\documentclass[review, nonatbib]{elsarticle}

\journal{Progress in Particle and Nuclear Physics}

\usepackage[T1]{fontenc}		
\usepackage[utf8]{inputenc}	
\usepackage{amsmath}
\usepackage{amssymb}
\usepackage{amsbsy}
\usepackage{amsfonts}
\usepackage[normalem]{ulem}

\usepackage{titlesec}
\usepackage{sectsty}
\titleformat{\section}{\normalfont\Large\bfseries\boldmath}{\thesection}{1em}{}
\titleformat{\subsection}{\normalfont\large\bfseries\boldmath}{\thesubsection}{1em}{}
\titleformat{\subsubsection}{\normalfont\normalsize\bfseries\boldmath}{\thesubsubsection}{1em}{}

\numberwithin{equation}{section}
\numberwithin{table}{section}
\numberwithin{figure}{section}

\usepackage[dvipsnames,table]{xcolor}

\usepackage{graphicx} 
\usepackage{float} 		 
\usepackage{subcaption}  
\usepackage[font=footnotesize,labelfont=bf]{caption} 

\usepackage{array} 
\usepackage{multirow}
\usepackage{hhline}
\usepackage{cases}
\usepackage{makecell} 

\usepackage{bm}	
\usepackage{dsfont}               
\usepackage{mathrsfs}             
\usepackage{slashed}              

\usepackage{physics} 
\usepackage{leftindex}

\usepackage{pifont} 	

\usepackage[final]{hyperref} 
\hypersetup{
 	colorlinks=true,    
 	linktoc=all,
 	linkcolor=cyan,      
 	citecolor=blue,      
 	filecolor=magenta,   
 	urlcolor=blue         
}

\makeatletter
\long\def\pprintMaketitle{\clearpage
	\iflongmktitle\if@twocolumn\let\columnwidth=\textwidth\fi\fi
	\resetTitleCounters
	\def\baselinestretch{1}%
	\printFirstPageNotes
	\begin{\elsarticletitlealign}%
		\thispagestyle{pprintTitle}%
		\def\baselinestretch{1}%
		\hfill\parbox[b]{4cm}{
	   		$\begin{gathered}\includegraphics[width=.15\textwidth]{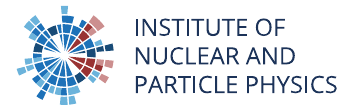}\end{gathered}$ \\
	   		$\begin{gathered}\includegraphics[width=0.05\textwidth]{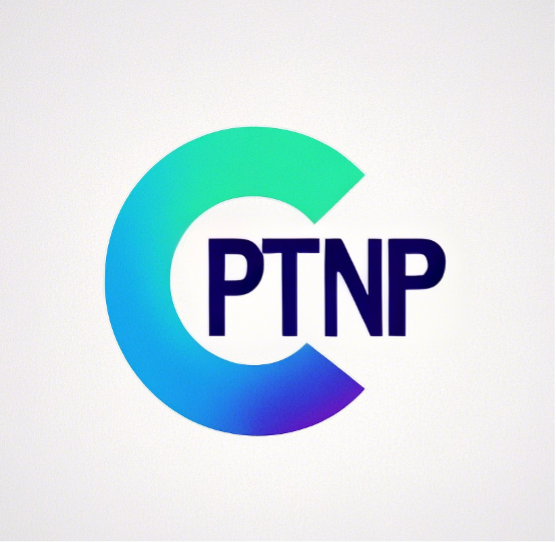}\end{gathered}$ CPTNP-2025-023}\vskip1cm
		{\Large\bfseries\@title}\par\vskip18pt%
		\ifx\@elsarticlenewpageafter\newpage@after@title
		\newpage
		\fi%
		\ifdoubleblind
		\vspace*{2pc}
		\else
		\normalsize\elsauthors\par\vskip10pt
		\footnotesize\itshape\elsaddress\par\vskip36pt
		\fi
		\ifx\@elsarticlenewpageafter\newpage@after@author
		\newpage
		\fi%
		\hrule\vskip12pt
		\ifvoid\absbox\else\unvbox\absbox\par\vskip10pt\fi
		\ifvoid\keybox\else\unvbox\keybox\par\vskip10pt\fi
		\hrule\vskip12pt
		\ifx\@elsarticlenewpageafter\newpage@after@abstract
		\newpage
		\fi%
	\end{\elsarticletitlealign}%
	\gdef\thefootnote{\arabic{footnote}}%
}

\makeatother

\usepackage{xspace} 

\definecolor{viable}{rgb}{0., 1., 0.5}
\definecolor{ruledout}{rgb}{0.97, 0.51, 0.47}
\definecolor{UVCruledout}{rgb}{0.6, 0.4, 0.8}
\definecolor{positive}{rgb}{0.2, 1, 0.4}
\definecolor{negative}{rgb}{0.3, 0.7, 1}
\definecolor{bothsigns}{rgb}{0.6, 0.4, 0.8}

\newcommand{\amu}{a_\mu\xspace}
\newcommand{\amuExp}{\amu^\text{Exp}\xspace}

\newcommand{\Damu}{\Delta a_\mu\xspace}

\newcommand{\DamuFinal}{\Delta a_\mu^\text{Exp$-$WP2025}\xspace}
\newcommand{\DamuOld}{\Delta a_\mu^\text{Exp$-$WP2020}\xspace}

\newcommand{\amuExpfinacc}{125 \text{ ppb}} 

\renewcommand{\vev}{\emph{vev}\xspace}
\newcommand{\La}{\mathcal{L}}
\newcommand{\PL}{\text{P}_\text{L}}
\newcommand{\PR}{\text{P}_\text{R}}

\newcommand{\SUc}{\text{SU}(3)_c}
\newcommand{\SUL}{\text{SU}(2)_L}
\newcommand{\UY}{\text{U}(1)_Y}
\newcommand{\GSM}{\SUc\times\SUL\times\UY}
\newcommand{\GEW}{\SUL\times\UY}
\newcommand{\GLEFT}{\text{SU}(3)_c\times\text{U}(1)_\text{em}}

\renewcommand{\O}{\mathcal{O}\xspace}
\newcommand{\F}{\mathcal{F}}
\newcommand{\G}{\mathcal{G}}

\newcommand{\QEDref}{Aoyama:2012wk,Volkov:2019phy,Volkov:2024yzc,
	Aoyama:2024aly,Parker:2018vye,Morel:2020dww,Fan:2022eto}

\newcommand{\EWref}{Czarnecki:2002nt,Gnendiger:2013pva,
	Ludtke:2024ase,Hoferichter:2025yih}

\newcommand{\HVPref}{RBC:2018dos,Giusti:2019xct,Borsanyi:2020mff,Lehner:2020crt,Wang:2022lkq,
	Aubin:2022hgm,Ce:2022kxy,ExtendedTwistedMass:2022jpw,RBC:2023pvn,Kuberski:2024bcj,Boccaletti:2024guq,
	Spiegel:2024dec,RBC:2024fic,Djukanovic:2024cmq,ExtendedTwistedMass:2024nyi,MILC:2024ryz,
	Bazavov:2024eou,Keshavarzi:2019abf,DiLuzio:2024sps,Kurz:2014wya}

\newcommand{\HLbLref}{Colangelo:2015ama,Masjuan:2017tvw,Colangelo:2017fiz,Hoferichter:2018kwz,
	Eichmann:2019tjk,Bijnens:2019ghy,Leutgeb:2019gbz,Cappiello:2019hwh,Masjuan:2020jsf,Bijnens:2020xnl,
	Bijnens:2021jqo,Danilkin:2021icn,Stamen:2022uqh,Leutgeb:2022lqw,Hoferichter:2023tgp,
	Hoferichter:2024fsj,Estrada:2024cfy,Ludtke:2024ase,Deineka:2024mzt,Eichmann:2024glq,Bijnens:2024jgh,
	Hoferichter:2024bae,Holz:2024diw,Cappiello:2025fyf,Colangelo:2014qya,Blum:2019ugy,Chao:2021tvp,
	Chao:2022xzg,Blum:2023vlm,Fodor:2024jyn}

\renewcommand*{\arraystretch}{1.2}

\newlength{\Cwidth}
\newcolumntype{C}{>{\centering\arraybackslash}p{\Cwidth}}

\makeatletter
\let\c@author\relax
\makeatother
\usepackage[style=numeric-comp, backend=bibtex, 
sorting=none, abbreviate=true, maxbibnames=99]{biblatex}
\begin{document}

\begin{frontmatter}
       
	\title{The Muon Magnetic Moment and Physics Beyond the Standard Model}
	
	\author[1,2]{Peter Athron}\ead{peter.athron@coepp.org.au}
	\author[3]{Kilian Möhling}\ead{kilian.moehling@tu-dresden.de}
	\author[3]{Dominik Stöckinger\corref{cor1}}\ead{dominik.stoeckinger@tu-dresden.de}					
	\author[3]{Hyejung Stöckinger-Kim}\ead{hyejung.stoeckinger-kim@tu-dresden.de} 	
	
	\address[1]{Institute of Theoretical Physics and Institute of
          Physics Frontiers and Interdisciplinary Sciences, Nanjing
          Normal University, Wenyuan Road, Nanjing, Jiangsu, 210023,
          China}

        \address[2]{Nanjing Key Laboratory of Particle Physics and
          Astrophysics, Nanjing, 210023, China}

	\address[3]{Institut für Kern- und Teilchenphysik, TU Dresden, 
				Zellescher Weg 19, 01069 Dresden, Germany}
				
	\cortext[cor1]{Corresponding author}

	\date{\today}
	
	\begin{abstract}

        We review the role of the anomalous
        magnetic moment of the muon $\amu$ as a powerful probe of physics beyond the
        Standard Model (BSM), taking advantage of the final result of
        the Fermilab $g-2$ experiment and the recently updated
        Standard Model value. This review provides 
        both a comprehensive summary of the current status, as well as
        an accessible entry point for phenomenologists with interests
        in dark matter, Higgs and electroweak or neutrino and flavour
        physics in the context of a wide range of BSM scenarios.  It
        begins with a qualitative overview of the field and a 
        collection of key properties and typical results. It then
        focuses on model-independent, generic formulas and classifies
        types of BSM scenarios with or without chiral enhancements.
        A strong emphasis of the review are the connections to a large
        number of other observables --- ranging from the muon
        mass and the muon--Higgs coupling and related dipole
        observables to dark matter, neutrino masses and high-energy
        collider observables.  Finally, we survey a
        number of well-motivated BSM scenarios such as dark photons,
        axion-like particles, the two-Higgs doublet model,
        supersymmetric models and models with leptoquarks, vector-like
        leptons or neutrino mass models. We discuss the impact of the
        updated Standard Model value for $\amu$ and of complementary
        constraints, exploring the phenomenology and identifying
        excluded and viable parameter regions.

\end{abstract}

\end{frontmatter}

\newpage	
\thispagestyle{empty}
\tableofcontents
	
\section{Introduction}

Magnetic dipole moments are among the most
precisely known observables in physics, with a long and rich history.
They represent simple and yet
characteristic particle properties reflecting details of the
fundamental interactions.
The anomalous magnetic moments of leptons $a_\ell=(g-2)_\ell/2$
are also among the most precisely
computed quantities in physics. Their values originate
exclusively from loop diagrams in renormalisable quantum field
theories and are thus  sensitive to quantum fluctuations
of all particles that can interact with leptons. For elementary particle physics, the
anomalous magnetic moment of the muon  $\amu=(g-2)_\mu/2$
is of particular interest.
Compared to the electron, the muon is much heavier and enables higher-energetic
quantum fluctuations, 
and  although the muon lifetime is finite,
it is long enough to allow measurements almost as precise as
for the electron, which is not the case for the even heavier
$\tau$-lepton.  Thus in combination $\amu$ is the most sensitive to 
heavy elementary particles and constitutes an excellent probe of the
fundamental laws of physics.

Recently, two landmark results were obtained for the experimental and
theoretical determination of $\amu$. The Fermilab muon $g-2$ experiment has obtained its 
final result, decreasing the uncertainty of the world average to $\amuExpfinacc$ \cite{Muong-2:2025xyk}. 
Meanwhile, the SM theory prediction has been scrutinised, 
and the immense effort of the community has led to two White Papers \cite{Aoyama:2020ynm,Aliberti:2025beg} detailing the involved calculations. 
These two White Papers differ in their evaluation of the
non-perturbative leading-order hadronic vacuum polarisation (HVP) 
contributions. The most recent prediction is based on lattice gauge
theory computations, which have become very precise and consistent and are
now cross-checked by several groups in multiple ways. The previous
prediction was instead based on data-driven evaluations of the HPV contribution using
experimental data for $e^+e^-\to$ hadrons together with dispersion relations.
However, at the time of Ref.~\cite{Aoyama:2020ynm} some tensions between these data sets were already present
and have subsequently become more severe with the appearance of additional measurements.
These tensions are under intense scrutiny and further
improvements on the SM prediction can be expected in the forthcoming
years.

The earlier SM
prediction based on Ref.~\cite{Aoyama:2020ynm} is lower than the
experimental value, favouring a sizeable and positive contribution
to $\amu$ from physics beyond the SM (BSM).
In contrast, the current SM prediction of
Ref.~\cite{Aliberti:2025beg} is in full agreement with the 
experimental result for $\amu$.  
The newly obtained agreement may mark the emergence
of $\amu$ as an important constraint on many BSM scenarios,
allowing only (positive or negative) contributions to $\amu$ of 
sufficiently small magnitude.

In this review we show the power and significance of $\amu$ as a
precision test for BSM physics, provide detailed insights into
calculations, describe  correlations with other
observables and discuss the phenomenological behaviour for many
models. The first sections of the review focus on generic statements and
relationships that can be described in a model-independent way or with
the help of simplified examples. The last section focuses on a broad
range of motivated BSM scenarios and presents up-to-date constraints
from $\amu$ and complementary observables. We  mostly  use  $\amu$ as
a constraint on new physics rather than as a signal, using the new
White Paper result and uncertainty. However the current SM prediction
is a snapshot of the present 
situation and progress on the SM prediction might lead to a resolution
of the tensions between lattice and data-driven results and could end
up pointing at a new deviation from the experimental value. In this
case, all the calculations and correlations we discuss would still
apply, and the phenomenological plots would still display the
preferred parameter regions. In either case, the present
review thus aims to provide
a timely overview of the impact of $\amu$ as a precision test
for new physics at the beginning of this new era for muon
$g-2$. 
\newline

In the following Section~\ref{sec:Motivation}
we provide a brief 
overview of the field and motivate $\amu$ as a probe of fundamental
physics with unique properties. In addition, we also
summarise the content of the review in more detail and preview
selected results in a qualitative manner. 
The remainder of this
Introduction  then contains
details on the quantum field theoretical definition of $\amu$ in
Sec.~\ref{sec:definition}, and  the status of the
experiment and SM theory are reviewed in
Secs.~\ref{sec:experiment}, \ref{sec:SMtheory}.
After this extended Introduction the review is structured into three
essentially model-independent, generic sections and one section
with detailed model-specific results.

In the model-independent, generic part of the review we
describe in detail the general properties of the observable $\amu$
from the perspective of BSM physics and correlations and
complementarities to other observables ranging from low-energy dipole
moments and the Higgs--muon interaction to neutrino masses and dark
matter.
We further analyse generically which classes of BSM scenarios
can lead to large or small contributions to $\amu$ and link
characteristic properties to fundamental motivations such as dark
matter or the origin of mass or flavour.
Specifically, 
generic formulas for $\amu$ together with their interpretation and applications
to simple and generic models are explained in Sec.~\ref{sec:Generic}.
Section \ref{sec:Observables} describes the relationships of $\amu$
with the muon mass, with other dipole observables, the muon--Higgs
interaction, 
electroweak precision observables and other collider observables as well as
with neutrino masses and dark matter. Section \ref{sec:HigherOrder}
provides additional technical details on model contributions including
higher orders both in Feynman diagrammatic and in effective field
theory approaches. 

The model-specific part of the review is Sec.~\ref{sec:Models}, where we  present an extensive
survey of up-to-date results on a large set of motivated BSM
scenarios. They cover models with light or heavy new particles, with
or without dark matter, and with a wide range of relevant
complementary observables. For
each model the theory and phenomenology, and particularly
the interplay of $\amu$ and complementary constraints is
explained. Viable model parameter regions are
characterised, and it is shown which scenarios could lead to sizeable
contributions to $\amu$ and which scenarios are preferred by the new
measurement and SM prediction. 

There exist already several reviews and surveys with full or partial focus on
$\amu$ and BSM, such as
Refs.~\cite{Czarnecki:2001pv,Stockinger:2006zn,Czarnecki:2009hvh,Stockinger:2009fns,Lindner:2016bgg,Athron:2021iuf}
and \cite{Melnikov:2006sr,Jegerlehner:2017gek,Jegerlehner:2009ry,Keshavarzi:2021eqa}. Although
there is some unavoidable overlap, the present review aims to be
complementary in several respects. In the past years, a combination of
general, model-independent analyses and explicit investigations in
specific models have led to a detailed understanding of how $\amu$ is
related to a number of other observables, and how properties of BSM
scenarios are relevant for such relationships. Here we  provide a
coherent and self-contained discussion of these insights with the goal
of serving as a reference and summary but also as a possible entry
point into the field. 
 
In addition, the review appears after the final result of the Fermilab
muon $g-2$ experiment and after significant progress on the SM theory
prediction, although further SM theory progress is expected. It takes
advantage of the current status and provides a detailed account of how
the recent developments affect the constraints on BSM models. In our
selection of models we are broad but focus particularly on a wide
range of models which are intrinsically motivated independently of
$\amu$ and explain how they are constrained by the combination of $\amu$ and
other appropriate observables.

\subsection{Motivation and overview}
\label{sec:Motivation}

The use of $\amu$ as a precision test of the SM and probe of new
physics follows many decades of investigations into magnetic moments
that have continuously helped shape our understanding of fundamental
laws of nature.  The discovery that the Dirac
equation predicts the gyromagnetic ratio $g=2$ for the electron \cite{Dirac:1928hu}, in
agreement with experiments at the time, was the first example of the
fruitful interplay between theory and experiment.

Key progress was made by Schwinger, who realised that quantum effects,
i.e.\ loop Feynman diagrams in QED lead to a deviation from $g=2$, the
anomalous magnetic moment $a=(g-2)/2$ \cite{Schwinger:1948iu,Schwinger:1951nm}. The agreement between the
one-loop result $a_e=\alpha/2\pi$ for the electron and experiments at
the time confirmed 
relativistic quantum field theory (QFT) as the correct framework for
fundamental interactions. The essential QFT property entering here is
the possibility for quantum fluctuations of particle number, in
modern language equivalent to the contributions of loop diagrams with
the exchange of virtual particles.

Subsequently over the decades $a_e$ and $\amu$ were measured and
computed at increasingly high levels of precision, finally arriving at
the latest Fermilab measurement and SM theory prediction
mentioned above.  With this precision, the comparison
between experiment and theory for $\amu$ is sensitive to all known
particles and interactions of the SM --- even the heaviest particles
such as $W$-, $Z$- and Higgs boson and the top quark enter the theory
prediction in a relevant way.
\newline

The SM is a mathematically rigorous, consistent, renormalisable QFT, and
it is in excellent agreement with a multitude of experimental
tests. Nevertheless it leaves open important questions and
hence cannot be the ultimate description of nature. It
cannot describe observational facts such as the evidence for dark
matter in the universe and the existence of neutrino oscillations and
neutrino masses. It also cannot explain the baryon-antibaryon
asymmetry of the universe unless one assumes ad hoc asymmetric
initial conditions. Beyond these observations, the structure of the
SM with the gauge group $\SUc\times\SUL\times\UY$, three
generations of quark and lepton doublets and singlets, and one scalar
Higgs doublet, suggests that a deeper understanding of the origin of
this structure might be possible: What is the origin of  this
particular gauge group or the number of three
generations? What is the origin of the Higgs field, the Higgs
potential and of electroweak symmetry breaking, and what sets the
interactions of quarks and leptons with the Higgs field and ultimately
the fermion masses?

Achieving progress on such questions requires the investigation of 
physics beyond the SM  in an interplay of theory and experiment.
Here the muon anomalous magnetic moment $\amu$  plays a special role.
Via quantum
fluctuations, $\amu$ could be influenced by any currently unknown BSM
particles as long as these have direct or indirect interactions with
muons. $\amu$ is thus sensitive to a large set of hypothetical BSM
scenarios, e.g.\ scenarios with extended scalar, fermion or gauge
boson sectors which may be motivated by the open questions listed
above. The new states may be light or heavy, strongly or weakly
interacting, they may be similar to the known SM leptons, quarks or
Higgs, or may have very different properties such as leptoquarks or
dark matter particles.  $\amu$ can also test elaborate constructions
such as supersymmetric models or extended non-abelian gauge
groups. Importantly, $\amu$ is complementary to high-energy
experiments at colliders such as the LHC, where detector acceptances
and SM backgrounds are relevant, and to many lower-energy observables
which correspond to symmetry violations such as charge-parity (CP) violating or
flavour-violating observables. 
\newline

In this Section~\ref{sec:Motivation} we  give an
extended motivation and an  overview of the field. We briefly describe the current
status and provide the experimental and SM numbers. We highlight the
distinctive properties of $\amu$ as a probe of 
fundamental interactions and constraint
on new physics and describe the important role of chirality flips. A
table at the
end of Sec.~\ref{sec:Motivation}  also collects useful generic formulas
and illustrative model-specific results for BSM contributions to
$\amu$. This section also describes the content of the
review in more detail, including  
forward references to later sections.

\subsubsection{Status of $\amu$ and current values}

The anomalous magnetic moment of the muon  $\amu$   can be measured
via the spin precession of the muon in a homogeneous, weak magnetic
field, and computed in quantum field theory in terms of the
one-particle irreducible muon--photon 3-point function. This
  definition is reviewed in Sec.~\ref{sec:definition}, including a
  discussion of suppressed  effects, e.g.~from non-linearities or
  inhomogeneities in the $B$-field or radiation effects, which are neglected in the quantum
  field theoretical derivation.

Measurements of $\amu$ have been carried out by
experiments at CERN \cite{CERNMuonStorageRing:1977bbe,CERN-Mainz-Daresbury:1978ccd}, 
Brookhaven \cite{Muong-2:2001kxu,Muong-2:2002wip,Muong-2:2004fok,Muong-2:2006rrc} and 
Fermilab \cite{Muong-2:2021ojo,Muong-2:2023cdq,Muong-2:2024hpx,Muong-2:2025xyk}
using the same experimental principle. All results are in full
agreement, and the final uncertainties are statistics dominated,
leading to a very satisfactory experimental status. The
recent final result of the Fermilab experiment, and the resulting
world average experimental value are
\begin{align}
	\amu^{\textrm{FNAL,Run-1--6}} &= 11\,659\,207.05(1.48)\times10^{-10}, \label{Eq:FNALfinalresult}\\
	\amuExp &= 11\,659\,207.15(1.45)\times10^{-10}. \label{eq:amuExp}
\end{align}
In Sec.~\ref{sec:experiment} we
provide further details on the experimental method and the results of
the individual measurements.

In parallel to the experimental progress, the SM theory prediction has
continuously been scrutinised. It can essentially be divided into four
contributions, from QED alone, from the hadronic vacuum polarisation
(HVP), from hadronic light-by-light scattering (HLbL), and from the
weak interactions and all remaining SM particles. The individual
definitions and values are given in Sec.~\ref{sec:SMtheory}. Already the
precision of the Brookhaven measurement required the evaluation of all
these SM contributions, and in preparation of the Fermilab experiment,
a worldwide theory initiative was formed to obtain the best
possible SM evaluation based on a coordinated, conservative and
inclusive community effort. The theory initiative has published two
extensive White Papers \cite{Aoyama:2020ynm,Aliberti:2025beg}, with
the results 
\begin{align}
	\amu^{\textrm{SM,WP2020}} &= 11\,659\,181.0(4.3)\times10^{-10}, \label{eq:amuSM2020}\\
	\amu^{\textrm{SM,WP2025}} &= 11\,659\,203.3(6.2)\times10^{-10}. \label{eq:amuSM2025}
\end{align}
The result from the year 2020 is based on original references on QED \cite{Aoyama:2012wk,Aoyama:2019ryr},
electroweak \cite{Czarnecki:2002nt,Gnendiger:2013pva}, HVP
\cite{Davier:2017zfy,Keshavarzi:2018mgv,Colangelo:2018mtw,Hoferichter:2019mqg,Davier:2019can,Keshavarzi:2019abf,Kurz:2014wya}
and HLbL contributions
\cite{Melnikov:2003xd,Masjuan:2017tvw,Colangelo:2017fiz,Hoferichter:2018kwz,Gerardin:2019vio,Bijnens:2019ghy,Colangelo:2019uex,
	Pauk:2014rta,Danilkin:2016hnh,Jegerlehner:2017gek,Knecht:2018sci,Eichmann:2019bqf,Roig:2019reh,Blum:2019ugy,Colangelo:2014qya}. The
2020 SM prediction  deviates substantially from the experimental
result. At the time of the Fermilab Run-1 publication
\cite{Muong-2:2021ojo} the significance was $4.2\sigma$,
leading to great enthusiasm in the community and many attempts to
explain the deviation in terms of BSM physics.

The SM prediction from
2025 is based on the following original references
for QED \cite{\QEDref},  electroweak \cite{\EWref}, HVP
\cite{\HVPref}, and HLbL contributions \cite{\HLbLref}. Its value is now very close to
the experimental result and in full agreement within uncertainties.

Sec.~\ref{sec:SMtheory} will explain this change in 
more detail; in brief,
crucial new results have appeared for the HVP
contributions.
The HVP contributions
are non-perturbative, and two possible evaluation strategies are based
on dispersion relations, the optical theorem and experimental
low-energy hadronic data on the one hand, and on lattice gauge theory
on the other hand. In recent years there was tremendous progress on
lattice calculations. A first milestone was the result of the BMW
collaboration \cite{Borsanyi:2020mff}, which was then confirmed by a 
number of further lattice results using a multitude of different
approaches.
In contrast, the dispersive analysis relies on measurements of
$e^+e^-\to$ hadrons at low energies. Different available data sets used
to be in sufficient, though not perfect agreement. Now the
recent CMD3 measurement of $e^+e^-\to$ hadrons
\cite{CMD-3:2023alj,CMD-3:2023rfe} is in stronger tension
with previous $e^+e^-$ data. At present no
concrete reason has been identified to dismiss any of 
the $e^+e^-$ data sets \cite{Aliberti:2025beg} and the origin of the disagreements is unknown.
The current tensions therefore do not allow a meaningful dispersive evaluation of the
HVP contributions. Hence, Ref.~\cite{Aliberti:2025beg} evaluates the LO HVP
correction based on lattice calculations only, which are, however, internally very consistent.

After these recent developments, the deviation between the
experimental world average and the SM prediction
is \cite{Aliberti:2025beg}
(all numbers are
rounded appropriately to the level of $10^{-11}$)
\begin{align}\label{eq:DamuFinal}
       \DamuFinal &= 3.8(6.3) \times 10^{-10},
\end{align}
where the uncertainties are summed in quadrature. This corresponds to
full agreement and shows no sign for physics
beyond the SM. Nevertheless, it corresponds to an important constraint
on BSM scenarios. The current result in Eq.~\eqref{eq:DamuFinal} will be the
basis of the phenomenological discussions in the present review.
In order to understand the development of the field over the past
years it is useful to
also record the discrepancy between the experimental result
\eqref{eq:amuExp} and the SM prediction based on the first
White Paper Eq.~\eqref{eq:amuSM2020},
\begin{align} \label{eq:DamuOld}
           \DamuOld &= 26.2(4.5) \times 10^{-10},
\end{align}
which would correspond to a $>5\sigma$ deviation.

Given the rapid progress and intense effort of the community, further
progress on the HVP contributions can be expected
\cite{Colangelo:2022jxc,Colangelo:2023rpc,Aliberti:2025beg}.
The result of such
progress is unknown at present, but we can highlight the following
logical possibilities from the point of view of BSM interpretations:
\begin{itemize}
\item
  The agreement (\ref{eq:DamuFinal}) is consolidated and sharpened
  with smaller uncertainty. In this case,  BSM contributions will
  be  allowed to be zero, positive, or negative, but their 
  magnitude must be sufficiently small.
\item
  The central value of $\Damu$ remains compatible with the
  $1\sigma$ band of Eq.~(\ref{eq:DamuFinal}) and thus much  smaller
  than the previous deviation $\DamuOld$, but the uncertainty goes down such that an
  indication for non-zero BSM contributions re-emerges. In this
  case, BSM contributions will be constrained to be non-zero and
  positive, but of a smaller magnitude $\lesssim10\times10^{-10}$.
\item
  The central value of $\Damu$ increases towards a value similar to
  the value after the 2020 White Paper, as
  Eq.~\eqref{eq:DamuOld}. In this case, there will be a 
  strong indication for BSM contributions to $\amu$, which will be
  constrained to be positive and approximately of the order $20\times10^{-10}$.
\end{itemize}
In the present review we will also comment on the impact of such possible
future developments. Mostly, however, we will explain the impact of
the dramatically shifted result (\ref{eq:DamuFinal}) on BSM physics. As a
first result, we record here the $2\sigma$ interval around the new
result,
\begin{align}
  \Damu(2\sigma)=[-8.8\ldots16.5]\times10^{-10}.
  \label{DamutwoSigma}
\end{align}
The interval (\ref{DamutwoSigma}) is now the allowed region for BSM
contributions to $\amu$.

\subsubsection{Sensitivity to fundamental  interactions and
   generic BSM contributions}
\label{sec:amuprobe}

Magnetic dipole moments are some of the most precisely measured
observables in physics and play a very important role in understanding
elementary particles and how they interact.
Their sensitivity to details of fundamental interactions becomes immediately apparent
by comparing the gyromagnetic ratios $g$ for the electron
\cite{ParticleDataGroup:2024cfk}, the muon (Eq.~\eqref{eq:amuExp}) and the proton
\cite{ParticleDataGroup:2024cfk},
\begin{subequations}\label{eq:g_e-mu-p}
	\begin{align}
		g_e^\text{exp} &= 2.002\,319\,304\,361\,24\, (24) , \label{gfactorElectron} \\
		g_\mu^\text{exp} &= 2.002\,3\mathbf{3}1\,\mathbf{8}4\mathbf{1}\,43\, (29) ,\label{gfactorMuon} \\
		g_p^\text{exp} &= 5.585\,694\;689\;26\,(164) .
	\end{align}
\end{subequations}
The electron $g$-factor is most precisely known. Its value is very
close to 2, the value predicted by the Dirac equation, and the
per-mille level deviation from the Dirac value is consistent with
Schwinger's calculation $a_e=\alpha/2\pi$. Section
\ref{sec:LeptonDipole} will provide details on the relationships
between the dipole moments of the muon and the electron, and further
important leptonic dipole observables. As described there, the full
value of $g_e^{\text{exp}}$ is in good agreement with
theory, confirming the electron as a fundamental elementary spin 1/2
particle of the SM.

From a low-energy and long-distance perspective, the proton seems no
less elementary than the electron. Its $g$-factor, however,
dramatically differs from the electron 
and from the Dirac values, establishing that the proton
behaves fundamentally differently than the electron. In this way the
$g$-factor, which corresponds to a long-distance/low-energy
observable, directly reflects the fact that the proton actually is not
elementary but a strongly interacting bound state.

For the muon, comparing the three $g$-factors immediately shows that
the muon behaves fundamentally similarly to the electron, in line with
the SM where the muon is the 2nd-generation sibling of the electron,
with identical interactions except for the Yukawa interaction to the
Higgs field and the resulting heavier muon
mass. Eq.~\eqref{gfactorMuon} highlights several digits in
boldface. At the 5th decimal place the muon and electron $g$-factors start to
differ due to the higher muon mass providing higher energy for 
quantum fluctuations. The major difference results from two-loop diagrams with electron loops
contributing to the muon versus muon loops contributing to the
electron. At the 7th decimal place the muon $g$-factor is influenced
by hadronic effects, i.e.\ by strongly interacting particles. At the
9th decimal place the muon $g$-factor is influenced by the weak
interactions and the heaviest SM particles. 

This discussion shows that the muon magnetic moment $\amu$ at the
current level of precision is sensitive
to all elementary particles and interactions we know. The digits of
the single number reflect our combined understanding of fundamental
physics. Clearly, in the same way $\amu$ can also be sensitive to possible
BSM particles depending on their mass and their interactions.

In the present review, Sec.~\ref{sec:Generic} describes the most
important general ways how BSM
particles can impact $\amu$. Starting from generic expressions
for one-loop Feynman diagrams of renormalisable QFTs,
it becomes apparent that different types of BSM scenarios can lead to
very different contributions. If only two gauge multiplets enter the
loop, the contributions have a very simple structure and are typically
small and automatically in agreement with the new result in
Eq.~(\ref{eq:DamuFinal}), though there are exceptions if BSM masses are
very small. Often the sign of
such contributions is fixed by the quantum 
numbers of the relevant gauge multiplets, and several tables in Sec.~\ref{sec:MinimalBSM} list
large classes of models of this kind, and their contributions to
$\amu$.

If three or more gauge multiplets enter the loop, so-called
chiral enhancements are possible. Corresponding BSM scenarios are often stringently
constrained by Eq.~(\ref{eq:DamuFinal}). The characteristic model properties
leading to chiral enhancements are discussed with the help of explicit
simplified examples and the mass-insertion approximation in
Sec.~\ref{sec:genericthreefield}. An instructive
qualitative account of the interplay between chirality, chirality
flips, the Higgs mechanism and $\amu$ will be given further below in
Sec.\ \ref{sec:ChiralityFlips}.

A very useful approach both for qualitative discussions and for
precision calculations is the one of effective field theories. These
can provide systematic low-energy approximations for more fundamental
theories involving heavy degrees of freedom. Two particularly
important effective field theories for $\amu$ are LEFT and SMEFT,
which extend QED/QCD or the full Standard Model by higher-dimensional
operators, respectively. These theories, and the description of $\amu$
within them, are explained in Sec.~\ref{sec:genericeft}.

A later section \ref{sec:HigherOrder} complements the one-loop
discussions by describing typical and important higher-order
contributions, both using Feynman diagrams and effective field theory
arguments.

\subsubsection{Key properties of $\amu$ and complementarities to other observables}
\label{sec:amuuniqueproperties}

Via quantum effects and loops, $\amu$ is an inclusive probe of many
candidate BSM scenarios. Unravelling
the existence and the properties
of potential new physics requires complementary experimental
information from high-energy colliders such as LHC, from astrophysics
and low-background measurements related e.g.\ to dark matter or
neutrinos, and from low-energy observables such as flavour-violating
rare decays and dipole moments.

The muon magnetic moment has a special role because it is sensitive to a large class of
models and it combines several properties in a unique way:
$a_\mu$ is
\begin{enumerate}[left=2cm]
	\item[\ding{114}]
	\emph{flavour conserving},
	\item[\ding{114}]
	\emph{CP conserving},
	\item[\ding{114}]
	\emph{loop induced},
	\item[\ding{114}]
	\emph{chirality flipping}.
\end{enumerate}
In contrast,  many other low-energy precision observables are CP or
flavour violating, and many high-energy collider observables at the
LHC  are not loop induced or not chirality flipping. The four properties
are thus the reason why $\amu$ is complementary to many
other important observables and why $\amu$ can help to disentangle the
nature of BSM physics. Sec.~\ref{sec:Observables} explores and
discusses such complementarities in detail, but still in a
model-independent way.

The last item in particular means that the QFT operator for $\amu$ connects
left-handed and right-handed muon fields, similar to the muon mass
operator.
Sec.~\ref{sec:ChiralityFlips} below will explain how this feature
helps to obtain a qualitative and quantitative understanding of
possible BSM contributions to $\amu$, and how investigating $\amu$ may
shed light on the origin of the muon mass. The
relationship between $\amu$ and the muon mass is then further explored
in Sec.~\ref{sec:MuonMass}, where it becomes clear that BSM scenarios
that change the way the muon receives its mass can be particularly
strongly constrained by $\amu$. Also special cases are considered such
as models with radiative muon mass generation and models with the 
maximum possible contributions to $\amu$ compatible with perturbative
unitarity.

$\amu$ has especially close relationships to other
dipole observables, which share the chirality-flipping nature. Among
these, the electron magnetic moment 
$a_e$ is however less sensitive to heavy BSM contributions, the
electric dipole moments $d_\mu$ and $d_e$ are CP-violating observables
and hence only sensitive to BSM effects which violate CP, and the
decay $\mu\to e\gamma$ is only sensitive to BSM effects which violate
lepton flavour. These relations are quantified in
Sec.~\ref{sec:LeptonDipole}. Via chirality flips $\amu$ is  also
related to the muon--Higgs interaction which itself is strongly
related to the muon mass and
which can be directly constrained at the LHC. The nature of this
relation and the resulting constraints on models are discussed in
Sec.~\ref{sec:muon-Higgs}. A difference is that $\amu$ is loop-induced
while the muon--Higgs
coupling arises at tree level. This has specific implications for the strength
of the correlation and the class of models with particularly large effects.

Clearly, $\amu$ has only weak or no correlations with many important
high-energy collider observables. This makes constraints on BSM from $\amu$
complementary to the ones from direct BSM particle searches at LHC or
high-precision measurements of electroweak interactions. It is
instructive to discuss this
complementarity in a model-independent way. This is done in
Secs.~\ref{sec:EWPO}, \ref{sec:Collider} on a qualitative level and with the
help of simplified models. In case of electroweak precision
observables discussed in Sec.~\ref{sec:EWPO} there exists one
interesting connection to $\amu$ via the quantity $\Delta\alpha$
and the hadronic vacuum polarisation,
which enters both kinds of observables in an important
way. Sec.~\ref{sec:Collider} on the other hand focuses on the reach of
direct collider searches for new particles and describes how this
reach compares to constraints from $\amu$ for different classes of
models.

Some of the most pressing open questions of particle physics are
related to the nature of neutrino masses and the nature of dark
matter. Perhaps surprisingly, $\amu$ has a stronger impact on many
candidate explanations of dark matter often more straightforwardly
than on many neutrino mass generation models. There are, however,
specific classes of neutrino mass models which potentially lead to
significant contributions to $\amu$ and which therefore are
constrained by $\amu$. These connections are explained in
Secs.~\ref{sec:DarkMatter} and \ref{sec:neutrino_masses}.

\subsubsection{Structure of BSM contributions and the role of chirality flips and electroweak gauge
  invariance}
\label{sec:ChiralityFlips}

Here we begin discussing quantitatively how BSM physics can contribute
to $\amu$. As a preview, the very simplest type of one-loop BSM
contributions take the generic form
\begin{align}
  \Damu & \sim
  \frac{|c|^2}{16\pi^2}
  \frac{m_\mu^2}{M^2}.
  \label{amusimplest}
\end{align}
The first factor involves a generic BSM coupling $c$ and a
one-loop suppression. The second factor provides a squared mass
suppression by the ratio of the muon and the heavier BSM masses $m_\mu$,
$M$. The symbol $\sim$ denotes equality up to order-one factors. The sign
can be positive or negative but is often fixed for a given model with
this type of contributions.

It turns out that the need for chirality flips mentioned above in
subsection \ref{sec:amuuniqueproperties} is
of pivotal importance for the BSM phenomenology of $\amu$
\cite{Czarnecki:2001pv,Stockinger:2009fns,Stockinger:2022ata}. 
Accordingly, in some BSM scenarios there can be so-called 
\emph{chiral enhancements}, leading to modified formulas such as
\begin{align}\label{amuchiralenhancement}
  \Damu & \sim
  R_\chi
    \times\frac{c_L c_R}{16\pi^2}
  \frac{m_\mu^2}{M^2},
\end{align}
where $R_\chi$ represents a dimensionless chiral enhancement factor,
and the formula also highlights that two different kinds of coupling
constants $c_{L,R}$ are in general required to invoke chiral
enhancements.
Phenomenologically, in such scenarios the contributions to $\amu$ can
often be large and involve intricate parameter dependencies, leading
to non-trivial constraints. In such contributions, often the sign of
the product $c_L c_R$, and sometimes also of the factor $R_\chi$, can
be positive or negative.

Here we explain how the basic structure of these or other model-specific
formulas emerges and discuss the role of chirality flips.
The discussion will make explicit a
fundamental connection to the muon mass and later allow us to elucidate
how different fundamental 
BSM mechanisms can contribute to $\amu$. It will also help us classify
BSM scenarios and understand parameter dependences, as well as
correlations to other observables.

First we note that massless fermions cannot have a non-vanishing
anomalous magnetic dipole moment, i.e.~$\amu=0$, if $m_\mu=0$. The
reason is that the spin of  massless
particles reduces to helicity and cannot undergo the continuous spin
precession as is needed and measured for an anomalous
magnetic dipole moment. Hence, non-vanishing $\amu$ requires
non-vanishing $m_\mu$, establishing a deep link.

Next, on a more technical level, 
the operators describing the mass and the anomalous magnetic
dipole moment of the muon
can be
written as
\begin{subequations}
  \begin{align}
  {\La}_{m}&=-m_\mu\ \left(\overline{\mu_L}\mu_R +
  \overline{\mu_R}\mu_L \right),
  \label{LDiracmLR}
  \\
  {\La}_{\amu}\  & = -\amu\frac{Qe}{4m_\mu}\,\left(\overline{\mu_L}
  \sigma^{\mu\nu}\mu_R + \overline{\mu_R}
  \sigma^{\mu\nu}\mu_L\right)\,F_{\mu\nu}.
  \label{LamuLR}
  \end{align}
\end{subequations}
Here
the spin-1/2 muon with  charge $Q$ is described  by a
Dirac spinor $\mu=\mu_L+\mu_R$, with chirality eigenstates
$\gamma_5\mu_{L,R}=\mp\mu_{L,R}$ and $F_{\mu\nu}$ denotes
the photon field strength tensor. More details
on the definition of $\amu$ and such Lagrangians will be provided in
Secs.~\ref{sec:definition} and \ref{sec:genericeft}.

These Lagrangians allow us to introduce the notions of \emph{chirality flips},
\emph{chiral transformations} and \emph{chiral symmetry}.
Interpreted as a Feynman rule, both the mass term and the $\amu$-term
correspond to an incoming right-handed and outgoing left-handed muon
$\mu_R\to\mu_L$ or vice versa. This is called a chirality flip.

In general, a chiral transformation is a transformation
where the left- and right-handed fields transform with
opposite phases. Specifically for the muon, a chiral transformation
reads 
\begin{align}\label{muchiralsym}
	\mu_R & \to e^{i\alpha}\mu_R &
	\mu_L & \to  e^{-i\alpha}\mu_L .
\end{align}
A theory is chirally symmetric if this transformation is  a symmetry
(possibly by assigning suitable transformation
laws for other fields).
Importantly, the  mass
term and the $\amu$-term in Eqs.~\eqref{LDiracmLR} and \eqref{LamuLR}
are never chirally symmetric.

The need for chirality flips and breaking of the chiral symmetry Eq.~(\ref{muchiralsym}) has very important consequences
especially in the SM and BSM scenarios defined at the weak scale or above. These 
theories must include electroweak interactions and respect the associated
$\GEW$ gauge invariance. In contrast to QED or QCD,
in such theories no gauge invariant Dirac mass
terms for leptons or quarks are possible. 
Specifically the left-handed
muon $\mu_L$ is part of an $\SUL$ doublet with hypercharge $-1/2$
while the right-handed muon $\mu_R$ is an $\SUL$ 
      singlet with hypercharge $-1$.
The only way to generate a muon mass in a theory with electroweak
gauge invariance (SM or beyond) is to have \emph{spontaneous 
electroweak symmetry breaking} (EWSB) and to couple the muon in a gauge
invariant way to the corresponding vacuum expectation value (\vev). In the
SM,  the muon mass generated at tree-level can be written as
\begin{align}
  m_\mu^{\text{SM, tree}}&=\frac{y_\mu
  v}{\sqrt2}.
\end{align}
Here $y_\mu$ is the strength of the muon--Higgs Yukawa coupling, which
acts as the breaking parameter of the muon-specific chiral symmetry, and 
$v$ is the EWSB Higgs \vev.

Analogously,  any
extension of the SM which has electroweak gauge invariance must have
one or several vacuum expectation values 
responsible for EWSB. These may be expectation
values of fundamental or of composite fields. The physical muon mass
$m_\mu$ will always be generated from (tree-level and/or
loop-induced) couplings to these \vev{}s.
And any such theory must contain
one or several parameters which break the relevant chiral symmetry
Eq.~(\ref{muchiralsym}).

\begin{figure}[t!]
\centering
\includegraphics[width=.8\textwidth]{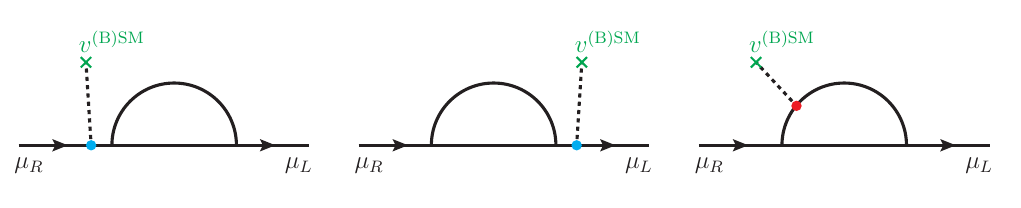}
\caption{One-loop Feynman diagrams illustrating
  Eqs.\ (\ref{mmugeneric},\ref{amugeneric}) and the possibility of
  chiral enhancements. Each contribution to the
  muon mass and to $\amu$ must involve one factor of some SM or BSM
  \vev   $v^{\text{(B)SM}}$ breaking electroweak symmetry, as well as
  a chirality flip between $\mu_L$ and 
  $\mu_R$. In order to obtain a contribution to $\amu$, a photon (not
  shown in the diagrams) needs
  to couple to any of the charged internal lines.
  In the first two diagrams, the chirality is flipped at the external
  line, producing an explicit factor $m_\mu$ in the computation of the
  diagrams. In these two diagrams, the loop only couples to $\mu_L$ or
  $\mu_R$, respectively. In the third diagram, the chirality is
  flipped via the loop, possibly via virtual BSM
  particles.}\label{fig:chirflipdiagrams}
\end{figure}

As a result, any contribution to the physical muon mass in any
such theory will involve the factors
\begin{align}
  \Delta m_\mu\sim
          \Big[\Big(\text{chirality-flipping parameter(s)}\Big)\times
          \Big(\text{EWSB \vev}\Big)\Big]
          {\times(\text{other factors})}.
            \label{mmugeneric}
\end{align}
This expresses that any contribution to $m_\mu$ must be proportional
to the combination of parameters
responsible for chiral symmetry breaking and to the vacuum expectation
values responsible
for EWSB; Fig.\ \ref{fig:chirflipdiagrams} shows illustrative Feynman
diagrams with different insertions of \vev{}s. The ``other factors'' can
involve any couplings of the theory in question which appear in appropriate
Feynman diagrams and do not have to be related to EWSB or chiral
symmetry breaking. They can also include numerical prefactors or
dimensionless loop
functions. 

Exactly the same kind of discussion applies to $\amu$ in view of the
analogous structure of Eqs.\ \eqref{LDiracmLR} and \eqref{LamuLR}.
We can therefore write the generic relationship
\begin{align}\label{amugeneric}
	{\Delta\amu} &\sim {m_\mu}\times{
        \Big[\big(\text{chirality-flipping parameter(s)}\big)\times
          \big(\text{EWSB \vev}\big)\Big]
      }
        {\times\frac{(\text{other factors})}
          {M^2}}
\end{align}
for any contribution to $\amu$ in any model with electroweak gauge invariance.
The factors in the square bracket reflect
the need for chirality flips and EWSB. Since the coefficient in the
Lagrangian Eq.~(\ref{LamuLR}) actually corresponds to $\amu/m_\mu$,
solving for $a_\mu$ produces
the explicit factor $m_\mu$ on the r.h.s.~of Eq.~(\ref{amugeneric}).
The denominator $M^2$ represents a
typical mass scale of the theory and must appear for dimensional
reasons.
The ``other factors'' have the same properties as the ones in
Eq.\ (\ref{mmugeneric}). We note that the separation into these
factors is not necessarily unique but always possible, as will also be
apparent from results in concrete models discussed later.

The formulas in Eqs.~\eqref{mmugeneric} and \eqref{amugeneric}
demonstrate the deep relationship between the muon mass and
$\amu$ and lead to the notion of models with or without
chiral enhancements as indicated at the beginning of this
subsection
in Eqs.~\eqref{amusimplest} and \eqref{amuchiralenhancement}.
In the SM, the factors in the square brackets  of
these formulas simply amount 
to the product of the SM Yukawa coupling and  Higgs \vev, i.e.\ to the
muon mass, $[\ldots]\to m_\mu$. There is a class of BSM scenarios
where the same is true, e.g.\ the ones illustrated by the first two
Feynman diagrams in Fig.~\ref{fig:chirflipdiagrams} where the
chirality is flipped at the external muon line, necessarily
producing the factor
$m_\mu$. Such contributions have the 
straightforward behaviour of
Eq.~(\ref{amusimplest}).

In another
class of BSM scenarios, however, the factors in the square brackets
can be very different and potentially larger. This
is illustrated by the third diagram in 
Fig.~\ref{fig:chirflipdiagrams} where the \vev is inserted at an
internal line possibly corresponding to a BSM particle. The new couplings relevant for chirality flips might be
generalised Yukawa couplings to  \vev{}s from 
extended Higgs sectors; they can also correspond to 
a suitable product of couplings which break chiral symmetry in
combination. 
With such new sources of symmetry breakings,
it can happen that some terms in the square brackets in
Eqs.~(\ref{mmugeneric}) and (\ref{amugeneric}) are enhanced,
$[\ldots]\gg m_\mu$. This is the case for the chiral enhancement
exemplified in Eq.~(\ref{amuchiralenhancement}).  In this case it is
always possible to define a characteristic quantity $R_\chi\gg1$ which is given by enhancement
factors appearing in $[\ldots]$ divided by the muon mass.  For example
the \vev insertion in the third diagram in
Fig.~\ref{fig:chirflipdiagrams} could come from the Yukawa coupling to
a heavier fermion in the loop with mass $m_F$, in which case one can
get an enhancement  factor $R_\chi=\frac{m_F}{m_\mu} = \frac{y_F}{y_\mu}$.  The
enhanced diagram in Fig.~\ref{fig:chirflipdiagrams} also illustrates
that two different couplings $c_{L,R}$ to left- and right-handed muon
are typically required, such that the generic formula
(\ref{amuchiralenhancement}) becomes valid.

\subsubsection{Overview of BSM scenarios}
\label{sec:overviewBSMscenarios}

\begin{table}
	\centering
	\renewcommand{\arraystretch}{1.5}
	\begin{tabular}{|>{\hspace{1cm}}p{8.5cm} l|}
		\hline
		\multicolumn{2}{|c|}{\textbf{Generic one-loop results}} \\ \hline\hline
		\multicolumn{2}{|l|}{(a) \emph{without} chiral enhancement, discussed in Secs.~\ref{sec:genericoneloop} and \ref{sec:MinimalBSM}} \\ 
		$\displaystyle\Damu\displaystyle\sim \pm\frac{|c|^2}{64\pi^2} \frac{m_\mu^2}{M^2}$ & $\displaystyle \approx\pm1.8 \times 10^{-9}  |c|^2 \Big(\frac{100~\text{GeV}}{M}\Big)^2$ \\
		\multicolumn{2}{|l|}{(b) \emph{with} chiral enhancement, discussed in Secs.~\ref{sec:genericthreefield} and \ref{sec:Collider}} \\
		$\displaystyle\Damu\sim \pm R_\chi\times\frac{c_Lc_R}{16\pi^2} \frac{m_\mu^2}{M^2} $
		& $\displaystyle \approx\pm 6.7 \times 10^{-8}  c_L c_R \frac{R_\chi}{100} \Big(\frac{1~\text{TeV}}{M}\Big)^2$ \\
		\multicolumn{2}{|l|}{(c) radiative muon mass generation ($m_\mu = \Delta m_\mu$). See Eq.~\eqref{amuradmmu} in Sec.~\ref{sec:MuonMass}} \\
		$\displaystyle \Damu\sim \pm\frac{16\pi^2 \Delta m_\mu}{\lambda_L\lambda_R m_\mu}\times\frac{\lambda_L\lambda_R}{16\pi^2}\frac{m_\mu^2}{M^2}  = \pm\frac{m_\mu^2}{M^2}$ &
		$\displaystyle \approx\pm2\times 10^{-9}  \Big(\frac{2.4~\text{TeV}}{M}\Big)^2$ \\ 
		\multicolumn{2}{|l|}{(d) maximal contribution compatible with perturbativity: $R_\chi,c_i\sim \sqrt{8\pi}$. See Eq.~\eqref{eq:amu-max-pert}} \\ 
		$\displaystyle \Damu\sim \pm\frac{\sqrt{8\pi} v}{m_\mu}\times\frac{1}{2\pi} \frac{m_\mu^2}{M^2} $ & 
		$\displaystyle \approx\pm2\times 10^{-9}  \Big(\frac{100~\text{TeV}}{M}\Big)^2$ 
		\\[.2cm]  \hline
		\multicolumn{2}{l}{{}} \\[-.2cm]
		\hline
		\multicolumn{2}{|c|}{\textbf{Model-specific results}} \\ \hline\hline
		\multicolumn{2}{|l|}{(e) \underline{Supersymmetry}: WHL one-loop contribution. See Eq.~\eqref{eq:SUSYWHL} in Sec.~\ref{sec:SUSYamu}} \\
		$\displaystyle\Damu\approx\tan\beta\times\frac{5 g_2^2}{192\pi^2} \frac{m_\mu^2}{M^2} $ & 
		$\displaystyle \approx+2.1\times 10^{-9}  \frac{\tan\beta}{40} \Big(\frac{500~\text{GeV}}{M}\Big)^2 $ \\
		
		\multicolumn{2}{|l|}{(f) \underline{2HDM}: two-loop Barr-Zee contribution. See Eqs.~\eqref{eq:FA2HDM-amu-estimate} and \eqref{eq:2HDM-amu-numerical}
			in Sec.~\ref{sec:2HDM}} \\ 
		$\displaystyle \Damu\approx -\zeta_l\zeta_u \frac{m_t^2}{v^2}\times\frac{N_c Q_t^2 \alpha}{8\pi^3} \frac{m_\mu^2}{M^2}  (\mathcal{F}_1 - \mathcal{F}_2)$ & 
		$\displaystyle \approx-7\times 10^{-10}  \frac{\zeta_l\zeta_u}{100} \Big(\frac{1~\text{TeV}}{M}\Big)^2$
		\\
		\multicolumn{2}{|l|}{(g) \underline{Leptoquarks}: $S_1$ + top one-loop contribution. See Eqs.~\eqref{DamuLQ} and \eqref{LQtopnumerical} in Sec.~\ref{sec:LQ} } \\
		$\displaystyle\Damu\approx\frac{m_t}{m_\mu}\times\frac{N_c \lambda_L^{32}\lambda_R^{32}}{16\pi^2} \frac{m_\mu^2}{M^2}  \mathcal{F}^\text{FS} $ & 
		$\displaystyle \approx+3.3 \times 10^{-9}  \frac{\lambda_L^{32}\lambda_R^{32}}{0.01} \Big(\frac{2~\text{TeV}}{M}\Big)^2 $\\
		\multicolumn{2}{|l|}{(h) \underline{Vector-like leptons}: $L\oplus E$ one-loop contribution. See Eq.~\eqref{eq:VLL-Damu-enhanced} in Sec.~\ref{sec:VLF}} \\
		$\displaystyle \Damu\approx -\frac{\bar\lambda_\Phi v}{m_\mu}\times\frac{\lambda_L\lambda_R}{16\pi^2}\frac{m_\mu^2}{M^2} $& 
		$\displaystyle \approx-4\times 10^{-10}  \frac{\lambda_L\lambda_R \bar\lambda_\Phi}{0.01} \Big(\frac{2~\text{TeV}}{M}\Big)^2$\\[.2cm] \hline
	\end{tabular}
	\caption{
		A collection of BSM contributions to $\Damu$ in different scenarios. The first column shows approximate analytical formulas and the second
		the corresponding seminumerical approximation. As indicated by
		the $\sim$ symbol, the generic results are given up to $\O(1)$ factors depending e.g.\ on the $\SUL$ 
		representations	and electric charges, while the model-specific
		formulas are approximations to the full results explained in the indicated sections.        
		The first factors on the left of the $\times$ symbols correspond to
		the chiral enhancement factors in the form of
		Eq.~\eqref{amuchiralenhancement} and discussed in
		Sec.~\ref{sec:ChiralityFlips}, see also
		Sec.~\ref{sec:Collider}.
		In case of the leptoquark and 2HDM
		we have suppressed the arguments of the loop-functions and
		neglected the resulting $\ln(M)$ dependence in the
		seminumerical results. In case of the radiative muon mass
		model, the chiral enhancement corresponds to an inverse
		one-loop expression.}
	\label{tab:estimates}
\end{table}

As explained above, the
chirality-flipping interactions have a key influence on the behaviour
of BSM contributions to $\amu$.
The instructive relations Eqs.~\eqref{mmugeneric} and \eqref{amugeneric}
for $m_\mu$ and $\amu$  highlight the relevance of
factors related to EWSB and to chirality flips. This insight
allows for a straightforward understanding of how
different models can contribute to $\amu$, including a distinction between
models with or without chiral    enhancements.

In the following we give a brief overview of important
types of contributions and important models. Table \ref{tab:estimates}
summarises corresponding analytical and numerical results for contributions to
$\amu$ in a selection of relevant models and generic model classes. 
First, there are many models without chiral enhancement, where EW and
chiral symmetry are broken by the SM Higgs \vev\ and the SM Yukawa coupling. Examples
include the SM itself, as well as many BSM scenarios such as dark
photon, dark $Z$, more general \emph{dark matter} (DM) and dark sector models, and
also several leptoquark models. Generally, 
models where the new states can couple either only to the
left-handed muon or only to the right-handed muon cannot lead to
chiral enhancement.
The contributions of such models to $\amu$ have a simple
structure and depend on the involved couplings and masses. As shown in
the table, they can only     lead to significant contributions to
$\amu$ if the masses are     smaller than around 100 GeV and/or
couplings are     very large. This is essentially only viable if the
models have very     weak or no interactions at colliders, i.e.\ with
electrons, photons,    or quarks. Hence models with DM
candidates or a larger dark sector with light particles can often lead to potentially
visible effects in $\amu$; conversely such models can be strongly
constrained by $\amu$. This will be explored in
Sec.~\ref{sec:LightDarkSector}.

There are also many well-motivated models with chiral enhancement. The
reason is that fundamental questions related to the Higgs sector and
electroweak symmetry breaking or to the origin of flavour and the
three generations motivate the study of modified Higgs sectors and
extended fermion or Yukawa sectors --- these often contain
new sources of breaking of the chiral symmetry
in Eq.~\eqref{muchiralsym}. Such models can lead to
modifications of the factors in square brackets in
Eq.~\eqref{amugeneric}, i.e.\ potentially to chiral enhancements.
Similarly, models with the goal to explain the origin of neutrino
masses often include new fields which as a by-product also contribute
to the muon mass and thereby also to $\amu$. Many such models can
potentially lead to large contributions to $\amu$, and their
parameter space can be significantly constrained by $\amu$.

Table \ref{tab:estimates} shows several well-known
examples. 
\emph{Supersymmetry} (SUSY) is one of the best motivated and most studied
ideas for BSM physics, and it has been known for a long time that the
parameter $\tan\beta$, the ratio of two Higgs vacuum expectation
values, can lead to a chiral enhancement which brings SUSY
contributions into a very interesting ballpark for SUSY mass ranges
which are motivated for other reasons. SUSY also provides very
promising potential explanations of dark matter. The current LHC
and dark matter limits together with $\amu$ constrain the SUSY
parameter space in a  complementary way. This will be discussed in
detail in Sec.~\ref{sec:SUSY}, focusing both on constrained scenarios
motivated by GUT- or Planck-scale assumptions and on phenomenological
SUSY with different ways to accommodate dark matter.

The \emph{Two-Higgs doublet model} (2HDM) is one of the simplest
extensions of the Higgs sector, but its leading contributions to
$\amu$ arise in a rather complicated way on the two-loop level from
so-called Barr-Zee diagrams where a second Higgs state connects the
muon to a loop of e.g.\ the top quark. Such contributions are chirally
enhanced if the additional Higgs field has enhanced Yukawa couplings
to the muon. However, there is a multitude of
constraints from collider, leptonic and quark flavour
observables
on the new Yukawa couplings and Higgs
masses. Depending on the type and flavour structure of the 2HDM, these
constraints complement constraints from $\amu$ in different ways. This
interplay and resulting viable 2HDM scenarios are discussed in
Sec.~\ref{sec:2HDM}. 

The existence of \emph{leptoquarks} (LQ), i.e.\ spin-0 or spin-1 particles which
directly connect leptons and quarks, can be motivated e.g.\ by attempts to
unify quarks and leptons. Specific versions of leptoquarks allow gauge
invariant couplings both to the left- and the right-handed muon and
the left- and right-handed top quark. Such models lead to a strong
chiral enhancement where e.g.\ the top- or charm-quark mass governs the chirality flip
instead of the muon mass. Even though there are strong lower mass limits on
leptoquarks from the LHC, such leptoquarks can very significantly
contribute to $\amu$ as well as to many quark flavour and lepton
flavour observables. This is discussed in Sec.~\ref{sec:LQ}.

The SM contains three generations of chiral quarks and leptons, and it
is known that no fourth chiral generation can exist. However,
additional vector-like leptons or quarks with TeV-scale masses are
possible. Specifically \emph{vector-like leptons} (VLL) can contribute to
the muon mass not only via loop corrections but also through mixing already at
tree level in a way similar to the seesaw mechanism for neutrino
masses. Because of this difference in loop order, the relationships
between $\amu$, the muon mass, and in particular also the muon--Higgs
coupling are of special interest in this class of
models. In addition collider and electroweak observables also yield
important complementary constraints. In Sec.~\ref{sec:VLF} we discuss how the combination of these 
observables can significantly constrain such VLL models and
we also characterise the remaining viable regions in parameter space.

Another open question, not addressed by the SM, is the origin of the \emph{neutrino masses}.
These tiny but non-zero masses were thoroughly established with the discovery of 
neutrino oscillation, but so far none of the proposed explanations could be confirmed
experimentally. The close connection between neutrinos and charged leptons via the 
electroweak symmetry might suggest that the answer could also have a profound impact on the
charged lepton sector. The simplest possible solutions discussed in Sec.~\ref{sec:neutrino_masses}
typically require extremely heavy new particles or very weak coupling, such that $a_\mu$ remains mostly
unaffected. However, there are nonetheless many models that simultaneously explain the tiny neutrino masses while
also resulting in a significant contribution on $a_\mu$. 
A wide range of such models will be discussed in detail in Sec.~\ref{sec:neutrino_mass_gm2}

Table \ref{tab:estimates} also includes several special model-independent cases of
chirally enhanced contributions. First it is possible that the
tree-level muon mass vanishes and the muon mass is instead radiatively
generated by BSM loop effects. In terms of
Eqs.~\eqref{mmugeneric} and \eqref{amugeneric} this means
that the chiral enhancement must be large enough to compensate all
loop suppression factors such that $\Delta m_\mu=m_\mu$ and $\Damu\sim
m_\mu^2/M^2$. This is possible in a variety of SUSY and non-SUSY
scenarios. Models with radiative muon mass generation lead to
$\Damu\sim10^{-9}$ for a relevant mass scale of around 3~TeV.

Finally, the maximum possible chiral enhancement from one-loop
diagrams where all appearing couplings are as large as allowed by
perturbative unitarity and the relevant \vev is the full SM
\vev. In
such a case $\Damu\sim10^{-9}$ can be obtained even for a mass scale
of order 100~TeV. Conversely speaking, the scale of 100~TeV is the
heaviest mass scale that can be constrained by $\amu$ in perturbative
models. 

\subsection{Definition of $a_\mu$}
\label{sec:definition}

Like mass and charge, the  magnetic dipole moment is
an intrinsic property of the muon that influences its
motion in the presence of an external classical electromagnetic field
$A_\mu^\text{ext}(x)$. Writing the magnitude of the magnetic moment
$\bm{\mu}$ in terms of the spin $\bm{s}$ and the total charge $q$ as 
\begin{align}
|\bm{\mu}|=  g_\mu \frac{q}{2m_\mu}|\bm{s}|,
\end{align}
the gyromagnetic ratio $g_\mu$ and the anomalous magnetic moment
$\amu=(g_\mu-2)/2$ is defined.
Interestingly, the value of $\amu$ always vanishes at tree level and
is a prediction arising from quantum effects in any renormalisable
quantum field theory. The derivation of $\amu$ from first principles
within QFT has been pioneered by Schwinger  \cite{Schwinger:1951nm}
and can be found in many textbooks and reviews. We reproduce it here,
particularly highlighting and quantifying  several of the
approximations that are made in the process. For an alternative
definition in the context of axiomatic QFT, see Ref.~\cite{Steinmann:2002fu}.

The physical situation relevant for the definition and measurement of
$\amu$ is in essence nonrelativistic. In its rest frame the muon only
interacts with the external quasiclassical electromagnetic field,
which is assumed to be weak compared to the intrinsic energy scale set
by the muon rest mass. 
The available energy is not sufficient to create new particles. Here,
photon radiation emitted off the accelerated charge is neglected, and the
muon decay is neglected as well. 

In QFT, this single-particle system in the presence of an external field
can be described as follows. The muon is described by a 4-component
Dirac spinor field $\mu(x)$ and its interaction with the photon field 
is given by the renormalisable Lagrangian term
\begin{align}
	\La \supset - Qe \bar{\mu} \slashed{A} \mu,
\end{align}
which follows from gauge invariance.
The quasiclassical external field
$A_\mu^{\text{ext}}$ can be introduced via a coherent quantum state
under which the quantum field operator $A_\mu$ develops the
expectation value $\langle A^\mu\rangle=A^{\text{ext}}_\mu$. 
Then, computations in the presence of the coherent state are 
simply equal to computations after the
variable shift $A_\mu\to A_\mu+A^{\text{ext}}_\mu$ in
the action \cite{PhysRev.138.B740,PhysRev.139.B1326}, see also
Refs.\ \cite{PhysRevD.97.016007,Fedotov:2022ely} for further
progress. The variable shift effectively leads to an additional
term in the interaction Lagrangian,
\begin{align}
	\La \supset - Qe \bar{\mu} \slashed{A}^\text{ext} \mu.
\end{align}
We
assume the field is weak and can be treated perturbatively. Then the interaction with the
external field is described by an additional Feynman given by 
\begin{align}\label{eq:Def:ext-Feynman-Rule}
	\begin{gathered}
		\includegraphics[scale=.8]{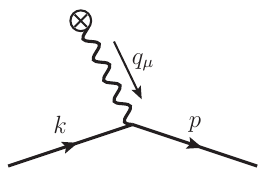}
	\end{gathered} 
	= -i Qe \gamma^\mu \tilde A_\mu^\text{ext}(q)
\end{align}
where $\tilde A_\mu^\text{ext}$ denotes the Fourier transform. For a
discussion of the Furry picture, where the external field is resummed
and incorporated in the unperturbed propagator, we refer to the recent
Refs.~\cite{Fedotov:2022ely,HernandezAcosta:2023msl}.

In this setup, the 2-point Green function  $G(x,y)=\langle
T\mu(x)\bar{\mu}(y)\rangle^{(A^{\text{ext}}_\mu)}$  computed in the
presence of the additional Feynman rule of
Eq.~(\ref{eq:Def:ext-Feynman-Rule})  thus describes the 
time-dependent motion of a fermion 
subject to the external field. In a perturbative treatment of the
external field it is useful to also define  
${G}^{\text{amp}}$ as the amputated Green function obtained by
dropping the trivial tree-level contribution and removing the external
unperturbed propagators. Assuming
time-independence of the external field and thus energy conservation, the
Green function only depends on the 7 arguments $\bm{x}$, $\bm{y}$
and $y^0-x^0$. It is useful to define its momentum-space
version $\tilde{G}^{\text{amp}}(\bm{p},\bm{k},k^0)$ as the 7-fold
Fourier transformation where $\bm{k}$ and $\bm{p}$ are incoming and outgoing
3-momenta and $k^0=p^0$ is enforced by energy conservation.

To later match the QFT description to a single-particle quantum theory
it is useful to also consider scattering of the muon off the
external field.
This is described by the amplitude
of an in-state with definite 4-momentum $k_\mu$ and polarisation
$s_\mu$ scattering into an out-state with four momentum $p_\mu$
and polarisation $s'_\mu$, where $(k,s)\ne(p,s')$,
\begin{align}\label{eq:Def:QFT-S-Matrix}
  \prescript{}{\text{out}}{\braket{p,s'}{k,s}_{\text{in}}} =
 - 2\pi i \delta(k_0-p_0) T_{ps';ks}.
\end{align}
The T-matrix element $T_{ps';ks}$ is also defined assuming energy
conservation and is
given by the LSZ formula\footnote{For simplicity, here we assume on-shell renormalisation such that 
the LSZ factor $\mathcal{Z}=1$.} in terms of the amputated two-point
function introduced before,
\begin{align}
  -iT_{ps';ks} =
  \bar{u}(p,s') G^{\text{amp}}(\bm{p},\bm{k},k^0)u(k,s).
\end{align}
As mentioned above, if the external field is strong, the Furry picture
provides a more appropriate starting point. In that context, 
the recent Refs.~\cite{Ilderton:2012qe,HernandezAcosta:2023msl,Fedotov:2022ely} 
have also considered the situation where infrared real radiation is
relevant. 

Generally, we use Dirac spinors $u(k,s)$ describing a free
muon with on-shell momentum $k$ and polarisation $s$ satisfying the
conditions 
\begin{align}
	(\slashed{k} - m) u(k,s) = 0, \qquad \text{and} \qquad \gamma^5\slashed{s} u(k,s) = u(k,s)
\end{align}
where the 4-vectors fulfil $k^2=m_\mu^2$, $s^2=-1$ and $k\cdot s=0$. Similarly, the negative frequency solution $v(k,s)$
obeying $(\slashed{k}+m)v(k,s)=0$ is also an eigenspinor of the helicity operator fulfilling $\gamma^5 \slashed{s}v(k,s) = -v(k,s)$.
Explicitly, the Dirac spinors are given by \cite{Bjorken:1965zz, Itzykson:1980rh,Peskin:1995ev,Quigg:2013ufa}
\begin{align}\label{eq:Def:Dirac-Spinors}
	\renewcommand*{\arraystretch}{.8}
	u(k,s) = \frac{\slashed{k}+m}{\sqrt{2(k_0+m)}} \begin{pmatrix} \xi(s) \\ \xi(s) \end{pmatrix}, \qquad
	v(k,s) = \frac{\slashed{k}-m}{\sqrt{2(k_0+m)}} \begin{pmatrix} -\xi(-s) \\ \xi(-s) \end{pmatrix},
\end{align}
in terms of 2-spinors $\xi$,
which allows to derive the following projection operators 
\begin{align}
	u(k,s) \bar{u}(k,s) &= \tfrac{1}{2}\big(\slashed{k}+m\big)\big(\mathds{1} + \gamma^5\slashed{s}\big) \\
	v(k,s) \bar{v}(k,s) &= \tfrac{1}{2}\big(\slashed{k}-m\big)\big(\mathds{1} + \gamma^5\slashed{s}\big).
\end{align}
From now on we assume a fixed polarisation axis, such that there are
only two discrete possible values $\pm s$ for the polarisation vector;
a corresponding completeness relation is $\sum_s 	u(k,s)
\bar{u}(k,s)=\slashed{k}+m$.

\begin{figure}[t]
	\centering
	\includegraphics[width=\textwidth]{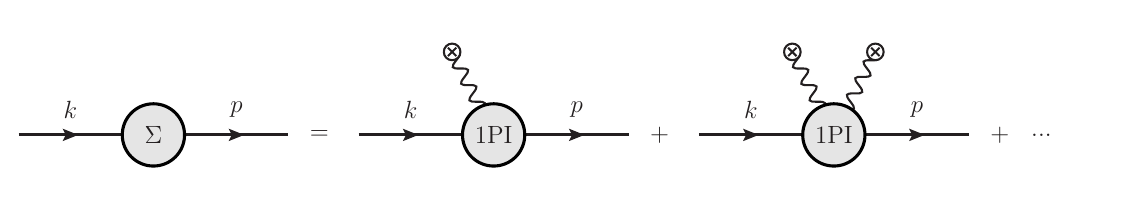}
	\caption{1PI on-shell self-energy of the muon. The crosses correspond to insertions of the interaction vertex with 
		the external field $A_\mu^\text{ext}$. Note that for
                off-shell momenta the r.h.s.~would contain an
                additional 1PI term without insertions of the external
        field.}
	\label{fig:Def:1PI-SE}
\end{figure}
\begin{figure}[t]
	\centering
	\includegraphics[width=\textwidth]{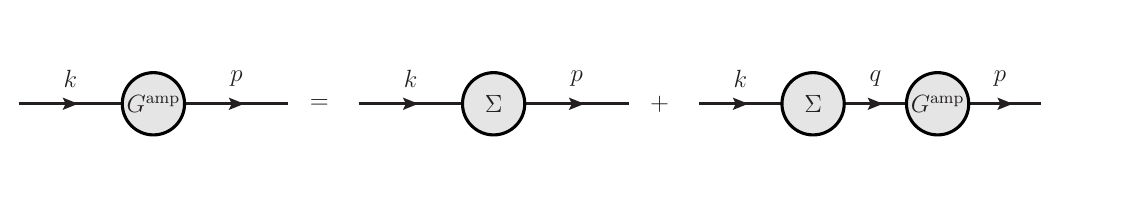}
	\caption{Dyson-Schwinger equation for the two-point Green
          function, corresponding to Eq.~(\ref{DSE}).}
	\label{fig:Def:Schwinger-Dyson}
\end{figure}

As pioneered by Schwinger, the Green function
$G^{\text{amp}}(\bm{p},\bm{k},k^0)$ and the T-matrix element
$T_{ps';ks} $   allow to read
off the value of the gyromagnetic ratio $g_\mu$
\cite{Schwinger:1951nm}. In order to proceed it is useful to derive a
Dyson-Schwinger like equation adapted to the non-relativistic
situation. 
Introducing the self-energy $\Sigma(\bm{p},\bm{k};k_0)$ as the sum of the
one-particle irreducible (1PI) diagrams in the presence of the external
field (see Fig.~\ref{fig:Def:1PI-SE})
allows us to write down the Dyson-Schwinger equation for
$G^{\text{amp}}$ described e.g.~in the review
Ref.~\cite{Roberts:1994dr} and illustrated in
Fig.~\ref{fig:Def:Schwinger-Dyson}, 
\begin{align}\label{DSE}
  G^{\text{amp}} (\bm{p},\bm{k};k_0) = i\Sigma(\bm{p},\bm{k};k_0) 
  + \int \frac{d^4q}{(2\pi)^4}  (2\pi)\delta(k_0-q_0)\,
  i\Sigma(\bm{p},\bm{q};q_0)
 \frac{i}{\slashed{q}-m_\mu+i\epsilon} G^{\text{amp}} (\bm{q},\bm{k};k_0).
\end{align}
After rewriting the tree-level fermion propagator in terms of
positive-time and negative-time contributions we arrive at
\begin{align}
	\begin{split}
		G^{\text{amp}}(\bm{p},\bm{k};p_0) = i\Sigma(\bm{p},\bm{k};p_0) - \sum_{r}\int\frac{d^3q}{(2\pi)^3 2\omega_q} \bigg\{&\Sigma(\bm{p},\bm{q};p_0)
		\frac{u(q,r)\bar u(q,r)}{p_0-\omega_q+i\epsilon} G^{\text{amp}}(\bm{q},\bm{k};p_0) \\
		&+
		\Sigma(\bm{p},-\bm{q};p_0)
		\frac{v(q,r)\bar v(q,r)}{p_0+\omega_q-i\epsilon} G^{\text{amp}}(-\bm{q},\bm{k};p_0)
		\bigg\},
	\end{split}
\end{align}
where $\omega_q=\sqrt{\bm{q}^2+m_\mu^2}$ corresponds to the
relativistic energy-momentum relation.
The first term in the brackets corresponds to the exchange of a
forward propagating muon and the second to a backward propagating
anti-muon.

At this point the non-relativistic limit can be used. In this limit,
the energies $p_0=k_0$ and $\omega_q$ are approximately equal to the rest mass
$m_\mu$. Hence the
forward propagator is unsuppressed while the denominator of the
backward propagator leads to a relative $1/m_\mu$ suppression. Up to $1/m_\mu$
corrections, this backward propagator term can be neglected.

The Dyson-Schwinger equation for the Green function implies a
similar equation for the T-matrix element. Defining
$T_{ps';ks}^{\text{1PI}} \equiv
-\bar{u}(p,s')\Sigma(\bm{p},\bm{k};p_0) u(k,s)$ for the
1PI part we can write the following iterative integral equation, 
\begin{align}\label{eq:Def:QFT-Lippmann-Schwinger}
  T_{ps';ks} &= T_{ps';ks}^{\text{1PI}}
  +
  \sum_r \int\frac{d^3q}{(2\pi)^3 2\omega_q} \frac{T^{\text{1PI}}_{ps';qr} T_{qr;ks}}{k_0-\omega_q+i\epsilon}.
\end{align}
This represents an approximate equation that includes the interaction
with the external fields to all orders, but where $1/m_\mu$-suppressed
terms were neglected.\footnote{%
The equation would become exact if the 1PI self energy were replaced
by a redefined self energy that is 1-particle irreducible only with
respect to the forward propagator $\sim \frac{1}{p_0 - \omega_q + i\epsilon}$.}

\begin{table}
	\centering
	\begin{tabular}{c @{\hskip 1em}|@{\hskip 1em} c}
		\begin{tabular}[t]{|c c c|}
			\multicolumn{1}{c}{} & \multicolumn{1}{c}{\textbf{S.I.}} & \multicolumn{1}{c}{\textbf{natural units}} \\ \hline
			$m_\mu$ & 0.113 u & 0.105 GeV \\ \hline 
			$\Gamma_\mu$ & $4.55\times 10^{5}$ Hz & $3\times 10^{-19}$ GeV \\ \hline\hline 
			$B_0$ & $1.45$ T & $2.68\times 10^{-15} \,\text{GeV}^2$  \\ \hline
			$R_0$ & $7.112$ m & $3.6\times 10^{16} \, \text{GeV}^{-1}$ \\ \hline
			$\frac{1}{R_0}\partial_\phi B_0|_{\text{max}}$ & $100\frac{\text{ppm}}{\text{deg}}\times \frac{B_0}{R_0}$ &
			$4\times 10^{-34} \,\text{GeV}^3$\\\hline\hline
			$\omega_c=\frac{eB_0}{\gamma m_\mu}$ & $4.21\times 10^7$ Hz & $2.77\times 10^{-17}\,\text{GeV}$  \\ \hline
			$\omega_a=a_\mu \frac{eB_0}{m_\mu}$ & $1.44\times
			10^6$ Hz & $9.47\times 10^{-19}\,\text{GeV}$
			\\ \hline
		\end{tabular}
		&
		\begin{tabular}[t]{|cc|}
			\multicolumn{1}{c}{\textbf{Ratios}} & \multicolumn{1}{c}{\textbf{numerical value}} \\\hline
			$\gamma\tau_\mu \omega_c$ & $2700$ \\ \hline
			$ v\approx R_0 \omega_c$ & 0.9994 \\ \hline\hline
			$\frac{B_0}{m_\mu^2}$ & $2.54\times 10^{-14}$ \\ \hline
			$\frac{\Gamma_\mu}{m_\mu}$ & $2.84\times 10^{-18}$  \\ \hline 
			$\frac{\partial_\phi B_0}{R_0 m_\mu^3}$ &
			$4\times10^{-31}$
			\\\hline\hline
			$\Delta_\text{synch}=\frac{4\pi\alpha\gamma^4}{3R_0m_\mu\gamma}$
			& $2\times10^{-13}$ 
			\\\hline
		\end{tabular}
	\end{tabular}
	\caption{Magnitude of  quantities relevant to $a_\mu$ and
		neglected contributions in the derivation. The left table
		contains dimensionful quantities in S.I.\ and natural units,
		the right table contains dimensionless combinations.
		Here $\Gamma_\mu$
		is the muon width, $B_0$ and $R_0$ are the magnetic
		field and the radius of the ring
		used in the Fermilab experiment, $\omega_{c,a}$ are the
		frequencies for the circular motion and the spin precession
		relative to the motion. The ``magic'' $\gamma$ used for the
		muons is $\gamma=29.3$. $\Delta_\text{synch}$ is the relative
		energy loss per turn caused by 
		synchrotron radiation, and $\frac{1}{R_0}\partial_\phi
		B_0|_{\text{max}}$ corresponds to the
		maximum field inhomogeneity in azimuthal direction. See Refs.~\cite{Muong-2:2021vma,Muong-2:2021ovs,Muong-2:2021xzz}.}
	\label{tab:exp-quantities}
\end{table}

This integral equation provides a starting point to match the full QFT
description to the one of non-relativistic single-particle quantum
mechanics, which then allows to extract the effective value of the magnetic moment.
There, the muon is
described by a 2-spinor wave function $	\psi(t,\bm{x};\bm{s}) = \braket{t,\bm{x};\bm{s}}{\psi}$ whose time-dependence is  
governed by the Schrödinger equation
\begin{align}
  i\tfrac{\partial}{\partial t}\psi(t,\bm{x};\bm{s}) = H(\bm{x}) \psi(t,\bm{x};\bm{s}).
\end{align}
Here $\bm{s}$ denotes the polarisation 3-vector and the Hamiltonian
$H$ is constrained by gauge invariance and
takes the general form
\begin{align}\label{eq:QM-Hamiltonian}
	H = \tfrac{1}{2 m_\mu} \big[\bm{\hat p}-
          q\bm{A}^{\text{ext}}\big]^2 + q \phi^{\text{ext}}  - 
	\bm{\mu} \cdot\bm{B} - \bm{d} \cdot \bm{E} + H_\text{eff}.
\end{align}
Here $\phi^{\text{ext}}$ and $\bm{A}^{\text{ext}}$ are the same external
fields as in the QFT description, and $\bm{B}$, $\bm{E}$ are the
corresponding physical magnetic and electric fields.
The first term of the Hamiltonian gives the Pauli kinetic term and the
coefficients $\bm{\mu}$ and $\bm{d}$ of the 
second and third term by definition correspond to the magnetic and
electric dipole 
moments, respectively. These dipole coefficients $\bm{\mu}$ and
$\bm{d}$ are of the order $1/m_\mu$.
The last term $H_\text{eff}$ represents an infinite tower of
additional effective contributions suppressed by higher powers of
$1/m_\mu$. These include e.g.~higher-order derivatives and non-linear terms in
the external fields.\footnote{%
Sample terms in $H_\text{eff}$ can be $
   \frac{e}{m_f^2}(\bm{\partial}\cdot\bm{E})$,
  $ \frac{e}{m_f^2}  (\bm{p}-Qe\bm{A}^{\text{ext}})\cdot\left(
  \bm{E}\times\bm{\sigma}\right)$, or
  $\frac{e^2}{m_f^3}\bm{E}^2
$ which are analogous to terms appearing in effective field theories
  such as NRQED or the leptonic equivalent of HQET
\cite{Caswell:1985ui,Kinoshita:1995mt,Manohar:1997qy,Hill:2012rh,Paz:2015uga},
although these field theories are still different from the
single-particle quantum theory described here.}

The single-particle quantum theory allows to describe the same
scattering of the non-relativistic particle off the external field. In
this approach, the full Hamiltonian  Eq.~\eqref{eq:QM-Hamiltonian}
is split into $H=H_0 + V$ with the free part $H_0=\frac{\bm{\hat p}^2}{2m_\mu}$ 
and the interaction potential $V$; the scattering T-matrix element
corresponding to
Eq.~(\ref{eq:Def:QFT-S-Matrix}) can be written as \cite{Weinberg:1995mt}
\begin{align}
  T_{ps';ks}^\text{QM}&
  = \leftindex^{\text{QM}}_{0}{} \bra{\bm{p};\bm{s}'}V\ket{\bm{k};\bm{s}}_\text{in}^\text{QM},
\end{align}
which corresponds to a matrix element of $V$ between
an asymptotically free in-state and a free momentum eigenstate
satisfying $H_0\ket{\bm{p};\bm{s}}_0^\text{QM} =
p_0\ket{\bm{p};\bm{s}}_0^\text{QM}$
(with the non-relativistic dispersion relation $p_0=\frac{\bm{p}^2}{2m_\mu}$), 
and the normalisation
$	\braket{t,\bm{x};\bm{s}'}{\bm{p};\bm{s}}_0^\text{QM} =
e^{-i(p_0t - \bm{px})} \delta_{ss'} \xi(\bm{s})$,
where the 2-spinors $\xi(\bm{s})$ are the same as those in
Eq.~\eqref{eq:Def:Dirac-Spinors} in the rest frame.

This T-matrix element fulfils 
the Lippmann-Schwinger equation
\begin{align}\label{eq:Lippmann-Schwinger-QM}
	T^\text{QM}_{ps';ks} = V_{ps';ks} + \sum_{r}\int\frac{d^3q}{(2\pi)^3} \frac{V_{ps';qr}T^\text{QM}_{qr;ks}}{k_0 - q_0 + i\epsilon},
\end{align}
where $V_{ps';ks}=
\leftindex^{\text{QM}}_{0}{}{\bra{\bm{p};\bm{s}'}V\ket{\bm{k};\bm{s}}_0^\text{QM}}$.
Obviously this Lippmann-Schwinger equation is an iterative integral
equation of the same form as 
Eq.~\eqref{eq:Def:QFT-Lippmann-Schwinger} which followed from the
Dyson-Schwinger equation of the QFT Green function.
Matching the QM
and the QFT T-matrix elements thus yields the equation
\footnote{the factor of $2p_0$ accounts for the relativistic normalisation of the QFT states}
\begin{align}\label{eq:Def:matching}
	T^{\text{1PI}}_{ps';ks} = 2 p_0 V_{ps';ks}.
\end{align}

To complete the matching, we are left with computing $T^{\text{1PI}}_{ps';ks}$
and interpreting its result in terms of contributions to
$V_{ps';ks}$, which can be rewritten as $ \xi(\bm{s}')^\dagger \tilde
V(\bm{p},\bm{k})\xi(\bm{s})$, where the Fourier transform of
the interaction potential is given by
\begin{align}\label{eq:Def:potential}
	\tilde V(\bm{p},\bm{k}) &= - \frac{q}{2m_\mu} \tilde{\bm{A}}(\bm{q})\cdot\bm{P} + q\tilde{\phi}(\bm{q})
	+ i \bm{\mu} \cdot \big(\bm{q}\times\tilde{\bm{A}}(\bm{q})\big) - i \bm{d}\cdot \bm{q} \tilde{\phi}(\bm{q}) + \ldots.
\end{align}
Here $\bm{q}=\bm{k}-\bm{p}$ and $\bm{P}=\bm{k}+\bm{p}$, and the dots
denote terms of higher order in $1/m_\mu$ that we neglect in the following.

As discussed above,  the field relevant for
the $a_\mu$ experiments is weak. Hence only terms of first order in
$A_\mu^\text{ext}$ have to be taken into account in the computation of
the self energy $\Sigma$ and to
$	T^{\text{1PI}}_{ps';ks}$; higher powers of the field would
contribute to terms in $H_{\text{eff}}$ suppressed by additional
factors $eA_\mu^{\text{ext}}/m_\mu$.\footnote{%
Note that the same would not be true for reducible diagrams, where
higher powers of the external field can appear with the weaker
suppression $A_\mu^{\text{ext}}/(\slashed{q}-m_\mu)$.}
Accordingly, in Fig.~\ref{fig:Def:1PI-SE}
only the first term on the r.h.s.\ is required. In this approximation
the 1PI T-matrix element can be written as
\begin{align}\label{eq:Def:vertex-function}
	\begin{gathered}
		\includegraphics{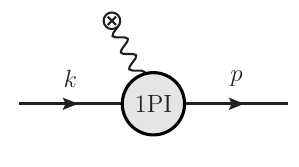}
	\end{gathered}
	\equiv -iQe \bar{u}(p,s') \Gamma^\mu(p,k) u(k,s) \times \tilde A_\mu^{\text{ext}}(q).
\end{align}
Equivalently, in this approximation we have the relationship
$\Sigma(\bm{p},\bm{k};p_0)=-Qe\Gamma^\mu(p,k)\tilde A_\mu^{\text{ext}}(q)$.
The computation of the vertex function $\Gamma^\mu$ is now independent
of the external field and can be done utilising ordinary Feynman rules
in vacuum. In general, $\Gamma^\mu$ can be written in terms of the
covariant decomposition 
\begin{align}
	\Gamma^\mu(p,k) = \gamma^\mu \Big[A_1(q^2) + A_2(q^2) \gamma^5 \Big] + \frac{P^\mu}{2m_\mu} \Big[A_3(q^2) + iA_4(q^2) \gamma^5\Big] + 
	\frac{q^\mu}{2m_\mu} \Big[A_5(q^2) + A_6(q^2) \gamma^5\Big],
\end{align}
where $q=p-k$ and $P=p+k$ is defined also for the 4-momenta. Because
the muon momenta are on-shell the prefactors $A_i$ are Lorentz
invariant and depend only on $q^2$.
Furthermore, gauge invariance for the external field implies the Ward identity
\begin{align}
	q_\mu \cdot \bar{u}(p,s') \Gamma^\mu(p,k) u(k,s) = 0
\end{align}
and requires $A_5(q^2)=0$ as well as $q^2 A_6(q^2) = -(2m_\mu)^2
A_2(q^2)$. The vertex function can therefore be written in terms of
only four form factors 
\begin{align}
	\Gamma^\mu(p,k) = \gamma^\mu A_1(q^2) + \Big[\gamma^\mu - \tfrac{2m_\mu}{q^2}q^\mu\Big] \gamma^5 A_2(q^2) + \frac{P^\mu}{2m_\mu} \Big[A_3(q^2) + iA_4(q^2) \gamma^5\Big]
\end{align}
or, using the Gordon identities to rewrite the spinor products,
equivalently as
\begin{align}\label{eq:photon-vertex-decomp}
	\Gamma^\mu(p,k) = \gamma^\mu F_E(q^2) + \frac{i\sigma^{\mu\nu}q_\nu}{2m_\mu} \Big[F_M(q^2) - i F_D(q^2)\gamma^5\Big] 
	+ \Big[\gamma^\mu - \tfrac{2m_\mu}{q^2}q^\mu\Big] \gamma^5 A_2(q^2),
\end{align}
where $F_E=A_1 + A_3$, $F_M=-A_3$ and $F_D=A_4$. Using the explicit results for the Dirac spinors Eq.~\eqref{eq:Def:Dirac-Spinors} we 
can expand the vertex function in the non-relativistic limit. To first order in $\bm{q}$ and $\bm{P}$ this gives
\begin{subequations}
	\begin{align}
		\bar{u}(p,s') \Gamma^0(p,k) u(k,s) &\approx \xi^\dagger(\bm{s}') \bigg\{ 2m F_E(0) + F_D(0) i \bm{q\sigma}\bigg\} \xi(\bm{s}) \\
		\bar{u}(p,s') \Gamma^k(p,k) u(k,s) &\approx \xi^\dagger (\bm{s}')
		\bigg\{ F_E(0) \Big(\bm{P}^k - i\epsilon^{kij}\bm{q}_i \sigma_j\Big) - F_M(0) i\epsilon^{kij}\bm{q}_i \sigma_j \bigg\} \xi(\bm{s}).
	\end{align}
\end{subequations}
We may then insert this expansion together with $\tilde A_\mu^\text{ext}(q) = (2\pi)\delta(q_0) (\tilde\phi(\bm{q}),-\tilde{\bm{A}}(\bm{q}))$ into
Eq.~\eqref{eq:Def:vertex-function} after which Eq.~\eqref{eq:Def:matching} reduces to
\begin{align}
	\tilde{V}(\bm{p},\bm{k}) \approx Qe F_E(0) \bigg(\tilde\phi(\bm{q}) - \frac{1}{2m_\mu} \tilde{\bm{A}}(\bm{q})\cdot\bm{P} \bigg)
	+ i \frac{Qe}{2m_\mu} \Big(F_E(0) + F_M(0)\Big) \bm{\sigma} \cdot \big(\bm{q}\times \tilde{\bm{A}}(\bm{q}) \big)
	+ i \frac{Qe}{2m_\mu} F_D(0) \bm{\sigma} \cdot \bm{q}\tilde\phi(\bm{q})
\end{align}
After comparing the coefficients with Eq.~\eqref{eq:Def:potential} we arrive at
\begin{subequations}
	\begin{align}
		q &= F_E(0) Qe \\
		\bm{\mu} &= 2\Big(F_E(0) + F_M(0)\Big) \frac{Qe}{2m_\mu} \frac{\bm{\sigma}}{2} \equiv g_\mu \frac{Qe}{2m_\mu} \frac{\bm{\sigma}}{2} \\\
		\bm{d} &= - 2F_D(0) \frac{Qe}{2m_\mu} \frac{\bm{\sigma}}{2}
	\end{align}
\end{subequations}
If the electric charge is on-shell renormalised, $F_E(0)=1$, and we obtain the usual simple expression for the anomalous magnetic moment as
\begin{align}
  \label{amuDefFM}
	\amu = \frac{g_\mu-2}{2} = F_M(0).
\end{align}
This result provides the starting point for quantum field theoretical
computations of $\amu$. We note that it also corresponds to the
effective Lagrangian already announced in Eq.~(\ref{LamuLR}) which was
used to explain the importance of chirality flips in
Sec.~\ref{sec:ChiralityFlips}.

The derived effective single-particle Pauli quantum theory is appropriate to describe the
setup of $g-2$ measurements both for the electron and the muon. In the
latter case, the muon and its spin  behave quasiclassically, and
the time-dependence of the spin expectation value is described by the classical
Bargmann-Michel-Telegdi equation \cite{Bargmann:1959gz}
	\begin{align}\label{eq:BMT}
		\frac{d \bm{s}}{dt} & = -\frac{Qe}{m_\mu} \bigg[\bigg(\frac{g_\mu}{2} + \frac{1-\gamma}{\gamma}\bigg)\bm{B} - 
		\frac{a_\mu \gamma}{1+\gamma}(\bm{v}\cdot\bm{B})\bm{v} - \bigg(\frac{g_\mu}{2} - \frac{\gamma}{1+\gamma}\bigg) \bm{v}\times \bm{E}\bigg]\times \bm{s}
	\end{align}
which can be derived starting from the quantum
theory neglecting the $1/m_f^2$-suppressed terms
as described in Ref.~\cite{LandauVol4}, $\S$41.

As indicated at the beginning, the derivation has involved several
approximations. Synchrotron radiation emitted by the accelerated muon and
the muon decay were ignored in setting up a single-particle
description, and terms with relative
$1/m_\mu$ suppression were neglected. As the derivation showed, the
$1/m_\mu$-suppressed terms can correspond to inhomogeneities in the
field $\sim\bm{\partial}\cdot\bm{B}$, higher orders in the field, or
relativistic corrections to the energy-momentum relation. In
Tab.~\ref{tab:exp-quantities} we collect a number of experimental
quantities, converted into natural units and combined to dimensionless
suppression factors. While a more complete evaluation of impact of
the neglected contributions on the definition of $\amu$ in
Eq.~(\ref{amuDefFM}) is beyond the scope of this review, the small
values of these factors confirm that 
the experimental measurements of $\amu$ are quite immune to such
corrections. 

\subsection{Status of the $a_\mu$ Measurement}\label{sec:experiment}

	In this section we summarise the history of the $g-2$ experiments and discuss the current
	status after the latest results of the final run of the Fermilab experiment \cite{Muong-2:2025xyk}.
	For a more complete historic overview and experimental details
        we refer to the reviews in
        Refs.~\cite{Miller:2012opa,Roberts:2018vsx,Keshavarzi:2021eqa}
        and to the original literature.

	The first experimental measurements of $a_\mu$ were performed as early as 1957 at Nevis \cite{Garwin:1957hc,Cassels:1957} in an effort
	to test the prediction of parity violation in the weak interaction by Lee and Yang a year earlier \cite{Lee:1956qn}.
	In their seminal paper Lee and Yang proposed several processes in which parity violation could be tested.
	In particular, they noted that parity violation and conservation of angular momentum in the pion decay 
	$\pi \to \mu + \nu$ as well as the subsequent decay of the muon $\mu \to e + \nu_\mu\bar{\nu}_e$ leads to a strong correlation between
	the momentum and polarisation of the muon (and electron). After the experimental confirmation \cite{Garwin:1957hc,Friedman:1957mz}
	this fact became crucial to control the initial polarisation and observe the final spin of the muons in the measurements of $a_\mu$.
	
	Initially, several experiments determined $g_\mu$ through the spin precession of a muon at rest in an external magnetic
	field $\bm{B}$. Here the spin precession is described by  \cite{Jackson:1998nia}
	\begin{align}
		\frac{d \bm{s}}{dt} = \bm{\omega}_s \times \bm{s},
	\end{align}
	and the precession rate (Larmor frequency) is given by $\bm\omega_s = -g_\mu (Qe/2m_\mu) \bm{B}$ and can thus be adjusted by
	varying the value of the external field. The experiments were
        set up to first produce polarised muons ($\mu^+$) from an incoming
	beam of pions ($\pi^+$) which were stopped in a target inside of a static magnetic field. A scintillator then counted the emitted positrons
	over a fixed interval, and the  precession frequency
        $\bm\omega_s$ could be measured for several values of $|\bm
        B|$, to extract the magnetic moment. By 1960 the results were
        precise enough to confirm Schwingers one-loop correction term $\alpha/2\pi$ \cite{Garwin:1960zz}
	\begin{align}
		a_\mu^\text{1960} = 0.00122(8),
	\end{align}
	and	by 1965 the first CERN experiment (CERN I) reached the precision of the QED two-loop contribution \cite{Charpak:1961mz,Charpak:1962zz,Charpak:1965zz},
	\begin{align}
		a_\mu^\text{CERN I} = 0.001162(5).
	\end{align}
	However, while very successful early on this approach was ultimately limited. For one, performing the experiment with negatively charged
	muons ($\mu^-$) was not possible as these would be captured by the target after stopping. In addition, the lifetime
	of the muon at rest $\tau_\mu \approx 2.2~\mu$s only allowed for a short measurement period and consequently limited statistics.
	
	\begin{figure}[t]
		\centering
		\begin{subfigure}{.45\textwidth}
			\centering
			\includegraphics[width=.6\textwidth]{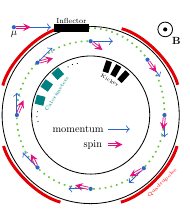}
			\caption{Illustration of the spin precession in storage ring}
			\label{fig:storage-ring}
		\end{subfigure}
		\begin{subfigure}{.45\textwidth}
			\centering
			\includegraphics[width=.9\textwidth]{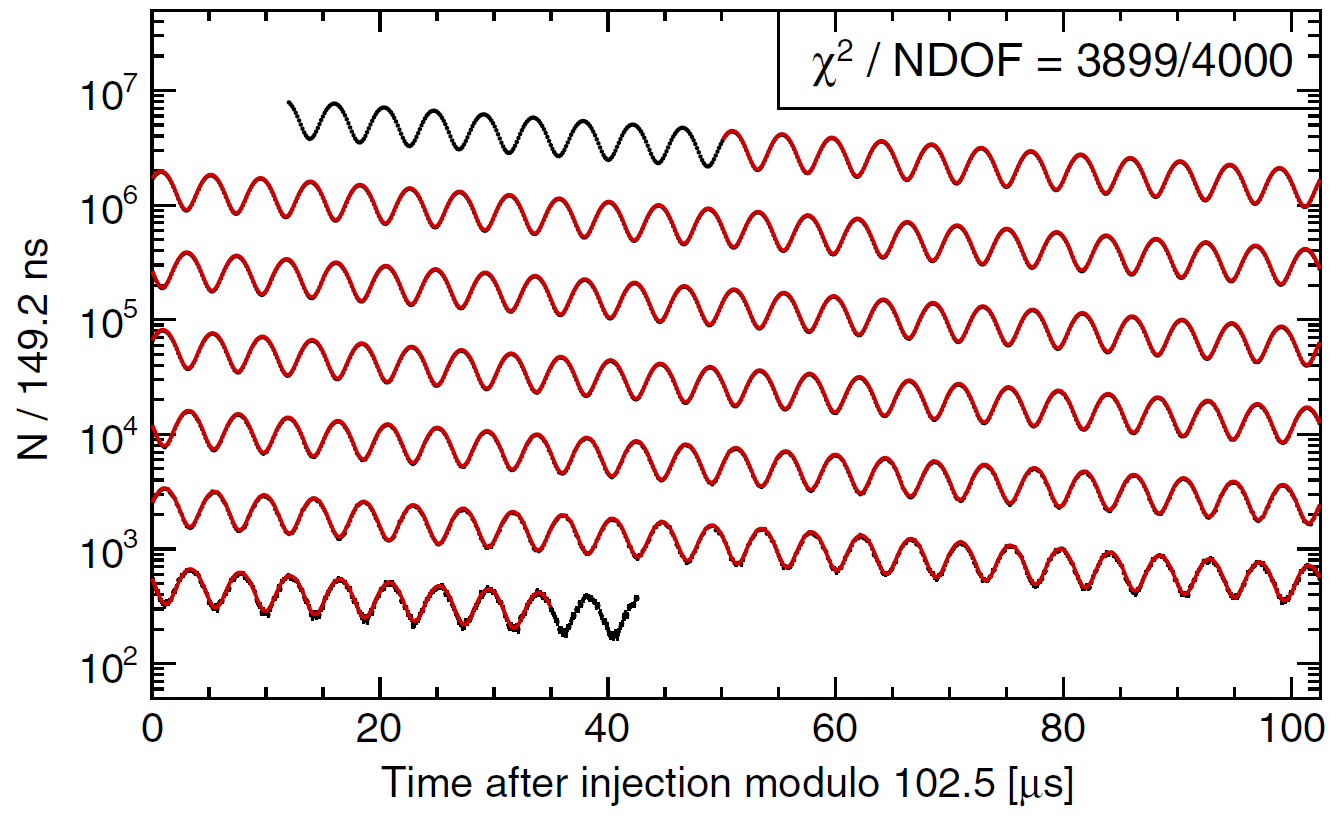}
			\caption{Fit of the positron count in FNAL run-1d \cite{Muong-2:2021vma}}
			\label{fig:FNAL-6}
		\end{subfigure}
		\caption{\textbf{(a)} Schematic of the storage ring and illustration of the muon spin precession in the constant magnetic field.
			In total, 24 calorimeters distributed evenly around the ring measure the positrons emitted by the decaying muon beam.
		\textbf{(b)} Fit of the positron count overlayed on top of the measured data in the FNAL run-1d \cite{Muong-2:2021vma}.
		}
	\end{figure}
	These issues prompted the development of a new method using relativistic muons resulting in longer lifetimes and thereby significantly
	extending the measurement period. This technique was pioneered by the second CERN experiment (CERN II) where, for the first time,
	a storage ring was used to contain the muon ($\mu^+$ or $\mu^-$) beam \cite{Combley:1974tw,CERNMuonStorageRing:1977bbe}. Because of the (relativistic) velocities of the muon, its
	spin precession frequency is modified by the
        Bargmann-Michel-Telegdi-equation \cite{Bargmann:1959gz}, see Eq.~(\ref{eq:BMT}),
	\begin{align}\label{eq:spin-precession}
		\bm\omega_s = -\frac{Qe}{m_\mu} \bigg[\bigg(\frac{g_\mu}{2} + \frac{1-\gamma}{\gamma}\bigg)\bm{B} - 
		\frac{a_\mu \gamma}{1+\gamma}(\bm{v}\cdot\bm{B})\bm{v} - \bigg(\frac{g_\mu}{2} - \frac{\gamma}{1+\gamma}\bigg) \bm{v}\times \bm{E}\bigg].
	\end{align}
	At the same time, the muon velocity also rotates
	\begin{align}
		\frac{d\bm{v}}{dt} = \bm{\omega}_c \times \bm{v} + \frac{Qe}{\gamma m_\mu} \frac{1}{\gamma^2-1} (\bm{v}\cdot \bm{E}) \bm{v},
	\end{align}
	with the cyclotron frequency
	\begin{align}\label{eq:velocity-precession}
		\bm\omega_c = -\frac{Qe}{m_\mu} \bigg[\frac{1}{\gamma} \bm{B} - \frac{\gamma}{\gamma^2-1} \bm v\times \bm E\bigg].
	\end{align}
	Because of the difference between Eq.~\eqref{eq:spin-precession} and
	Eq.~\eqref{eq:velocity-precession} over time the polarisation of the muon beam  
	deviates from the direction of the velocity as sketched in Fig.~\ref{fig:storage-ring}.
	This difference can be tracked through the positrons (or electrons) emitted from the muon beam
	and measured by detectors arranged around the inside of
        the ring. Up to experimental corrections, the number of positrons varies with time as 
	\begin{align}
		N(t) = N_0 e^{-t/\gamma\tau_\mu} \Big[1 + \mathcal{A} \cos\big(|\bm\omega_a| t + \phi_0\big)\Big],
	\end{align}
	where $N_0$ and $\phi_0$ denote the initial beam density and average initial polarisation angle, 
	$\mathcal{A}$ stems from the weak decay asymmetry and dictates the amplitude of the oscillation around the
	average exponential decay with the anomalous precession frequency
	\begin{align}\label{eq:omega_a}
		\bm\omega_a = \bm\omega_s - \bm\omega_c=-\frac{Qe}{m_\mu} \bigg[a_\mu \bm{B} - \frac{a_\mu\gamma}{1+\gamma} (\bm{v}\cdot\bm{B})\bm{v}
		- \Big(a_\mu - \frac{1}{\gamma^2-1}\Big) \bm v\times \bm E\bigg].
	\end{align} 
	The magnetic field is typically measured using nuclear magnetic resonance (NMR) probes that yield $|\bm{B}|=\omega_p/2\mu_p$
	in terms of the proton Larmor frequency and magnetic moment $\omega_p$ and $\mu_p$.
	After extracting $|\bm\omega_a|$ from fitting the positron
        count, subtracting experimental corrections
        and writing $e=4 m_e \mu_e / g_e$ in terms of the electron mass $m_e$ and magnetic moment $\mu_e$, $a_\mu$ is given by
	\begin{align}
		a_\mu = \frac{\omega_a}{\omega_p} \frac{\mu_p}{\mu_e} \frac{m_\mu}{m_e} \frac{g_e}{2}.
	\end{align} 
	Here the ratios $\mu_p/\mu_e$ and $m_\mu/m_e$ as well as $g_e$
        are taken from independent experiments and are sufficiently
        well known.
		
	CERN II relied solely on weak magnetic focusing, such that no external electric field was present ($\bm E=0$) and the 
	Lorentz factor $\gamma\approx 12$ was set by the ring geometry ($r=2.5$ m) and the external magnetic field $|\bm{B}|\approx 1.71$ T.
	This first measurement using a relativistic muon beam
        \cite{Combley:1974tw}  already
        improved the precision by a factor of 16 compared to CERN I 
	\begin{align}
		a_\mu^\text{CERN II} = 0.00116610(31).
	\end{align}
	However, there were a number of drawbacks with the first design, the most severe stemming from the inclusion of the pion
	production target as a muon kicker inside of the storage ring. While this avoided the technical challenge of producing and
	injecting the polarised muon beam into the storage ring, it also resulted in a large hadronic background and decreased
	statistics due to the small number of muons kicked into the correct orbit.
	To improve on this result a cleaner muon beam with longer
        lifetime was required.

        This was achieved at the CERN III experiment,
	where the $g-2$ experiment in its modern form was implemented for the first time \cite{CERN-Mainz-Daresbury:1978ccd}.
	Instead of the weak magnetic focusing, electric quadrupoles were used to provide vertical stability. According to Eq.~\eqref{eq:omega_a},
	this additional electric field would enter the frequency
        $\bm{\omega}_a$, potentially leading to uncertainties. However, it was noted that
	this effect could be suppressed by choosing the magic $\gamma=\sqrt{1+1/a_\mu}\approx 29.3$.
	To achieve this with a comparable magnetic field
        $|\bm{B}|\approx 1.45$ T, a new ring with radius $r=7.112$ m
	needed to be constructed. Further upgrades were the injection of a pion instead of proton beam, the development of
	an inflector magnet and the uniform magnetic field allowing for a simpler averaging over the muon distribution.
	In combination, CERN III yielded another factor of 36
        improvement over the previous results \cite{CERN-Mainz-Daresbury:1978ccd},
	\begin{align}
		a_\mu^\text{CERN III} = 11\,659\,240(85)\times10^{-10}.
	\end{align}
Famously, this measurement was precise enough to verify the relevance
of the hadronic vacuum polarisation contributions in the SM theory
prediction, which contribute at the level of $7\times10^{-7}$.

	A decade after CERN III the next $g-2$ experiment E821 \cite{Muong-2:2001kxu} at Brookhaven National Laboratory (BNL) was approved.
	The basic measurement principle from the CERN collaboration was kept, however with a number of further upgrades,
	most notably a new superconducting storage ring magnet (that would also be reused at Fermilab)
	and the injection of a much more intense beam. The magnetic field was closely monitored by a number 
	of NMR probes installed around the ring as well as an NMR trolley that
	could map the magnetic field in the ring when the muon beam was inactive.
	A new muon kicker allowed for the injection of a much cleaner muon beam, which significantly
	decreased the number of remaining pions compared to CERN.
	Altogether these upgrades yielded another factor of 15
        improvement in $a_\mu$
        \cite{Muong-2:2002wip,Muong-2:2004fok,Muong-2:2006rrc} (the
        number was updated in Ref.~\cite{Muong-2:2021vma} to reflect
        changes of CODATA values for external inputs)
	\begin{align}\label{amuBNL}
	a_\mu^\text{BNL} &= 11\,6 59\,2 09.1(6.3)\times10^{-10}.
	\end{align} 
	This was precise enough to come within reach of the
        electroweak SM contribution, which contributes at the
        $10^{-9}$ level; in addition, this result differed by around
        $3\sigma$ from the SM prediction at the time, motivating
        considering BSM contributions to $\amu$ as well as further
        experimental scrutiny.
	
	\begin{figure}[t]
		\centering
		\includegraphics[width=.6\textwidth]{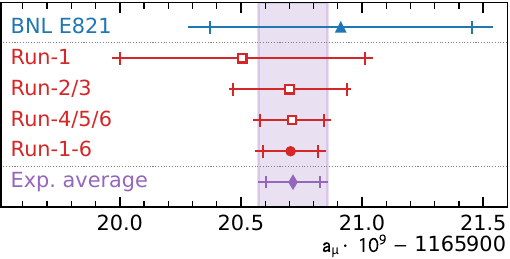}
        \caption{Reprinted from Ref.~\cite{Muong-2:2025xyk}.
        	Summary plot of the experimental values of $a_\mu$ obtained from
        	the FNAL run-1, run-2/3 and run-4/5/6 measurements (red squares) 
        	as well as the final BNL E821 result (blue triangle). The FNAL final value Eq.~\eqref{Eq:FNALfinalresult}
        	is indicated by the red circle and the current world average Eq.~\eqref{eq:amuExp} by the
        	purple diamond. The additional results can be found in
        Eqs.~\eqref{amuBNL} and \eqref{amuExpAdditional}. The
          vertical bars indicate the statistical uncertainties, the
          horizontal lines the total uncertainties.}
        \label{fig:amu-Exp}
	\end{figure}
	
	In order to scrutinise this result, the follow-up experiment E989 at the Fermilab National Accelerator Laboratory (FNAL)
	was commissioned with the goal to improve the overall uncertainty by another factor of 4 \cite{Muong-2:2015xgu}. 
	The central part, the 14 m ring magnet, was moved from Brookhaven to the Fermilab campus, but (apart from the inflector)
	essentially all of the electronics were replaced. The new calorimeters provided a much higher spatial and timing resolution, and an additional monitoring system was used to stabilise the energy measurement. 
	An in-vacuum tracking system
	was developed in order to measure the beam profile. Another major improvement
	was the much longer muon delivery path from the production target to the injector, that allowed for most of the 
	excess pions to decay before arriving at the storage ring. In addition a delivery ring was constructed that
	helped separating remaining protons from the muon bunches.
	With these upgrades, already the results from Run-1 published in 2021 \cite{Muong-2:2021ojo,Muong-2:2021ovs,Muong-2:2021vma,Muong-2:2021xzz}
	reached the precision of the BNL experiment, and together with the Run-2 and 3 results published
	in 2023 \cite{Muong-2:2023cdq,Muong-2:2024hpx} further improved on this result by more than a factor of $2$. 

	The final analysis of Run-4 to 6 was presented earlier this
        year \cite{Muong-2:2025xyk}. Within uncertainties, it is in full
        agreement with all previous measurements at Fermilab and
        Brookhaven. The total Run-1--6 result has surpassed its design goal and reduced the
        statistical and systematic uncertainties to
        {$(1.14)^{\text{stat}}(0.91)^{\text{syst}}\times10^{-10}$} , in
          total to better than 4 times the precision of the Brookhaven
          measurement.\footnote{%
          {Experimental numbers which cannot be found in the
            references cited in the main text are from the internal
            Muon $g-2$ collaboration document \#GM2-doc-32016-v10,
            revised 11th June 2025, by A.\ Lusiani.}
          }
        In total, the results of the Fermilab muon $g-2$ experiment,
        and the resulting current world average are (taking into
        account updates reflecting experimental corrections that were
        applied after the original 
        publications)
        \begin{subequations}
          \label{amuExpAdditional}
	\begin{align}
a_\mu^\text{FNAL,Run-1} &= 11\,6 59\,2 05.1\ \,(5.4)\ \times10^{-10},\\
a_\mu^\text{FNAL,Run-2,3} &= 11\,6 59\,2 07.0\ \,(2.5)\ \times10^{-10},\\
a_\mu^\text{FNAL,Run-4,5,6} &= 11\,6 59\,2 07.10(1.62)\times10^{-10},\\
a_\mu^\text{FNAL,Run-1--6} &= 11\,659\,207.05(1.48)\times10^{-10},\\ 
	\amuExp &= 11\,659\,207.15(1.45)\times10^{-10}. 
	\end{align}
        \end{subequations}
        These numbers are also illustrated in Fig.~\ref{fig:amu-Exp}.

	Finally, let us give a brief outlook for the planned $a_\mu$ experiment at J-PARC 
	\cite{Mibe:2010zz,Iinuma:2011zz,Mibe:2011zz,Saito:2012zz,Abe:2019thb}. This
        experiment aims to be complementary in that it minimises the
        electric-field term in Eq.~\eqref{eq:omega_a} not by choosing a magic
        $\gamma$ but instead by vanishing electric field. 
	A key requirement for this is a lower emittance of the muon
        beam compared to the Fermilab and Brookhaven experiments.
	To achieve this, the polarised muon beam produced by the pion decay is  cooled
	and then re-accelerated before injection into storage ring. 
	Such cooling will eliminate the need for additional electric
        focusing, and thereby the constraint to the {magic} $\gamma$, and generally significantly reduce the systematics associated with the muon beam.
	The novel technology required for the cooling was recently demonstrated for the first time
	\cite{Aritome:2024jiv}.

\subsection{Status of $a_\mu$ in the Standard Model}\label{sec:SMtheory}

Here we describe the SM theory prediction  and explain
the origin of the results obtained in the Theory Initiative White Papers
\cite{Aoyama:2020ynm,Aliberti:2025beg} and given in
Eqs.~(\ref{eq:amuSM2020},\ref{eq:amuSM2025}). Generally, the theory
  prediction is given in terms  1PI Feynman
  diagrams with external on-shell muons and an external photon, see
  Eq.~(\ref{amuDefFM}). As
  mentioned before, for the SM prediction all SM
particles and interactions matter in a relevant way. The hadronic
contributions are responsible for the largest part of the theory
uncertainty; for this review it is also of interest to highlight
certain aspects of the QED and electroweak contributions which  are
similar to many potential BSM contributions.

Traditionally, the full SM theory prediction for $\amu$ is split into
\begin{align}\label{amuSMsplit}
  \amu^{\textrm{SM}} &=
  \amu^{\textrm{QED}} + \amu^{\textrm{EW}} +
  \amu^{\textrm{HVP}} + \amu^{\text{HLbL}},
\end{align}
where the abbreviations denote QED,
electroweak, hadronic vacuum polarisation and hadronic light-by-light
contributions, respectively.

\begin{figure}[t]
	\centering
	\begin{subfigure}{.32\textwidth}
		\centering
		\includegraphics[width=.6\textwidth]{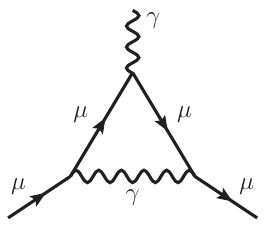}
		\caption{Schwinger one-loop diagram}
		\label{fig:amu-Schwinger}
	\end{subfigure}
	\begin{subfigure}{.32\textwidth}
		\centering
		\includegraphics[width=.6\textwidth]{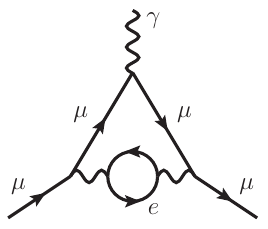}
		\caption{Electron-loop VP correction}
		\label{fig:amu-QED-2l-e}
	\end{subfigure}
	\begin{subfigure}{.32\textwidth}
		\centering
		\includegraphics[width=.6\textwidth]{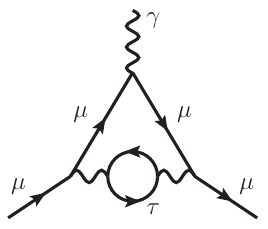}
		\caption{$\tau$-loop VP correction}
		\label{fig:amu-QED-2l-tau}
	\end{subfigure}
	\caption{One- and two-loop QED contributions to $a_\mu$ in the Standard Model.
		Diagram \textbf{(a)} corresponds to the well-known Schwinger correction, while diagrams \textbf{(b)} and
		\textbf{(c)} show two-loop leptonic vacuum-polarisation (VP) corrections from an electron and $\tau$ loop
		inserted into the photon propagator.}
	\label{fig:amu-QED}
\end{figure}

The QED and electroweak (EW) contributions are very accurately
known. Their uncertainties were already negligible in
Ref.~\cite{Aoyama:2020ynm}, and their values have changed only by
tiny amounts in Ref.~\cite{Aliberti:2025beg}. The QED contributions are
given by Feynman diagrams involving only photons and muons or other
leptons. They have been evaluated including the 5-loop level to \cite{\QEDref,Aoyama:2019ryr,Aliberti:2025beg}
\begin{align}\label{amuSMQEDnew}
  \amu^{\textrm{QED}} &= 11\,658\,471.88(2)\times10^{-10}.
\end{align}
Three QED diagrams of particular interest are shown in
Fig.~\ref{fig:amu-QED}.
The diagram in Fig.~\ref{fig:amu-Schwinger} is the well-known
QED one-loop diagram leading to Schwinger's result
$\amu^{\textrm{QED,1L}}=\alpha/2\pi$ \cite{Schwinger:1951nm}. The
diagrams in Fig.~\ref{fig:amu-QED-2l-e} and \ref{fig:amu-QED-2l-tau} illustrate
important features of mass-dependent contributions. They are two-loop
corrections where either an electron loop or a $\tau$-lepton loop is
inserted into the photon propagator. The electron mass $m_e$ is
lighter than the muon mass, leading to a logarithmic enhancement
\begin{align}
  \amu^{\textrm{QED,Fig.~\ref{fig:amu-QED-2l-e}(b)}} &=
  \left(\frac{\alpha}{\pi}\right)^2
  \left[-\frac{25}{36}+\frac{1}{3}\ln(\frac{m_\mu}{m_e})\right]
    \sim6\times10^{-6}.
\end{align}
This logarithmic enhancement is the main reason why the muon magnetic
moment $\amu$ is larger than the electron magnetic moment $a_e$, see
also the comparison in
Eqs.~(\ref{gfactorElectron},\ref{gfactorMuon}). The $\tau$-lepton mass
in contrast is larger than the muon mass, leading to a quadratic mass
suppression
\begin{align}
  \amu^{\textrm{QED,Fig.~\ref{fig:amu-QED-2l-tau}(c)}} &=
  \left(\frac{\alpha}{\pi}\right)^2
  \left[\frac{1}{45}\frac{m_\mu^2}{m_\tau^2}+{\cal
      O}\left(\frac{m_\mu^4}{m_\tau^4}\right)\right]
  \sim4\times10^{-10}. 
\end{align}
Such a quadratic mass suppression is typical of the contributions of
heavy particles also in the context of BSM physics.

\begin{figure}[t]
	\centering
	\begin{subfigure}{.32\textwidth}
		\centering
		\includegraphics[width=.6\textwidth]{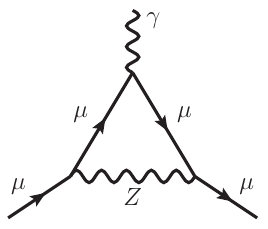}
		\caption{$Z$-boson contribution}
	\end{subfigure}
	\begin{subfigure}{.32\textwidth}
		\centering
		\includegraphics[width=.6\textwidth]{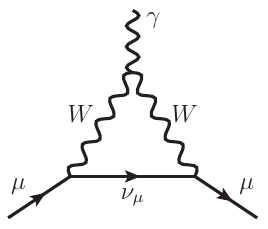}
		\caption{$W$-boson contribution}
	\end{subfigure}
	\begin{subfigure}{.32\textwidth}
		\centering
		\includegraphics[width=.6\textwidth]{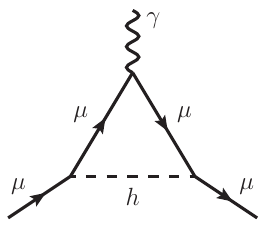}
		\caption{Higgs contribution}
	\end{subfigure}
	\caption{Complete electroweak one-loop contribution to $a_\mu$ in unitary gauge.
	Diagrams \textbf{(a)} and \textbf{(b)} correspond to the leading $Z$- and $W$-boson correction,
	while diagram \textbf{(c)} gives the suppressed Higgs correction. 
	In other gauges (like $R_\xi$) additional contributions from the appropriate Goldstone bosons
	need to be included.}
	\label{fig:SMEW1}
\end{figure}

The EW contributions are given by all Feynman diagrams which contain
any of the non-QED/QCD particles, i.e.\ the heavy $W$, $Z$ or Higgs boson
or neutrinos. The one-loop EW diagrams are shown in
Fig.~\ref{fig:SMEW1}.
For example, the W-boson contribution is essentially determined by
\begin{align}
  \label{eq:amuew1approx}
  \amu^{\textrm{W}} \propto
  \frac{g_2^2}{16\pi^2  }
  \frac{m _\mu ^2}{M _W ^2} \sim 10^{-9},
\end{align}
where $g_2^2/16\pi^2$ corresponds to a loop factor involving the
weak $\SUL$ gauge coupling, and 
$m_\mu ^2 / M _W ^2 \sim 10 ^{-6}$ is again a mass suppression factor arising
from the heavy $W$-boson. This shows that the EW contributions are six
orders of magnitude smaller than the QED contributions and affect the
total $g$-factor of the muon at the 9th decimal place, as announced in
Eq.~(\ref{gfactorMuon}) in Sec.~\ref{sec:amuprobe}. Nevertheless, the EW contributions are
relevant at the current level of precision. In detail, 
the EW contributions can be split into
\begin{align}
\label{EWsplitup}
a_\mu^{\rm EW} &=
a_\mu^{\rm EW(1)}
+ {a_{\mu;\text{bos}}^{\rm EW(2)}}
+
{a_{\mu;\text{ferm}}^{\rm EW(2)}}
+ a_\mu^{\rm EW(\ge3)},
\end{align}
corresponding to one-loop diagrams, two-loop diagrams without/with
fermion loop, and $\ge3$-loop diagrams. The analytic evaluation of the
diagrams in Fig.~\ref{fig:SMEW1} (together with the corresponding Goldstone boson contributions) gives
\begin{align}\label{eq:amu-EW-1}
	a_\mu^{\rm EW(1)} = \frac{G_F}{\sqrt{2}} \frac{m_\mu^2}{24\pi^2} \bigg\{  
	\underbrace{\Big(1-4 s_W^2\Big) - 5 }_{Z} \,~
	\underbrace{ + \,~ 10\vphantom{\Big(}}_{W}
	\,~ \underbrace{- ~\, \tfrac{m_\mu^2}{M_h^2} \Big[7 + 6\ln(\tfrac{m_\mu^2}{M_h^2})\Big]}_{h} + \O\Big(\tfrac{m_\mu^2}{M_Z^2}\ln(\tfrac{m_\mu^2}{M_Z^2})\Big)  \bigg\}
\end{align}
The one-loop and bosonic
two-loop contributions
\cite{Czarnecki:1995sz,Heinemeyer:2004yq,Gribouk:2005ee,Gnendiger:2013pva,Ishikawa:2018rlv}
essentially depend on the heavy $W$, $Z$, and Higgs 
boson masses and amount to $a_\mu^{\rm EW(1)}=19.479(1)$ and 
${a_{\mu;\text{bos}}^{\rm EW(2)}}=-1.9962(3)$ in units of $
10^{-10}$, respectively. The large, negative two-loop correction is
mainly due to large QED-like logarithms and appears also in the
context of BSM contributions \cite{Degrassi:1998es,Czarnecki:2002nt}, see also Sec.~\ref{sec:photonic}.
The fermionic two-loop contributions \cite{Czarnecki:1995wq,Czarnecki:2002nt,Gnendiger:2013pva,Ludtke:2024ase,Hoferichter:2025yih} together with the $\ge3$-loop
contributions amount to $-2.04(4)\times10^{-10}$,
where about half of this value originates from loops of the
3rd-generation fermions $\tau$-lepton, top- and bottom-quarks and the
uncertainty is mainly caused by 2nd-generation hadronic contributions
\cite{Hoferichter:2025yih} and by unknown 
$\ge3$-loop contributions \cite{Czarnecki:2002nt}. In total,
therefore,
\begin{align}
a_\mu^{\rm EW}&= 15.44(4)\times10^{-10}.
\label{amuEWNew}
\end{align}

\begin{figure}[t]
	\centering
	\begin{subfigure}{.32\textwidth}
		\centering
		\includegraphics[width=.6\textwidth]{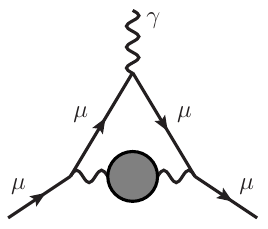}
		\caption{LO HVP}
		\label{fig:amu-HVP}
	\end{subfigure}
	\begin{subfigure}{.32\textwidth}
		\centering
		\includegraphics[width=.6\textwidth]{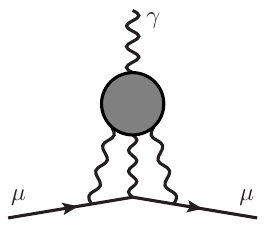}
		\caption{HLbL}
		\label{fig:amu-HLBL}
	\end{subfigure}
	\caption{Non-perturbative hadronic contributions to $a_\mu$ in the SM.
		Diagram \textbf{(a)} shows the leading-order hadronic
                vacuum polarisation  (HVP) correction and 
		diagram \textbf{(b)} the hadronic light-by-light
                (HLbL) scattering correction.}
	\label{fig:amu-hadronic}
\end{figure}

The hadronic vacuum polarisation (HVP) contributions correspond to
diagrams such as Fig.~\ref{fig:amu-hadronic} and appropriate higher-order
corrections, where a hadronic vacuum polarisation is inserted into a
QED one-loop diagram. They are almost two orders of magnitude larger than the EW
contributions and enter the muon $g$-factor at the 7th digit in
Eq.~(\ref{gfactorMuon}). This can be understood by noting the
diagrammatic similarity to the electron-loop
and $\tau$-loop diagrams of Fig.~\ref{fig:amu-QED-2l-e} and \ref{fig:amu-QED-2l-tau}
together with the value of the lightest hadron masses.

These HVP contributions are the cause of the
largest SM theory uncertainty and are subject to intense current
scrutiny. They require non-perturbative techniques, and two approaches are
established.
The more traditional approach is based on the optical
theorem and dispersion relations and relates $\amu^{\textrm{HVP}}$ to
cross sections for $e^+e^-\to\gamma^*\to$~hadrons processes, which are
experimentally measurable. The result used in the first White Paper
\cite{Aoyama:2020ynm} is based on this
approach. Ref.~\cite{Aoyama:2020ynm} obtained
\begin{align}\label{amuHVPWP1}
  \amu^{\textrm{HVP,LO,\cite{Aoyama:2020ynm}}} &= 693.1(4.0)\times10^{-10}
\end{align}
for the most important leading-order HVP contributions.
This result is based on original references
\cite{Davier:2017zfy,Keshavarzi:2018mgv,Colangelo:2018mtw,Hoferichter:2019mqg,Davier:2019can},
and it needs to be combined with known next- and next-to-next-to-leading
order corrections \cite{Keshavarzi:2019abf,Kurz:2014wya}.
In evaluating 
Eq.~(\ref{amuHVPWP1}), the most
critical hadronic channel is $e^+e^-\to\pi^+\pi^-$ at energies between
$0.6\ldots0.9$ GeV. This
had been measured most
precisely by KLOE \cite{KLOE-2:2017fda}, Babar
\cite{BaBar:2012bdw},
CMD-2 \cite{CMD-2:2001ski,CMD-2:2003gqi}, SND \cite{Achasov:2006vp,SND:2020nwa} and BESIII \cite{BESIII:2015equ}. However,
there were significant tensions between these data sets. For instance,
evaluating $\amu$ by using just one of the two individually most
precise data sets KLOE and Babar in this energy region leads to
results that differ by $10\times10^{-10}$. The tensions were taken
into account by locally and globally rescaling uncertainties to obtain
the result (\ref{amuHVPWP1}) and its uncertainty. For detailed
discussions we refer to
Refs.~\cite{Davier:2017zfy,Keshavarzi:2018mgv,Colangelo:2018mtw,Hoferichter:2019mqg,Davier:2019can},
Refs.~\cite{Aoyama:2020ynm,Aliberti:2025beg} and
references therein. 

The second approach to evaluate the HVP contributions is based on
lattice gauge theory, which allows a direct non-perturbative
computation without relying
on experimental data. When Ref.~\cite{Aoyama:2020ynm} appeared,
the lattice results were not yet sufficiently consolidated to be
used. However, around the same time the
Budapest-Marseille-Wuppertal (BMW) lattice collaboration
\cite{Borsanyi:2020mff} published the first lattice determination of
the HVP contribution with an uncertainty
estimate that is small enough to be competitive with the data-driven
estimates.  This BMW prediction is not in agreement with the data-driven
results used in  Ref.~\cite{Aoyama:2020ynm} but is
significantly larger.  The disagreement is
strengthened with the recent update  \cite{Boccaletti:2024guq} which obtains
\begin{equation} \label{BMWupdate}
    a_{\mu}^\textrm{HVP,LO,\cite{Boccaletti:2024guq}} = 714.1(3.3)\times10^{-10}.
\end{equation}
Recent further lattice groups have found agreement with the
BMW calculation, and full HVP,LO results were obtained by the BMW
\cite{Borsanyi:2020mff,Boccaletti:2024guq}, Mainz/CLS
\cite{Ce:2022kxy,Kuberski:2024bcj,Djukanovic:2024cmq} and RBC/UKQCD collaborations \cite{RBC:2018dos,RBC:2023pvn,RBC:2024fic}. Further important
results, partially for intermediate/long- or short-distance Euclidean
time windows where specific optimisations are possible, are given in Refs.~\cite{Giusti:2019xct,Lehner:2020crt,Wang:2022lkq,Aubin:2022hgm,
ExtendedTwistedMass:2022jpw,Spiegel:2024dec,ExtendedTwistedMass:2024nyi,MILC:2024ryz,Bazavov:2024eou}.
All these lattice results are consistent with each other and uniformly show  a tension between lattice and
the data-driven results used in Ref.~\cite{Aoyama:2020ynm}.

The situation is further
complicated by experimental data from the recent CMD-3 experiment \cite{CMD-3:2023alj} which gives
systematically larger cross-sections for $e^+e^-\rightarrow
\pi^+\pi^-$ than the previous  CMD-2, SND, Babar, KLOE and BESIII
measurements.
The  reasons for the disagreement between the
HVP contributions based on earlier $e^+e^-\to\pi\pi$ data, based on CMD-3 data,
or based on lattice gauge theory are under intense scrutiny
\cite{Aliberti:2025beg,Colangelo:2022vok,Davier:2023fpl,Davier:2023cyp,Benton:2023dci,Masjuan:2023qsp}.
Currently, the reasons are unknown, and as discussed in detail in the
second White Paper \cite{Aliberti:2025beg},
there are no scientific grounds on which any of the
$e^+e^-$ data sets could be disregarded. However, the increased tensions between the different data
sets make it currently impossible to obtain a reliable update of the HVP result
using the  dispersive and data-based approach.

\begin{figure}
	\centering
	\includegraphics[width=.7\textwidth]{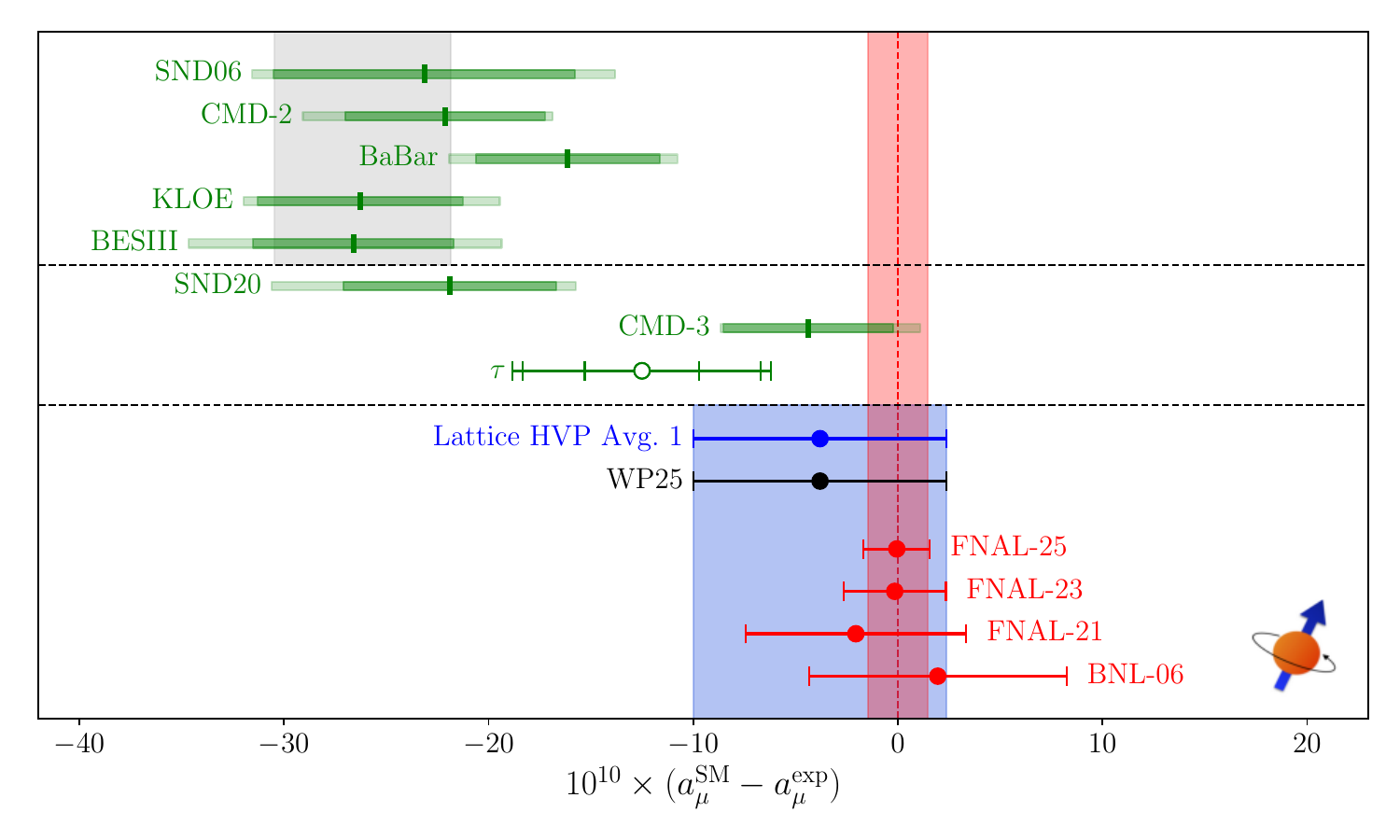}
	\caption{Reprinted from Ref.~\cite{Aliberti:2025beg}. 
		Summary plot of the SM predictions for $a_\mu$, based on different evaluations of the LO HVP.
		The black point and blue band correspond to the new WP2025 value Eq.~\eqref{eq:amuSM2025}
		obtained from the lattice-QCD results,
		while the grey band shows the old WP2020 value Eq.~\eqref{eq:amuSM2020}, derived from the
		$e^+ e^-$ data sets collected above the upper dashed line.
		The values between the dashed lines were obtained from the SND20 and CMD-3 data sets that appeared
		after the WP2020 \cite{Aoyama:2020ynm} as well as a reevaluation of the LO HVP based on $\tau$-decay data
		and are not included in either WP value for $a_\mu$.
		The red points and band	correspond to the experimental values already shown in Fig.~\ref{fig:amu-Exp}.
	}
	\label{fig:amu-SM}
\end{figure}

Notwithstanding these complications, the lattice gauge theory results
on their own are now sufficiently precise, obtained by several
independent groups, fully consistent and cross-checked in multiple
ways. Hence the second Theory Initiative White Paper
Ref.~\cite{Aliberti:2025beg} evaluates the HVP,LO contributions based on
the lattice results from Refs.~\cite{Borsanyi:2020mff,Boccaletti:2024guq,Ce:2022kxy,Kuberski:2024bcj,Djukanovic:2024cmq,
	RBC:2018dos,RBC:2023pvn,RBC:2024fic,Giusti:2019xct,Lehner:2020crt,Wang:2022lkq,Aubin:2022hgm,
	ExtendedTwistedMass:2022jpw,Spiegel:2024dec,ExtendedTwistedMass:2024nyi,MILC:2024ryz,Bazavov:2024eou} 
and obtains
\begin{align}
	\label{eq:WP25-amuHVP-LO}
	\amu^{\textrm{HVP,LO,\cite{Aliberti:2025beg}}} &=   713.2(6.1)\times10^{-10}.
\end{align}
Here, the value (\ref{BMWupdate})  has not been
used since it employs a combination of lattice (95\%) and $e^+e^-$
data (5\%), but both values agree very well.
Including higher orders, the full HVP result reads \cite{Aliberti:2025beg}
\begin{align}
  	\label{eq:WP25-amuHVP}
	\amu^{\textrm{HVP,\cite{Aliberti:2025beg}}} &=   704.5(6.1)\times10^{-10}.
\end{align}
Figure \ref{fig:amu-SM} from
Ref.~\cite{Aliberti:2025beg} summarises the different evaluations of
the HVP contributions and compares the full SM prediction to the
experimental value (\ref{eq:amuExp}).

Finally we briefly comment on the hadronic light-by-light
contributions. They are given by diagrams of the form given in
Fig.~\ref{fig:amu-HLBL} where a non-perturbative hadronic light-by-light
scattering subdiagram is attached to the muon line. They are the
smallest of the four classes of SM contributions in
Eq.~(\ref{amuSMsplit}) but still highly relevant. Significant progress
has been achieved in their evaluation in the past decades. After
earlier computations via hadronic models summarised in
Refs.~\cite{Prades:2009tw,Jegerlehner:2009ry}, more recent
phenomenological approaches based on dispersion relations have been
developed in Refs.~\cite{Melnikov:2003xd,Masjuan:2017tvw,Colangelo:2017fiz,Hoferichter:2018kwz,Gerardin:2019vio,Bijnens:2019ghy,Colangelo:2019uex,Pauk:2014rta,Danilkin:2016hnh,Jegerlehner:2017gek,Knecht:2018sci,Eichmann:2019bqf,Roig:2019reh,Colangelo:2014qya}
and \cite{Holz:2024diw,Holz:2024lom,Estrada:2024cfy,Leutgeb:2022lqw,Bijnens:2024jgh,Bijnens:2022itw}, and also lattice gauge theory
computations have become feasible \cite{Blum:2019ugy,Chao:2021tvp,Blum:2023vlm,Fodor:2024jyn}. All
approaches lead to compatible results, and the White Paper
Ref.~\cite{Aliberti:2025beg} obtains the value
\begin{align}\label{amuSMHLbLnew}
  \amu^{\textrm{HLbL}} &= 115.5(9.9)\times10^{-11}.
\end{align}

In total, adding up the values given in
Eqs.~(\ref{amuSMQEDnew},\ref{amuEWNew},\ref{eq:WP25-amuHVP},\ref{amuSMHLbLnew})
yields the full SM theory prediction for $\amu$ given already in
Eq.~(\ref{eq:amuSM2025}). The difference between the values obtained in
the two White Papers Refs.~\cite{Aoyama:2020ynm,Aliberti:2025beg}
and shown in Eqs.~(\ref{eq:amuSM2020},\ref{eq:amuSM2025}) and
Fig.~\ref{fig:amu-SM} is mainly caused 
by the new result for the HVP contributions as discussed above.

While the current SM prediction agrees very well 
with the experimental determination of $\amu$, significant progress
and updates can be expected. Specifically the leading-order HVP
prediction is under intense scrutiny
\cite{Colangelo:2022vok,Davier:2023fpl,Davier:2023cyp,Benton:2023dci,Masjuan:2023qsp,Aliberti:2025beg}.
The understanding of $e^+e^-$ data can improve via new
measurements, new analyses also including rigorous validation of
theory ingredients such as higher-order calculations and
Monte Carlo tools \cite{Aliberti:2024fpq}. Ideally,  the current tensions among the $e^+e^-$
data can be resolved. In addition, the
inclusion of $\tau$-decay data
\cite{Alemany:1997tn,Davier:2023fpl,Aliberti:2025beg}
and the MUonE experiment
\cite{CarloniCalame:2015obs,MUonE:2016hru,MUonE:2019qlm,Banerjee:2020tdt,Abbiendi:2022oks,Ignatov:2023wma} measuring the
vacuum polarisation in the space-like instead of the time-like
region can provide further cross-checks and independent
input.\footnote{%
The MUonE experiment can also be used to study physics beyond the SM,
as discussed in Refs.~\cite{Dev:2020drf,Masiero:2020vxk,Asai:2021wzx,GrillidiCortona:2022kbq,Atkinson:2022qnl,Le:2023ceg}.}

Forthcoming lattice computations by many groups will employ a wide
range of methods. This progress
promises
even more
precise and consolidated results for the full SM prediction of $\amu$,
further sharpening its role as a test of all aspects of the SM and a
unique constraint on BSM physics.

\newpage\section{Generic BSM scenarios}
\label{sec:Generic}

Physics beyond the SM can contribute to $\amu$ via loop Feynman
diagrams. The contributions have an anatomy governed by the need
for chirality flips and electroweak symmetry breaking, and there are
classes of models with or without chiral enhancements. In this section
we provide technical details and connect them to such qualitative
discussions. 
Sec.~\ref{sec:genericoneloop}
provides  generic one-loop results which are
applicable to all renormalisable quantum field theories. The form of
one-loop Feynman diagrams that can contribute   to $\amu$
is restricted by the requirement that there must be couplings to
a photon and two muons, in addition to the usual restrictions from
gauge and Lorentz invariance and renormalisability.  As a result the possible
one-loop contributions to muon $g-2$ are  well known
and depend only on the couplings, masses and topology of
the diagrams.
The formulas of Sec.~\ref{sec:genericoneloop}
allow mass eigenstates with arbitrary couplings to left-handed and
right-handed
muons.

If the particles in the loop are assumed to coincide with
gauge eigenstates or with mixtures of only two gauge eigenstates, the
possible gauge quantum numbers are quite restricted.
Such  specific BSM extensions are discussed in Secs.~\ref{sec:MinimalBSM}
and \ref{sec:genericthreefield}, where only one, two, or three new BSM
fields with specific gauge quantum numbers and with gauge-invariant
interactions are added to the SM.
The set of such
extensions is quite
limited and the possible models can be systematically
categorised, and
immediate consequences can be drawn regarding the possible signs and magnitudes of
the BSM contributions to $\amu$. The comparisons of the generic
mass eigenstate formulas  with the behaviour of two-field and
three-field models also sheds additional light on the role of
chirality flips and illustrates the usefulness of mass-insertion
approximations. Finally, 
Sec.~\ref{sec:genericeft}
introduces the description of dipole moments in important effective
field theories.
These frameworks are useful to study general properties of $\amu$ and
correlations to other observables in model-independent ways, although
renormalisability is given up. The LEFT
framework can be used for observables below the electroweak scale, and
the SMEFT framework is appropriate for observables above the
electroweak scale as long as no BSM particles exist at the considered
mass scale. The effective theory description will also be used later
to obtain higher-order corrections that are applicable to wide classes
of models.

\subsection{General one-loop formulas for $\Damu$}\label{sec:genericoneloop}

\begin{figure}[tb]
	\centering
	\begin{subfigure}[t]{.3\textwidth}
		\centering
		\includegraphics[width=.49\textwidth]{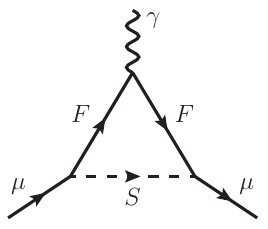}
		\includegraphics[width=.49\textwidth]{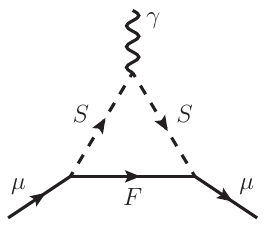}
		\caption{FS-type diagrams}
		\label{fig:amu-FS}
	\end{subfigure}\hfill
	\begin{subfigure}[t]{.3\textwidth}
		\centering
		\includegraphics[width=.49\textwidth]{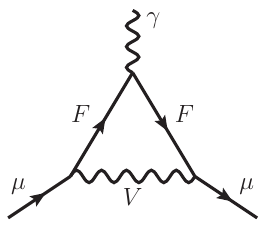}
		\includegraphics[width=.49\textwidth]{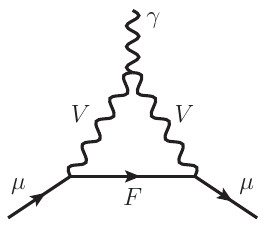}
		\caption{FV-type diagrams}
		\label{fig:amu-FV}
	\end{subfigure}\hfill
	\begin{subfigure}[t]{.3\textwidth}
		\centering
		\includegraphics[width=.49\textwidth]{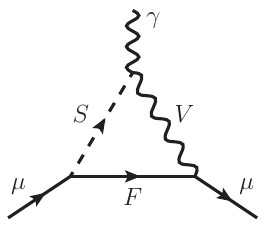}
		\includegraphics[width=.49\textwidth]{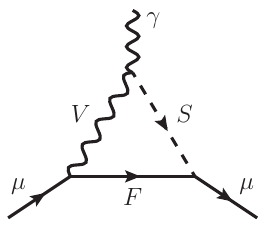}
		\caption{FSV-type diagrams}
		\label{fig:amu-FSV}
	\end{subfigure}
	\caption{All possible one-loop diagrams contributing to $\amu$ in generic renormalisable QFTs, with
		internal scalars (S), spin 1/2 fermions (F) or vector-bosons (V). }
	\label{fig:amu-one-loop}
\end{figure}

In this section we collect the general formulas for $\Damu$ at
one-loop order in generic renormalisable relativistic  quantum field theories.
Such theories may contain scalar
fields, spin 1/2 fermion fields and vector fields which interact via
operators of dimension 4 or less. The generically possible one-loop
Feynman diagrams that contribute to $\amu$ can be classified into
FS-type contributions shown in Fig.~\ref{fig:amu-FS}, FV-type
contributions shown in Fig.~\ref{fig:amu-FV} and FSV-type
contributions shown in Fig.~\ref{fig:amu-FSV}.
All diagrams can be computed in a model-independent way based on a
generic Lagrangian.

Details and discussions of the calculation have been presented in the literature many times in the past.
Specifically, results for unitary gauge can be found in Refs.~\cite{Leveille:1977rc,Jegerlehner:2009ry,Lindner:2016bgg,Yu:2021suw,Hue:2023rks}
and for t'Hooft Feynman gauge in
Refs.~\cite{Lynch:2001zs,Lavoura:2003xp,Athron:2021iuf}.

To list the results, we start from the generic interaction Lagrangian in the mass basis after EWSB. The relevant terms for the interaction between the muon 
and the additional fields read
\begin{align}\label{eq:L_SFV_muon}
	\La_{\mu}^\text{SFV} &
	= \overline{F}\big(c_L \PL + c_R \PR\big)\mu S^\dagger + \overline{F} \gamma^\nu \big(g_L \PL + g_R \PR\big) \mu V_\nu^\dagger + h.c.,
\end{align}
while the interaction terms involving the photon are given by
\begin{align}\label{eq:L_SFV_photon}
	\begin{split}
		\La^\text{SFV}_\gamma &= e \bar\mu \slashed{A} \mu - Q_F e \bar F \slashed{A} F - i Q_S e \big(S^\dagger \partial_\mu S - S \partial_\mu S^\dagger \big) A_\mu\\
		&\hspace{.43cm} -iQ_V e \big(F^{\mu\nu} V_\mu V^\dagger_\nu -  V^{\mu\nu} A_\mu V_\nu^\dagger + V_{\mu\nu}^\dagger A^\mu V^\nu \big)
		- \mu_0 e A^\mu \big(V_\mu^\dagger S + V_\mu S^\dagger\big).
	\end{split}
\end{align}
Here $F=F_L+F_R$ is a new fermion field, $S$ and $V_\nu$ are new
scalar and vector fields, and $\mu=\mu_L+\mu_R$ is the muon field. The
muon electric charge is set to $Q=-1$, and for non-vanishing
contributions to $\amu$, charge conservation implies
$Q_S=Q_V=-1-Q_F$. In  Fig.~\ref{fig:amu-one-loop} the
charges always flow from left to right.
Further, $V_{\mu\nu}=\partial_\mu V_\nu - \partial_\nu V_\mu$ and $\mu_0$ denotes a model-dependent mass parameter, e.g. for Goldstone bosons $\mu_0 = Q_V m_V$.
The appearing fields are
assumed to be mass eigenstates with masses $m_{F,S,V}$, but are not
required to be gauge eigenstates with well-defined gauge quantum
numbers. 
The Lagrangian of any concrete model can
be mapped to these Lagrangians, leading to specific values of the
coupling prefactors.
In general, the full contribution to $\amu$ must be gauge
independent. However, propagator and vertex Feynman rules and
individual Feynman diagrams can depend on the
choice of gauge. Here we present the results in unitary gauge, where
the vector boson propagators take a specific form and Goldstone-boson contributions
vanish. We note that, while 
absent in the SM, physical FSV-type contributions can exists in
extensions to the SM (e.g. in the type-II see-saw mechanism). 

After the charge $Q_F$ is eliminated, the physical results corresponding
to the generic diagrams Fig.~\ref{fig:amu-FS}, \ref{fig:amu-FV} and \ref{fig:amu-FSV} can be written compactly as 
\begin{subequations}\label{eq:Damu-oneloop-results}
	\begin{align}
		\Damu^\text{FS} &= \frac{1}{16\pi^2} \frac{m_\mu^2}{m_S^2} \bigg(
		\Re{c_L^* c_R}\, \frac{m_F}{m_\mu}  \F^\text{FS}(x_\mu, x_F;Q_S)
		+ \big\{|c_L|^2+|c_R|^2 \big\} \G^\text{FS}(x_\mu,x_F;Q_S)	\bigg),
		\label{eq:Damu-FS}
		\\ \label{eq:Damu-FV}
		\Damu^\text{FV} &= \frac{1}{16 \pi^2} \frac{m_\mu^2}{m_V^2} \bigg(
		\Re{g_L^* g_R} \frac{m_F}{m_\mu} \F^\text{FV}(z_\mu,z_F;Q_V)
		+ \big\{|g_L|^2+|g_R|^2 \big\} \G^\text{FV}(z_\mu,z_F;Q_V) \bigg), 
		\\ \label{eq:Damu-FSV}
		\Damu^\text{FSV}&= \frac{\mu_0}{16\pi^2}\frac{m_\mu}{m_V^2} \bigg( \Re{g_R^* c_L + g_L^* c_R} \F^\text{FSV}(z_\mu,z_F,z_S)
		+ \Re{g_R^* c_R + g_L^* c_L} \frac{m_F m_\mu}{m_V^2} \G^\text{FSV}(z_\mu,z_F,z_S) \bigg),
	\end{align}
\end{subequations}
where $x_i\equiv m_i^2/m_S^2$ and $z_i\equiv m_i^2/m_V^2$. The fully general
expressions for the loop functions are listed at the end of this
subsection.\footnote{%
We mention that one-loop contributions from non-renormalisable models
with spin-3/2 particles have been considered in
Ref.~\cite{Criado:2021qpd} and from spin-2 Kaluza-Klein excitations of
gravitons in Ref.~\cite{Graesser:1999yg}.
}

\begin{figure}
	\centering
	\includegraphics[width=.7\textwidth]{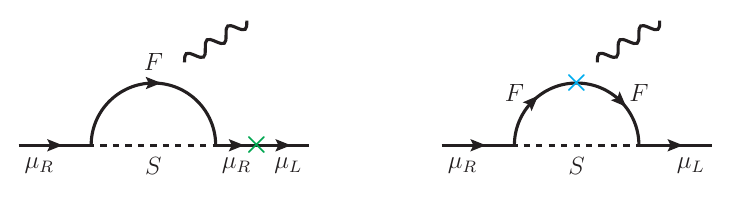}
	\caption{\label{fig:twoVsthree} Illustration of the chirality flips in the FS-type one-loop contribution
		 Eq.~\eqref{eq:Damu-FS} to $\Damu$. The photon can
                 couple to any charged internal line. The left diagram corresponds to a chirality flip on the external line,
		 producing the $|c_{L,R}|^2 m_\mu$ term, while the right diagram corresponds to an internal chirality flip
		 resulting in the $c_L^* c_R m_F$ term. There exist
                 similar diagrams with the roles of $\mu_{L,R}$
                 exchanged, and the FV- and FSV-type contributions can be
                 understood analogously.
		  }
		\label{fig:amu-chirality-flip}
\end{figure}

All of the formulas contain two different kinds of terms,
corresponding to different chirality combinations and involving
different fermion mass combinations.
As explained in Sec.~\ref{sec:ChiralityFlips},
any
contribution to $\amu$ requires a $L\leftrightarrow R$ flip of the
muon chirality. This flip can happen at
the external muon line or via the internal loop.
In the FS- and FV-type diagrams the  terms involving $|c_{L,R}|^2$
or $|g_{L,R}|^2$
are exemplified by the diagram in Fig.\ \ref{fig:amu-chirality-flip} (\emph{left}),
where only the right-handed (or similarly only the left-handed) muon couples to the loop and the chirality
is flipped at an external muon line. 
In the diagrammatic computation
this corresponds to an application of the Dirac equation $\slashed{p}
u(p)=m_\mu u(p)$ on the external muon spinor, and the resulting
contributions to $\amu$ are suppressed by $m_\mu^2/M^2$, where $M$ is
the potentially heavy internal scale.
In terms of the discussion of Sec.~\ref{sec:ChiralityFlips} and
Eqs.\ \eqref{mmugeneric} and \eqref{amugeneric}, the factors in 
square brackets  simply amount to
$[\ldots]\to m_\mu$, corresponding to no chiral enhancement.

The other kind of terms of the FS- and FV-type results
are illustrated by the Feynman
diagram of Fig.\ \ref{fig:amu-chirality-flip} (\emph{right}), where
one of the vertices connects to  a right-handed muon, the other vertex
to a left-handed muon, and where the chirality is flipped via the loop. 
In the computation
of the diagram, the fermion mass $m_F$ arises via the propagator
of the fermion $F$ in the loop, leading to the relative, potential
enhancement factor $m_F/m_\mu$.
In terms of Eqs.\ \eqref{mmugeneric} and \eqref{amugeneric}, the factors 
in the square brackets $[\ldots]$ correspond to the combinations
$c_L^*c_R m_F$ and $g_L^*g_R m_F$.
In order to enable these potentially chirality-flip enhanced
contributions, there must be
mixing, i.e.\ the particles in the 
loop must be linear combinations of states with different gauge
quantum numbers. This mixing affects the potential enhancement since
it requires the explicit or implicit appearance of a factor of an EWSB \vev
as illustrated by the \vev insertion in
Fig.\ \ref{fig:chirflipdiagrams} (right).
For example the fermion mass $m_F$ may arise from an EWSB
\vev, or the vertex factors $c_{L,R}$, $g_{L,R}$ might involve mixing
matrix elements which effectively again depend on some EWSB \vev.
The role of the two kinds of terms in the FSV-type diagrams is
analogous but reversed compared to the cases of the FS- and FV-type diagrams.

The numerical behaviour of all these terms and their appearance in
concrete models will be discussed at length in subsequent sections.
As a first illustration, already the EWSM one-loop diagrams shown in Fig.~\ref{fig:SMEW1}
are of all the above types. The $W$-boson diagram only involves
left-handed muon couplings, whereas the $Z$- and Higgs-boson diagrams
lead to terms of both chirality combinations. However, in these
diagrams the internal fermion is a muon, hence the factor
$m_F/m_\mu=1$ is no enhancement.  In contrast, in many BSM scenarios,
the terms with a chirality flip on the internal line are potentially
enhanced, see also Tab.~\ref{tab:estimates}.

The generic formulas in Eq.~(\ref{eq:Damu-oneloop-results}) depend on
loop functions of the appearing masses. We first provide these loop
functions in approximations for important limits. Very often, BSM
particles are heavy and the muon mass can be neglected. In the
corresponding limits $x_\mu,z_\mu\to0$, the loop functions for FS-type
diagrams depend only on the single ratio $y\equiv m_F^2/m_S^2$ and are given by
\footnote{\label{footnoteloopfunctions}
In the literature the loop functions are often given for the
individual diagrams of Fig.~\ref{fig:amu-one-loop} instead of the
combined contributions in Eq.~\eqref{eq:Damu-oneloop-results}. 
It is straightforward to translate between these results. For example
the FFS, FSS, FFV and FVV functions defined in
Ref.~\cite{Jegerlehner:2009ry,Fargnoli:2013zia,Athron:2021iuf}
correspond to setting either $Q_S=0$, $Q_F=0$, or $Q_V=0$ and can be
recovered as
\begin{align}
  F^C_2(y)&=F(y)=\frac{3}{2}\F^\text{FS}(0,y,0),
  &
  F^N_2(y)&=C(y)=-3\F^\text{FS}(0,y,-1),\\
F^C_4(y)&=\frac{2}{3}C(y)+\frac{y}{3}F(y)=\frac{1}{2}\F^\text{FV}(0,y,0),
&
F^N_4(y)&=2K(y)+\frac{y}{6}C(y)=-\frac{1}{2}\F^\text{FV}(0,y,-1).
\end{align}
Further useful loop functions are obtained by taking differences,
\begin{align}
  F_a(x,y)&=-\frac{F_2^C(x)-F_2^C(y)}{3 (x-y)},&
  F_b(x,y)&=-\frac{F_2^N(x)-F_2^N(y)}{6 (x-y)}.
\end{align}
}
\begin{subequations}
	\begin{align}
		\F^\text{FS}(0,y;Q_S) &= -\frac{3-4y+y^2+2\ln(y)}{(1-y)^3} - \frac{2Q_S(1-y+\ln(y))}{(1-y)^2}, \\
		\G^\text{FS}(0,y;Q_S) &= \frac{2+3y-6y^2+y^3+6y\ln(y)}{6(1-y)^4} + \frac{Q_S(1-y^2+2y\ln(y))}{2(1-y)^3},
	\end{align}
\end{subequations}
for the FV-type diagrams \ref{fig:amu-FV} the loop functions depend on $y\equiv m_F^2/m_V^2$ and are
given by\footnote{In t'Hooft Feynman gauge ($\xi=1$), the loop functions 
	obtained from Fig.~\ref{fig:amu-FV} are instead given in terms of the definitions in Ref.~\cite{Athron:2021iuf} by  
	$\F^\text{FV}_{\xi=1}(0,y;Q_V)=\frac{4}{3}(1+Q_V)C(y) + 3Q_V K(y)$
	and $\G^\text{FV}_{\xi=1}(0,y;Q_V)=-\frac{(1+Q_V)}{3} M(y)-\frac{Q_V}{6}J(y)$.
	The physical result Eq.~\eqref{eq:amu-FV-LF} is recovered once the Goldstone boson contributions are added.
	
}
\begin{subequations}\label{eq:amu-FV-LF}
	\begin{align}
		\F^\text{FV}(0,y;Q_V) &= \frac{4-3y-y^3+6y\ln(y)}{(1-y)^3} + \frac{2Q_V(4-5y+y^2+3y\ln(y))}{(1-y)^2},\\
		\G^\text{FV}(0,y;Q_V) &= -\frac{8-38y+39y^2-14y^3+5y^4-18y^2\ln(y)}{6(1-y)^4}
		-\frac{3Q_V(2-7y+6y^2-y^3-2y^2\ln(y))}{2(1-y)^3},
	\end{align}
\end{subequations}
and for the FSV-type diagrams \ref{fig:amu-FSV}  the loop functions become
\begin{subequations}
	\begin{align}
		\F^\text{FSV}(0,y,z) &= \frac{y^2(1-z)-2(y-z)}{2(1-y)(1-z)(y-z)} - \frac{(1-z)y^3\ln(y)}{2(1-y)^2(y-z)^2} + \frac{[2(y-z)+y(1-z)]z^2\ln(z)}{2(1-z)^2(y-z)^2},\\
		\G^\text{FSV}(0,y,z) &= \frac{2yz(1-y)-y(3+y)(y-z)}{6(1-y)^2(y-z)^2} + \frac{[y(1+z)(3z-y)-z^2(3+y^2)]y\ln(y)}{3(1-y)^3(y-z)^3}
		+\frac{z^3\ln(z)}{3(1-z)(y-z)^3}.
	\end{align}
\end{subequations}

Generally, the loop functions take values of the order
one, and it is of
interest to collect further limiting cases. Equally heavy internal
masses correspond to $y=1$, 
very heavy boson mass $m_{S,V}\gg m_F\gg m_\mu$ corresponds to $y\to0$
and very heavy fermion mass corresponds to $y\to\infty$. In case of
FSV, equal boson masses $m_S=m_V$ corresponds to $z=1$. In these
limits, the loop functions become
\begin{subequations}\label{LoopfunctionFSlimits}
	\begin{alignat}{5}
		\F^\text{FS}(0,y;Q_S) &\overset{y\to 1}{=} \frac{2}{3} + Q_S && \G^\text{FS}(0,y;Q_S) &&\overset{y\to1}{=} \frac{1+2Q_S}{12} \\
		\F^\text{FS}(0,y;Q_S) &\overset{y\to 0}{\simeq} -2(1+Q_S)\ln(y) - 3 - 2Q_S \qquad && \G^\text{FS}(0,y;Q_S) &&\overset{y\to0}{=} \frac{2+3Q_S}{6}\\
		\F^\text{FS}(0,y;Q_S) &\overset{y\to\infty}{=} \frac{1+2Q_S}{y} && \G^\text{FS}(0,y;Q_S) &&\overset{y\to\infty}{=} \frac{1+3Q_S}{6y},
		\intertext{and}
		\F^\text{FV}(0,y;Q_V) &\overset{y\to1}{=} 2 + 5Q_V \hspace{1.5cm} && \G^\text{FV}(0,y;Q_V) &&\overset{y\to1}{=} -\frac{13}{12}-\frac{5Q_V}{2} \\
		\F^\text{FV}(0,y;Q_V) &\overset{y\to0}{=} 4+8Q_V \hspace{1.5cm} && \G^\text{FV}(0,y;Q_V) &&\overset{y\to0}{=} -\frac{4+9Q_V}{3} \hspace{1.5cm} \\
		\F^\text{FV}(0,y;Q_V) &\overset{y\to\infty}{=} 1+2Q_V, &&\G^\text{FV}(0,y;Q_V) &&\overset{y\to\infty}{=} -\frac{5+9Q_V}{6},
		\intertext{and}
		\F^\text{FSV}(0,y,1)&\overset{y\to 1}{=} -\frac{1}{2} && \G^\text{FSV}(0,y,1)&&\overset{y\to 1}{=} \frac{1}{12} \\
		\F^\text{FSV}(0,y,1) &\overset{y\to 0}{=} - \frac{1}{2} &&  \G^\text{FSV}(0,y,1) &&\overset{y\to 0}{=} \frac{1}{3} \\
		\F^\text{FSV}(0,y,1) &\overset{y\to \infty}{=} - \frac{1}{2}, && \G^\text{FSV}(0,y,1) &&\overset{y\to \infty}{=} \frac{1}{6y}.
	\end{alignat}
\end{subequations}
It is worth noting that none of the above functions are singular in the limits $y,z\to 1,\infty$ (or $z\to y$),
and only $\F^\text{FS}$ develops a logarithmic enhancement for
$y\to 0$ in the term proportional to the fermion charge
$Q_F=-1-Q_S$. This enhancement is present e.g.\ in the EWSM
Higgs-boson one-loop contribution shown in Fig.~\ref{fig:SMEW1}, where
the internal muon is much lighter than the internal Higgs boson,
however due to the small Yukawa couplings that diagram is nevertheless
negligible.

\paragraph{General expressions of the loop functions}
Finally, we list the general results for the loop functions in Eq.~\eqref{eq:Damu-oneloop-results} valid for arbitrary masses of the muon and internal particles.
We make use of the following abbreviation
\begin{align}
	\tilde\Lambda(x,y,z) \equiv \frac{\sqrt{\lambda(x,y,z)}}{x} \ln(\frac{\sqrt{\lambda(x,y,z)} - x + y + z}{2\sqrt{yz}})
\end{align}
where $\lambda(x,y,z)=(x-y-z)^2-4yz$ denotes the the K\"all\'en function.
For the FS-diagrams Fig.~\ref{fig:amu-FS} the loop functions are given by
\begin{subequations}\label{FSloopfunctionfull}
	\begin{align}
		\begin{split}
			\F^\text{FS}(x,y;Q) &= \frac{2}{x} + \frac{\ln(y)}{x^2}\Big[1+x-y+Qx\Big] + \frac{2\tilde\Lambda(x,y,1)}{x\lambda(x,y,1)} \Big[1-2y+(x-y)^2  + Q x (1+x-y) \Big]
		\end{split} \\[.4cm]
		\begin{split}
			\G^\text{FS}(x,y;Q) &= - \frac{1+2Q}{2x} - \frac{1-y}{x^2} - \frac{\ln(y)}{2x^3} \Big[(1 - y)^2 - y x + Q x (1-y)\Big] \\
			& - \frac{\tilde\Lambda(x,y,1)}{x^2\lambda(x,y,1)} \Big[(1-y)^3 - x (1-y)(1+2y) - y x^2 + Q x\big\{(1-y)^2-x(1+y)\big\} \Big],
		\end{split}
	\end{align}
\end{subequations}
for the FV Fig.~\ref{fig:amu-FV} diagrams by
\begin{subequations}\label{FVloopfunctionfull}
	\begin{align}
		\begin{split}
			\F^\text{FV}(x,y;Q) = &\; 4(1+Q) - \frac{2(2+y)}{x} - \frac{\ln(y)}{x^2} \Big[2-(x+y)-(x-y)^2 + Q x(2-x+y)\Big] \\
			& - \frac{2\tilde\Lambda(x,y,1)}{x\lambda(x,y,1)} \Big[2-3(x+y) - (x-y)^3
			+ Q x \Big\{ 2(1-x) - (x + y) - (x-y)^2\Big\}\Big]
		\end{split} \\[.4cm]
		\begin{split}
			\G^\text{FV}(x,y;Q)= &- \frac{Q(2+x+y)}{x} + \frac{y(2-3x+2y)-(2-x)^2}{2x^2} - \frac{\ln(y)}{2x^3} \Big[2-3(x+y) + y\big(2x+(x-y)^2\big) \\
			& + Q x \Big\{2-3x - y(1-x+y)\Big\} \Big] - \frac{\tilde\Lambda(x,y,1)}{x^2\lambda(x,y,1)} \Big[2 - 5x(1-x) - 5y(1-y) \\
			&- (x-y)^2\big(2-y(1+x-y)\big) + Q x \Big\{ 2(1-x) - 3 (x+y) + 3 x^2 + y \big(x +(x-y)^2\big) \Big\}\Big],
		\end{split}
	\end{align}
\end{subequations}
and for the FSV diagrams Fig.~\ref{fig:amu-FSV} by
\begin{subequations}\label{FSVloopfunctionfull}
	\begin{align}
		\begin{split}
			\F^\text{FSV}(x,y,z) &= \frac{x-y+z^2}{3x(z-1)} - \frac{2+2x+y}{6x} - \frac{(1-z)\ln(y)}{12 x^2}\Big[2-3(x+y)+2z\Big] 
			+ \frac{\ln(z)}{12x^2(1-z)^2} \Big[(x-y)^3(1+z)\\
			&\hspace{-1cm}-3z(x+y)(2x-2y+3z-z^2)+2z^3(2-z)\Big] + \frac{\tilde\Lambda(1,\frac{1}{x},\frac{y}{x})}{6x(1-z)^2} \Big[(x-y)^2(1+z) + (x+y)(1-5z)+ 2(2z-1)\Big] \\
			&\hspace{-1cm}- \frac{\tilde\Lambda(1,\frac{z}{x},\frac{y}{x})}{6x(1-z)^2} \Big[(x-y)^2(1+z) + z(x+y)(z-5) + 2z^2(2-z)\Big]
		\end{split} \\[.4cm]
		\begin{split}
			\G^\text{FSV}(x,y,z) &= - \frac{2+x-4y+2z}{6x^2} - \frac{\ln(y)}{6x^3}\Big[1-3y(1+x-y+z)+z(1+z)\Big] - \frac{\ln(z)}{6x^3(1-z)} \Big[(x-y)^3\\
			&\hspace{-1cm} -3yz(x-y+z)+z^3\Big] - \frac{\tilde\Lambda(1,\frac{1}{x},\frac{y}{x})}{3x^2(1-z)} \Big[(x-y)^2 + 1+x-2y\Big] + \frac{\tilde\Lambda(1,\frac{z}{x},\frac{y}{x})}{3x^2(1-z)} \Big[(x-y)^2 + z(x-2y+z)\Big].
		\end{split}
	\end{align}
\end{subequations}

\newcommand{\FieldP}{$({\bf 1},{\bf 1}, 1)$}
\newcommand{\FieldN}{$({\bf 1},{\bf 1}, 0)$}
\newcommand{\FieldC}{$({\bf 1},{\bf 1},-1)$}
\newcommand{\FieldD}{$({\bf 1},{\bf 2},-1/2)$}
\newcommand{\FieldDp}{$({\bf 1},{\bf 2},1/2)$}
\newcommand{\FieldY}{$({\bf 1},{\bf 2},-3/2)$}
\newcommand{\FieldA}{$({\bf 1},{\bf 3}, 0)$}
\newcommand{\FieldT}{$({\bf 1},{\bf 3},-1)$}
\newcommand{\FieldDalt}{$({\bf 1},{\bf 2},-3/2)$}

\newcommand{\FieldTp}{$({\bf 1},{\bf 3},1)$}
\newcommand{\FieldTpp}{$({\bf 1},{\bf 3},-2)$}
\newcommand{\FieldDpp}{$({\bf 1},{\bf 2},3/2)$}

\newcommand{\OneFieldP}{\FieldP\@\xspace}
\newcommand{\OneFieldN}{\FieldN\@\xspace}
\newcommand{\OneFieldC}{\FieldC\@\xspace}
\newcommand{\OneFieldD}{\FieldD\@\xspace}
\newcommand{\OneFieldDp}{\FieldDp\@\xspace}
\newcommand{\OneFieldY}{\FieldY\@\xspace}
\newcommand{\OneFieldA}{\FieldA\@\xspace}
\newcommand{\OneFieldT}{\FieldT\@\xspace}

\newcommand{\CombinedFieldsSNFC}{\FieldN$_0$ &-- \FieldC$_{1/2}$\@\xspace}
\newcommand{\CombinedFieldsSCFN}{\FieldC$_0$ &-- \FieldN$_{1/2}$\@\xspace}
\newcommand{\CombinedFieldsSDFN}{\FieldD$_0$ &-- \FieldN$_{1/2}$\@\xspace}
\newcommand{\CombinedFieldsSNFD}{\FieldN$_{0}$ &-- \FieldD$_{1/2}$\@\xspace}
\newcommand{\CombinedFieldsSDFC}{\FieldD$_{0}$ &-- \FieldC$_{1/2}$\@\xspace} 
\newcommand{\CombinedFieldsSDpFC}{\FieldD$_{0}$ &-- \FieldC$_{1/2}$\@\xspace}
\newcommand{\CombinedFieldsSCFD}{\FieldC$_0$ &-- \FieldD$_{1/2}$\@\xspace} 
\newcommand{\CombinedFieldsSCFDp}{\FieldC$_0$ &-- \FieldDp$_{1/2}$\@\xspace}
\newcommand{\CombinedFieldsSDFD}{\FieldD$_0$ &-- \FieldD$_{1/2}$\@\xspace}
\newcommand{\CombinedFieldsSDFA}{\FieldD$_{0}$ &-- \FieldA$_{1/2}$\@\xspace}
\newcommand{\CombinedFieldsSDpFA}{\FieldDp$_{0}$ &-- \FieldA$_{1/2}$\@\xspace}
\newcommand{\CombinedFieldsSDFT}{\FieldD$_{0}$ &-- \FieldT$_{1/2}$\@\xspace} 
\newcommand{\CombinedFieldsSDpFT}{\FieldDp$_{0}$ &-- \FieldT$_{1/2}$\@\xspace}
\newcommand{\CombinedFieldsSAFD}{\FieldA$_{0}$ &-- \FieldD$_{1/2}$\@\xspace} 
\newcommand{\CombinedFieldsSAFT}{\FieldA$_{0}$ &-- \FieldT$_{1/2}$\@\xspace}
\newcommand{\CombinedFieldsSTFD}{\FieldT$_{0}$ &-- \FieldD$_{1/2}$\@\xspace}
\newcommand{\CombinedFieldsSTpFD}{\FieldTp$_{0}$ &-- \FieldD$_{1/2}$\@\xspace}
\newcommand{\CombinedFieldsSTFA}{\FieldT$_{0}$ &-- \FieldA$_{1/2}$\@\xspace}
\newcommand{\CombinedFieldsFCVN}{\FieldC$_{1/2}$ &-- \FieldN$_{1}$\@\xspace}
\newcommand{\CombinedFieldsFDVN}{\FieldD$_{1/2}$ &-- \FieldN$_{1}$\@\xspace}
\newcommand{\CombinedFieldsFDVA}{\FieldD$_{1/2}$ &-- \FieldA$_{1}$\@\xspace}
\newcommand{\CombinedFieldsFNVP}{\FieldN$_{1/2}$ &-- \FieldP$_{1}$\@\xspace}
\newcommand{\CombinedFieldsFDVC}{\FieldD$_{1/2}$ &-- \FieldC$_{1}$\@\xspace}
\newcommand{\CombinedFieldsFTVA}{\FieldT$_{1/2}$ &-- \FieldA$_{1}$\@\xspace}

\newcommand{\CombinedFieldsFDFN}{\FieldD$_{1/2}$ &-- \FieldN$_{1/2}$\@\xspace}
\newcommand{\CombinedFieldsFDFC}{\FieldD$_{1/2}$ &-- \FieldC$_{1/2}$\@\xspace}
\newcommand{\CombinedFieldsFDFT}{\FieldD$_{1/2}$ &-- \FieldT$_{1/2}$\@\xspace}
\newcommand{\CombinedFieldsFDFA}{\FieldD$_{1/2}$ &-- \FieldA$_{1/2}$\@\xspace}
\newcommand{\CombinedFieldsFDaltFC}{\FieldDalt$_{1/2}$ &-- \FieldC$_{1/2}$\@\xspace}
\newcommand{\CombinedFieldsFDaltFT}{\FieldDalt$_{1/2}$ &-- \FieldT$_{1/2}$\@\xspace}

\newcommand{\CombinedFieldsSTFDpp}{\FieldT$_0$ &-- \FieldDpp$_{1/2}$\@\xspace}
\newcommand{\CombinedFieldsSTFDalt}{\FieldT$_0$ &-- \FieldDalt$_{1/2}$\@\xspace}
\newcommand{\CombinedFieldsSDppFT}{\FieldDpp$_0$ &-- \FieldT$_{1/2}$\@\xspace}
\newcommand{\CombinedFieldsSDaltFTp}{\FieldDalt$_0$ &-- \FieldTp$_{1/2}$\@\xspace}
\newcommand{\CombinedFieldsSYFDp}{\FieldY$_0$ &-- \FieldDp$_{1/2}$\@\xspace}
\newcommand{\CombinedFieldsSDpFY}{\FieldDp$_0$ &-- \FieldY$_{1/2}$\@\xspace}
\newcommand{\CombinedFieldsSTppFTp}{\FieldTpp$_0$ &-- \FieldTp$_{1/2}$\@\xspace}
\newcommand{\CombinedFieldsSTpFTpp}{\FieldTp$_0$ &-- \FieldTpp$_{1/2}$\@\xspace}

\newcommand{\TwoFieldsSNFP}{\FieldN$_0$ -- \FieldP$_{1/2}$\@\xspace}
\newcommand{\TwoFieldsFNVP}{\FieldN$_{1/2}$ -- \FieldP$_1$\@\xspace}

\newcommand{\TwoFieldsSNFC}{\FieldN$_0$ -- \FieldC$_{1/2}$\@\xspace}
\newcommand{\TwoFieldsFNVC}{\FieldN$_{1/2}$ -- \FieldC$_{1}$\@\xspace}

\newcommand{\TwoFieldsSNFD}{\FieldN$_{0}$ -- \FieldD$_{1/2}$\@\xspace}
\newcommand{\TwoFieldsSNFDp}{\FieldN$_{0}$ -- \FieldDp$_{1/2}$\@\xspace}
\newcommand{\TwoFieldsFNFD}{\FieldN$_{1/2}$ -- \FieldD$_{1/2}$\@\xspace}
\newcommand{\TwoFieldsFNVD}{\FieldN$_{1/2}$ -- \FieldD$_{1}$\@\xspace}

\newcommand{\TwoFieldsSNFA}{\FieldN$_{0}$ -- \FieldA$_{1/2}$\@\xspace}
\newcommand{\TwoFieldsFNFA}{\FieldN$_{1/2}$ -- \FieldA$_{1/2}$\@\xspace}
\newcommand{\TwoFieldsFNVA}{\FieldN$_{1/2}$ -- \FieldA$_{1}$\@\xspace}

\newcommand{\TwoFieldsSCFN}{\FieldC$_0$ -- \FieldN$_{1/2}$\@\xspace}
\newcommand{\TwoFieldsFCVN}{\FieldC$_{1/2}$ -- \FieldN$_{1}$\@\xspace}

\newcommand{\TwoFieldsSCFD}{\FieldC$_{0}$ -- \FieldD$_{1/2}$\@\xspace}
\newcommand{\TwoFieldsFCFD}{\FieldC$_{1/2}$ -- \FieldD$_{1/2}$\@\xspace}
\newcommand{\TwoFieldsFCVD}{\FieldC$_{1/2}$ -- \FieldD$_{1}$\@\xspace}

\newcommand{\TwoFieldsSYFC}{\FieldY$_{0}$ -- \FieldC$_{1/2}$\@\xspace}
\newcommand{\TwoFieldsFYFC}{\FieldY$_{1/2}$ -- \FieldC$_{1/2}$\@\xspace}
\newcommand{\TwoFieldsFYVC}{\FieldY$_{1/2}$ -- \FieldC$_{1}$\@\xspace}

\newcommand{\TwoFieldsSYFT}{\FieldY$_{0}$ -- \FieldT$_{1/2}$\@\xspace}
\newcommand{\TwoFieldsFYFT}{\FieldY$_{1/2}$ -- \FieldT$_{1/2}$\@\xspace}
\newcommand{\TwoFieldsFYVT}{\FieldY$_{1/2}$ -- \FieldT$_{1}$\@\xspace}

\newcommand{\TwoFieldsSDFN}{\FieldD$_0$ -- \FieldN$_{1/2}$\@\xspace}
\newcommand{\TwoFieldsFDFN}{\FieldD$_{1/2}$ -- \FieldN$_{1/2}$\@\xspace}
\newcommand{\TwoFieldsFDVN}{\FieldD$_{1/2}$ -- \FieldN$_1$\@\xspace}

\newcommand{\TwoFieldsSDFC}{\FieldD$_{0}$ -- \FieldC$_{1/2}$\@\xspace}
\newcommand{\TwoFieldsFDFC}{\FieldD$_{1/2}$ -- \FieldC$_{1/2}$\@\xspace}
\newcommand{\TwoFieldsFDVC}{\FieldD$_{1/2}$ -- \FieldC$_{1}$\@\xspace}

\newcommand{\TwoFieldsSDFD}{\FieldD$_{0}$ -- \FieldD$_{1/2}$\@\xspace}
\newcommand{\TwoFieldsFDFD}{\FieldD$_{1/2}$ -- \FieldD$_{1/2}$\@\xspace}
\newcommand{\TwoFieldsFDVD}{\FieldD$_{1/2}$ -- \FieldD$_{1}$\@\xspace}

\newcommand{\TwoFieldsSDFA}{\FieldD$_{0}$ -- \FieldA$_{1/2}$\@\xspace}
\newcommand{\TwoFieldsFDFA}{\FieldD$_{1/2}$ -- \FieldA$_{1/2}$\@\xspace}
\newcommand{\TwoFieldsFDVA}{\FieldD$_{1/2}$ -- \FieldA$_{1}$\@\xspace}
    
\newcommand{\TwoFieldsSDFT}{\FieldD$_{0}$ -- \FieldT$_{1/2}$\@\xspace}
\newcommand{\TwoFieldsFDFT}{\FieldD$_{1/2}$ -- \FieldT$_{1/2}$\@\xspace}
\newcommand{\TwoFieldsFDVT}{\FieldD$_{1/2}$ -- \FieldT$_{1}$\@\xspace}

\newcommand{\TwoFieldsSAFD}{\FieldA$_{0}$ -- \FieldD$_{1/2}$\@\xspace}
\newcommand{\TwoFieldsFAFD}{\FieldA$_{1/2}$ -- \FieldD$_{1/2}$\@\xspace}
\newcommand{\TwoFieldsFAVD}{\FieldA$_{1/2}$ -- \FieldD$_{1}$\@\xspace}

\newcommand{\TwoFieldsSAFT}{\FieldA$_{0}$ -- \FieldT$_{1/2}$\@\xspace}
\newcommand{\TwoFieldsFAFT}{\FieldA$_{1/2}$ -- \FieldT$_{1/2}$\@\xspace}
\newcommand{\TwoFieldsFAVT}{\FieldA$_{1/2}$ -- \FieldT$_{1}$\@\xspace}

\newcommand{\TwoFieldsSTFD}{\FieldT$_{0}$ -- \FieldD$_{1/2}$\@\xspace}
\newcommand{\TwoFieldsFTFD}{\FieldT$_{1/2}$ -- \FieldD$_{1/2}$\@\xspace}
\newcommand{\TwoFieldsFTVD}{\FieldT$_{1/2}$ -- \FieldD$_{1}$\@\xspace}

\newcommand{\TwoFieldsSTFA}{\FieldT$_{0}$ -- \FieldA$_{1/2}$\@\xspace}
\newcommand{\TwoFieldsFTFA}{\FieldT$_{1/2}$ -- \FieldA$_{1/2}$\@\xspace}
\newcommand{\TwoFieldsFTVA}{\FieldT$_{1/2}$ -- \FieldA$_{1}$\@\xspace}

\subsection{Contributions from minimal extensions of the Standard Model}
\label{sec:MinimalBSM}

The one-loop diagrams discussed in the previous section can provide
new contributions to $\amu$ from physics beyond the standard model when BSM
particles enter these loops.  There are many possibilities and some
popular BSM theories have many particles that can contribute. Here
however we want to consider minimal contributions from a limited
number of new fields. The key difference to the generic diagrams
considered above is that here each new field is required to have
specific quantum numbers under the SM gauge group and gauge invariant
interactions. In contrast, the generic Lagrangian in
Eq.~(\ref{eq:L_SFV_muon}) was not 
required to be gauge invariant by itself and the appearing
mass-eigenstate fields could be mixtures of fields with different
gauge quantum numbers.

The scenarios considered here 
represent minimal BSM extensions that give non-trivial effects to
$\amu$, which can be relevant on their own or which can be part of
larger theories when only a few particles 
are light and constitute building blocks for contributions from
theories with many new particles. Their understanding provides a
useful basis for the general discussion of BSM contributions to $\amu$.

We therefore systematically categorise the simplest extensions of the
standard model that give new physics contributions to the anomalous
magnetic moment of the muon. 
Drawing on similar systematic investigations in the literature
\cite{Freitas:2014pua,Queiroz:2014zfa,Biggio:2014ela,Biggio:2016wyy,Kowalska:2017iqv,Calibbi:2018rzv,Athron:2021iuf}
we focus here on one-field and two-field extensions, defined as BSM
scenarios where the SM is extended by exactly one or two new fields
that have renormalisable and gauge invariant interactions involving
the muon.

In both the one-field and two-field cases only
specific gauge quantum numbers of such fields allow couplings to the
muon and thus one-loop contributions to $\amu$. For example, if a
single scalar field $\phi$ is added to the SM, it must be paired with
an internal SM fermion to get a contribution via one of the diagrams
in Fig.~\ref{fig:amu-one-loop}. Hence at least one of the field
combinations $\phi\, \mu_{L}\, f$ and $\phi\, \mu_R\, f$ must allow a
gauge invariant operator, where $f$ can be any of the SM fermion
multiplets $f\in\{l_{Li},e_{Ri},q_{Li},u_{Ri},d_{Ri}\}$, possibly involving hermitian
conjugation. Using such arguments leads to
Tab.~\ref{tab:onefieldmodels}. This table contains all possibilities
for one-field extensions with possible one-loop contributions to
$\amu$. Similar considerations gives Tab.~\ref{tab:twofieldmodels}
containing two-field extensions where two new BSM states of different spins enter the
one-loop contributions to muon $g-2$.

To be specific on the considered class of models, we use the following
requirements: The added fermions are 4-component Dirac spinors, matching
conventions in Ref.\ \cite{Freitas:2014pua}.
For the  one-field models, each proposed new field must
have renormalisable and gauge invariant couplings to the muon,
together with some other SM field. This excludes models such as a
scalar singlet from the table, where the singlet could contribute to
$\amu$ via mixing with the SM Higgs doublet. For the two-field models,
we require that both new fields together have direct couplings to the
muon and can form a loop without any further SM fields. This excludes
models with two different vector-like leptons or so-called bridge
models \cite{Guedes:2022cfy} with one vector-like lepton and a scalar
singlet. Many models that go beyond the tables will be discussed in
later sections.

\begin{table}
\begin{center}
\begin{tabular}{|c|c|c|c|c|c|} \hline
Model & Spin & $\GSM$ & Sign($\Delta
a^\text{BSM}_\mu$) & Muon interactions &  Chiral enhancement?   \\ \hline
 1 &     0 &                 $({\bf 1},{\bf 1},1)$ & \cellcolor{negative} - & L &  \cellcolor{negative} No \\
 2 &     0 &                 $({\bf 1},{\bf 1},2)$ & \cellcolor{negative} - & R & \cellcolor{negative} No \\
 3 &     0 &          $({\bf 1},{\bf 2},-1/2)$ &\cellcolor{bothsigns} $\pm$   & L--R &  \cellcolor{bothsigns} only with FV  \\ 
 4 &     0 &                $({\bf 1},{\bf 3},-1)$ & \cellcolor{negative} - & L & \cellcolor{negative} No   \\
 5 &     0 &   $({\bf \overline{3}},{\bf 1}, 1/3)$ & \cellcolor{bothsigns} $\pm$   & L and R &   \cellcolor{positive} Yes    \\
 6 &     0 &   $({\bf \overline{3}},{\bf 1}, 4/3)$ & \cellcolor{negative} - & R & \cellcolor{negative} No   \\
 7 &     0 &   $({\bf \overline{3}},{\bf 3}, 1/3)$ & \cellcolor{positive} +  & L & \cellcolor{negative} No \\
 8 &     0 &              $({\bf 3},{\bf 2}, 7/6)$ & \cellcolor{bothsigns} $\pm$ & L and R & \cellcolor{positive}Yes    \\
 9 &     0 &              $({\bf 3},{\bf 2}, 1/6)$ & \cellcolor{negative} -  & L &\cellcolor{negative} No \\
10 & $1/2$ &                 $({\bf 1},{\bf 1},0)$ & \cellcolor{negative} -  & L &  \cellcolor{negative} No  \\
11 & $1/2$ &                $({\bf 1},{\bf 1},-1)$ & \cellcolor{bothsigns} $\pm$  & L & \cellcolor{negative} No  \\
12 & $1/2$ &          $({\bf 1},{\bf 2},-1/2)$ & \cellcolor{positive} +   & R &  \cellcolor{negative} No   \\
13 & $1/2$ &          $({\bf 1},{\bf 2},-3/2)$ & \cellcolor{negative} -  & R &  \cellcolor{negative} No   \\
14 & $1/2$ &                 $({\bf 1},{\bf 3},0)$ & \cellcolor{negative} - & L & \cellcolor{negative} No \\
15 & $1/2$ &                $({\bf 1},{\bf 3},-1)$ & \cellcolor{negative} -  & L &\cellcolor{negative} No \\
16 &   $1$ &                 $({\bf 1},{\bf 1},0)$ & \cellcolor{bothsigns} $\pm$ & L and R &\cellcolor{bothsigns} only with FV  \\
17 &   $1$ &              $({\bf 1},{\bf 2},-3/2)$ & \cellcolor{positive} + & L--R & \cellcolor{bothsigns} only with FV  \\ 
18 &   $1$ &                $({\bf 1},{\bf 3}, 0)$ & \cellcolor{positive} + &L &  \cellcolor{negative} No \\
19 &   $1$ & $({\bf \overline{3}}, {\bf 1}, -2/3)$ & \cellcolor{bothsigns} $\pm$ & L and R & \cellcolor{positive} Yes \\
20 &   $1$ & $({\bf \overline{3}}, {\bf 1}, -5/3)$ & \cellcolor{positive}  +  & R & \cellcolor{negative} No \\
21 &   $1$ & $({\bf \overline{3}}, {\bf 2}, 5/6)$  & \cellcolor{bothsigns} $\pm$ & L and R &\cellcolor{positive} Yes\\
22 &   $1$ & $({\bf \overline{3}}, {\bf 2}, -1/6)$ & \cellcolor{negative} - & L  &\cellcolor{negative} No \\
23 &   $1$ & $({\bf \overline{3}}, {\bf 3}, -2/3)$ & \cellcolor{positive} +   & L  &\cellcolor{negative} No  \\
\hline
\end{tabular}
\caption{Gauge invariant single field extensions with one-loop
  contributions to the anomalous magnetic moment of the muon.  Column
  two  specifies the spin of the state where we consider spin 0, 1/2
  and 1.  Column three specifies the representations under SM gauge
  groups $\SUc$ and $\SUL$ and the hypercharge.  The fourth
  column shows the sign of the one-loop contribution to $a_\mu$, while
  the fifth column shows which kind of gauge invariant and renormalisable couplings to
  the muon are possible: ``L'', ``R'' denote couplings to the
  left-handed or right-handed muon (and some other unspecified SM field), respectively, and ``L--R'' denotes
  a Lagrangian term involving a product of left- and right-handed
  muon.  The sixth column states whether there is a chirality flipping enhancement in the one-loop diagrams allowing large new physics contributions to $a_\mu$.  Special cases where a chirality enhancement at one-loop is only possible by introducing lepton flavour violating couplings are marked as ``only with FV''.  Note these models were originally grouped together in \cite{Athron:2021iuf} which itself drew upon previous systematic investigations and classifications in Ref.\ \cite{Freitas:2014pua,Queiroz:2014zfa,Biggio:2014ela,Biggio:2016wyy}. \label{tab:onefieldmodels} 
}
\end{center}
\end{table}

\begin{table}
	\begin{center}
		\setlength\tabcolsep{1.5pt}
		\begin{tabular}{|c|rl|c|c|c|c|} \hline
			Model &	\multicolumn{2}{|c|}{{\scriptsize $(\GSM)_\textrm{spin}$}}  &
			sign$(\Delta a^\text{BSM}_\mu)$  & Muon interactions &
			Chiral enhancement? \\ \hline
			24 & \CombinedFieldsSNFC & \cellcolor{positive}+ &  R   & \cellcolor{negative} No \\  
			25 &	\CombinedFieldsSCFN & \cellcolor{negative}- & R & \cellcolor{negative} No \\ 
			26 &	\CombinedFieldsSDFN & \cellcolor{negative}- & L &  \cellcolor{negative} No \\ 
			27 &	\CombinedFieldsSNFD & \cellcolor{positive}+  & L  & \cellcolor{negative} No \\
			28 &	\CombinedFieldsSDpFC & \cellcolor{positive}+ & L  & \cellcolor{negative} No \\
			29 &	\CombinedFieldsSCFDp & \cellcolor{negative}- & L & \cellcolor{negative} No \\
			30 &	\CombinedFieldsSDFD & \cellcolor{bothsigns}$\text{sign}(M_S-M_F)$ & R  &\cellcolor{negative} No \\
			31 &	\CombinedFieldsSDpFA & \cellcolor{positive}+ & L &\cellcolor{negative} No \\
			32 &	\CombinedFieldsSDpFT & \cellcolor{positive}+  & L & \cellcolor{negative} No\\
			33 &	\CombinedFieldsSAFD & \cellcolor{negative}-  & L & \cellcolor{negative} No  \\
			34 &	\CombinedFieldsSAFT & \cellcolor{positive}+  & R  &\cellcolor{negative} No \\
			35 &	\CombinedFieldsSTpFD & \cellcolor{negative}-  & L & \cellcolor{negative} No \\ 
			36 &	\CombinedFieldsSTFA & \cellcolor{negative}- & R & \cellcolor{negative} No \\ 
			37 &    \CombinedFieldsSTFDalt    &  \cellcolor{positive}+ & L  &  \cellcolor{negative} No \\
			38 &    \CombinedFieldsSDaltFTp    & \cellcolor{negative}- & L  &  \cellcolor{negative} No \\
			39 &    \CombinedFieldsSYFDp     & \cellcolor{negative}-   & R  &   \cellcolor{negative} No \\
			40 &    \CombinedFieldsSDpFY     &  \cellcolor{positive}+    & R  &   \cellcolor{negative} No \\
			41 &    \CombinedFieldsSTppFTp   &  \cellcolor{negative}-  & R  &   \cellcolor{negative} No \\
			42 &    \CombinedFieldsSTpFTpp   &  \cellcolor{positive}+           & R  &   \cellcolor{negative} No \\\hline
			43 &	\CombinedFieldsFCVN & \cellcolor{negative}-   & R &  \cellcolor{negative} No \\ 
			44 &	\CombinedFieldsFDVN & \cellcolor{negative}-   & L &  \cellcolor{negative} No \\ 
			45 &	\CombinedFieldsFDVA &
                        \cellcolor{positive}+ & L &\cellcolor{negative} No \\ 
			46 &	\CombinedFieldsFNVP & \cellcolor{positive}+ & R &\cellcolor{negative} No \\  
			47 &	\CombinedFieldsFDVC & \cellcolor{positive}+ & L &\cellcolor{negative} No   \\ 
			48 &	\CombinedFieldsFTVA &\cellcolor{negative}-   & R & \cellcolor{negative} No  \\ \hline
		\end{tabular}
		\caption{\label{tab:twofieldmodels}Gauge invariant two-field extensions with one-loop contributions to the anomalous magnetic moment of the muon where both new fields enter the loop diagrams.  Column two shows the representations under SM gauge groups $\SUc$ and $\SUL$ and the hypercharge of each state inside brackets and the spin of the states is shown as the subscript.  Column three shows the sign of the one-loop contribution to $a_\mu$, while the fourth column shows which kind of gauge invariant couplings to
			the muon are possible: ``L'', ``R'' denote couplings to the
			left-handed or right-handed muon, respectively.  The fifth and final
			column states whether there is a chirality flipping enhancement in
			the one-loop diagrams allowing large new physics contributions to
			$a_\mu$. Note most of these models were originally grouped together
			in \cite{Athron:2021iuf} which itself drew upon previous systematic
			investigations and classifications in
			Ref.\ \cite{Freitas:2014pua,Kowalska:2017iqv,Calibbi:2018rzv}.  
		}
	\end{center}
\end{table}

We first focus on the models with the table entry ``Chiral enhancement=No''
in Tabs.~\ref{tab:onefieldmodels} and \ref{tab:twofieldmodels}. This
set includes most one-field models and all two-field models with two
fields of different spin. A prominent feature of most of these
models is that the sign of the contributions to $\amu$ is fixed to be
positive or negative definite, as given 
in the table in column 4 of Tab.~\ref{tab:onefieldmodels} and
column 3 of Tab.~\ref{tab:twofieldmodels}.
In terms of the formulas of the previous subsection \ref{sec:genericoneloop}, the reason
is that these models allow either
only left-handed or only right-handed couplings to the muon, as
indicated in column 5 of Tab.~\ref{tab:onefieldmodels} and
column 4 of Tab.~\ref{tab:twofieldmodels}. In such
cases with only single chiralities, only the $|c_{L,R}|^2$-type terms of Eqs.~(\ref{eq:Damu-oneloop-results}) play a
role. These terms have essentially a fixed sign which depends on the loop
functions and the particle charges as can be seen e.g.\ in the limits
in Eqs.~(\ref{LoopfunctionFSlimits}). 
Exceptions can arise in models with larger gauge multiplets where
several Feynman diagrams with loop functions with different signs and
different mass-dependencies can contribute.

For example, the one-field model with a single scalar triplet $\phi_T$ in line 4 of
  Tab.~\ref{tab:onefieldmodels} with quantum
  numbers $(1,3,-1)$ allows only the gauge invariant  term $\phi_T^\dagger l_L
  l_L$; hence it only interacts with left-handed muons. The charges of
  the appearing physical scalars are $Q_S=0,-1,-2$, and the
  overall sign of the contribution to $\amu$ is always
  negative. Similarly, the one-field model 20 is a spin-1 leptoquark which
  couples to the right-handed muon and to right-handed up-type
  quarks. The relevant vector boson charge is $Q_V=-5/3$, and in line
  with Eq.~\eqref{LoopfunctionFSlimits} the contribution to $\amu$ is
  positive.
For all two-field models with two fields of different spin, the
situation is the same. Whatever the quantum numbers of the new states are, at most either
the coupling to the left-handed muon  or to the right-handed muon can be gauge invariant,
but not both. As a result in most cases the sign of the contribution
to $\amu$ is fixed.

Such models where there are only gauge invariant couplings to either left-
or right-handed muons cannot have muon $g-2$ contributions with a
chiral enhancement by a factor $m_F/m_\mu$, since the muon chirality
must be the same at each vertex.  However, assuming flavour conserving
couplings, there are also several examples which can couple to both but
still do not get the chiral enhancement.  For example Model 16 adds a
new gauge boson that can have couplings to both left- and right-handed
muons (as indicated in the gauge invariant muon interactions column of
Table \ref{tab:twofieldmodels}), but the only possibility for
flipping the chirality is through a muon mass insertion which therefore does not
lead to a chiral enhancement, even if it is done at an internal line.  Another
case is given by e.g.\ model 3 where we have a second scalar doublet
that can have Yukawa couplings connecting left- and right-handed muons
(written as L--R in the respective column). This Yukawa coupling does flip the chirality, but since Yukawa
coupling must appear at both vertices one has an even number of
chirality flips.  Thus there is no net change in chirality from the
one-loop diagram and no chiral enhancement at one-loop order.  

Thus for the models with only single muon chiralities as well as the
exceptional additional models without chiral enhancement discussed above the magnitude of the muon 
$g-2$ contributions are rather limited. For all models with the table entry ``Chiral
enhancement=No'', the contributions to $\amu$ are of the order
\begin{align}
  \Delta\amu \sim \frac{|c|^2(1+2Q_S)}{192\pi^2}\frac{m_\mu^2}{M^2} \sim \frac{|c|^2}{64\pi^2}\frac{m_\mu^2}{M^2}
  \approx|c|^2\left(\frac{100\,\text{GeV}}{M}\right)^2\times18\times10^{-10},
\label{nonenhancedoneloop}
\end{align}
where the FS-type result Eq.~(\ref{eq:Damu-FS}) in
the equal-mass limit of Eq.~(\ref{LoopfunctionFSlimits})  has been
used as a reference; in the second equation  $Q_S=1$ has been used as
an example to obtain a numerical estimate of the typical magnitude of
the contribution. Note that the charges can be negative, where
$Q_F+Q_S=-1$, so the sign of the contributions depends on the values of
the particle charges appearing in the loop. This illustrates how the
different sign predictions visible in the tables emerge.

This result allows a general understanding of the $\amu$
phenomenology which applies to many of the one- and two-field models.
If the BSM masses are sufficiently heavy, $M\gtrsim200$ GeV, and
couplings at most $\O(1)$, the
contributions to $\amu$ are typically smaller than
$\sim5\times10^{-10}$ and hence in agreement with the new
result $\DamuFinal$. Previously, such scenarios could not explain the
large deviation $\DamuOld$. Now, because of the automatic agreement
such scenarios are not nontrivially constrained by $\amu$.

For smaller masses, the models can potentially contribute sizeably to
$\amu$ but are also constrained by complementary limits
e.g.~from collider searches. Such complementary limits are very model
dependent. For example, leptoquarks as light as $100\ldots200$
GeV are strongly excluded, while a light neutral boson can be viable.
If complementary constraints allow light masses, the models can be
constrained by $\amu$.

Here the new result changes the
conclusions significantly. Previously, $\DamuOld$ preferred a non-zero positive
result, which excludes many models and forces models with positive
contributions into parameter regions with small masses and rather
large couplings. Now, $\DamuFinal$ is compatible with positive or
negative $\Damu$ contributions. However the contributions must be
sufficiently small, resulting in upper limits on couplings in such
models.

These remarks will be refined for specific models in later
sections. Here we highlight two special classes of models.
First, there is a  class of models with
only electrically and colour neutral BSM particles. Within the minimal models
considered here, these are in particular the one-field models with gauge
quantum number $({\bf 1},{\bf 1},0)$ such as a $Z'$, dark photon or
neutral scalar. Such new particles easily evade
collider limits and hence can be rather light. They could be part of a
dark sector and related to dark matter. Due to their lightness,
$\amu$ can place significant constraints on their parameter
space. Such light dark sector particles are the subject of
Sec.~\ref{sec:LightDarkSector}.
Similarly, 
there are certain   two-field models 
where compressed spectra provide more challenging scenarios for the
LHC, allowing  gaps in the exclusion at lower
masses. Thus special scenarios may be found where larger $\Damu$ contributions
generated by light masses are not ruled out.  Again, such models are  also
interesting in view of dark matter,  and a more detailed discussion
will be given in
Sec.~\ref{sec:DarkMatter}.

Second, the one-field model 3 corresponds to the
two-Higgs doublet model, where very important two-loop Barr-Zee
diagrams
can dominate the $\Damu$ contributions and strongly alter the
one-loop discussion here. Such two-loop diagrams will be
discussed in Sec.~\ref{sec:Barr-Zee}, and the resulting phenomenology
of the two-Higgs doublet model will be discussed in
Sec.~\ref{sec:2HDM}.

Finally we discuss the models with chiral enhancement, i.e.\ models
where the $m_F/m_\mu$-terms in Eq.~(\ref{eq:Damu-oneloop-results}) can
contribute. As is clear from the previous discussion and visible
in the tables, a necessary condition for such an enhancement is the
existence of gauge invariant couplings both to left-handed muons and
to right-handed muons.
The crucial common feature of all these chirally enhanced
contributions is that not two, but three different gauge multiplets
appear inside the loop contributing to $\amu$. 

Since the mechanism of chiral enhancements and the models
with chiral enhancements are of particular importance, the following
subsection \ref{sec:genericthreefield} is devoted to an analysis of
generic 3-field models representing minimal models with chiral
enhancements, and Sec.~\ref{sec:Models} contains detailed discussions
of concrete important models. Here we briefly explain how one- and
two-field models can combine with SM fields to get three gauge
multiplets in the loop.

Indeed, in the one-field case most models with chiral enhancement are
leptoquark models where the leptoquark couples left-handed muons to
quarks of one chirality and right-handed muons to quarks of the
opposite chirality. Then, in the $\amu$ Feynman diagrams of
Fig.~\ref{fig:amu-one-loop} the internal fermion is a quark and the
chirality can be flipped from a left-handed to a right-handed quark.
The three relevant fields are therefore the single BSM leptoquark and
two SM multiplets, a quark singlet and a quark doublet.  In addition
to the leptoquark examples there are also special cases where the
factor $m_F/m_\mu$ is only large when the new BSM state has lepton
flavour-violating couplings, allowing the internal lepton to be the
$\tau$-lepton.  These exceptions are model 3 which can be interpreted
as the two-Higgs doublet model, model 16 which can be interpreted as a
Z' model, and model 17 which could be embedded in a 331 model \cite{Pisano:1992bxx}.

Though excluded from our Table \ref{tab:twofieldmodels}, there are
classes of two-field models which allow chiral enhancements. Examples
are models with two different new vector-like fermions  which can
combine e.g.\ with
the SM Higgs field in a one-loop diagram. Such models will be
discussed later in
  Sec.\ \ref{sec:VLF} and Eq.\ \eqref{Eq:2fieldVLLModels} provides a
  complete list.
Another type of examples are bridge
models \cite{Guedes:2022cfy} where one vector-like lepton and a scalar
can combine e.g.\ with a SM lepton to give a one-loop contribution to
$\amu$.

In general we see that extending the SM by only one or two fields
typically leads to small contributions to muon $g-2$, though we have
highlighted special cases where they can be larger and are thus
constrained by $\DamuFinal$.  In the next
sections we will show that when we consider extensions with three BSM
fields, many more models with an internal chirality flip are possible
allowing a much wider variation in the size of the contributions.

\subsection{Three-field extensions  with chiral enhancements and mass-insertion
  approximations}
\label{sec:genericthreefield}

We now focus in detail on a set of simple and generic models which lead to chirally
enhanced contributions to $\amu$. 
The key point is that the potential chiral enhancement via the $m_F$-terms in the generic formulas \eqref{eq:Damu-oneloop-results}
requires a chirality flip on one of the internal lines in the one-loop diagrams. It must correspond to an 
additional source of chiral symmetry breaking besides $m_\mu$ as 
discussed in Sec.~\ref{sec:ChiralityFlips} and illustrated in Eqs.~\eqref{amuchiralenhancement} and \eqref{amugeneric}.
To achieve this within a minimal setup, three distinct
gauge-eigenstate fields are needed that allow for additional gauge invariant
couplings to the left- and right-handed muons as well as to the SM Higgs.

The models with chiral enhancements discussed in
Sec.~\ref{sec:MinimalBSM} combine one BSM field with two SM fields,
and later in Sec.~\ref{sec:VLF} we will discuss models where one SM
field and two BSM fields are combined to give chiral enhancements. The
generic models discussed here illustrate such cases, but they also
cover the case of three BSM fields. This last case of 3-field models
represents a wide class of very 
important BSM scenarios in a simplified and generic way.
Here we specifically focus on models with either two new fermions 
and one additional scalar or one new fermion and two additional scalars,
which have often been used to illustrate the generic behaviour in scenarios with chiral enhancement
\cite{Baker:2020vkh,Baker:2021yli,Kowalska:2017iqv,Calibbi:2018rzv,Capdevilla:2021rwo,Crivellin:2021rbq}.

We essentially follow the conventions of Refs.~\cite{Baker:2020vkh,Baker:2021yli} and introduce 
three classes of 3-field models. Class I and II contain two new Dirac fermions $\psi$ and $\chi$ and one scalar
field $\phi$, while Class III instead contains a single new Dirac fermion $\psi$ and two scalar fields $\phi$ and $\eta$\footnote{
	In the literature, Class I and III correspond to FLR and SLR of
	Ref.~\cite{Calibbi:2018rzv}, Class II and III were considered in
	Refs.~\cite{Baker:2020vkh,Baker:2021yli}, the other references cited above
	consider very similar models corresponding to combinations of the
	classes, and Ref.~\cite{Kowalska:2017iqv} also considers
        additional	simple models. Similar models are also studied
        in Ref.~\cite{Kawamura:2020qxo}, and several of the references
        also study dark matter in the context of these models. We note that the same kind of
        three-field models can also be used as a starting point to
        discuss the constraints of asymptotic safety on values of
        couplings and the magnitude of contributions to $\amu$, and
        corresponding results are obtained in
        Refs.~\cite{Hiller:2019mou,Hiller:2020fbu,Kowalska:2020zve}}.  
Both Dirac fermions are formed from left-handed and right-handed
fields of equal gauge quantum numbers, such that explicit Dirac
masses, denoted by $m_\psi$ and $m_\chi$, are allowed by gauge invariance.
Similarly, the scalar masses are written as $m_\phi$ and $m_\eta$. 
The interaction Lagrangians connecting the new fields to the SM are defined as\footnote{For simplicity we assume all couplings to be real.}
\begin{subequations}\label{eq:3-field-La}
	\begin{alignat}{4}
		&\text{Class I:} \qquad &&\La &&\supset -\lambda_L\, \overline{l_L} \psi_R \phi^\dagger
			-\lambda_R\, \overline{\chi_L}\mu_R\phi -\lambda_\Phi \overline{\psi_R}\chi_L\Phi &&+ h.c., \label{eq:LaClassI}
		\\
		&\text{Class II:} \qquad &&\La &&\supset  -\lambda_L\, \overline{l_L}\psi_R\phi^\dagger
			-\lambda_R\, \overline{\chi_L}\mu_R\phi -\bar{\lambda}_\Phi \overline{\psi_L}\chi_R\Phi &&+ h.c., \label{eq:LaClassII}
		\\
		&\text{Class III:}\qquad &&\La &&\supset -\lambda_L\, \overline{l_L}\psi_R\phi^\dagger
			-\lambda_R\, \overline{\psi_L}\mu_R\eta - \,a_\Phi\, \eta^\dagger\Phi\phi &&+ h.c., \label{eq:LaClassIII}
	\end{alignat}
\end{subequations}
where $l_L, \mu_R$ and $\Phi$ denote the left-/right-handed muon and
Higgs fields respectively (cf.\ \ref{sec:Conventions})
and we have used 4-spinor notation for the fermion fields.
In order to ensure $\UY$ invariance the hypercharges of the Class I/II
fields $(\phi^\dagger,\psi,\chi)$ and Class III fields $(\psi,\phi^\dagger,\eta^\dagger)$
need to be of the form $Y=(X,-\frac{1}{2}-X,-1-X)$, where $X$ can be an arbitrary constant. 
Possible $\SUL$ quantum number assignments of $
(\psi,\phi^\dagger,\eta^\dagger)$ and $(\phi^\dagger,\psi,\chi)$
include the combinations $(\bm1,\bm2,\bm1)$, $(\bm2,\bm1,\bm2)$, $(\bm2,\bm3,\bm2)$, $(\bm3,\bm2,\bm3)$, (where $\bm1, \bm2, \bm3$ denote singlets,
doublets and triplets). 
Note that the $\SUL$ contractions in Eq.~\eqref{eq:3-field-La}
  have been suppressed but are, in general, non-trivial in order to
  ensure gauge invariance.
For further details and more explicit Lagrangians see e.g.\ Refs.~\cite{Crivellin:2021rbq,Calibbi:2018rzv,Kowalska:2017iqv}.
To be definite, we focus on the $(\bm1,\bm2,\bm1)$ case for the explicit calculations and refer to the literature for further results. The $\SUc$ representations are left implicit, i.e.\
the new fields may also carry appropriate colour charges.

In each model class, $\lambda_{L,R}$ are Yukawa couplings
of the the new fields to the left-/right-handed muons, and the coupling of the new fields
to the SM Higgs is denoted either by $\lambda_\Phi$ and $\bar{\lambda}_\Phi$
in case of two fermions or by the dimensionful coupling $a_\Phi$ in case of two new scalars.
The difference between Class I and II is the
chirality of the fermions interacting with the Higgs field.  In Class
I, only $\psi_R$ and $\chi_L$ interact with the muon and the Higgs,
whereas $\psi_L$ and $\chi_R$ only have gauge interactions that play
no role for the current discussion.  In Class II the muon interactions are unchanged, 
but the opposite chiralities of the new fields $\psi_L$ and $\chi_R$ interact with the Higgs. 
Although the chirality is different, in both Class I and II the third term implies a 
mass mixing between $\psi$ and $\chi$ after EWSB. Similarly in Class III
a mass mixing between the scalars is induced by their coupling to the Higgs boson.  

As mentioned in the beginning, the chirally enhanced models in Tab.~\ref{tab:onefieldmodels}
are connected to the classes introduced above by identifying two of the fields $\psi,\chi,\phi$ and $\eta$
with SM fields. For example, the 1-field leptoquark models 5 and 8 discussed in Sec.~\ref{sec:LQ} correspond to Class I where the fermions $\psi_R$ and $\chi_L$
are identified with the SM quark singlets and doublets while $\phi$ constitutes the leptoquark.
Similarly,
the two-Higgs doublet model (model 3 in Tab.~\ref{tab:onefieldmodels}) is recovered from Class I
identifying $\psi_R$ and $\chi_L$ with the lepton singlets and
doublets (and dropping the Dirac masses) and $\phi$ with the second Higgs doublet. 
In this case the parameter $\lambda_\Phi$ corresponds to a lepton Yukawa coupling and is thus $ \propto m_l$, such that chiral enhancement is  present if the fermions are identified with
the $\tau$ and the Yukawa couplings to $\phi$ are lepton-flavour violating. For a detailed discussion of the 2HDM see Sec.~\ref{sec:2HDM}.
The 1-field models with a BSM spin-1 particle and chiral enhancements appearing in
Tab.~\ref{tab:onefieldmodels} have no correspondence to the 3-field models considered here, but their behaviour is similar
and has been discussed in the literature (see in particular Ref.~\cite{Biggio:2016wyy}).
Furthermore, many 2-field models with chiral enhancement, like the vector-like leptons discussed in Sec.~\ref{sec:VLF}, 
also correspond to special cases of the Class I, II or III models.

\begin{figure}
	\centering
	\begin{subfigure}{.24\textwidth}
		\centering
		\includegraphics[width=\textwidth, clip, trim=0 10 0 0]{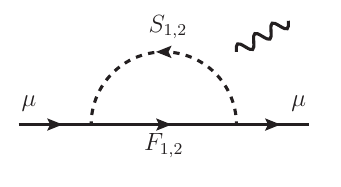}
		\caption{full one-loop result}
	\end{subfigure}
	\begin{subfigure}{.24\textwidth}
		\centering
		\includegraphics[width=\textwidth, clip, trim=0 10 0 0]{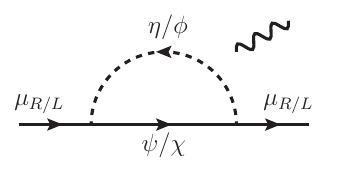}
		\caption{no mass insertion}
		\label{fig:zeroMIA}
	\end{subfigure}
	\begin{subfigure}{.24\textwidth}
		\centering
		\includegraphics[width=\textwidth, clip, trim=0 10 0 0]{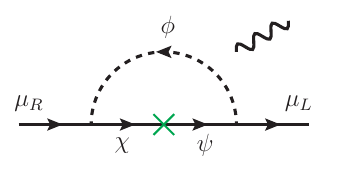}
		\caption{Class I and II mass insertion}
		\label{fig:oneMIA-I+II}
	\end{subfigure}
	\begin{subfigure}{.24\textwidth}
		\centering
		\includegraphics[width=\textwidth, clip, trim=0 10 0 0]{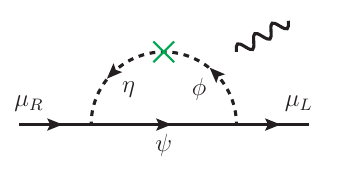}
		\caption{Class III mass insertion}
		\label{fig:oneMIA-III}
	\end{subfigure}
	\caption{Diagrams generating $\Delta a_\mu$ at one-loop in Class I, II and III. Diagram \textbf{(a)}
		corresponds to the full calculation in the mass-basis, \textbf{(b)} to the zero-order mass-insertion 
		and diagrams \textbf{(c)} and \textbf{(d)} to the single mass-insertion in Class I/II and III, respectively.}
	\label{fig:MIAClasses}
\end{figure}

We will now analyse the models and exhibit their essential new features related to $\amu$, chirality flips and the impact of mixing in several complementary ways. We will begin with the
mass-insertion approximation method and connect it to the discussion of
chirality flips of Sec.~\ref{sec:ChiralityFlips}, then we 
describe the explicit calculation of $\amu$ via 
results of Sec.~\ref{sec:genericoneloop}.

The mass-insertion approximation method (M.I.A.) is a generally useful tool
to obtain insight into the behaviour of Feynman diagrams in case of
mixing. It allows to qualitatively understand the dominant parameter
dependences, and can also be used for quantitatively accurate
computations, see
e.g.\ Refs.~\cite{Moroi:1995yh,Stockinger:2006zn,Dedes:2015twa,Crivellin:2018mqz}. Its main idea is to treat
bilinear Lagrangian terms which mix different fields and are $\propto v$  as
small interaction terms giving bilinear vertices in
Feynman diagrams. In the models considered here, the ${\lambda}_\Phi v$-,
$\bar{\lambda}_\Phi v$- and the $a_\Phi v$-terms arising after inserting the
Higgs \vev $\ev{\Phi} = (0,v/\sqrt2)^T$ into the Lagrangian, are of this kind.

In general, these bilinear terms from couplings to the Higgs \vev\ give rise to mass mixing between the fields.
For example, in the $(\bm1,\bm2,\bm1)$ representation the mass matrices for the gauge-eigenstate fields are given by
\begin{align}\label{eq:3-field-mass-matrices}
	\text{Class I:}\quad
	\bordermatrix{ & \psi_{2R} & \chi_R \cr 
	\overline{\psi_{2L}} & m_\psi & 0 \cr
	\overline{\chi_L} & \lambda_\Phi \frac{v}{\sqrt{2}} & m_\chi } \qquad\quad
	\text{Class II:}\quad
	\bordermatrix{ & \psi_{2R} & \chi_R \cr 
		\overline{\psi_{2L}} & m_\psi & \bar\lambda_\Phi \frac{v}{\sqrt{2}} \cr
		\overline{\chi_L} & 0 & m_\chi } \qquad\quad
	\text{Class III:}\quad
	\bordermatrix{ & \phi_2 & \eta \cr 
		\phi_2^\dagger & m_\phi^2 & a_\Phi \frac{v}{\sqrt{2}} \cr
		\eta^\dagger & a_\Phi \frac{v}{\sqrt{2}} & m_\chi },		
\end{align}
where $\psi_2$ and $\phi_2$ denote the lower-components of the respective doublets.

Fig.~\ref{fig:MIAClasses} compares the full diagram with several mass-insertion diagrams where either zero or one of the off-diagonal entries are inserted as a bilinear vertex. The mass-insertion 
diagrams thus correspond to a Taylor expansion in $v$; the exact
result would be obtained by resumming diagrams with any number of insertions,
however often  the leading contributions are sufficient.
In this way one can avoid explicitly diagonalising the mass matrices
and obtain  a useful analytic approximation of the main behaviour.

The zeroth-order M.I.A.\ diagram Fig.~\ref{fig:zeroMIA} is present in all three classes. 
It can only contain two out of the three fields of each class and thus behaves like the
one- or two-field models without chiral enhancement discussed in the previous subsection.
These diagrams contribute to $\amu$ only via the non-enhanced $|c_{L,R}|^2$-terms in
Eq.~\eqref{eq:Damu-FS}, and the magnitude of this contribution can hence be estimated as (cf. Eq.~\eqref{nonenhancedoneloop})
\begin{align}\label{eq:noMIAcontribution}
	|\Damu^\text{no insertion}| \sim \frac{|\lambda_{L,R}|^2}{64\pi^2}\frac{m_\mu^2}{M^2},   
\end{align}
where $M$ is the typical mass scale of the particles in the loop.

The diagrams \ref{fig:oneMIA-I+II} and \ref{fig:oneMIA-III} with single mass insertions 
are the crucial new contribution present in the three-field models. 
They involve all three of the new couplings, resulting in a chirality flip inside of the loop. 
A straightforward estimate based on the structure of the diagrams gives 
\begin{align}\label{eq:MIAcontribution}
  |\Delta\amu^\text{one insertion}| \sim \frac{\lambda_L\lambda_R}{16\pi^2} \frac{m_\mu v}{\sqrt2 M^2}
  \left\{ \lambda_\Phi, \; \bar\lambda_\Phi \frac{m_\psi m_\chi}{M^2},\; \frac{ m_\psi a_\Phi }{M^2}\right\}. 
\end{align}
for Class I, II and III respectively.
Here the coupling factors in the numerator correspond to the three vertices of
the diagrams. The different powers of the internal fermion masses arise because the numerator 
of each fermion propagator (each of the form  $(\slashed{k}+m) $) is sandwiched
between specific combinations of the projectors ${P}_{L/R}$, the factor $v/\sqrt{2}$ comes from the
mass insertions and the muon mass $m_\mu$ stems from the definition of $\amu$. 
The mass in the denominator is then fixed by the dimensionality and again corresponds to the typical mass scale of the loop.

It is instructive to relate these explicit results to the generic
discussion of chirality flips and gauge invariance in
Sec.~\ref{sec:ChiralityFlips}. First, the single mass insertions in
Fig.~\ref{fig:MIAClasses} correspond to the coupling to the EWSB \vev
in Fig.~\ref{fig:chirflipdiagrams} (right) and to the ``(EWSB \vev)'' factor in Eq.~\eqref{amugeneric}. 
Since we assume the additional scalars to not develop vacuum expectation values,
the only relevant \vev is the one of the SM Higgs. Diagrams with several mass
insertions would correspond to terms of higher power in the mixing and to higher-order
corrections to formulas such as Eq.~\eqref{amugeneric}. 

As a second connection, the different combinations of couplings appearing in Eq.~\eqref{eq:MIAcontribution}
can be understood in terms of new sources of (muon) chiral symmetry breaking present in these models.
The chiral transformation of Eq.~\eqref{muchiralsym} may be extended to the new fields as
\begin{align}
	\psi_{L/R}\to e^{-i\mathcal{C}_{L/R}^\psi\alpha}\psi_{L/R}, \qquad \chi_{L/R}\to e^{-i\mathcal{C}^\chi_{L/R}\alpha}\chi_{L/R}, \qquad
	\phi\to e^{-i\mathcal{C}_\phi\alpha}\phi, \qquad \text{and}\qquad  \eta\to e^{-i\mathcal{C}_\eta\alpha}\eta
\end{align}
with initially unspecified chiral charges $\mathcal{C}_i$ for the new fields. 
These transformations would leave the Lagrangians in Eq.~\eqref{eq:3-field-La} invariant if the following equations 
\begin{align}
	\text{Class I/II:}\quad 0 = \begin{cases}
		1 + \mathcal{C}_\phi - \mathcal{C}^\psi_R \\
		1 + \mathcal{C}^\chi_L - \mathcal{C}_\phi \\
		\mathcal{C}^\psi_{R/L} - \mathcal{C}^\chi_{L/R}
	\end{cases}, \qquad\qquad
	\text{Class III:}\quad 0 = \begin{cases}
		1 + \mathcal{C}_\phi - \mathcal{C}^\psi_R \\
		1 + \mathcal{C}^\psi_L - \mathcal{C}_\eta \\
		\mathcal{C}_\eta - \mathcal{C}_\phi
	\end{cases}
\end{align}
could hold. However, in Class I there is no simultaneous solution, and in 
Class II and III
the only possible solutions require $\mathcal{C}_L^{\psi,\chi} \neq \mathcal{C}_R^{\psi,\chi}$.
The latter inequality is then inconsistent with the invariance of the Dirac mass terms in the respective models.
Note however that dropping any one of the equations does yield consistent solutions in all three classes.
Consequently, this implies that the order parameters of chiral symmetry breaking in Class I, II and III respectively are the following specific combinations,
\begin{align}\label{3fieldchiralenhancements}
	\lambda_L \lambda_\Phi \lambda_R, \qquad \lambda_L \bar\lambda_\Phi \lambda_R m_\psi m_\chi, \qquad \text{and} \qquad
	\lambda_L \lambda_R a_\Phi m_\psi.
\end{align}
It is precisely those combinations that appear in Eq.~\eqref{eq:MIAcontribution} and lead to the 
chiral enhancement.\footnote{%
	\label{note:3field}
	Some of the additional models of Ref.~\cite{Kowalska:2017iqv} provide an interesting negative illustration of
	  chiral symmetry breaking as a prerequisite for large chiral enhancements. 
	  While e.g.~their Models 3 or 9 are similar to our Classes I, II and lead to chiral enhancements, 
	  their Model 6 does not. Model 6 corresponds to combination of Class I and Class II but with a crucial
	  modification where the replacement $\lambda_L \overline{l_L}\phi^\dagger\psi_R\to\lambda_L \overline{l_L} \phi\psi_R$ 
	  is applied to the Lagrangian. After this difference, the Lagrangian always is chirally symmetric under 
	  $\mu_R \to e^{i\alpha}\mu_R$, $l_L \to e^{-i\alpha}l_L$, and $\phi\to e^{-i\alpha}\phi$. Hence, the
	  product of the couplings $\lambda_L\lambda_R\lambda_\Phi$ in this case
	  would not break chiral symmetry and could not contribute to $\amu$, as also pointed out in Ref.~\cite{Kowalska:2017iqv}.  }
In this way they represent an explicit realisation of the ``(chirality flipping parameters)'' in 
Eqs.~\eqref{mmugeneric} and \eqref{amugeneric}, while the ``(other factors)'' here correspond to the $1/16\pi^2$
loop suppression as well as group factors depending on the specific representations.
The essential relative chiral enhancement between Eq.~\eqref{eq:noMIAcontribution} and the single-insertion diagrams is 
(assuming $\lambda_L\sim\lambda_R$ and $m_\psi,m_\chi\sim M$)
\begin{align}
	\frac{|\Damu^\text{one insertion}|}{|\Damu^\text{no insertion}|} \sim \frac{v}{m_\mu}
	\left\{ \lambda_\Phi , \bar{\lambda}_\Phi,\frac{a_\Phi}{m_\psi}\right\} .
\end{align}
As mentioned in Sec.~\ref{sec:genericoneloop}, the true chiral enhancement is therefore not just the factor $m_F/m_\mu$ as
might have been guessed from the formulas in Eq.~\eqref{eq:Damu-oneloop-results}, but the behaviour is more
complicated due to the presence of mixing.
\newline

With this qualitative understanding we now proceed with the exact computation of the one-loop contributions to $\amu$ in the model and 
elucidate how it reproduces the above estimates. Although the calculations proceed similarly for all
representations, we again take $(\bm1,\bm2,\bm1)$ as the explicit example.
To calculate the full diagram in Fig.~\ref{fig:MIAClasses} and employ the generic one-loop formulas (\ref{eq:Damu-oneloop-results}), we first introduce the mass-basis fields $F_i$ and $S_i$  in the different
classes as
\begin{subequations}
	\renewcommand*{\arraystretch}{.7}
	\begin{alignat}{4}
		&\text{Class I/II:} \quad
		&&\begin{pmatrix}
			F_1 \\ F_2
		\end{pmatrix}_{L/R} &&= 
		\begin{pmatrix}
		\cos\theta_{L/R} & -\sin\theta_{L/R} \\ \sin\theta_{L/R} & \cos\theta_{L/R}
 		\end{pmatrix} \begin{pmatrix}
		\psi_2 \\ \chi
		\end{pmatrix}_{L/R}, \qquad && S = \phi, \\
		&\text{Class III:} \quad
		&&\begin{pmatrix}
			S_1 \\ S_2
		\end{pmatrix} && = 
		\begin{pmatrix}
			\cos\theta & -\sin\theta \\ \sin\theta & \cos\theta
		\end{pmatrix}
		 \begin{pmatrix}
			\phi_2 \\ \eta
		\end{pmatrix}, && F = \psi,
	\end{alignat}
\end{subequations}
where the rotation matrices are chosen to diagonalise the mass matrices \eqref{eq:3-field-mass-matrices}, 
such that the corresponding terms in the Lagrangians become
\begin{alignat}{2}
	&\text{Class I/II:} \qquad &&\La \supset - m_1 \bar{F}_1F_1 - m_2 \bar{F}_2 F_2 - m_\phi^2 |S|^2, \\
	&\text{Class III:} \qquad  &&\La\supset - m_1^2 |S_1|^2 - m_2^2 |S_2|^2 - m_\psi \bar{F}F.
\end{alignat}
In Class I and II two independent transformations of the left- and right-handed fields with mixing angle $\theta_{L,R}$ are required
and fulfil the following useful relation \cite{Baker:2021yli}
\begin{align}\label{couplingsClassII}
	\sin\theta_L\cos\theta_R&=
	\cos\theta_L\sin\theta_R \bigg\{\frac{m_1}{m_2},\frac{m_2}{m_1}\bigg\}
	=\frac{1}{m_2^2-m_1^2} \frac{v}{\sqrt2} \bigg\{m_1\lambda_\Phi,m_2\bar{\lambda}_\Phi \bigg\}.
\end{align}
In Class III only a single mixing angle is required which also fulfils a similar relation
\begin{align}
	\sin\theta \cos\theta &= \frac{a_\Phi v}{\sqrt2 (m_2^2-m_1^2)}.
\end{align}
The scalar mass-eigenstate(s) coupling to the muon in Class I and II
(Class III) have the electric charge $Q_S = X$ ($Q_S = -1-X$),
and the relevant products of mass-basis Yukawa couplings appearing in Eq.~\eqref{eq:Damu-FS} are given by
\begin{align}
	\text{Class I/II:} \quad \begin{cases}
		F_1 : \quad c_L^* c_R = - \lambda_L\lambda_R \sin\theta_L\cos\theta_R \\
		F_2 : \quad c_L^* c_R = + \lambda_L\lambda_R \sin\theta_R\cos\theta_L \\
	\end{cases} \qquad
	\text{Class III:}
	\quad \begin{cases}
		S_1 : \quad c_L^* c_R = - \lambda_L\lambda_R \sin\theta\cos\theta \\
		S_2 : \quad c_L^* c_R = + \lambda_L\lambda_L \sin\theta\cos\theta. \\
	\end{cases}
\end{align}
Similar expressions hold for the non-enhanced $|c_{L/R}|^2$ terms which we neglect for the discussion here.
With this, the enhanced contributions to $\Damu$ read
\begin{subequations}\label{eq:3field-amu}
	\begin{alignat}{8}
		&\text{Class I:} \qquad &&\Damu^\text{enhanced} = \frac{\lambda_L\lambda_R}{16\pi^2} \frac{ \lambda_\Phi v m_\mu}{\sqrt{2} m_\phi^2} &&\times
		\frac{1}{x_2-x_1}&&\bigg[x_2 &&\mathcal{F}^\text{FS}(0,x_2;Q_S) &&- x_1 &&\mathcal{F}^\text{FS}(0,x_1;Q_S) &&\bigg],
		\\
		&\text{Class II:} \qquad &&\Damu^\text{enhanced} = \frac{\lambda_L\lambda_R}{16\pi^2} \frac{ \bar\lambda_\Phi v m_\mu}{\sqrt{2} m_\phi^2} &&\times
		\frac{\sqrt{x_1x_2}}{x_2-x_1}&&\bigg[&&\mathcal{F}^\text{FS}(0,x_2;Q_S) &&\;- &&\mathcal{F}^\text{FS}(0,x_1;Q_S) &&\bigg], \\
		&\text{Class III:} \qquad &&\Delta\amu^\text{enhanced} = \frac{\lambda_L\lambda_R}{16\pi^2} \frac{ a_\Phi v m_\mu}{\sqrt{2} m_\psi^3} &&\times
		\,\frac{z_1 z_2}{z_1 - z_2}&&\bigg[ z_2 &&\mathcal{F}^\text{FS}(0,z_2;Q_S) &&- z_1 &&\mathcal{F}^\text{FS}(0,z_1;Q_S) &&\bigg],
	\end{alignat}
\end{subequations}
where $x_i =m_i^2/m_\phi^2$ and $z_i=m_\psi^2/m_i^2$. For small mixing the mass eigenvalues $m_{1,2}$ appearing in these exact results correspond to
$m_{\psi,\chi}$ in Class I/II and $m_{\phi,\eta}$ in Class III. Since the loop-functions are of order one this result is in full agreement with
the naive estimate Eq.~\eqref{eq:MIAcontribution}. This is further illustrated by the following simple limits.
Taking $x_i = z_i \equiv m_F^2 / m_S^2 \equiv x$, the limits $m_F\ll m_S$, $m_F=m_S$ and $m_F\gg m_S$ are of the form
\begin{subequations}\label{eq:3-field-limits}
	\begin{align}
		\Damu^\text{enhanced} = \frac{\lambda_L \lambda_R}{16\pi^2} \frac{m_\mu v}{\sqrt{2} m_S^2}
		\Big\{\lambda_\Phi, \bar\lambda_\Phi, \frac{a_\Phi}{m_F}  \Big\}
		\begin{cases}
			\big\{1,1,x\big\}[a_1 + b_1 \ln(x)] & m_F \ll m_S\\
			\frac{1}{6} a_2  & m_F = m_S \\
			\big\{\frac{1}{x^2},\frac{1}{x},\frac{1}{x}\big\} [a_3 + b_3 \ln(x)] & m_F \gg m_S
		\end{cases}
	\end{align}
\end{subequations}
where the entries in the brackets correspond to Class I, II and III respectively. The coefficients $a_i$ and $b_i$ depend on the representation
of the field multiplets as well as the hypercharge $X$ and are listed in Tab.~\ref{tab:3-field-limits}.
There are interesting relations between the Class I and III coefficients, namely $a_{1,2,3}^\text{I}=a_{3,2,1}^{\text{III}}$ and $b_{1,3}^\text{I}=-b_{3,1}^{\text{III}}$.
Furthermore, for Class II, $b_i=0$ for all listed representations, i.e.\ in this case none of the limits results in logarithmic enhancement.
On the other hand, both Class I and III develop logarithmic enhancements not only in the limit $x\to 0$ as expected from Eq.~\eqref{LoopfunctionFSlimits},
but also for $x\to\infty$. This is another consequence of the mixing between the gauge-eigenstates and once again confirms 
that this mixing is crucial to understand the chirally enhanced contribution.

\begin{table}[t]
	\centering
	\begin{tabular}{|c|c||c|c|c|c|c|}
		\hline
		Rep. & Class & $a_1$ & $b_1$ & $a_2$ & $a_3$ & $b_3$ \\ \hline\hline
		\multirow{3}{*}{\rotatebox[origin=c]{90}{$(\bm1,\bm2,\bm1)$}} 
		& I   & $-5-4X$ & $-2-2X$ & $1+2X$ & $1-4X$ & $2X$ \\ \cline{2-7}
		& II  & $-2-2X$ & 0 & $-3-4X$ & $-1-2X$ & 0 \\ \cline{2-7}
		& III & $1-4X$ & $-2X$ & $1+2X$ &$-5-4X $ & $2+2X$ \\ \hline\hline
		\multirow{3}{*}{\rotatebox[origin=c]{90}{$(\bm2,\bm1,\bm2)$}} 
		& I   & $3+4X$ & $1+2X$ & $ -2X$ & $-3+4X$ & $1-2X$  \\ \cline{2-7}
		& II  & $1+2X$ & 0 & $1+4X$ & $2X$ & 0 \\ \cline{2-7}
		& III & $-3+4X$ & $-1+2X$ & $-2X$ &$3+4X$ & $-1-2X$ \\ \hline\hline
		\multirow{3}{*}{\rotatebox[origin=c]{90}{$(\bm2,\bm3,\bm2)$}} 
		& I   & $-17-12X$ & $-7-6X$ & $4+6X$ & $1-12X$ & $1+6X$ \\ \cline{2-7}
		& II  & $-7-6X$ & 0 & $-11+12X$ & $-4-6X$ & 0 \\ \cline{2-7}
		& III & $1-12X $ & $-1-6X$ & $4+6X$ & $-17-12X$ & $7+6X$ \\ \hline\hline
		\multirow{3}{*}{\rotatebox[origin=c]{90}{$(\bm3,\bm2,\bm3)$}} 
		& I   & $-7-12X$ & $-2-6X$ & $-1+6X$ & $11-12X$ & $-4+6X$ \\ \cline{2-7}
		& II  & $-2-6X$ & 0 & $-1-12X$ & $1-6X$ & 0 \\ \cline{2-7}
		& III & $11-12X$ & $4-6X$ & $-1+6X$ & $-7-12X$ & $2+6X$ \\ \hline
	\end{tabular}
	\caption{Coefficients of the limits in Eq.~\eqref{eq:3-field-limits} for different $\SUL$ representations of the
		Class I/II fields $(\phi^\dagger,\psi,\chi)$ and Class III fields $(\psi,\phi^\dagger,\eta^\dagger)$.
		As described in the text, their respective hypercharges are given in terms of the arbitrary constant 
		$X$ as $(X,-\frac{1}{2}-X,-1-X)$.  }
	\label{tab:3-field-limits}
\end{table}

As mentioned, the behaviour found in these simple 3-field models is
characteristic for a large number of BSM scenarios which involve
new couplings to both left-handed and right-handed muons. Examples are
supersymmetric models,
vector-like leptons, certain leptoquarks and the flavour-violating
2HDM and more general models. We will also use the 3-field models later to
illustrate connections to other observables such as the muon mass and
the muon--Higgs coupling.

\subsection{Effective field theory description}\label{sec:genericeft}

\begin{figure}
	\centering
	\begin{subfigure}{.19\textwidth}
		\centering
		\includegraphics[width=.8\textwidth]{Diagrams/fig.amu-Schwinger.pdf}
		\caption{dimension-4}
                \label{fig:LEFT-amudim4}
	\end{subfigure}
	\begin{subfigure}{.38\textwidth}
		\centering
		\includegraphics[width=.8\textwidth]{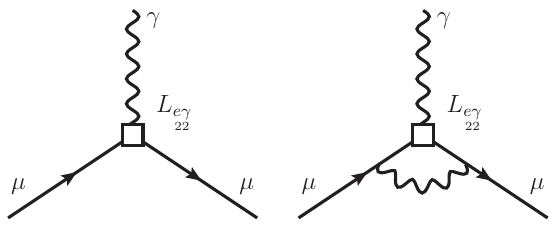}
		\caption{dimension-5}
                \label{fig:LEFT-amudim5}
	\end{subfigure}
	\begin{subfigure}{.38\textwidth}
		\centering
		\includegraphics[width=.8\textwidth]{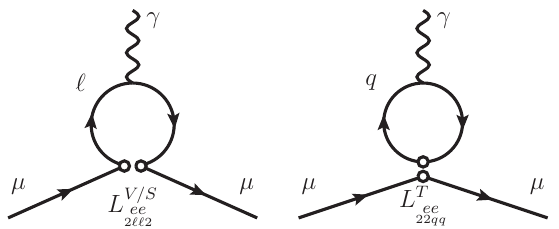}
		\caption{dimension-6}
                \label{fig:LEFT-amudim6}
	\end{subfigure}
	\caption{Sample diagrams of contributions to $a_\mu$ in the
          LEFT. The dimension-4 contribution is just the usual QED
          Schwinger diagram already shown in Fig.~\ref{fig:amu-Schwinger},
	the dimension-5 contribution stems solely from the muon dipole
        operator in Eq.~(\ref{eq:LEFT-dipole-operator}),  and the
        dimension-6 four-fermion contributions come from 
	scalar or vector operators defined in
        Eq.~(\ref{eq:LEFT-Z-matching}) involving the leptons and
        tensor operators involving the heavy quarks. 
	}
	\label{fig:LEFT-amu}
\end{figure}

To conclude the section on generic BSM scenarios we discuss $\Damu$ in the context of effective field theories (EFT).
Over the past decades, EFT has become one of the most important tools for studying new physics at or above the electroweak scale,
particularly as it allows for model independent analyses of BSM effects on a wide range of observables.
EFT is especially potent for low-energy observables, like $\Damu$, where the expansion in $E/M_\text{BSM}$ converges rapidly.
For  recent reviews we refer to Refs.~\cite{Isidori:2023pyp,
  Allwicher:2021jkr, Aebischer:2021uvt}.

The first appropriate effective theory of this kind
is known as low-energy EFT (LEFT) \cite{Jenkins:2017jig} or weak
effective theory (WET) \cite{Buchalla:1995vs}. Here
heavy states like the $W$-, $Z$- or Higgs boson and potential heavy
BSM fields are integrated out and the resulting EFT is an
$\GLEFT$ gauge theory which contains QED and QCD, the light
SM fields and non-renormalizable terms suppressed by inverse powers of 
the electroweak scale in the Lagrangian.
It is a valid description of phenomena below the electroweak scale as
long as there is no light BSM physics.

The Lagrangian of this low-energy EFT contains the 
dipole operator (we follow the conventions of Ref.~\cite{Jenkins:2017jig})
\begin{align}\label{eq:LEFT-dipole-operator}
	\La_\text{LEFT} \supset L_{\underset{22}{e\gamma}}
        \Big(\overline{\mu_L} \sigma^{\alpha\beta} \mu_{R}\Big)
        F_{\alpha\beta} + h.c.
\end{align}
where $L_{\underset{22}{e\gamma}}$ is a Wilson coefficient assumed to
scale as $\propto M^{-1}$, where $M$ is the
relevant heavy scale. The Wilson coefficient is defined in
the $\overline{\text{MS}}$ scheme, $\mu_{L,R}$ are the left- and
right-handed muon fields and $F_{\mu\nu}$ denotes the photon
field-strength tensor defined in
Eq.~\eqref{eq:QED-covariant-derivative}.
Like the mass and Yukawa terms, this dimension-5 term connects the left- and right-handed muons
resulting in an additional breaking of the muon chiral symmetry (cf. Sec.~\ref{sec:ChiralityFlips}).
The dipole operator yields a contribution to $\amu$ already at tree-level,
\begin{align}\label{eq:LEFT-amu-tree}
	\amu^\text{LEFT,tree} = \frac{4m_\mu}{e} \Re L_{\underset{22}{e\gamma}}.
\end{align}
Further contributions, shown in Fig.~\ref{fig:LEFT-amu},
arise at the one-loop level e.g.\ from four-fermion operators, but also the usual QED corrections.
The total $\amu$ at one-loop order in the $\overline{\text{MS}}$ renormalised LEFT is given by \cite{Aebischer:2021uvt}
\begin{align}\label{eq:LEFT-amu-one-loop}
  \amu^\text{LEFT} = \frac{\alpha}{2\pi} +
  \frac{4 m_\mu}{e} \Re L_{\underset{22}{e\gamma}}(\mu) \bigg\{ 1 - \frac{\alpha}{4\pi} \bigg[2 + 5 \ln(\frac{\mu^2}{m_\mu^2})\bigg] \bigg\}
	+  \Damu^{4l} + \Damu^{2l2q} + \O(L_{e\gamma}^2).
\end{align}
Here,  $e\equiv e(\mu)$ and $L_{\underset{22}{e\gamma}}$ denote the
QED gauge-coupling and the Wilson coefficient in the
$\overline{\text{MS}}$ scheme,
and $m_\mu$ is the physical muon
pole mass.\footnote{%
For more details on the scheme definition of the running
quantities we refer to the original literature, and to Ref.~\cite{Gnendiger:2017pys}
for an overview of variants of dimensional regularisation.} The first term corresponds to Fig.~\ref{fig:LEFT-amudim4}
and to Schwinger's one-loop QED 
result, which is part of the LEFT prediction for $\amu$. The second
term corresponds to the tree-level and one-loop corrections of
Fig.~\ref{fig:LEFT-amudim5} involving the dimension-5 dipole
operator. Specifically the logarithmic term is a higher-order
correction to Eq.~(\ref{eq:LEFT-amu-tree}).\footnote{
If $m_\mu$ is also taken in the $\overline{\text{MS}}$ scheme at the
scale $\mu$, the coefficient of the leading one-loop logarithm in Eq.~\eqref{eq:LEFT-amu-one-loop}
is modified. For a detailed discussion of the logarithmic terms see Sec.~\ref{sec:photonic}.}

The one-loop corrections $\Damu^{4l}$ and $\Damu^{2l2q}$ illustrated
in Fig.~\ref{fig:LEFT-amudim6} arise from the following dimension-6 four-fermion operators involving two muons and either two leptons
or two quarks,
\begin{align}
	\begin{split}
		\La_\text{LEFT} &\supset L_{\underset{2ij2}{ee}}^{V,LR} \big(\overline{\mu_L}\gamma^\mu e_{Li}\big)\big(\overline{e_{Rj}}\gamma_\mu \mu_R\big)
		+ L_{\underset{2ll2}{ee}}^{S,RR} \big(\overline{\mu_L} e_{Ri}\big)\big(\overline{e_{Lj}} \mu_R\big) \\
		&+ L_{\underset{22ij}{eu}}^{T,RR} \big(\overline{\mu_L}\sigma^{\mu\nu} \mu_R\big)\big(\overline{u_{Lj}}\sigma_{\mu\nu} u_{Rj}\big)
		+ L_{\underset{22ij}{ed}}^{T,RR} \big(\overline{\mu_L}\sigma^{\mu\nu} \mu_R\big)\big(\overline{d_{Lj}}\sigma_{\mu\nu} d_{Rj}\big),
	\end{split}
\end{align}
and result in the contributions
\begin{subequations}
	\begin{align}
		\Damu^{4l} &= \frac{m_\mu}{4\pi^2} \sum_{i} m_{e_i}
                \bigg\{  \ln\bigg(\frac{\mu^2}{m_{e_i}^2}\bigg) \Re
                L_{\underset{2ii2}{ee}}^{S,RR}(\mu)  - \Re
                L_{\underset{2ii2}{ee}}^{V,LR}(\mu)   \bigg\}
                \label{eq:LEFT-amu4l}\\
		\Damu^{2l2q} &= \frac{m_\mu}{\pi^2} \sum_{q_j,i} N_c Q_{q_j} m_{q_j} \ln\bigg(\frac{\mu^2}{m_{q_j}^2}\bigg) \Re L_{\underset{22ii}{eq}}^{T,RR}(\mu),
	\end{align}
\end{subequations}
where the sum runs over the heavy quarks $q_j$ active the scale $\mu$ (i.e.\ usually $c$ and $b$). The light quark contributions, not included here, 
have to be computed non-perturbatively and are discussed in
Refs.~\cite{Dekens:2018pbu,Aebischer:2021uvt}.

In the LEFT result Eq.~(\ref{eq:LEFT-amu-one-loop}),
the dipole operator contributes  at tree-level while the four-fermion
operators contribute first at one-loop order.
When applying this result to concrete renormalizable theories it is
important to keep in mind that this loop counting might not reflect
the loop counting of the underlying theories.
For example, in many UV completions $L_{ee}$ and $L_{eq}$ arise at tree-level
while $L_{e\gamma}$ is generated radiatively, such that their overall
contributions will be of similar size.

It is instructive to illustrate the application of LEFT to the EWSM
contributions where the above comment is relevant.
For example,  integrating out the $Z$ and $W$-boson from the SM yields (in the
$\overline{\text{MS}}$ scheme with naive dimensional regularisation) 
\begin{align}\label{eq:LEFT-Z-matching}
	L_{\underset{2222}{ee}}^{V,LR}(M_Z) = \frac{4G_F}{\sqrt{2}}\big(1-2 s_W^2\big) s_W^2, \qquad \text{and} \qquad
	L_{\underset{22}{e\gamma}}(M_Z) = \frac{em_\mu}{48\pi^2} \frac{G_F}{\sqrt{2}} \big(3+8 s_W^2 - 16 s_W^4\big)
\end{align}
and the SM one-loop contribution Eq.~\eqref{eq:amu-EW-1} 
discussed in Sec.~\ref{sec:SMtheory} is recovered once both Wilson coefficients are inserted into 
Eq.~\eqref{eq:LEFT-amu-one-loop}.
This agreement includes the absence of a large logarithm of the form
$\log(M_Z/m_\mu)$ in the $Z$-boson contribution.
Applying the same procedure to the SM Higgs one-loop contribution
yields a non-vanishing tree-level result for $L_{ee}^{S,RR}$, and
inspecting Eqs.~(\ref{eq:LEFT-amu-one-loop}) and (\ref{eq:LEFT-amu4l})
shows that this Higgs contribution contains a large logarithm, in line
with the limit of the relevant loop function in
Eq.~(\ref{LoopfunctionFSlimits}). 

Similarly, certain BSM scenarios such as many one-field models of
Tab.~\ref{tab:onefieldmodels} can yield non-vanishing Wilson
coefficients of all these types. There are, however, also many BSM
scenarios such as the  models with two fields of different spin in
Tab.~\ref{tab:twofieldmodels} which contribute to $\amu$ only via the
dipole operator. For all such models there can be no large logarithm
at the one-loop level, but the logarithm present explicitly in
Eq.~(\ref{eq:LEFT-amu-one-loop})  constitutes an important two-loop 
correction.
This universal logarithmically enhanced QED correction will be
discussed in more detail in Sec.~\ref{sec:photonic}.

The LEFT description is valid up to the scale $\mu_\text{max} \sim \min(v,M_\text{BSM})$. If the new physics is significantly heavier than $v$,
it may be appropriate to first integrate out the BSM modes before going to the LEFT. Generically, this yields a non-renormalizable 
$\GSM$ gauge theory referred to as
Standard Model EFT (SMEFT), 
whose operators and Lagrangian have been systematically analysed in Refs.~\cite{Buchmuller:1985jz,Grzadkowski:2010es}.
Similar to Eqs.~\eqref{eq:LEFT-dipole-operator} and \eqref{eq:LEFT-amu-tree}
the SMEFT contains dipole operators that contribute to $a_\mu$ at
tree-level. In the conventions of Ref.~\cite{Grzadkowski:2010es} the
relevant part of the Lagrangian reads
\begin{align}\label{eq:SMEFT-dipole-operator}
	\La_\text{SMEFT} \supset C_{\underset{22}{eB}} \big(\overline{l_{L2}} \sigma^{\mu\nu} e_{R2}\big) \Phi B_{\mu\nu} +
	C_{\underset{22}{eW}} \big(\overline{l_{L2}} \sigma^{\mu\nu} e_{R2}\big) \sigma^a \Phi W^a_{\mu\nu},
\end{align}
where the $C_i$ are SMEFT Wilson coefficients with the assumed scaling
$\propto M_\text{BSM}^{-2}$. Here we slightly differ from the conventions of Ref.~\cite{Grzadkowski:2010es}
and absorb the BSM scale in the dimensionful Wilson coefficients $C_i$. As an
alternative parametrisation of SMEFT Wilson coefficients we will use
\begin{align}
  C_i&\equiv\frac{c_i}{M_\text{BSM}^2}
\end{align}
with dimensionless coefficients $c_i$. 
Like the dimension-4 Yukawa interaction, these dipole operators connect the left-handed lepton doublet
with the right-handed singlet.
Thus, similar to LEFT, the corresponding Wilson coefficients both individually break the muon chiral symmetry.

Note that in this case gauge invariance results in the appearance of the Higgs fields and the
electroweak field strength tensors 
defined in Eq.~\eqref{eq:SM-field-strengths}, rather than
$F_{\mu\nu}$ directly. The lowest-order contribution to $\amu$ from
the SMEFT dipole operators is then
\begin{align}\label{SMEFTDamutree}
	\begin{split}
		\Damu^{\text{SMEFT,tree}} &= \frac{4m_\mu v}{e\sqrt{2}} \Re C_{\underset{22}{e\gamma}},
	\end{split}
\end{align}
where $C_{e\gamma} \equiv c_W C_{eB} - s_W C_{eW}$. The result can be
compared to its LEFT counterpart and related to the need for
chirality flips.
In contrast to Eq.~\eqref{eq:LEFT-dipole-operator}, the SMEFT dipole
operators are of dimension 6 and consequently their contribution to $\amu$ is suppressed by a factor of $1/M_\text{BSM}^2$.
Because of the appearance of the Higgs field $\Phi$, this contribution scales like $\Damu\propto m_\mu v / M_\text{BSM}^2$.
This SMEFT behaviour corresponds to the discussion of chirality
flips and electroweak symmetry breaking leading to the structure
(\ref{amugeneric}) for contributions to $\amu$. In contrast, LEFT
ignores electroweak gauge invariance and hence leads to a simpler
scaling of $\Damu\propto m_\mu/M_{\text{BSM}}$. However, the two
behaviours coincide when $M_\text{BSM}\sim v$, which is the heaviest
possible scale in LEFT.

More generally, the LEFT dipole operator emerges from Eq.~\eqref{eq:SMEFT-dipole-operator} after EWSB as can be seen after decomposing the SMEFT 
operators in terms of the gauge-boson mass eigenstates
\begin{align}\label{SMEFTphotondipole}
	(\overline{l_{L2}} \sigma^{\mu\nu} e_{R2}) \Big[C_{\underset{22}{eB}} B_{\mu\nu} + C_{\underset{22}{eW}} \sigma^a W_{\mu\nu}^a\Big]\Phi
	\quad \overset{\text{EWSB}}{\longrightarrow} \quad 
	\tfrac{v}{\sqrt{2}} C_{\underset{22}{e\gamma}} (\overline{\mu_L} \sigma^{\mu\nu} \mu_R) F_{\mu\nu} + ...
\end{align}
where $C_{e\gamma}$ has been defined above and the dots denote
additional terms involving the $W$, $Z$ and Higgs bosons. 
Results for the complete matching of SMEFT to LEFT at tree-level and one-loop order have been presented in Refs.~\cite{Jenkins:2017jig,Dekens:2019ept}.
Similar to above, a BSM UV theory will typically give rise to
correlations between the SMEFT Wilson coefficients and the related observables, several examples of which will be discussed in Sec.~\ref{sec:Observables}.

If we ignore such correlations and assume coefficients of equal size, the largest SMEFT contribution to $\Damu$ stems from the two dipole operators in 
Eq.~\eqref{eq:SMEFT-dipole-operator} as well as the following four-fermion operator
\begin{align}
	\La_\text{SMEFT} \supset C_{\underset{2233}{lequ}}^{(3)} \big(\bar l^a_{L2} \sigma^{\mu\nu} e_{R2}\big) \epsilon_{ab} 
	\big(\bar{q}_{L3}^b \sigma_{\mu\nu} u_{R3} \big)
\end{align}
involving the left- and right-handed muon and top-quark multiplets. Together their leading contributions are given by \cite{Buttazzo:2020ibd,Allwicher:2021jkr}
\begin{align}\label{SMEFTDamu}
	\begin{split}
		\Damu^{\text{SMEFT}} &= \frac{4m_\mu v}{e\sqrt{2}} \Re \Big[ c_W  C_{\underset{22}{eB}}(\mu) - s_W C_{\underset{22}{eW}}(\mu) \Big]
		- \frac{2m_\mu m_t}{\pi^2} \ln(\frac{\mu^2}{m_t^2}) \Re  C_{\underset{2233}{lequ}}^{(3)}(\mu).
	\end{split}
\end{align}
For a more elaborate seminumerical version of this equation that takes
into account higher-order corrections from RGE running we refer to
Ref.~\cite{Aebischer:2021uvt}. Similarly,
Ref.~\cite{Cirigliano:2021peb} has obtained the analogous result in an
extended EFT, $\nu$SMEFT, where neutrino masses are taken into account
and where additional operators can contribute.

\begin{figure}
	\centering
	\includegraphics[width=.4\textwidth]{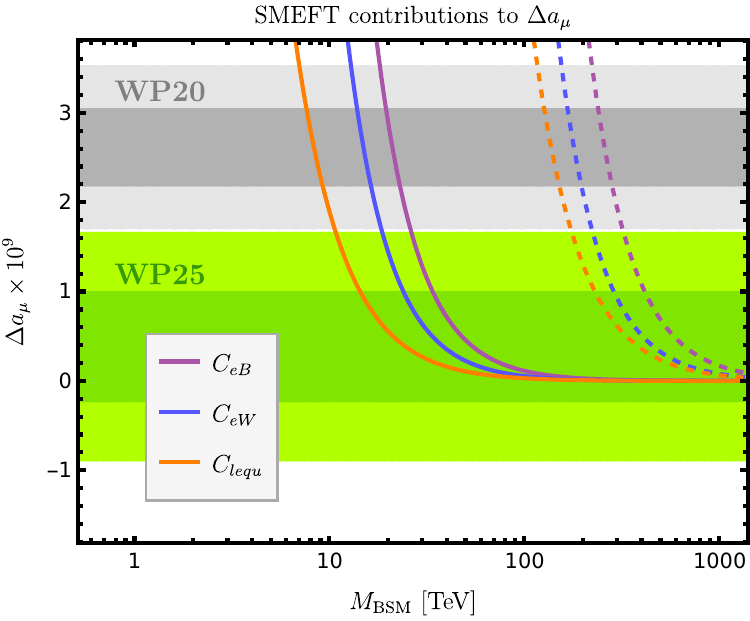}
	\caption{
		Contribution of the SMEFT Wilson coefficients $C_{eB}, C_{eW}$ and $C_{lequ}$ to $\Damu$ as a function of the BSM scale $M_\text{BSM}$.
		The continuous (dashed) lines correspond to
                coefficients generated by weakly (strongly) coupled
                new physics as described in the text.
        The green bands indicate the new result
                  $\DamuFinal$ and the grey bands the previous
                $\DamuOld$ (at the 1$\sigma$ and $2\sigma$ levels).}
	\label{fig:SMEFT-damu-scale}
\end{figure}

In Fig.~\ref{fig:SMEFT-damu-scale}
the  contributions to $\amu$ from the individual SMEFT Wilson coefficient are
plotted, illustrating how a non-zero deviation 
like $\DamuOld$ restricts the possible BSM mass scale from above,
while the current result $\DamuFinal$ restricts the possible mass
scale from below.
We show the two cases where either $C_i=1/M_\text{BSM}^2$,
corresponding to strongly interacting new physics (dashed), or
$C_i=1/(16\pi^2M^2_\text{BSM})$, corresponding to weakly interacting new physics
with large chiral enhancement (solid).

The plot shows the general behaviour in the simplest cases. Clearly,
the quantitative mass dependence in actual models depends also on
other model parameters and can differ from the ones shown in the plot. 
It is instructive to connect the SMEFT framework to the results of explicit
models. The simple 1-field and 2-field models without chiral
enhancement lead to
\begin{align}
  C_{\underset{22}{e\gamma}}\sim\frac{y_\mu |c|^2}{16\pi^2}\frac{e}{M^2_\text{BSM}}
\end{align}
in line with the behaviour of $\Damu$ in Eq.~(\ref{nonenhancedoneloop}). As an example of a model with chiral
enhancement, the Class I model with a common mass scale $m_F, m_S \sim M_\text{BSM}$ leads to (cf. Eq.~(\ref{eq:MIAcontribution}))
\begin{align}
  C_{\underset{22}{e\gamma}}\sim\frac{\lambda_L\lambda_R\lambda_\Phi}{16\pi^2}\frac{e}{M^2_\text{BSM}}.
\end{align}
Refs.~\cite{Buttazzo:2020ibd,Allwicher:2021jkr} have used the SMEFT framework to classify
plausible values of $\Damu$ from very heavy new physics in general:
  For strongly interacting new physics with large chiral enhancement,
  the relevant couplings can be 
  assumed to compensate the $1/16\pi^2$ suppression,\footnote{%
  The additional factor $e$ in this and the following formula compared
  to the generic formulas used for all Wilson coefficients in
Fig.~\ref{fig:SMEFT-damu-scale} is natural for couplings to the photon
but inconsequential for the following arguments.}
 such that
  \begin{align}
    C_{\underset{22}{e\gamma}}&\sim\frac{e}{M^2} &
    \Damu&\sim \frac{m_\mu v}{M^2}\sim
    \left(\frac{100\,\text{TeV}}{M}\right)^2\times 20\times10^{-10}.
  \end{align}
    This is arguably the largest plausible  magnitude of $\Damu$ for
  any given mass scale $M$, and it shows that measurable contributions
  to $\amu$ can be generated even by new physics as heavy as
  100\,TeV, without violating constraints from unitarity
  \cite{Allwicher:2021jkr}. 
Similarly, for weakly coupled new physics with large chiral enhancement, a
  suitable estimate is
  \begin{align}
    C_{\underset{22}{e\gamma}}&\sim\frac{1}{16\pi^2}\frac{e}{M^2} &
    \Damu&\sim \frac{m_\mu v}{16\pi^2 M^2}\sim
    \left(\frac{10\,\text{TeV}}{M}\right)^2\times 15\times10^{-10}.
  \end{align}
For each of these cases the SMEFT framework then allows to evaluate
the implication of a sizeable $\Damu$  on other observables. It turns
out that the same
new physics producing a deviation around $3\times10^{-9}$, i.e.\ as large as
the previous deviation (\ref{eq:DamuOld}), would inevitably lead to observable effects at a future
muon collider  \cite{Buttazzo:2020ibd,Bhardwaj:2022qtk}, see also
Ref.~\cite{Fortuna:2024rqp} for an analogous study in a related EFT HEFT.

\section{Relationships to other observables}\label{sec:Observables}

The observable $\amu$ is  flavour and CP conserving,
loop induced and chirality flipping. This set of properties implies
similarities,  differences and complementarities  to many other
observables. Such relationships are explored and described in this
section. We work on generic, model-independent levels and use
qualitative arguments, and we also use simple models for quantitative
details and technical explanations. Sec.~\ref{sec:MuonMass} continues
the discussion of Sec.~\ref{sec:ChiralityFlips} and focuses on the
connection to the muon mass implied by the common need for chirality
flips and electroweak symmetry breaking. It also explains special
cases such as models with radiative muon mass
generation. Sec.~\ref{sec:LeptonDipole} discusses differences and
relations with other dipole observables such as $a_e$, electric dipole
moments and flavour-violating decays. The muon--Higgs coupling also
shares many properties with $\amu$, though it appears at tree-level
and is measured at the LHC. Its connection with $\amu$ is explored in
Sec.~\ref{sec:muon-Higgs}. Electroweak precision observables such as
$M_W$ or $\sin^2\theta_W$ depend on quite different properties such as
custodial symmetry, but they share the importance of the hadronic
vacuum polarisation with $\amu$. The consequences are described in
Sec.~\ref{sec:EWPO}. Sections \ref{sec:neutrino_masses} and
\ref{sec:DarkMatter} focus neutrino masses and dark matter --- two
phenomena which both require BSM physics. These sections describe how
models explaining neutrino masses or dark matter can contribute to
$\amu$ and can be constrained by it. Finally, Sec.~\ref{sec:Collider}
focuses on the general complementarity of $\amu$ with collider
constraints, particularly from the LHC.

\subsection{Muon mass and generic estimates of
  possible $\amu$ contributions}
\label{sec:MuonMass}

The analysis of chirality flips, electroweak gauge invariance and
muon-chiral symmetry has revealed a deep connection between the muon
mass and the muon magnetic moment. Here we provide further details and
phenomenological conclusions for generic classes of BSM scenarios, as
already illustrated in Tab.~\ref{tab:estimates}.
The relationship between the two observables is expressed by the patterns
Eqs.~\eqref{mmugeneric} and \eqref{amugeneric} for any contribution to
$m_\mu$ and $a_\mu$ in the SM or any BSM extension.
These patterns can also be summarised
by the following expressions for  contributions to $\amu$ and $m_\mu$ in any given BSM scenario as (see also
Refs.\ \cite{Czarnecki:2001pv,Stockinger:2009fns,Stockinger:2022ata}),
\begin{align} \label{eqn:GeneralGM2Contribution}
  \Damu^{\text{BSM}} &=
  C_{\text{BSM}}\frac{m_\mu^2}{M_{\text{BSM}}^2} ,
  &
  \frac{\Delta m_\mu^{\text{BSM}}}{m_\mu} &=
  {\mathcal O}(C_{\text{BSM}}),
\end{align}
where $M_{\text{BSM}}$ is the relevant mass scale.
This equation highlights that a BSM
scenario contributing to $a_\mu$ generically also contributes to
$m_\mu$ and the contributions are related. The complicated origin of
the muon mass, the necessity for EWSB and chiral symmetry breaking are
all encapsulated in the dimensionless coefficient $C_{\text{BSM}}$. 
This coefficient summarises the
interesting chirality-flipping and EWSB factors in the square brackets in
Eqs.\ \eqref{mmugeneric} and \eqref{amugeneric} together with the
``other factors'', and is normalised to the full muon mass $m_\mu$.
This factor  reflects interesting dynamical
details of the model in question and is thus very
model-dependent.
In exceptional cases it can happen that the contributions to either
$m_\mu$ or $\amu$ in a model vanish or dominant terms arise at
different loop orders. In such cases the ${\mathcal O}(1)$ factors
between the two formulas in Eq.~(\ref{eqn:GeneralGM2Contribution})
correspond to zero or infinity, and in such BSM scenarios,
Eq.~(\ref{eqn:GeneralGM2Contribution}) does not capture the
dominant behaviour.

A qualitative conclusion of the relationship is that 
$a_\mu$ can provide a window to the muon mass generation mechanism:
According to the SM the muon mass is generated by EWSB generating a
Higgs \vev 
and the Yukawa interaction which couples the muon to the Higgs
\vev. 
But many open questions remain.
Hence many BSM scenarios  modify the Higgs sector and/or the Yukawa sector.
Technically, the muon mass generation mechanism is reflected by the
factors in square brackets in
Eqs.\ \eqref{mmugeneric} and \eqref{amugeneric}, hence many BSM scenarios
substantially modify these factors. The three-field models discussed
in Sec.~\ref{sec:genericthreefield} provide simple examples of such
modifications, and many models discussed in
Sec.~\ref{sec:Models} provide  further examples. 

For this reason, the $\amu$ phenomenology of  such
BSM scenarios with modified
Higgs/Yukawa sectors is particularly rich. They can lead to strong chiral
enhancements
and thereby provide potentially large and highly model-dependent
contributions to $a_\mu$. Such models were often considered promising
explanations  of a deviation as large as
$\DamuOld$. In general, the parameter spaces of such models are
severely constrained by $\amu$, and  the new result
$\DamuFinal$ sets an upper limit on chirally enhanced contributions
which  in individual models translates into limits on the modified
Higgs, Yukawa and related sectors.

On a technical level, the validity of
Eq.~(\ref{eqn:GeneralGM2Contribution}) in a wide range 
of scenarios can be illustrated by juxtaposing generic one-loop
formulas for $\Damu$ 
and for the muon mass 
correction $\Delta m_\mu$. Using the same conditions and notations as
in  Sec.~\ref{sec:genericoneloop}, and specialising for concreteness
to the case of the FS type contributions, we have
\begin{subequations}\label{oneloopexamples}
	\begin{alignat}{2}
		\Delta m_\mu &= \frac{1}{16\pi^2}\, m_\mu\, \bigg( 
		\big\{|c_L|^2+|c_R|^2\big\} \frac{1}{2}B_1
		&&-\frac{m_{F}}{m_\mu} \text{Re}\big\{c_L^*c_R\big\} B_0
		\bigg),
		\\
		\Delta\amu&=\frac{1}{16\pi^2}\frac{m_\mu^2}{m_S^2}
		\bigg(
		\big\{|c_L|^2+|c_R|^2\big\}~\G^\text{FS}
		&&+\frac{m_{F}}{m_\mu} \text{Re}\big\{c_L^*c_R\big\} \F^\text{FS} \bigg).
  \end{alignat}
\end{subequations}
Here the
loop functions $B_{0,1}\equiv B_{0,1}(m_\mu^2,m_F,m_S)$ are standard
Passarino-Veltman functions, and the $\Damu$ result corresponds to
Eq.~(\ref{eq:Damu-FS}) with suppressed arguments. As an illustration, in the
 limit $m_S=m_F\gg m_\mu$ the loop functions reduce to
 $B_0=-2B_1=1/\bar{\epsilon}+\ln\left(\mu^2/m_S^2\right)$ and
 $\G^\text{FS}=(1+2Q_S)/12$, $\F^\text{FS}=(2+3Q_S)/3$,
 where $\mu$ is the renormalisation scale and where $1/\bar{\epsilon}$
 is the dimensional regularisation parameter which is set to zero in the
$\overline{\text{MS}}$-renormalisation scheme. Hence in general all appearing loop
functions have values of ${\mathcal O}(1)$.
The analogous structure of $\Delta m_\mu$ and $\Delta\amu$ is
apparent. In both cases there are terms without chiral enhancement
governed by $|c_{L,R}|^2$ and a term with potential chiral enhancement governed
by $c_L^*c_R$. For each term, the contributions to $\amu$ and $m_\mu$
differ in the relative factor
$m_\mu/m_S^2$ and in the ${\mathcal O}(1)$ coefficients and loop
functions. Hence each term reflects the general relationship
stipulated by Eqs.~\eqref{mmugeneric} and \eqref{amugeneric}.

Relations such as Eq.~(\ref{eqn:GeneralGM2Contribution}) and the
discussion of chiral enhancements allow to obtain a qualitative
understanding of the behaviour of $\amu$ in many BSM scenarios, and
Tab.~\ref{tab:estimates} already presented a collection of estimates.
Here we specifically provide details on three classes of BSM scenarios
with particularly large contributions to $\Delta m_\mu$ and
$\Damu$. All of these can be exemplified by the three-field models of
Sec.~\ref{sec:genericthreefield} but are also relevant beyond this
simple setup. The classes are
\begin{itemize}
\item
  models with radiative muon mass generation,
\item
  models with ``finetuning in the muon mass'',
\item
  models saturating perturbative unitarity for chiral enhancements.
\end{itemize}
Models with radiative muon mass generation realise the idea that the
tree-level muon mass vanishes and instead the entire muon mass arises
via BSM loop effects, such that
\begin{align}\label{muonmassradmmu}
  \Delta m_\mu^{\text{BSM}} &= m_\mu.
\end{align}
The idea is appealing since it offers a chance
to explain the smallness of the muon mass; it might be accompanied by
radiative generation of the electron mass or other fermion masses
\cite{Czarnecki:2001pv,Baker:2020vkh}.
As emphasised in Ref.~\cite{Czarnecki:2001pv} (and highlighted in Tab.~\ref{tab:estimates} (c)), 
combining Eq.~(\ref{muonmassradmmu}) with Eq.~(\ref{eqn:GeneralGM2Contribution})
shows that models with radiative muon mass lead to a specific, large
value of $\Damu$,
\begin{align}\label{amuradmmu}
  \Damu &= {\mathcal O}(1)\frac{m_\mu^2}{M_{\text{BSM}}^2}.
\end{align}
In Refs.~\cite{Czarnecki:2001pv,Baker:2021yli,Yin:2021yqy} the impact of radiative muon mass
generation on $\amu$ has been investigated using several simple
examples, in
Refs.~\cite{Chiang:2021pma,Chiang:2022axu,Bonilla:2023wok} the idea is
expanded to more elaborate models which explain the lepton masses
radiatively and study connections to neutrino and dark matter
physics. In the context of supersymmetry, the tree-level muon mass 
can be set to zero either by setting one of two Higgs \vev{}s to zero,
or by setting the muon Yukawa coupling to zero, and the corresponding
impact on $\amu$ has been investigated in Refs.~\cite{Bach:2015doa}
and \cite{Borzumati:1999sp,Crivellin:2010ty,Thalapillil:2014kya}, respectively. Here we follow
Ref.~\cite{Baker:2021yli} and illustrate radiative muon mass
generation using the three-field models of Class II, III in
Sec.~\ref{sec:genericthreefield} in the special case where the muon
Yukawa coupling is set to zero; later in Sec.~\ref{sec:HeavySUSY} we
discuss it in the context of supersymmetry. In 
the  two three-field models, setting $y_\mu=0$ restores
muon-specific chiral symmetry on the dimension-4 level, while the BSM
parameters made explicit in Eq.~(\ref{3fieldchiralenhancements}) break chiral symmetry only
softly on the dimension-3 level. Hence in these cases, the loop
contribution $\Delta m_\mu$ is
guaranteed to be finite and unaffected by renormalisation. In
contrast, setting $y_\mu=0$ in the three-field Class I model does not restore
chiral symmetry on the dimension-4 level.

For example, in the Class II model, the chirally enhanced muon mass
correction and $\Damu$ can be 
written as
\begin{subequations}
	\begin{alignat}{2}
		\frac{\Delta m_\mu^{\text{Class II}}}{m_\mu} &= C^{\text{II}}~\times
		&&\Big[B_0(0,m_1,m_\phi)-B_0(0,m_2,m_\phi)\Big]
		,\\
		\Damu^{\text{Class II}} &= C^{\text{II}}\frac{m_\mu^2}{ m_\phi^2}
		&&\Big[\mathcal{F}^\text{FS}(0,x_2;Q_S)-\mathcal{F}^\text{FS}(0,x_1;Q_S) \Big],	
	\end{alignat}
\end{subequations}
using the common prefactor
\begin{align}
	C^{\text{II}} &=
	\frac{\lambda_L\lambda_R}{16\pi^2} \frac{ \bar\lambda_\Phi v }{\sqrt{2}m_\mu} \times
	\frac{m_1m_2}{m_2^2-m_1^2}.
\end{align}
Here the generic result Eq.~(\ref{oneloopexamples}) has been
specialised to the Feynman rules given around
Eq.~(\ref{couplingsClassII}) and where $x_{i}=m_{i}^2/m_\phi^2$ as in
Eq.~(\ref{eq:3field-amu}). By requiring radiative muon mass generation,
$m_\mu=\Delta m_\mu^{\text{Class II}}$, the prefactor can be
eliminated, and the resulting $\Damu$ can be expressed  via
a ratio of loop functions,
\begin{align}\label{DamuClassIIrad}
  \Damu^{\text{Class II, rad $m_\mu$}} &= \frac{m_\mu^2}{
    m_\phi^2}\left[
    \frac{\mathcal{F}^\text{FS}(0,x_2;Q_S)
      -\mathcal{F}^\text{FS}(0,x_1;Q_S)
    }{B_0(0,m_1,m_\phi)-B_0(0,m_2,m_\phi)}
    \right].
\end{align}
It is further illuminating to record the result for the simple case
where all masses are similar, $m_1\approx m_2\approx
m_\phi$. In this case, the muon mass and $\Damu$ can be written as
\begin{align}\label{ClassIIradmmusimple}
m_\mu^{\text{Class II, rad $m_\mu$}} &
  \approx
  \frac{\lambda_L\lambda_R}{32\pi^2} \frac{ \bar\lambda_\Phi v }{\sqrt{2}}
,&
  \Damu^{\text{Class II, rad $m_\mu$}} &
  \approx
  \frac{m_\mu^2}{
    m_\phi^2}\left[\frac{1}{3}+\frac{4Q_F}{3}\right].
\end{align}
These equations are all in line with the
discussion above. The contributions to the muon mass and to $\amu$ are
related in the expected way, the model-dependent prefactors are the
same, and the final $\Damu$ behaves as expected in
Eq.~(\ref{amuradmmu}). The ${\cal O}(1)$ prefactor of that equation corresponds to
the ratio of dimensionless loop functions in the square brackets in
Eqs.~(\ref{DamuClassIIrad}) or (\ref{ClassIIradmmusimple}). Interestingly, this prefactor 
can be positive or negative, and in exceptional cases it can also
vanish. For the important cases where the charges of the BSM fermion
is either $Q_F=-1$ or $Q_F=0$, the prefactor
becomes $-1$ or $+1/3$, respectively.

Similarly, for the Class III model, the results are
\begin{subequations}
	\begin{alignat}{2}
		\frac{\Delta m_\mu^{\text{Class III}}}{m_\mu} &= C^{\text{III}}~\times&&\Big[B_0(0,m_1,m_\psi)-B_0(0,m_2,m_\psi)\Big]
		,\\
		\Damu^{\text{Class III}} &=   C^{\text{III}}\frac{m_\mu^2}{ m_\psi^2}
		&&\Big[z_2\mathcal{F}^\text{FS}(0,z_2;Q_S)
		-z_1\mathcal{F}^\text{FS}(0,z_1;Q_S) \Big],
	\end{alignat}
\end{subequations}
using the common prefactor
\begin{align}
  C^{\text{III}} &=
  \frac{\lambda_L\lambda_R}{16\pi^2} \frac{ a_\Phi v }{\sqrt{2}m_\mu} \times
  \frac{m_\psi}{m_2^2-m_1^2},
\end{align}
the result for the case of radiative muon mass generation is
\begin{align}
  \Damu^{\text{Class III}} &=  \frac{m_\mu^2}{
    m_\psi^2}\bigg[\frac{z_2\mathcal{F}^\text{FS}(0,z_2;Q_S)
      -z_1\mathcal{F}^\text{FS}(0,z_1;Q_S)}{B_0(0,m_1,m_\psi)-B_0(0,m_2,m_\psi)}\bigg],
\end{align}
and the approximate results in the equal-mass limit are
\begin{align}\label{ClassIIIradmmusimple}
  m_\mu^{\text{Class III}} & \approx
  \frac{\lambda_L\lambda_R}{32\pi^2} \frac{ a_\Phi v }{\sqrt{2}m_\psi} 
  ,&
  \Damu^{\text{Class III}} &    \approx
  \frac{m_\mu^2}{
    m_\psi^2}\left[\frac13 +\frac{2Q_F}{3}\right].
\end{align}
The results for these two explicit models are representative for the general class of
radiative muon mass generation models. In all cases, there is a strong
chiral enhancement that at first depends on the chirality-flipping
model parameters. These model parameters can be eliminated, leaving a
prediction for $\Damu$ that involves the typical factor
$m_\mu^2/M_{\text{BSM}}^2$ and where the complicated model parameter
dependence has cancelled. Most prominently, the overall sign of
$\Damu$ is essentially fixed and does not depend on the sign of the
model parameters. Numerically, the contributions can be large even for
BSM masses in the multi-TeV region. Conversely, $\amu$ constitutes a
strong constraint on radiative muon mass models.

In the general case with non-zero tree-level muon mass it is possible
to consider the situation
\begin{align}\label{muonmassfinetuning}
  |\Delta m_\mu^{\text{BSM}}| &> m_\mu.
\end{align}
Here the radiative corrections are  larger than required by radiative
muon mass generation, and the total muon mass
arises from a cancellation between tree-level and higher-order
effects. Accordingly, the contributions to $\amu$ can be larger than
in Eq.~(\ref{amuradmmu}), i.e.
\begin{align}
|    \Damu| &\gtrsim\frac{m_\mu^2}{M_{\text{BSM}}^2}.
\end{align}
It is clear from the above that
already the simple three-field models considered above allow this to
happen if the model couplings are larger than what is required by 
Eqs.~\eqref{ClassIIradmmusimple} and \eqref{ClassIIIradmmusimple}.
Nevertheless, the possible values can be limited from above by two relevant
considerations.

First, a cancellation between contributions to the muon mass might be
considered as undesirable finetuning. If no cancellation at all is
allowed, Eq.~(\ref{amuradmmu}) constitutes an upper limit on $\Damu$;
conversely, a value of $\Damu\sim10^{-9}$ can only be reached for
$M_{\text{BSM}}\lesssim3$\,TeV. If a finetuning to a degree of
e.g.\ 10\% \cite{Calibbi:2020emz} or 
1\% \cite{Capdevilla:2021rwo} is allowed, the maximum value of $\Damu$
increases accordingly.

A second consideration is perturbative unitarity.
While finetuning constraints can be considered optionally, unitarity
constraints are quantum field theoretically stricter. Such unitarity
constraints can be derived within explicit models such as the
three-field models defined in Sec.~\ref{sec:genericthreefield}
\cite{Capdevilla:2020qel,Capdevilla:2021rwo}.
In this context, the
unitarity constraints are equivalent to the absence of
non-perturbative new physics at the energy scale of consideration, and e.g.\ for the three-field models of
Classes I, II, Refs.~\cite{Capdevilla:2020qel,Capdevilla:2021rwo} obtain the unitarity limits
$|\lambda_{L,R}|<\sqrt{8\pi}$ and
$|\lambda_\Phi,\bar\lambda_\Phi|<\sqrt{4\pi}$; the limits for the Class III
model are similar. Thus perturbative unitarity implies an upper limit
on $\Damu$ for any given mass scale. However this upper limit is huge;
for example even in models where
$M_{\text{BSM}}\lesssim100$\,TeV, $\Damu\sim20\times10^{-10}$ is
possible \cite{Capdevilla:2020qel,Capdevilla:2021rwo}, confirming EFT
results described in Sec.~\ref{sec:genericeft}.
In general, the results obtained in these references can be summarised
\begin{align}\label{eq:amu-max-pert}
  \Damu^{\text{max,perturbative unitarity}} & \sim
  \frac{1}{16\pi^2}
  \frac{m_\mu[(8\pi)^{3/2} v]}{M^2}
  &&
  \approx 
  \left(\frac{100\text{\, TeV}}{M}\right)^2\times20\times10^{-10}.
\end{align}
Hence even such
heavy BSM scenarios can be constrained by $\DamuFinal$.  At the same
time, the results of these references and the related
Refs.~\cite{Buttazzo:2020ibd,Paradisi:2022vqp} imply that muon colliders have a strong
sensitivity to essentially all high-scale models which involve
sizeable corrections to $\amu$.

\subsection{Other leptonic dipole observables and
  charged lepton flavour violation}
\label{sec:LeptonDipole}

Next to $\amu$, there is a number of further related dipole observables in the lepton sector.
This includes the anomalous magnetic moment of the electron $a_e$,
the electric dipole moments (EDM) of the electron and muon $d_{e,\mu}$
as well as several charged lepton flavour violating (CLFV) processes such as $\mu\to e\gamma$.
The magnetic dipole moments $a_e$ and $\amu$  are CP- and flavour-conserving
observables, but correspond to different generations, and comparing them as in Eq.~\eqref{eq:g_e-mu-p} 
provides a high precision test of the universality of the lepton interactions.
The other dipole observables violate CP and flavour conservation and hence test different aspects of fundamental interactions. At the same time, 
all mentioned observables arise via very similar operators and are all sensitive to chirality flips in the sense discussed in Sec.~\ref{sec:ChiralityFlips}. For this reason,
BSM contributions to all dipole observables could be related. Here we explore  relationships between these and further related CLFV observables
in a generic, model-independent way.

We begin with a collection of the SM predictions and experimental results.
For the electron $a_e$, the current experimental measurement reads \cite{Fan:2022eto,ParticleDataGroup:2024cfk} 
\begin{align}
	a_e^{\text{Exp}} &= 1\,159\,652\,180.62(12)\times10^{-12}\ \text{\cite{ParticleDataGroup:2024cfk}} \label{aeExp},
\end{align}
and the SM prediction is given by
\begin{subnumcases}{a_e^\text{SM}=}
	1\,159\,652\,181.61(23)\times10^{-12} & for $\alpha_{\text{Cs}}$, \,\cite{Parker:2018vye} \label{aeSM2018} \\
	1\,159\,652\,180.252(95)\times10^{-12} & for $\alpha_{\text{Rb}}$. \cite{Morel:2020dww} \label{aeSM2020}
\end{subnumcases}
The two SM values were compiled in Ref.\ \cite{Keshavarzi:2021eqa} and
are based on a long series of high-precision theory calculations
reviewed in Ref.\ \cite{Aoyama:2019ryr} and
two different measurements of the fine-structure constant
$\alpha$ using atomic interferometry via either caesium or rubidium
\cite{Parker:2018vye,Morel:2020dww}. These $\alpha$ measurements
differ by more than 5 standard deviations; hence at present there is
no clear evidence for BSM physics in  $a_e$.\footnote{%
\label{footnoteae}
At the time the result (\ref{aeSM2018}) was published, the difference
between experiment and the SM prediction of $\Delta a_e$ was negative,
and a variety of studies analysed to what extent this deviation could
be explained by BSM physics, in particular in conjunction with the
positive $\Damu$ at the time. For an overview we refer to
Ref.~\cite{Crivellin:2018qmi}, further studies will be mentioned in
Sec.~\ref{sec:Models}.} 
Clearly, though, the magnitude of BSM contributions to $a_e$ is
bounded to be below $|\Delta a_e^{\text{BSM}}|\lesssim 10^{-12}$.

The EDMs $d_\ell$ are CP-violating observables.
So far, the only known source of CP
violation (CPV) stems from the CKM matrix in the quark sector of the SM.
The lepton EDMs resulting from this CPV are first generated at the 4-loop level yielding \cite{Booth:1993af,Pospelov:2005pr,Yamaguchi:2020eub,Yamaguchi:2020dsy}
\begin{align}
	d_{e,\mu,\tau}^\text{SM} \sim 10^{-40} ... 10^{-38}~ e\cdot \text{cm}.
\end{align}
Similarly, LFV has so far only been established firmly in the neutral sector through the discovery of neutrino oscillation.
The required (tiny) neutrino masses induce CLFV through radiative corrections, however, this is again very strongly suppressed
\cite{Petcov:1976ff,Marciano:1977wx, Lee:1977tib}
\begin{align}
	\text{BR}(\mu\to e\gamma)_\text{SM} \sim 10^{-50}.
\end{align}
Consequently, a measurement of non-vanishing electric dipole moments (within the projected future sensitivities)
or observation of CLFV processes would pose an unambiguous signal for physics beyond the Standard Model.

For the CP-violating EDMs $d_e$ and $d_\mu$, so far no nonzero signals have been found. 
The most sensitive searches for the electron and muon EDM have been obtained 
using   electrons confined in HfH$^{+}$ ions and thus subjected to huge
intramolecular fields and in the Brookhaven
Muon g-2 experiment, respectively,
\begin{subequations}\label{eq:EDM-bounds}
	\begin{alignat}{3}
		&|d_e| &&< 4.1\times10^{-30} ~e\cdot\text{cm} \quad  &&\text{\cite{Roussy:2022cmp}},
		\label{deBound}\\
		&|d_\mu| && < 1.9 \times 10^{-19} ~e\cdot\text{cm} \quad &&\text{\cite{Muong-2:2008ebm}}.
		\label{dmuBound}
	\end{alignat}
\end{subequations}
Similarly, the CLFV decays $\ell_i\to\ell_j\gamma$ have not yet been observed.
The best available limits have been obtained by the MEG experiment at PSI and the BaBar and Belle collaborations
\begin{subequations}\label{eq:ltolgamma-bounds}
	\begin{alignat}{3}
	  &\text{BR}(\mu\to e\gamma) &&<  3.1 \times 10^{-13}\quad &&\text{\cite{MEGII:2023ltw}}, \label{MuegBound} \\
		&\text{BR}(\tau\to e\gamma) &&< 3.3\times 10^{-8}  &&\text{\cite{BaBar:2009hkt}}, \\
		&\text{BR}(\tau\to \mu\gamma) &&< 4.2 \times 10^{-8} &&\text{\cite{Belle:2021ysv}}.
	\end{alignat}
\end{subequations}
The absence of
signals in these experiments puts strong constraints on the possible CP and
flavour textures of potential BSM scenarios. Still, there is good potential that such signals could be seen in future experiments.
On the one hand, it has been known for a long time that the CPV of the SM is insufficient to address
the baryon-antibaryon asymmetry in the early universe. Many models introduced to resolve
this issue also induce lepton EDMs significantly above the SM value.
On the other hand, lepton flavour is an accidental symmetry of the SM that is generically broken by higher-dimensional terms like the four-fermion
and, in particular, dipole operators. Thus new physics scenarios above
the electroweak scale often induce new sources of CPV and/or CLFV,
leading to
enhanced rates potentially observable in the future, for reviews see e.g.~\cite{Pospelov:2005pr,Roberts:2009xnh,Lindner:2016bgg,Vives:2025clr}.

Since all observables considered here are defined at low energies, it
is appropriate to describe them using an effective Lagrangian. The appropriate framework is the one of LEFT, 
discussed in Sec.~\ref{sec:genericeft}. In particular,
the dipole operator introduced in Eq.~\eqref{eq:LEFT-dipole-operator}
can be generalised to
\begin{align}\label{eq:LEFT-dipole-LFV}
	\La \supset L_{\underset{ij}{e\gamma}} \, \bar{e}_{Li} \sigma^{\mu\nu} e_{Rj} F_{\mu\nu} + h.c.
\end{align}
The diagonal Wilson coefficients  induce the dipole moments
	\begin{align}
	\Delta a_i & = \frac{4 m_i}{e} \Re L_{\underset{ii}{e\gamma}},
	\\
	d_i &= 2 \Im L_{\underset{ii}{e\gamma}},
	\end{align}
while the off-diagonal Wilson coefficients contribute to the CLFV processes, most notably
\cite{Crivellin:2013hpa, Pruna:2014asa, Lindner:2016bgg}
\begin{align}
	\text{BR}(\ell_i\to \ell_j\gamma) = \frac{m_i^3}{4\pi\Gamma_i} \Big(|L_{\underset{ij}{e\gamma}}|^2 + |L_{\underset{ji}{e\gamma}}|^2\Big).
\end{align}

These equations make the connections between the dipole observables manifest.
Although the different coefficients $L_{\underset{ij}{e\gamma}}$ are in general independent, in many concrete BSM scenarios 
they arise simultaneously from common origins and, as a consequence, are correlated \cite{Giudice:2012ms,Lindner:2016bgg,Crivellin:2018qmi,Vives:2025clr}. 
The nature of these relationships is of course model-dependent and can manifest in different ways.
For a general, qualitative discussion we follow Refs.~\cite{Giudice:2012ms,Crivellin:2018qmi}. As e.g.~in Ref.~\cite{Giudice:2012ms} we first introduce
generic CPV phases $\phi_i$ and effective 
	flavour mixing angles $\theta_{ij}$ (for $i\neq j$) as
	\begin{align}\label{Paradisiparametrisation}
		\tan \phi_i \equiv \frac{\Im L_{\underset{ii}{e\gamma}} }{ \Re L_{\underset{22}{e\gamma}}}, \qquad \text{and} \qquad
		\theta_{ij} \equiv \frac{L_{\underset{ij}{e\gamma}} }{L_{\underset{22}{e\gamma}}},
	\end{align}
        where we will assume $\theta_{ij}$ to be real for
        simplicity. 
It is instructive to use these generic parameters to express several potential BSM properties as
\begin{subequations}
	\begin{alignat}{6}
		\text{no CPV}: \quad
		&&|\tan \phi_i| &= 0 \qquad\qquad
		&\text{no LFV}:\quad
		&&|\theta_{ij}| &= 0,
		\qquad
		\\
		\text{weak CPV}:\quad
		& &|\tan \phi_i| & \ll 1 \qquad\qquad
		&\text{weak LFV}:\quad
		&&|\theta_{ij}| &\ll 1,
		\qquad
		\\
		\text{generic CPV}:\quad
		& &|\tan \phi_i| & \sim 1 \qquad\qquad
		&\text{generic LFV}:\quad
		&&|\theta_{ij}| &\sim 1  .
	\end{alignat}
\end{subequations}
In the absence of any fundamental mechanism, one may expect that BSM contributions generically violate both CP and lepton flavour; however, certain fundamental BSM properties might guarantee that these symmetries are only weakly or not at all violated.

Using the parametrisation (\ref{Paradisiparametrisation})  we can express generic and qualitative relationships
between the observables in semi-numerical form. For the EDMs we can write
\begin{subequations}\label{eq:EDM-semi-num}
	\begin{alignat}{2}
		d_e &\approx \Big(\frac{\Delta a_e}{10^{-12}}\Big)
                \Big(\frac{\tan\phi_e}{2\times 10^{-7}}\Big) &&\ 4.1 \times 10^{-30} ~ e\cdot\text{cm}, \label{eq:e-EDM-num}\\
		d_\mu &\approx \Big(\,\frac{\Delta a_\mu}{10^{-9}}
                \,\Big) ~\Big(\frac{\tan\phi_\mu}{2000}\Big) &&\ 1.9 \times 10^{-19} ~ e\cdot\text{cm}. \label{eq:mu-EDM-num}
	\end{alignat}
\end{subequations}
A similar expression for $\mu\to e\gamma$ reads
	\begin{align}\label{eq:mueg-semi-num}
		\text{BR}(\mu\to e\gamma) \approx 
		\Big(\frac{\Delta a_\mu}{10^{-9}}\Big)^2 	
                \Big( \frac{\theta_{12}^2 + \theta_{21}^2}{(4.3\times 10^{-5})^2} \Big)
		\Big(1 + \tan^2\phi_\mu\Big)^2\ 3.1\times 10^{-13}.
	\end{align}

These three equations illuminate the relationships between the dipole observables in various ways.
First, comparing to the experimental bounds in
Eqs.~\eqref{eq:EDM-bounds} and \eqref{eq:ltolgamma-bounds} the product of the brackets on the r.h.s.\ must be smaller than 1.
In addition, the $\Delta a_i$ are normalised to their current
experimental sensitivities, such that these terms must also be $\lesssim 1$. In the electron sector,
Eq.~\eqref{eq:e-EDM-num} for the EDM $d_e$ therefore implies that BSM
contributions to $a_e$ are either far below the current experimental
sensitivity, or that they must be essentially CP conserving with
$\tan\phi_e\lesssim10^{-7}$. In contrast, Eq.~\eqref{eq:mu-EDM-num} for
the muon EDM $d_\mu$ still allows for a sizeable $\Damu$ even in the
presence of large CPV $\tan\phi_\mu\gg 1$.

Similarly, Eq.~\eqref{eq:mueg-semi-num} shows that sizeable contributions to the muon
magnetic moment $\Damu$ at the level of the current sensitivity are only possible in models with $\theta_{12,21}\lesssim 10^{-5}$, i.e.\
which are essentially flavour conserving. 
Such a suppression typically requires some additional symmetries or underlying mechanisms that impose
specific flavour structures. This is illustrated e.g.\ in Ref.~\cite{Lindner:2016bgg}
where a number of simplified models have been explored in the context
of CLFV and $\amu$. In addition, Refs.~\cite{Calibbi:2021qto,Lopez-Ibanez:2021yzu,} 
consider the connection between $\amu$ and CLFV in models with
discrete flavour symmetries, and Ref.~\cite{Isidori:2021gqe} considers
the generic situation also including renormalisation-group running,
which further strengthens the conclusion that sizeable BSM
contributions to $\amu$ are only viable in the presence of non-trivial
flavour symmetry properties of the underlying BSM mechanisms.

Here we will consider two particular general patterns of BSM contributions that appear in a variety of scenarios.
First, if the SM Yukawa couplings are the only source           
of chiral symmetry breaking, the discussion of
Sec.~\ref{sec:ChiralityFlips} shows that the contributions are typically related by
\begin{align}
	\frac{L_{\underset{11}{e\gamma}}}{
              L_{\underset{22}{e\gamma}} } \sim \frac{m_e}{m_\mu}.
\end{align}
This behaviour is often called \emph{naive scaling} (n.s.) \cite{Giudice:2012ms}. It also arises in BSM scenarios when additional chirality breaking BSM parameters are present 
but are forced to be proportional to the SM Yukawa couplings by an underlying mechanism. 
If naive scaling holds, the coefficient $C_{\text{BSM}}$ introduced in Eq.~(\ref{eqn:GeneralGM2Contribution}) 
is generation-independent. A discussion on the related minimal flavour violation in the lepton sector 
can be found in Ref.~\cite{Cirigliano:2005ck}.

In case of naive scaling, additional and tighter relationships between the electron and muon dipole moments emerge,
\begin{subequations}
	\begin{align}
		\Delta a_e^{(\text{n.s.})} &\sim \frac{m_e^2}{m_\mu^2} \Delta a_\mu \approx
		    2.3\times 10^{-5} \Delta a_\mu,\\
		d_e ^{(\text{n.s.})} &\sim \frac{m_e}{m_\mu} d_\mu \approx 4.8 \times 10^{-3}
		    d_\mu .
	\end{align}
\end{subequations}
As a consequence, the currently strongest experimental bounds on $\Damu$ and $d_e$
imply mutually stronger constraints of the orders
\begin{align}\label{eq:Dipole-NS-Bounds}
	|\Delta a_e^{(\text{n.s.})}| & \lesssim 10^{-13}, \qquad \text{and} \qquad
	| d_\mu^{(\text{n.s.})}| \lesssim 10^{-27}~ e\cdot\text{cm} 
\end{align}
on the dipole moments of the opposite generations.
In particular, combined with Eq.~\eqref{eq:EDM-semi-num} we arrive at
a relation between $d_e$ and $\Damu$,
\begin{align}
	d_e ^{(\text{n.s.})}&\approx \Big(\,\frac{\Delta a_\mu}{10^{-9}} \,\Big)
        ~\Big(\frac{\tan\phi_\mu}{10^{-5}}\Big) \ 4.1 \times
        10^{-30} ~ e\cdot\text{cm}.
\end{align}
Together with Eq.~(\ref{eq:e-EDM-num}), this  requires either $\Delta\amu$ to be well below current experimental sensitivities, 
or CP conservation to an excellent degree in both the muon and the electron sector, $\tan\phi_{\mu,e} < 10^{-5}$.

As a second special pattern of BSM contributions we consider the relations implied by a \emph{single particle contribution} (s.p.c.) \cite{Crivellin:2018qmi}.
This corresponds to the special case of a single BSM particle which couples to leptons with
generation-dependent couplings $c_i$ and produces a contribution to the dipole operators. In this case, the Wilson coefficients scale as
$L_{\underset{ij}{e\gamma}}\propto c_i c_j$ and are thus related as
\begin{align}\label{Crivellinsingleparticle}
	L_{\underset{ij}{e\gamma}}
	\sim
	L_{\underset{ji}{e\gamma}}
	\sim
	\sqrt{L_{\underset{ii}{e\gamma}}
	L_{\underset{jj}{e\gamma}} }.
\end{align}
Under this assumption, BR$(\mu\to e\gamma)$ can be written as
\begin{align}\label{LFVsingleparticleresult}
		\text{BR}(\mu\to e\gamma)^{(\text{s.p.c.})}& \approx 
		\Big(\frac{\Delta a_e}{10^{-12}}\Big)\Big(\frac{\Delta
                  a_\mu}{10^{-9}}\Big) \Big(2.2\times 10^8\Big) 
		\Big(1 + \tan^2\phi_\mu\Big)^2 \ 3.1\times 10^{-13} .
\end{align}
In the case of such simple scenarios  the existing $\mu\to e\gamma$
bounds thus exclude 
simultaneously sizeable contributions to $\Delta a_e$ and
$\Damu$ at the level of the current sensitivities
\cite{Crivellin:2018qmi}. This observation 
was of interest particularly before the result (\ref{aeSM2020}) became
available. As mentioned in footnote \ref{footnoteae}, the slight discrepancy
between Eqs.\ (\ref{aeExp}) and (\ref{aeSM2018}) sparked numerous
attempts to explain the negative $\Delta a_e$ simultaneously with
positive $\Delta a_\mu$, which are constrained by Eq.~(\ref{LFVsingleparticleresult}).

\begin{table}[t]
	\centering
	\begin{tabular}{|c||c|c|c|}
		\hline
		Observable & Future sensitivity & Experiment \\ \hline\hline
		$\text{BR}(\mu\to e\gamma)$   & $6\times10^{-14}$ &
                MEG II \cite{MEGII:2018kmf} \\\hline
		$\text{BR}(\mu\to 3e)$  & $2\times10^{-15}$ & Mu3e
                (phase 1) \cite{Mu3e:2020gyw} \\ \hline
		$\text{BR}(\mu\to 3e)$  & $10^{-16}$ & Mu3e
                (phase 2) \cite{Mu3e:2020gyw} \\ \hline
		$\text{BR}(\mu \text{Al}\to e \text{Al})$ &
                $7\times10^{-15}$ & COMET (phase 1) \cite{COMET:2018auw} \\ \hline
		$\text{BR}(\mu \text{Al}\to e \text{Al})$ &
                $3\times10^{-17}$ & COMET (phase 2) \cite{COMET:2018auw}, Mu2e \cite{Mu2e:2014fns} \\ \hline\hline
		$\text{BR}(\tau\to \mu\gamma)$  & $10^{-9}$ &
                \multirow{5}{*}{Belle II \cite{Belle-II:2018jsg} (Figure~189)}
                \\ \cline{1-2}
                $\text{BR}(\tau\to e\gamma)$  & $3\times10^{-9}$ &
                \\ \cline{1-2}                
                $\text{BR}(\tau\to 3\mu)$  & $3.5\times10^{-10}$ &
                \\ \cline{1-2}                
                 $\text{BR}(\tau\to \mu \pi)$  & $4.5\times10^{-10}$ &
                \\ \cline{1-2}
                 $\text{BR}(\tau\to \mu ee)$  & $3\times10^{-10}$ &
                \\ \hline
	\end{tabular}
	\caption{Future experimental sensitivities to a selection of
          CLFV observables. The first set involves $\mu\to e$
          transitions, the second set involves $\tau$-lepton decays
          into final states involving muons and/or electrons. The
          decays of the type $\ell_i\to \ell_j\gamma$ are generated by
        dipole operators, the other decays can be generated also by
        other operators.}
	\label{tab:futureCLFV}
\end{table}

In the following we focus more broadly on CLFV observables. Table
\ref{tab:futureCLFV} collects a selection of promising observables and
the sensitivities of planned experiments.
In the $\mu$--$e$ sector, there are two particularly important
CLFV processes complementary to $\mu\to e\gamma$. These are the decay
$\mu^+\to e^+ e^- e^+$ ($\mu\to 3e$ for short) and $\mu\to e$
conversion $\mu\text{N}\to e\text{N}$, where  a muon bound in the
field of a nucleus N converts into an electron, where the nucleus
absorbs the recoil. 
These processes receive contributions not only from the 
dipole terms in Eq.~\eqref{eq:LEFT-dipole-LFV} but also from chirality-preserving four-fermion operators.
Such observables are therefore particularly relevant in scenarios without chiral enhancements, where
the bounds from $\mu\to e\gamma$ and the dipole moments are much less
stringent, and Refs.~\cite{Ardu:2023yyw,Ardu:2024bua} provide an overview of the interplay
between various CLFV observables.

For our purposes in connection with $\amu$ it is interesting to estimate the impact of the additional
muon decay observables in the case of
\emph{dipole dominance} (d.d.), i.e. 
in the case where the non-enhanced BSM contributions are absent or can be neglected. In this case
the expressions for the branching ratios reduce to \cite{Kuno:1999jp,Kitano:2002mt,Crivellin:2013hpa,Lindner:2016bgg,Crivellin:2017rmk}
\begin{align}\label{dipoledominancemu3e}
	\text{BR}(\mu\to 3e)^{\text{(d.d.)}} &\simeq \frac{\alpha m_\mu^3}{48\pi^2 \Gamma_\mu} 
	\bigg[8\ln(\frac{m_\mu}{m_e}) - 11\bigg] \times \Big(|L_{\underset{12}{e\gamma}}|^2 + |L_{\underset{21}{e\gamma}}|^2\Big)
	\approx \text{BR}(\mu\to e\gamma) \ 6\times 10^{-3}
        \end{align}
and
\begin{align}\label{dipoledominancemue}
	\text{BR}(\mu N\to e N)^{\text{(d.d.)}} &\simeq \frac{m_\mu^3 D^2_N}{4\Gamma_\text{capt}^N} \Big(|L_{\underset{12}{e\gamma}}|^2 + |L_{\underset{21}{e\gamma}}|^2\Big)
	\approx \text{BR}(\mu\to e\gamma) \ \begin{cases}
		2.7\times 10^{-3} & (\text{Al}), \\
		3.9\times 10^{-3} & (\text{Au}),
	\end{cases}
\end{align}
where the result in the second case depends on the nucleus $N$.
Thus in case of dipole dominance all these processes are strictly
correlated by fixed, known factors; the additional suppression factors appearing here
arise from an additional phase space factor and from nuclear overlap
integrals, respectively.  
The branching ratios dictated by dipole dominance  are therefore
roughly a factor of $1000$ smaller than for $\mu\to e\gamma$. Comparing
to the bounds set by SINDRUM and SINDRUM II, 
\begin{subequations}
	\begin{alignat}{2}
		\text{BR}(\mu\to 3e) &< 10^{-12} \qquad &&\text{\cite{SINDRUM:1987nra}}, \\
		\text{BR}(\mu \text{Au}\to e\text{Au}) &< 7\times 10^{-13} \qquad &&\text{\cite{SINDRUMII:2006dvw}},
	\end{alignat}
\end{subequations}
and to Eq.~(\ref{MuegBound}), shows that currently these processes do
not constitute additional constraints beyond $\mu\to e\gamma$ on BSM
scenarios with dipole dominance.
However, as Tab.~\ref{tab:futureCLFV} shows, the sensitivities in the
measurements are expected to improve 
by a factor of more than $10^4$ in the upcoming 
Mu3e experiment at PSI  for $\mu\to 3e$ as well as
 the COMET and Mu2e
experiments at J-PARC and
Fermilab  for $\mu\to e$ conversion in the presence of
an Al-nucleus.
These future experiments will have a sensitivity to CLFV that
can surpass the one of $\mu\to e\gamma$ even in case of dipole
dominance, and they additionally have a unique sensitivity to
non-dipole operators.

Finally we also provide qualitative comments on CLFV observables
involving $\tau$ decays. Table~\ref{tab:futureCLFV} lists a small
selection of $\tau$-decay modes that can be investigated at the
Belle II experiment.
Specifically the so-called ``golden modes'' \cite{Belle-II:2018jsg}
$\tau\to\mu\gamma$ and $\tau\to3\mu$ behave analogously to $\mu\to
e\gamma$ and $\mu\to 3e$. Similarly to the discussion above in the
$\mu$--$e$ sector, the correlation
of $\tau\to\mu\gamma$ to $\amu$ and the correlation between
$\tau\to\mu\gamma$ and $\tau\to3\mu$ in case of dipole dominance can
be analysed in 
analogous ways. The $\tau$-lepton sector offers a wide range of
additional observables. There are observables with $\tau$--$e$
transitions or even observables that involve all three lepton flavours
(such as $\tau\to \mu ee$ in the table), and observables with leptons
and hadrons in the final state  (illustrated by $\tau\to \mu\pi$ in
the table). The latter may be sensitive to special
classes of BSM physics, e.g.\ to leptoquarks.
For a brief account of how the $\tau$-decay observables provide complementary
constraints on BSM physics and on EFT operators we refer to section
15.2.1 of Ref~\cite{Belle-II:2018jsg}.

To summarise, the three  observables with $\mu\to e$ transitions
provide the highest-sensitivity probes of
CLFV, and their interplay can specifically test the hypothesis of dipole dominance,
which in turn relates to the discussion of $\amu$. In general,
the constraints from all CLFV observables including muons or
$\tau$-leptons are sensitive to different properties of BSM 
physics and are therefore complementary. Positive or negative signals
combined with limits from magnetic and electric dipole moments provide key
information on the chiral and flavour structure of BSM physics.

\subsection{Muon--Higgs coupling}\label{sec:muon-Higgs}
In the previous sections \ref{sec:MuonMass} and
\ref{sec:LeptonDipole}
we have considered several cases where large BSM effects in $\Damu$ also leave an 
imprint in other low-energy chirality-flipping observables. Here we extend this discussion
into the high-energy regime by exploring the connection between $\Damu$ and the muon--Higgs coupling $\lambda_{\mu\mu}$ 
probed at the LHC.
Like the magnetic moment and the muon mass, the Higgs coupling
$\lambda_{\mu\mu}$ is flavour conserving and chirality flipping. Much
of the discussion of chirality flips in Sec.~\ref{sec:ChiralityFlips}
can also be applied to this quantity, suggesting that large chiral enhancements present in
$\Damu$ generically translate into similar enhancements in
$\lambda_{\mu\mu}$. This has been observed and evaluated in the
context of concrete models such as vector-like lepton models
\cite{Kannike:2011ng,Dermisek:2013gta}, leptoquark models
\cite{Crivellin:2020tsz,Crivellin:2020mjs,Fajfer:2021cxa}, the
two-Higgs doublet model \cite{Fajfer:2021cxa} and generalised within
EFT frameworks
\cite{Fajfer:2021cxa,Crivellin:2021rbq,Dermisek:2022aec,Dermisek:2023nhe}.

Over the past decade many of the Higgs decay channels have been measured with impressive accuracy.
In particular, the current world average for the di-muon signal strength is given by \cite{ATLAS:2020fzp,CMS:2022dwd,ParticleDataGroup:2024cfk}
\begin{align}\label{eq:Rmumu-exp-bound}
	\mu(H\to \mu\mu) = \frac{\sigma\cdot\text{BR}(H\to\mu\mu)}{[\sigma\cdot\text{BR}(H\to\mu\mu)]_\text{SM}} = 1.21 \pm 0.35.
\end{align}
In many BSM scenarios the modifications to both the production cross section $\sigma$ and the total decay width of the Higgs are 
negligible, such that this measurement provides a direct constraint on
the effective
interaction strength between muon and the Higgs, the muon--Higgs
coupling $\lambda_{\mu\mu}$, 
\begin{align}
	\mu(H\to\mu\mu) \simeq \frac{\Gamma(H\to\mu\mu)}{\Gamma(H\to\mu\mu)_\text{SM}} = \bigg|\frac{\lambda_{\mu\mu}}{\lambda_{\mu\mu}^\text{SM}}\bigg|^2.
\end{align}
The LHC result Eq.~(\ref{eq:Rmumu-exp-bound}) is consistent with the
SM prediction but also allows BSM contributions of more than $20\%$ to $\lambda_{\mu\mu}$.

\begin{figure}
	\centering
	\includegraphics[width=.7\textwidth]{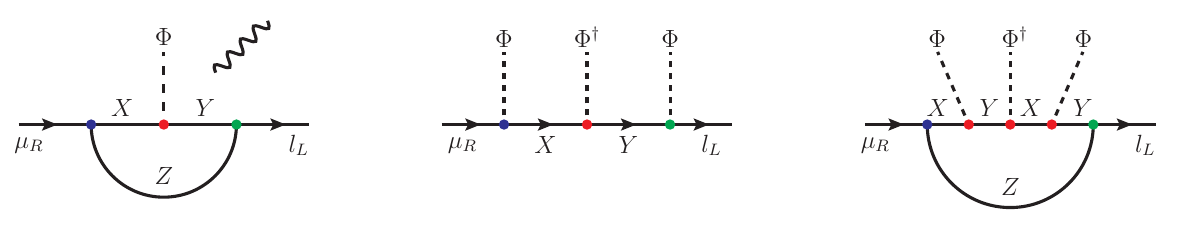}
	\caption{Generic diagrams from Ref.~\cite{Dermisek:2022aec} contributing to the muon--Higgs vertex or SMEFT dipole operator (\emph{left})
		as well as to $C_{e\Phi}$ at tree-level (\emph{middle}) and one-loop (\emph{right}). Rather than the genuine 1PI one-loop topologies displayed here
		in some models there are instead only one-\emph{light}-particle-irreducible contributions, like e.g. obtained by connecting two of the Higgs lines in the middle diagram.}
	\label{fig:SMEFT-muon-Higgs}
\end{figure}

The relationship between the muon--Higgs coupling and $\amu$ is illustrated by the generic diagrams 
in Fig.~\ref{fig:SMEFT-muon-Higgs} (left) where the same chiral
structure generates both a correction to the muon Yukawa coupling and
thus the muon mass (without external photon) and also to the magnetic
moment $\Damu$ (with external photon).
This diagram is similar to the ones illustrating the relationship
between $\amu$ and
the muon mass in Fig.~\ref{fig:chirflipdiagrams}
and to the mass-insertion diagrams for three-field models in
Fig.~\ref{fig:MIAClasses}, but it highlights the possible appearance
of a physical Higgs boson, not only of the Higgs \vev.
As Fig.~\ref{fig:SMEFT-muon-Higgs}  shows, the couplings responsible
for the left diagram can also generate effective interactions between
the muon and three Higgs fields, either at tree level (middle) or at
one-loop order (right). We note that the left and middle diagrams
involve the same power of couplings, however of different
couplings. Hence depending on the model, the tree-level diagram might
vanish or might be of the same order as the diagram on the left. In
contrast, the left and right diagrams involve the same type of
couplings, however in different powers.

The relationships
between contributions of these kinds to $\amu$, $m_\mu$ and
$\lambda_{\mu\mu}$ can be understood well by using the framework of
SMEFT introduced in Sec.~\ref{sec:genericeft}.
Here we follow particularly 
Refs.~\cite{Dermisek:2022aec,Dermisek:2023nhe}, which considered all
types of contributions in  Fig.~\ref{fig:SMEFT-muon-Higgs}.
In this context, the left diagram in Fig.~\ref{fig:SMEFT-muon-Higgs}  gives
rise to a matching correction to the fundamental muon Yukawa coupling
$y_\mu$, which then differs from its SM value, and to the dipole
operator, and the other two diagrams induce
the dimension-6 mass operator $\bar{l}_L \mu_R \Phi |\Phi|^2$. The
relevant resulting  effective Lagrangian is
\begin{align}
	\La \supset - y_\mu \big(\bar{l}_L \mu_R \Phi\big) + C_{\underset{22}{e\Phi}} \big(\bar{l}_L \mu_R \Phi\big) |\Phi|^2 + C_{\underset{22}{e\gamma}} \big(\bar{l}_L\sigma^{\mu\nu}\mu_R\big) \Phi F_{\mu\nu} + h.c.,
\end{align}
where the first term is the usual dimension-4 muon--Higgs Yukawa
coupling, the second term is a dimension-6 muon--Higgs coupling with
three powers of the Higgs doublet, the third term is the muon dipole
operator. The first two terms contribute to the muon mass and the
physical muon--Higgs coupling $\lambda_{\mu\mu}$, the third term to
$\amu$ and $d_\mu$. The Lagrangian can be easily augmented
by few four-fermion operators to form a closed system under
renormalisation-group equations \cite{Fajfer:2021cxa}; here it has
been simplified and  written in
terms of the photon dipole by using
Eq.~(\ref{eq:SMEFT-dipole-operator}), which is sufficient for the
following discussion.

The diagrammatic relationship in Fig.~\ref{fig:SMEFT-muon-Higgs}
translates into a correlation between the two dimension-6 Wilson coefficients 
that can be parametrised in terms of a model-dependent factor $\kappa$ \cite{Dermisek:2022aec,Dermisek:2023nhe},
\begin{align}\label{eq:SMEFT-WC-correlation}
	C_{\underset{22}{e\Phi}} = \frac{\kappa}{e} C_{\underset{22}{e\gamma}}.
\end{align}
In general, the factor $\kappa=|\kappa|e^{i\phi_\kappa}$ is complex and depends
on the BSM masses and couplings. It plays a similar role as the
model-dependent ${\cal O}(1)$ factor 
between $\amu$ and $\Delta m_\mu/m_\mu$ implicit in
Eq.~(\ref{eqn:GeneralGM2Contribution}).
In contrast, however, $\kappa$ does not have to be ${\cal
  O}(1)$. Instead its order of magnitude is determined 
by the relative loop suppression between the contributions to the dipole and dimension-6 mass operator. Typically
the dipole operator is induced at one-loop, while $\O_{e\Phi}$ can be
induced at tree-level (by diagrams as in
Fig.~\ref{fig:SMEFT-muon-Higgs} (middle) if they exist)   or at one-loop level (by diagrams as in
Fig.~\ref{fig:SMEFT-muon-Higgs} (right)). Hence, typical values can be
\begin{align}
	|\kappa| \sim \begin{cases}
	16\pi^2 & \text{if tree-level diagrams for
          $C_{\underset{22}{e\Phi}}$ exist,} \\ 
	|\lambda_{XY}|^2 & \text{if one-loop is leading order,}
	\end{cases}
\end{align}
where $\lambda_{XY}$ are the additional coupling factors present in
Fig.~\ref{fig:SMEFT-muon-Higgs} (right).
To see how Eq.~\eqref{eq:SMEFT-WC-correlation} translates into a correlation between the Higgs decay and $\Damu$
we consider the muon mass and effective Higgs coupling generated in
the SMEFT (at dimension 6) after EWSB. At leading order,
these are given by
\begin{subequations}\label{eq:SMEFT-muon-Higgs}
	\begin{alignat}{3}
		m_\mu &= \tfrac{1}{\sqrt2}\big(y_\mu v &&- \tfrac{1}{2} C_{\underset{22}{e\Phi}} v^3\big), \\
		\lambda_{\mu\mu} &= \tfrac{1}{\sqrt2}\big(y_\mu &&- \tfrac{3}{2} C_{\underset{22}{e\Phi}} v^2\big).
	\end{alignat}
\end{subequations}
The important factor 3 arises from the expansion $(v+h)^3= v^3 +
3v^2h+...$ and results in a modification of the correlation between
the muon mass and the muon--Higgs coupling. While the SM correlation
is simply $ \lambda_{\mu\mu}^\text{SM}=m_\mu / v$, the above equations
can be combined by eliminating the SMEFT Yukawa coupling $y_\mu$,
yielding the correlation
\begin{align}
		\lambda_{\mu\mu} &= \frac{m_\mu}{v}-C_{\underset{22}{e\Phi}} \frac{v^2}{\sqrt2}.
\end{align}
In this approximation the modification is thus caused solely by the
dimension-6 mass operator involving three Higgs fields; further,
subleading effects are discussed in Ref.~\cite{Dermisek:2023nhe}.

Using Eq.~\eqref{eq:SMEFT-WC-correlation}, the SMEFT expressions for the muon dipole moments discussed in 
sec.~\ref{sec:genericeft} and sec.~\ref{sec:LeptonDipole} give
\begin{subequations}
	\begin{align}
		 \Re\big\{e^{-i\phi_\kappa}C_{\underset{22}{e\Phi}}\big\}&= \frac{|\kappa|\Damu}{2\sqrt{2}v m_\mu} \\
		 \Im\big\{e^{-i\phi_\kappa}C_{\underset{22}{e\Phi}}\big\} &= -\frac{|\kappa| d_\mu}{\sqrt{2}v e}.
	\end{align}
\end{subequations}
After eliminating EFT Yukawa coupling $y_\mu$ in favour of $m_\mu$ and
expressing $C_{e\Phi}$ in terms of the above equations, we
arrive at the following correlation between the muon--Higgs coupling and dipole moments
\begin{align}\label{eq:Rmumu-ellipse}
    \abs{\frac{\lambda_{\mu\mu}}{\lambda_{\mu\mu}^\text{SM}}}^2 =
    \bigg(\cos(\phi_\kappa)-
     \frac{|\kappa|}{728} \Big(\frac{\Damu}{10^{-9}}\Big)\bigg)^2
     + \bigg(\sin(\phi_\kappa) -
     \frac{|\kappa|}{684}\Big( \frac{d_\mu}{10^{-22}\,e\cdot\text{cm}}\Big)\bigg)^2,
\end{align}
  where the numerical prefactors correspond to the factors
  $\frac{v^2}{4m_\mu^2}$ and $\frac{v^2}{2 m_\mu}$, respectively.

\begin{figure}
	\centering
	\includegraphics[width=.24\textwidth]{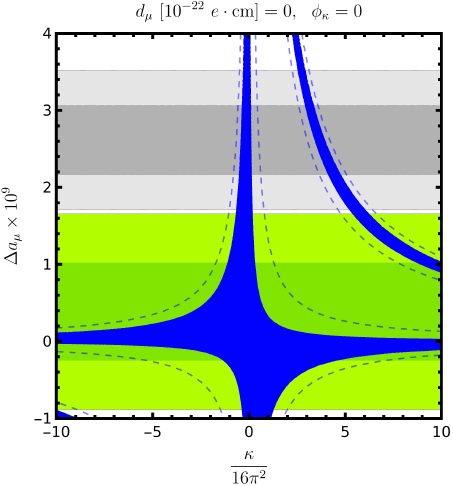}
	\includegraphics[width=.24\textwidth]{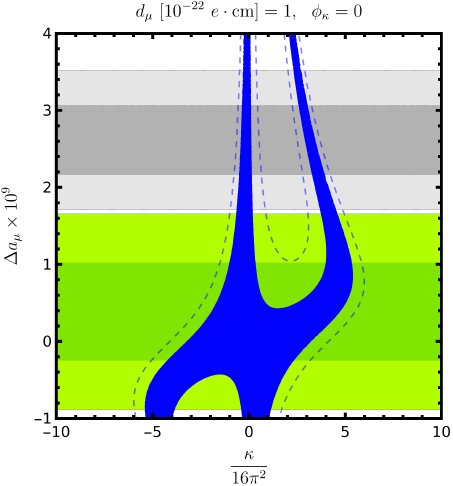}
	\includegraphics[width=.24\textwidth]{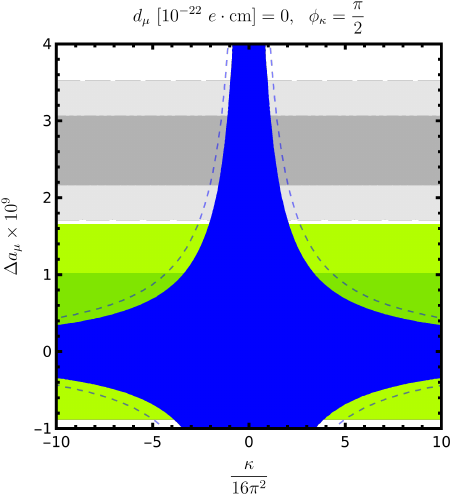}
	\includegraphics[width=.24\textwidth]{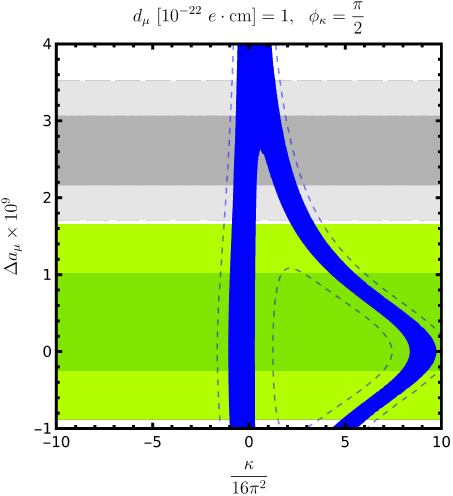}
	\caption{Contour plots for the correlation between $\Damu$ and
          the muon--Higgs coupling, in the $\kappa$ -- $\Damu$ plane,
          assuming the tree-level correlation
          Eq.~\eqref{eq:Rmumu-ellipse}. The  current experimental
          bounds Eq.~\eqref{eq:Rmumu-exp-bound} on the muon--Higgs
          coupling at 1$\sigma$ (2$\sigma$) are shown in blue (dashed blue). 
		The green bands indicate the new result
                  $\DamuFinal$ and the grey bands the previous
                $\DamuOld$ (at the 1$\sigma$ and $2\sigma$ levels), and
		negative $\kappa$ correspond to the case $\phi_\kappa\to \phi_\kappa + \pi$.
		}
	\label{fig:kappa-amu-contours}
\end{figure}

A shift of the muon--Higgs coupling of a relevant magnitude can
therefore be expected if e.g.~the
 dipole moments are significant, $\Damu\sim10^{-9}$ or
$d_\mu\sim10^{-22}\,e\cdot\text{cm}$, and if $|\kappa|\sim \O(100)$.
Here the value $d_\mu=10^{-22} ~e\cdot\text{cm}$ is of interest since
it is in reach of the future PSI experiment described in
Ref.~\cite{PSImuEDM:2023dsd}, and a value of $|\kappa|\sim \O(100)$ is
particularly possible in models where the tree-level diagram
Fig.~\ref{fig:SMEFT-muon-Higgs} (middle) is generated.  

In Refs.~\cite{Dermisek:2022aec,Dermisek:2023nhe} it was stressed that
for any given model with fixed $\kappa$, the correlation in
Eq.~(\ref{eq:Rmumu-ellipse}) defines an ellipse in the
$\Damu$--$d_\mu$ plane. Here we use the correlation to study allowed
values for the model factor $\kappa$ for different values of $\Damu$
and $d_\mu$.
Specifically, Fig.~\ref{fig:kappa-amu-contours}  shows the 
$\kappa$--$\Damu$ regions allowed by Eq.~\eqref{eq:Rmumu-exp-bound}
and Eq.~\eqref{eq:Rmumu-ellipse} for 
four example scenarios with either $d_\mu=0$ or $d_\mu=10^{-22}
~e\cdot\text{cm}$ and with real or imaginary $\kappa$.
The first case with $d_\mu=0$ and $\phi_\kappa=0$ is particularly
simple but interesting in its own right. Here in the correlation
formula the second bracket on the r.h.s.~vanishes, and the relation
can be simplified to
  \begin{align}\label{eq:Rmumu-ellipse-real}
\left|\frac{\lambda_{\mu\mu}}{\lambda_{\mu\mu}^\text{SM}}\right| =
\left|1-
     \frac{|\kappa|}{728} \Big(\frac{\Damu}{10^{-9}}\Big)\right|.
  \end{align}
The LHC constraints on the muon--Higgs coupling act only on
absolute values and require the left-hand side of this equation to be
close to one. Hence there are two distinct 
ways to fulfil these LHC constraints:
Either $\lambda_{\mu\mu}\approx+\lambda_{\mu\mu}^\text{SM}$, which
implies small $\kappa$ and/or small $\Damu$. Or
$\lambda_{\mu\mu}\approx-\lambda_{\mu\mu}^\text{SM}$, which requires a
very specific value of the product $\kappa\Damu$ where both factors
are rather large.
Both cases are visible in the first plot of
Fig.~\ref{fig:kappa-amu-contours}. The large cross-like regions
centred around the origin corresponds to 
$\lambda_{\mu\mu}\approx+\lambda_{\mu\mu}^\text{SM}$. It contains the
two limits where $\Damu=0$ (and $\kappa$ is unconstrained) and where
$\kappa=0$ (and $\Damu$ is unconstrained). The allowed region towards larger
$\kappa$ corresponds to
$\lambda_{\mu\mu}\approx-\lambda_{\mu\mu}^\text{SM}$.

The second plot of the same figure shows the case with 
non-zero $d_\mu$, where the general correlation
Eq.~(\ref{eq:Rmumu-ellipse}) applies. Here the two allowed regions
merge at small $\Damu$, and there is an upper limit on
$|\kappa|$ even at small $\Damu$. The third and fourth plots of 
Fig.~\ref{fig:kappa-amu-contours} show the case
of imaginary $\kappa=i|\kappa|$. Here, Eq.~\eqref{eq:Rmumu-ellipse} is
purely quadratic in $\Damu$ and its 
effect therefore suppressed. Consequently, in the case of $d_\mu=0$ the allowed ranges of $|\kappa|$ are larger and the flipped-sign
region is absent. For non-zero $d_\mu$ the scenario is essentially opposite to the one with real $\kappa$;
now $\sin(\phi_\kappa) =1$ and for small $\Damu$ there are two
distinct regions where the second bracket in
Eq.~\eqref{eq:Rmumu-ellipse} is close to
either $+1$ or $-1$; these two regions merge as $\Damu$ becomes large
where the $|\kappa|^2$ term in Eq.~\eqref{eq:Rmumu-ellipse} starts to dominate.

As expected, in all scenarios the constraints are relevant  for large
$|\kappa|\sim\O(16\pi^2)$, which are expected only in
models where the tree-level diagram Fig.~\ref{fig:SMEFT-muon-Higgs} (middle) exists.
In this case the allowed quantum numbers of the new fermions are strongly constrained and $\kappa$ can be written as 
\begin{align}
	\kappa = \frac{64\pi^2}{\mathcal{Q}}
\end{align}
where the $\O(1)$ values of $\mathcal{Q}$ are listed in Tab.~\ref{tab:VLL-coefficients}.
These scenarios will be discussed in more detail later in Sec.~\ref{sec:VLF}.

In contrast, in a much wider class of models,  $C_{e\Phi}$ is
generated at one-loop order.
Suitable examples are again the generic
three-field extensions with chiral enhancements
discussed in Sec.~\ref{sec:genericthreefield}, and a corresponding
detailed analysis has been carried out in 
Refs.~\cite{Crivellin:2021rbq,Dermisek:2023nhe}.
These references have computed the relevant
and the relevant matching conditions for the Wilson coefficients. The corresponding values
of $\kappa$ are necessarily real, and translated to our conventions
for the Class I, II, III models
they can be written as (for simplicity we set $m_{\phi,\eta}=m_{\psi,\chi}\equiv M$)
\begin{align}\label{eq:kappa-classes}
	\kappa^{I} = \frac{|\lambda_H|^2}{\mathcal{Q}}, \qquad \kappa^{II} = \frac{|\bar\lambda_H|^2}{\mathcal{Q}}, \qquad 
	\kappa^{III} = \frac{|a|^2}{\mathcal{Q} M^2}.
\end{align}
The coefficients $\mathcal{Q}$ depend on the gauge quantum numbers of
the BSM fields, and their values for the different classes and
representations are listed in Tab.~\ref{tab:kappa-coeff}. 
In these models, $\kappa$ is thus typically small, and in this case the muon--Higgs coupling
currently provides no relevant constraint on $\Damu$ via the correlation in
Eq.~\eqref{eq:Rmumu-ellipse},
though this could change with increased precision that can be reached
at the future FCC-hh \cite{Benedikt:2022kan}.

In exceptional cases, however, the hypercharges $X$ are such that
$\mathcal{Q}$ vanishes. This
 corresponds to scenarios with a cancellation of the 
chirally enhanced contributions to $\Damu$. For values of $X$ close to
this point, $\kappa$ is strongly enhanced and the constraints from the
muon--Higgs coupling become more relevant \cite{Crivellin:2021rbq}.

\begin{table}[t]
	\centering
	\begin{tabular}{|c|c|c|c|c|}
		\hline
		$\SUL$ rep. & $(\bm1,\bm2,\bm1)$ & $(\bm2,\bm1,\bm2)$ & $(\bm3,\bm2,\bm3)$ & ($\bm2,\bm3,\bm2$) \\ \hline \hline
		Class I   & $-\frac{1}{6}(1+2X)$ & $-\frac{1}{3}X$ & $\frac{1}{30}(1-6X)$ & $-\frac{1}{15}(2+3X)$ \\ \hline
		Class II  & $-\frac{1}{2}(3+4X)$ & $-\frac{1}{2}(1+4X)$ & $-\frac{1}{10}(1+12X)$ & $-\frac{1}{10}(11+12X)$ \\ \hline
		Class III & $-\frac{1}{2}(1+2X)$ & $-X$ & $\frac{1}{10}(1-6X)$ & $-\frac{1}{5}(2+3X)$ \\ \hline
	\end{tabular}
	\caption{Values of $\mathcal{Q}$ in Eq.~\eqref{eq:kappa-classes} for the class I, II and III models and different 
		possible representations of $(\phi^\dagger,\psi,\chi)$ and $(\psi,\phi^\dagger,\eta^\dagger)$. 
		Here, $X$ denotes the hyper-charge of $\phi^\dagger$ (class I or II) or $\psi$ (class III).
	}
	\label{tab:kappa-coeff}
\end{table}

\subsection{Electroweak precision observables}\label{sec:EWPO}

Electroweak precision observables (EWPO)  provide  stringent tests of
the electroweak sector of the SM at the quantum level as well as strong bounds on new physics scenarios.
Key electroweak observables are the electroweak boson masses
$M_{W,Z,h}$, the muon life time and the related Fermi constant $G_F$,
and  effective weak mixing angles $\sin^2\theta_{\text{eff}}^f$ for
the interaction of a fermion $f$ with the $Z$ boson at the
$Z$-pole. Like $\amu$, the EWPOs are strongly sensitive to loop
effects. But while $\amu$ is related to chirality flips and can be
chirally enhanced, EWPOs instead are sensitive to so-called custodial
symmetry, an approximate  SU(2)$_\text{L+R}$ symmetry of the
electroweak vacuum \cite{Sikivie:1980hm}.\footnote{In the SM this symmetry
becomes exact in the limit $g_1 \to 0$ and $y_u=y_d$.}
In short, the phenomenology of EWPOs is quite independent and
complementary to the one of $\amu$, but there is one important common
ingredient, which we will discuss in the following.

The EW sector of the SM at tree level is parametrised by the $\GEW$  gauge couplings $g_2$ and $g_1$ as well as
the quartic Higgs coupling $\lambda$ and the mass parameter $\mu^2$ that characterise the Higgs potential and govern the EWSB scale.
In precision calculations, these four parameters are typically traded
for the finestructure constant $\alpha$, $G_F$, $M_Z$ and $M_h$ which is the minimal
set of experimental input parameters with the smallest relative
uncertainty. To illustrate the behaviour of higher orders and compare
with $\amu$, we list the leading shifts to two important EWPOs, the
$W$-boson mass and the effective weak mixing angle (for a review we
refer to Ref.~\cite{ALEPH:2005ab}),
\begin{align}\label{DeltaMWshift}
  \frac{\Delta M_W^2}{M_W^2} &=
  \frac{s_W^2}{s_W^2-c_W^2} \Delta\alpha(M_Z^2)
  +
  \frac{c_W^2}{c_W^2-s_W^2} \Delta\rho + \ldots ,\\
  \frac{\Delta \sin^2\theta_{\text{eff}}^f}{\sin^2\theta_{\text{eff}}^f} &=
  \frac{c_W^2}{c_W^2-s_W^2} \Delta\alpha(M_Z^2)
  +
  \frac{c_W^2}{s_W^2-c_W^2} \Delta\rho + \ldots .
\end{align}
Here $\Delta\rho$ is the universal correction to the $\rho$ parameter
defined via the neutral- to charged-current ratio at low energies. It
is also related to the ratio
$	\frac{M_W^2}{c_W^2 M_Z^2}$, which is equal to
$\rho^{\text{tree}}=1$ in the SM at tree level because of custodial symmetry, but
$\Delta\rho$ arises from breakings of this symmetry such as the large
top-bottom mass splitting and similar properties of BSM
extensions \cite{Veltman:1977kh,Peskin:1991sw,Sikivie:1980hm}. The behaviour of $\Delta\rho$
depends on very different physical mechanisms and is therefore
typically quite unrelated to the behaviour of $\amu$.

The intriguing connection to $\amu$ arises via the
quantity $\Delta\alpha(q^2)$  appearing in the above corrections
to the EWPOs. This quantity corresponds to 
the relative shift of the finestructure constant between its value in
the Thomson limit in the on-shell scheme, $\alpha$, and the effective
$\alpha(q^2)$ at higher energies. This shift arises from the photon
vacuum polarisation, which also plays an important role in the
evaluation of $\amu$.

While the leptonic contributions to $\Delta\alpha$ are uncontroversial
and precisely known, the hadronic corrections are of particular
interest here.
The hadronic vacuum polarisation (HVP) can be defined as
\begin{align}
	\begin{gathered}
		\includegraphics[scale=.8, clip, trim=0 4 0 0]{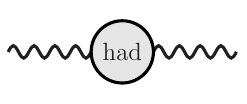}
	\end{gathered} \equiv -i \Pi_\text{had}(q^2) \Big(q^2 g^{\mu\nu} - q^\mu q^\nu\Big).
\end{align}
In case of $a_\mu$ the leading HVP contribution (cf. Sec.~\ref{sec:SMtheory}) generated by Fig.~\ref{fig:amu-HVP} can be expressed
in terms of the following dispersion integral \cite{Durand:1962zzb,Gourdin:1969dm,Aoyama:2020ynm}
\begin{align}\label{eq:amu-HVP}
	a_\mu^\text{HVP, LO} = \frac{\alpha}{\pi^2} \int_{s_{th}}^{\infty} \frac{ds}{s} \, K(s) \Im{\Pi_\text{had}(s)},
\end{align}
where  the integration starts at the $\pi^0 \gamma$ production threshold $s_{th}=m_{\pi^0}^2$ and the kernel function is given by
\begin{align}\label{eq:HVP-kernel}
	K(s) = \int_0^1 dz \frac{(1-z) z^2}{z^2 + (1-z)\frac{s}{m_\mu^2}}.
\end{align}
This is the basic origin of $a_{\mu}^\textrm{HVP,LO}$. As
discussed in Sec.~\ref{sec:SMtheory}, non-perturbative techniques are
required to obtain accurate values for $\Pi_{\text{had}}(s)$, though currently
there are unresolved tensions between different
data-driven
evaluations based on experimental $e^+e^-\to\gamma^*\to$ hadrons measurements, and between data-driven evaluations and lattice simulations.

This hadronic vacuum polarisation also enters the EWPO
through the hadronic correction $\Delta \alpha^{(5)}_\text{had}(q^2)$
to the effective fine structure constant, given by
\begin{align}
  \Delta \alpha^{(5)}_\text{had}(q^2) & = - \Re \Big(\Pi_{\text{had}}(q^2) - \Pi_{\text{had}}(0)\Big).
\end{align}
This hadronic correction can be expressed in terms of a dispersion integral similar to Eq.~(\ref{eq:amu-HVP}),
\begin{align}
	\Delta \alpha^{(5)}_\text{had}(M_Z^2) = \frac{M_Z^2}{\pi}~ \text{p.v.}\int_{s_{th}}^\infty ds \, \frac{\Im{\Pi_\text{had}(s)}}{s(M_Z^2 - s)}.
\end{align}
however with a qualitatively different kernel function.
While $K(s)$ decreases monotonically as $1/s$ and thus gives the greatest weight to
the low-energy region around $s\gtrsim s_{th}$, the kernel for $\Delta \alpha^{(5)}_\text{had}(M_Z^2)$ remains nearly
constant for $s\ll M_Z^2$ resulting in an increased sensitivity to the contributions also from $s\gg s_{th}$.

Notably the HVP contributions to $a_\mu^\text{HVP,LO}$ and $\Delta \alpha^{(5)}_\text{had}(M_Z^2)$ are positively correlated, that is, 
an increase of $\Im\Pi_\text{had}$ 
(e.g.\ due to a larger measured $e^+ e^-$ cross section in case of the
data-driven extraction)
leads to an increase in both of these quantities, and a resulting
decrease of the prediction for $M_W$, see Eq.~(\ref{DeltaMWshift}) \cite{Passera:2008jk}.
After the Brookhaven measurement, the SM prediction for $\amu$ was
about $3\sigma$ lower than the experimental value, and the SM
prediction for $M_W$ was slightly lower than the measurement at the
time.
The above correlation means that
assuming an increased HVP to reduce the discrepancy for $\amu$ 
inevitably worsens the discrepancy for $M_W$ and thus generally worsens the fit
of EWPOs to experiment  \cite{Passera:2008jk}.

\begin{figure}[t]
	\centering
	\begin{subfigure}{.48\textwidth}
		\centering
		\includegraphics[width=\textwidth]{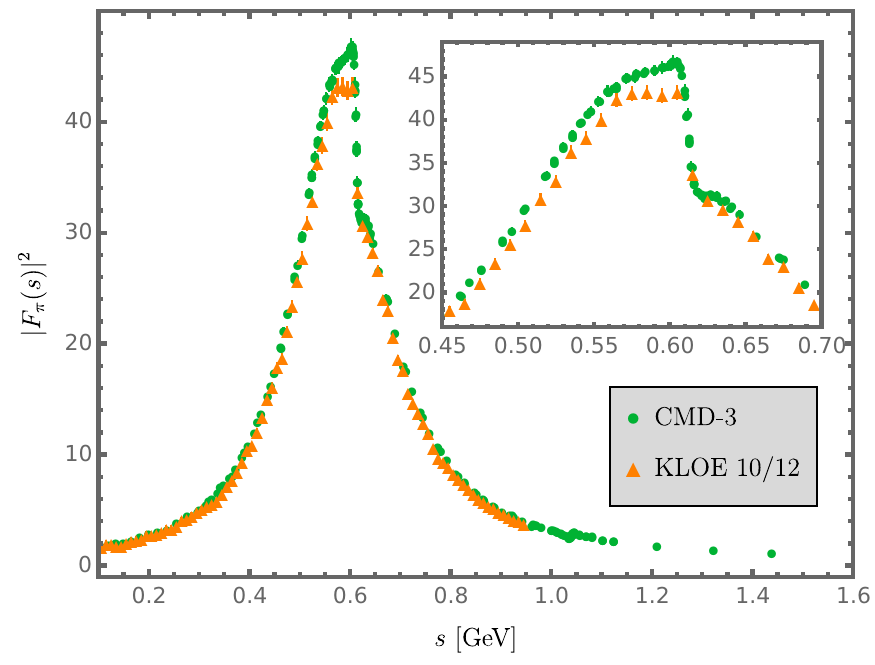}
		\caption{Comparison between $|F_\pi(s)|^2$ determined by KLOE and CMD-3}
		\label{fig:pion-FF}
	\end{subfigure}\hfill
	\begin{subfigure}{.48\textwidth}
		\centering
		\includegraphics[width=\textwidth]{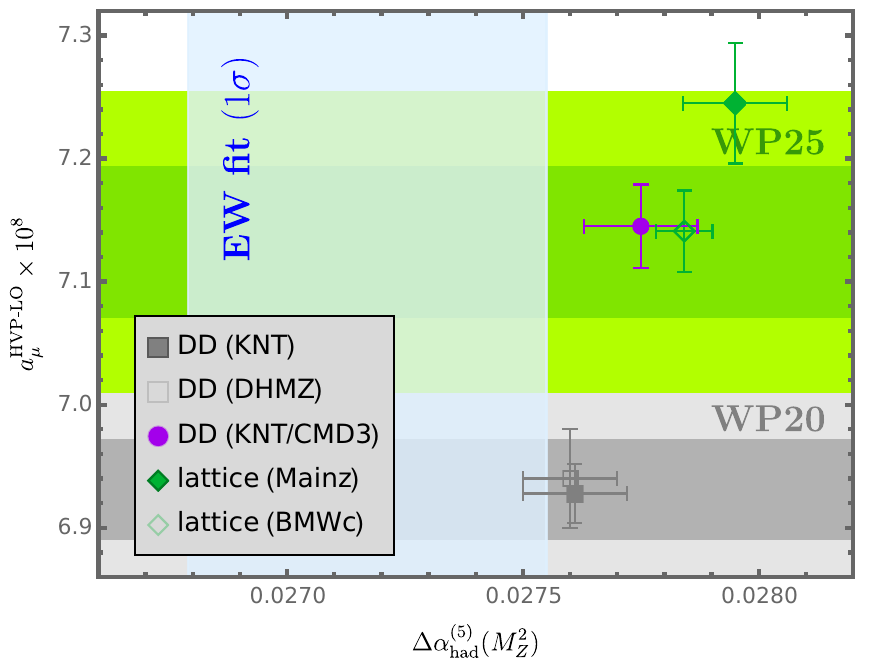}
		\caption{Compilation of $\Delta\alpha_\text{had}^{(5)}(M_Z^2)$ and $a_\mu^\text{HVP,LO}$ from different references}
		\label{fig:Daplha-vs-amu}
	\end{subfigure}
	\caption{\textbf{(a)} Comparison between the pion form factor $F_\pi(s)$ obtained from the KLOE 10/12 \cite{KLOE:2010qei,KLOE:2012anl}
		and CMD-3 \cite{CMD-3:2023alj,CMD-3:2023rfe} measurements of $\sigma(e^+ e^- \to \pi^+\pi^-(\gamma))$ for c.o.m.\ energies
		around the $\rho$-peak.
		\textbf{(b)} Compilation of values for $\Delta\alpha_\text{had}^{(5)}(M_Z^2)$ and $a_\mu^\text{HVP,LO}$ obtained from the
		data driven	KNT \cite{Keshavarzi:2019abf,Keshavarzi:2020bfy} (grey square), DHMZ \cite{Davier:2019can} 
		(empty grey square) and the KNT analysis with CMD-3 data substituted in the available energy range \cite{DiLuzio:2024sps} (purple circle)
		as well as the lattice results from the Mainz \cite{Djukanovic:2024cmq} (green diamond) and BMW \cite{Boccaletti:2024guq} (empty green diamond) collaborations 
		where the corresponding $\Delta\alpha_\text{had}^{(5)}(M_Z^2)$ was taken from Ref.~\cite{Erler:2023hyi}.
		The blue band shows the 1$\sigma$ region for $\Delta\alpha_\text{had}^{(5)}(M_Z^2)$ determined from the EW fit \cite{Haller:2018nnx}
		and the grey and green bands correspond to
                $a_\mu^\text{HVP,LO}$ obtained in the WP20
                \cite{Aoyama:2020ynm} and WP25
                \cite{Aliberti:2025beg}, respectively.
			}
\end{figure}

However, because of the significantly different energy dependence of the two kernel functions the impact on the EW fit 
depends on the energy scale at which $\Im\Pi_\text{had}$ is
modified. After Ref.~\cite{Passera:2008jk} appeared, several major developments took place. 
First, the discovery of the Higgs boson and measurement of $M_h$
removed an important source of uncertainty.
Secondly, subsequent updates of the analysis \cite{Crivellin:2020zul,Keshavarzi:2020bfy,Malaescu:2020zuc,Athron:2022qpo},
particularly in light of the first lattice QCD results \cite{Borsanyi:2020mff} for $a_\mu^\text{HVP,LO}$, as well as further theoretical
considerations \cite{deRafael:2020uif,Colangelo:2020lcg} and more detailed comparisons between data driven and lattice evaluations
\cite{Davier:2023cyp,Colangelo:2022vok} confirmed that a modification
of  $\Im\Pi_\text{had}$ at
very low energies (e.g.\ around the $\rho$ peak at $s\sim
0.6~\text{GeV}^2$) are actually consistent with EWPO within
uncertainties. And finally, as described in 
Sec.~\ref{sec:SMtheory} there is now strong evidence from many lattice
results that   $\Im\Pi_\text{had}$  actually is indeed larger than
previous $e^+e^-$-data based determinations at the time of Refs.~\cite{Passera:2008jk} or
\cite{Aoyama:2020ynm} suggested. This is also supported by the more recent CMD-3 
measurement of the pion form factor \cite{CMD-3:2023alj,CMD-3:2023rfe}
(see Fig.~\ref{fig:pion-FF}), although despite extensive efforts
\cite{Campanario:2019mjh,BaBar:2023xiy,Davier:2023fpl,Aliberti:2024fpq}
so far no causes of the discrepancies to earlier $e^+e^-$-data based results have been
identified.\footnote{
The persisting discrepancies between $e^+e^-$-data based and
lattice-based results have led to speculations whether BSM effects
could be the cause. We will discuss such ideas in 
Sec.~\ref{sec:HVP-models}.}
Finally, the fit of the EWPO to data has also itself further improved e.g.~through
the known Higgs boson mass and
recent LHC measurements of the $W$-boson mass,
\begin{subequations}\label{MWATLASCMS}\begin{align}
  M_W^{\text{ATLAS}} &= 80.3665(159) \text{ GeV \cite{ATLAS:2024erm}},\\
  M_W^{\text{CMS}} &= 80.3602(99) \text{ GeV \cite{CMS:2024lrd}}.
\end{align}
\end{subequations}
These recent measurements are the two of the most precise $M_W$ measurements
and are both in full agreement with the SM
prediction.\footnote{%
This is in contrast to the CDF result $80.4335(94)\text{ GeV}$
\cite{CDF:2022hxs}, which 
is in significant tension with the SM and with other measurements, leading to a poor fit if combined with the other measurements \cite{LHC-TeVMWWorkingGroup:2023zkn}, and
is therefore excluded from the global average,
as discussed in
Ref.~\cite{ParticleDataGroup:2024cfk}.
}

In Fig.~\ref{fig:Daplha-vs-amu} we show a comparison between results for $\Delta \alpha_\text{had}^{(5)}(M_Z^2)$ 
and $a_\mu^\text{HVP,LO}$ obtained by different groups as well as (for reference) the value preferred by (extracted indirectly from) the
EW fit of
Ref.~\cite{Haller:2018nnx}\footnote{We use Ref.~\cite{Haller:2018nnx}
since it provides $\Delta\alpha_\text{had}^{(5)}(M_Z^2)$, however it
appeared before the recent measurements of $M_W$
\cite{ATLAS:2024erm,CMS:2024lrd}, both of which would improve the
agreement or the EW fit with data. As an example, the 
 fit result of Ref.~\cite{Haller:2018nnx} for
$M_W$ without using the experimental $M_W$ input is $80.354(7)$ GeV,
 in excellent agreement with Eqs.~(\ref{MWATLASCMS}).}
\begin{align}\label{eq:Dalpha-EWfit}
	\Delta \alpha_\text{had}^{(5)}(M_Z^2)[\text{EW fit}] = 0.02716(39).
\end{align}
The data points are from $e^+e^-$-data driven evaluations of
$\Im\Pi_{\text{had}}$ without
\cite{Keshavarzi:2019abf,Keshavarzi:2020bfy,Davier:2019can} and with
the recent CMD-3 data \cite{DiLuzio:2024sps} 
as well as the lattice results from the Mainz
\cite{Djukanovic:2024cmq} and BMW \cite{Boccaletti:2024guq}
collaborations. For details see the figure caption. The plot                
illustrates how the increased low-energy contributions found by CMD-3 and the lattice evaluations resolve
the $a_\mu$ discrepancy without significantly affecting the EW fit.
It is worth pointing out that the uncertainty of Eq.~\eqref{eq:Dalpha-EWfit} is driven mainly by the uncertainties
of the experimental input and thus further experimental improvements could eventually establish strong tensions within the EW fit. 
Furthermore, additional observables like $a_e$ and the muonium hyper-fine splitting
also appear promising for future tests of the hadronic contributions \cite{DiLuzio:2024sps}.

\newcommand{\noredcolor}[1]{{#1}}
\newcommand{\nobluecolor}[1]{{#1}}

\subsection{Neutrino masses }
\label{sec:neutrino_masses}
One of the important breakthroughs of the past decades in particle physics is the observation of neutrino oscillations~\cite{Super-Kamiokande:1998qwk,SNO:2001kpb}, which has confirmed the existence of nonvanishing neutrino masses. 
A number of measurements have resulted in precise determinations of
the mass-squared differences $\Delta m_{ij}^2 \equiv m_{i} ^2 -
m_{j}^2$ between mass eigenvalues
as well as the entries of the Pontecorvo-Maki-Nakagawa-Sakata (PMNS)
matrix, relating the mass-eigenstates to the flavour eigenstates
$\nu_{e,\mu,\tau}$.
This PMNS matrix can be expressed in terms of three mixing angles
$\theta _{ij}$ for $i,j = 1,2,3$ and one CP violating phase, $\delta
_{CP}$. In total, the
 current combined experimental constraints for the case of normal ordering (inverted ordering) are \cite{Esteban:2024eli}
\footnote{see NuFIT, \url{www.nu-fit.org} .}

\begin{align}
	\Delta m_{21} ^2 &= 7.49_{-0.19} ^{+0.19}   & (&7.49_{-0.19} ^{+0.19})    &\times 10^{-5} \text{eV}^2,\\
	\Delta m_{3l} ^2 &= 2.513_{-0.019} ^{+0.021} & (&-2.484_{-0.020} ^{+0.020}) &\times 10^{-3} \text{eV}^2,
\end{align}

\begin{align}
	\sin ^2 \theta _{12} &= 0.308_{-0.011} ^{+0.012}  & (&0.308 _{-0.011} ^{+0.012}),\\
	\sin ^2 \theta _{23} &= 0.470 _{-0.013} ^{+0.017} & (&0.550 _{-0.015} ^{+0.012}),\\
	\sin ^2 \theta _{13} &= 0.02215 _{-0.00058} ^{+0.00056}    &
        (&0.02231 _{-0.00056} ^{+0.00056}),
\end{align}
\begin{align}
  \delta _{CP}/^\circ &= 212 _{-41} ^{+26}
  & (&274) _{-25}
        ^{+22} .
\end{align}

For normal ordering (NO) the mass hierarchy and the largest mass splitting are $m_1 < m _2 < m_3$ and $\Delta m_{3l} ^2 = \Delta m_{31}^2$, and for  inverted ordering (IO) they are $m_3 < m _1 < m_2$ and $\Delta m_{3l} ^2 = \Delta m_{32} ^2$. This observation implies at least two massive neutrinos and three different mass eigenvalues.

In addition there are absolute upper limits on the maximum neutrino mass from cosmology (see Ref.~\cite{Esteban:2024eli} and references therein, and see Ref.~\cite{Brieden:2022lsd} for further discussions of  BOSS, eBOSS and Planck CMB data),

\begin{align}
  \label{data:nmasscosm}
  m_{\text{tot}} &= \sum m_{\nu} < 3 \times 10^{-10}\text{ GeV},
\end{align}
and on the effective electron (anti-)neutrino mass from the KATRIN experiment~\cite{Esteban:2024eli,KATRIN:2021uub},
\begin{align}
  \label{data:nmassKATRIN}
  m_{\nu_{e}} ^{\text{(eff)}}
  &< 4 \times 10^{-10}\text{ GeV}.
\end{align}

One important fact is that all experiments so far observed only interactions of left-handed neutrinos (or right-handed antineutrinos), corresponding to maximal parity violation. Accordingly the Standard Model (SM) contains only left-handed neutrino fields as part of the left-handed lepton doublets. Therefore, Yukawa interaction terms of neutrinos to the Higgs field, which would generate Dirac neutrino masses similarly to the masses of all other charged SM fermions, are not allowed. Hence neutrinos remain massless in the Standard Model and BSM physics is needed to generate and explain neutrino masses.
It is natural to ask whether neutrino masses can be linked to charged lepton observables such as $\amu$. Here we will briefly review ideas to explain the origin of neutrino masses and give generic arguments on how the mechanism to generate neutrino masses could contribute to $\amu$.

\subsubsection{Overview of neutrino mass terms}

Since neutrinos are electromagnetically neutral they can be either Dirac or Majorana fermions, i.e.\ fermions which are equal to their antiparticles. Dirac masses  are compatible with lepton number conservation and lead to the existence of a neutrino magnetic dipole moment~\cite{Fujikawa:1980yx}.
Majorana masses imply lepton number violation by $|\Delta L| = 2$, and consequently processes such as neutrinoless double $\beta$-decay ($0\nu 2\beta$) are possible --- these are intensely searched for in various experiments~\cite{KamLAND-Zen:2022tow,GERDA:2020xhi}.
There are many attractive ways to generate Majorana neutrino masses, such as the seesaw mechanism~\cite{Minkowski:1977sc,Gell-Mann:1979vob,Yanagida:1979as,Schechter:1980gr} discussed below.
Currently it is unknown how neutrino masses arise and whether neutrinos are of Majorana or Dirac type.

For a brief overview of neutrino mass terms we allow both Dirac and Majorana masses, and we assume the existence of $m + n$  neutrino gauge eigenstate fields, combined into $\omega_R \equiv (\nu_{L\,1} ^C, \cdots, \nu_{L\,m} ^C, \noredcolor{N_{R\,1}}, \cdots, \noredcolor{N_{R\,n}})^T$, where the $\nu_{L\,i}$ contain the  $m=3$ generations of SM neutrino fields and where \noredcolor{$N_{R\,i}$} denotes the additional neutral right-handed fermions, which can be called right-handed neutrinos or sterile neutrinos or neutral leptons, depending on the context.
We can define a combined Majorana field $\Omega \equiv \omega_R + \omega_R ^C$ and write the general form of the neutrino mass Lagrangian in terms of these neutrino gauge eigenstate fields as

\begin{align}
  \label{nmassLagrangianOmega}
  \La_{\nu -\text{mass}} &=
  -\frac{1}{2}{\bar{\Omega}} \mathcal{M}_\nu \Omega,
\end{align}
where the mass matrix takes a block form
\begin{align}
  \label{eq:nmassMatrixOmega}
  \mathcal{M}_\nu &= 
  \begin{pmatrix}
    \nobluecolor{\delta\mathcal{M}_L} & \mathcal{M}_D\\
    \mathcal{M}_D ^T & \noredcolor{\mathcal{M}_R}
  \end{pmatrix}. 
\end{align}

We do not consider CP violation here and keep $\mathcal{M}_\nu$ real for simplicity. Depending on the model, the number $n$ of additional neutrino fields might be zero or non-zero, and some of the blocks might be zero or non-zero. Generally, $\mathcal{M}_D$ denotes Dirac neutrino masses which can be generated from  Yukawa-type interactions e.g.\ if right-handed SU(2)$_L$ singlet neutrinos $N_{R}$ exist. The block $\mathcal{M}_R$ denotes the Majorana mass of the additional right-handed neutrinos. Finally, $\delta\mathcal{M}_L$ is a potential Majorana mass of the SM-like left-handed neutrinos.

We can use this general structure to classify the BSM scenarios which aim to explain the neutrino mass generation. 
There are four general possibilities:

\begin{enumerate}
\item[(a)] $ \noredcolor{\mathcal{M}_R}=0$, $\mathcal{M}_D \neq 0$
  obtained e.g.~from Yukawa interaction terms (e.g.~pure Dirac masses),
\item[(b)] $\mathcal{M}_D = \noredcolor{\mathcal{M}_R} = 0$
  (e.g.~seesaw type II, Zee and Babu models),
\item[(c)] $\noredcolor{\mathcal{M}_R}\neq 0$, and
  $\mathcal{M}_D \neq 0$ (e.g.~seesaw type I, III),
\item[(d)] $\mathcal{M}_D = 0$, $\noredcolor{\mathcal{M}_R} \neq
  0$, non-zero $\nobluecolor{\delta\mathcal{M}_L}\neq 0$ obtained
  from loop diagrams (e.g.~Ma/scotogenic model).
\end{enumerate}
In all cases where $\nobluecolor{\delta\mathcal{M}_L}$ is not specified, it may be generated either at the tree-level or at the loop level, depending on model details.

Scenario (a) together with $\nobluecolor{\delta\mathcal{M}_L}=0$ corresponds to the only scenario with Dirac masses and lepton number conservation. In the simplest case, the neutrino masses can arise in the same way as the mass generation of the up-type quarks of the SM. This requires at least three generations of right-handed singlet neutrinos $N_{R\,i}, (i = 1,2,3,\ldots)$ and the additional Yukawa interaction terms
\begin{align}
  \label{eq:nmassDirac}
  \mathcal{L}_{\text{$\nu$-Dirac mass}} &=  -y_\nu ^{ij}
          {\overline{l_{Li}}}{\tilde{\Phi}}\noredcolor{N_{R\,j}} + {\text{ h.c.}}.
\end{align}
In this case the neutrinos obtain (matrix-valued) Dirac masses $m_{\nu} = y_{\nu} v/\sqrt2$. The required Yukawa coupling values are however extremely tiny, $|y_\nu|\lesssim \O(10^{-12})$. 

Before exemplifying the other  scenarios (b,c,d) we mention that neutrino masses can also be described using the approach of low-energy effective theories. Especially if SMEFT is the correct EFT, see Sec.~\ref{sec:genericeft}, then neutrino masses can arise via the following dimension-5 operator~\cite{Weinberg:1979sa,Grzadkowski:2010es}\footnote{We use the convention of Ref.~\cite{Grzadkowski:2010es}. The general expression of the dimension-5 operator for more than one Higgs doublet is $Q_{\underset{ab}{\nu\nu}} = f_{abrp}{\overline{l_{jaL}^{C}}} l_{kbL} \phi_{r} ^{m} \phi_{p} ^{n}\epsilon_{jm}\epsilon_{kn} + f_{abrp}'{\overline{l_{jaL} ^C}} l_{kbL} \phi _{r} ^{m}\phi _{p} ^{n} \epsilon_{jk}\epsilon_{mn}$~\cite{Weinberg:1979sa}.}
\begin{align}
  \label{eq:nmassDim5}
  Q_{\underset{ab}{\nu\nu}} &= \epsilon_{jk}\epsilon_{mn}\Phi ^{j}\Phi ^{m}(l_{La} ^{k})^T C l_{Lb} ^{n} \nonumber\\
  &= \epsilon _{jk} \epsilon _{mn} \Phi ^{j} \Phi ^{m} {\overline{(l_{La} ^k)^C}}l_{Lb} ^n \,,
\end{align}
where $a$ and $b$ denote the flavour (generation) indices. 
This operator violates lepton number by $|\Delta L|=2$~\cite{Weinberg:1979sa,Grzadkowski:2010es}, and upon inserting the Higgs \vev it generates  Majorana masses for neutrinos
\begin{align}
  \label{mneudim5}
  (M_\nu)_{ab} &= c_{\underset{ab}{\nu\nu}} \frac{v^2}{2M_{\Lambda}},
\end{align}
where the suppression scale $M_{\Lambda}$ has been pulled out of the Wilson coefficient $C_{\underset{ab}{\nu\nu}}=c_{\underset{ab}{\nu\nu}} /M_{\Lambda}$.

This neutrino mass term has the structure and effect of the block matrix $\nobluecolor{\delta\mathcal{M}_L}$ in Eq.~\eqref{eq:nmassMatrixOmega}, so one may also say that this block matrix often arises in a low-energy EFT upon integrating out heavy fields in the fundamental theory.

\subsubsection{Seesaw mechanism and radiative mass generation}

\begin{table}
  \begin{center}
    \begin{tabular}{ |c| c c |l| }
      \hline
      \textbf{Model} & \textbf{Field} & $(\mathcal{G}_\text{EW})_\text{spin}$ & \textbf{Parameters}
      \\\hline\hline
      Seesaw Type I
      &
      $N_{R\,i}$ & $(\bm1,0)_{\frac{1}{2}}$
      &
      $(\mathbf{y}_{\nu})_{ij},~M_{R\,i}$
      \\\hline
      Seesaw Type II
      &
      $\Delta$ & $(\bm3,1)_0$
      & $(\mathbf{y}_{\nu})_{ij},~M_{\Delta},~\boldsymbol{\mu}$
      \\\hline
      Seesaw Type III
      &
      $\Sigma_{R\,i}$ & $(\bm3,0)_{\frac{1}{2}}$
      &
      $(\mathbf{y}_{\nu})_{ij},~M_{R\,i}$
      \\\hline
    \end{tabular}
    \caption{Field content of three tree-level seesaw models. The respective spins and representations with respect to $\mathcal{G}_\text{EW}=\GEW$ are given in the brackets. The indices $i,j$ are generation indices, while SU(2)$_L$ indices are suppressed. In Type-I and III the number of fields corresponds to the number of neutrino masses generated at tree-level. All fields are singlets under $\SUc$.}
    \label{table:seesaw}
  \end{center}
\end{table}

\begin{figure}[t]
	\centering 
	\begin{subfigure}{0.3\textwidth}
		\centering
		\includegraphics[width=.8\textwidth]{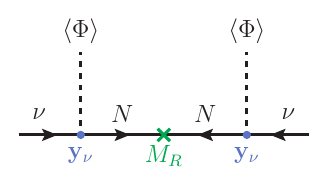}
		\subcaption{Type-I}
	\end{subfigure}
	\begin{subfigure}{0.3\textwidth}
		\centering
		\includegraphics[width=.8\textwidth]{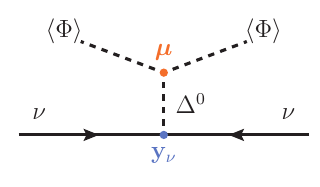}
		\subcaption{Type-II}
	\end{subfigure}
	\begin{subfigure}{0.3\textwidth}
		\centering
		\includegraphics[width=.8\textwidth]{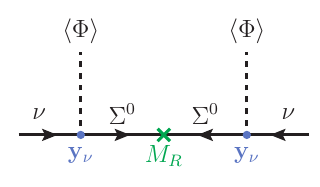}
		\subcaption{Type-III}
	\end{subfigure}
	\caption{\label{fig:seesawtyp123diagrams}
		Diagrams generating neutrino masses and
		the effective operator Eq.~\eqref{eq:nmassDim5} at  tree-level in the type I, II and III seesaw mechanism.}
\end{figure}

Here we briefly survey six different ways to generate neutrino masses, exemplifying cases (b,c,d) above. All these models lead to Majorana neutrino masses described at low energies by the formula~\eqref{mneudim5}. 
In this way the small neutrino masses are explained either by very heavy fundamental masses $M_{\Lambda} \approx 10 ^{15}$ GeV with $c_{\underset{ab}{\nu\nu}} \approx {\mathcal{O}}(1)$, or by tiny lepton-number violating couplings leading to a very small coefficient $c_{\underset{ab}{\nu\nu}}$.

The seesaw mechanism is a particularly attractive and simple way to generate neutrino masses. Here, new fields are added to the SM such that a mass matrix of the form~\eqref{eq:nmassMatrixOmega} arises and effectively neutrino masses of the form~\eqref{mneudim5} are generated at tree-level. The three types that allow this are illustrated in Fig.~\ref{fig:seesawtyp123diagrams}. The additional fields for each seesaw types are displayed in Table~\ref{table:seesaw}. 
Type I~\cite{Minkowski:1977sc,Gell-Mann:1979vob,Yanagida:1979as,Schechter:1980gr} involves right-handed neutrino singlets $N_R$ with heavy Majorana mass term and gauge invariant Yukawa interaction with the SM lepton doublets and Higgs field. Type III~\cite{Foot:1988aq} is similar to type I but adds right-handed fermion triplets $\Sigma_{R\,i}$ to the SM.
Type II~\cite{Mohapatra:1980yp} instead adds a triplet scalar field $\Delta$ with gauge invariant interactions of the form $ll\Delta$ and $\Phi\Phi\Delta^\dagger$. The essential Lagrangian terms can be written as
\begin{subequations}
  \begin{alignat}{4}
    \label{seesawI}
    \text{type I:}& \qquad
    &&\La \supset - \Big[(\mathbf{y}_\nu)_{ij} \, \overline{l_{Li}}\tilde{\Phi}N _{R\,j} &&+\text{h.c.}\Big]
    &&-\frac12 M_{R\,i} \overline{N^{C}_{R\,i}}N _{R\,i},
    \\
    \label{seesawIII}
    \text{type III:}& 
    &&\La\supset - \Big[(\mathbf{y}_{\nu})_{ij} \overline{l_{Li}}\Sigma^{a}_{R\,j}\sigma^a\tilde{\Phi}
      &&+\text{h.c.}\Big]
    &&-\frac12 M_{R\,i} \overline{\Sigma^{C}_{R\,i}}\Sigma_{R\,i}  ,
    \\
    \label{seesawII}
    \text{type II:}& 
    &&\La \supset -\Big[(\mathbf{y}_\nu)_{ij} \overline{l^C_{Li}}\Delta^a\sigma^a l_{Lj}&&+ \text{h.c.}\Big]
    &&-M_{\Delta}^2|\Delta|^2
    -{\boldsymbol{\mu}}{\Phi}^\dagger (\Delta^a\sigma^a)\tilde{\Phi},
  \end{alignat}
\end{subequations}
The tree-level mass matrices $\mathcal{M}$ arising from replacing the Higgs \vev
into the Lagrangians have the structures
\begin{subequations}
  \label{neutrinomassmatricesseesaw}
  \begin{align}
    \text{type I, III: }&&
    \nobluecolor{\delta\mathcal{M}_L} &= 0
    &
    \mathcal{M}_D&\sim \mathbf{y}_\nu v
    &
    \noredcolor{\mathcal{M}_R} &\sim M_R
    \\
    \text{type II: }&&
    \nobluecolor{\delta\mathcal{M}_L} &= \frac{v^2 \boldsymbol{\mu}\mathbf{y}_\nu}{2M_\Delta^2}
    &
    \mathcal{M}_D &=0
    &
    \noredcolor{\mathcal{M}_R} &=0.
  \end{align}
\end{subequations}
They correspond to the above scenarios (c) and (b),
respectively. After integrating out all heavy BSM fields, the
resulting values of the dimensionful coefficients of the dimension-5
operator  have the form\footnote{\label{seesawradiativecorrections}%
We note that in these seesaw models, it is possible that some neutrinos remain massless at tree level. In such a case, loop corrections can still generate viable, non-vanishing neutrino masses. For instance in the seesaw type I case, Ref.~\cite{Pilaftsis:1991ug} shows that specific textures of the mass matrices $\mathcal{M}_R$ and $\mathcal{M}_D$ lead to vanishing tree-level neutrino masses. Non-vanishing masses and effective contributions to $\delta \mathcal{M}_L$ then arise from loop diagrams with $Z$- or Higgs-boson exchange.
In the seesaw type III case, it is also possible to generate three massive active neutrinos with just a single $\Sigma_R$ field. Here, there can be only a single massive light neutrino at tree level, but the other light neutrino masses can be generated by two-loop diagrams involving two $W$-bosons~\cite{Babu:1988ig}. 
}
\begin{align}
  \label{cneutrinomassesseesaw}
  C_{\underset{}{\nu\nu}} & \sim \frac{\mathbf{y}_\nu^2}{M_R} \text{ (type I, III)}
  &
  C_{\underset{}{\nu\nu}} & \sim\frac{\boldsymbol{\mu} \mathbf{y}_\nu}{M_\Delta^2} \text{ (type II)},
\end{align}
suppressing flavour indices.

In type II, the mass matrix arises from a triplet \vev $v_\Delta$ which arises automatically as $v_\Delta\sim \mu v^2/M_\Delta^2$ because the potential receives a linear term $\sim\mu v^2\Delta$ after EWSB.
The light eigenvalues of the mass matrices in Eq.~\eqref{neutrinomassmatricesseesaw} agree with the neutrino masses obtained via $v^2 C_{\underset{ab}{\nu\nu}}$ from Eq.~\eqref{cneutrinomassesseesaw}.

In all cases the neutrino masses follow the general pattern in Eq.~\eqref{mneudim5}, but the detailed parameter dependences differ between the types. Of course, the total particle spectrum and wider phenomenology of the three types are very different.
We note in passing that the general idea of generating small neutrino
masses at tree-level via the exchange of BSM states can be extended in
many ways. Two illuminating examples are the inverse seesaw 
and the linear seesaw mechanisms, proposed in
Refs.~\cite{Mohapatra:1986bd,Gonzalez-Garcia:1988okv} and
Refs.~\cite{Barr:2003nn,Malinsky:2005bi}, respectively. Both cases may
be regarded as extensions of the Type I seesaw mechanism and
likewise correspond to our above scenario (c). Compared to Type I seesaw,
the number of BSM neutrinos is doubled, and our matrices
$\mathcal{M}_R$ and $\mathcal{M}_D$ become $2\times2$ and $1\times2$
block matrices.
These block matrices and the resulting active neutrino masses  have the structure 
\begin{align}
  \text{inverse seesaw: }&
&\mathcal{M}_R&\sim   \begin{pmatrix}
    0 & {M}\\
    {M} ^T & \mu
  \end{pmatrix}
  &
  \mathcal{M}_D&\sim
  \begin{pmatrix}
    \mathbf{y}_\nu v & 0
  \end{pmatrix}
  &
  M_\nu&  \sim 
  \mu (\mathbf{y}_\nu v )^2/M^2,
  \\
  \text{linear seesaw: }&
&  \mathcal{M}_R&\sim   \begin{pmatrix}
    0 & {M}\\
    {M} ^T & 0
  \end{pmatrix}
  &
  \mathcal{M}_D&\sim
  \begin{pmatrix}
    \mathbf{y}_\nu v & M_L
  \end{pmatrix}
  &
  M_\nu&\sim
  M_L \mathbf{y}_\nu v  /M.
\end{align}
  In the inverse seesaw case, $M$ is a Dirac mass in the BSM neutrino sector
  and $\mu$ is a lepton-number violating
  parameter. The smallness of the active neutrino masses is then
  explained by small $\mu$ without the need for very large $M$. In the
  linear seesaw case, $M_L$ acts as a lepton-number violating mass
  parameter, and the active neutrino masses turn out to depend only
  linearly on the Yukawa couplings.

\begin{table}
  \centering
  \begin{tabular}{| c | c | c c | c |}
    \hline
    \textbf{Model} & \textbf{Symmetry} & \textbf{Fields} & \textbf{Rep.} & \textbf{Parameters}\\\hline\hline  
    Zee Model 	& $\mathcal{G}_\text{EW}$ & \makecell[ct]{$h^+$ \\ $\Phi_2$}  & \makecell[lt]{$(\bm1,1)_0$ \\ $(\bm2,\frac{1}{2})_0$} & $\mathbf{f}_{ij},~\mathbf{y}^\alpha_{ij},~ \mathbf{M}_{\alpha\beta}$ \\ \hline
    Babu Model 	& $\mathcal{G}_\text{EW}$ & \makecell[ct]{$h^+$ \\ $k^{++}$} & \makecell[lt]{$(\bm1,1)_0$ \\ $(\bm2,2)_0$} & $\mathbf{f}_{ij},~\mathbf{h}_{ij},~ \bm\mu$ \\\hline
    \makecell[tc]{Ma Model \\ (scotogenic)} & $\mathcal{G}_\text{EW};~\mathds{Z}_2$ & \makecell[ct]{$N_{R\,i}$ \\ $\eta$} & \makecell[lt]{$(\bm1,0;-)_{\frac{1}{2}}$ \\ $(\bm2,\frac{1}{2};-)_0$} & 
    $M_{R\,i},~\mathbf{h}_{ij},~ \boldsymbol{\lambda}_5$\\\hline
  \end{tabular}
  \caption{One-loop neutrino mass models. The respective spins and representations with respect to $\mathcal{G}_\text{EW}=\GEW$ (or $\mathds{Z}_2$) are given in the brackets. All fields are singlets under $\SUc$.}
  \label{table:ZBM}
\end{table}

A very attractive alternative to the three seesaw types mentioned
above or their tree-level extensions is the idea of radiative neutrino mass generation, where $\delta\mathcal{M}$ is effectively obtained from loop diagrams, providing additional suppression. 
We illustrate the idea again by briefly reviewing three well-known models by Zee~\cite{Zee:1980ai}, Babu~\cite{Babu:1988ki}, and by Ma~\cite{Ma:2006km}, respectively. In the Zee and Babu models only the scalar sector is enlarged whereas in the scotogenic model (or Ma model) the fermion and scalar sectors are enlarged with Majorana fermions and an inert Higgs doublet. The field contents in each model are listed in Table~\ref{table:ZBM}.
The Zee model can be understood as an extended two Higgs doublet model (2HDM), of which the scalar sector is enlarged by one more charged scalar singlet $h^+$. The Babu model is built on the SM by adding two charged scalar singlets: one singly charged scalar $h^+$ and one doubly charged scalar $k^{++}$. The scotogenic model can be considered an extended inert two Higgs doublet model where  three right-handed neutrinos $N_{R\,1,2,3}$ are included in addition to the inert Higgs doublet $\eta$.
All three models are based on the SM gauge invariance, but in the scotogenic model also a discrete $\mathbb{Z}_2$ invariance is imposed. The additional fields are odd under $\mathbb{Z}_2$ and the SM fields even. 
As no additional Majorana fermions are added, $\mathcal{M}_R$ and $\mathcal{M}_D$ vanish in the Zee and Babu models, corresponding to scenario (b) above, while the scotogenic model represents scenario (d). 

The additional Lagrangians of the three models can be written as

\begin{subequations}
  \begin{alignat}{5}
    \text{Zee-model:}& &&\La \supset \mathbf{f}_{ij}\,\overline{l_{Li}^{C}} l_{Lj} h^{+}  
    &&-\mathbf{y}^\alpha_{ij}\,\overline{l_{Li}}\Phi_\alpha e_{Rj} 
    &&+ \mathbf{M}_{\alpha\beta}\tilde\Phi^\dagger_{\alpha} \Phi_{\beta} h^{+}
    &&+\text{h.c.}, \label{LZee} \\
    \text{Babu-model:}& \qquad&&\La \supset \mathbf{f}_{ij}\, \overline{l _{Li}^C} l _{Lj} h ^{+} 
    &&+ \mathbf{h}_{ij}\, \overline{e_{Ri}^C}e_{Rj} k ^{++} 
    &&+ \boldsymbol\mu \left( h^+ h^+ k^{--} \right)\, 
    &&+ \text{h.c.}, \label{LBabu} \\
    \text{Ma-model:}& &&\La \supset \tfrac12 M_{R\,i} \overline{N^{C}_{R\,i}} N_{R\,i}
    &&- \mathbf{h}^\dagger_{ij} \overline{l_{Li}}N_{R\,j} \tilde{\eta} 
    &&- \tfrac{1}{2}\boldsymbol\lambda_5 (\eta^\dagger\Phi)(\eta^\dagger\Phi) 
    &&+ \text{h.c.} \label{Lscotogenic}
  \end{alignat}
\end{subequations}

In the Zee  and Babu models the scalar singlet $h^+$ has charge $Q=+1$ and Lepton number $L=2$; the Higgs doublet(s) in the Zee model are called $\Phi_a$, $a \ge 2$ (the SM Higgs doublet is denoted as $\Phi\equiv\Phi_1$ in this context). As the \vev of $h^+$ vanishes it does not contribute the gauge boson masses. 

In the Zee model the cubic interaction term $\sim M_{ab}$ is responsible for the one-loop Majorana mass generation for the neutrinos.   Similarly, in the Babu model, the triple scalar coupling term $\sim\mu$ is responsible for the two-loop generation of Majorana neutrino masses.

In the Ma/scotogenic model the parameter $\boldsymbol\lambda_5 $ is lepton-number violating and responsible for the loop generation of Majorana neutrino masses. Note that the parameter $\mathbf{h}$ in the scotogenic model has a different meaning than the parameter $\mathbf{h}$ in the Zee and Babu models.

\begin{figure}[t]
	\begin{center}
		\begin{subfigure}{0.3\textwidth}
			\includegraphics{{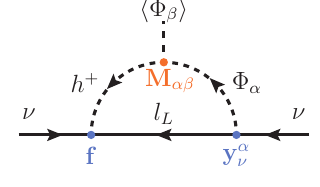}}
			\subcaption{Zee model}
			\label{fig:nmassZee}
		\end{subfigure}
		\begin{subfigure}{0.3\textwidth}
			\includegraphics{{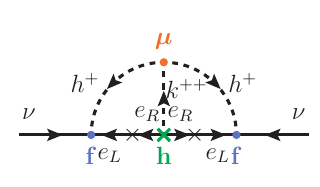}}
			\subcaption{Babu model}
			\label{fig:nmassBabu}
		\end{subfigure}
		\begin{subfigure}{0.3\textwidth}
			\includegraphics{{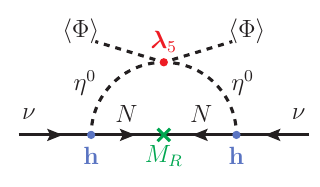}}
			\subcaption{Ma model (scotogenic model)}
			\label{fig:nmassScoto}
		\end{subfigure}
	\end{center}
	\caption{\label{fig:radiativeneutrinomassloops}Lowest-order one-loop or two-loop diagrams contributing to the dimension-5 operator Eq.~\eqref{eq:nmassDim5} in three well-known radiative neutrino mass scenarios, the Zee-, Babu- and Ma/scotogenic models. See Eqs.~\eqref{radiativeneutrinomassresults}. The lepton-number violating vertices are highlighted in orange.}
\end{figure}

The lowest-order contributions to the dimension-5 operator is then given by the one- and two-loop diagrams in Fig.~\ref{fig:radiativeneutrinomassloops}.
Like in the seesaw case, the neutrino masses are then generated via the effective dimension-5 operator with the schematic form
\begin{subequations}\label{radiativeneutrinomassresults}
  \begin{alignat}{2}
    \text{Zee-model}:& \qquad  &&C_{\underset{}{\nu\nu}} \sim
    \frac{1}{16\pi^2}
    \frac{\mathbf{f} ^{} y ^{} v m_\ell}{M_h ^2}\mathbf{M}_{}\ln(M_h ^2/M_\Phi ^2) \,,\\
    \text{Babu-model}:& &&C_{\underset{}{\nu\nu}} \sim
    \frac{1}{16 \pi^2}
    \frac{\mathbf{f} ^{}_{} m_\ell \mathbf{h}^{}_{} m _\ell
      \mathbf{f}^\dagger}{2 \pi^2 M_{k^{++}} ^2}\boldsymbol\mu \ln ^2 \left( \frac{M_{h^+} ^2 + M _{k^{++}} ^2}{M_{h^{+}} ^2}\right)\,,\\
    \text{Ma-model}:& &&C_{\underset{}{\nu\nu}} \sim
    \frac{1}{16\pi^2} 2 {\boldsymbol\lambda} _5 
    \frac{\mathbf{h} ^{}_{}\mathbf{h} ^{}_{} M_R}{m_0 ^2 - M_R^2}
    \left(1 - \frac{M_R ^2}{m _0 ^2 - M_R ^2} \ln\frac{m_0^2}{M_R ^2} \right)\,,
    \label{eq:cdim5ZBscoto}
  \end{alignat}
\end{subequations}
where all indices corresponding to flavour or the number of the Higgs doublet have been suppressed and where $m_\ell$ denotes a lepton mass of any generation. In the scotogenic model, $m_0$ denotes the new scalar mass scale.
The loop-generated masses are suppressed by the loop factor $\frac{1}{16 \pi ^2} $ which allows lower BSM masses for small neutrino masses. 

The six sketched models represent basic ways how neutrino masses can
be generated. They can of course be extended and generalised. For
example, tree-level and loop generation can be combined as mentioned
in footnote \ref{seesawradiativecorrections} or as in Ref.~\cite{Grimus:1989pu}, where additional scalar doublets and right-handed singlet neutrinos are added; mass generation may arise only at higher loop orders, or models may use an extended gauge group.
Finally it is possible that the dimension-5 operator in Eq.~\eqref{eq:nmassDim5} does not provide an appropriate description because higher-dimension operators or a different EFT should be used. We will discuss a variety of modified models in Sec.~\ref{sec:neutrino_mass_gm2}, together with their impact on $\amu$.

\subsubsection{Example contributions of neutrino mass generation mechanisms to $\amu$}

We can now return to the question how neutrino masses can be related to $\amu$. In essence, not the light neutrinos themselves, but the BSM particles which are responsible to generate the neutrino masses can often cause potentially significant contributions to $\amu$.\footnote{%
For similar connections between $\amu$ and neutrino transition moments see Ref.~\cite{Babu:2021jnu}.
}

Two simple considerations illustrate this point. First, nonvanishing neutrino masses directly modify the value of the EWSM diagram with W-$\nu$-loop in Fig.~\ref{fig:SMEW1}, see also Eq.~\eqref{eq:amu-EW-1}. Technically, the second argument of the appearing $\F^{\text{FV}}, \G^{\text{FV}}$ loop function is then $z_\nu=m_{\nu}^2/M_W^2$ instead of zero.
Given the absolute neutrino mass bound, the variable $z_\nu\lesssim10^{-24}$ and the absolute impact of neutrino masses on this diagram is 
\begin{align}
  \Delta\amu^W({\nu\text{-mass}}) \lesssim10^{-33}
  \label{Eq:amu_from_nu_bound}
\end{align}
and thus totally negligible.
\footnote{Here we do not yet consider the effects of any mixing or contributions from new states that may accompany neutrino masses.} 
Second we can consider the EFT point of view and compare the dimension-5 term Eq.~\eqref{eq:nmassDim5} for neutrino masses with the dimension-6 dipole operator term Eq.~\eqref{SMEFTphotondipole}.
The neutrino masses and $\amu$ contributions are then parametrised as
\begin{align}
  m_\nu&\sim c_{\underset{}{\nu\nu}}\times\frac{v^2}{M_\Lambda}
  &
  \Delta\amu &\sim c_{e\gamma} \times\frac{vm_\mu}{M_{\Lambda}^2},
\end{align}
again pulling out appropriate powers of $M_\Lambda$ to obtain dimensionless Wilson coefficients. Here the neutrino masses have only a single $M_\Lambda$ suppression, while $\Delta\amu$ has a suppression $vm_\mu/M_{\Lambda}^2$. In addition, both contributions involve Wilson coefficients $c_{\underset{ab}{\nu\nu}}$, $c_{e\gamma}$ which can in principle be independent. 

In the simplest case where $c_{\underset{ab}{\nu\nu}}\sim1$ and the small neutrino masses require $M_\Lambda\sim10^{15}$ GeV, we again obtain entirely negligible values of $\Delta\amu$. However, it is possible that the BSM scale $M_\Lambda$ is significantly smaller and the smallness of neutrino masses is instead explained by a strongly suppressed $L$-violating parameter $c_{\underset{}{\nu\nu}} \ll c_{e\gamma}$. In this case, the above relation implies the numerical relationship
\begin{align}
  \label{genericamuneutrinorelation}
  \Delta\amu \sim
   10^{-9}
  \left(\frac{\text{1 TeV}}{M_\Lambda}\right)
  \left(\frac{c_{e\gamma}}{c_{\underset{}{\nu\nu}}} \times 10^{-8}\right).
\end{align}
Here a neutrino mass of the order $0.05$ eV, slightly below the current upper limit, has been assumed. The formula shows that in scenarios where the neutrino masses arise from underlying BSM physics not far above the TeV scale, and where the $\Delta L=2$ operators receive an additional suppression, significant $\Delta\amu$ can arise from the same underlying physics. Such neutrino mass models are constrained by $\Delta\amu$.

\begin{figure}[t]
  \centering
  \begin{subfigure}{.49\textwidth}
    \centering
    \includegraphics[width=.5\textwidth]{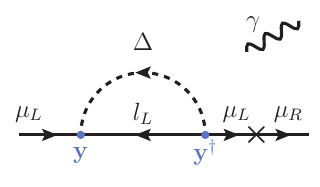}
    \caption{$\Damu$ in seesaw type II}
    \label{fig:amutypeII}
  \end{subfigure}
  \begin{subfigure}{.49\textwidth}
    \centering
    \includegraphics[width=.5\textwidth]{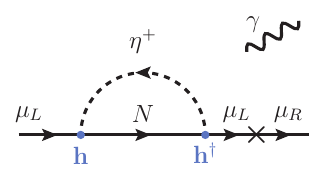}
    \caption{$\Damu$ in the Ma/scotogenic model}
    \label{fig:amuscotogenic}
  \end{subfigure}
  \caption{One-loop contributions to $\amu$ in two neutrino mass models. See Eqs.~(\ref{amutypeII},\ref{amuscotogenic}). The chirality flip is  shown explicitly on the external line; similar diagrams with the chirality flip on the other line exist as well.}
  \label{fig:amuneutrinomodels}
\end{figure}

The possible connections between $\amu$ and neutrino masses thus depend strongly on details of the underlying mechanisms. Here we illustrate the possible relationships with the help of few simple and well-motivated models as a background for further discussions in later sections. We begin with  two examples pointed out in Ref.~\cite{Ma:2001mr}, which first asked the question whether neutrinos and contributions to $\amu$ could be related. The first example is the seesaw type II model (\ref{seesawII}) with scalar triplet $\Delta$ and two mass parameters $\mu$, $M_\Delta$ and the neutrino mass scale given via Eq.~\eqref{cneutrinomassesseesaw} as $\mu y_\nu v^2/M_\Delta^2$. The small neutrino masses can be well explained by a low seesaw scale $M_\Delta\sim$ 1 TeV, if the lepton-number violating parameter $\mu$ is tiny. A one-loop diagram for $\amu$ is obtained via two muon--triplet--anti-lepton vertices governed by the Yukawa coupling $y_\nu$, as shown in Fig.~\ref{fig:amutypeII}.
The contributions correspond to the one-field model 4 in Tab.~\ref{tab:onefieldmodels} discussed in Sec.~\ref{sec:MinimalBSM}. They cannot contain a chiral enhancement, and the sign is fixed to be negative. Hence,  $\Damu$ in the seesaw type II model is of the order
\begin{align}
  \Delta&\amu\sim-\frac{|\mathbf{y}_\nu|^2}{64\pi^2}\frac{m_\mu^2}{M_\Delta^2}.
  \label{amutypeII}
\end{align}
The  statement on the negative sign of $\Damu$ has been generalised to all seesaw type I and III models in Ref.~\cite{Biggio:2008in}.

Another example is the scotogenic model (\ref{Lscotogenic}) with loop-generated neutrino masses via loops of $\mathbb{Z}_2$-odd right-handed neutrinos and Higgs doublet $\eta$. Again, the formula for the resulting neutrino masses allows a low seesaw scale $M_R$ if the lepton-number violating coupling ${\boldsymbol\lambda} _5$ is tiny. 
Contributions to $\amu$ correspond to model 26 in Tab.~\ref{tab:twofieldmodels}; they are given by a diagram of the same kind as in the seesaw type II case, see Fig.~\ref{fig:amuscotogenic}.
The result is thus of the same form, with the appropriate coupling of the lepton to the right-handed neutrino and the new Higgs doublet $\eta$,
\begin{align}
  \Delta&\amu\sim-\frac{|\mathbf{h} ^{}|^2}{64\pi^2}\frac{m_\mu^2}{M_\Delta^2}.
  \label{amuscotogenic}
\end{align}

Both of these examples, together with the neutrino masses in Eqs.~\eqref{cneutrinomassesseesaw} and~\eqref{eq:cdim5ZBscoto}, are concrete illustrations of the relationship given in Eq.~\eqref{genericamuneutrinorelation}. The role of the BSM scale $M_\Lambda$ is taken by the triplet or right-handed neutrino masses, and the role of $c_{\underset{ab}{\nu\nu}}/c_{e\gamma}$ is taken by either the small ratio $\mu/M_\Delta$ or by the small parameter ${\boldsymbol\lambda} _5 $, respectively. Note that $\mu$ and ${\boldsymbol\lambda} _5 $ are both lepton-number violating and can therefore be assumed to be small in a technically natural way.  
Due to the negative signs, none of the tree-level models are candidates to explain large positive deviations like $\DamuOld$. Beyond that, Ref.~\cite{Ma:2001mr} showed that the correlations to lepton flavour violation of the kind discussed in Sec.~\ref{sec:LeptonDipole} are strong and essentially exclude sizeable magnitudes of contributions to $\amu$ in both models.
Hence the new result $\DamuFinal$ does not exclude additional parameter regions.

Though $\Damu$ in these two examples turns out to be ultimately small, the examples illustrate how sizeable $\Damu$ may arise in neutrino mass models and how such models may be constrained by $\amu$ results. In Sec.~\ref{sec:neutrino_mass_gm2} we will discuss further concrete neutrino mass models. The models show a great diversity of ideas how to explain neutrino masses and their smallness, and some are motivated also by other phenomenological goals. Many of them are elaborate constructions extending the scotogenic/Ma model, see Tab.~\ref{table:scotoextension}, but several others are distinct, see Tab.~\ref{table:moreextension}. Sec.~\ref{sec:neutrino_mass_gm2} will also discuss the contributions to $\amu$ in these models.
In many of them, chirally enhanced contributions to  $\amu$ arise of the kind present in the three-field models of Classes I/II/III in Sec.~\ref{sec:genericthreefield}. For this reason, these models are non-trivially constrained by $\amu$.

\subsection{Dark Matter}\label{sec:DarkMatter}
Dark Matter (DM) is one of the primary motivations for physics beyond
the standard model, and is evidenced by galaxy rotation curves (see
e.g.\ \cite{Rubin:1970zza, Rubin:1980zd}), gravitational lensing
(e.g.\ \cite{Clowe:2006eq,Gavazzi:200933G}) and the cosmic microwave
background \cite{Aghanim:2018eyx}.  Furthermore cold dark matter is
one of the postulates of the standard model of cosmology and is
required to reproduce the structure formations of
galaxies.  See Refs. \cite{Bertone:2016nfn,Cirelli:2024ssz,Balazs:2024uyj} for detailed, pedagogical reviews of the evidence and further references to the original work.

To contribute to the dark matter inferred from astrophysics
observations a particle dark matter candidate should be
electromagnetically neutral\footnote{The possibility of milli-charged
  dark matter has also been considered in the literature, see
  e.g.\ \cite{Goldberg:1986nk,Cheung:2007ut,Huh:2007zw}, and remains
  possible as long as the charges are small enough
  \cite{McDermott:2010pa}, but we will not discuss it further here.}
and non-baryonic and is thus expected to interact with the SM at most
through weak or gravitational interactions or some new interaction
between the SM and the dark sector.  DM can influence the anomalous
magnetic moment of the muon ($a_\mu$) when the stable dark matter
particle enters the loops, and/or from loops involving other new
states that are introduced to make a consistent gauge invariant theory
explaining dark matter.  This can lead to a positive or negative shift
away from SM prediction of $a_\mu$ and means that a measurement of
$a_\mu$ with a large deviation from the SM prediction could be
experimental evidence of particle dark matter, while measurements
consistent with the standard model prediction serve as an important
constraint on theories that explain dark matter.

The possible topologies for one-loop diagrams that can influence
$a_\mu$ are shown in Fig.\ \ref{fig:amu-one-loop} in
Sec.\ \ref{sec:genericoneloop} where they are discussed in detail and
the form of the corrections they imply may be found.  Since it should
be electromagnetically neutral the dark matter particle cannot couple
to the photon, and since it should also be stable it must not have
kinematically allowed channels through which it decays. Thus in
one-loop corrections to $a_\mu$, the dark matter candidate may then only play the role of the particle connecting the muons in the loop and must therefore have couplings between the DM candidate, the muon and the charged particle in the loop.  These properties  significantly  constrain the possible quantum numbers, and can also constrain the mass of the DM candidate, since decays through this coupling must be kinematically forbidden.

In the following we discuss the possible impact
  of the new $\amu$ result on models for dark matter. We begin with
  general discussions of dark matter properties and
  possible contributions of dark matter
  candidates to $\amu$, then we systematically discuss dark matter
  models with one, two, or three new fields.

\subsubsection{The relic density of dark matter, WIMPs and the detection of dark matter}
 Any particle with the properties described above is a dark matter
 candidate and {\it may} contribute to the observed dark
 matter. However, this is not sufficient to explain dark matter
 observations, one also needs to check that the correct abundance of dark matter can be generated in order to agree with these observations.   More precisely the relic density of dark matter should be in agreement with the value measured by the Planck collaboration \cite{Aghanim:2018eyx},
\begin{align} \Omega_{\text{DM}} h^2 = 0.120\pm0.001. \label{Eq:PlanclRelicDensityDM}\end{align}
Here,
\begin{align} \Omega_{\text{DM}} = \frac{\rho_{\text{DM}}}{\rho_{\text{c}}} = \frac{n_{\chi, 0} m_{\chi}}{\rho_{\text{c}}} \label{Eq:relic_densiy_def}  \end{align} is the DM density normalised to the critical value for the total energy density for a flat Universe $\rho_{\text{c}} := 3 H_0^2 / (8 \pi G)$ where $G$ is Newton's Gravitational constant and $H_0$ is the present day value of the Hubble parameter.  In Eq.\ \eqref{Eq:PlanclRelicDensityDM} the measured value is presented by convention with the square of the dimensionless Hubble rate $h = H_0 / (100\, \text{km s}^{-1}\text{Mpc}^{-1}) = 0.6736\pm 0.0054$ \cite{Aghanim:2018eyx}. On the right-hand side of Eq.\ \eqref{Eq:relic_densiy_def} we have written the density of dark matter in terms of the present day number density of the dark matter $n_{\chi,0}$ and the mass of the dark matter candidate $m_\chi$, as is appropriate for particle dark matter, and in the case where the dark matter is composed of multiple components with different masses this should be interpreted as a sum over the components $\chi$.   

Satisfying the relic density provides an important test for standard model extensions that provide dark matter candidates and typically places strong restrictions on the parameter space or even excludes scenarios entirely.   Even if one does not require an explanation of  all the dark matter within the model, justifying this by assuming some or most of the dark matter is made of other essentially orthogonal new physics\footnote{For example when considering a model with a  WIMP candidate one could allow for the possibility that there is also an axion dark matter candidate that can contribute.}, this still provides an important test or constraint on new physics. This is because it should still be checked that the model does not predict an over abundance of dark matter, i.e.\ the above Planck measurement should still be used to give an upper bound on $\Omega h^2$.

The most commonly considered mechanism for generating the DM is the freeze-out mechanism \cite{Roszkowski:2017nbc,Feng:2022rxt} where the DM is thermally produced. When there are very high temperatures in the early universe the particle DM candidates and SM particles are in thermal equilibrium, such that the production and annihilation rates of the DM candidates are equal. When the temperature drops below the mass of the dark matter, production is suppressed and the dark matter is no longer in thermal equilibrium with the SM states and annihilation of the dark matter then drives a rapid decrease in their number densities.  As the Universe then cools down and expands, reducing the number densities,  we reach a point where the DM can no longer annihilate.  The relic abundance of the dark matter freezes out at this time/temperature and subsequently DM number densities evolve according to only the expansion of space, leading to a prediction for the relic density for the dark matter today.

The number density of the dark matter evolves during this process
according to the Boltzmann equation,
\begin{equation}
  \frac{dn_\chi}{dt} = - 3 H n_\chi - \langle \sigma_{\text{eff}}\, v\rangle (n_\chi^2 - n_{\chi\,\text{eq}}^2)\label{Eq:n_chiBolzmann}
\end{equation}
where $\langle \sigma_\text{eff}\, v\rangle$ is the thermal average of
the effective annihilation cross-section which is computed for the
specific particle physics model and $n_{\chi\,\text{eq}}$ is the number density at
equilibrium.  The first term in Eq.\ \eqref{Eq:n_chiBolzmann} is
present because the number density is affected by the expansion of
space as well as by the annihilations.   When freeze out occurs $n_\chi / a^3$ becomes constant, where $a$ is the scale factor for the expansion of space.  

The Boltzmann equations
should be solved numerically, as is done in public software packages, e.g.\ MicrOMEGAs and DarkSUSY \cite{Belanger:2001fz,Gondolo:2004sc,Arbey:2018msw}. However rough approximations can also be used to give crude estimates of the relic density \cite{Feng:2022rxt},
\begin{equation}
  \Omega_{\text{DM}} \approx \frac{x_f T_0^3}{\rho_c M_\text{Pl}  \langle \sigma_{\text{eff}}\, v\rangle } \approx \frac{x_f T_0^3}{\rho_c M_\text{Pl}} \frac{16 \pi^2m_\chi^2}{g^4},
  \label{Eq:crude_rel_den}
\end{equation}
where $x_f = \frac{m_\chi}{T_f}$ is the ratio of dark matter mass to
the freeze-out temperature $T_f$, $T_0$ is the present day temperature
of the Universe and $M_\text{Pl}$ is the Planck mass. In the last step
we have used a simple ansatz for the thermally averaged
annihilation cross-section based on the dimensionality and assuming
weak interactions, with $m_\chi$ being the only mass scale involved.
Eq.\ \eqref{Eq:crude_rel_den} shows that higher annihilation
cross sections lead to later freeze-out and hence reduced relic
density. The resulting sensitivity to
the mass $m_\chi$ and coupling $g$ via the ratio $m_\chi^2/g^4$ illustrates that the
  freeze-out mechanism can easily produce too much or too little dark
  matter, depending on the parameter choices.  In fact since collider experiments are typically placing lower limits on masses and/or upper limits on interactions, avoiding overabundance can be the biggest challenge with the freeze-out mechanism and is also the constraint which is hardest to argue away, as discussed above.       

In Eq.\ \eqref{Eq:crude_rel_den} we have used $g$ to denote the
coupling to indicate the connection to the weak gauge coupling. While
there can be other possibilities where the dark matter (or more
generally dark sector) interacts with the standard model via new
couplings, if we assume it is the weak gauge coupling then it implies
the full relic density is explained when $m_\chi\sim 100$ GeV -- $1$
TeV, which is known as the WIMP miracle.  The apparent miracle here is
it just so happens that a weak gauge coupling and weak scale mass give
the annihilation cross-section needed to predict the observed relic
density of dark matter.  This led to the common Weakly Interacting
Massive Particle (WIMP) paradigm, with WIMPs being the most popular
dark matter candidate. Now, the term WIMP is generally used for a dark
matter candidate that is a thermal relic, generating the relic
abundance via the freeze-out mechanism, with a mass around the weak
scale.  For example singlet Higgs portal dark matter couples to the
standard model through quartic interactions with the Higgs,
e.g.\ $\lambda |H|^2SS$ where $S$ is the dark matter candidate.
Although the mass and coupling $\lambda$ are typically close to those
of the Weak scale and Weak force, the DM candidate is a singlet under
the SM gauge group and does not interact via the Weak force.
Nonetheless for such candidates the freeze-out mechanism generates the
relic density and they are typically referred to as WIMPs.

\begin{figure}
  \begin{center}
    \includegraphics[width=0.5\textwidth]{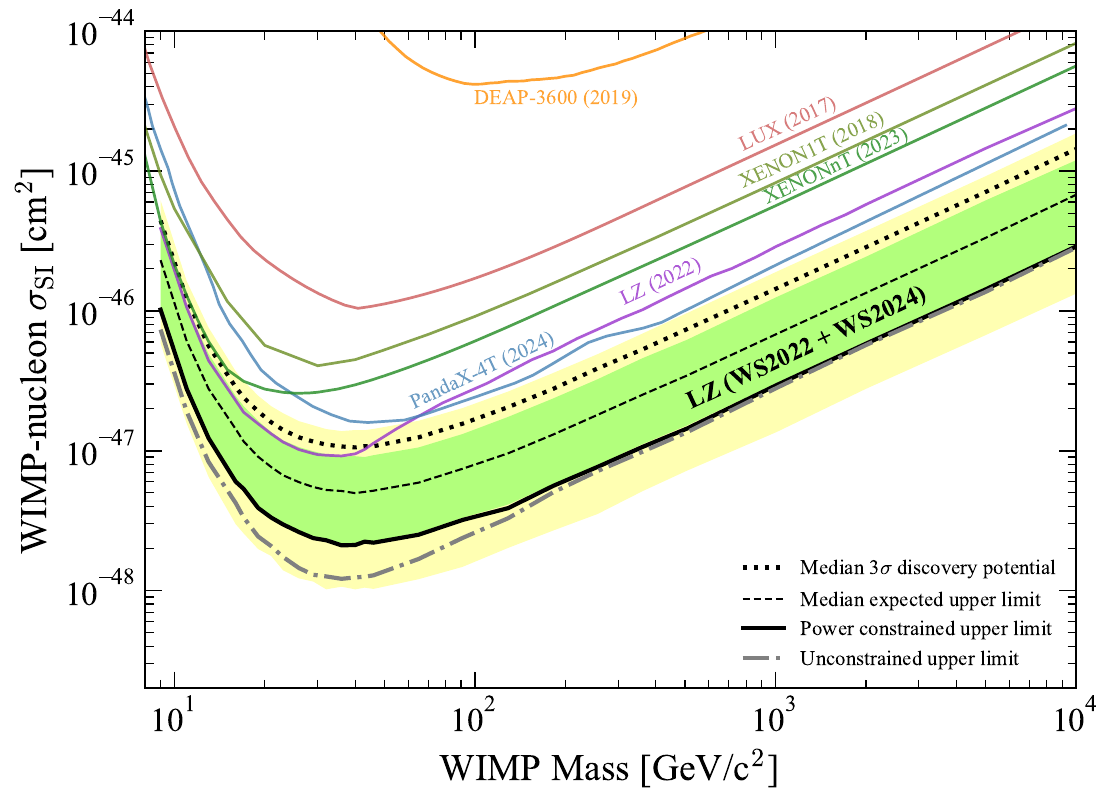}          
  \end{center}
  \caption{\label{LZ-DDlimits} The latest direct detection limits on WIMP-nucleon cross-section, plotted against the mass of the dark matter candidate, from the LZ collaboration \cite{LZ:2024zvo} which is shown as the solid black line.  Also shown for comparison are $1 \sigma$ limits from their 2022 release \cite{LZ:2022lsv} and from other collaborations: PandaX-4T \cite{PandaX-4T:2021bab}, LUX \cite{LUX:2016ggv}, XENONnT \cite{XENON:2023cxc}, XENON1T \cite{XENON:2018voc} and DEAP-3600 \cite{DEAP:2019yzn}, while the green and yellow regions show the range of expected upper limits based that would be seen $68\%$ and $95\%$ of the time respectively assuming background only.  This figure is taken from Ref.\ \cite{LZ:2024zvo} where more details on the analysis and precise meaning of the limits can be found. }
\end{figure}

However in recent years the WIMP paradigm has come under increased
pressure from non-observation in both direct and indirect detection
experiments as well as LHC searches, leading to a reduction in the
favouritism towards WIMPs \cite{Arcadi:2017kky}.  While LHC searches
(see Sec.\ \ref{sec:Collider}) focus on the production of the DM from
SM particles, direct detection experiments instead try to detect the
recoil of the nucleus in ordinary matter from the scattering of
incident dark matter particles with the nucleons.  The first direct
detection limits on DM appeared in 1987 \cite{Ahlen:1987mn}, and since
then many experiments have successively increased sensitivity over the
years, producing strong constraints in the plane of the mass of the DM
candidate and (spin-independent and spin-dependent) WIMP-nucleon cross
section. Fig.\ \ref{LZ-DDlimits} shows a compilation of experimental
results including the most recent results of the LZ experiment
\cite{LZ:2024zvo}, which provide the strongest direct detection constraints on
WIMPs \cite{LZ:2024zvo}.  To obtain these limits the flux
of the DM particles passing through the earth has to be estimated. In
dark matter direct detection experiments this estimate is typically
based on the Standard Halo Model, see Ref.\ \cite{Baxter:2021pqo} for
recommended values (used by LZ) and the references that these value
are based upon. To apply these limits to a new physics model the
WIMP-nucleon cross-section for the new physics model must be
computed. This cross-section depends on what interactions the DM
candidate has with the SM and as a result so does the impact of direct
detection.  DM interacting via the weak force may interact with the
nucleons via tree level $Z$ boson exchange or, if the DM candidate is
a gauge multiplet with a heavier charged state, through inelastic
upscattering to the charged state mediated by the $W$ boson.  The
Higgs boson may also act as a mediator if the DM couples with the
Higgs. Alternatively the mediator may be a BSM state and/or the DM may
only couple to nucleons at the loop level. 

Indirect detection experiments instead look for excesses in cosmic rays  made up of known particles (charged particles, photons, neutrinos) which could result from dark matter annihilation in regions where the dark matter density is large enough, or dark matter decay products if unstable states are present\footnote{This applies if one considers dark matter that is not perfectly stable, but rather decays on time scales longer than the cosmological scales needed to explain the data.}\cite{Slatyer:2017sev,Cirelli:2010xx}. In the former case, if the annihilation cross-section determines the abundance via the freeze-out mechanism, observation of the annihilation products would probe the mechanism for generating the observed relic abundance.  In contrast to the well-controlled low background direct detection experiments, indirect detection has significant uncertainties due to the astrophysical backgrounds and the distribution of the dark matter in the galaxy. Furthermore, unknown or poorly modelled astrophysical sources can make assessing signals and limits challenging. Nonetheless indirect detection experiments using gamma-ray telescopes (e.g.\ Fermi-LAT \cite{Fermi-LAT:2015att}, MAGIC \cite{MAGIC:2016xys}, VERITAS \cite{VERITAS:2017tif}), neutrino detectors 
(e.g.\  IceCube \cite{IceCube:2016dgk}, ANTARES \cite{ANTARES:2016xuh} )  
and cosmic rays (e.g.\ AMS-02 \cite{AMS:2016oqu}, PAMELA \cite{PAMELA:2013vxg}), looking for signals from  sources such as dwarf spheroidal galaxies and the galactic centre can be an important source of potential signals and constraints on new physics \cite{Gaskins:2016cha}.

Here we note that the above discussion of WIMPs is rather crude, and in
specific models the actual results for cross-sections and dark matter
annihilation can be quite different. These differences can matter
especially for the correlation with $\amu$ or other particle physics
observables.
For example,
mechanisms to boost the dark matter annihilation cross-section have been explored in various WIMP scenarios where experimental constraints otherwise push the mass limits beyond those where 
the relic density can be naturally explained. 
Secondly, it is easy to vary the coupling and mass simultaneously in
Eq.\ \eqref{Eq:crude_rel_den} such that the cross-section does not change. 
Thirdly, besides DM production via freeze-out other possibilities, like the freeze-in mechanism \cite{Hall:2009bx}, can be considered. 
Finally it is worth noting that the relic density can also be affected by non-standard cosmologies (see e.g.\ \cite{Arbey:2018msw}) and explanations 
through exotic macroscopic objects like primordial black holes remain possible \cite{Carr:2020gox,Carr:2021bzv}. Such explanations, however, also require the existence of new physics.

Thus there is a vast range of mass scales (ranging from $10^{-21}$ eV
to $10^{73}$ eV) and many different types of dark
matter candidates to consider.  However since the dark matter could
not enter the diagrams for $a_\mu$ with strong interactions, dark
matter candidates with masses far in excess of the TeV scale can be
expected to contribute negligibly.  In Sec.\ \ref{sec:LightDarkSector}
we will discuss $a_\mu$ predictions for light dark sectors, while in
the remainder of the present section we will briefly discuss the simplest possibilities for fitting both dark matter and $a_\mu$, focusing mainly on the standard freeze-out mechanism and WIMPs.

\subsubsection{Single-field extensions of the SM}
As we have discussed in Sec.\ \ref{sec:MinimalBSM}, it is possible for an extension of the standard model by a single field to contribute to $a_\mu$, with the possibilities listed in Table \ref{tab:onefieldmodels}.  At the same time there are a number of proposed explanations of dark matter that only extend the standard model by a single field, including scalar singlet dark matter \cite{Silveira:1985rk,McDonald:1993ex}, inert Higgs doublet dark matter \cite{LopezHonorez:2006gr} and minimal dark matter \cite{Cirelli:2005uq}. Nonetheless these two sets of models are mostly disjoint since to be a DM candidate the particle should be stable on cosmological time scales, while to enter the one-loop corrections to $a_\mu$ it must couple to the muon and another SM state, introducing a potential decay channel. Therefore unlike multi-field extensions, typically single field DM models do not contribute to $a_\mu$ and thus can neither be discovered nor constrained by this observable. 

More systematically, requiring colour singlets with an
electromagnetically (EM) neutral component cuts the one-field models
in Table \ref{tab:onefieldmodels} down to: scalar (3, 4), fermion (10,
12, 14, 15), vector (16, 18).  Avoiding decays then restricts us to
very light DM in all cases. At the same time as argued earlier the
freeze-out mechanism implies DM with SU$(2)_L$ interactions should fit the observed relic density when the mass is ${\cal O}(100-1000 \text{GeV})$.\footnote{For example Model 3 is a scalar SU$(2)_L$ doublet, which could be interpreted as inert Higgs doublet dark matter \cite{LopezHonorez:2006gr} and it will have the annihilation cross-section needed to deplete the relic density to the observed value (Eq.\ \eqref{Eq:PlanclRelicDensityDM}) when its mass is about 600 GeV \cite{LopezHonorez:2006gr}.}  This is clearly too large for the erstwhile DM candidate to avoid decaying into muons through the Yukawa coupling that allows the inert Higgs doublet to contribute to $a_\mu$ at the one-loop level. Furthermore SU$(2)_L$ doublets and triplets also contain EM charged components and 
interact via $Z$- or $W$-exchange. In addition lower bounds on their masses from collider limits, which we will discuss in Sec.\ \ref{sec:Collider}, imply they must be too heavy to avoid decays into SM states.

Thus we are left with singlet fermion and vector states that must be very light.\footnote{There is no scalar singlet in Table \ref{tab:onefieldmodels} since it does not have direct couplings to the muon.}
Thanks to the Pauli exclusion principle, the lower bound on the mass
of fermion dark matter is ${\cal O}(100\, \text{eV})$
\cite{Tremaine:1979we,Alvey:2020xsk}.\footnote{This limit could be
significantly reduced by introducing very large numbers of different
fermion species \cite{Davoudiasl:2020uig} but this would be very far
from the single field extensions we consider here.} Furthermore Model
10 can be interpreted as a right-handed neutrino or a vector-like
lepton (VLL; specifically the VLL model with singlet $N$ state shown
in later Sec.\ \ref{sec:VLF}).  For the right handed neutrino, the Yukawa coupling is constrained by the seesaw mechanism and upper limits on neutrino masses, so only negligible contributions arise from providing a mass to the neutrino (see Eq.\ \eqref{Eq:amu_from_nu_bound} in Sec.\ \ref{sec:neutrino_masses}) or from mixing effects. At the same time the VLL can decay virtually through heavier SM states into neutrinos, and in order to suppress these decays sufficiently to allow it to explain DM, the couplings would have to be so small that it also contributes negligibly to muon g-2. 

This leaves only Model 16, a new  spin-1 vector boson that is a
singlet under the SM gauge group, as a potential single-field dark
matter model which can contribute to $\amu$.  One realisation of Model
16 is the dark photon
\cite{Holdom:1985ag,Pospelov:2007mp,Arkani-Hamed:2008hhe,Pospelov:2008zw}
(see also Sec.\ \ref{sec:DarkPhoton}) where the couplings to SM states come only from  gauge kinetic mixing.  If the dark photon has a mass that is lighter than two times the lightest neutrino mass then it could be an example of very light dark matter. In this case the DM relic is not produced via the freeze-out mechanism as discussed earlier, but non-thermal mechanisms that have been developed for light dark matter including  vacuum misalignment \cite{Nelson:2011sf,Arias:2012az}, quantum fluctuations during inflation \cite{Graham:2015rva} or abundance transfer from e.g. axions \cite{Agrawal:2018vin} or the inflation \cite{Bastero-Gil:2018uel} via tachyonic instabilities. See also Ref.\ \cite{Cline:2024qzv} for a recent brief review on dark photons.  This model may also be extended in various ways, e.g. by including direct mass mixing with the $Z$ boson, referred to as a dark $Z$ \cite{Davoudiasl:2012ag,Davoudiasl:2012qa,Davoudiasl:2012ig,Davoudiasl:2014kua,Cadeddu:2021dqx}.
Direct couplings to the SM are possible through the charge assignments of the new $U(1)$ gauge group. Importantly, these assignments are restricted by the anomaly cancellation conditions.
Without additional matter fields besides the SM leptons and quarks, the only allowed possibilities are U(1)$_{B-L}$, U(1)$_{L_e-L_\mu}$, U(1)$_{L_e -L_\tau}$ and U(1)$_{L_\mu - L_\tau}$. 
Again, in order for the $Z^\prime$ to be a stable dark matter candidate it must be light enough so that it cannot decay into the SM particles. 
The U(1)$_{L_\mu - L_\tau}$ is special because it avoids couplings to the electron (which are strongly constrained) and, compared to the other possibilities, can give the largest contributions to $\Damu$.  
It is worth noting that these models can also be part of a wider dark matter sector where the new gauge boson may not be the stable dark matter but instead acts as a mediator between the visible and dark sectors.
Such scenarios and their implications for $\Damu$ are discussed in more detail in Sec.\ \ref{sec:LightDarkSector}.

In that section we also discuss another interesting exception to the arguments given above against scalar candidates: the axion or axion-like particles (ALPs) \cite{Chadha-Day:2021szb}.  The axion is a pseudoscalar state introduced to solve the strong CP problem \cite{Peccei:1977hh,Wilczek:1977pj,Weinberg:1977ma}, and as briefly mentioned earlier it can be a viable dark matter candidate. ALPs are a generalisation of the axion that do not necessarily solve the strong CP problem, but interact with SM states through shift-invariant (but suppressed) effective couplings. As such, they do not really satisfy the selection criteria for our Table \ref{tab:onefieldmodels}, 
but can nonetheless have important implications for both DM and $a_\mu$.\footnote{Similarly one can also add a real scalar singlet that couples via effective couplings or by mixing with the Higgs field.  The latter can only affect the one loop result through mixing and the impact should be small, but the former is could be very light dark matter similar to the axion.}  In this case the DM relic abundance can be produced through the vacuum misalignment mechanism \cite{Preskill:1982cy,Abbott:1982af,Dine:1982ah} or from topological defects such as cosmic strings or domain walls, see e.g.\ Ref.\ \cite{Kawasaki:2014sqa} and references therein.
The $a_\mu$ contributions from axions are discussed in detail in Sec.\ \ref{sec:ALPs}.

\subsubsection{Two-field extensions}
\label{sec:DMtwofield}
Compared to the one-field extensions, the situation where two new
fields enter the loop  corrections to $\Damu$ is very different.  In
particular, there are far more DM models of this nature that may
contribute non-negligibly to $a_\mu$. In models with two new fields
of different spin, both neutral and charged states can enter $\Damu$
via the diagrams in Fig.\ \ref{fig:amu-one-loop}.  The neutral state
provides a dark matter candidate as long as it is lighter than the
charged state and also does not have further interactions through
which it can decay.
The latter can be
easily realised within a specific model by e.g.\ imposing a
$\mathbb{Z}_2$ symmetry.

If the two new fields have the same spin then they can
only enter the same one-loop diagram via mixing, and there must also
be a SM state in the loop that couples to the photon. Therefore the
possibilities for dark matter models of this kind with significant
contributions to $\Damu$ are restricted in a similar way to
the one-field extensions as discussed above.  Thus although models like this are of interest because they can get a large chiral enhancement (see e.g.\ Sec.\ \ref{subsec:VLL})
they are not particularly interesting for dark matter explanations and
we will not discuss them further here. 

Therefore we now focus on the case of two fields with different
spin. Models of this nature are collected in Table
\ref{tab:twofieldmodels} in Sec.~\ref{sec:MinimalBSM} and, as
discussed there, these models cannot result in chirally enhanced
contributions to $\Damu$.  Instead for all of these models the
couplings at the vertices of the one-loop diagram must either be to
right handed muons only (written as `R' in Table
\ref{tab:twofieldmodels}) or to left handed muons only (written as `L'
in Table \ref{tab:twofieldmodels}).  This means the $a_\mu$
contributions are roughly given by the simple formula in
Eq.\ \eqref{nonenhancedoneloop} or Tab.~\ref{tab:estimates} (a), and
even before looking at the details, we already understand that while
we are no longer pushed towards very light dark matter, the
contributions to the anomalous magnetic moment from these states are
limited.  Large contributions beyond the 2$\sigma$ range corresponding
to $\DamuFinal$ (see Eq.~\eqref{DamutwoSigma}) are only possible for
states with ${\cal O}(1)$ couplings with ${\cal O}(100)$ GeV masses or
lighter. Nonetheless these models appear to be the simplest class of
models that can generically offer a dark matter candidate and provide
non-negligible contributions to the anomalous magnetic moment of the
muon.

In the following we consider models with a new scalar and a new Dirac
fermion, which have been studied in Refs.\ \cite{Calibbi:2018rzv,
  Athron:2021iuf,Kowalska:2017iqv}, see also
Ref.~\cite{Agrawal:2014ufa} for an earlier study.
Following Ref.\ \cite{Calibbi:2018rzv} we will refer to these as RR or
LL models respectively and construct the Dirac fermion to be composed
of two Weyl spinors from conjugate representations, i.e.\ we consider
vector-like fermions for which we can write an explicit
gauge invariant mass term.  The form of a generic Lagrangian, without
specifying field representations, can be written in terms of only
left-handed Weyl spinors as
\begin{align} \label{eqn:Min2FieldsLL}
  {\cal L}_{\text{LL}} &= \Big(\lambda_L l_{L2} \psi \phi - M_{\psi} \psi^c \psi + h.c.\Big) - M_{\phi}^2 |\phi|^2, \\
  \label{eqn:Min2FieldsRR}
	{\cal L}_{\text{RR}} &= \Big(\lambda_R  \mu_R^\dagger \phi \psi - M_{\psi} \psi^c \psi + h.c.\Big) - M_{\phi}^2 |\phi|^2,
\end{align}
where $l_{L2}$ is the SU$(2)_L$ doublet containing left-handed Weyl spinors for the muon and muon-neutrino, while $\mu_R^\dagger$ is a left-handed Weyl spinor that is the hermitian conjugate of the Weyl spinor for the right-handed muon. For the LL models SU$(2)_L$ indices for contractions between $l_{L2}$ and the new fermion $\psi$  or new scalar $\phi$ (depending on the particular representations of the model) are suppressed and similarly for the RR models any SU$(2)_L$ indices  for contractions between the new scalar and fermion are also suppressed.  The possible quantum numbers for one-loop contributions from fermion-scalar loops
(i.e.\ the FS type diagrams in Fig.\ \ref{fig:amu-one-loop}) where
both the fermion and scalar are BSM states are systematically listed in Table 1 of Ref.~\cite{Calibbi:2018rzv}. 
These models are also possible realisations\footnote{Alternatively, for example, Ref.\ \cite{Freitas:2014pua} included realisations of Models 24 to 36 without the $Z_2$ symmetry.} 
of Models 24 to 42 included in our Table \ref{tab:twofieldmodels}.

For simplicity, like Refs.\ \cite{Calibbi:2018rzv, Athron:2021iuf,Kowalska:2017iqv} we assume muon-philic couplings, which
avoids both severe constraints from direct detection of dark matter that would be induced by tree-level couplings to quarks as well as constraints from flavour violating processes. 
Note however, that the latter in general also play an important role in constraining models that can both
explain dark matter and provide large contributions to $\Damu$. For further details on the LFV constraints 
see Sec.~\ref{sec:LeptonDipole} as well as the model-specific discussions in Sec.~\ref{sec:Models}.

Now that we have specified the models we can go beyond the simple estimate of Eq.\ \eqref{nonenhancedoneloop}, and in particular obtain expressions showing the dependence on multiplicities from the dimension of the representations. This allows us to further discriminate between the different LL or RR models.  For LL models the dimension of the SU$(2)_L$ representations of the fermion $N_\psi$ and scalar $N_\phi$ must be related by $N_\phi= N_\psi \pm 1$, while for the RR models $N_\phi=N_\psi$.   Neglecting mass splittings amongst the components of the SU$(2)_L$ multiplets, the one-loop $\Damu$ results can be expressed as,  
\begin{align}
\Delta a_\mu^\text{LL} &=  -\frac{|\lambda_L|^2m_\mu^2}{16\pi^2 M_\phi^2}  \frac{2 N_\psi (N_\psi \pm 1)}{2 N_\psi - (1\mp 1) } \left[  f^S_\text{LL} + \left(  Y_S +\frac16(\pm N_\psi -4) \right)(f^S_\text{LL} + f^F_\text{LL}) \right], \label{Eq:amuLL} \\
\Delta a_\mu^\text{RR} &= - \frac{N_\psi|\lambda_R|^2m_\mu^2}{8 \pi^2 M_\phi^2} \left[ f^S_\text{LL} + Y_{F_L}(f^S_\text{LL} +   f^F_\text{LL}) \right] ,
\end{align}
where the sum over all $N_\psi$ multiplet components has been carried
out with the hypercharges $Y_{F_L,S}$ corresponding to the average
electric charges of the multiplets; after this summation the result is
conveniently expressed in terms of loop functions satisfying
$\mathcal{G}^\text{FS}(0,x;Q_S) = 2f_{LL}^F(x)
+2Q_S(f_\text{LL}^S(x)+f_\text{LL}^F(x))=2Q_S
f_\text{LL}^S(x)+2(1+Q_S)f_\text{LL}^F(x)$ in terms of the loop
function defined in
Sec.~\ref{sec:genericoneloop}.	\footnote{Generally, the
loop-functions used in Ref.~\cite{Calibbi:2018rzv}  are related to the
ones of Sec.~\ref{sec:genericoneloop} as:
		\begin{alignat*}{4}
				&f_{LL}^F(x) &&= +\frac{1}{2} \mathcal{G}^\text{FS}(0,x;0), \qquad &&f_{LR}^F(x) &&= +\frac{1}{2} \mathcal{F}^\text{FS}(0,x;0), \\
				&f_{LL}^S(x) &&= -\frac{1}{2} \mathcal{G}^\text{FS}(0,x;-1), \qquad &&f_{LR}^S(x) &&= -\frac{1}{2} \mathcal{F}^\text{FS}(0,x;-1). \\
			\end{alignat*}
		}

For models that couple only to left-handed muons, or only to
right-handed muons, the sign of the new physics contributions is often
fixed. 
More precisely for RR models Ref.\ \cite{Calibbi:2018rzv} show that the sign of $\Damu$ depends on the hypercharge of the new fermion $Y_F$, as follows,
\begin{align}
	\text{RR sign rules:} \quad \begin{cases}
		Y_F \geq -1/3  &\qquad\Rightarrow\qquad \text{sign}(\Damu) = -1 \\
		Y_F \leq - 2/3 &\qquad\Rightarrow\qquad \text{sign}(\Damu) = +1 \\
		Y_F = -\frac12 &\qquad\Rightarrow\qquad \text{sign}(\Damu) = \text{sign}(M_\phi -M_\psi),
	\end{cases}
\end{align} 
where $M_\phi$ is the mass of the scalar and $M_\psi$ is the mass of the fermion.  Note here that $Y_F$ is the hypercharge of the left-handed Weyl spinor in Eqs.\ \eqref{eqn:Min2FieldsRR} and matches the hypercharge assignments given in Ref.\ \cite{Calibbi:2018rzv}, which can differ from those given in Table \ref{tab:twofieldmodels}.

For LL models the rule they found is a little more complicated.  The
sign of $a_\mu$ depends on the dimension of the SU$(2)_L$ representation
of the fermion field, $N_F$, and the hypercharge of the new scalar,
$Y_S$.  Then for models where the the dimension of the SU$(2)_L$
representation of the scalar is $N_S = N_F \pm 1$ the sign of $\Damu$
is determined as follows, 
\begin{align}
	\text{LL sign rules:}\quad \begin{cases}
		Y_S \geq (\mp N_F + 2)/6 &\qquad\Rightarrow\qquad  \text{sign}(\Damu) = -1 \\
		Y_S \leq \mp N_F/6 		 &\qquad\Rightarrow\qquad  \text{sign}(\Damu) = +1 \\
		Y_S = (\mp N_f +1) / 6   &\qquad\Rightarrow\qquad  \text{sign}(\Damu) = \text{sign}(M_\phi -M_\psi).
	\end{cases}
\end{align}
Note that similar to above, here the $Y_S$ matches the hypercharge
assignments given in Ref.\ \cite{Calibbi:2018rzv}, which can differ
from those given in Table \ref{tab:twofieldmodels} due to differing
conventions. Nonetheless these rules are reflected in signs shown in
Table \ref{tab:twofieldmodels}, which we have explicitly checked by
translating between conventions.

\begin{figure}[t]
	\centering
	\begin{subfigure}{.45\textwidth}
		\centering
		\includegraphics[width=\textwidth]{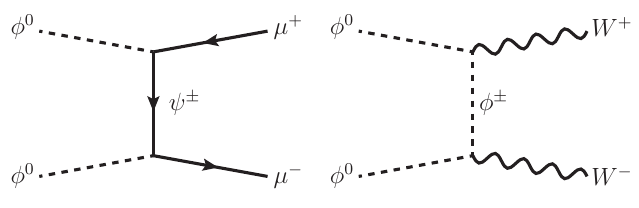}
		\caption{Scalar DM}
	\end{subfigure}\hfill
	\begin{subfigure}{.45\textwidth}
		\centering
		\includegraphics[width=\textwidth]{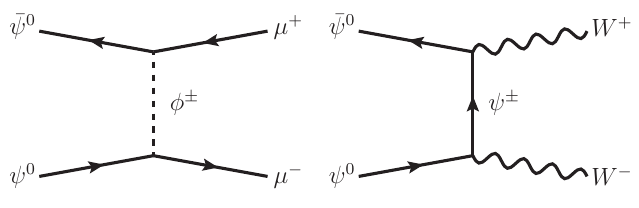}
		\caption{Fermionic DM}
	\end{subfigure}
	\caption{\label{fig:DM-ann_2fields} The dominant annihilation channels for scalar \textbf{(a)} and fermionic \textbf{(b)} dark matter in two-field extensions of the SM discussed in the text.}
\end{figure}

These RR and LL models may contain fermionic or scalar dark matter with a
relic density generated by the freeze-out mechanism discussed earlier.  For both fermionic and scalar dark
matter, the dark matter predominantly annihilates into muons or
$W^\pm$ gauge bosons before the relic density freezes out, as indicated in
Fig.\ \ref{fig:DM-ann_2fields} where the dominant
annihilation diagrams are shown.
Note that the annihilation into gauge bosons involves only the
states from a single SU$(2)_L$ multiplet, and depends only on the hypercharge of that multiplet as well as the SM gauge couplings $g_{1,2}$. 
The thermally averaged annihilation cross section for this channel is given by results obtained for minimal dark matter \cite{Cirelli:2005uq},
\begin{align}
	\langle\sigma v\rangle^{\phi\phi\rightarrow WW} &\approx \frac{1}{128\pi M_\phi^2 N_\phi} 
	\left[(N_\phi^4 - 4 N_\phi^2 + 3) + 16 Y_\phi^2 g_1^2 + 8 g_2^2 g_1^2 Y_\phi^2 (N_\phi^2 -1)\right], \label{Eq:sigmaannSSVV}\\
	\langle\sigma v\rangle^{\psi\psi\rightarrow WW} &\approx \frac{1}{512 \pi M_\psi^2 N_\psi} 
	\left[(2N_\psi^4 +17N_\psi^2 -19) + 4 Y_\psi^2 g_1^2(41+8Y_\psi^2) + 16 g_2^2 g_1^2(N_\psi^2 -1)\right],   \label{Eq:sigmaannFFVV}
\end{align}  
where we assume here a complex scalar $\phi$ and a Dirac fermion $\psi$, but the result for a real scalar or Majorana fermion is trivially obtained by multiplying with a factor 2. 
This annihilation channel is active even if the $a_\mu$ contribution
is zero.  The annihilation channels into muons, on the other hand,
depend on the couplings $\lambda_L$ or $\lambda_R$. Neglecting
velocity suppressed terms and assuming real couplings for simplicity,
these channels are given by \cite{Calibbi:2018rzv},
\begin{align}
	\langle\sigma v\rangle^{\phi\phi\rightarrow \mu^+\mu^-} &\approx  \frac{\lambda_{L,R}^4}{16\pi}\frac{m_\mu^2}{(M_\psi^2 + M_\phi^2)^2}, \label{Eq:AnhCS-Cscalar} \\
	\langle\sigma v\rangle^{\psi\psi\rightarrow \mu^+\mu^-} &\approx \frac{\lambda_{L,R}^4}{32\pi} \frac{M_\psi^2}{(M_\phi^2 + M_\psi^2)^2},\label{Eq:AnhCS-Dfermion} 
\end{align}
where we again have given the result for a complex scalar and Dirac fermion. In this case 
the result for a real scalar is obtained by multiplying with a factor of 4, while for the cross-section for a Majorana fermion additional factor of $m_\mu^2 / M_\psi^2$ must be included\footnote{If one includes velocity suppressed terms there are additional differences in both cases, as can be seen from the expressions in Ref.\ \cite{Calibbi:2018rzv}.}. 
Although not considered in Ref.\ \cite{Calibbi:2018rzv} for scalar dark matter, depending on the quantum numbers of the particular states, couplings from the scalar potential may also play a role. 
For example the Higgs portal (i.e.\ mixed quartic) coupling plays an important role in the scalar singlet dark matter, see e.g. Ref.\ \cite{Kowalska:2017iqv}.

To maximise the $\Delta a_\mu$ contribution one can set
$\lambda_{L,R}$ to the perturbative limit and this also means that
both channels for DM annihilations will be active.  However note that the annihilation cross-section in 
Eqs.\ \eqref{Eq:sigmaannSSVV} and \eqref{Eq:sigmaannFFVV} increases
with the dimension of representation;  schematically for
$N=\{N_\psi,N_\phi\}$ and $M=M_{\psi,\phi}$ if the annihilation into
$W^+W^-$ dominates, the scaling is
\begin{align}
  \langle\sigma v\rangle \sim \frac{N^3}{M^2}.
\end{align}
Thus for higher dimensional representations the relic density will be
lower for the same masses and couplings,
\begin{align}
 \Omega \sim \frac{M^2}{N^3}.
\end{align}
To adjust for this the mass can be increased, such that the observed
relic density will be explained fully in higher dimensional models at
higher masses (assuming fixed $\lambda_{L,R}$), with a scaling
\begin{align}
  M^2 (\Omega_\text{Planck}) \sim N^3,
\end{align}
where $\Omega_\text{Planck}$ refers to the measured value given in
Eq.~\eqref{Eq:PlanclRelicDensityDM}.  However increasing the mass will reduce $\Delta a_\mu$.  Although
$\Delta a_\mu$ also increases with the dimension of the
representation, the reduction dominates and  we have         
\begin{align}
  \Delta a_\mu \sim \frac{N}{M^2} \sim N^{-2},
\end{align}
Thus as $N$ is increased the
maximum $\Delta a_\mu$ that is consistent with the model explaining
all the dark matter reduces.  

The above argument was originally
presented in Ref.\ \cite{Calibbi:2018rzv}.  There they were interested
in explaining the large anomaly from the BNL measurement and data
driven estimates of the SM prediction and therefore they subsequently
applied a more careful analysis to derive an upper limit on the relic
density that can be obtained when explaining this large deviation
within $2\sigma$.  This ruled out most models of this type from
providing dual solutions, i.e.\ simultaneously explaining dark matter
and giving such a large contribution to $a_\mu$, with Models 24, 27
and 28 being the only three models that were capable of the dual
explanation.  These three models have the lowest dimensional
representations of SU$(2)_L$ that could give a positive contribution to
$a_\mu$.

In the present context based on $\DamuFinal$ in
Eq.~(\ref{eq:DamuFinal}) where the SM theory
prediction and experimental result are consistent within errors,
restricting us to only small deviations, it is these lowest
dimensional models where constraints from $a_\mu$ will be strongest
and have the most impact.  For higher dimensional representations muon
$g-2$ predictions are suppressed further, due to the heavier masses
needed to fulfil the total measured relic density of dark matter.
Thus for higher dimensional representations that explain dark matter,
the maximum $\Delta a_\mu$ contributions they can predict are even
smaller than the simple estimate using Eq.~\eqref{nonenhancedoneloop}.

  In summary, two-field models with a dark matter candidate(s) cannot
  have a chirality flipping enhancement; hence the contributions to
  $\amu$ can be sizeable at most in the region of large couplings and
  small masses. Dark matter constraints however require sufficiently
  high masses, and large multiplicities further strengthen this
  requirement. Hence for most of these DM explanations the current
  $\amu$ bounds do not provide strong complementary parameter
  constraints. The few cases where small masses and sizeable $\Damu$
  are allowed by DM constraints are constrained by collider data. This
  will be discussed in Sec.\ \ref{sec:Collider}.  However before
  moving onto that section we will first discuss another class of
  models where explaining dark matter does not place strong
  restrictions of the size of $a_\mu$ contributions. In these models,
  constraints from $\DamuFinal$ or signals from a large deviation like
  $\DamuOld$, may go beyond the reach of the LHC and will be a very
  important test of models explaining DM.

\subsubsection{Three-field extensions}

Solutions to DM with much larger new physics contributions to $\amu$
are possible when there is both a chiral enhancement and a stable dark
matter candidate.  The minimum number of new fields where this is
possible is three.\footnote{ Interestingly a similar result including
  also flavour physics was derived in Ref.~\cite{Arcadi:2021cwg},
  where the requirements of sizeable $\Damu$ and explanation of dark
  matter were combined with accommodating $B$-physics anomalies
  present at the time, and at least four BSM fields were found to be
  necessary.}  For models with two fermions and one scalar or two
scalars and one fermion the relevant Lagrangian interactions allowing
the chirality flip are given in Sec.\ \ref{sec:genericthreefield},
Eqs.\ (\ref{eq:LaClassI},\ref{eq:LaClassII},\ref{eq:LaClassIII}). As
described there, these models introduce two particles of the same spin
that can mix, allowing a chirality flip in the internal loop diagram.
At the same time they also contain a particle of different spin to
enter the loop diagram and as a result the new physics states may have
couplings that only involve the muon and another BSM state.  Thus
there is no longer any requirement that new physics states must decay
into SM states and if the lightest of the new states is neutral it can
be a dark matter candidate as long as no other couplings through which
it decays are present in the model (this could be achieved by imposing
a $\mathbb{Z}_2$ symmetry for example).

Models of this nature have been studied in
Refs.\ \cite{Calibbi:2018rzv, Athron:2021iuf,Kowalska:2017iqv}.   All
of these papers (like  Sec.\ \ref{sec:genericthreefield}) consider
classes of models involving  fermions and scalars.   The models in
Ref.\ \cite{Calibbi:2018rzv, Athron:2021iuf} correspond to the
three-field models of Class I and Class III given in
Sec.\ \ref{sec:genericthreefield}, while Ref.\ \cite{Kowalska:2017iqv}
also considers two models that correspond to a combination of Class I and Class II.

 Ref.\ \cite{Calibbi:2018rzv} systematically  constructed all models of Class I and Class III types in dimensions of the SU$(2)_L$ representations, up to triplets.  We reproduce this classification of the models in Tables
\ref{tab:FLRmodels} and \ref{tab:SLRmodels}. We again follow
Refs.\ \cite{Calibbi:2018rzv,Athron:2021iuf,Kowalska:2017iqv} and show the Lagrangian here in terms of left-handed Weyl spinors, and we label the BSM left-handed Weyl spinors $\psi$ and $\chi$. 
	\begin{align} 
		{\cal L}_{\text{2F1S}} &= \bigg(\lambda_\Phi \Phi \chi \psi + \lambda_L l_{L2} \phi \psi + \lambda_R \chi \mu_R^\dagger \phi^\dagger 
		- M_{\psi} \psi^c \psi - M_{\chi} \chi^c \chi + h.c.\bigg) - M_{\phi}^2 |\phi|^2, \label{eqn:Min3FieldsFLRS} \\
       {\cal L}_{\text{2S1F}} &= \begin{pmatrix}a_\Phi \Phi\cdot\phi_a \phi_b + \lambda_L \phi_a l_{L2} \psi + \lambda_R \phi_b \mu \psi^c - M_{\psi} \psi^c \psi_s + h.c.\end{pmatrix} - M_{\phi_a}^2 |\phi_a|^2 - M_{\phi_b}^2 |\phi_b|^2.   \label{eqn:Min3FieldsXLRF}        
	\end{align}
Note that although Ref.\ \cite{Calibbi:2018rzv} writes the generic Lagrangians in this form, additional couplings allowed by gauge invariance may be included for specific models where they perform detailed numerical investigations.

\begin{table}
\centering
\begin{tabular}{|c|c|c|c|c c c|} \hline
 \textbf{Model}  & $\psi$     & $\chi$     &  $\phi$    & \multicolumn{3}{c|}{\textbf{DM candidates (DD channel)}}  \\ \hline\hline
 FLR1   & \FieldN    & \FieldD    & \FieldDp   & $\psi$ ($H$-ex), & $\chi^0$ ($H$-ex), & \cellcolor{ruledout} $\phi^0$ ($Z$-ex) \\
 FLR2   & \FieldP    & \FieldDalt & \FieldD    & -- & -- & \cellcolor{ruledout}{$\phi^0$ ($Z$-ex)} \\
 FLR3   & \FieldD    & \FieldN    & \FieldP    & $\psi^0$ ($H$-ex), & $\chi$ ($H$-ex), & -- \\
 FLR4   & \FieldD    & \FieldA    & \FieldTp   & $\psi^0$ ($H$-ex), & $\chi^0$ ($H$-ex), & \cellcolor{ruledout} $\phi^0$ ($Z$-ex) \\
 FLR5   & \FieldDp   & \FieldC    & \FieldN    & \cellcolor{ruledout} $\psi$ ($Z$-ex), & -- & $\phi$ ($H$-ex) \\
 FLR6   & \FieldDp   & \FieldT    & \FieldA    & \cellcolor{ruledout} $\psi^0$ ($Z$-ex), & \cellcolor{ruledout}$\chi^0$ ($Z$-ex), & $\phi^0$ ($H$-ex) \\
 FLR7   & \FieldDpp  & \FieldTpp  & \FieldT    & -- & -- & \cellcolor{ruledout} {$\phi^0$ ($Z$-ex)} \\
 FLR8   & \FieldT    & \FieldDp   & \FieldDpp  & \cellcolor{ruledout} {$\psi^0$ ($Z$-ex),} & \cellcolor{ruledout} {$\chi^0$ ($Z$-ex),} & -- \\
 FLR9   & \FieldA    & \FieldD    & \FieldDp   & $\psi^0$ ($H$-ex), & $\chi^0$ ($H$-ex), & \cellcolor{ruledout} {$\phi^0$ ($Z$-ex)} \\
 FLR10  & \FieldTp   & \FieldDalt & \FieldD    & \cellcolor{ruledout}{$\psi^0$ ($Z$-ex),} & -- & \cellcolor{ruledout}{$\phi^0$ ($Z$-ex)} \\
 \hline
\end{tabular}
\caption{Models with dark matter candidates that have a chiral
  enhancement with precisely two new Dirac fermions and one new scalar
  field. For each field from the Lagrangian (\ref{eqn:Min3FieldsFLRS}) the representations under $\GSM$ are shown in columns 2,3 and 4. The last
  column shows the possible dark matter candidates together with the direct detection channel expected to be most constraining for each of these DM candidates. 
  Here, $H-ex$ stands for tree-level Higgs exchange and $Z-ex$ stands for tree-level $Z$-exchange.  
  If all DM candidates are constrained by tree-level $Z$-exchange the entry is coloured red reflecting the fact that constraint is strong enough to exclude the model, 
  as discussed in the text.  This table is based on similar tables presented in Ref. \cite{Calibbi:2018rzv}.   \label{tab:FLRmodels}}
\end{table}

\begin{table}
\begin{center}
\begin{tabular}{|c|c|c|c|c c c|} \hline
 \textbf{Model}  & $\psi$     & $\phi_a$   & $\phi_b$    & \multicolumn{3}{c|}{\textbf{DM candidates (DD channel)}}  \\ \hline\hline
 SLR1   & \FieldP    & \FieldD    & \FieldN    & -- & $\phi_a^0$ ($H$-ex), & $\phi_b$ ($H$-ex) \\
 SLR2   & \FieldN    & \FieldDp   & \FieldC    & \cellcolor{ruledout} $\psi$ ($Z$-ex), & $\phi_a^0$ ($H$-ex), & --   \\
 SLR3   & \FieldDp   & \FieldN    & \FieldD    & \cellcolor{ruledout} $\psi^0$ ($Z$-ex), & $\phi_a$ ($H$-ex), & $\phi_b^0$ ($H$-ex)  \\
 SLR4   & \FieldDp   & \FieldA    & \FieldD    & \cellcolor{ruledout} $\psi^0$ ($Z$-ex), & $\phi_a^0$ ($H$-ex), & $\phi_b^0$ ($H$-ex)  \\
 SLR5   & \FieldDpp  & \FieldT    & \FieldDp   & -- & \cellcolor{ruledout} $\phi_a^0$ ($Z$-ex), & \cellcolor{ruledout} $\phi_b^0$ ($Z$-ex) \\
 SLR6   & \FieldD    & \FieldP    & \FieldDalt & \cellcolor{ruledout} $\psi^0$ ($Z$-ex), & -- & --  \\
 SLR7   & \FieldD    & \FieldTp   & \FieldDalt & \cellcolor{ruledout} $\psi^0$ ($Z$-ex), & \cellcolor{ruledout} $\phi_a^0$ ($Z$-ex), & --   \\
 SLR8   & \FieldTp   & \FieldD    & \FieldA    & \cellcolor{ruledout} $\psi^0$ ($Z$-ex), & $\phi_a^0$ ($H$-ex), & $\phi_b^0$ ($H$-ex)  \\
 SLR9   & \FieldA    & \FieldDp   & \FieldT    & $\psi^0$ ($H$-ex), & \cellcolor{ruledout} $\phi_a^0$ ($Z$-ex), & \cellcolor{ruledout} $\phi_b^0$ ($Z$-ex)  \\
 SLR10  & \FieldT   & \FieldDpp & \FieldTpp    & \cellcolor{ruledout} $\psi^0$ ($Z$-ex), & -- & --  \\
 \hline
\end{tabular}
\caption{Models with dark matter candidates that have a chiral enhancement with precisely one new Dirac fermions and two new scalar fields.  
	For each field from the Lagrangian (\ref{eqn:Min3FieldsFLRS}) the representations under $\GSM$ are shown in columns 2,3 and 4. 
	The last column shows the DM candidates and direct detection channel as described in the caption for Table \ref{tab:FLRmodels}.  \label{tab:SLRmodels}}
\end{center} 
\end{table}

The dominant channels for dark  matter annihilations that deplete the
relic density are typically the same as described in the previous
Sec.\ \ref{sec:DMtwofield}, though the coupling to the Higgs means
additional co-annihilation processes are possible and at low masses
near resonant annihilation through the Higgs can also be important.
The cross-sections for annihilations into $W$ Bosons are still given
by Eqs.\ \eqref{Eq:sigmaannSSVV} and \eqref{Eq:sigmaannFFVV}, while
the annihilations into muons are significantly modified with a new
contribution that requires both $\lambda_L$ and $\lambda_R$ to be non
zero,
\begin{align}
	\langle\sigma v\rangle^{\phi\phi\rightarrow \mu^+\mu^-} &\approx  \frac{1}{16\pi}\left( (\lambda_L^4 + \lambda_R^4)\frac{m_\mu^2}{M_\psi^2} + 4 \lambda_L^2\lambda_R^2 \right)\frac{M_\psi^2}{(M_\psi^2 + M_\phi^2)^2}, \label{Eq:AnhCS-Cscalar3F} \\
	\langle\sigma v\rangle^{\psi\psi\rightarrow \mu^+\mu^-} &\approx \frac{(\lambda_L^2 + \lambda_R^2)^2 }{32\pi} \frac{M_\psi^2}{(M_\phi^2 + M_\psi^2)^2}.\label{Eq:AnhCS-Dfermion3F} 
\end{align}
As before, some modifications for the case of a real scalar and for a Majorana fermion are required, with the full expressions given in Ref.\ \cite{Calibbi:2018rzv}.  As discussed in Ref.\ \cite{Athron:2021iuf} in certain scenarios where this latter channel is dominant, the $m_\mu / M_F$ suppression of the first term in Eq.~\eqref{Eq:AnhCS-Cscalar3F} can lead to the relic density following a similar parabolic curve in the $\lambda_L$--$\lambda_R$ plane to contours of fixed $\Damu$.  

With direct couplings to the quarks avoided by the muon-philic choice of couplings, the most important remaining direct detection channels for these models are tree-level $Z$ exchange, tree-level Higgs exchange induced by the new coupling to the Higgs and loop-level contributions that play an important role in constraining lepto-philic dark matter \cite{Bell:2014tta}.  The tree-level scattering via $Z$ exchange applies to models where the dark matter is in a multiplet where the hyper-charge is non-zero ($Y\neq 0$)  and if the scattering is elastic this provides very strong exclusions essentially ruling out them out as WIMP dark matter, with cross-sections far above the limits set by direct detection experiments across their full mass range.  This limit cannot be avoided by adjusting BSM couplings, since it proceeds through gauge interactions. However if there is a small mixing this suppresses or forbids the elastic scattering and therefore can be evaded, see e.g. \cite{Cirelli:2005uq,Bottaro:2022one}.    When this is not active, Higgs exchange will typically provide the dominant direct detection  limits where the impact depends on the Higgs mixing parameter ($a_H$ or $\lambda_H$) and the masses, but this can also potentially exclude states up to multi TeV masses. The dominant direct detection channel is indicated in Tables \ref{tab:FLRmodels}, \ref{tab:SLRmodels}.

If the interactions with the Higgs and either $\lambda_L$ or
$\lambda_R$ are set to zero we return to the situation discussed in
the previous subsection for two-field
  models. \footnote{Introducing non-zero couplings to the Higgs boson introduces tree-level Higgs exchange as an important channel for direct detection of dark matter via nuclei recoil,  as well as allowing the new (co-)annihilation channels mentioned above for the relic density.  Thus the effect would be to just further constrain scenarios in comparison to the situation in Sec.\ \ref{sec:DMtwofield}, and in parameter regions where the new (co-)annihilation channels are active, further increasing the mass at which the dark matter can be fully explained and thus suppressing $\Damu$ further. If the mixing with the Higgs is zero but both $\lambda_L$ and $\lambda_R$ are non-zero then one effectively gets an extra contribution to $\Damu$, and there is some modification to the DM annihilation cross-sections, but the situation is qualitatively unchanged from that described in Sec.\ \ref{sec:DMtwofield}.}
Therefore here we take the natural assumption that all the new couplings are ${\cal O}(0.1$--$1)$, implying a large chiral enhancement to $\Damu$ which can generate much larger contributions to the anomalous magnetic moment of the muon.   Even with all of these couplings non-zero though, in principle a similar argument to that presented above in Sec.\ \ref{sec:DMtwofield} regarding the dimensionality of these models applies here too.  That is, as the dimensionality of the representation increases, the maximum size of the muon $g-2$ contributions that are consistent with explaining the observed relic density of dark matter decreases.  However due to the chirality enhancement this maximum value is much larger for a given representation in this class of models than in the models discussed in Sec.\ \ref{sec:DMtwofield}. As a result, as long as one does not consider huge representations, far above the triplet representations categorised here\footnote{Ref.\ \cite{Calibbi:2018rzv} found that large contributions consistent with $\DamuOld$ were possible for representations smaller than ${\cal O}(20)$.  }, large $a_\mu$ corrections well in excess of the $2\sigma$ bound on $\Damu$ are possible for scenarios that fully explain the observation of DM by fitting the measured relic density.   

Dark matter models must also respect complementary constraints
e.g.~from dark matter direct detection (DMDD) experiments shown in
Fig.~\ref{LZ-DDlimits}. In fact, in such three-field DM scenarios with
large chiral enhancements there is a complex interplay between muon
$g-2$, the relic density and direct detection constraints that must be
studied carefully to understand whether or not it is possible to
explain dark matter in these models.  For example a model with a
Lagrangian of the form of Eq.~\eqref{eqn:Min3FieldsXLRF} is studied in
detail in Ref.\ \cite{Athron:2021iuf}.  There it is shown that
parameter regions exist where the DMDD constraints are fulfilled while
$\Damu$ is as large as the old deviation $\DamuOld$. But there are
also parameter regions where $\Damu$ is small while the DMDD
constraints are violated. Presently, with the small value $\DamuFinal$
both such parameter regions are excluded, demonstrating the
complementarity of $\amu$ and DMDD constraints in probing DM models.
The reason for this complementarity is that the two couplings
$\lambda_{L,R}$ and the two masses $M_{\phi,\psi}$ enter the relic
density, $\Damu$ and the DMDD cross section in different ways.  Later
in Sec.~\ref{sec:SUSY} we will encounter an analogous situation in the
context of supersymmetric models.

Away from these equal mass scenarios other possibilities include scenarios where the relic density and $\Damu$ contours are orthogonal, such that fitting the combination can dramatically shrink the parameter space or where they are roughly parallel contours allowing muon g-2 constraints to exclude fitting the relic abundance entirely in these planes.  In general the direct detection limits typically depend on $a_H$,  a small mass splitting between the two states that are mixed via the Higgs and the general mass scale, whereas constraints from muon g-2 strongly depend also on $\lambda_L$ and $\lambda_R$, and the relic density is typically more sensitive to $\lambda_L$ and $\lambda_R$ away from special regions where the co-annihilations between the scalar and fermion are important (i.e.\ when the fermion and scalar mass are very close) or for very light masses tuned such that resonant annihilation via the Higgs boson is an active channel.  As a result it is clear that to truly understand the allowed parameters space, one needs to perform global fits including the relic density, direct detection and muon g-2.

The situation is similar for other models of this type given in Table \ref{tab:SLRmodels} and for models with a Lagrangian of the form of Eq.~\eqref{eqn:Min3FieldsFLRS} given in Table \ref{tab:FLRmodels}, as has been illustrated by detailed studies of specific cases in Refs.\ \cite{Calibbi:2018rzv,Athron:2021iuf,Kowalska:2017iqv}.  Thus $a_\mu$ and direct detection both play important roles in constraining such explanations of dark matter, and may both play a role in further testing or even discovering such scenarios in the future and the inclusion of muon g-2 in global fits of models of these types (or similar) should be mandatory.

\subsection{Collider observables}\label{sec:Collider}
Searches for new particles produced in collider experiments provide the most direct probes of new physics and are applicable to a very wide variety of new physics scenarios.  This both competes with and complements precision measurements on new physics.  In this section we discuss this complementarity and competition between the muon $g-2$ measurement and collider searches and discuss the impact collider limits have on possible NP signals
in $\Damu$.  

Determining the limits from the Large Hadron Collider (LHC) can be
highly model dependent and should be done by identifying possible
collider signatures of the specific model, finding the collider
searches that can be relevant for this and reinterpreting their
results. This is most easily done through publicly available
reinterpretation software,
e.g.\ \cite{GAMBIT:2017qxg,Kraml:2013mwa,Tattersall:2016nnx}.
Nonetheless, as a generality, the LHC typically provides much stronger
constraints on new physics with strong interactions, since the LHC collides
protons and can these particles through strong interactions between
the constituent quarks and gluons.  Constraints on
strongly interacting fundamental new physics tend to reach into the
multi-TeV range, with examples being simplified model limits on
supersymmetric states like the gluinos and first and second generation squarks reaching up to around 2.4 TeV \cite{ATLAS:2022ckd,ATLAS:2022ihe,CMS:2021beq,CMS:2019ybf} and 1.8 TeV \cite{ATLAS:2018nud,CMS:2019zmd} respectively. Similarly for scalar leptoquarks the mass reaches from 2 to 5 TeV are possible depending for some specific states and couplings \cite{ATLAS:2022wcu,CMS:2024bnj,CMS:2023bdh,CMS:2025iix}.  Note however these limits can be weaker depending on the specific model.  For example the limits on stops are weaker than on the light squarks, with current limits reaching around 1.3 TeV at most \cite{ATLAS:2020dsf,CMS:2021eha} but also varying significantly depending on the specifics of the simplified model.

In contrast to this, typical limits on weakly interacting states from
the LHC do not reach so high and tend to be lower than around 1
TeV.\footnote{An important exception to this are $Z'$ bosons, that tend
  to have multi-TeV limits \cite{CMS:2024nmz,CMS:2022uul,CMS:2011ckb} because through its couplings to quarks it can be
  produced resonantly and its decay into two leptons is a very clean
  signal with low backgrounds. }  For example the LHC limits on
sleptons reach at most 700 GeV \cite{ATLAS:2019lff}, while for
charginos they can extend at most to around 1.1 TeV
\cite{ATLAS:2018ojr}.  As stated above these limits are very model
dependent, for example it has been previously shown that while in
certain simplified models of neutralinos and charginos the LHC was
placing quite tough limits on these states, when a careful
reinterpretation for the MSSM was applied there was no general
exclusions in the mass plane of the lightest neutralino and the
lightest chargino \cite{GAMBIT:2018gjo}. When the all relevant degrees
of freedom for realistic MSSM scenarios are tested, for every point in
this plane there are at least some realistic scenarios with those
masses that escape the LHC limits. Since then ATLAS and CMS have
produced studies that can close various holes in the exclusions, but
this result still emphasises how the collider limits depend on models
and specific parameter values that influence decay
kinematics.

\subsubsection{Mass limits from Colliders and $\Damu$}

As indicated in Tables \ref{tab:onefieldmodels} and \ref{tab:twofieldmodels} (as well as Tables \ref{tab:FLRmodels} and \ref{tab:SLRmodels} in Sec.\ \ref{sec:DarkMatter}), typically states that contribute to muon $g-2$ at one-loop have $\SUL$ and/or electromagnetic interactions, but only a few examples have strong interactions.  Thus for the models of interest here we are typically dealing with an LHC reach that does not extend much beyond 1 TeV and can in many cases be much weaker than this depending on model-specific details.   However if the model contains charged states then there should at least be very robust limits from LEP of around 100 GeV of the mass of the charged state (e.g.\ Refs.\ \cite{L3:2001xsz, ALEPH:2001oot}).  Therefore we are typically comparing limits from the precision measurement of $a_\mu$ to collider limits that vary between ${\cal O} (100$--$1000)$ GeV.        
The limits from muon $g-2$ can be broadly understood by looking at the
schematic form of the one-loop new physics contributions
discussed in Sec.\ \ref{sec:ChiralityFlips}, where it is pointed out
that typically the one-loop new physics contributions can be brought into the form
 \begin{equation}
  \Damu = R_\chi \times\frac{c_L c_R}{16\pi^2}\frac{m_\mu^2}{M^2}.
 \label{Eq:genericBSMDamReproduuce}
\end{equation}
 Here $R_\chi$ is a chiral enhancement factor, $c_{L,R}$ are the
 relevant couplings and $M$ is the  mass scale of the new physics, and
 there can be additional factors from particle multiplicities and an
 order one loop-function.  Since the production of new particles needs
 sufficiently large couplings and sufficiently low masses,
 non-observation of new physics at colliders typically places limits
 on large couplings and small masses.  If the couplings are fixed or
 one requires that the couplings at least stay perturbative, this
 generally implies a lower limit on the masses.  Using Eq.\ (\ref{Eq:genericBSMDamReproduuce}) the 2$\sigma$ limits
 on $\Damu$ from Eq.\ (\ref{DamutwoSigma}) translate into 

\begin{align}\label{Eq:MLim_amu}
	M \gtrsim \big|R_\chi c_L c_R\big|^{\frac{1}{2}} \begin{cases}
		206~\text{GeV} & \quad \text{for} \quad \Damu > 0 \\
		282~\text{GeV} & \quad \text{for} \quad \Damu < 0
	\end{cases}.
\end{align}
  This can be used to illustrate the generic behaviour of mass (and coupling) limits on new physics scenarios that are set from chirally enhanced muon $g-2$ contributions in the new physics model.  
  Note that without chiral enhancement ($R_\chi=1$) and order one couplings $c_{L,R}\sim\O(1)$ this limit excludes masses below $\lesssim 200$ GeV, which is already stronger than the
  generic LEP bound of 100 GeV for heavy charged and neutral particles \cite{L3:2001xsz}. On the other hand, this limit lies well below the possible reach of $\sim \O(1~\text{TeV})$ for 
  production of weakly interacting states at the LHC. 
  Thus, in such cases a careful (re)interpretation of the LHC data is typically necessary and often leads to competing or stronger exclusion limits.
 
\begin{figure}[tb]
	\centering
	\includegraphics[width=0.95\textwidth]{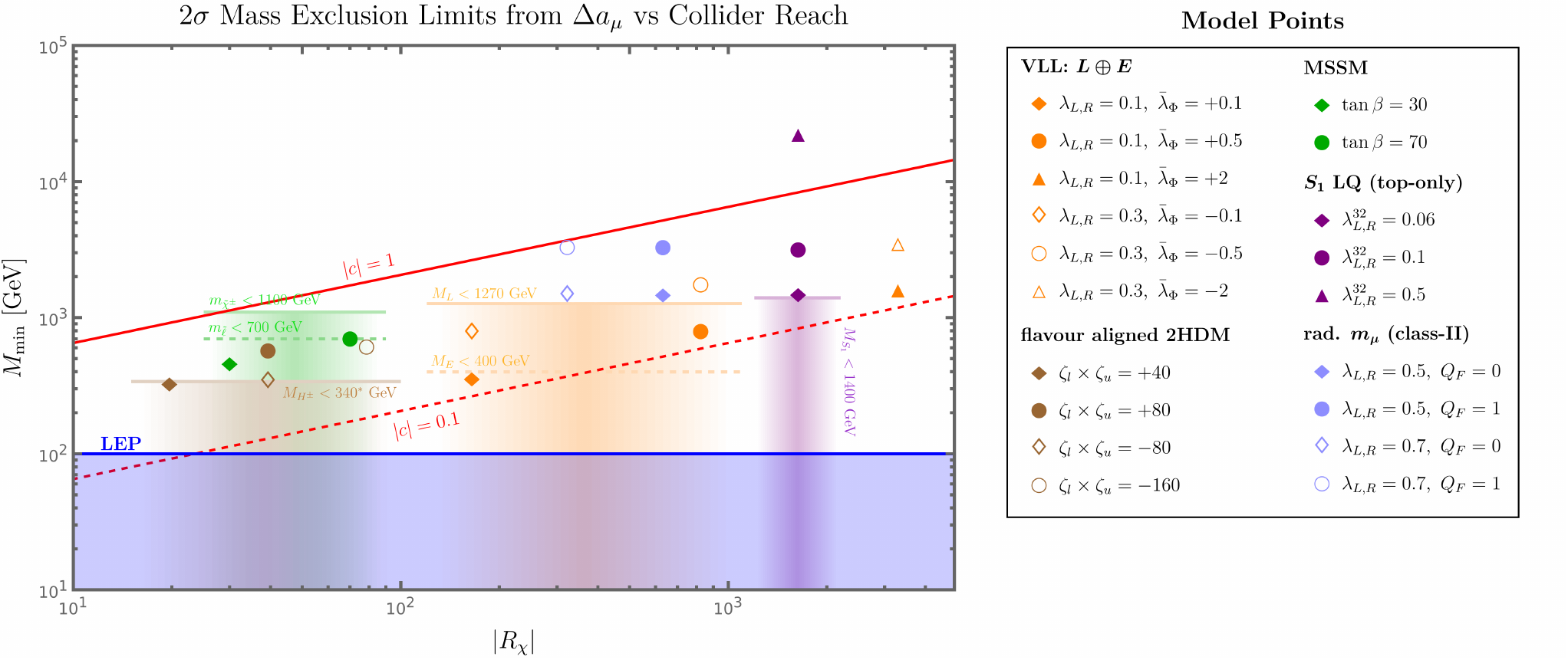}
	\caption{Comparison of mass limits on new physics coming from
          the measurement of the anomalous magnetic moment of the
          muon, and the mass reach and limits of the LHC and LEP. The
          LEP limit is indicated by the solid blue line and blue
          shading. The mass reach from the LHC is shown for charginos (solid green) and   sleptons
          (dashed green) from simplified
          models of supersymmetry, vector-like leptons $L$ (solid
          orange) and $E$ (dashed orange), $S_1$ scalar leptoquarks
          coupling to the top and the muon (purple) and finally the
          flavour aligned 2HDM (brown).  The limits on sleptons and
          charginos should be interpreted as the maximum masses that
          could be constrained by current LHC data, but are not robust
          limits that exclude all lighter possibilities. Similarly the
          relevant VLLs and leptoquarks should be safe when heavier
          than the LHC reaches indicated, while it may be possible for
          lighter states to avoid the bound if the branching ratio is
          suppressed. On the other hand the solid brown line is not a
          maximum mass reach, but rather a robust limit on the charged
          Higgs mass in a specific parameter region of the 2HDM discussed in Sec.~\ref{sec:FA2HDMpheno} and the LHC does also restrict masses heavier than this.   These collider constraints can be compared to both the red lines and the points.  The solid and dashed red lines show the mass limits from the generic one-loop contributions given in Eq.\ \eqref{Eq:genericBSMDamReproduuce} ( with couplings $c := \sqrt{c_Lc_R} =1$ and $c=0.1$ respectively), which also appears in Sec.\ \ref{sec:ChiralityFlips} and Table \ref{tab:estimates}.  
		These are just rough generic expectations and the actual limits may deviate substantially from these as demonstrated by diamonds, circles and triangles, where the colour coding indicating the model matches that of the collider limits.  The exact choice of couplings are given in the key beside the plot and all $\Damu$ used to obtain these results are computed from the fit formulae in Table \ref{tab:estimates}.    
		\label{Fig:mass_reach_gm2VScoll}
	}
\end{figure}

However, if the NP contributions yield a large chiral enhancement ($R_\chi\gg1$), the mass exclusion range from $\Damu$ is substantially increased
and can easily extend beyond the typical reach of the LHC.
This is illustrated in Fig.\ \ref{Fig:mass_reach_gm2VScoll} where we plot the mass exclusion contour from Eq.~\eqref{Eq:MLim_amu} (assuming $\Damu>0$)
as a function of $R_\chi$ for fixed $c_{L,R} \equiv c = 1$ and $0.1$.
This generic limit is compared to various collider mass bounds on specific new particles, 
shown as solid or dashed lines with shading indicating the corresponding models.  
Besides the 100 GeV LEP limit shown in blue, we have included the reach for sleptons $m_{\tilde\ell}>700$ GeV and charginos $m_{\chi^\pm} > 1.1$ TeV in the MSSM (in green),
the bounds on VLL singlets $M_E>400$ GeV and doublets $M_L>1270$ GeV (in orange), the $S_1$ leptoquark bound $M_{S_1}>1400$ GeV for $\mu t$ final states (in purple)
as well as a generic bound applicable to the charged scalar in the 2HDM-X $M_{H^\pm}>340$ GeV (in brown).

As will be discussed in Sec.\ \ref{sec:SUSY}, the slepton and chargino bounds do not represent strict exclusion limits, but rather indicate the upper reaches 
beyond which MSSM scenarios are generally safe from current LHC searches. Nevertheless, significant room for scenarios with masses below these reaches
remains viable.

In contrast, the limits on VLL from searches at ATLAS \cite{ATLAS:2015qoy,ATLAS:2023sbu,ATLAS:2024mrr} and CMS \cite{CMS:2019hsm,CMS:2023dhv,CMS:2024bni}
are much more robust and essentially completely exclude scenarios with lighter masses. The main reason for the stronger limit in case of the VLL doublet
is the larger production cross section resulting from the additional $\SUL$ interactions. Consequently, limits for similar states like e.g.\ VLL triplets 
are expected to be comparable or even stronger. For more details see Sec.~\ref{sec:VLF}.

Similarly, many searches for scalar leptoquarks with a variety of possible final states have been performed at the LHC in recent years 
\cite{CMS:2018ncu,CMS:2018lab,CMS:2018oaj,CMS:2020wzx,ATLAS:2019ebv,ATLAS:2019qpq,ATLAS:2022wcu,ATLAS:2021jyv,ATLAS:2023uox,ATLAS:2021oiz, ATLAS:2020dsk,ATLAS:2020xov,ATLAS:2023prb, ATLAS:2020dsf,ATLAS:2021yij,CMS:2023bdh,CMS:2025iix,CMS:2024bnj}. The resulting limits typically range between $1$ to $2$ TeV \cite{ATLAS:2022wcu,CMS:2024bnj},
and even as far $3$ to $5$ TeV in some specific scenarios \cite{CMS:2023bdh,CMS:2025iix}. In the context of $a_\mu$, the largest chiral enhancement is achieved for 
LQs coupling to the muon and top-quark (discussed in Sec.~\ref{sec:LQ}) for which the ATLAS and CMS searches \cite{CMS:2022nty,ATLAS:2022wcu} apply, resulting in the comparatively modest 
bound shown in Fig.~\ref{Fig:mass_reach_gm2VScoll}. In fact, this bound may be slightly weakened if additional decay modes are allowed and are dominant.
Such scenarios again require a more dedicated recasting of the LHC results.

Finally we note that limits on additional Higgs bosons are very model dependent. Here, the limit $M_{H^\pm}>340$ GeV stems from
a CMS search for $\tilde\tau_L$ \cite{CMS:2022syk}, that can be directly reinterpreted as a bound on the type-X 2HDM
assuming BR$(H^\pm\to \tau^\pm\nu)\simeq 100\%$. This limit again differs if additional decay channels are allowed, however, in this
case the bounds resulting from a dedicated recasting are typically more stringent. For a detailed discussion on the current constraints
on the 2HDM see Sec.~\ref{sec:FA2HDMpheno}.

While we have only shown limits from a few selected searches here, it is worth noting that many LHC analyses often obtain limits in a vast array of models that can be very 
different from the original interpretation or motivation given for the search. 
Therefore the limits presented here can also be considered as representative examples of the kind of limits that may be set in other models for which no dedicated LHC searches have been preformed.  
Comparing these to the generic one-loop chirally enhanced contribution shown by the solid and dashed red lines provides a rough indication of how mass limits from the $a_\mu$ measurement and the direct collider limits compete.    

However specific models do not match the generic one-loop result precisely. Often there are further significant factors that can arise from e.g.\ colour or logarithmic 
enhancements in the loop functions.  
Therefore, for a more accurate comparison, in Fig.\ \ref{Fig:mass_reach_gm2VScoll} we also a show mass limits from $\Damu$ in specific models.  
We compute these using the expressions for $R_\chi$ given in Table~\ref{tab:estimates}, including the full mass-dependent loop functions.
These points are shown with the same colour coding as the corresponding collider limits.

For the MSSM we have $R_\chi = \tan\beta$ and we show examples for $\tan \beta=30$ (green diamonds) and $\tan\beta=70$ (green circle) where the Wino, Higgsino and left smuon and smuon neutrino are the light states and give the dominant contribution.   All these points lie below the chargino collider reach indicated for the MSSM,  but nonetheless as discussed  above and later in Sec.\ \ref{sec:SUSY}, many viable scenarios can be found well below these limits.  In these cases the $a_\mu$ measurement can be stronger than the direct collider limit and plays an important role in exclusions and determining the allowed region.  Note that in supersymmetric models the couplings to gauginos are fixed by the gauge couplings, while couplings between muons and Higgsinos are given by Yukawa couplings.  
As a result of this and the specific loop factors they lie well below the solid red line for the generic one-loop contribution with the couplings set to 1, 
as could be anticipated by comparing the expression in table \ref{tab:estimates}.       

Next to the MSSM, we show several points in the flavour aligned 2HDM where the chiral enhancement factor $R_\chi=-\zeta_l \zeta_u m_t^2 / v^2$  is also typically of $\O(10)$.
The 2HDM is unique compared to the other examples shown, because here the largest correction to $\Damu$ comes from two-loop Barr-Zee diagrams (see also Sec.~\ref{sec:Barr-Zee}).
While this contribution is suppressed by an additional loop factor, this suppression is easily compensated by the logarithmic enhancement in the loop function
as well as further combinatoric and colour factors. 
The brown symbols indicate specific values of $\zeta_l \zeta_u = 40,80, -80,-160$, corresponding to solid diamond, solid circle, empty diamond and empty circle respectively.  While  the chiral enhancement and the mass limit increases with $|\zeta_l \zeta_u|$, the sign is also important as can be seen by comparing the solid circle and empty diamond. The latter effect is entirely due to the different limits on positive and negative $\Damu$ shown in Eq.\ (\ref{Eq:MLim_amu}). The phenomenology and the impact of the $a_\mu$ measurement on 2HDM models is discussed in more detail in Sec.\ \ref{sec:2HDM}.

For vector-like lepton models which the chiral enhancement is given by $R_\chi = -\bar\lambda_\Phi v / m_\mu$. 
Here $\bar\lambda_\Phi$ is a free parameter that can be varied to increase the size of the chirality flip with different choices represented by the orange diamonds, circles and triangles showing how increasing $|\bar\lambda_\Phi|$ increases the chiral enhancement and the mass limit. 
At the same time the mass limit increases with $|\lambda_L \lambda_R|^{\frac{1}{2}}$ and again becomes more severe if the sign is chosen so that $\Damu <0$, as indicated by comparing the filled and unfilled symbols in orange.  These points demonstrate how the mass limit obtained from the $a_\mu$ measurement can be stronger than collider limits for $\O(1)$ values of the couplings. Note, however, that in some cases additional constraints apply which are discussed in detail in Sec.~\ref{sec:VLF}.     

In case of the $S_1$ scalar leptoquarks coupling to the top-quark and muon the resulting chiral enhancement $R_\chi = m_t / m_\mu \approx 1700$ is fixed, but rather large. 
Additionally, there is again a colour factor and sizeable loop function, such that the actual limits for these models easily exceed the generic one-loop bounds indicated by the red lines.  
Since the chiral enhancement is fixed by the measured SM fermion masses, the mass limit only varies with the couplings to left- and right-handed muons.  
For example, the current collider limit on $M_{S_1}$ is surpassed for couplings $\sqrt{\lambda_L^{32}\lambda_R^{32}} > 0.06$, 
as indicted by comparing to the purple diamond in Fig.\ \ref{Fig:mass_reach_gm2VScoll}. 
The mass limit from the $a_\mu$ measurement for larger couplings of $0.1$ and $0.5$ are indicated by the purple circle and triangle respectively, 
with the latter showing that even masses well above 10 TeV can be excluded by the $a_\mu$ measurement emphasising the power of this observable.

Finally, we have also shown some points corresponding to the radiative muon mass scenarios listed in Table~\ref{tab:estimates} and discussed in more detail
in Sec.~\ref{sec:MuonMass}. Here the chiral enhancement is given by $R_\chi\sim 16\pi^2/(\lambda_L\lambda_R)$, however, since the couplings
are always adjusted to produce the correct muon mass, the resulting mass limits are independent of the model parameters and instead depend only on the representation and charges.
In turn, the resulting bounds are quite rigid and now typically exclude masses in such scenarios to below the TeV scale. 

\subsubsection{Two-field models and compressed spectra}

As we have seen muon $g-2$ can be a very powerful observable for constraining models with large chiral enhancements. On the other hand generically for models without any enhancement, direct collider limits on masses are higher than the limit from the $a_\mu$ measurement.  However precision measurements of muon $g-2$ still play an important role in the phenomenology of models without an internal chirality flip.  
As discussed above the collider bounds are very model dependent and can often be evaded.  
The limits depend significantly on very specific signal regions that can change in realistic and well motivated models where multiple states may be light or due to other factors influencing the kinematics.  In these cases rather than competing, collider searches and the precision $a_\mu$ measurement can provide complementary constraints and cross-checks.     

We will illustrate this complementarity in the context of models that extend the SM by two new fields with different spin, which we discussed in  Sec.\ \ref{sec:MinimalBSM} and classified in in Table \ref{tab:twofieldmodels}.  As already discussed there these models can only achieve large $\Damu$ contributions, exceeding the $2\sigma$ bound, for masses that are well within the reach of the LHC searches.  These models are particularly interesting as they are the simplest extensions that can both explain dark matter and contribute non-negligibly to $a_\mu$ and they were also discussed in this context in Sec.\ \ref{sec:DMtwofield}. There it was shown that the DM relic density can only be explained for such low masses for the lowest dimensional representations.  Nonetheless this does not exclude lower mass scenarios, as it is possible the dark matter candidate provides only a small fraction of the full relic density of dark matter or in fact for the model to have no dark matter candidate at all if one does not assume a $\mathds{Z}_2$ symmetry (models like this were considered in Ref.\ \cite{Freitas:2014pua}). However both collider constraints and the precision $a_\mu$ measurement are very relevant for these masses in all cases, i.e.\ for the models that can explain dark matter, for models that predict an under abundance of DM and for models that have no DM candidate at all.   

In particular we will consider two models with a scalar $\phi$ and Dirac fermion $\psi$, one that couples only to left-handed muons, which we will call Model L here, and one that couples to only right-handed muons, which we will call  Model R.  Model L and Model R have the Lagrangian's given Eq.\ \eqref{eqn:Min2FieldsLL} and \eqref{eqn:Min2FieldsRR}, respectively,  in Sec.\ \ref{sec:DMtwofield}. In both models the scalar and fermion are $\SUc$ singlets and for the $\SUL$ representation we choose from the models of this type with the lowest dimensional representations to improve the chances of explaining dark matter.  In Model L $\phi$ is a scalar singlet and $\psi$ is an $\SUL$ doublet, while in Model R both the scalar and the fermion are $\SUL$ singlets, with the scalar having hypercharge zero such that it is a dark matter candidate and the fermion has hypercharge $-1$ such that it is a charged state that can couple to the photon in the one-loop diagram from $\Damu$. 

\begin{figure}[tb]
		\centering
		\includegraphics[width=0.4\textwidth]{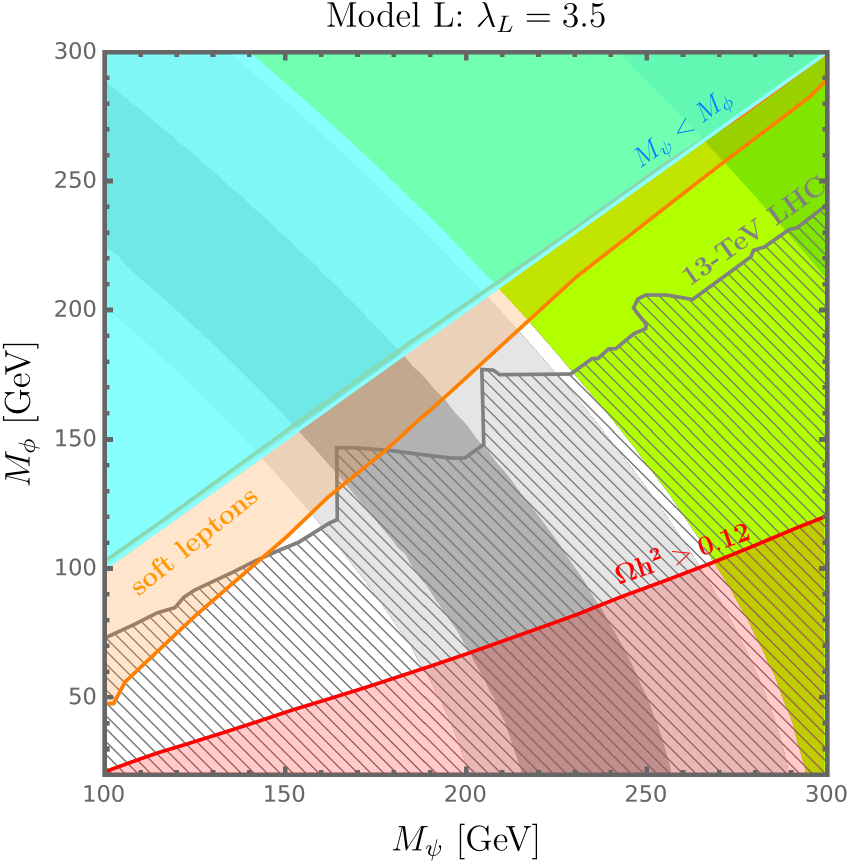}\hspace{1.5cm}
        \includegraphics[width=0.4\textwidth]{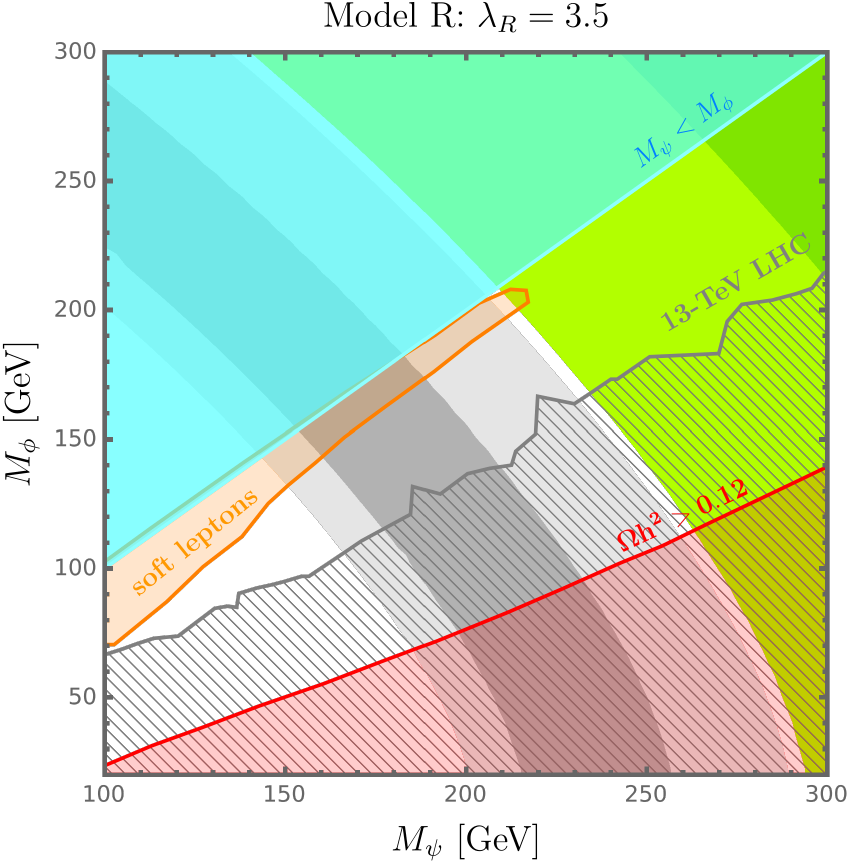}
		\caption{LHC, dark matter and $\Damu$ exclusion contours in the $M_\psi$ -- $M_\phi$
			plane of the Model L (\emph{left}) and R (\emph{right}) discussed in the text. In both cases the scalar $\phi\sim(\bm1,0)$ is a singlet with hyper-charge 0 under $\GEW$,
			while in Model L $\psi\sim(\bm2,\frac{1}{2})$ and in Model R $\psi\sim(\bm1,-1)$. The (light) green contour shows the 1$\sigma$ (2$\sigma$) region allowed by $\DamuFinal$,
			while the (light) grey contour shows the 1$\sigma$ (2$\sigma$) region allowed by the previous deviation $\DamuOld$.
			Compared to Ref.\ \cite{Athron:2021iuf} which took into account higher-order corrections to $\Damu$, here we have only used the leading one-loop contribution.
			The resulting difference is, however, negligible for the present discussion. 
			The exclusion limits from 13-TeV LHC (grey hatched region), soft lepton searches (orange shading) and dark matter overabundance $\Omega h^2>0.12$ (red shading)
			have been directly adapted from Ref.\ \cite{Athron:2021iuf}.
			The region with $M_\psi < M_\phi$ (cyan contour) is excluded by the fact that the lightest state is electrically charged.
			}
		\label{fig:Min2FieldsLLSlice}
	\end{figure}

Even in these relatively simple models, with only two new physics states, there is a generic challenge in constraining so-called compressed spectra where the mass of the fermion and scalar are close in size. For a sufficiently small mass splitting the kinetic energy of the decay products is too low, resulting in soft leptons or soft jets that can be below the detection threshold.  In the past such gaps in the collider exclusion have been interpreted as a possible way that models without a chiral enhancement could explain a large deviation in muon $g-2$, as illustrated in Fig.\ \ref{fig:Min2FieldsLLSlice} which we reproduce from Ref.\ \cite{Athron:2021iuf}.             

        There we see a substantial gap between the exclusion from LHC searches and the $M_\phi = M_\psi$ line, which borders the cyan region that is excluded because it has a stable charged particle.  This is nonetheless further constrained by dedicated searches for compressed spectra \cite{CMS:2018kag}.\footnote{Since Ref.\ \cite{Athron:2021iuf} was released there have been further papers tightening constraints on compressed spectra, see e.g.\ Refs.\ \cite{ATLAS:2022hbt,ATLAS:2019lng,ATLAS:2025evx,CMS:2019zmn,CMS:2024gyw,CMS:2021knz}.  Nonetheless things do not change qualitatively as there always remains a mass gap in the LHC exclusions.}   Despite these efforts, gaps in the exclusions remain for small enough mass splittings.

By contrast contours of $\Damu$ cross the whole $M_\phi-M_\psi$ plane
with no gaps. This is illustrated by the light and dark, grey and
green shaded contours in Fig.\ \ref{fig:Min2FieldsLLSlice}.  Here the
light (dark) grey contours show the region compatible within $2\sigma$
($1\sigma$) with the deviation $\DamuOld$. 
Previously the mass gaps discussed above and plots such
as those in Fig.\ \ref{fig:Min2FieldsLLSlice} were used to provide a
way in which these models could explain the then large deviation between
experiment and SM predictions for $a_\mu$ and the grey contours show where that was
possible. However in the present context where the 2025 White Paper
prediction is very close to the measured value, such plots should be
reinterpreted.  Therefore we have added the the light (dark) green
contours that show the current $2\sigma$ ($1\sigma$) limit for these
models on top of the results from Ref.\ \cite{Athron:2021iuf}. The
precision $a_\mu$ measurement then excludes the entire plane below the
light green region, including the previously preferred region in grey.

In this way although the mass reach of the FNAL muon $g-2$ experiment does not currently extend to masses large enough to escape the LHC sensitivity, the precision $a_\mu$ measurement can exclude various gaps in the collider constraints at lower masses where colliders lack sensitivity. Thus muon $g-2$ experiments provide a complementary test on light new physics that can close holes in the exclusion limits from colliders.  While we have only discussed two particular examples above, this behaviour is typical of the whole class of models extending the SM by a Dirac fermion and a scalar singlet that we have also discussed in previous sections.  Furthermore when one considers models with many new states at low energies, as is predicted by some models based on fundamental principles, such as supersymmetry, there are many more ways for holes in the collider limits to be introduced.  Therefore the precision measurement of $a_\mu$ provides a  complementary constraint on new physics masses of ${\cal O}(100)$ GeV that is of great importance to particle physics phenomenology.

\section{Higher-order contributions to $a_\mu$}
\label{sec:HigherOrder}
In this section we discuss several techniques and results for calculations of $\Delta\amu$
in BSM scenarios beyond the one-loop level. 
We first explain in Sec.~\ref{sec:TraceProjectors} how the scalar form factors can be obtained from the tensor amplitude
using trace projectors. This methods helps to avoid the tedious reductions of the Dirac covariants via Gordon Identities
and, in particular for higher order calculation, drastically reduces the computational complexity.
Section \ref{sec:Barr-Zee} then focuses on the important class of Barr-Zee two-loop diagrams, 
which are relevant or even dominant in a variety of different models, including e.g. the 2HDM or vector-like fermion extensions.
Finally, in Sec.~\ref{sec:photonic} we discuss
logarithmically enhanced higher-order corrections arising e.g.\ from QED. 
There, the origin and structure of such contributions is explained using EFT
methods and the impact on various BSM scenarios is sketched.

\subsection{Form factor trace projectors}\label{sec:TraceProjectors}
On-shell amplitudes like Eq.~\eqref{eq:photon-vertex-decomp}  typically consist of a small number
of relevant covariants that define the physical form factors.
However, especially at higher loop orders, many additional Dirac
structures can appear at intermediate steps that require further simplifications in order to bring the result
in the desired form. It is therefore often much more efficient to
project out the relevant terms already at the amplitude level by using trace projectors.
This technique was first developed in 4-dimensions \cite{Brodsky:1966mv,Aldins:1970id,Barbieri:1972as,Barbieri:1974nc,Barbieri:1975qf}
and later adapted to dimensional regularisation \cite{Czarnecki:1996if,Melnikov:2006sr,Jegerlehner:2009ry,Jegerlehner:2017gek}, for which 
we collect the results here.
In case of the photon vertex, we denote the projectors as $\mathcal{P}^\mu_i$ such that the corresponding form factors are given by
\begin{align}\label{eq:FF-trace}
	F_i(q^2) = \Tr\Big\{\mathcal{P}_i^\mu(p,p') \cdot \Gamma_{\mu}(p,p')\Big\}.
\end{align}
After moving the trace to the integrand level only scalar integrals remain on the right-hand side which directly yield the desired 
contribution to the respective form factors.
The main purpose of the projectors is to enforce the Dirac equation,
that is, to remove terms of the form $...(\slashed{p}-m_\mu)$ and $(\slashed{p}'-m_\mu)...$ from $\Gamma^\mu$.
This is achieved by the following ansatz,
\begin{align}
	\mathcal{P}_i^\mu = (\slashed{p} + m_\mu) 
	\bigg[ \gamma^\mu \Big(C_1 + C_2 \gamma^5\Big)  + \frac{P^\mu}{2m_\mu}\Big(C_3 + i C_4 \gamma^5\Big) + \frac{q^\mu}{2m_\mu} \Big(C_5 + C_6 \gamma^5\Big) \bigg]
	(\slashed{p}' + m_\mu),
\end{align}
where the momenta are taken on-shell, $p^2=p'^2=m_\mu^2$. The terms vanishing due to the Dirac equation then drop out automatically.
The coefficients $C_i$ can be determined by inserting this ansatz together with the generic vertex function Eq.~\eqref{eq:photon-vertex-decomp} into
Eq.~\eqref{eq:FF-trace} and collecting the coefficients in front of
the form factors. Solving the resulting system of linear equations
assuming $D=4-2\epsilon$ space-time dimensions
yields the coefficients
\begin{alignat}{2}
	\mathcal{P}^\mu_E: \qquad & C_1=\frac{1}{2(D-2)(q^2-4m_\mu^2)} \qquad 
	&&C_3 = \frac{2(D-1)m_\mu^2}{(D-2)(q^2 - 4 m_\mu^2)^2} ,\\
	\mathcal{P}^\mu_M: \qquad & C_1=\frac{-2m_\mu^2}{(D-2)q^2(q^2-4m_\mu^2)} \qquad 
	&&C_3 = \frac{-2m_\mu^2((D-2)q^2 + 4 m_\mu^2)}{(D-2)q^2(q^2 - 4 m_\mu^2)^2} ,\\
	\mathcal{P}^\mu_D: \qquad &C_4 = \frac{2m_\mu^2}{q^2(q^2-4m_\mu^2)}.
\end{alignat}
Note that the projectors for $F_M(q^2)$ and $F_D(q^2)$ are both
divergent in the limit $q^2\to 0$. Although  this divergence
cancels after the evaluation of the loop integrals, it constitutes
an inconvenience for the calculation of $a_\mu$.
For this, it is instead desirable to obtain dedicated projectors that immediately yield the form factors at $q^2=0$.
A simple way to achieve this is by expanding the amplitude in $q_\mu$ after setting $p=(P-q)/2$ and $p'=(P+q)/2$. Since the dipole
covariants are linear in $q_\nu$ it is sufficient to expand to first order
\begin{align}
	\Gamma^\mu(P,q) \simeq \Gamma^\mu(P,0) + \frac{\partial}{\partial q_\nu} \Gamma^\mu(P,q)\Big|_{q=0} q_\nu  \equiv V^\mu + T^{\mu\nu} q_\nu.
\end{align}
We can then average $\mathcal{P}_i\cdot\Gamma$ over the spatial directions of $q_\mu$ which leaves the trace unchanged but allows us to replace
\cite{Czarnecki:1996if,Jegerlehner:2017gek}
\begin{align}
	\overline{q^\mu},~\overline{q^\mu q^\nu q^\rho} \to 0 \qquad \text{and} \qquad \overline{q^\mu q^\nu} \to \frac{q^2}{D-1} \Big(g^{\mu\nu} - \frac{P^\mu P^\nu}{P^2}\Big).
\end{align}
The resulting expressions for the trace projectors are given by
\begin{subequations}
	\begin{align}
		\begin{split}
			F_M(0) &= \frac{1}{4(D-1)m_\mu^2} \Tr\bigg\{ \Big[m^2_\mu\gamma^\alpha - (D-1) p^\alpha m_\mu - Dp^\alpha\slashed{p}\Big]  V_\alpha(p) \bigg\} \\
			&\quad + \frac{1}{8(D-2)(D-1)m_\mu} \Tr\bigg\{ (\slashed{p}+m_\mu)[\gamma^\alpha,\gamma^\beta](\slashed{p}+m_\mu) T_{\alpha\beta}(p)  \bigg\},
		\end{split}\\
		\begin{split}
			F_D(0) &= \frac{i p^\alpha}{4(D-1)m_\mu^2} \Tr\bigg\{  (\slashed{p}+m_\mu)\gamma^\alpha \gamma^5(\slashed{p}+m_\mu) T_{\alpha\beta}(p)  \bigg\} 
			-\frac{i p^\alpha}{16m_\mu} \Tr\bigg\{ \gamma^5 V_\alpha(p) \bigg\}.
		\end{split}
	\end{align}
\end{subequations}
Here we have substituted $P=2p$ which holds for $q\to 0$. Since both $V$ and $T$ depend only on $p$, the
computation of $F_M(0)$ and $F_D(0)$ reduces to an evaluation of on-shell self-energy amplitudes rather than the full three-point function. 
Furthermore, for $\amu$ only the anti-symmetric part of $T_{\mu\nu}$ contributes due to the appearance of the commutator.

\subsection{Two-loop Barr-Zee diagrams}\label{sec:Barr-Zee}

\begin{figure}[t]
	\centering
	\begin{subfigure}{.44\textwidth}
		\centering
		\includegraphics[width=.8\textwidth]{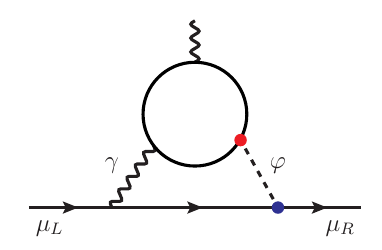}
		\caption{Generic Barr-Zee diagram}
		\label{fig:BZ-generic}
	\end{subfigure}
	\begin{subfigure}{.53\textwidth}
		\centering
		\includegraphics[width=.9\textwidth]{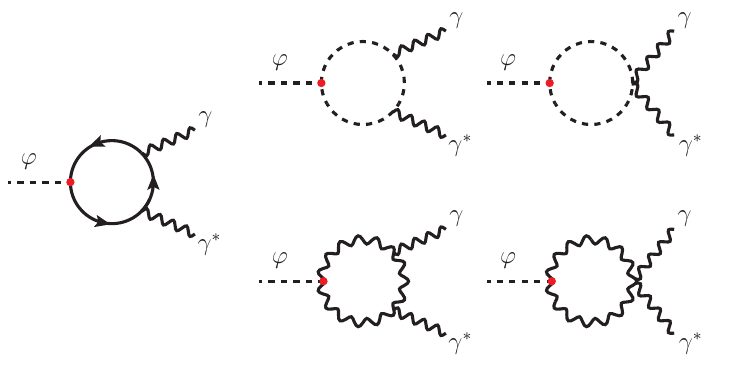}
		\caption{One-loop subdiagrams}
		\label{fig:BZ-sub-loops}
	\end{subfigure}
	\caption{\textbf{(a)}: Generic two loop Barr-Zee diagram with internal photon and real scalar. The blue vertex denotes the small Yukawa coupling of the scalar to 
		the muon while the red vertex denotes the (potentially enhanced) coupling of the scalar to the virtual loop.
		\textbf{(b)}: one-loop contributions of fermions, scalars and gauge bosons to the $\varphi\gamma\gamma$ subdiagram.}
	\label{fig:Barr-Zee-diagrams}
\end{figure}
One important class of two-loop diagrams (some of which are shown in Fig.~\ref{fig:Barr-Zee-diagrams}) with potentially large contributions to 
$\Damu$ was first identified by Barr and Zee in
Ref.~\cite{Barr:1990vd}.
In these diagrams, an inner fermion or boson loop generates an
effective $\gamma$--$\gamma$--$\varphi$ vertex, which then couples to the
muon via a second loop. $\varphi$ can be the SM Higgs or a new scalar.
Barr and Zee initially studied possibility of CP violation mediated by
an extended Higgs sector, where $\varphi$ corresponds  to a BSM
Higgs. The diagrams can then generate important effects on the EDM of
the electron, but the same enhancement mechanism can also lead to 
a large contribution to $\Damu$. 
The main reason for this enhancement is that, compared to a one-loop
diagram of the kind in Fig.~\ref{fig:amu-FS} with a virtual $\varphi$, one of the scalar vertices is lifted from the muon line into the virtual loop. 
On the one-hand, this exchanges one of the muon Yukawa couplings
$y_\mu$ for the (potentially) much larger coupling of the scalar
$\varphi$ to the particle running in the inner loop and
on the other hand, no further chirality flip on the muon line is
necessary, removing a factor of $m_\mu$. The compensating mass
parameter 
is given either by the triple boson coupling or, in case of a fermion
loop, by an additional required chirality flip on one of the internal
heavy fermion lines.
In both cases this results
in a significant potential enhancement factor, which can often
overcome the additional loop suppression.

To specify the diagrams generically, we denote the generic Yukawa and
scalar interaction terms in a mass-basis Lagrangian by
\begin{align}\label{BarrZeeLagrangian}
	\La_Y = y^\varphi_f \bar{f} \PR f \varphi + h.c., \qquad\quad
	\La_{\varphi S^2} = - \lambda_{\varphi S^2} \varphi |S|^2, \qquad\quad
	\La_{\varphi V^2} = g_{\varphi V^2} \varphi V_\mu^\dagger V^\mu,
\end{align}
while the photon interactions are taken from
Eq.~\eqref{eq:L_SFV_photon}.  We assume
fermion flavour changing  
couplings to be absent and $\varphi$ to be a real scalar.

With these couplings, the above discussion can be made concrete. The
one-loop diagram with virtual $\varphi$ and muon has the 
chirality-flipping and EWSB factors $[\ldots]\to |y^\varphi_\mu|^2
m_\mu$ in the notation of Eq.~(\ref{amugeneric}). The two-loop diagram
with fermion loop instead involves the factors $y^\varphi_\mu
y^\varphi_f m_f\alpha/4\pi$, hence it is relatively enhanced by
\begin{align}
\label{eq:Barr-Zee-scaling}
	\bigg|\frac{\Delta\amu^\text{BZ}}{\Delta\amu^\text{FS}}\bigg|
        \sim \frac{\alpha}{4\pi} \frac{m_fy^\varphi_f}{m_\mu
          y^\varphi_\mu}.
\end{align}
In the typical case where $m_f\propto y^\varphi_f$, the two-loop
Barr-Zee diagram is bigger than the one-loop FS-type diagram if the
mass $m_f\gtrsim$few GeV. The Barr-Zee type diagrams with internal
boson loops behave similarly.

Thus, particularly in BSM scenarios with new scalars (such as the 2HDM
discussed in Sec.~\ref{sec:2HDM} or extensions thereof) the Barr-Zee diagrams often
provide very substantial or even dominant contributions to $\amu$.

In the following we collect some of the results previously presented in the literature. We focus particularly on the photonic Barr-Zee 
diagrams shown in Fig.~\ref{fig:Barr-Zee-diagrams}. In principle, there are further Barr-Zee type diagrams where the internal photon is exchanged for a massive
gauge boson. Although these will have a relative suppression due to the massive propagator, they can still give significant contributions e.g.\
in case of a charged scalar or as a consequence of other
enhancements. For further discussion and details on the calculation
for $\Damu$ we
refer to Refs.~\cite{Chang:2000ii,Cheung:2001hz} for the fermionic
loops and Ref.~\cite{Ilisie:2015tra} for more general
Barr-Zee diagrams.

The calculation of the diagrams is conveniently split into two steps. First, the sub-loops depicted in Fig.~\ref{fig:BZ-sub-loops}
are evaluated for off-shell momenta $p$ and $k$ of the scalar and one of the photons, while the external photon momentum $q$ can be treated on-shell. 
The resulting effective vertex has the following relevant gauge invariant structure
\begin{align}
	i\Gamma^{\mu\nu}(k,q) = i\Gamma(k,q) \big[g^{\mu\nu} k\cdot q - k^\mu q^\nu\big] + i\Gamma^5(k,q)\epsilon^{\mu\nu\alpha\beta} k_\alpha q_\beta.
\end{align}
Note that gauge-variant covariants like $k^\mu k^\nu$ can also appear during the calculation,  
but they do not contribute to the final result and can therefore be dropped \cite{Abe:2013qla}.
After inserting this effective vertex back into the two-loop amplitude, the remaining integral can be evaluated,
yielding the contributions to the dipole moments. It is also worth pointing out that the sub-loop is finite and the effective vertex has 
the structure of a propagator. Therefore also the full amplitude is finite and does not require sub-renormalisation.

With the couplings defined in the Lagrangian (\ref{BarrZeeLagrangian}), the results of the diagrams in Fig.~\ref{fig:Barr-Zee-diagrams} (neglecting higher orders in $m_\mu$) are given by\footnote{The
vector contribution is generalised from the 2HDM diagrams with $W$-boson loops computed in Ref.~\cite{Ilisie:2015tra}
and corresponds to the physical result including appropriate Goldstone boson contributions.}
\begin{align}
	\Delta\amu^{\text{BZ},f} &= \frac{\alpha}{8\pi^3} N_c^f Q_f^2 
	\frac{m_\mu m_f}{m_\varphi^2}
        \bigg[\Re\big\{y_f^\varphi\big\}\Re{y_\mu^\varphi}
          \mathcal{F}_1\Big(\tfrac{m_f^2}{m_\varphi^2}\Big) +
          \Im\big\{y_f^\varphi\big\}\Im{y_\mu^\varphi}
          \mathcal{F}_2\Big(\tfrac{m_f^2}{m_\varphi^2}\Big)\bigg],
\label{BZfermionic}
        \\
	\Delta\amu^{\text{BZ},S} &= \frac{\alpha}{16\pi^3} N_c^S Q_S^2 \frac{m_\mu \lambda_{\varphi S^2}}{m_\varphi^2} \Re{y_\mu^\varphi}\F_3\Big(\tfrac{m_S^2}{m_\varphi^2}\Big),\\
	\Delta\amu^{\text{BZ},V} &= \frac{\alpha}{32\pi^3} Q_V^2 \frac{m_\mu g_{\varphi V^2}}{M_V^2} \Re\big\{y_\mu^\varphi\big\} \F_4\Big(\tfrac{m_V^2}{m_\varphi^2}\Big),
\end{align}
where the loop functions are normalised slightly differently compared
to Ref.~\cite{Ilisie:2015tra} and are given by
\begin{align}\label{eq:BZ-loop-functions}
	\mathcal{F}_i(x) &= \int_0^1 du \frac{\mathcal{N}_i(u;x)}{x-u(1-u)} \ln(\frac{x}{u(1-u)}),
\end{align}
and the numerators for the respective integrals read
\begin{subequations}
	\begin{align}
		\mathcal{N}_1(u;x) &= 2u(1-u) -1 \\
		\mathcal{N}_2(u;x) &= 1 \\
		\mathcal{N}_3(u;x) &= u(1-u) \\
		\mathcal{N}_4(u;x) &= u(1-u) - x u [3u(4u-1) + 10].
	\end{align}
\end{subequations}
The loop functions depend on the mass ratios
$x=m_{f,S,V}^2/m_\varphi^2$ between the two involved
internal masses. Often the two
masses both correspond to heavy states and are of similar order, but
sometimes the cases of light $\varphi$, $x\to\infty$, or heavy
$\varphi$, $x\to0$, are of interest. Table \ref{tab:BZ-asymptotes}
  collects all these limits.

\begin{table}[t]
	\centering
	\begin{tabular}{|c|c|c|c|}
		\hline
		& $x\to 0$ & $x\to\infty$ & $x\to 1$ \\ \hline\hline
		\rule{0pt}{2.5ex}$\F_1$ & $-\ln(x)^2-2\ln(x)-4-\frac{\pi^2}{3}$ & $-\frac{6\ln(x)+13}{9x}$ & $-1.66$ \\ \hline
		\rule{0pt}{2.5ex}$\F_2$ & $\ln(x)^2 + \frac{\pi^2}{3}$ & $ \frac{\ln(x) + 2}{x} $ & $2.34$ \\ \hline
		\rule{0pt}{2.5ex}$\F_3$ & $-\ln(x) - 2$ & $\frac{3\ln(x)+5}{18x}$ & $0.34$ \\ \hline
		\rule{0pt}{2.5ex}$\F_4$ & $-\ln(x) - 2$ & $-7\ln(x) - \frac{89}{6} $ & $-16.76$ \\ \hline
	\end{tabular}
	\caption{Asymptotic expressions for the Barr-Zee loop functions.}
	\label{tab:BZ-asymptotes}
\end{table}

For phenomenological applications not only the magnitude but also the
signs of the contributions are of high interest. Taking again the
corresponding one-loop diagrams with virtual $\varphi$ and muon as a
reference, its result in the limit of large $m_\phi$ can be written as (see Sec.~\ref{sec:genericoneloop})
\begin{align}
	\Damu^\text{one-loop} \approx \frac{1}{4\pi^2} \frac{m_\mu^2}{m_\varphi^2} \ln(\frac{m_\varphi}{m_\mu}) \Big[\Re{y^\varphi_\mu}^2 - \Im{y^\varphi_\mu}^2\Big]
\end{align}
in the notation of the Lagrangian (\ref{BarrZeeLagrangian}).
For a scalar, the couplings are real and the one-loop contribution is
positive; for a pseudoscalar the one-loop contribution is negative. In
contrast, e.g.\ the fermionic Barr-Zee diagrams of Eq.~(\ref{BZfermionic}) have
exactly the opposite behaviour. The was first observed and used for the 2HDM
in Refs.~\cite{Chang:2000ii,Cheung:2001hz,Wu:2001vq,Krawczyk:2002df}.

\subsection{Leading logarithms and EFT resummation}\label{sec:photonic}

\begin{figure}
	\centering
	\includegraphics[width=.9\textwidth]{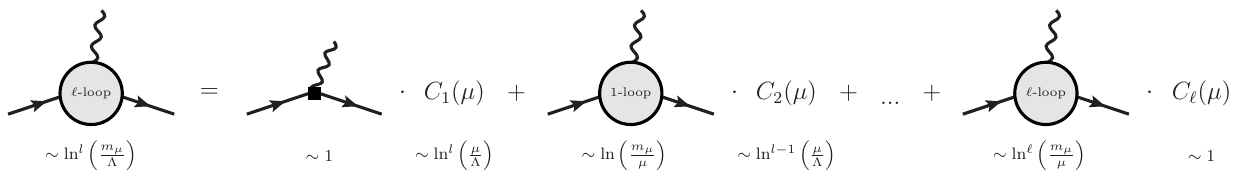}
	\caption{Illustration of the decomposition of an $\ell$-loop contribution to $\Damu$ into EFT (loop) diagrams
	and Wilson coefficients with the corresponding logarithmic scaling.}
	\label{fig:EFT-amu-Logs}
\end{figure}

Another class of important higher-order corrections can arise in BSM
scenarios with heavy masses of the order of a scale $\Lambda\gg m_\mu$. In
such models, large logarithms $\ln(\Lambda/m_\mu)$ involving the two physical
scales can appear, where
e.g.~for $\Lambda=1$ TeV,
\begin{align}
  \ln\left(\frac{\Lambda}{m_\mu}\right) \approx 9.2.
\end{align}
For example, the electroweak SM contributions involve heavy particles
and behave similar to many BSM scenarios. The one-loop $W$- and
$Z$-boson contributions involve no large logarithm, but the one-loop
Higgs-boson contribution involves a large logarithm $\ln(M_h/m_\mu)$,
see Eq.~(\ref{eq:amu-EW-1}) in Sec.~\ref{sec:SMtheory}. As discussed
there, the two-loop electroweak corrections reduce the one-loop result
by about 20\% and are dominated by large
logarithms. These and further logarithms were analysed and computed in
Ref.~\cite{Degrassi:1998es}, and a prediction valid in many BSM
scenarios was derived,
\begin{align}\label{DGlogs}
  \Damu^{\text{2L,leading log}}=\Damu^{\text{1L}}\left(1-\frac{4\alpha}{\pi}  \ln\left(\frac{\Lambda}{m_\mu}\right)\right)
,
\end{align}
i.e.\ the one-loop contributions are reduced by a large log, and the
reduction amounts to about 9\% for $\Lambda\sim1$ TeV. This holds in
scenarios where the one-loop results are governed by purely heavy
loops and are proportional to $m_\mu^2$,
i.e.\ where the necessary chirality flip (see
Sec.~\ref{sec:ChiralityFlips}) is proportional to the muon Yukawa
coupling or some generalisation. The analysis of SM logarithms was
refined in Ref~\cite{Czarnecki:2002nt}, the result (\ref{DGlogs}) was
rederived together with non-logarithmic corrections by Feynman
diagrammatic computations in Ref.~\cite{vonWeitershausen:2010zr}, and
more general results were obtained in Ref.~\cite{Aebischer:2021uvt}.

Our aim in this subsection is to  show how the techniques of
effective field theory (EFT) and renormalisation group equations (RGE)
allow to compute such logarithms and discuss the corresponding
results.
While we focus here on applications to $\Damu$, we stress that the
method is much more general and has been applied e.g.~in the context
of quark flavour physics \cite{Buchalla:1995vs} and can be applied to
a large number of other observables using full one-loop
RGEs available for both LEFT \cite{Jenkins:2017dyc} and SMEFT  
\cite{Jenkins:2013zja,Jenkins:2013wua,Alonso:2013hga}.

\subsubsection{The EFT renormalisation group equation}
\label{sec:EFTRGE}

In sec.~\ref{sec:genericeft} we have already discussed how EFT can be applied to analyse BSM contributions to
$\amu$ in a model-independent way. A prominent feature of this approach is the separation
of the BSM (or electroweak) scale $\Lambda$ and the muon mass scale $m_\mu$. The BSM scale only enters in the Wilson coefficients 
after matching to the UV theory, while the muon mass (and other light particle masses) appears only in the EFT diagrams.
As an important consequence, large \emph{physical} logarithms
$\ln(m_\mu/\Lambda)$ appearing in the 
full result of the UV theory are split
into $\ln(m_\mu/\mu)$ contributions from EFT (loop) 
diagrams
and  $\ln(\mu/\Lambda)$ contributions from running Wilson coefficients
(WCs).
In this way, the physical logarithms are connected to the unphysical logarithms $\ln(\mu)$ of the EFT 
renormalisation scale $\mu$.
In particular, as illustrated in Fig.~\ref{fig:EFT-amu-Logs}, 
the highest powers of $l$-loop logarithms $\ln^l(m_\mu/\Lambda)$ (leading logarithms) 
are contained entirely in the tree-level Wilson coefficient.
This connection can be exploited to obtain the leading (or subleading)
higher-order corrections from the
 RGEs within the EFT which govern the WCs,
rather than from explicit multi-loop integrals involving several scales.

As a setup we consider first a  generic EFT Lagrangian in terms of the bare composite operators $\O_i^0$
and corresponding Wilson coefficients $C_i^0$
\begin{align}
	\La = \sum_i C_i^0 \O^0_i.
\end{align}
In dimensional regularisation with dimension $D=4-2\epsilon$ we denote
the operator mass dimension by $[\O_i^0] = n_i -(2+\varsigma_i)\epsilon$ (with $n_i=1,2,...$), 
such that $[C_i^0] = 4-n_i + \varsigma_i \epsilon$. The bare quantities are given in terms of the renormalised ones as
\begin{align}
	C_i^0 = \mu^{\varsigma_i\epsilon} \Big(C_i(\mu) + \delta C_i(\mu)\Big), \qquad \O_i^0 = Z_{\O_i}(\mu) \O_i(\mu),
\end{align}
where $Z_{\O_i}$ is the product of composite field wave-function renormalisations and
the renormalisation scale $\mu$ was introduced in such that the
renormalised Wilson coefficients are integer dimensional (in the
following we suppress the argument). In the $\overline{\text{MS}}$-scheme 
the renormalisation constants can be written as
\begin{align}
	\delta C_i = \sum_{n=1}^\infty \frac{\delta C_i^{[n]}}{\epsilon^n}, \qquad Z_{\O_i} = 1 + \sum_{n=1}^{\infty} \frac{Z_{\O_i}^{[n]}}{\epsilon^n}.
\end{align}
The fact that the bare coupling parameters are independent of $\mu$ and that the ($\overline{\text{MS}}$) renormalisation constants $\delta C_i^{[n]}$
depend on $\mu$ only through the renormalised couplings results in the following RGE
\begin{align}\label{eq:EFT-general-RGE}
	\mu\frac{d C_i}{d\mu} = - \Big(\delta_{ij} + \frac{\partial\delta C_i}{\partial C_j}\Big)^{-1} \varsigma_j \epsilon \big(C_j + \delta C_j\big)
	\overset{\epsilon\to 0}{=} \frac{\partial \delta C_i^{[1]}}{\partial C_j} \varsigma_j C_j - \varsigma_i \delta C_i^{[1]}.
\end{align}
While this equation is generally valid for all couplings, it can often
be  significantly simplified for the 
WCs of the non-renormalisable operators:
\begin{itemize}
	\item  It is often sufficient to consider effects only up to $1/\Lambda^2$. 
	In this case only terms linear in the dimension-6 coefficients and (at most) quadratic 
	in the dimension-5 coefficients have to be taken into account.	
	\item In addition, the dimension-5 coefficients are usually generated radiatively, such
	that two insertions of the dimension-5 operator are loop suppressed by the matching condition.
	In this case also the $\sim C^2$ terms can be dropped.	
\end{itemize}
Under these assumptions, the renormalisation constants of the WCs are then given by $\delta C_i = Z_{ij} C_j$ 
where the matrix $Z_{ij}$ depends only on the renormalisable (dimension$\leq$4) couplings (denoted $\alpha_k$ in the following), and 
the RGE Eq.~\eqref{eq:EFT-general-RGE} takes the simple form
\begin{align}\label{eq:ADM}
	\mu \frac{d C_i}{d\mu} = (\gamma^T)_{ij} C_j \qquad \text{where} \qquad
	\gamma_{ji} = (\varsigma_j - \varsigma_i) Z_{ji}^{[1]} + \frac{\partial Z_{ij}^{[1]}}{\partial \alpha_k} \varsigma_k \alpha_k.
\end{align}
This simplified RGE can be solved iteratively giving
\begin{align}\label{eq:ADM-RGE}
	\bm{C}(\mu) = 
	\sum_{n=0}^\infty \int_{\Lambda}^{\mu}\frac{dt_1}{t_1}... \int_{\Lambda}^{t_{n-1}} \frac{dt_n}{t_n} \gamma^T(t_1)...\gamma^T(t_n) \bm{C}(\Lambda)
	\equiv \mathcal{T}\bigg\{ e^{\int_\Lambda^\mu \frac{dt}{t}\gamma^T(t)  }\bigg\} \bm{C}(\Lambda).
\end{align}
The coefficient matrix $\gamma$ is therefore also called anomalous dimension matrix (ADM) since the RGE running changes 
the naive scaling $C_i \sim \Lambda^{4-n_i}$ to (symbolically) $C_i\sim \Lambda^{4-n_i-\gamma}$.
The ADM depends on $\mu$ only through the renormalisable
couplings $\alpha_k(\mu)$. In principle, the RGEs for these couplings
also receive corrections from the higher-dimensional operators, however, under the above assumptions these corrections
can be neglected in the RGE of the Wilson coefficients. The running
couplings are therefore given by the usual results in QED and QCD (if
the EFT is LEFT, see Sec.~\ref{sec:genericeft})
or the SM (for SMEFT). In particular, for the following discussion we
will need the one-loop RGEs for LEFT given by
\begin{subequations}\label{eq:LEFT-gauge-RGE}
	\begin{alignat}{2}
		\mu\frac{d e}{d\mu} &= \frac{e^3}{12\pi^2} \Big(n_e +
                \tfrac{4}{3} n_u + \tfrac{1}{3} n_d\Big) &&\equiv
                \frac{\beta_e e^3}{16\pi^2} , \\
		\mu\frac{d g_s}{d\mu} &= \frac{g_s^3}{16\pi^2}
                \Big(\tfrac{2}{3}(n_u + n_d) - 11\Big) &&\equiv
                \frac{\beta_s g_s^3}{16\pi^2} ,
	\end{alignat}
\end{subequations}
where $n_e=n_d=3$ and $n_u=2$ count the number of charged fermions in the LEFT. In addition we also need the running fermion masses
and expressions for the wave-function renormalisations
\begin{align}\label{eq:LEFT-mmu-RGE}
	\mu\frac{d m_f}{d\mu} = -\frac{3m_f}{8\pi^2} \Big(Q_f^2 e^2 + C^f_F g_s^2\Big) = 6 m_f Z_f^{[1]}, 
\end{align}
where $C_F^e = 0$ and $C_F^{u,d}=\frac{4}{3}$.

\subsubsection{Photonic logarithms for purely heavy BSM scenarios}

\begin{figure}
	\centering
	\includegraphics{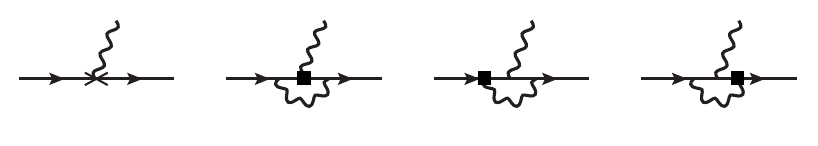}
	\caption{Diagrams contributing to the $\O_{e\gamma}$ self-mixing at one-loop order in LEFT.}
	\label{fig:LEFT-dipole-mixing}
\end{figure}

We first focus on a simple but important case, where new physics
contributions to $\amu$ arise from purely heavy one-loop diagrams. This is the
case for many 2- and 3-field models as well as e.g.~in supersymmetric
or leptoquark+top-loop diagrams. Counterexamples are contributions
from heavy $Z'$ or leptoquark+charm-loop diagrams. In this purely heavy
case, the dipole operator is the single relevant operator, and only
the corresponding diagonal entry of the ADM matters.
The corresponding important one-loop diagrams in the EFT are shown in
Fig.~\ref{fig:LEFT-dipole-mixing} and correspond to the dipole
self-mixing via photon loops. In the full BSM theory they correspond
to two-loop diagrams with a heavy one-loop $\amu$ subdiagram and an
additional photon loop, plus the appropriate renormalisation
\cite{vonWeitershausen:2010zr}, and the following computation will
yield the leading two-loop photonic logarithms.

Including the counterterm diagram, the total amplitude corresponding to Fig.~\ref{fig:LEFT-dipole-mixing} gives
\begin{align}
	i\Gamma^\mu = - L_{\underset{22}{e\gamma}} \bigg[ Z_{L_{e\gamma}L_{e\gamma}}^{[1]} + Z_{\O_{e\gamma}}^{[1]} - \frac{\alpha}{\pi} \bigg] \frac{2}{\epsilon} \sigma^{\mu\nu}q_\nu + \text{finite terms}.
\end{align}
According to Sec.~\ref{sec:EFTRGE}, the operator renormalisation $Z_{\O_{e\gamma}}=Z_\mu\sqrt{Z_A}=Z_\mu Z_e^{-1}$ is given by
\begin{align}
	Z_{\O_{e\gamma}}^{[1]} = Z_\mu^{[1]} - Z_e^{[1]} = - \frac{\alpha}{8\pi} \Big(2 + \beta_e\Big),
\end{align}
hence the resulting diagonal renormalisation constant and ADM element of the Wilson
coefficient must be (the complete dimension-6 ADM results  are listed in Ref.~\cite{Jenkins:2017dyc})
\begin{align}
	Z_{L_{e\gamma}L_{e\gamma}}^{[1]} = \frac{\alpha}{8\pi} \Big(10 + \beta_e\Big) \qquad \implies \qquad 
	\gamma_{\O_{e\gamma}\O_{e\gamma}} = \frac{5\alpha}{2\pi} +\frac{\alpha}{4\pi} \beta_e.
\end{align}
Integrating this equation gives an expression of the running
Wilson coefficient $L_{\underset{22}{e\gamma}}(Q)$ as a function of
the high-scale  $L_{\underset{22}{e\gamma}}(\Lambda)$ at the
leading-logarithmic level. Using  $\Damu = 4m_\mu / e(m_\mu)
L_{e\gamma}(m_\mu)$, we obtain the leading-logarithmic result
\begin{align}\label{eq:photoniclogs-Legamma}
          \Damu &=
m_\mu\left[4 \frac{L_{\underset{22}{e\gamma}}(\Lambda)}{e(\Lambda)}\right]
 \bigg\{ 1 + \frac{5\alpha}{2\pi}  \ln(\frac{m_\mu}{\Lambda}) + \O(\alpha^2)\bigg\},
\end{align}
where the running of $e(m_\mu)$ has cancelled the $\beta_e$ appearing
in the running of $L_{\underset{22}{e\gamma}}(m_\mu)$. We note in
passing that values of the ADM matrix and Wilson coefficients in
general can depend on the regularisation scheme \cite{Ciuchini:1993fk}
and we refer to the original literature for the scheme specifications
and Ref.~\cite{Gnendiger:2017pys} for a review. Physical results and the dipole coefficient
$L_{\underset{22}{e\gamma}}(\Lambda)$ in the purely heavy models, however, are
scheme independent.

This formula
can be applied e.g.~to the contributions of a heavy
leptoquark+top-loop to $\amu$, where the one-loop result can be
sketched as (see also the discussion in
Sec.~\ref{sec:overviewBSMscenarios} and Tab.~\ref{tab:estimates}) 
\begin{align}
  \Damu^{\text{LQ,1L}}  &= c\frac{m_\mu[m_t\lambda_L\lambda_R]}{M^2}
\end{align}
with a dimensionless constant $c$ which depends on details such as
loop functions and loop factors. Apart from the factor $m_\mu$ this
defines the matching value for the square bracket in
Eq.~(\ref{eq:photoniclogs-Legamma}), such that the leading-logarithmic
two-loop correction is
\begin{align}
  \Damu^{\text{LQ,1L+2LLL}}  &=
  \left.\Damu^{\text{LQ,1L}}\right|_{\text{$\overline{\text{MS}}$}}
 \bigg\{ 1 + \frac{5\alpha}{2\pi}  \ln(\frac{m_\mu}{\Lambda}) + \O(\alpha^2)\bigg\},
\end{align}
where the one-loop term, except the factor $m_\mu$, needs to be
evaluated in the ${\text{$\overline{\text{MS}}$}}$-scheme at the high
scale $\Lambda$.

A useful, slightly different way to write the same result has been
obtained in Ref.~\cite{Degrassi:1998es}. There, the WC
is normalised differently by pulling out a factor of the running muon
mass, effectively $C_{H_\mu}(\mu) = - \frac{16\pi^2}{e(\mu)m_\mu(\mu)}
\Re L_{\underset{22}{e\gamma}}(\mu)$. The running of this WC differs
from Eq.~(\ref{eq:photoniclogs-Legamma}) by the anomalous dimension of
the muon mass, such that
\begin{align}\label{eq:photoniclogs-CHmu}
          \Damu &=
-\frac{m_\mu^2}{4\pi^2}\left[C_{H_\mu}(\Lambda)\right]
 \bigg\{ 1 + \frac{4\alpha}{\pi}  \ln(\frac{m_\mu}{\Lambda}) + \O(\alpha^2)\bigg\}.
\end{align}
This formula is particularly naturally applicable in models, where a
factor $m_\mu^2$ arises in the one-loop 
computation of $\Damu$. An example is the MSSM, where the one-loop
contributions can be sketched as
\begin{align}
  \Damu^{\text{MSSM,1L}}  &= c\frac{m_\mu^2\tan\beta}{M^2}
\end{align}
with a dimensionless constant $c$ which also depends e.g.~on gauge
couplings and further model details. Then, the result  including
leading two-loop logarithms is the one
announced in Eq.~(\ref{DGlogs}), i.e.
\begin{align}
  \Damu^{\text{MSSM,1L+2LLL}}  &=
  \left.\Damu^{\text{MSSM,1L}}\right|_{\text{$\overline{\text{MS}}$}}
 \bigg\{ 1 + \frac{4\alpha}{\pi}  \ln(\frac{m_\mu}{\Lambda}) + \O(\alpha^2)\bigg\},
\end{align}
where now the one-loop contribution, except the factor $m_\mu^2$, must
be 
evaluated in the ${\text{$\overline{\text{MS}}$}}$-scheme at the high
scale $\Lambda$. 
These leading photonic two-loop logarithms have also been obtained in
an explicit diagrammatic two-loop computation in
Ref.~\cite{vonWeitershausen:2010zr}, where 
also the dependence on the renormalisation scheme was discussed and
additional non-logarithmic photonic corrections were obtained.

\subsubsection{More general logarithms}

In the more general case, not only the dipole operator but also
further LEFT operators are relevant, corresponding to 4-fermion
operators involving leptons and quarks and to 4-lepton operators.
The general  RGE for $L_{e\gamma}$ reads
\begin{align}
	\begin{split}
		\mu\frac{d }{d\mu} L_{\underset{22}{e\gamma}} = \frac{\alpha}{4\pi} \Big(10 + \beta_e \Big) L_{\underset{22}{e\gamma}}
		- \frac{e Q_d }{2\pi^2} N_c m_{d_i} L^{T,RR}_{\underset{22ii}{ed}} - \frac{e Q_u}{2\pi^2} N_c m_{u_i} L^{T,RR}_{\underset{22ii}{eu}}
		- \frac{e m_{e_i}}{8\pi^2} L^{S,RR}_{2ii2} + \O(L^2).
	\end{split}
\end{align}
Integrating this equation from the high scale $\Lambda$ to a scale $m_\mu$ gives the leading-logarithmic corrections 
\begin{align}\label{eq:LEFT-Legamma-running}
	\begin{split}
		L_{\underset{22}{e\gamma}}(m_\mu) &\simeq L_{\underset{22}{e\gamma}}(\Lambda) \bigg\{ 1 + \frac{\alpha}{4\pi} \Big(10+\beta_e\Big) \ln(\frac{m_\mu}{\Lambda}) + \O(\alpha^2)\bigg\} \\
		&\quad - \frac{e}{8\pi^2} \bigg\{ 4 N_c Q_d m_{d_i} L^{T,RR}_{\underset{22ii}{ed}}(\Lambda) + 4 N_c Q_u L^{T,RR}_{\underset{22ii}{eu}}(\Lambda)
		+ m_{e_i} L^{S,RR}_{2ii2}(\Lambda) + \O(\alpha) \bigg\} \ln(\frac{m_\mu}{\Lambda}) .
	\end{split}
\end{align}
It is instructive to compare this calculation to the  full LEFT
one-loop result including non-logarithmic terms from
Ref.~\cite{Aebischer:2021uvt}, which was discussed 
in Sec.~\ref{sec:genericeft}. 
The logarithmic contributions (up to logarithms between the light
scales) from Eq.~\eqref{eq:LEFT-amu-one-loop} are reproduced here by
identifying 
$\Lambda=\mu$. As in the context of
Eq.~(\ref{eq:photoniclogs-Legamma}), the running of $e$ cancels the
appearance of $\beta_e$ in Eq.~\eqref{eq:LEFT-amu-one-loop}.
If desired, Eq.~(\ref{eq:LEFT-Legamma-running}) can be extended to
include higher orders. In this way, Refs.~\cite{Degrassi:1998es,Czarnecki:2002nt} evaluated leading
logarithms for the EW SM contributions up to the 3-loop
level. Similarly, the formalism explained here can also be used to
understand the structure of the logarithms appearing in the two-loop 
Barr-Zee diagrams of Sec.~\ref{sec:Barr-Zee}, see Tab.~\ref{tab:BZ-asymptotes}.
If the particles running in the sub-loop are heavy ($x\to\infty$), the dimension-5 $\phi F_{\mu\nu}F^{\mu\nu}$ operator
obtained from integrating out the heavy particles mixes directly into $\O_{e\gamma}$ at one-loop.
Combined with Eq.~\eqref{eq:ADM-RGE} this leads to the two-loop $\ln^1$ terms.
On the other hand if the scalar is heavy ($x\to 0$), integrating out $\phi$ leads to a tree-level matching
condition for operators like $\bar{\mu}\mu |S|^2$, $\bar{\mu}\mu \bar{f}f$ and $\bar{\mu}\mu |V|^2$.
These either mix directly into $\O_{e\gamma}$ at two-loop leading again to a two-loop $\ln^1$ term or,
only in case of a light fermion loop, first into $\O^T_{ee}$ which then mixes into the dipole operator leading to
a two-loop $\ln^2$ term.

\section{Specific BSM scenarios}
\label{sec:Models}

In this section we focus on a large set of concrete and well-motivated
BSM scenarios. For each of them we explain the motivation, the theory
and the contributions to $\amu$, survey the literature and discuss
complementary constraints and viable parameter space. We begin with
models with light new particles in Sec.~\ref{sec:LightDarkSector}, which can be connected to
dark matter and larger dark sectors. The considered models include
dark photon, $Z'$ and ALPs models as well as models trying to explain
the existing discrepancies related to the hadronic vacuum
polarisation.

Section \ref{sec:SUSY} focuses on supersymmetry, both in its minimal
form and in alternative realisations  that change the
phenomenology. Supersymmetric models contain dark matter candidates
and lead to chiral enhancements in $\Damu$, hence there is an
important interplay between constraints from $\amu$, from dark matter,
and from the LHC, with significant recent progress.

The two-Higgs doublet model is discussed in Sec.~\ref{sec:2HDM},
allowing the well-known types I II, X, Y but also more general variants. For
$\amu$ this model is special since the main contributions arise
typically at the two-loop level, and recent LHC and flavour
observables significantly narrow down the viable parameter space.

Leptoquark models have been discussed frequently in the context of
$B$-physics anomalies, but also in the context of $\amu$ because of
potentially very large chiral enhancements. In
Sec.~\ref{sec:LQ} we describe cases with and without chiral
enhancement and determine parameter constraints from $\amu$ and
flavour physics.

Section \ref{sec:VLF} focuses on models with new fermions. Especially
vector-like leptons constitute a special class of models with
particularly strong connections between $\amu$ and the muon--Higgs
coupling. The constraints from the interplay between these and further
collider observables is explained for the complete set of such models
and some of their generalisations.

Finally, Sec.~\ref{sec:neutrino_mass_gm2} discusses a wide range of
neutrino mass models. The neutrino masses are generated at the
tree-level or the loop level, up to three-loop order.
These models involve many ideas and
elements discussed in other sections such as new gauge groups, new
fermions or new scalars; the new particles can be vector-like leptons,
leptoquarks, a second Higgs doublet, or dark matter
candidates. In many models the contributions to $\amu$ are strongly
chirally enhanced, such that $\amu$ places an important complementary
parameter constraint on the models. 

The phenomenology of $\amu$ has also been studied in additional
  BSM scenarios. In most cases, the general remarks made in
  Sec.~\ref{sec:Generic} on models with one, two, or three new fields apply 
and the relationships to other observables explained in
  Sec.~\ref{sec:Observables} provide a good understanding of many
  properties, but a dedicated model-specific discussion is beyond the scope of 
  this review. Examples are
  heavy new gauge bosons
  \cite{Huang:2001zx,
Gninenko:2001hx,
Murakami:2001cs,
Heo:2008dq,
Belanger:2015nma,
    Dasgupta:2023zrh,
    Kriewald:2022erk,
    Biswas:2021dan,
    Ashry:2022maw,
    Dcruz:2023mvf}
  including the 331-model
  \cite{    Pinheiro:2021mps,
    Li:2021poy,
    Hue:2021xap,Hue:2021zyw,
    Hong:2022xjg,
    Cherchiglia:2022zfy,
    CarcamoHernandez:2021tlv,
    Binh:2024lez,
    Hernandez:2021mxo,
      Hong:2024yhk,
      Doff:2024cap,
},
  studies of specific models with heavy dark matter particles
  beyond the discussions in
   Secs.~\ref{sec:MinimalBSM}, \ref{sec:DarkMatter} 
  \cite{Chen:2020tfr,
    Horigome:2021qof,
      Borah:2023dhk,
      Chowdhury:2021tnm,
      Bickendorf:2022buy,
  },
  models with extra dimensions, technicolour or compositeness
  \cite{Park:2001uc,Kim:2001rc,Agashe:2001ra,Calmet:2001si,
    Xiong:2001rt,Calmet:2001dc,Dai:2001vv,Yue:2001db,Appelquist:2001jz, 
    Das:2004ay,Tabbakh:2006zy,Blanke:2007db,
Hektor:2008xu,Hundi:2012uf,
Doff:2015nru,Hong:2016uou,Megias:2017dzd,
  Anchordoqui:2021lmm,Anchordoqui:2021vrg,
      Cacciapaglia:2021gff,
      Xu:2022one},
  non-commutative geometry, Lorentz violation and spin-2 particles
  \cite{Crivellin:2022idw,
      Lin:2021cst,
      Huang:2022zet,Huang:2022zop}. For detailed discussions we refer to the literature.

\subsection{Light Dark Sectors}\label{sec:LightDarkSector}

Some of the big open questions in fundamental physics are related to
the nature of dark matter and of neutrino masses. Answers could
possibly be related to light BSM particles which interact very weakly
with ordinary matter. In particular the evidence for dark matter may
not only point towards the existence of a single kind of dark matter
particle but also indicate a richer dark sector, possibly with a
number of different states and non-trivial interactions between
them. Such states may contain scalars, fermions and there may be dark
gauge sectors with dark vector bosons. The weakness of their
interactions with detectors allows their masses to be light,
i.e.\ below the GeV-scale. There is growing interest and activity in
detailed investigations of such light dark sectors, as reviewed in
Refs.~\cite{Jaeckel:2010ni,Essig:2013lka,Beacham:2019nyx,Agrawal:2021dbo,Antel:2023hkf,MartinCamalich:2025srw}. 

In connection with $\amu$ the possible existence of such light dark sectors is of
high interest. Even without chirality-flip enhancement, see
  Sec.\ \ref{sec:ChiralityFlips}, significant BSM contributions are possible provided
dark sector particles have some interactions with muons. Particular
classes of light dark sectors can therefore be candidates to explain a potential
large $\Delta\amu$ effect, and generally $\amu$ provides relevant
constraints on such dark sector parameter spaces. Similarly, electric
dipole moments provide constraints on dark sectors which violate CP \cite{Ardu:2024bxg}.

In general, dark sector states and their interactions are often
classified according to so-called ``portal interactions'' with the SM,
which are couplings of dark sector operators to gauge invariant
SM operators. In the notation of Sec.\ 1.2
of Ref.\ \cite{Agrawal:2021dbo},
\begin{align}
  \label{kineticportal}
  B^{\mu\nu}F'_{\mu\nu}&&&
  \text{kinetic mixing portal}
  \\
  \label{Higgsportal}
  (\Phi^\dagger \Phi)(AS+\lambda S^2) &&&
  \text{Higgs portal}
  \\
  \label{neutrinoportal}
  (\bar{l}\Phi)N&&&
  \text{neutrino portal}
  \\
  \label{gaugeportal}
  (\bar{\psi}\gamma^\mu\psi)A'_\mu&&&
  \text{generic gauge portal}
  \\
  \label{axionportal}
  (\bar{\psi}\gamma^\mu\psi)\partial_\mu a/f_a, (F\tilde{F})a/f_a &&&
  \text{axion-like portal}.
\end{align}
Here, the first factors are always gauge invariant SM operators,
$\psi$ is a generic SM fermion and $F$, $\tilde{F}$ a generic SM field
strength/dual field strength tensor. The dark sector fields are an abelian
gauge field $A'_\mu$, a neutral scalar $S$, a neutral fermion $N$, an axion-like
pseudoscalar $a$, and $A$, $\lambda$, $f_a$ are parameters.

Only specific dark sector particles and interactions can contribute to
$\amu$ in one-loop diagrams involving only light fields. New neutral
bosons of spin 0 or spin 1 can contribute if they couple to muons,
possibly in a lepton flavour-violating way with couplings of the form
boson--$\mu$--$e/\mu/\tau$; in contrast, new neutral fermions on their
own cannot give non-negligible contributions.  Here we will focus on several cases
of particular interest: the dark photon and kinetic mixing, a light
U(1)$_{L_\mu-L_\tau}$ gauge boson and the gauge portal, and axion-like
particles. We will briefly also comment on extensions and generalised
scenarios.
For  general studies of many neutral singlet particles with either spin
0 or spin 1 and the interplay of contributions to $\amu$ with
collider constraints and
we also refer to
Ref.~\cite{Capdevilla:2021kcf}, and for the interplay with quark
flavour physics to Ref.~\cite{Darme:2021qzw}.

\subsubsection{Analytical results for light-particle contributions}

\begin{figure}
	\centering
	\begin{subfigure}{.3\textwidth}
		\centering
		\includegraphics[width=.7\textwidth]{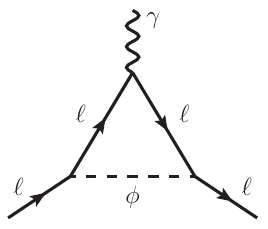}
		\caption{Scalar exchange}
	\end{subfigure}\hspace{2cm}
	\begin{subfigure}{.3\textwidth}
		\centering
		\includegraphics[width=.7\textwidth]{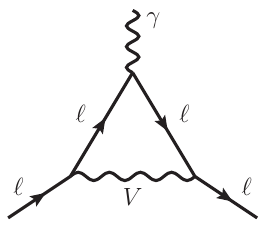}
		\caption{Vector exchange}
	\end{subfigure}
	\caption{Light dark sector one-loop contribution to $a_\ell$ from a neutral scalar $\phi$ or vector boson $V$.}
	\label{fig:lightbosondiagrams}
\end{figure}

Here we provide basic one-loop formulas for $\amu$ and the dipole
moments the other leptons for the case of light neutral bosons. We
specifically focus on the two kinds of 
diagrams in Fig.~\ref{fig:lightbosondiagrams}, with the exchange of
either a neutral vector boson $V^\mu$ with vector-like couplings or a
neutral real scalar $\phi$ with scalar and pseudoscalar couplings,
with masses $M_{V,\phi}$ respectively. 

We take the Lagrangians for the interactions with a SM lepton
$\ell=e,\mu,\tau$ as 
\begin{align}
 \La_{\text{int}}^V &= - gV^\mu\bar{\ell}\gamma_\mu \ell, \\
 \La_{\text{int}}^\phi &= - \lambda_S^{\ell}\phi\bar{\ell}\ell-i\lambda_P^{\ell}\phi\bar{\ell}\gamma_5 \ell.
\end{align}
The resulting contributions to $a_\ell$ corresponding to Fig.~\ref{fig:lightbosondiagrams} are given by\footnote{
In terms of the generic loop functions defined in Sec.~\ref{sec:genericoneloop} the results can be expressed as
\begin{align*}
	\Delta a_\ell^\text{V} &= 
	\frac{g^2}{16\pi^2} z_\ell \bigg\{ \mathcal{F}^\text{FV}(z_\ell,z_\ell;0) + 2 \mathcal{G}^\text{FV}(z_\ell,z_\ell;0) \bigg\} \\
	\Delta a_\ell^\phi &=  \frac{1}{16\pi^2} x_\ell \bigg\{ \Big[ \lambda_S^{\ell\,2} - \lambda_P^{\ell\,2}\Big] 
	\mathcal{F}^\text{FS}(x_\ell,x_\ell;0) + \Big[\lambda_S^{\ell\,2} + \lambda_P^{\ell\,2}\Big] 2\mathcal{G}^\text{FS}(x_\ell,x_\ell;0) \bigg\}
\end{align*}
where $x_\ell = m_\ell^2 / M_\phi^2$ and $z_\ell = m_\ell^2 / M_V^2$.}
\begin{subequations}
	\begin{align}
	    \label{eq:amulightneutralvector}
	    \Delta a_\ell^V &= \frac{g ^2}{8 \pi ^2}
	    \int_0 ^1 d x \frac{2 x (1-x)^2}{(1-x)^2 + r x}, \\
	    \label{eq:amulightneutralscalar}
	    \Delta a_\ell^\phi &= \frac{1}{8 \pi ^2}
	    \int_0 ^1 d x \frac{(1-x)^2}{(1-x)^2 + r x} \Big[ \lambda_S^{\ell\,2} (1+x)
	    - \lambda_P^{\ell\,2} (1-x)\Big],
	\end{align}
\end{subequations}
where $r = M_{V,\phi}^2 / m_\ell^2$ for the vector and scalar case respectively. 
Note that the vector and the scalar contributions are always positive while the
pseudoscalar contributions are always negative.
Analytic formulas for more
general cases with e.g.\ lepton-flavour violating couplings, or with
couplings of a muon to two different light new particles can be
readily obtained from Sec.~\ref{sec:genericoneloop}, see also
Ref.\ \cite{Lindner:2016bgg}, see also Ref.~\cite{Gamboa:2023mhv} for
possible additional correction terms appearing in certain cases.

It is instructive to consider the limiting cases
\begin{subequations}
	\begin{alignat}{2}
		\Delta a_\ell^V &= \,\frac{g^2}{8\pi^2} &&\begin{cases}
			1 & m_\ell \ll M_V \\
			\frac{2}{3 r} & m_\ell \gg M_V
		\end{cases} \label{eq:amulightneutralvectorlimit}\\
		\Delta a_\ell^\phi &= \frac{1}{16\pi^2} &&\begin{cases}
			3 \lambda_S^{\ell\,2} - \lambda_P^{\ell\,2} & m_\ell \ll M_\phi \\
			\frac{\lambda_S^{\ell\,2}}{3r} \big[6\ln(r)-7\big] - \frac{\lambda_P^{\ell\,2}}{3r} \big[6\ln(r) - 11\big] & m_\ell \gg M_\phi
		\end{cases} \label{eq:amulightneutralscalarlimit}
	\end{alignat}
\end{subequations}
The limit of heavy $M_{V,\phi}$ is similar to the limits considered in
Sec.\ \ref{sec:genericoneloop} and can be relevant for light particles
with masses around 1 GeV, while the
opposite limit is of interest for very light BSM particles. Here the behaviour differs, and the
contributions to $a_\ell$ approach mass-independent constants. The
light-vector limit in Eq.~(\ref{eq:amulightneutralvectorlimit})
corresponds to Schwinger's result $\alpha/2\pi$. A noteworthy general property of $\Damu$ in this limit 
is that naive scaling between the different lepton generations in the sense discussed in Sec.~\ref{sec:LeptonDipole} is typically violated. 
Concretely, in the case where the couplings are
lepton-generation independent, the ratio of contributions to the
electron and muon behave as
\begin{subequations}\label{Lightratio}
	\begin{numcases}{
		\frac{\Damu}{\Delta a_e} \simeq }
		1 & $M_{V,\phi}\ll m_e$ \label{Lightequalratio}\\
		\tfrac{m_\mu^2}{m_e^2}
		& $M_{V,\phi}\gg m_\mu$. \label{Lightnaivescaling}
	\end{numcases}
\end{subequations}
Here the behaviour corresponds to naive scaling for sufficiently heavy
masses; but the ratio becomes 1 for very small masses, and
it interpolates  between these two limiting values in the mass range
between the electron and muon masses.

In general, models with very light particles and
generation-independent couplings behaving as Eq.~\eqref{Lightequalratio} are
severely constrained by the electron $a_e$ value \eqref{aeExp} and therefore cannot
lead to observable contributions to $\amu$ within the current sensitivities. 
In contrast, models with particle masses between $m_e$ and $m_\mu$ or slightly larger than
$m_\mu$ can lead to sizeable effects, and their parameter space can be
constrained by $\amu$. 

\subsubsection{Dark Photons}
\label{sec:DarkPhoton}

A particularly simple and attractive kind of dark sector particle is
the dark photon. This term refers to a BSM vector boson $A'_\mu$ which
in a low-energy EFT only interacts via kinetic mixing \cite{Holdom:1985ag}
with the photon, as
\begin{align}
  \La_{\text{dark photon}} &= -\frac14 F^{\mu\nu}F_{\mu\nu}
  -\frac{\epsilon}{2} F^{\mu\nu}F'_{\mu\nu}
  -\frac14 F'{}^{\mu\nu}F'_{\mu\nu}-J^\mu_{\text{em}}A_\mu
  + \frac{M_{A'}^2}{2}A'{}^\mu A'_\mu.
\end{align}
Here $A_\mu$ and $A'_\mu$ are the ordinary photon and the new vector
field, the field strength tensors are defined accordingly, and
$J^\mu_{\text{em}}$ is the electromagnetic current to which the photon
couples. $M_{A'}^2$ is the dark photon mass parameter, which may arise
due to some Higgs mechanism associated with the dark gauge group. As
mentioned above, the only interaction between the dark sector and the
SM happens via the kinetic mixing term, which is a product of two
different gauge invariant abelian field strength tensors. Direct
couplings of SM quarks and leptons to $A'_\mu$ are assumed to be absent, i.e.~the
SM fermions are uncharged under the dark gauge group.

Via  the non-unitary  transformation $A_\mu\to A_\mu+\epsilon A'_\mu$,
the Lagrangian can be transformed to
\begin{align}
  \La_{\text{dark photon}} &= -\frac14 F^{\mu\nu}F_{\mu\nu}
  -\frac{1+\epsilon^2}{4}
  F'{}^{\mu\nu}F'_{\mu\nu}-J^\mu_{\text{em}}(A_\mu+\epsilon A'_\mu)
  + \frac{M_{A'}^2}{2}A'{}^\mu A'_\mu.
\end{align}
Thus we get, up to ${\cal O}(\epsilon^2)$-terms, canonically
normalised kinetic terms and no mixing between $A_\mu$ and
$A'_\mu$. After the redefinition, the $A'_\mu$ corresponds to
a  physical
mass-eigenstate which couples to the electromagnetic current with the
suppression factor $\epsilon$. I.e.\ it inherits all couplings of the
ordinary photon but multiplied by $\epsilon$.

The dark photon especially couples to muon and electron with the
equal coupling strength $\epsilon Q e$, leading to the behaviour
displayed in Eq.~(\ref{Lightratio}) and the resulting pattern of
naive-/non-naive scaling depending on $M_{A'}$.

Dark photon contributions to $g-2$ of the muon and to $g-2$ of the
electron have been first discussed in Ref.\ \cite{Pospelov:2008zw} and
further investigated in
Refs.\ \cite{Endo:2012hp,Davoudiasl:2012ag,Davoudiasl:2012qa,Davoudiasl:2012ig,Davoudiasl:2014kua}.

\begin{figure}[t]
	\centering
	\includegraphics[width=0.75\textwidth]{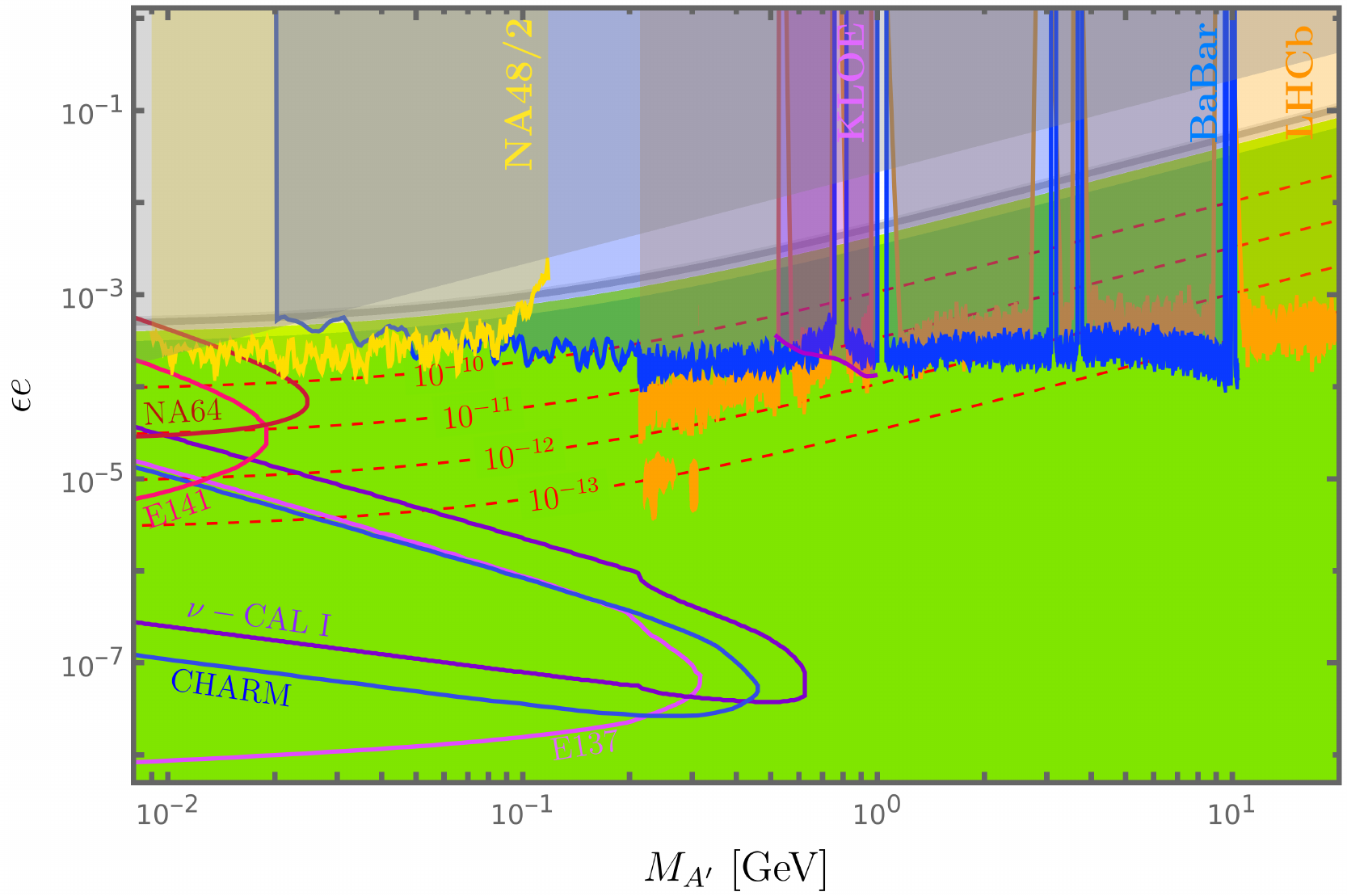}
	\caption{Constraints on the dark photon parameter space based on the recasting 
		performed in Ref.~\cite{Ilten:2018crw} (Fig.~4) and the accommodating code \texttt{DarkCast}.
		The 1$\sigma$ (2$\sigma$) contour of $\DamuFinal$ is shown in (light) green, while
		the (light) grey contour shows the 1$\sigma$ (2$\sigma$) range of $\DamuOld$ for reference.
	}
  \label{fig:darkphotonresult}
\end{figure}

Here we focus on the simplest original dark photon case and discuss
its current status. Fig.~\ref{fig:darkphotonresult} was
obtained using public data and code of
Ref.~\cite{Ilten:2018crw},
and it displays the two-dimensional parameter space $(\epsilon e,
M_{A'})$ of the dark photon and shows a set of constraints. The
constraints have been derived from a multitude of experiments as
indicated in the plot (for original references we refer to
Ref.~\cite{Ilten:2018crw}). We first focus only on $\amu$ and $a_e$.
The triangular shaded region at the top is
excluded by $g-2$ of the electron $a_e$. The shape is due to
the simple $m_e^2/M_{A'}^2$ behaviour of $\Delta a_e$ in this regime. The
contours for $\Damu$ interpolate between the $m_\mu^2/M_{A'}^2$
behaviour at high masses and approaching a constant at low masses. As
a result, and as explained after Eq.~(\ref{Lightratio}), the bounds on
$a_e$ automatically exclude sizeable contributions $\Damu$ for low
masses. For example, in the plot $\Damu>5\times10^{-10}$ is excluded by $a_e$ for
$M_{A'}\lesssim10$ MeV. Only higher dark photon masses are of
relevance for discussing $\amu$.

As the plot shows, other constraints now restrict the parameter space
more strongly than $a_e$. Values of $\Damu$ from dark photons
as large as the previous
deviation $\DamuOld$ were ruled already for some time.
Nevertheless, a wedge-like
region around $M_{A'}=10\ldots100~\text{MeV}$ and $\epsilon e=\O(10^{-4})$ is not
excluded. This is the region that allows the largest $\epsilon e/M_{A'}$ ratio
and hence the overall largest viable contribution to $\amu$ in the
dark photon scenario.

This maximum contribution, obtained from
Eq.~(\ref{eq:amulightneutralvector}) with $g=\epsilon e$, is
\begin{align}
  \label{darkphotonamumax}
  \Delta\amu^{\text{dark photon, }\text{max}} \approx 5\times 10^{-10} \qquad
  \text{ for } \qquad 
  M_{A'}\approx 20\,\text{MeV}\quad \text{and} \quad \epsilon e\approx 25\times10^{-5}.
\end{align}
Especially the NA48/2 and NA64 searches
 limit this parameter region. Smaller masses are excluded by NA64 \cite{NA64:2018lsq}, which searched for visible
$A'\to e^+ e^-$ decays with a particular $A'$ lifetime and decay
length  in beam dump
bremsstrahlung events. Larger couplings are excluded by  the NA48/2
experiment \cite{NA482:2015wmo}, which searched for the decay chain
$\pi^0\to\gamma A'$, $A'\to e^+e^-$, assuming that the dark photon
decays promptly, i.e.\ at the production point.

In most of the remaining parameter space the total contribution of the dark photon to 
$\amu$ is extremely small and thus fully compatible with the current value $\DamuFinal$ in Eq.~\eqref{eq:DamuFinal}.
Presently $\Damu$ therefore does not yield a complementary
constraint on the dark photon parameter space.
Conversely, if a smaller but non-zero deviation within experimental sensitivities should 
emerge from future progress on the SM prediction, $\Damu$ will provide a strong lower
bound on both $M_{A'}$ and $\epsilon e$ and the viable parameter regions will be tiny.

Dark photon contributions to $\amu$ have also been considered in
Refs.~\cite{Endo:2012hp,Davoudiasl:2012ag,Davoudiasl:2012qa,Davoudiasl:2012ig,Davoudiasl:2014kua,Chen:2015vqy,Mohlabeng:2019vrz,Tsai:2019buq,Ge:2021cjz,Datta:2022zng,Wojcik:2022woa,Abdallah:2023pbl,Gamboa:2023mhv,Abdullahi:2023tyk,Harigaya:2023uhg,Zhevlakov:2023jzt},
partly with the same results, partly in combinations with further BSM
states by which bounds can be weakened, $\Damu$ increased, or other
observables become relevant.

The straightforward dark photon scenario discussed so far can be
changed in several ways.
Most importantly, the dark photon is motivated in the
context of dark sectors involving also dark matter candidates. In this
context it is possible that the dark photon decays mainly invisibly,
\begin{align}
  A'\to \chi\chi
\end{align}
into a dark matter particle $\chi$.
This additional decay channel can significantly weaken the collider bounds shown in 
Fig.~\ref{fig:darkphotonresult} \cite{Davoudiasl:2014kua}. However, in this case other searches become
applicable and tend to yield even stronger exclusion limits. Most recently, 
a dedicated search for invisible $A'$ decays via missing-energy signatures was performed 
at the NA64 experiment \cite{NA64:2023wbi}. In the relevant mass range the constraints
on the dark photon coupling are actually stronger than the ones shown
in Fig.~\ref{fig:darkphotonresult}, hence the possible values of
$\Delta\amu$ for dark photons with invisible decays are even smaller
than Eq.~\eqref{darkphotonamumax}.

In general, the experimental bounds depend on the details of the dark sector.
For example, Refs.~\cite{Tsai:2019buq,Abdullahi:2020nyr,Abdullahi:2023tyk} found scenarios with extended dark sectors 
under which the bounds from both visible and invisible decays of the
dark photon can be circumvented, and where also neutrino masses can be
described. 
One possibility includes inelastic dark matter, where the dark photon decays dominantly
into two different dark-sector particles $A'\to \chi_1 \chi_2$ (with very similar masses) the heavier of which 
subsequently decays to $\chi_2\to\chi_1 e^+e^-$.
In this case the $A'$ decay is semivisible and the bounds on the parameter space are weaker.
Consequently, larger contributions to $\amu$ are possible and $\DamuFinal$ provides a complementary constraint in 
such models.

Similarly, in case the dark photon has more general couplings,
specifically if there is also a dark fermion $F$ which allows the
vertex $A'$--$F$--$\mu$, the contribution to $\amu$ from an $A'$--$F$
loop could be chirally enhanced by the potentially larger $F$-mass
\cite{Wojcik:2022woa} as
in the three-field models of
Sec.~\ref{sec:genericthreefield}.\footnote{%
Note that the current discussion is at the level of an effective
low-energy theory where only QED is relevant; in this case the
discussion of chirality flips simplifies compared to the one of
Sec.~\ref{sec:genericthreefield}.}

Another generalisation of the dark photon scenario is to allow mass
mixing via a $2\times2$ mass matrix involving the SM $Z$ boson and the
dark gauge boson, now called dark $Z$ in this setup. Such a mass mixing
can arise e.g.\ if the SM Higgs carries a dark charge and thus its
\vev contributes to the dark $Z$ mass
\cite{Davoudiasl:2012ag,Davoudiasl:2012qa,Davoudiasl:2012ig,Cadeddu:2021dqx}. 
The main effect is that the dark $Z$ couples to a linear combination of
currents $\epsilon J^\mu_{\text{em}}+\epsilon_Z J^\mu_{\text{NC}}$,
where $J^\mu_{\text{NC}}$ is the neutral current which couples to the
SM $Z$ boson. Again, this scenario is viable but does not lead to
significant changes in the conclusion for $\amu$ compared to
Fig.~\ref{fig:darkphotonresult}.

In light of the new $\DamuFinal$ value, the simplest dark photon model 
remains an appealing candidate for light new physics with viable parameter space as shown in
Fig.~\ref{fig:darkphotonresult}, 	while $\Damu$ now enters as an
important constraint in  	models with a larger dark sectors
where complementary constraints can be weakened.

\subsubsection{$Z'$ and U(1)$_{L_\mu-L_\tau}$}
\label{sec:LmuMinusLtau}

Now we turn to a wider class of a new neutral spin-1 gauge
bosons, generically called  $Z'$. In distinction 
from the previous case of the dark photon or dark $Z$, we now allow that
the SM fermions can be charged under the new gauge group. Hence
 there
can be direct couplings of the $Z'$ to fermions, i.e.\ couplings of the
form
(\ref{gaugeportal}) corresponding to the generic gauge portal, with
prefactors given by the assignments of the new quantum number
$X$. Kinetic mixing or mass mixing remains allowed and possible but is
typically subdominant.

The phenomenology depends on the gauge group and the $X$ charges. For
a large variety of cases, the couplings of electron and muon to $Z'$ are equal or at least of the same order. 
A prime example of a gauge group with similar couplings to all SM fermions
is U(1)$_{B-L}$ corresponding to the gauged baryon minus lepton number that often arises in Grand Unified
Theories (GUTs).
In such models the phenomenology is similar
to the one of the dark photon discussed in the previous section. In particular, the
correlation between $\amu$ and $a_e$ combined with the bounds from experiments sensitive to $Z'\to e^+e^-$
leads to similar constraints on the parameter space as
Fig.~\ref{fig:darkphotonresult}. Models of this kind have been studied
already in Ref.~\cite{Gninenko:2001hx}. More recent investigations
for $Z'$ with couplings to quarks and leptons can be found in
Refs.~\cite{Kang:2020gfi,Cen:2021ryk,Chowdhury:2022jde,Ferreira:2023buj,Ghosh:2023dgk},
where different charge assignments were used and different
complementary low-energy observables were studied. Similarly,
Refs.~\cite{Lee:2014tba,Nomura:2020dzw,Hooper:2023fqn,Verma:2024zav} studied $Z'$ models with couplings to muons and
electrons, which can also be interesting in view of neutrino physics,
and Ref.~\cite{Greljo:2022dwn} gave a general overview of such models.

The similarity between models with couplings to electrons, muons and
quarks of the same order and the dark photon scenario is
particularly demonstrated in Ref.~\cite{Ilten:2018crw}, where the same experimental
constraints are applied to a variety of models including the original
dark photon model as well as $Z'$ models based on various quantum number assignments. 
Here we focus on  $Z'$ models that differ from such scenarios, 
i.e.\ in particular where the couplings to the electron and muon are noticeably different.

In general, the choice of the quantum number $X$ is constrained by the absence of
gauge anomalies. It turns out that this requirement is very
restrictive. For instance, U(1)$_{B-L}$ only becomes anomaly free once
three generations of right-handed 
neutrinos are added to the SM fermion content. Without adding BSM
fermions, all viable options for U(1)$_X$ involve generation-dependent
quantum numbers \cite{Foot:1990mn}. It has been found that viable options are
the quantum numbers
\begin{align}
  X\quad=\quad L_\mu-L_\tau,\quad L_\mu-L_e\quad \text{or} \quad L_e-L_\tau,
\end{align}
where $L_i$ corresponds to lepton number of generation $i$. Such gauge groups are also of interest in
view of flavour model building, and they are of particular interest
here, in view of their distinct properties.

Among these, $L_\mu-L_\tau$ is special since it implies no direct
$Z'$-coupling to electrons and quarks, such that many of the usual experimental constraints
are evaded. Similar models with slightly different gauge groups which
nevertheless only allow couplings to muons and $\tau$-leptons were
studied in Refs.~\cite{Araki:2015mya,Altmannshofer:2016brv}.
In the following we therefore focus on the U(1)$_{L_\mu-L_\tau}$
gauge group and the corresponding $Z'$ contributions to $\amu$. These
were first pointed out to be potentially significant in
Refs.~\cite{Baek:2001kca,Ma:2001md,Heeck:2011wj}, where also appropriate parameter
constraints and links to neutrino masses were discussed. The
relevant Lagrangian is given by
\begin{align}
  \La_{\text{int, U(1)$_{L_\mu-L_\tau}$}} &\supset
  -g_{Z'}Z'_\mu\left(
  \bar{\mu}\gamma^\mu\mu-\bar{\tau}\gamma^\mu\tau+\bar{\nu}_\mu\gamma^\mu\nu_\mu
  -\bar{\nu}_\tau\gamma^\mu\nu_\tau
  \right)
  +\frac{M_{Z'}^2}{2}Z'{}^\mu Z'_\mu
  ,
\end{align}
where the $Z'$ mass  originates as
$M_{Z'}=g_{Z'}v'$ from some unspecified Higgs mechanism
which spontaneously breaks U(1)$_{L_\mu-L_\tau}$. 
The contribution to $\amu$ is given by Eq.~\eqref{eq:amulightneutralvector} and simplifies to the following expression
for masses $M_{Z'}\gg m_\mu$,
\begin{align}
  \Damu^{Z'} &=
  \frac{1}{12\pi^2}\frac{g_{Z'}^2m_\mu^2}{M_{Z'}^2}
  =
  \frac{1}{12\pi^2}\frac{m_\mu^2}{v'{}^2}
  ,
\end{align}
where the \vev $v'$ remains as the only relevant model parameter.

\begin{figure}
	\centering
	\includegraphics[width=.3\textwidth]{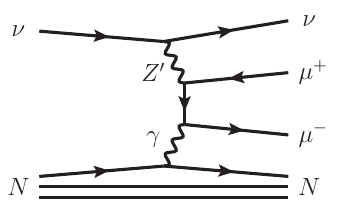}
	\caption{Tree-level $Z'$ contribution to the \emph{neutrino trident} production of a $\mu^+\mu^-$ pair
	in the scattering of neutrinos on nucleons.	}
	\label{fig:neutrinotridentdiagram}
\end{figure}

Refs.~\cite{Altmannshofer:2014cfa,Altmannshofer:2014pba} identified
the so-called neutrino trident process as a very important constraint
on this model. The process is illustrated by the diagram in
Fig.~\ref{fig:neutrinotridentdiagram} and corresponds to
nucleon--neutrino scattering where the $Z'$ interaction produces a
$\mu^+\mu^-$ pair. In the large-mass region $M_{Z'}\gg m_\mu$, the correction
to the trident production cross section results in the simple formula \cite{Altmannshofer:2014cfa}
\begin{align}
  \frac{\sigma(\nu N \to \nu N + \mu^+\mu^-)}{\sigma(\nu N \to \nu N + \mu^+\mu^-)_{\text{SM}}}
    &\approx \frac{1+\left(1+4s_W^2+2v^2/v'{}^2\right)^2}{1+\left(1+4s_W^2\right)^2}.
\end{align}
This process has been measured in several experiments including Charm-II \cite{CHARM-II:1990dvf}, CCFR \cite{CCFR:1991lpl}
and NuTeV \cite{NuTeV:1998khj}, yielding the combined
limit $\sigma_{\text{Exp}}/\sigma_{\text{SM}}=0.83\pm0.18$ obtained in Ref.~\cite{Altmannshofer:2014cfa}.
This limit results in a strong lower bound on $v'$ given by
\begin{align}\label{Zpvplimit}
  v'\gtrsim 750~\text{GeV}.
\end{align}
Even though there has been discussion about the applicability of
various  neutrino trident experiments \cite{Altmannshofer:2014pba}
it is clear that this bound severely limits the possible $Z'$ contributions
to $\amu$. If $M_{Z'}\gg m_\mu$ (such that Eq.~(\ref{Zpvplimit})
holds), the contributions are as small as
\begin{align}\label{DamuZpmax}
  \Damu^{Z'} &< 1.7\times10^{-10}
\end{align}
However, for smaller $Z'$ masses the behaviour of the neutrino trident
cross section is different (see Ref.~\cite{Altmannshofer:2014pba} for the full calculation),
and larger contributions to $\amu$ become possible.

\begin{figure}[t]
	\centering
	\includegraphics[width=.75\textwidth]{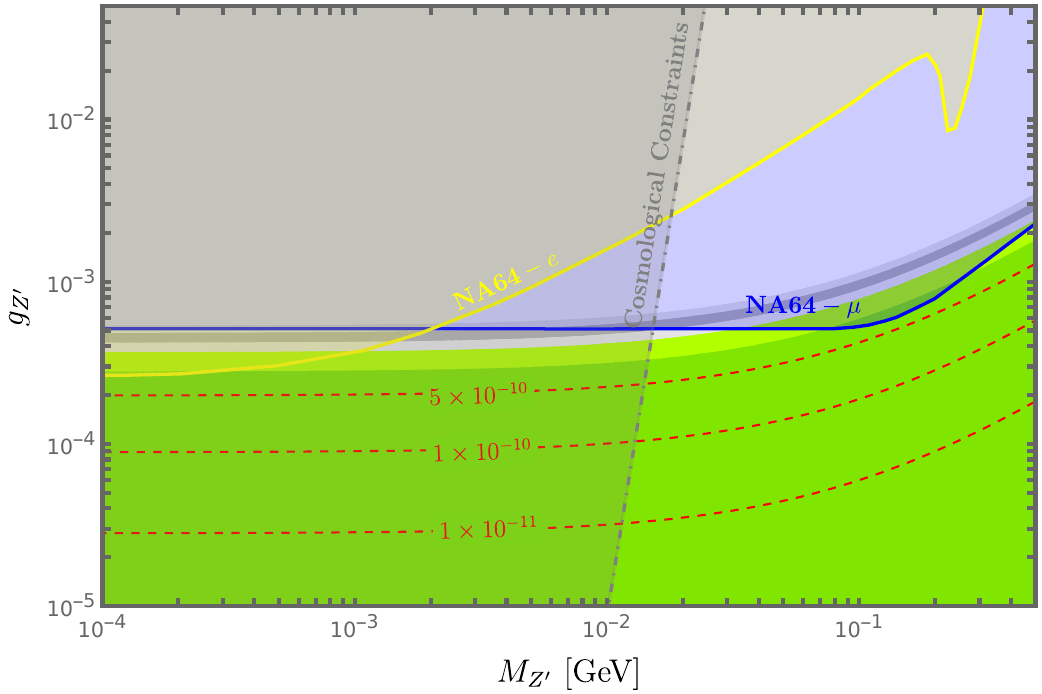}
	\caption{ The $M_{Z'}$ -- $g_{Z'}$ plane of the
          U(1)$_{L_\mu-L_\tau}$ model. Regions excluded by NA64-e,
          NA64-$\mu$ and cosmological bounds are shown in yellow, blue
          and light grey, respectively.
		The (light) green area shows the 1$\sigma$ (2$\sigma$) range of $\DamuFinal$, while the (light) grey band
		shows the 1$\sigma$ (2$\sigma$) range of the old deviation $\DamuOld$.}
	\label{fig:LmuminusLtau}
\end{figure}

After Refs.~\cite{Heeck:2011wj,Altmannshofer:2014cfa,Altmannshofer:2014pba},
the region of lower $M_{Z'}$ has been investigated,
a multitude of further constraints on the model have been identified
and dedicated experiments have been carried out. An overview is given
in Ref.~\cite{Bauer:2018onh}.

The resulting parameter space for $M_{Z'}\lesssim 1$ GeV is shown in
Fig.~\ref{fig:LmuminusLtau}, together with the strongest current exclusion limits. 
One important constraint comes from searches for invisible $Z'$ decays via missing-energy
signatures. Even without extended dark sectors, invisible decay channels like $Z'\to \nu\bar\nu$ to
muon- or tau-neutrinos are guaranteed to have a large branching ratio. The two direct search bounds obtained by NA64-$e$ \cite{Andreev:2024lps} 
and NA64-$\mu$ \cite{NA64:2024nwj} are shown in Fig.~\ref{fig:LmuminusLtau}.
These experiments measured the invisible decay rates of $Z'$ bosons produced as bremsstrahlung 
in high-energy scattering of electron- or muon-beams off a fixed target.
The resulting bounds apply similarly to a
wide range of $Z'$ scenarios with invisible decays. 
In the region with light $M_{Z'}$ shown in the plot these bounds are stronger than the ones
obtained from the neutrino trident process discussed above.
In combination these experiments already excluded most of the light
$Z'$ parameter space with large $\Damu$,
leaving only a small region around $M_{Z'}\sim \O(1~\text{MeV})$ where the deviation could have
been as large as $\DamuOld$.

However, this region is mostly excluded by a number of constraints
stemming from cosmology and astrophysics. 
Due to its properties, a light $Z'$ can have a significant impact on both
the thermal evolution of the universe as well as many astrophysical phenomena 
that have been observed and measured with great precision over the past decades.

For instance, the temperature of white dwarfs decreases over time resulting in characteristic 
luminosity functions that have been extracted from the Sloan Digital Sky Survey \cite{Harris:2005gd,DeGennaro:2007yw}.
The white dwarf cooling is well understood and simulations based on the SM are in good agreement with
the data. The $Z'$ contributes to this cooling through the decay of plasmons (electron excitations in the white dwarf plasma)
into neutrinos \cite{Dreiner:2013tja}. 
This constraint is in particular relevant for small masses \cite{Bauer:2018onh,Foldenauer:2024cdp}. 
For derivations and surveys of astrophysical limit from e.g. solar neutrinos, stellar cooling or the supernova 1987A neutrino burst 
we refer to Refs.~\cite{Amaral:2020tga,Amaral:2021rzw,Escudero:2019gzq,Manzari:2023gkt}. 

Another very strong cosmological constraint on $Z'$ bosons with masses in the MeV-range 
stems from its interactions with the neutrinos in the early universe.
In this case the $Z'$ is still present in the thermal bath around the time
of neutrino decoupling and can have a significant impact on the evolution of
the radiation energy density. This density is usually parametrised in terms of
the effective number of neutrinos $N_\text{eff}$ which is determined from observations of the
cosmic microwave background \cite{Planck:2018vyg} and given by
\begin{align}
	N_\text{eff} = 2.99 \pm 0.17
\end{align}
The $Z'$ results in a small positive correction $\Delta N_{\text{eff}}$ 
that in turn translates into a stringent lower bound on $M_{Z'}\gtrsim 10$ GeV 
\cite{Escudero:2019gzq}\footnote{Interestingly, for small shifts around $\Delta N_\text{eff} \sim 0.2\cdots0.5$ this correction might 
help alleviate the Hubble tension \cite{Escudero:2019gzq,Drees:2021rsg,DiValentino:2021izs}, but larger $\Delta N_{\text{eff}}$ are excluded.}.
In Fig.~\ref{fig:LmuminusLtau} this constraint excludes most of the remaining parameter space with a deviation
as large as $\DamuOld$. We refer to Refs.~\cite{Gninenko:2014pea,Altmannshofer:2016oaq,Patra:2016shz,Biswas:2016yjr,Gninenko:2018tlp,Carpio:2021jhu,Hapitas:2021ilr,Drees:2021rsg,Cheng:2021okr,Nagao:2022osm,Ganguly:2022qxs,Wang:2023pnd,Paul:2024dsh,Asai:2024pzx} for further
phenomenological studies of $\amu$ in the U(1)$_{L_\mu-L_\tau}$-model, partially
including also additional fields to describe dark matter.

Taking into account the above constraints the maximum value of $\Damu$ allowed in the plot is
\begin{align}\label{DamuLmuLtaumax}
	\Delta\amu^{Z',\text{max}} \approx 20\times10^{-10}
	\qquad \text{for} \qquad
	M_{Z'}\approx20\,\text{MeV}\quad \text{and} \quad  g_{\mu\tau}\approx5\times10^{-4}.
\end{align}
Smaller values of the mass $M_{Z'}$ are excluded by cosmology, for the
relevant coupling strength; for larger masses than shown in the plot the neutrino
trident and further collider constraints take over, see
Eq.~(\ref{DamuZpmax}) and Refs.~\cite{Huang:2021nkl,Brown:2024jah}
for further collider investigations of the model.

The plot confirms that in this U(1)$_{L_\mu-L_\tau}$ model the
contributions to  $\amu$ can be larger than for the dark
photon in the mass region below $m_\mu$. The new result $\DamuFinal$
therefore now disfavours the 
parameter region given in Eq.~(\ref{DamuLmuLtaumax}) and thus constitutes
an important novel constraint on the model.
Still, the parameter region where such large $\Damu$ is possible is very
narrow and specific.  For all
other values of the $Z'$ mass, the constraints from NA64-$e,\mu$ and
from cosmology are stronger than the one from $\DamuFinal$.

The plot and its discussion cover the basic (\emph{vanilla})
$L_\mu-L_\tau$ model. As stated in the beginning, the gauge group and
its properties are appealing, and the model can be extended in many
motivated ways. In such a larger framework, the interplay between the
various bounds can also change. Like for dark photons, an obvious
extension is to include a dark matter candidate particle $\chi$ to the
theory, where $\chi$ can interact with the $Z'$. Here constraints arise from
the need to have sufficiently high dark matter annihilation
cross section to avoid overabundance of dark matter while avoiding too
large $\Damu$. 
A naive estimate of the scaling for the dark matter annihilation cross section (see Sec.~\ref{sec:DarkMatter}) 
and $\Damu$ gives
\begin{align}
  \ev{\sigma v}\propto \frac{g_\chi^4}{m_\chi^2} \qquad \text{while} \qquad
  \Damu\propto g_\mu^2
\end{align}
where $g_{\mu,\chi}$ denote the $Z'$ couplings to the muon and the dark matter particle and $m_\chi$
the dark matter mass. If one argues that in a gauge theory it is natural that $g_{\mu,\chi}$
are of the same order of magnitude, typically the cross section is too
low or $\Damu$ too high
\cite{Foldenauer:2018zrz,Holst:2021lzm}. Hence special regions of
parameter space, and/or special dark matter models are needed. In
Ref.~\cite{Holst:2021lzm}, resonant annihilation is considered, which
happens in the specific case where $M_{Z'}\approx 2m_\chi$. In
Ref.~\cite{Qi:2021rhh} co-annihilation of two scalars with similar
masses is considered. And Ref.~\cite{Figueroa:2024tmn} accepts strongly non-universal couplings 
$g_\chi\gg g_\mu$ while allowing arbitrary dark matter masses. The
cosmological bounds on the scenario with dark matter have also been
evaluated \cite{Drees:2021rsg}, however, there is a subtle
dependence on the value of the kinetic mixing which leads to small but
non-zero couplings of the $Z'$ to electrons \cite{Hapitas:2021ilr}.

Interestingly, the fact that $\DamuFinal$ is now compatible
with zero and places a stringent upper limit  on the
coupling $g_\mu$ (while $\DamuOld$ preferred a non-vanishing value)
exacerbates this tension between dark matter  and $\amu$ in this kind
of models. For a viable explanation of dark matter in this context,
the construction of special scenarios like in the mentioned
references has therefore become even more important.
Such a more involved scenario is considered in
Ref.~\cite{Kawamura:2020qxo}, where dark matter is self-interacting
and the viable parameter space opens up.

It is also possible to extend the $L_\mu-L_\tau$ model to include
neutrino masses \cite{Heeck:2011wj} in the parameter region of
sizeable $\Damu$
\cite{Biswas:2016yan,Asai:2018ocx,Greljo:2021npi,Costa:2022oaa}, and such extensions
lead also to additional constraints
\cite{Borah:2020jzi,Borah:2020swo,Borah:2021khc,Borah:2021jzu,Borah:2021mri,Singh:2022tvz,Eijima:2023yiw,Ibe:2025rwk}.
Here,
Refs.~\cite{Asai:2018ocx,Costa:2022oaa} propose variants of the
tree-level seesaw mechanism to generate neutrino masses, where
partially Majorana mass terms of the kind $N_e N_\mu$ must be obtained
from a \vev of a U(1)$_{L_\mu-L_\tau}$-breaking new Higgs field which
also gives mass to the $Z'$. These setups find no tension between
explaining neutrino masses and $\Damu$. In contrast,
Ref.~\cite{Borah:2021khc} combines the $Z'$ with the scotogenic model
for neutrino masses and finds a tension between $\mu\to e\gamma $ and
$\Damu$ of the kind discussed in Sec.~\ref{sec:LeptonDipole},
specifically the single-particle scenario.

\subsubsection{Axion-like particles and other light scalar particles}
\label{sec:ALPs}

Here we discuss the phenomenology of light new spin-0 particles in
view of $\amu$. Such spin-0 particles could couple to muons via
renormalisable couplings of the form described in
Sec.~\ref{sec:Generic}, such that the general statements of that section
apply. Here we mainly focus on axion-like particles (ALPs),
which are a particularly well motivated type of light dark sector
particles with different kinds of couplings, and we will come back to
generic scalars at the end of this section.

ALPs are excitations of a neutral pseudoscalar field $a$ which has only effective 
(nonrenormalisable) interactions of the axion-portal type \eqref{axionportal} with the SM. 
The properties are modelled after the original axion proposed to solve the strong CP problem by
dynamically generating a vanishing coefficient in front of the $G_{\mu\nu}\tilde{G}^{\mu\nu}$ 
term in the effective Lagrangian via a potential for the axion \cite{Peccei:1977hh,Wilczek:1977pj,Weinberg:1977ma}. 
Since this proposal, many phenomenological models have been constructed and gained attention also
in the context of $\Damu$.

A crucial ingredient in the original formulation was the shift symmetry $a(x)\to a(x)+\text{const}$.
The interaction terms allowed in the general ALP Lagrangian  are required to
have the same shift symmetry, even though the ALPs mass and coupling
strengths are not required to solve the strong CP problem.
Using the conventions of e.g. Refs.~\cite{Buen-Abad:2021fwq,Neubert:2024jal}, the relevant low-energy 
Lagrangian can then be written as 
\begin{align}\label{ALPsLagrangian}
	\La_{\text{ALP}} &\supset
	\frac{c_{ii}}{2}
	\frac{\partial_\mu a}{f_a}
	\left(\bar{e}_i\gamma^\mu\gamma_5 e_i\right)
	+
	c_{\gamma\gamma}\frac{\alpha}{4\pi}
	\frac{a}{f_a}
	F_{\mu\nu}\tilde{F}^{\mu\nu}
	-\frac{m_a^2 a^2}{2} 
	+\ldots.
\end{align}
Here $F^{\mu\nu}$ and $\tilde{F}^{\mu\nu}=\frac12\epsilon^{\mu\nu\rho\sigma}F_{\rho\sigma}$ 
are the photon field strength tensor and its dual.
The hallmark of such theories are effective $a\gamma\gamma$
interaction vertices and derivative interactions with the
leptons. Similar terms involving other fermions or the gluon field
strength tensor $G^{\mu\nu}$ exist but are omitted here. All these terms correspond to non-renormalisable
dimension-5 operators. By convention, the Wilson coefficients are
written via dimensionless constants $c_{\gamma\gamma}$ and $c_{ii}$,
pulling out a mass  scale $f_a$. Further, a loop factor has been
pulled out of  $c_{\gamma\gamma}$, taking into account that in fundamental theories the
corresponding term will typically be loop-suppressed
compared to the $c_{ii}$ term, even though enhancements are possible
\cite{Bauer:2020jbp}. Here we have neglected the possibility of lepton flavour
violating axion couplings which we will come back to at the end of this section.

The Lagrangian in Eq.~\eqref{ALPsLagrangian} can be simplified by using
equations of motion of the renormalisable part of the theory \cite{Buen-Abad:2021fwq,Bauer:2017ris},
such that the interaction terms can equivalently be written as
\begin{align}
	\label{ALPsLagrangiansimple}
	\La_{\text{int, ALP}} &=
	-{c_{ii}}
	\frac{m_i }{f_a}
	a\left(\bar{e}_ii\gamma_5 e_i\right)
	+
	\tilde{c}_{\gamma\gamma}\frac{\alpha}{4\pi}
	\frac{a}{f_a}
	F_{\mu\nu}\tilde{F}^{\mu\nu}
	+\ldots,
\end{align}
where the shifted $a\gamma\gamma$ coupling parameter is given by
$\tilde{c}_{\gamma\gamma}=c_{\gamma\gamma}+\sum_i Q_i^2c_{ii}$ \cite{Neubert:2024jal}. The sum here 
extends over all SM 
fermions, although in the following we assume the quark couplings to vanish. 
The appearance of the lepton mass prefactors in the 
ALPs--lepton couplings arises naturally from the shift symmetry of the
fundamental Lagrangian, combined with the equations of motion.
Interestingly for discussing $a_\mu$ and $a_e$, the conventions are
such that naive scaling simply corresponds to flavour-independent
Wilson coefficients $c_{\mu\mu}\approx c_{ee}$. 

\begin{figure}
	\centering
	\begin{subfigure}{.19\textwidth}
		\centering
		\includegraphics[width=.9\textwidth]{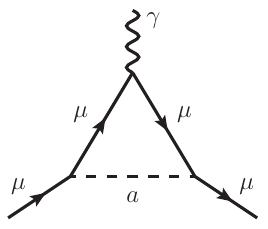}
		\caption{}
	\end{subfigure}\hfill
	\begin{subfigure}{.19\textwidth}
		\centering
		\includegraphics[width=.9\textwidth]{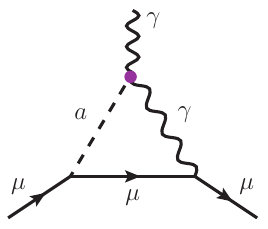}
		\caption{}
	\end{subfigure}\hfill
	\begin{subfigure}{.19\textwidth}
		\centering
		\includegraphics[width=.9\textwidth]{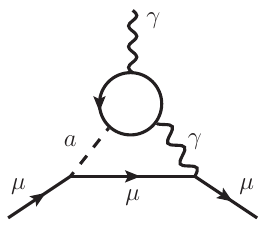}
		\caption{}
	\end{subfigure}\hfill
	\begin{subfigure}{.19\textwidth}
		\centering
		\includegraphics[width=.9\textwidth]{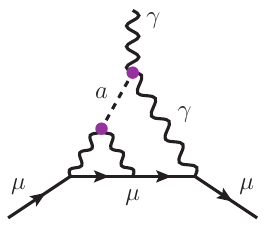}
		\caption{}
	\end{subfigure}\hfill
	\begin{subfigure}{.19\textwidth}
		\centering
		\includegraphics[width=.9\textwidth]{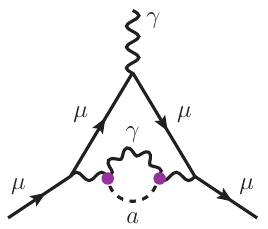}
		\caption{}
	\end{subfigure}\hfill
	\caption{ALP one- and two-loop contributions to $\amu$. 
		The purple vertex corresponds to the effective $a\gamma\gamma$ coupling.}
	\label{fig:ALPsdiagrams}
\end{figure}

ALPs generally contribute to $\amu$ via the diagrams of
Fig.~\ref{fig:ALPsdiagrams}.
Such contributions were  discussed in
Ref.~\cite{Marciano:2016yhf}, where diagrams (a,b,d,e) were
computed. The calculations were then refined in
Refs.\ \cite{Bauer:2017ris,Bauer:2020jbp,Buen-Abad:2021fwq,Bauer:2021mvw,Galda:2023qjx,Neubert:2024jal}. 
Note that the  EFT analyses used for the definition of the Lagrangian in Eq.~\eqref{ALPsLagrangiansimple}
also suggested that the $c_{\gamma\gamma}$-term is loop suppressed; hence
diagrams (b,c) should be treated of equal order and diagrams (d,e) as loop suppressed in comparison. 
The result for diagram (a) can be obtained 
from Eq.~\eqref{eq:amulightneutralscalar}, and the result for diagram
(b) reads (using the form of Refs.\ \cite{Marciano:2016yhf,Buen-Abad:2021fwq})
\begin{align}
  \Delta\amu^{a\gamma\gamma} &=
  -  \frac{m_\mu^2 c_{\mu\mu}\tilde{c}_{\gamma\gamma}\alpha}{8\pi^3 f_a^2}
  \left(\ln\frac{\Lambda^2}{m_\mu^2}
+\frac72 -h_2(r)\right)
  ,
  \intertext{where the loop function is given in terms of
    $r=m_a^2/m_\mu^2$ as}
  h_2(r) &=
1+\frac{r^2}{6}
  \ln r
  -\frac{r}{3}
  -\frac{r+2}{3}\sqrt{r(r-4)}
  \ln\left(\frac{\sqrt{r}+\sqrt{r-4}}
  {2}\right).
\end{align}
Since the $a\gamma\gamma$ coupling is nonrenormalisable, the diagram
is divergent. The result shown here is obtained using a momentum cutoff
$\Lambda$, which is assumed to be of the order TeV. In a proper EFT
setup, the logarithmic dependence can be understood via
renormalisation-group methods \cite{Neubert:2024jal,Galda:2023qjx}. 
The divergence is cancelled by the renormalisation of the dipole
operator corresponding to $\amu$, and the cutoff-dependence is
replaced by a dependence on an unknown further Wilson coefficient for
the dipole operator \cite{Bauer:2017ris}.

The diagram (c) is a two-loop Barr-Zee diagram which however needs to be
evaluated in the limit of very light $a$, which cannot be readily
obtained from the formulas in Sec.~\ref{sec:Barr-Zee}. The
results are provided in Refs.~\cite{Buen-Abad:2021fwq,Neubert:2024jal}.
Since diagrams (b) and (c) are formally of the same order, it is
expected that two-loop diagram is important as well. This is indeed
confirmed by the explicit calculations performed in Refs.~\cite{Buen-Abad:2021fwq,Neubert:2024jal,Galda:2023qjx}.
However, the main effect of the
Barr-Zee diagram (c) can be easily understood. In fact, the contribution of the Barr-Zee diagram (c) can essentially be absorbed in diagram (b) 
by replacing the $a\gamma\gamma$ coupling $\tilde{c}_{\gamma\gamma}$ with an effective coupling
$\tilde{c}_{\gamma\gamma}^{\text{eff}}$ that takes into account the inner
fermion-loop correction to the $a\gamma\gamma$ vertex~\cite{Neubert:2024jal}. The shift is not negligible and has been quantified in the same reference as a function of model parameters.

The interplay between the different contributions to $\amu$ is important. 
The phenomenological behaviour of the diagrams (a) and (b+c)
can be concisely summarised as \cite{Cornella:2019uxs,Buen-Abad:2021fwq}
\begin{align}\label{ALPsoverview}
  \Damu^{a\mu\mu} & \propto-\frac{c_{\mu\mu}^2}{16\pi^2},
  &
  \Delta\amu^{a\gamma\gamma}
  &\propto-\frac{c_{\mu\mu}\tilde{c}^{\text{eff}}_{\gamma\gamma}\alpha}{
    16\pi^3} .
\end{align}
As mentioned below Eq.~\eqref{eq:amulightneutralvector} the pseudoscalar FS-type one-loop diagram (a) 
with two ALP--muon couplings is always negative, whereas
the diagrams (b) and (c) with effective $a\gamma\gamma$-coupling can have either sign depending on the relative sign of the two couplings.
In particular, to obtain a positive contributions to $\amu$ diagrams (b+c) involving the non-renormalisable 
$\tilde{c}^{\text{eff}}_{\gamma\gamma}$-term must be positive and dominate, which is possible for large
logarithmic enhancements from $\Lambda\gg m_\mu$ and large couplings $|\tilde{c}_{\gamma\gamma}|\gtrsim |c_{\mu\mu}|$.

The general scaling in Eq.~\eqref{ALPsoverview} means that a large and
positive contribution $\Damu$ as motivated in the past by $\DamuOld$
is not trivial to achieve. In contrast, when the negative deviation
$\Delta a_e$ corresponding to Eq.~\eqref{aeSM2018} was reported around 2018,
one-loop ALP contributions were considered one of the plausible explanations. 
Combining this with positive contributions to $\amu$ was possible via diagram (b) or via
lepton-flavour violating couplings
\cite{Davoudiasl:2018fbb,Cornella:2019uxs,Bauer:2019gfk,Buttazzo:2020vfs}.

\begin{figure}
	\centering
	\includegraphics[width=.8\textwidth]{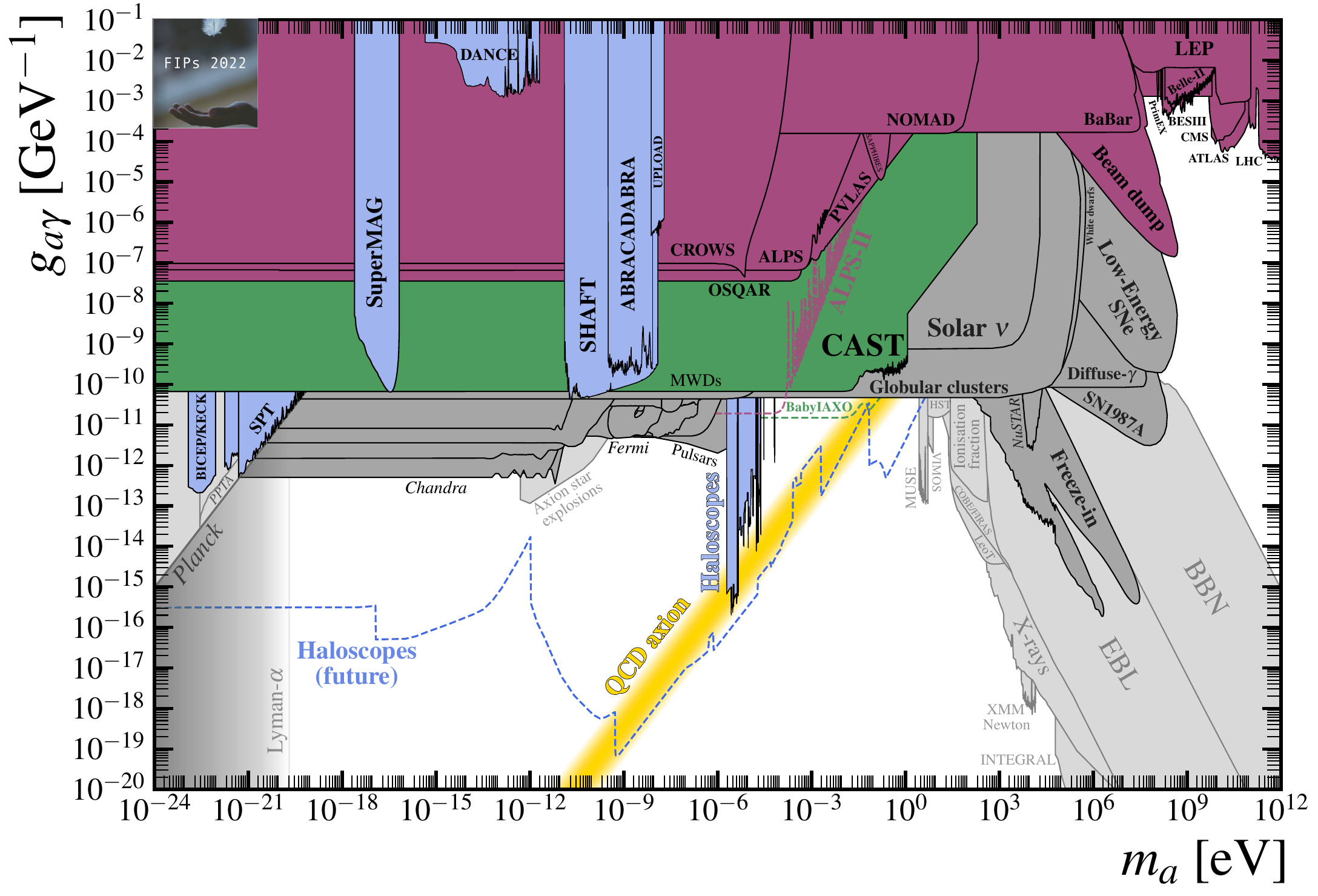}
	\caption{Fig.~50 from Ref.~\cite{Antel:2023hkf},
	          displaying the bounds on the ALPs--photon coupling
	          $g_{a\gamma}$ as  a function of the ALPs mass up to
	          the TeV-scale. The coupling is related to the parameter $c_{\gamma\gamma}$ as
	          $g_{a\gamma}= - \frac{{c}_{\gamma\gamma}}{f_a}\frac{\alpha}{\pi}$.
	          For original references and
	          explanations we refer to Ref.~\cite{Antel:2023hkf}.}
	\label{fig:FIPSALPs}
\end{figure}

Now we turn to the phenomenology of ALP contributions to $\amu$, focusing on present constraints. 
For earlier studies for similar scenarios we refer to Refs.~\cite{Marciano:2016yhf,Bauer:2017ris,Cornella:2019uxs,Buen-Abad:2021fwq,Ge:2021cjz,Keung:2021rps,Brdar:2021pla,Ganguly:2022imo,Pustyntsev:2024ygw,Sharma:2025vsh},
and for scenarios including dark matter to Ref.~\cite{Darme:2020sjf}.
Similarly to the dark photon and the U(1)$_{L_\mu-L_\tau}$ scenarios, a large number of experimental
constraints restrict the viable parameter space
$(m_a,\tilde{c}_{\gamma\gamma},c_{\mu\mu})$. Fig.~\ref{fig:FIPSALPs}
from Ref.\ \cite{Antel:2023hkf} shows constraints on
the $a\gamma\gamma$ coupling as a function of the mass $m_a$, where
the coupling is normalised as
\begin{align}\label{eq:gagammagamma}
  g_{a\gamma}&= - \frac{{c}_{\gamma\gamma}}{f_a}\frac{\alpha}{\pi}
  =- \frac{\tilde{c}_{\gamma\gamma}-\sum_i Q_i^2c_{ii}}{f_a}\frac{\alpha}{\pi}.
\end{align}
For low masses,
cosmological bounds essentially exclude ALPs unless the coupling is
extremely tiny with negligible impact on $\Damu$. For large $m_a$, collider
constraints become restrictive. Intriguingly there is a window for
$m_a$ close to the muon mass $m_\mu$ up to few GeV, where the largest
couplings are viable.

In this mass region, the BESIII result \cite{BESIII:2022rzz} provides
a strong constraint on $g_{a\gamma}$, based on the process $J/\Psi\to\gamma^* \to \gamma a$ \cite{Merlo:2019anv}.
If the ALP also couples to the Higgs and the decay $h\to a a$ is allowed, another important bound
stems from the recent ATLAS search \cite{ATLAS:2023ian} for $h\to aa\to 4\gamma$. In fact, for sizeable couplings to 
the Higgs this search excludes most of the parameter region with large $\Damu$. Equivalently, large $g_{a\gamma}$ translate
into a strong bound on the ALP--Higgs coupling. In the following we
will not use this bound but only consider the bounds that directly
depend on the ALPs couplings to photons and leptons, i.e.\ from the
lepton dipole moments and BESIII.
 
\begin{figure}
	\centering
	\includegraphics[width=.65\textwidth]{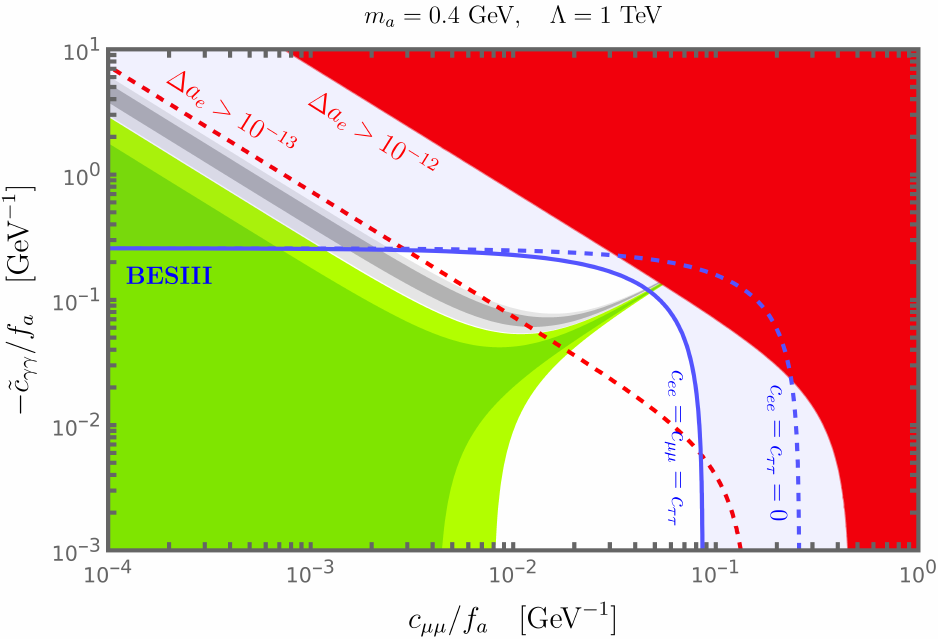}
	\caption{Plot of the 1$\sigma$ (2$\sigma$) ranges for $\Damu$
          in the ALP $c_{\mu\mu}$ -- $\tilde c_{\gamma\gamma}$ plane 
		for $\DamuFinal$ in (light) green and for $\DamuOld$ in (light) grey. The ALP mass is fixed to $m_a=0.4$ GeV
		and the cutoff scale to $\Lambda=1$ TeV.
		The red contour shows the region excluded by $\Delta a_e$ (assuming $c_{ee}=c_{\mu\mu}$),
		while the blue contours show the BESIII bound on $g_{a\gamma}$ for equal couplings $c_{ee}=c_{\mu\mu}=c_{\tau\tau}$ to all leptons (solid)
		or muon specific couplings $c_{ee}=c_{\tau\tau}=0$ (dashed).}
	\label{fig:ALPsresult}
\end{figure}

In Fig.~\ref{fig:ALPsresult} we illustrate the
current status of $\amu$ contributions from an ALP $a$, following
Ref.~\cite{Pustyntsev:2024ygw}.\footnote{%
While finalising the review, Ref.~\cite{Pustyntsev:2025nwm} appeared,
which constitutes an update of Ref.~\cite{Pustyntsev:2024ygw}.
}
We fix $m_a=0.4$ GeV as a representative
example in the mass window allowing largest couplings. 
For this particular ALP mass the BESIII measurements exclude couplings 
$g_{a\gamma} > 6\times10^{-4}~\text{GeV}^{-1}$ which translate into a correlated
bound on $\tilde{c}_{\gamma\gamma}$ and $c_{\mu\mu}$ due to Eq.~\eqref{eq:gagammagamma}. 
In this translation we assume the contribution of $N_\ell$ equal lepton generations. Then notably the bound on $c_{\mu\mu}$ depends on the number of leptons $N_\ell$ coupling to 
the ALP and is weakest in the muon-specific scenario.

The $\Damu$ contour lines show a characteristic spike at large $c_{\mu\mu}$ values, where the contributions of diagram (a)
and (b) are both large but cancel, leaving a small fine-tuned region where $\Damu$ lies within
the experimental bounds. Above this region the $\tilde c_{\gamma\gamma}c_{\mu\mu}$ term dominates
and $\Damu \gg 10^{-9}$ while below the $c_{\mu\mu}^2$ term dominates resulting in a negative $\Damu < -10^{-9}$ which is also excluded.
For smaller $c_{\mu\mu}$ the region allowed by $\DamuFinal$ widens.
We also show the bound from $a_e$, which is generally important in case of light new physics with couplings to electrons. Here, it applies only in the case $N_\ell=3$ where $c_{ee}=c_{\mu\mu}=c_{\tau\tau}$, while it does not apply in the muon-specific case.
The experimental bound $\Delta a_e\lesssim 10^{-12}$ is weaker than the BESIII exclusion limit, while a stronger potential future bound
$\Delta a_e \lesssim 10^{-13}$ would rule out most of the remaining region with large $\Damu$.

We note that, similarly to Ref.~\cite{Pustyntsev:2024ygw}, 
both $\Damu$ and $\Delta a_e$ in Fig.~\ref{fig:ALPsresult} were computed 
taking into account only the contributions from diagram (a) and (b) in Fig.~\ref{fig:ALPsdiagrams},
while the Barr-Zee diagram (c) has been neglected. 
As mentioned above, this contribution can essentially be absorbed into $\tilde c_{\gamma\gamma}^\text{eff}$,
such that the $\Delta a_\ell$ contour lines in terms of this effective coupling would stay the same.
On the other hand, the same loop corrections to the $a\gamma\gamma$ vertex (see Eq.~(3.1) in Ref.~\cite{Neubert:2024jal})
that enter into the two-loop Barr-Zee diagram for off-shell ALP momenta also enter $g_{a\gamma}$ measured at BESIII.
However, since this vertex is measured via $a\gamma$ final states, the loop function must be evaluated for
on-shell momenta of the ALP. Consequently, the effective coupling $g_{a\gamma}$ constrained by BESIII 
corresponds to a different combination of the $c_{ii}$ compared to $\tilde c_{\gamma\gamma}^\text{eff}$.
It is plausible that in such a more detailed analysis the relation between the BESIII constraint and $\amu$
shown in  Fig.~\ref{fig:ALPsresult} would remain approximately the
same, such a study is however beyond the scope of the present review.

Besides the current bounds discussed here, several experimental proposals to further scrutinise the ALP
parameter space in the future have been put forward \cite{Cui:2021dkr,Cheung:2021mol,Davoudiasl:2021haa,Cheung:2022umw,Biekotter:2022ovp,Calibbi:2022izs,Agrawal:2022wjm,Liu:2022tqn,Knapen:2023iwg,Sharma:2025vsh}.

We have highlighted the interplay between the ALPs couplings to photons
  and to leptons, expressed in the relationships
  $c_{\gamma\gamma}$$\leftrightarrow$$\tilde{c}_{\gamma\gamma}$$\leftrightarrow$$\tilde{c}^{\text{eff}}_{\gamma\gamma}$,
  which all differ by terms governed by $c_{ii}$. In light of this interplay
one can consider more minimal setups, where the fundamental $a\gamma\gamma$ coupling vanishes,
$c_{\gamma\gamma}=0$, as (partially) done in Ref.~\cite{Cornella:2019uxs}. 
Even in this case, all contributions of Eq.~(\ref{ALPsoverview}) are present, and sizeable and positive
contributions to $\amu$ are possible. Similarly,
Ref.~\cite{Armando:2023zwz} effectively sets
$\tilde{c}_{\gamma\gamma}=0$ but takes into account the genuine
2-loop Barr-Zee diagrams of Fig.~\ref{fig:ALPsdiagrams} (c). In particular, 
this reference shows that this setup is compatible with explaining dark
matter by a light fermion which couples to $a$ similarly as the leptons.

Finally, we comment on variations of the basic ALPs setup considered so far. 
While we have mainly focused on the case without lepton flavour
violation, there is no fundamental necessity for the absence of CLFV
ALPs couplings such as $c_{\mu e}$, $c_{\mu\tau}$ or $c_{e\tau}$. 
In the case these couplings are non-zero, new effects appear which show interesting differences
to some of the generic cases discussed in Sec.~\ref{sec:LeptonDipole}. 
Within ALP scenarios with both
quark and lepton flavour violation, also past $B$-physics anomalies could be accommodated
together with large $\Damu$ \cite{Bhattacharya:2021shk,Bonilla:2022qgm}.
Generally, lepton flavour-violating ALPs couplings result in 
additional one-loop diagrams to $\amu$ or $a_e$. If the loops contain an 
internal $\tau$-lepton there is strong chiral enhancement, which
could lead to too large effects. Hence, there are upper limits on the
related couplings, of the order \cite{Calibbi:2024rcm}
\begin{align}
	\left|\frac{c_{\mu\tau}(m_\mu+m_\tau)}{2f_a}\right|\lesssim10^{-3\ldots-2}
	\qquad \text{where} \qquad 
	m_a=1\ldots100\text{ GeV},
\end{align}
as well as similar limits on $c_{e\tau}$.
In contradistinction, in an
internal electron loop contributes to $\amu$ there is no chiral
enhancement, but the loop function changes sign \cite{Bauer:2019gfk} 
and positive $\Delta a_e$ can result from the
diagram type (a). In general, the current null results for
deviations in $\amu$ and $a_e$ then
lead to stringent constraints on such CLFV ALPs couplings
\cite{Cornella:2019uxs,Bauer:2019gfk,Buttazzo:2020vfs,Calibbi:2024rcm}.

Regarding CLFV observables, ALPs scenarios with CLFV couplings typically do
not follow the patterns of naive scaling and dipole dominance
mentioned in Sec.~\ref{sec:LeptonDipole}
\cite{Cornella:2019uxs,Bauer:2019gfk,Calibbi:2020jvd}. Instead,
e.g.\ the ratio between $\mu\to e$ conversion to $\mu\to e\gamma$ is
not given by Eq.~(\ref{dipoledominancemue}) but can vary in the interval
\cite{Cornella:2019uxs}
\begin{align}	\text{BR}(\mu \text{Au}\to e \text{Au}) &\simeq
(1\ldots4)\times10^{-3} \times \text{BR}(\mu\to e\gamma),
\end{align}
if the coupling $c_{\gamma\gamma}$ is varied from small to large
values; for large ALP masses $m_a\gtrsim10$ GeV the dipole dominance
value $4\times10^{-3}$ is typically recovered unless the diagrams of type
(b,c) are tuned to zero. In addition, new decay modes such as
$\mu\to e a$ can be possible, followed by $a\to 
e^+e^-$ or $a\to\gamma\gamma$ \cite{Bauer:2019gfk}. In this way, CLFV decays such as
$\mu\to 3e$ might be dominant CLFV decay modes, unrelated to
$\mu\to e\gamma$.

As announced at the beginning of this section, we will now comment on
the case of light, dark spin-0 particles which are not ALPs but which
may have renormalisable couplings to muons or other SM particles.
Such particles were studied in the light of recent experimental data
in Refs.~\cite{Liu:2018xkx,
Abdallah:2020biq,
Escribano:2020wua,Jia:2021mwk,Caputo:2021rux,Capdevilla:2021kcf,Balkin:2021rvh,Gninenko:2022ttd,Herms:2022nhd,Ema:2022afm}.\footnote{%
We mention that also the case of scalars that  couple to muons only
via dimension-6 operators have been considered in the context of
$\amu$ and spin-1 dark matter \cite{Ghorbani:2021yiw}.}
Particularly Refs.~\cite{Escribano:2020wua,Capdevilla:2021kcf} provide
very general analyses of wide classes of models with light scalars and
relevant experimental constraints.

We highlight here that the new result $\DamuFinal$
impacts  potential explanations of dark matter related to
    such light scalars. For
    instance, Refs.~\cite{Jia:2021mwk,Herms:2022nhd} consider light fermionic dark
    matter, where a mediator particle is a light new Higgs doublet,
    which is shown to be viable. Similarly to the case of
    U(1)$_{L_\mu-L_\tau}$ discussed earlier, there is a tension
    between $\Damu$ and the explanation of dark matter.
    Viable
    dark matter tends to require large annihilation cross sections
    which --- in case of dark matter annihilating into muons --- would
    lead to an even larger $\Damu$ value than the previous deviation
    $\DamuOld$ in most of the parameter space.\footnote{%
    In a similar investigation where a light CP-even scalar with
    lepton-flavour violating couplings is assumed to couple to
    electrons and muons, and to dark matter particles
    \cite{Gninenko:2022ttd}, no such tension was found, but the
    parameter region viable for dark matter is very tightly constrained.
    }
    An appealing way out of
    this tension is to assume that dark matter only annihilates into
    $\tau$-leptons. In this case, $\Damu$ vanishes, which is now
    compatible with the new $\DamuFinal$ and allows large couplings
    such that dark matter can be well accommodated in such models.

\subsubsection{Models modifying $\sigma(e^+e^-\to\text{had})$}\label{sec:HVP-models}

As discussed in Secs.~\ref{sec:SMtheory} and \ref{sec:EWPO}, the
SM hadronic vacuum polarisation contributions $a_\mu^\text{HVP}$
are currently under intense scrutiny.
Traditionally \cite{Aoyama:2020ynm}, the leading order HVP
contributions were obtained from experimental $e^+e^-\to$~hadrons data,
leading to $\DamuOld$, while
the current SM prediction for HVP,LO is purely based on computations in lattice gauge theory \cite{Aliberti:2025beg}. 
This caused a dramatic shift of the SM value, leading to $\DamuFinal$.
Between the two Theory Initiative White Papers
Refs.~\cite{Aoyama:2020ynm,Aliberti:2025beg} several new lattice results appeared, 
which are internally very consistent and well cross-checked. 
Furthermore also the new $e^+e^-$ measurement by CMD-3 appeared,
 which is more consistent with current lattice results than with earlier $e^+e^-$ measurements.

While the new result $\DamuFinal$ shows full
agreement between the experimentally measured $\amu$ and the SM
prediction, it
remains unknown why the  
previous determinations of the HVP,LO contributions show such large
discrepancies in comparison.
Extensive analyses were performed in an effort to clarify the origin
of inconsistencies between details of the various measurements
and to produce the new SM prediction
\cite{Campanario:2019mjh,BaBar:2023xiy,Davier:2023fpl,Aliberti:2024fpq,Aliberti:2025beg}.
However, so far no clear sources of the discrepancies have been
identified, and Ref.~\cite{Aliberti:2025beg} concluded that 
no scientific grounds are known on which one may dismiss any of the
data sets. Therefore, further scrutiny is required in
order to satisfyingly resolve the current situation. 

\begin{figure}
	\centering
		\centering
		\includegraphics[width=.33\textwidth]{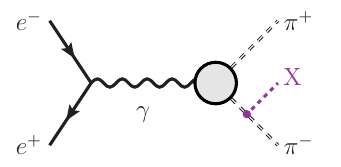}
		\includegraphics[width=.33\textwidth]{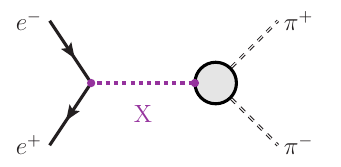}
		\caption{Direct BSM contributions to $e^+ e^- \to \pi^+ \pi^-$.}
		\label{fig:HVP-NP-direct}
\end{figure}

Meanwhile, this situation has motivated to ask whether physics beyond
the SM could be the cause of the observed discrepancies. Most
plausibly, light new particles
of the kind discussed in the present section, with masses similar to
hadron masses could be relevant. They might impact the $e^+ e^-$ cross
section measurements, by 
contaminating the data or by affecting the data analysis
\cite{DiLuzio:2021uty,Darme:2021huc,Darme:2022yal,Coyle:2023nmi,Agashe:2024apf}.
In order to
discuss this, it is useful to summarise the current situation in a
coarse-grained way (for details see Sec.~\ref{sec:SMtheory} and
Refs.~\cite{Aoyama:2020ynm,Aliberti:2025beg}):
\begin{enumerate}
  \item The $e^+e^-\to$~hadrons data from before
    Ref.~\cite{Aoyama:2020ynm} yield low values
    for the cross section and the resulting $a_\mu^\text{HVP,LO}$.
    \begin{enumerate}
    \item These data were taken by energy-scan experiments 
    SND and CMD-2 as well as using initial-state radiation at KLOE,
    BaBar and BESIII.
    \item There are some internal discrepancies particularly  between KLOE and BaBar.
    \end{enumerate}
  \item
    The  $e^+e^-\to$~hadrons data from the CMD-3 experiment yield a
    higher cross section and higher $a_\mu^\text{HVP,LO}$.
  \item
    The lattice results yield a higher $a_\mu^\text{HVP,LO}$.
\end{enumerate}
To see how BSM physics could cause some of these discrepancies, we
schematically write the experimentally determined cross section
$\sigma_{\text{had}}$ as
\begin{align}\label{eq:had-cs-exp}
	\frac{d\sigma_\text{had}}{ds} = \frac{N_\text{had} -
          N_\text{bg}}{\epsilon(s)L(s)}
        ,
\end{align}
where $N_\text{had}$ is the measured number of hadronic events, $N_\text{bg}$ the estimated number of background events,
$\epsilon$ the detection efficiency and $L(s)$ the luminosity for final states at c.o.m.\ energies $\sqrt{s}$.
In particular the following two scenarios were considered:
\begin{enumerate}[(i)]
	\item BSM directly affects the hadronic production rates,
          lowering $N_\text{had}$
          \cite{DiLuzio:2021uty,Coyle:2023nmi}. This could cause all
          $e^+e^-$-data based results to be systematically too low.
	\item BSM enters indirectly through other processes used to
          estimate $N_\text{bg}$, $\epsilon(s)$ or $L(s)$
          \cite{Darme:2021huc,Darme:2022yal}. This could cause
          different measured values depending on the details of the experiments.
\end{enumerate} 
In both cases the HVP contribution to $a_\mu$ extracted from the data corresponds to
\begin{align}
	a_\mu^\text{HVP,LO} = \frac{1}{4\pi^3} \int_{s_{th}}^\infty ds \, K(s) \Big(\sigma_\text{had}^\text{SM}(s) + \Delta \sigma_\text{had}^\text{NP}(s)\Big).
\end{align}
Thus, if the NP contribution $\Delta \sigma_\text{had}^\text{NP}(s) < 0$, the resulting value of $a_\mu^\text{HVP,LO}$ is decreased leading to
a deviation from the experiment.
Quantitatively, the correlation to the running finestructure constant
$\Delta\alpha$ discussed in Sec.~\ref{sec:EWPO}, and detailed
comparisons with lattice computations show that the cross section
should be modified mostly at low energies $\sqrt{s}\lesssim1$ GeV, and
there the shift of the cross section must be of the order of a few percent.
For simplicity in the following we
focus on the $2\pi$ channel that gives the dominant contribution to
$a_\mu^\text{HVP,LO}$.\footnote{%
Although most likely the discrepancies cannot be explained by
exclusively changing the $2\pi$ channel at very low energies
\cite{Colangelo:2020lcg,Crivellin:2022gfu}. In this context we note
that the comparison between lattice computations and data-based
results needs to be made on the level of integrals, since the cross
section $\sigma_{\text{had}}(s)$ with time-like momentum $s>0$ is not
accessible on the lattice. A detailed comparison is facilitated by
windowed integrals, isolating long-distance, short-distance, or
intermediate contributions. These windows can be applied both to
lattice computations \cite{RBC:2018dos} and to $e^+e^-$ data \cite{Colangelo:2022vok}.}

Fig.~\ref{fig:HVP-NP-direct} illustrates how $N_\text{had}$ can be
affected directly by
initial- or final-state radiation (ISR,FSR) of a BSM particle, or by a new
tree-level exchange diagram where the photon is replaced by a new
particle $X$. 
As pointed out in Ref.~\cite{DiLuzio:2021uty}, modifications of ISR or FSR alone are too small or too strongly constrained to account for the required few-percent
effect. However, if the $X$-particle couples  to both electrons and
pions, the new $s$-channel diagram could in principle lead to a
sizeable interference. Specifically, destructive interference is
required to get a negative BSM contribution. Scalar,
pseudoscalar, and axial vectors cannot destructively interfere with
the photon diagram (or the interference is suppressed by the electron
mass); hence the only plausible candidate is given by a light $Z'$
boson with vector couplings to both the electron and light quarks
\cite{DiLuzio:2021uty}, and a model Lagrangian can be written as
\begin{align}\label{eq:La-Zprime}
	\La \supset \big(g_e \bar{e} \gamma^\mu e + g_u \bar{u}\gamma^\mu u + g_d \bar{d}\gamma^\mu d\big) Z'_\mu.
\end{align}
The interference then induces the following tree-level correction to
the pion production cross section \cite{DiLuzio:2021uty},
\begin{align}\label{DiLuzioresult}
	\Delta \sigma_{\pi\pi}^\text{NP}(s) = \sigma_{\pi\pi}^{\text{SM}}(s) 
	\frac{2s(s-M_{Z'}^2) \epsilon + s^2 \epsilon^2}{(s-M_{Z'}^2)+M_{Z'}^2 \Gamma_{Z'}^2}, \qquad \text{where} \qquad
	\epsilon = \frac{g_e(g_u - g_d)}{e^2}.
\end{align}
This model therefore represents an explicit realisation of the
scenarios discussed in the
Refs.~\cite{Passera:2008jk,Keshavarzi:2020bfy,Crivellin:2020zul}
mentioned in Sec.~\ref{sec:EWPO}. It changes the extraction of the
photon vacuum polarisation from $e^+e^-$ data but itself does not
contribute to $\amu$ if the $X$ particle does not couple to the
muon. 
The required increase in the hadronic cross section then implies
$M_{Z'} < 1$ GeV and $|\epsilon|\approx 0.01$.

This basic 
idea has been investigated in detail in
Refs.~\cite{DiLuzio:2021uty,Coyle:2023nmi} and further scrutinised in
Ref.~\cite{Crivellin:2022gfu}. The relevant parameter space is already
strongly constrained, in particular by LEP measurements of $e^+e^-\to
q\bar{q}$  which limit the overall coupling strengths
as well as strong isospin-breaking observables like $\Delta m_\pi=m_{\pi^\pm}^2 -
m_{\pi^0}^2$ which limit the difference $|g_u-g_d|$.
According to the calculations of Ref.~\cite{DiLuzio:2021uty}, these
constraints fully exclude the scenario. The LEP constraint was
evaluated in more detail in
Refs.~\cite{Crivellin:2022gfu,Coyle:2023nmi}, finding slightly relaxed
bounds. The $\Delta m_\pi$-constraint involves hadronic uncertainties
and was evaluated using various arguments in all these
references. Furthermore, the tree-level calculation leading to
Eq.~(\ref{DiLuzioresult}) can be improved in case of a large $Z'$
width, leading to amplified effects \cite{Crivellin:2022gfu}. Despite
this, Ref.~\cite{Crivellin:2022gfu} confirms that the scenario is
excluded, while the evaluation of Ref.~\cite{Coyle:2023nmi} finds the
constraints on the scenario with $M_{Z'}\sim700$ MeV to be strong but
tolerable.

In total, the idea is strongly constrained and barely viable, even in
generalised forms where 
the $Z'$ can also decay invisibly or couple to muons and directly
affect $\amu$  \cite{Coyle:2023nmi}. Furthermore it is at odds with the more
recent discrepancy between the CMD-3 and other $e^+e^-$ measurements.

\begin{figure}
	\centering
		\includegraphics[width=.33\textwidth]{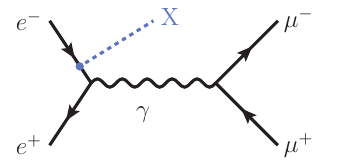}
		\includegraphics[width=.33\textwidth]{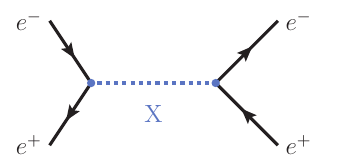}
		\caption{Indirect BSM contributions to the
                  $\sigma_{\text{had}}$ determination via normalisation processes.}
		\label{fig:HVP-NP-indirect}
\end{figure}

The second way BSM physics could affect the HVP discrepancies is via
indirect effects, as illustrated in Fig.~\ref{fig:HVP-NP-indirect}. 
If a BSM particle $X$  does not couple to hadrons but to electrons and
muons, the additional diagrams could still lead to a relevant modification of
the hadronic cross section determined via Eq.~\eqref{eq:had-cs-exp} by 
modifying the luminosity measurement \cite{Darme:2021huc,Darme:2022yal}.  
In detail, the luminosity for the
KLOE08 and KLOE10 analyses \cite{KLOE:2008fmq,KLOE:2010qei} was determined
by measuring $e^+e^-$ final states and comparing with the SM
prediction for Bhabha scattering. Assuming a dark photon can be
resonantly produced at KLOE and decays into $e^+e^-$ pairs, the
measured $e^+e^-$ rate is higher than the actual SM Bhabha rate and the
determined luminosity therefore larger than the true luminosity. 
Such a scenario could thus account for
an underestimate of the hadronic cross section $\sigma_{\text{had}}$ by 
KLOE08 and KLOE10.

In the  measurements by KLOE12 \cite{KLOE:2012anl} 
and BaBar \cite{BaBar:2012bdw}, the luminosity dependence in
Eq.~(\ref{eq:had-cs-exp}) is effectively replaced by measuring the
ratio of $\pi^+\pi^-(\gamma)$ versus $\mu^+\mu^-(\gamma)$ events, using SM
theory predictions for the $\mu^+\mu^-(\gamma)$ cross sections. These measurements
are therefore highly sensitive to additional $\mu^+\mu^-$ final states. 
The details depend on the experimental setup and are discussed in the
literature; for instance KLOE12 reconstructed photons from the
missing energy, hence
any additional $\mu^+\mu^- X$ events like in Fig.~\ref{fig:HVP-NP-indirect} 
would be included in the measured event number $N_{\mu\mu\gamma}$, decreasing
the value of $\sigma_{\pi\pi}$ estimated from
Eq.~\eqref{eq:had-cs-exp}. This again could happen e.g.\ in case of a dark photon or $Z'$
at a mass close to $1$ GeV.
Refs.~\cite{Darme:2021huc,Darme:2022yal} construct a working example
of a dark photon of mass around 1 GeV with kinetic mixing and hence equal but weak couplings
to electrons and muons. The scenario is accompanied by the idea of
inelastic dark matter, such that the dark photon bounds are weakened
as discussed in Sec.~\ref{sec:DarkPhoton}.

Because of the differing experimental setups, this scenario affects all ISR-based $e^+e^-$ 
measurements differently.
For example, the BaBar measurements are performed at a higher centre-of-mass energy compared to KLOE. 
Consequently a resonantly produced $Z'$ leading to large effects at one of the
experiments would be suppressed at the other. In turn this can alleviate 
some of the tensions between the ISR data sets.
In contrast, the scenario does not affect the energy scan measurements. 
In this way, the tensions between KLOE and BaBar on the one hand, and lattice
and CMD-3 on the other hand, can be significantly reduced. However,
the tensions between CMD-3 and previous energy scan based measurements like 
CMD-2 and SND and minor tensions between KLOE and BaBar remain \cite{Darme:2022yal}.

In summary, this scenario has not been excluded by other constraints,
but it does not fully resolve the convoluted situation regarding the
HVP determinations. An interesting lesson of this discussion is that
BSM generally might not only affect the observables of primary
interest like $\amu$ but also experimental quantities such as
luminosity measurements. In case of dark photons and similar
particles, the impact on $N_{\mu\mu\gamma}$ will actually be rather
generic \cite{Darme:2021huc}.

\newcommand{\fs}{\texttt{FlexibleSUSY}}
\newcommand{\sarah}{\texttt{sarah}}
\newcommand{\softsusy}{\texttt{SOFTSUSY}\@\xspace}
\newcommand{\ddcalc}{\texttt{DDCalc}\@\xspace}
\newcommand{\DDCalc}{\ddcalc}
\newcommand{\micromegas}{\texttt{micrOMEGAs}\@\xspace}
\newcommand{\calchep}{\texttt{CALCHEP}\@\xspace}
\newcommand{\pythia}{\texttt{PYTHIA}\@\xspace}
\newcommand{\checkmate}{\texttt{CheckMate}\@\xspace}
\newcommand{\madanalysis}{\texttt{MadAnalysis}\@\xspace}
\newcommand{\smodels}{\texttt{SModelS}\@\xspace}
\newcommand{\colliderbit}{\texttt{ColliderBit}\@\xspace}
\newcommand{\darkbit}{\texttt{DarkBit}\@\xspace}
\newcommand{\specbit}{\texttt{SpecBit}\@\xspace}
\newcommand{\GAMBIT}{\texttt{GAMBIT}\@\xspace}
\newcommand{\gambit}{\GAMBIT}
\newcommand{\GMTCalc}{\texttt{GM2Calc}\@\xspace}
\newcommand{\multinest}{\texttt{MultiNest}\@\xspace}

\subsection{Supersymmetry}
\label{sec:SUSY}

\newcommand{\amuWHL}{\Delta a_{\mu}^{1{\rm L}}(\text{WHL})} 
\newcommand{\amuWHnu}{\Delta a_{\mu}^{1{\rm L}}(\text{WHL(cha)})} 
\newcommand{\amuWHmuL}{\Delta a_{\mu}^{1{\rm L}}(\text{WHL(neu)})} 
\newcommand{\amuBHmuL}{\Delta a_{\mu}^{1{\rm L}}(\text{BHL})}
\newcommand{\amuBHmuR}{\Delta a_{\mu}^{1{\rm L}}(\text{BHR})}
\newcommand{\amuBmuLmuR}{\Delta a_{\mu}^{1{\rm L}}(\text{BLR})} 

\newcommand{\txl}{\text{L}}
\newcommand{\txr}{\text{R}}
\newcommand{\txw}{\text{W}}

Here we consider supersymmetry (SUSY), one of the most promising ideas for extending the SM. SUSY is the unique symmetry that relates fermions and bosons in relativistic quantum field theories. SUSY at the TeV scale offers potential explanations  of dark matter (DM) as the lightest SUSY particle (LSP), it explains the observed value of the Higgs-boson mass and improves electroweak naturalness. It is compatible with grand unification of gauge couplings, and local SUSY contains gravity and general relativity and can thus provide a link to ultra-high scale physics \cite{Nilles:1983ge,Haber:1984rc,Martin:1997ns}.

SUSY can be realised in different renormalizable models that are
viable extensions of the SM. The best known option is the Minimal
Supersymmetric Standard Model (MSSM), alternatives of interest for
$\amu$ include the Next-to-MSSM (NMSSM) and the Minimal R-symmetric
Supersymmetric Standard Model (MRSSM). All these models contain the SM
and enlarge the particle spectrum by superpartners (sparticles) to all
SM fields whose quantum numbers and interactions are governed by
SUSY. SUSY is explicitly broken by so-called soft-SUSY breaking terms
\cite{Girardello:1981wz}, which are free parameters governing
sparticle masses and mixings. The MSSM contains the minimal viable
number of fields and extends the SM by a second Higgs doublet and all
the superpartners. The MRSSM contains additional fields but also a
higher degree of symmetry and hence a smaller number of free
parameters, leading to dramatic changes of the $\amu$ predictions. The
NMSSM extends the MSSM by a scalar singlet and its superpartner,
leading to interesting changes e.g.\ in dark matter phenomenology
which can be correlated with $\amu$.  

It is assumed that the soft-SUSY breaking terms are not fundamental but originate from some underlying mechanism involving spontaneous SUSY breaking. Corresponding setups typically involve spontaneous SUSY breaking in a hidden sector which is then mediated to the visible sector, effectively leading to soft-SUSY breaking terms at TeV-scale energies. Example mechanisms include gravity-mediated SUSY breaking, gauge-mediated SUSY breaking, anomaly-mediated SUSY breaking or gaugino-mediated SUSY breaking, see Ref.~\cite{Martin:1997ns} and references therein. Especially in their simplest implementations, such as the Constrained MSSM (CMSSM) or minimal gauge-mediated SUSY breaking (mGMSB), they predict rigid patterns for the soft-SUSY breaking terms, often calculable as a function of few fundamental parameters. In this way models such as the CMSSM or mGMSB lead to specific predictions for $\amu$ and strong correlations to other observables. 

For the last two decades SUSY contributions to $\amu$ were considered
very promising explanations of the significant deviation $\DamuOld$ in
Eq.~(\ref{eq:DamuOld}). Especially in the MSSM that
deviation could be explained for values of parameters such as
$\tan\beta \sim 50$ and
$M_{\text{SUSY}} \sim 500$ GeV, which were attractive also for other reasons.
For early reviews  we refer to
Refs.~\cite{Czarnecki:2001pv,Stockinger:2006zn}; a large number of
phenomenological studies of SUSY and $\amu$ were inspired by the first 
Brookhaven $\amu$ measurement in 2001 \cite{Arnowitt:2001be,Czarnecki:2001pv,Baltz:2001ts,Everett:2001tq,Feng:2001tr,Chattopadhyay:2001vx,Komine:2001fz,Hisano:2001qz,Ibrahim:2001ym,Ellis:2001yu,Choi:2001pz,Kim:2001se,Martin:2001st,Komine:2001hy,Baek:2001nz,Carvalho:2001ex,Baer:2001kn,Chacko:2001xd,Baek:2001kh,Chen:2001kn,Arhrib:2001xx,Enqvist:2001qz,Cerdeno:2001aj,Kim:2001eg,Cho:2001nfa,Arnowitt:2001pm, 
    Belanger:2001am,deBoer:2001in,Roszkowski:2001sb,Adhikari:2001sf,deBoer:2001nu,Endo:2001ym,Cho:2001hx},
  and in the following years by improvements in the determination of
  $\amu$ and in complementary measurements
  \cite{Huang:2002dj,Baek:2002cc,Chattopadhyay:2002jx,Byrne:2002cw,Kim:2002cy,Baek:2002wm,Martin:2002eu,Heinemeyer:2003dq,Heinemeyer:2004yq,Marchetti:2008hw,Domingo:2008bb,Cheung:2009fc,Hofer:2009xb,Crivellin:2010ty,vonWeitershausen:2010zr,Babu:2014lwa}. 

Already before the dramatically
shifted SM contributions led to the new value  $\DamuFinal$,
SUSY scenarios  have become constrained
by other
measurements, most importantly from dark matter searches and from LHC.
In this review we first focus on the MSSM, its contributions to $\amu$
and complementary constraints from dark matter, LHC and other
considerations. We discuss several important standard scenarios and
their current status, then provide a detailed account of the
phenomenology of the general MSSM. Finally we also comment on
alternative realisations of SUSY such as the MRSSM, the NMSSM and
scenarios with unstable LSP.

\subsubsection{Brief account of the MSSM}
\label{sec:MSSMDefinition}
We begin with a brief introduction to the MSSM. We introduce the MSSM
notation as well as the aspects of key importance for the
phenomenological discussion. The MSSM contains the SM extended by a
second Higgs doublet of opposite hypercharge, and by superpartner
fields.
The field
content is illustrated in Tab.~\ref{tab:MSSM}. The SM quarks and
leptons of both chiralities receive spin-0 squark and slepton
superpartners; the SM gauge and Higgs fields receive spin-1/2 gaugino
and Higgsino superpartners. The two Higgs doublets are denoted ${\cal
  H}_{u,d}$, and supersymmetry only allows the tree-level coupling of
${\cal H}_{u}$ to up-type quarks and of ${\cal H} _{d}$ to down-type
quarks and leptons. This corresponds to the two-Higgs doublet type II
Yukawa structure, but in contrast to the non-SUSY two-Higgs doublet
model, this Yukawa structure is modified by loop corrections.

\begin{table}
	\centering
	\begin{tabular}{c |c|c c|c c|}
		\multicolumn{1}{c}{} & \multicolumn{1}{c}{\textbf{MSSM:}} & \multicolumn{2}{c}{Fermions} & \multicolumn{2}{c}{SUSY partners} \\ \cline{2-6}
		& Rep. & Field & Spin & Field & Spin \\ \hhline{~=====}
		\multirow{2}{*}{\rotatebox[origin=c]{90}{(s)leptons}} & $(\bm1,\bm2,-\frac{1}{2})$ & 
		\raisebox{.5\totalheight}{\vphantom{$\Big($}}$\renewcommand*{\arraystretch}{.6}\mqty(\nu_{Li} \\ e_{Li})$ & $\frac{1}{2}$ &
		\raisebox{.5\totalheight}{\vphantom{$\Big($}}$\renewcommand*{\arraystretch}{.6}\mqty(\tilde\nu_{Li} \\ \tilde e_{Li})$ & $0$ \\
		& $(\bm{1},\bm{1},-1)$ & ${e_{Ri}}$ & $\frac{1}{2}$ & $\tilde e_{Ri}$ & $0$ \\[.1cm] \cline{2-6}
		\multirow{3}{*}{\rotatebox[origin=c]{90}{(s)quarks}} & $(\bm3,\bm2,\frac{1}{6})$ & 
		\raisebox{.5\totalheight}{\vphantom{$\Big($}}$\renewcommand*{\arraystretch}{.6}\mqty(u_{Li} \\ d_{Li})$ & $\frac{1}{2}$ &
		\raisebox{.5\totalheight}{\vphantom{$\Big($}}$\renewcommand*{\arraystretch}{.6}\mqty(\tilde u_{Li} \\ \tilde d_{Li})$ & $0$ \\
		& $(\bar{\bm3},\bm1,\frac{2}{3})$ & ${u_{Ri}}$ & $\frac{1}{2}$ & $\tilde u_{Ri}$ & $0$\\
		& $(\bar{\bm3},\bm1,-\frac{1}{3})$ & ${d_{Ri}}$ & $\frac{1}{2}$ & $\tilde d_{Ri}$ & $0$\\ \cline{2-6}
	\end{tabular}\hfill
	\begin{tabular}{c |c|c c|c c|c}
		\multicolumn{1}{c}{} & \multicolumn{1}{c}{\textbf{MSSM:}} & \multicolumn{2}{c}{Bosons} & \multicolumn{2}{c}{SUSY partners} & \multicolumn{1}{c}{} \\ \cline{2-6}
		& Rep. & Field & Spin & Field & Spin & \\ \hhline{~=====}
		\multirow{2}{*}{\rotatebox[origin=c]{90}{Higgs(ino)}} & $(\bm1,\bm2,-\frac{1}{2})$ & 
		\raisebox{.5\totalheight}{\vphantom{$\Big($}}$\renewcommand*{\arraystretch}{.6}\mqty(\mathcal{H}_d^0 \\ \mathcal{H}_d^-)$ & $0$ &
		\raisebox{.5\totalheight}{\vphantom{$\Big($}}$\renewcommand*{\arraystretch}{.6}\mqty(\tilde{H}_d^0 \\ \tilde{H}_d^-)$ & $\frac{1}{2}$ &
		\multirow{4}{*}{$\left\}\rule{0pt}{12ex}\right.$
			\rotatebox[origin=c]{90}{
			\begin{minipage}{3.5cm}
					\centering
					$h,H,A,H^\pm,\gamma,Z,W^\pm$ \\ $\chi^0_{1,2,3,4},~\chi^\pm_{1,2}$
			\end{minipage}} }	 \\
		& $(\bm1,\bm2,\frac{1}{2})$ & 
		\raisebox{.5\totalheight}{\vphantom{$\Big($}}$\renewcommand*{\arraystretch}{.6}\mqty(\mathcal{H}_u^0 \\ \mathcal{H}_u^-)$ & $0$ &
		\raisebox{.5\totalheight}{\vphantom{$\Big($}}$\renewcommand*{\arraystretch}{.6}\mqty(\tilde{H}_u^0 \\ \tilde{H}_u^-)$ & $\frac{1}{2}$ & \\[.2cm] \cline{2-6}
		\multirow{3}{*}{\rotatebox[origin=c]{90}{Gauge(ino)}} & $(\bm1,\bm1,0)$ & $B_\mu$ & $1$ & $\tilde B$ & $\frac{1}{2}$ & \\
		& $(\bm1,\bm3,0)$ & $W_\mu^i$ & $1$ & $\tilde W^i$ & $\frac{1}{2}$ & \\ 
		& $(\bm8,\bm1,0)$ & $G_\mu^a$ & $1$ & $\tilde g^a$ & $\frac{1}{2}$ & \\\cline{2-6}
	\end{tabular}
	\caption{The field/particle content of the MSSM. The table shows the gauge-eigenstate fields and their quantum numbers; next to the table the mass
  eigenstates corresponding to the electroweak gauge
  and Higgs bosons and their superpartners are indicated.}
\label{tab:MSSM}
\end{table}

Via electroweak symmetry breaking, the two Higgs doublets receive \vev{}s $v_{u,d}$, and five physical Higgs fields $h,H,A,H^\pm$ emerge, where $h$ is assumed to be SM-like. 
The $\UY$, $\SUL$ and $\SUc$ gauginos, called Bino ${\tilde B}$, Winos ${\tilde W}^i$ and gluinos ${\tilde g} ^a$, mix with the Higgsinos ${\tilde H}_{u,d}$ to neutralino ${ \chi}^0 _i$ and chargino ${ \chi} ^\pm _i$ mass eigenstates, and also sfermions mix. 

Two central MSSM parameters that are of particular importance for $a_\mu$ are related to the two Higgs doublets. The first of these is the ratio of the two vacuum expectation values,
\begin{eqnarray}
\tan\beta &=& \frac{v_u}{v_d}.
\end{eqnarray}
As mentioned above, only the doublet ${\cal H}_d$ gives masses to
down-type quarks and leptons such as the muon. As a result, the
tree-level muon mass is given by $m_\mu=y_\mu v_d/\sqrt{2}$, and the
tree-level muon Yukawa coupling $y_\mu$ is enhanced compared to the
SM-value by a factor $1/\cos\beta$, which becomes approximately an
enhancement by $\tan\beta$ for large $\tan\beta$. Similarly the
up-type Yukawa couplings are enhanced by $1/\sin\beta$. In order to
avoid non-perturbative values of all Yukawa couplings, $\tan\beta$ is
commonly restricted to the range between about 1 and 50, though this
can be relaxed, which is discussed below.  
High values $\tan\beta={\cal O}(50)$ lead to similar top and bottom
Yukawa couplings and are therefore favoured by the idea of top--bottom
Yukawa coupling unification \cite{Ananthanarayan:1991xp}, see also
Ref.~\cite{Aboubrahim:2021phn} for a recent study in connection with
$\amu$. 

The second important parameter connecting the two Higgs doublets is the $\mu$-parameter, which appears in the MSSM Lagrangian in the terms
\begin{align}
\La ^{\text{MSSM}} &\ni \mu\tilde{H}_d\tilde{H}_u-\mu F_{H_d}{\cal H}_u-\mu F_{H_u}{\cal H}_d + \text{h.c.} .
\label{mudef}
\end{align}
The first term describes a Higgsino mass term, while in the other terms $F_{H_{d,u}}$ are auxiliary fields whose elimination gives rise to interactions of ${\cal H}_{d,u}$ with sfermions of the opposite type compared to the Yukawa couplings, producing e.g.\ mixing mass terms proportional to ${v}_d\tilde{t}_L\tilde{t}_R^\dagger$ and ${v}_u\tilde{\mu}_L\tilde{\mu}_R^\dagger$.  

In addition the MSSM contains a large number of parameters that
parametrise soft SUSY breaking.
  The relevant soft parameters are
\begin{align}
   M_1, M_2, M_3; m^2_{\tilde{q},\tilde{u},\tilde{d},\tilde{l},\tilde{e}}; A_{u,d,e},
\end{align}
where we follow the conventions of
Refs.~\cite{Allanach:2002nj,Aguilar-Saavedra:2005zyz}, which are
similar as e.g.\ in Refs.\ \cite{Martin:2001st,Stockinger:2006zn,Cho:2011rk}. Here $M_{1,2,3}$ are (Bino, Wino,
gluino) gaugino mass parameters, $m^2_{\tilde{i}}$ are $3\times3$ hermitian
matrices in generation space corresponding to the scalar masses in the
sector $i$; $A_{i}$ are   $3\times3$ general
matrices in generation space corresponding to trilinear interactions
between Higgs doublets and sfermion doublets and singlets. Further
parameters in the Higgs sector can be eliminated in favour of $\tan\beta$
and e.g.~the mass $M_{H^\pm}$ of the charged Higgs boson.
Except where explicitly stated we will restrict the number of the
parameters by neglecting generation mixing and complex phases in the sfermion sectors,
i.e.\ by assuming the corresponding matrices to be diagonal and real.
Allowing off-diagonal or complex sfermion mass parameters would lead to sizeable
predictions for flavour-violating processes and electric dipole moments in the quark and lepton
sectors, and specifically the correlations between $\amu$, CLFV
processes, and EDMs  follow the general pattern described in
Sec.~\ref{sec:LeptonDipole} as will be discussed below. 

Further, for the following phenomenological discussion it is common
practice to use the term ``sleptons'' only for smuons and selectrons,
while the ``staus'' are discussed separately. The reason is that in
many respects, especially for LHC phenomenology, the staus behave very
differently from the sleptons of the first two
generations. Accordingly, we will use the following simplified
notation for the slepton mass parameters,
\begin{align}
  m_\txl &\equiv m_{\tilde{l},11}=m_{\tilde{l},22},
  &m_{\tilde{\tau}_L} &\equiv m_{\tilde{l},33},\\
  m_\txr &\equiv m_{\tilde{e},11}=m_{\tilde{e},22},
  &m_{\tilde{\tau}_R} &\equiv m_{\tilde{e},33}.
\end{align}

Similarly, unless explicitly noted, we will assume that the MSSM conserves $R$-parity, where
all SM-like states have positive $R$-parity and all SUSY partner
states have negative $R$-parity. $R$-parity conservation implies that
SUSY particles must be produced in pairs, and the decay of each SUSY
particle ends up in a number of SM particles plus the LSP, and the LSP
is stable.

To fix our notation for the most important parameters we provide the mass matrices of the charginos, the neutralinos and the smuons. They are given as
\begin{align}
  X &= 
  \left(\begin{array}{cc}
    M_2 & M_W\sqrt2 \sin\beta\\
    M_W \sqrt2 \cos\beta & \mu
  \end{array}
  \right)
  ,
  \\
  Y &=
  \left(\begin{array}{cccc}
    M_1 & 0   & -M_Z s_\txw \cos\beta & M_Z s_\txw \sin\beta \\
    0   & M_2 & M_Z c_\txw \cos\beta & -M_Z c_\txw \sin\beta \\
    -M_Z s_\txw \cos\beta &M_Z c_\txw \cos\beta & 0 & -\mu \\
    M_Z s_\txw \sin\beta & -M_Z c_\txw \sin\beta& -\mu & 0
  \end{array}
  \right),\label{NeutralinoMassmatrix}
  \\
  M_{\tilde{\mu}}^2 &= \left(\begin{array}{lr}
    m_\mu^2 + m_\txl ^2 + M_Z ^2 \cos2\beta\left(-\frac12+s_\txw ^2\right)
    &
    m_\mu (-\mu\tan\beta+A_\mu^*)
    \\
    m_\mu (-\mu^*\tan\beta+A_\mu)
    &
    m_\mu^2 + m_{\txr}^2 + M_Z ^2\cos2\beta\ (-s_\txw ^2)
  \end{array}
  \right),
  \label{smuonmassmatrix}
\end{align}
where the fundamental SUSY parameters appearing in these expressions are $\tan\beta$ and the two gaugino (Bino and Wino) mass parameters $M_{1,2}$, the Higgsino mass $\mu$ and the
left-/right-handed smuon masses $m_{\txl, \txr}$. The smuon mass
matrix also involves the trilinear soft SUSY-breaking parameter
$A_\mu\equiv A_{e,22}$, which however plays a subdominant role compared to the $\mu\tan\beta$ term. The other appearing parameters are the SM parameters $m_\mu,M_{W,Z}$ and $s_\txw =\sqrt{1-c_\txw ^2}$.

\subsubsection{MSSM contributions to $\amu$}
\label{sec:SUSYamu}

The SUSY 
contributions to $\amu$ are mainly generated
by one-loop diagrams involving a smuon or
sneutrino, together with a neutralino or chargino, i.e.~the SUSY
partners of
the
electroweak SM gauge and Higgs bosons. The diagrams are of the generic
FS-type shown in Fig.~\ref{fig:amu-FS}.

These one-loop contributions have been
systematically and comprehensively studied for the MSSM in
Ref.\ \cite{Moroi:1995yh}, for 
reviews see e.g.~Refs.\ \cite{Martin:2001st,Stockinger:2006zn,Cho:2011rk}.

The MSSM one-loop contributions to $\amu$ feature a very important
chiral enhancement, which brings the possible values into the ballpark
tested by the Fermilab experiment.
Here we first focus on this dominant behaviour, using the
mass-insertion approximation, which allows us to directly read off the main parameter
dependences similarly to
Sec.~\ref{sec:genericthreefield}; then we comment on the full
prediction in the MSSM.\footnote{%
We note that this chiral enhancement is
present also in many extensions of the  MSSM, but e.g.~not in the
MRSSM, as will be discussed further below.}

\begin{figure}
  \centering
  \begin{subfigure}{0.3\textwidth}
    \centering \includegraphics{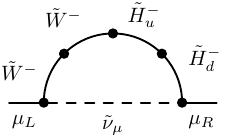}
    \caption{WHL (chargino)} \vspace{0.05\textwidth}
    \label{fig:HWsneu}
  \end{subfigure}
  \begin{subfigure}{0.3\textwidth}
    \centering \includegraphics{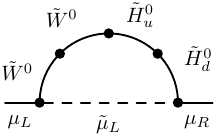}
    \caption{WHL (neutralino)} \vspace{0.05\textwidth}
    \label{fig:HWL}
  \end{subfigure}
  \begin{subfigure}{0.3\textwidth}
    \centering \includegraphics{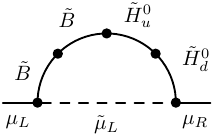}
    \caption{BHL} \vspace{0.05\textwidth}
    \label{fig:HBL}
  \end{subfigure}
  \begin{subfigure}{0.3\textwidth}
    \centering \includegraphics{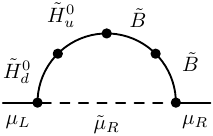}
    \caption{BHR}
    \label{fig:HBR}
  \end{subfigure}
  \begin{subfigure}{0.3\textwidth}
    \centering \includegraphics{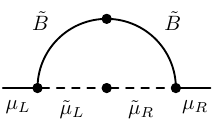}
    \caption{BLR}
    \label{fig:BLR}
  \end{subfigure}
  \caption{Mass-insertion diagrams of the MSSM contributing to $\amu$
  and to the muon self
  energy, see Eqs.~(\ref{eq:SUSYMIapprox}) and
  (\ref{eq:sigmamuparts}). For contributions to $\amu$, an additional
  external photon couples to any of the 
  charged particles in the loop.}
  \label{fig:MIAMSSM}
\end{figure}

The  relevant mass-insertion diagrams are shown in
Fig.~\ref{fig:MIAMSSM}.
Following the form given e.g.\ in
Refs.\ \cite{Cho:2011rk,Fargnoli:2013zia,Bach:2015doa} the corresponding
contributions can be written as
\begin{align}
\Damu^{\text{1L SUSY}} &\approx \amuWHL + \amuBHmuL \nonumber + \amuBHmuR +
\amuBmuLmuR, 
\end{align}
with
\begin{subequations}
\label{eq:SUSYMIapprox}
\begin{align}
  \amuWHL =& + m_\mu\left[\frac{y_\mu v_u}{\sqrt2}\right]
  \frac{g_{2}^{2}}{8 \pi ^{2}} \frac{ M_{2}\mu}{m_\txl ^{4}} \, 
F_{a}\left(\frac{M_{2}^{2}}{m_\txl ^{2}},\frac{\mu^{2}}{m_\txl
  ^{2}}\right)
\nonumber\\* 
&-m_\mu\left[\frac{y_\mu v_u}{\sqrt2}\right]
\frac{g_{2}^{2}}{16 \pi ^{2}} \frac{ M_{2}\mu}{m_\txl ^{4}} \,
F_{b}\left(\frac{M_{2}^{2}}{m_\txl ^{2}},\frac{1}{m_\txl ^{2}}\right)
&&\approx\pm21\times10^{-10} \left(\frac{500\text{ GeV}}{M_{\text{SUSY}}}\right)^2\frac{\tan\beta^{\text{eff}}}{40}
\label{eq:SUSYWHL}
,\\*
\amuBHmuL =& + m_\mu\left[\frac{y_\mu v_u}{\sqrt2}\right]
\frac{g_{1}^{2}}{16 \pi ^{2}}\frac{ M_{1}\mu}{m_\txl ^{4}}\,
F_{b}\left(\frac{M_{1}^{2}}{m_\txl ^{2}},\frac{\mu^{2}}{m_\txl
  ^{2}}\right)
&&\approx\pm1.2\times10^{-10}\left(\frac{500\text{
    GeV}}{M_{\text{SUSY}}}\right)^2\frac{\tan\beta^{\text{eff}}}{40}
,\\* 
\amuBHmuR =& -
m_\mu\left[\frac{y_\mu v_u}{\sqrt2}\right]
\frac{g_{1}^{2}}{8 \pi ^{2}} \frac{ M_{1}\mu}{m_\txr ^{4}}\, 
F_{b}\left(\frac{M_{1}^{2}}{m_\txr ^{2}},\frac{\mu^{2}}{m_\txr
  ^{2}}\right)
 &&\approx\mp2.4\times10^{-10}\left(\frac{500\text{ GeV}}{M_{\text{SUSY}}}\right)^2\frac{\tan\beta^{\text{eff}}}{40}
, \\* 
\amuBmuLmuR =& +m_\mu\left[\frac{y_\mu v_u}{\sqrt2}\right]
\frac{g_{1}^{2}}{8 \pi ^{2}} \frac{\mu}{M_{1}^{3}}\, 
F_{b}\left(\frac{m_\txl ^{2}}{M_{1}^{2}},\frac{m_\txr
  ^{2}}{M_{1}^{2}}\right)
&&\approx\pm2.4\times10^{-10}\left(\frac{500\text{ GeV}}{M_{\text{SUSY}}}\right)^3\frac{\tan\beta^{\text{eff}}}{40}\frac{\mu}{500\text{ GeV}}.
\end{align} 
\end{subequations}
The appearing loop functions are normalised as $F_a(1,1)=1/4$,
$F_b(1,1)=1/12$ and can be found in footnote \ref{footnoteloopfunctions}.
On the far right we provide numerical approximations valid if all SUSY
mass parameters are equal, where the signs are given by the sign of
the products $M_{1,2}\mu$ and where
$\tan\beta^{\text{eff}}=\tan\beta+$higher order corrections as
explained further below. 
The letters B,W,H,L, and R are the abbreviations for Bino, Winos,
Higgsinos, and Left- and Right-handed 
smuons/sneutrinos, respectively. As indicated and as visible in the Feynman
diagrams, each contribution depends on three of these states and their
respective masses. The factors $y_\mu v_u$ in square brackets directly reflect the
need for chirality flips and electroweak symmetry breaking discussed
in Sec.~\ref{sec:ChiralityFlips}, and generally the formulas are
written analogously to Eq.~(\ref{amugeneric}).
As an even simpler reference we provide an approximation if all
relevant sparticle masses are equal to a common scale
$M_{\text{SUSY}}$,
\begin{align}\label{DamuSUSYequalmass}
  \Damu^{\text{SUSY}}\approx 
22\times10^{-10}\left(\frac{500\text{
    GeV}}{M_{\text{SUSY}}}\right)^2\frac{\tan\beta^{\text{eff}}}{40}. 
\end{align}

There is a crucial chiral
enhancement by the factor $\tan\beta$ because the 
tree-level muon mass is given via the down-type \vev $v_d$, while
the up-type \vev $v_u$ can be larger, such that
\begin{align}\label{muonmassMSSMtree}
  \left[\frac{y_\mu v_u}{\sqrt2}\right] \to m_\mu^{\text{tree}}\tan\beta
\end{align}
at lowest order.
Diagrammatically, the chiral enhancement arises as follows. In the
WHL, BHL and BHR contributions, the muon chirality is flipped at the Yukawa coupling
to a Higgsino; the Higgsino is then converted to a gaugino via a Higgs
vacuum expectation value. The appearance of the enhanced VEV is always
accompanied by a factor of the Higgsino mass $\mu$ which converts
$\tilde{H}_d$ into $\tilde{H}_u$. This explains the
appearance of one Higgsino mass $\mu$ in the numerators of all
contributions. A  gaugino  (Bino or Wino) mass parameter $M_{1,2}$
must also appear in each numerator. These contributions are very
similar to the three-field models of Class II in
Sec.~\ref{sec:genericthreefield}, although the relevant Higgs doublet is
different.\footnote{%
Comparing e.g.~the  WHL case to the Class II behaviour in
Eq.~(\ref{eq:MIAcontribution}), the couplings 
$\lambda_{L}\lambda_R\bar{\lambda}_H$ would correspond 
to $g_2 y_\mu g_{2}$, and the masses $m_\psi m_\chi$
to $M_2\mu $.
}

The BLR
contribution in the last line is special, and it behaves like the
three-field model of Class III.
Here the muon chirality is flipped at the smuon line via the insertion
of a smuon-left-right flip, which is also governed by a product of the
Higgsino mass $\mu$ and the enhanced VEV in the smuon mixing
matrix whose origin is explained below Eq.~(\ref{mudef}).
In contrast to the other contributions, the BLR contribution is
approximately linearly enhanced by the Higgsino mass $\mu$ since this
parameter does not appear as the mass of a virtual Higgsino. The signs
of all contributions are 
determined by the signs of $\mu$ and the gaugino masses
$M_{1,2}$. Unless noted explicitly, in the following we will only
consider positive signs of 
all these parameters, leading to positive MSSM contributions to $\amu$.

Specific constraints on this BLR-contribution have
been very thoroughly investigated in Ref.\ \cite{Endo:2013lva,Chigusa:2022xpq,Chigusa:2023mqy}. 
Most importantly, vacuum stability requires that staus, the
superpartners of $\tau$-leptons, do not receive
a charge-breaking vacuum expectation value, and this provides a
constraint on the relation between the off-diagonal and diagonal
elements of the stau mass matrix similar to
Eq.\ (\ref{smuonmassmatrix}). As a quantitative example,
Ref.\ \cite{Endo:2013lva} finds that in case of universal left- and
right-handed stau masses
$m_{\tilde{\tau}_\txl}=m_{\tilde{\tau}_\txr}\equiv m_{\tilde{\tau}}$, the Higgsino mass has an upper limit,
specifically 
\begin{align}\label{VacStabEndo}
  \frac{\mu}{\text{1 TeV}}\frac{\tan\beta}{40}
  &\lesssim
  \frac{m_{\tilde{\tau}}}{\text{300 GeV}}.
\end{align}
An analogous limit from the smuon mass matrix exists but is weaker
because of the smaller muon mass; for a more accurate and general
treatment we refer to  Refs.\ \cite{Endo:2013lva,Chigusa:2022xpq,Chigusa:2023mqy}.  

It is illuminating to discuss a particular kind of higher-order
corrections to $\amu$ with the help of the same mass-insertion
diagrams in Fig.~\ref{fig:MIAMSSM}. As explained in
Secs.~\ref{sec:ChiralityFlips} and \ref{sec:MuonMass} there is a deep
connection between contributions to $\amu$ and to the muon mass
$m_\mu$. Interpreting the mass-insertion diagrams as contributions to
the muon mass, they correspond to one-loop corrections to the
relationship (\ref{muonmassMSSMtree}) between the muon mass and the
muon Yukawa coupling. The decisive feature is that the loops generate
a coupling of the muon to the larger \vev $v_u$. The result for the
one-loop corrected muon mass can be
written as
\begin{align}
  m_\mu =\frac{y_\mu v_d}{\sqrt{2}} + \frac{y_\mu v_u}{2}\Delta_\mu^{\text{red}},
\label{eq:mmufromSE}
\end{align}
where the one-loop quantity $\Delta_\mu^{\text{red}}$ can be
approximated as \cite{Bach:2015doa}
\begin{subequations}
\label{eq:sigmamuparts}
\begin{align}
  \Delta_\mu^{\text{red}}(\text{WHL})&=
    {-}\frac{g_2^2}{16 \pi^2} \,M_2 \mu \,I(M_2,\mu,m_L)\nonumber\\
    &\quad\
    {-}\frac{g_2^2}{32 \pi^2} \,M_2 \mu \,I(M_2,\mu,m_L),
    \label{eq:sigmamupartsHWL}\\
  \Delta_\mu^{\text{red}}(\text{BHL})&=
    \phantom{-}\frac{g_1^2}{32 \pi^2} \,M_1 \mu \,I(M_1,\mu,m_L),\\
  \Delta_\mu^{\text{red}}(\text{BHR})&=
    {-}\frac{g_1^2}{16 \pi^2} \,M_1 \mu \,I(M_1,\mu,m_R),\\
  \Delta_\mu^{\text{red}}(\text{BLR})&=
    \phantom{-}\frac{g_1^2}{16 \pi^2} \,M_1 \mu \,I(M_1,m_L,m_R),
\end{align}
\end{subequations}
where the loop function is given by
\begin{align}
  I(a,b,c)
  =\frac{a^2\,b^2\ln\frac{a^2}{b^2}+b^2\,c^2\ln\frac{b^2}{c^2}
  +c^2\,a^2\ln\frac{c^2}{a^2}}{\left(a^2-b^2\right)\left(b^2
  -c^2\right)\left(a^2-c^2\right)},
\end{align}
with a scaling behaviour  $I\sim1/M^2$,
where $M$ denotes the largest of the three mass arguments.

The parallel structure of the one-loop contributions to  $\amu$ and
for $m_\mu$ via $\Delta_\mu^{\text{red}}$ is apparent. This does not
only illustrate and confirm the general discussions of
Secs.~\ref{sec:ChiralityFlips} and \ref{sec:MuonMass}. It also
explains the possibility of an all-order resummation of large
$\tan\beta$-enhanced effects, first analysed in
Ref.~\cite{Marchetti:2008hw} and further studied in
Refs.~\cite{Dobrescu:2010mk,Altmannshofer:2010zt,Bach:2015doa}.
The solution of Eq.~(\ref{eq:mmufromSE}) leads to the improved
replacement for the Yukawa coupling $y_\mu$,
\begin{align}\label{muonmassMSSM1L}
  \left[\frac{y_\mu v_u}{\sqrt2}\right] \to
  \frac{  m_\mu v_u}{v_d + v_u\Delta_\mu^{\text{red}}}
  \equiv m_\mu\tan\beta^{\text{eff}},
\end{align}
where also an effective higher-order corrected
$\tan\beta^{\text{eff}}=\tan\beta/(1+\tan\beta\Delta_\mu^{\text{red}})$
has been defined. If this replacement instead of the tree-level 
Eq.~(\ref{muonmassMSSMtree})  is applied to the result for
$\amu$,  all $n$-loop corrections enhanced by $\tan^n\beta$ are taken
into account \cite{Marchetti:2008hw}. The resummed formula also allows
the limit $v_d\to0$, $\tan\beta\to\infty$
\cite{Dobrescu:2010mk,Altmannshofer:2010zt,Bach:2015doa}, in which
case the muon mass 
vanishes at tree level and is generated radiatively as in the examples
of Sec.~\ref{sec:MuonMass}.

Now we turn to the best available MSSM prediction for $\amu$. Given
the importance of the model, significant effort has been invested in
obtaining an accurate and reliable computation. The one-loop
diagrams for $\amu$ have been computed exactly
in the general MSSM in Ref.~\cite{Moroi:1995yh}, and the explicit
formulas can be found there and in all other references mentioned
above. In addition, a wide range
of higher-precision
calculations of 2-loop contributions is
available. The known
SUSY contributions (i.e.\ the difference between MSSM
and SM contributions) to $\amu$ can 
be summarised as
\newcommand{\amuSUOL}{\Damu^{\rm 1L\, SUSY}}
\newcommand{\amuFSf}{\Damu^{{\rm 2L,} f\tilde{f}}}
\newcommand{\amuTLa}{\Damu^{\rm 2L(a)}}
\newcommand{\amuphotonic}{\Damu^{\rm 2L,\ photonic}}
\begin{align}
\Damu^{\text{SUSY}} &= \left[\amuSUOL  +\amuTLa
+
\amuphotonic 
 +\amuFSf\right]_{t_\beta\text{-resummed}}.
\label{amuSUSYdecomposition}
\end{align}
Here  $\amuTLa$ corresponds  2-loop diagrams in which a SUSY loop is inserted into a
SM-like loop, including  Barr-Zee
diagrams discussed in Sec.~\ref{sec:Barr-Zee} where charginos or
sfermions run in the inner loop. These were computed and evaluated in
Refs.\ \cite{Arhrib:2001xx,Chen:2001kn,Heinemeyer:2003dq,Heinemeyer:2004yq,Cheung:2009fc},
and reviewed in detail in Ref.~\cite{Stockinger:2006zn};
given today's experimental constraints these diagrams must be
negligibly
small.  The full 2-loop QED corrections $\amuphotonic$, see
Refs.\ \cite{Degrassi:1998es,vonWeitershausen:2010zr}, include the 
leading QED-logarithms discussed in Sec.~\ref{sec:EFTRGE}.
Refs.\ \cite{Fargnoli:2013zda,Fargnoli:2013zia} computed
genuine SUSY 2-loop corrections $\amuFSf$ to the SUSY 1-loop diagrams which include
non-decoupling effects from e.g.\ heavy squarks. Finally, the
inclusion of a $\tan\beta$-resummation, i.e.~of  $n$-loop higher-order
terms enhanced by $(\tan\beta)^n$, was developed in
Refs.\ \cite{Marchetti:2008hw,Bach:2015doa}, where also the muon mass
corrections $\Delta_\mu^{\text{red}}$ are given without mass-insertion approximation.
Interestingly, each of
these three kinds of higher-order corrections can shift the 1-loop contributions by
around 10\%.
All these 1-loop and 2-loop contributions
are
implemented in the code GM2Calc \cite{Athron:2015rva}, which is used
in many phenomenological analyses in the recent
literature. Ref.~\cite{Athron:2015rva} also contains a compact collection of
explicit formulas of all contributions.\footnote{%
  For higher-order calculations in extensions of the MSSM see 
  Refs.\ \cite{Zhao:2014dxa,Yang:2018guw,Dong:2019iaf,Su:2020lrv,Liu:2020nsm,Zhao:2021eaa}.}

\begin{figure}[t]
  \includegraphics[height=.4\textwidth]{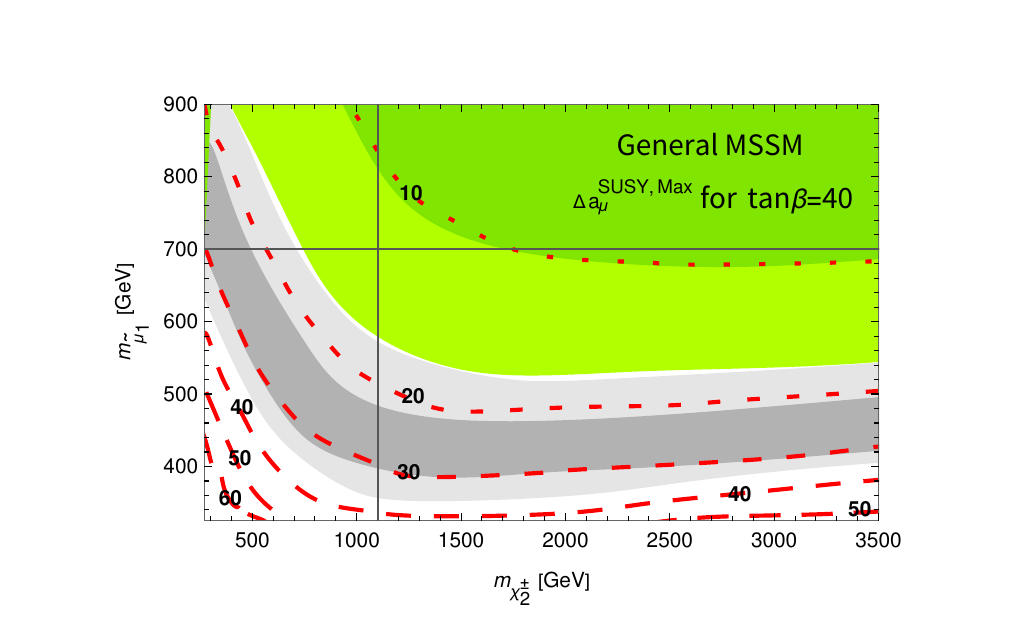}
  \includegraphics[height=.4\textwidth]{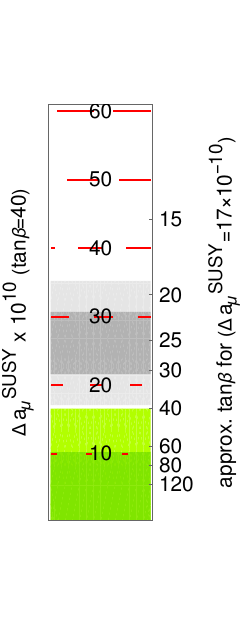}
  \caption{\label{fig:briefsurveyplot}
    The theoretical maximum MSSM contribution $\amu^{\text{SUSY,Max}}$ for
    $\tan\beta=40$ in
    the plane of the heaviest chargino and the lightest smuon
    mass. (For each point in the plane, the actual value of the MSSM
    contribution can take any value between $0$ and $\pm
    \Damu^{\text{SUSY,Max}}$, depending on the signs of parameters and
    details such as other masses and mixings.) The (light) green
    coloured regions show where
    $\Damu^{\text{SUSY,Max}}$ (for $\tan\beta=40$) is within the
    $1\sigma$ ($2\sigma$) bands corresponding
    to 
    $\DamuFinal$, see Eq.\ \eqref{eq:DamuFinal}. The (light) grey
    regions correspond to the value $\DamuOld$   in
    Eq.~\eqref{eq:DamuOld}. 
    The red dashed contour lines can be interpreted in two
    ways. Firstly, they directly correspond  to certain values of
    $\Damu^{\text{SUSY,Max}}$ for $\tan\beta=40$, as indicated in the left
    axis of the legend plot. Secondly, thanks to the approximate
  linearity in $\tan\beta$, each contour can be used to estimate the
  $\tan\beta$ value for which $\Damu^{\text{SUSY,Max}}$ just reaches
  the upper limit of the $2\sigma$ band around Eq.~(\ref{eq:DamuFinal}) (keeping other input
  parameters fixed). These $\tan\beta$ values can be
  read off from the right axis of the legend plot (the values are
  approximate since the linearity is not exact).
      As an example of the reinterpretation we take the point  $m_{\chi^\pm_2} = 1100$ GeV and
    $m_{\tilde{\mu}_1}= 400$ GeV.  For $\tan\beta =40$ we get
      $\Damu^{\text{SUSY,Max}} =30 \times 10^{-10}$. The legend plot
      then shows that $\Damu^{\text{SUSY,Max}} <17 \times
      10^{-10}$ if $\tan\beta\lesssim23$.
    The
      results for $\Damu^{\text{SUSY,Max}}$ were obtained from a scan using
      GM2Calc \cite{Athron:2015rva} in which all relevant SUSY masses
      are varied independently between $100$\,GeV and 4\,TeV.
      The black lines
    indicate the approximate maximum LHC reach for charginos and sleptons of
    $1100$ and $700$ GeV.
  }
\end{figure}

To provide a first overview, Fig.\
\ref{fig:briefsurveyplot} shows the theoretical maximum
of the SUSY   contributions  $\Damu^{\text{SUSY,Max}}$
in the plane of $m _{\chi ^\pm _2}$ and $m_{{\tilde{\mu}} _1}$, where $\chi ^\pm _2$ is the heavier 
chargino and $\tilde{\mu} _1$ the lighter smuon (a version of this
plot was also shown in Ref.~\cite{Athron:2021iuf}, and for similar plots in
different planes see
Refs.\ \cite{Byrne:2002cw,Stockinger:2006zn,Badziak:2014kea}).
It fixes $\tan\beta=40$  and allows all SUSY masses to vary
independently between $100$\,GeV and 
4\,TeV.\footnote{%
  We note that the choice of the mass interval is not crucial because the
  maximum $\Damu^{\text{SUSY}}$ is obtained in the bulk of the mass range, not at
  its boundary. This is a reflection of the fact that the leading
  contributions approximated in Eq.\ (\ref{eq:SUSYMIapprox}) are
  suppressed if Bino, Wino or Higgsino masses are too large or too small.}
The red dashed lines and the green/grey coloured regions correspond
to the indicated values of $\Damu^{\text{SUSY,Max}}$; specifically the
green and grey colours correspond to the $1/2\sigma$ regions for the
current result $\DamuFinal$ and the previous value $\DamuOld$ from
Eqs.\ (\ref{eq:DamuFinal}) and (\ref{eq:DamuOld}),
respectively.
As indicated on the right of the legend plot, an alternative
interpretation of the contour lines and regions is possible. Thanks to
the approximate linearity in $\tan\beta$, each value for $\Damu^{\text{SUSY}}$ with
$\tan\beta=40$ can be translated into approximate $\tan\beta$-values
for which  other values of $\Damu^{\text{SUSY}}$ can be obtained. In the legend
plot this translation is done for $\Damu^{\text{SUSY}}=17\times10^{-10}$,
the new $2\sigma$ upper limit.
For further details on the plot see the caption.

The figure and the previous formulas show that large contributions are
in principle possible, but these are now strongly constrained by the
new result $\DamuFinal$. Using the plot we can read off how the current
constraint can be satisfied if  $\tan\beta=40$:
\begin{itemize}
\item
Masses
$\gtrsim900\ldots1000$ GeV tend to be in the green $1\sigma$ region,
which is ``safe'', i.e.~in this region $\Damu^{\text{SUSY}}$ with
$\tan\beta=40$ is
guaranteed to be sufficiently small.
\item
Smaller masses below the green region can still be possible, since 
 $\Damu^{\text{SUSY,Max}}$ is shown, but they are strongly
constrained and their viability depends on further details of SUSY
masses and parameters.
\end{itemize}
Using the legend plot we can read off further viable parameter regions:
\begin{itemize}
\item
For any value of the masses, the legend plot shows which $\tan\beta$
is ``safe'' (to the extent that the rescaling is exact), i.e.~it shows the value of $\tan\beta$ for
which $\Damu^{\text{SUSY,Max}}$ is definitely below the current
$2\sigma$ upper limit. E.g.\ for $m_{\chi ^\pm _2}=m_{{\tilde{\mu}} _1}=400$ GeV, 
$\tan\beta\le14$ is ``safe''.
\item Again,
  larger values of $\tan\beta$ can be possible, depending on details
of SUSY masses and parameters.
\end{itemize}
The plot also shows in which mass region larger values of $\Damu^{\text{SUSY}}$ are 
possible.
E.g.\
$\Damu^{\text{SUSY}}>20\times10^{-10}$  can 
only occur (for $\tan\beta=40$ and $\mu\le4$\,TeV) if 
\begin{itemize}
\item either both chargino masses are lighter than around 1.1\,TeV 
(vertical black line in Fig.~\ref{fig:briefsurveyplot})
\item
  or one smuon is lighter than around 700\,GeV (horizontal black line  in Fig.~\ref{fig:briefsurveyplot}).
\end{itemize}
While this mass region is of interest also for other reasons, it is in
   potential
  tension with LHC and other data, as discussed below. This tension
  was analysed in  a large part of the literature in the past years.

To explain the behaviour of Fig.~\ref{fig:briefsurveyplot} in more
detail, we note that the WHL contributions (\ref{eq:SUSYWHL}) are by far
most important in the largest part of parameter space. These are
generically suppressed if any of the SUSY masses become large.
However for large $\mu$, the BLR contribution can become sizable and
even dominant due to its approximately linear $\mu$-dependence. In Fig.\ \ref{fig:briefsurveyplot} the slight rise of
$\Damu^{\text{SUSY,Max}}$ with increasing $m_{\chi^\pm_2}$ is due to the
BLR contribution. In contrast, the BHR and
particularly the BHL 
contributions are not important for the maximum value
$\Damu^{\text{SUSY,Max}}$, and in general they are  subdominant unless there
are extremely large mass splittings between left- and right-handed
smuons.\footnote{%
  Specifically the BHR contribution can be decisive if
  $\tan\beta\gg50$ and if $m_\txl\gg m_\txr$ \cite{Bach:2015doa}. For further
  dedicated investigations focusing on parameter situations in which
  the BHL or BHR contributions dominate we refer to
  Refs.\ \cite{Endo:2017zrj,Chakraborti:2022vds}.\label{footnoteBHR}}

\subsubsection{Complementary constraints on MSSM parameters}
\label{sec:SUSYconstraints}

The above formulas, in particular
Eq.~(\ref{DamuSUSYequalmass}), and Fig.~\ref{fig:briefsurveyplot}
illustrate that SUSY contributions to
$\amu$ can be sizeable if the relevant SUSY masses are in the
few-hundred GeV range and $\tan\beta$ is large. This general parameter
region is motivated also independently as a realisation of
SUSY. Accordingly, it has been investigated 
extensively, and the possibility of sizeable $\Damu$ in the MSSM has been
scrutinised using complementary constraints on the parameter
space. Earlier studies of the MSSM phenomenology of $\amu$ of the LHC
run-I 
era can be found in
Refs.\ \cite{Cho:2011rk,Endo:2011gy,Endo:2011xq,Endo:2011mc,Kim:2012fc,Ibe:2012qu,Zhang:2013hva,Bhattacharyya:2013xma,Akula:2013ioa,
	Evans:2013uza,Ibe:2013oha,Endo:2013lva,Endo:2013bba,Iwamoto:2014ywa,Kersten:2014xaa,Badziak:2014kea,Gogoladze:2014cha,Bach:2015doa,
	Ajaib:2015yma,Kowalska:2015zja,Khalil:2015wua,Gogoladze:2015jua,Flores-Baez:2015qdl,Harigaya:2015jba,Harigaya:2015kfa,Chakrabortty:2015ika,
	Padley:2015uma,Chowdhury:2015rja,Li:2016ucz,Okada:2016wlm,Belyaev:2016oxy,Kobakhidze:2016mdx,Belanger:2017vpq,Fukuyama:2016mqb},
more recent studies using detailed recasting of LHC run-II and/or dark
matter data have been done in 
Refs.\ \cite{Endo:2017zrj,Hagiwara:2017lse,Bagnaschi:2017tru,Choudhury:2017acn,Wang:2018vxp,Cox:2018qyi,Abdughani:2019wai,Endo:2020mqz,
	Chakraborti:2020vjp,Chakraborti:2021kkr,Cox:2021nbo,Athron:2021iuf,Chakraborti:2021dli,Chakraborti:2021mbr,Chakraborti:2022vds,
	Chakraborti:2023pis,Endo:2021zal,VanBeekveld:2021tgn,He:2023lgi}. In
particular, the (series of) works in Refs.~\cite{Endo:2013bba,Endo:2020mqz,Endo:2021zal,Chakraborti:2022vds},
Refs.~\cite{Chakraborti:2020vjp,Chakraborti:2021kkr,Chakraborti:2021dli,Chakraborti:2021mbr,Chakraborti:2023pis},
Ref.~\cite{Athron:2021iuf} have systematically analysed the general
MSSM parameter space coherently against LHC and dark matter data.
These latter works are complementary in that they organise the
discussions differently, according to dominant $\Damu$ contributions
\cite{Endo:2013bba,Endo:2020mqz,Endo:2021zal,Chakraborti:2022vds},
according to dark matter
scenarios \cite{Chakraborti:2020vjp,Chakraborti:2021kkr,Chakraborti:2021dli,Chakraborti:2021mbr,Chakraborti:2023pis},
or according to sparticle mass patterns, allowing all possible dark
matter realisations \cite{Athron:2021iuf};
Refs.~\ \cite{Endo:2013bba,Endo:2020mqz,Endo:2021zal,Chakraborti:2022vds,Athron:2021iuf,Chakraborti:2023pis}
allow $\tilde{\tau}$ masses different from other slepton masses,
Refs.~\cite{Endo:2013bba,Endo:2020mqz,Endo:2021zal,Chakraborti:2022vds,Chakraborti:2020vjp,Chakraborti:2021kkr,Chakraborti:2021dli,Chakraborti:2021mbr,Chakraborti:2023pis}
allow different left-handed and right-handed slepton
masses.

It
has turned out that although the LHC strongly constrains the relevant
parameter space, the dark matter constraints are at least as
important. The latter have further strengthened through recent experimental
progress, evaluated first in Ref.~\cite{He:2023lgi}. Here we briefly
summarise the most important constraints. We begin with
dark matter and LHC constraints, then we briefly mention constraints
from the Higgs-boson mass, EDMs and CLFV and related observables.

Dark matter provides a strong motivation to consider SUSY. As long as
there is no R-parity violation  the
lightest SUSY particle (LSP) is stable. If the LSP is a neutralino it
constitutes a viable dark matter candidate, but constraints arise both
from the dark matter relic density 
(DMRD) of $  0.120$, see Eq.~(\ref{Eq:PlanclRelicDensityDM}), and from
dark matter direct detection (DMDD) experiments.
The DMDD experiments place upper limits on the possible cross section
for dark matter--nucleon scattering, and the limits have substantially
improved in recent years in a series of experiments such as PandaX, LUX,
XENON1T and most recently LZ, see Fig.~\ref{LZ-DDlimits} and
particularly Ref.~\cite{LZ:2024zvo}.

Like in the general
discussion of Sec.~\ref{sec:DarkMatter}, the strong constraints on
dark matter and thus on neutralinos
significantly impact the parameter space of interest for
$\amu$. 
Neutralino dark matter has been extensively reviewed in
Ref.~\cite{Roszkowski:2017nbc}, and
Refs.~\cite{He:2023lgi,Martin:2024ytt} provide recent updates relevant for discussing
$\amu$, see also Ref.~\cite{Bisal:2023iip} for a computation of
higher-order corrections for dark matter direct detection. Here we provide a brief summary of the impact of the dark matter
constraints on the SUSY parameter space of interest.
Depending on the 
order of the gaugino and Higgsino masses $M_{1,2}$, $\mu$ entering the
neutralino mass matrix in Eq.~(\ref{NeutralinoMassmatrix}), the nature
of the LSP can be dominantly Bino-like, Wino-like or Higgsino-like,
with sizeable or negligible subdominant admixtures. In the following
we briefly comment on these cases.
\begin{itemize}
\item
  Bino-like LSP:
  In the case of a Bino-like LSP in the considered mass range, the
  relic density is typically too large unless a specific mechanism
  acts to enhance the dark matter annihilation and to suppress the
  relic density. In the mass range of Bino masses of around
  200\ldots600 GeV there are three promising possibilities: stau-coannihilation,
  other slepton-coannihilation, and Wino-coannihilation
  \cite{Roszkowski:2017nbc,Ellis:1998kh,Ellis:1999mm,Nihei:2002sc,Harigaya:2014dwa}.
  For higher masses, a Bino-like LSP can also explain the correct DMRD
  if there is coannihilation with coloured sparticles, squarks or
  gluinos \cite{Boehm:1999bj,Roszkowski:2017nbc}.
  These coannihilation mechanisms are active if the
  masses between the LSP and the coannihilation partner are nearly
  degenerate.

  If a Bino-like LSP explains the observed relic density, the most
  recent DMDD experiments provide very strong upper limits on
  neutralino--nucleon scattering. A Bino-like LSP can interact
  with nucleons via its Higgsino admixture. Hence strong gaugino--Higgsino mixing of the LSP is
  not viable, except in ``blind spots'' which are characterised by
  particular ratios $\mu/m_{\text{LSP}}$, require negative $\mu$ and depend on $\tan\beta$ and
  the CP-odd Higgs boson mass $M_A$\ \cite{Huang:2014xua,Baum:2021qzx,Baum:2023inl}.

  For
  positive $\mu$ we find the approximate bound
  \begin{align}\label{DMDDmulimit}
    \mu\gtrsim 520\text{ GeV}+ M_1 ,
  \end{align}
  based on the analytical approximations of Ref.~\cite{He:2023lgi} and the
  latest LZ result \cite{LZ:2024zvo} in the parameter space where
  $\tan\beta\approx40$ and the extra Higgs masses are large (the
  dependence of the bound on these parameters is mild). This bound
  supersedes the bounds found in Refs.~\cite{Endo:2013bba,Endo:2020mqz,Endo:2021zal,Chakraborti:2022vds,Chakraborti:2020vjp,Chakraborti:2021kkr,Chakraborti:2021dli,Chakraborti:2021mbr,Chakraborti:2023pis,Athron:2021iuf} and is used in the
  following plots, though it is not based on full simulations and thus
  has an intrinsic uncertainty.
\item
  Higgsino- and Wino-like LSP:
   If the LSP is Higgsino-like, there are automatically three almost
   degenerate states $\chi^0_{1,2}$ and $\chi^\pm_1$. Without any
   other  states close in mass, a Higgsino-like LSP generates the correct
   relic density for $M_{\text{LSP}}\approx\mu\approx1$ TeV. In case of Wino-like LSP
   without any other degenerate states, the relic density can be
   explained for $M_{\text{LSP}}\approx M_2\approx2\ldots3$
   TeV. Higgsino-like or Wino-like LSPs with smaller masses typically lead to a
   relic density which is too small. For relevant discussions, including exceptions due to
   coannihilation effects, and for additional dark matter candidates
   we refer to
   Refs.~\cite{Bramante:2015una,Roszkowski:2017nbc,Chakraborti:2017dpu,Kowalska:2018toh,Baer:2018rhs,Beneke:2020vff}.

   Scenarios with underabundant DMRD are viable (if additional
   dark matter candidates from beyond the MSSM are assumed), but still the
   DMDD constraints limit the gaugino--Higgsino mixing. For 
   the particularly interesting case of Higgsino-like LSP without additional
   coannihilation or non-thermal effects, the results of
   Ref.~\cite{Martin:2024ytt} imply the approximate bounds (derived for
   large $\tan\beta=50$, again with mild $\tan\beta$-dependence)
\begin{subequations}
\label{DMDDMlimit}
\begin{align}
\text{for $M_1\gg M_2>\mu$:}&&  M_2&\gtrsim
  1420\text{\,GeV}+3\mu ,\\
\text{for $M_2\gg M_1>\mu$:}&&  M_1&\gtrsim1100\text{\,GeV}+2.3\mu.
   \end{align}
   \end{subequations}
\end{itemize}

Now we turn to collider constraints on the MSSM. Particularly the
non-observation of sparticles at  
the LHC excludes large parts of the MSSM parameter space. Here we summarise the major
conclusions of the studies mentioned at the beginning of this
subsubsection.
The relevant LHC constraints can be grouped into ``standard'' searches
for electroweak sparticles and
searches optimised for compressed spectra.
The standard searches assume the production of heavy electroweak SUSY particles
(charginos/neutralinos or sleptons), followed by decay into 
SM particles and the LSP, which corresponds to missing energy.
For instance, the strongest chargino/neutralino mass limits are obtained from the
pair production channel $pp\to{\chi} ^\pm _1 {\chi} ^0
_2$ with subsequent decay via on-shell sleptons into three charged
leptons and two LSPs. Here the produced charginos/neutralinos could
be Wino-like, while the LSP is a significantly lighter Bino-like
neutralino and the slepton mass is in between. In simplified-model
interpretations, in which 100\%\ decay branching ratios are assumed
and the slepton mass is halfway between the LSP- and the chargino
mass, the limits extend up to  
\begin{align}\label{maximumChamasslimit}
  m_{\chi^\pm_1}(\text{max.~limit \cite{ATLAS:2018ojr,CMS:2017moi}}) \approx1100\text{
  GeV},
\end{align}
which is shown in Fig.~\ref{fig:briefsurveyplot} as
an illustration of the maximum LHC reach.
Direct searches for slepton pairs
  $\tilde{l}\tilde{l},(\tilde{l}=\tilde{e},\,\tilde{\mu})$ with
  subsequent decay into leptons plus LSP have  a
  mass reach up to
  \begin{align}
    m_{\tilde{l}}(\text{max.~limit \cite{ATLAS:2019lff,ATLAS:2018ojr,CMS:2018eqb}}) \approx 700\text{ GeV},
  \end{align}
which is also shown in the Figure.

Further LHC limits arise on  charginos/neutralinos from decays into
other final states which can be available if decays into sleptons are
impossible or subdominant.
  Searches dedicated to compressed spectra, i.e.\ the case where mass
  splittings are less than around 100 GeV exist for a variety of
  scenarios, such as pairs of charginos/neutralinos, or a pair of
  neutralino and slepton. The mass reach of these searches, however,
  extends only up to around 200 GeV; they are thus only relevant for
  very light sparticle masses. We refer to the original literature,
  particularly Refs.~\cite{Endo:2020mqz,Endo:2021zal,Chakraborti:2022vds,Chakraborti:2020vjp,Chakraborti:2021kkr,Chakraborti:2021dli,Chakraborti:2021mbr,Chakraborti:2023pis,Athron:2021iuf}
for
further discussions, references, and the recasting tools used to
  obtain LHC limits for realistic spectra.\footnote{%
For an up-to-date list of LHC SUSY searches and results see the web
pages
\url{https://twiki.cern.ch/twiki/bin/view/AtlasPublic/SupersymmetryPublicResults}
and
\url{https://twiki.cern.ch/twiki/bin/view/CMSPublic/PhysicsResultsSUS}.
}
\footnote{The plots with MSSM results shown in the following are based on
updates of plots in
    Ref.~\cite{Athron:2021iuf}, which in turn uses
  the \texttt{MSSMEFTHiggs\_mAmu} spectrum generator, created with {\fs}
  \cite{Athron:2014yba,Athron:2016fuq,Athron:2017fvs,Kwasnitza:2020wli,Athron:2021kve,Khasianevich:2024hpv}.  \fs also uses some numerical routines
  originally from \cite{Allanach:2001kg,Allanach:2013kza} and uses
  \sarah 4.14.1
  \cite{Staub:2009bi,Staub:2010jh,Staub:2012pb,Staub:2013tta}. It  is
  incorporated in the  \gambit-1.3 framework
  \cite{GAMBIT:2017yxo,GAMBITModelsWorkgroup:2017ilg,GAMBIT:2017qxg,GAMBITDarkMatterWorkgroup:2017fax,Martinez:2017lzg,GAMBITFlavourWorkgroup:2017dbx};
  for LHC recasting, \colliderbit within \gambit \cite{GAMBIT:2017qxg}
    was used. The fundamental
    SUSY parameters are treated as running $\overline{\text{DR}}$-parameters at
    the scale 1 TeV.
}
There are three general ways to evade the LHC constraints in MSSM
parameter space.
\begin{enumerate}
\item
  Sufficiently heavy masses. Clearly, in this case the bounds are
  trivially evaded without considering further details. Heavy SUSY
  with e.g.~$  m_{\chi^\pm_1}\gtrsim1100$ GeV and
  $m_{\tilde{l}}\gtrsim700$ GeV can be attractive for several reasons,
  including dark matter explanations via Higgsino- or Wino-like LSP,
  or in view of certain more fundamental models of SUSY breaking.
\item
  Compressed spectra. If the mass splitting between two relevant
  sparticles is too small typically below around 100 GeV, the standard searches become
  insensitive.  In this case,
  dedicated searches for scenarios with compressed spectra are
  important.
  Interestingly, the coannihilation mechanisms for dark matter, which
  rely on small mass splittings, are not strongly constrained by
  LHC. The mentioned references
  find that in the
  region of 100\ldots500 GeV for dark matter masses, the LHC
  constraints are easily evaded, dark matter can be explained and
  sizeable contributions to $\amu$ are possible.
\item
  Different decay modes. If the SUSY particle in question decays into
  final states that are hard to detect at the LHC, the LHC limits
  weaken. An important example of this kind are Higgsino-like
  charginos. Even in case of the mass hierarchy assumed by the search
  of Refs.~\cite{ATLAS:2018ojr,CMS:2017moi}, where the chargino
  is much heavier than the LSP 
  and the slepton mass is in between, the Higgsino is essentially
  invisible and the Higgsino mass is unconstrained by this search.
\end{enumerate}
In total, the discussion shows that a number of attractive MSSM
scenarios emerge, where dark matter constraints and LHC constraints
can be simultaneously satisfied and, partially, sizeable contributions
to $\amu$ are possible. They are summarised in
Tab.~\ref{tab:MSSMscenarios}.

\begin{table}[t]
\centerline{  \begin{tabular}{|c|c|c|c|c|}
    \hline
    Scenario & Number
    & Mass range (GeV)
    & $\Damu>5\times10^{-10}$ possible
    & Comments
    \\
    \hline\hline
    $\tilde{B}$--$\tilde{W}$-coann. &1
    &
    100\ldots500
    &
    yes
    &
    $\mu$: DMDD, (vacstab)
    \\
\hline
    $\tilde{B}$--$\tilde{l}$-coann.&2
    &
    100\ldots500
    &
    yes
    &
    $\mu$: DMDD, (vacstab)
    \\
\hline
    $\tilde{B}$--$\tilde{\tau}$-coann.&3
    &
    100\ldots500
    &
    yes
    &
    $\mu$: DMDD, vacstab
    \\
\hline
    $\tilde{B}$--various&4
    &
    up to several TeV
    &
    no
    &
    $\mu$: DMDD, vacstab
    \\
\hline
    $\tilde{H}$-LSP full DM&5
    &
    1000
    &
    no
    &
    $M_{1,2}$: DMDD
    \\
\hline
    $\tilde{H}$-LSP partial DM&6
    &
    $<1000$
    &
    yes
    &
    $M_{1,2}$: DMDD
    \\
\hline
    $\tilde{W}$-LSP full DM&7
    &
    $2000\ldots3000$
    &
    no
    &
    $\mu$: DMDD, (vacstab)
    \\
\hline
    $\tilde{W}$-LSP partial DM&8
    &
    $<2000$
    &
    yes
    &
    $\mu$: DMDD, (vacstab)
    \\
\hline
  \end{tabular}
  }
  \caption{%
    Example MSSM scenarios of particular interest for $\amu$, with
    characteristic mass patterns which are 
    generically compatible with constraints from dark matter and
    LHC. For Bino-like LSP the coannihilation partner is indicated
  where appropriate,
    and for Higgsino- or Wino-like LSP the two cases where the dark
    matter relic density is fully or partially explained are
    distinguished.
    In the given mass ranges it is easy to satisfy all
    constraints; the last column indicates whether constraints from
    dark matter direct detection are relevant for gaugino or Higgsino
    masses, and whether the constraint Eq.~(\ref{VacStabEndo}) from vacuum
    stability is relevant. The latter constraint is most relevant in
    case of light staus, which is optionally the case in some
    scenarios. Exceptions to the estimate for $\Damu$ can arise in
    cases with ultra-large $\tan\beta$ or $\mu$, as discussed in Sec.~\ref{sec:HeavySUSY}.
  }
  \label{tab:MSSMscenarios}
\end{table}

Another crucial constraint on the SUSY parameter space results
from the LHC measurement of the SM-like Higgs boson mass
\cite{ParticleDataGroup:2024cfk} 
\begin{align}\label{MSSMHiggsExp}
  M_H&=125.20(11)\text{ GeV}.
\end{align}
The constraint is in general unrelated to $\amu$, except in the 
important class of scenarios where a fundamental mechanism is assumed
to cause a correlation between SUSY parameters of the different
strongly and weakly interacting sectors. In the MSSM, the  prediction
for the SM-like Higgs-boson mass has been developed to a high
precision; for details and a
summary of recent efforts on high-precision computations we refer to
Refs.~\cite{Draper:2016pys,Slavich:2020zjv}. For the purposes here it
can be approximated in terms of only leading tree-level and one-loop
effects for $\tan\beta\gg1$. The leading one-loop quantum corrections are
governed by top-quarks and their stop superpartners  and dominantly depend on the 
mass scale of the stops $m_{\tilde{t}}$ and the left-right stop mixing
parameter $x_t=(A_t-\mu^*/\tan\beta)/m_{\tilde{t}}$. The approximation reads
\begin{align}\label{MSSMHiggsTh}
  M_h^2 &\approx M_Z^2 + \frac{1}{8\pi^2} \frac{m_t^4}{v^2}
  \left[24\ln\left(\frac{M_S}{m_t}\right)
    + x_t^2\left(12-x_t^2\right)\right].
\end{align}
First, the formula
illustrates a unique hallmark of SUSY theories in that the Higgs-boson
mass becomes a prediction instead of an input parameter, and the
prediction is of the order of $M_Z$, in agreement with
observation. Second, in order to obtain quantitative agreement between
experiment and the MSSM prediction, the tree-level value $M_Z$ must be
shifted by very large quantum corrections.
It turns out that
Eq.~(\ref{MSSMHiggsTh}), improved using state-of-the-art calculations, 
reproduces the measurement 
in Eq.~(\ref{MSSMHiggsExp}) e.g.~if either the stop masses are as
heavy as ${\cal O}(10\text{\,TeV})$ leading to a large logarithm, or
if the stop masses are approximately
${\cal O}(3\text{\,TeV})$ and there is large stop mixing,
$|x_t|\approx\sqrt6$, maximising the polynomial in $x_t$.
In specific models of SUSY breaking there exist relations between
smuon masses and stop masses. In such models, the constraints on
the Higgs mass can imply lower mass limits on smuons and thus impact the
possible values of
$\amu$. We will discuss examples e.g.~in
Secs.~\ref{sec:BaselineSUSY}, \ref{sec:HeavySUSY}
below. 

Observables which are more directly connected to $\amu$ are 
leptonic dipole observables  such as
electric dipole moments and CLFV processes like $\mu\to e\gamma$. 
In the MSSM, the relationship between $\amu$ and these other dipole
observables is a
prime example of the general discussion of
Sec.~\ref{sec:LeptonDipole}: either $\Damu^{\text{SUSY}}$ must be tiny
or the SUSY parameters must be CP and flavour conserving to a high
degree. For this reason we will set such CPV an CLFV parameters to
zero in most of the subsequent discussions but provide a short summary here.

In the MSSM, the dipole operators are chirally enhanced and 
satisfy naive scaling unless the
slepton mass terms are very hierarchical \cite{Giudice:2012ms,Crivellin:2018qmi}.
In principle, CP can be violated by complex phases of the Higgsino and
gaugino masses $\mu$, $M_{1,2,3}$ in a flavour-independent way, and
lepton flavour can be violated by off-diagonal elements of the mass
matrices $m^2_{\tilde{l},\tilde{e}}$ or the trilinear interactions
$A_e$.
The MSSM correlations between $\amu$ and the electron EDM and $\mu\to
e\gamma$ have been investigated in detail in
Refs.~\cite{Pospelov:2005pr,Ellis:2008zy,Cheung:2009fc,Han:2021ify} and  
\cite{Graesser:2001ec,Chacko:2001xd,Isidori:2007jw,Kersten:2014xaa,Calibbi:2014yha,Calibbi:2015kja},
and the results are fully in line with the general discussion 
of Sec.~\ref{sec:LeptonDipole}: e.g.~Ref.~\cite{Han:2021ify} finds 
(using a slightly weaker EDM bound than
the one in Eq.~(\ref{deBound})) the implication
\begin{align}
  \text{if }\Damu^{\text{SUSY}}&\approx10\times10^{-10}
&  \text{ then $d_e$ and Eq.~(\ref{deBound}) imply }&
&\left|  \text{arg}(\mu M_{1,2})\right|
&\lesssim 10^{-5}
\intertext{
under rather wide conditions. Similarly,
e.g.~Fig.~13 of Ref.~\cite{Kersten:2014xaa} can be used to read off the constraint}
  \text{if }\Damu^{\text{SUSY}}&\approx10\times10^{-10}
&  \text{ then  $\mu\to e\gamma$ and Eq.~(\ref{MuegBound}) imply }&
&\left|  \frac{m^2_{\tilde{l},12}}{\sqrt{m^2_{\tilde{l},11}
    m^2_{\tilde{l},22}}}\right|
&\lesssim2\times10^{-4}
\end{align}
on the left-handed slepton mixing matrix,
provided all relevant SUSY masses are approximately within a factor 2
of each other.
Similarly rescaled results hold for other values of
$\Damu^{\text{SUSY}}$, and complex phases and flavour mixing of ${\cal
    O}(1)$ are only possible if $\Damu^{\text{SUSY}}$ is many orders
of magnitude below the experimental sensitivity.

Similar analyses have been applied to the specific case where the MSSM
is extended to include neutrino masses and mixings,  and slepton
mixing is correlated to neutrino mixing
\cite{Han:2020exx,Nagai:2020xbq,Nakai:2021mha,Gomez:2024dts}. These references
find 
concrete SUSY models where neutrino mixing and bounds require only
small slepton mass mixing and small CP violating phases. Nevertheless,
they also show that in many particularly straightforward scenarios the
non-observation of CLFV implies small $\Damu$, hence
the current result
$\DamuFinal$ significantly opens up the parameter space of such
models.

Finally we briefly mention  that several studies
\cite{Badziak:2019gaf,Endo:2019bcj,Ali:2021kxa,Frank:2021nkq,Li:2021xmw,Li:2021koa,Cao:2023juc,Cao:2021lmj,Jia:2023xpx}
investigated
whether the MSSM can accommodate simultaneously the significant
previous deviation $\DamuOld$ and the deviation
in $a_e$ between Eqs.~(\ref{aeExp}) and (\ref{aeSM2018}) present at
the time before the appearance of
Ref.~\cite{Morel:2020dww}. Explaining
both is challenging in the MSSM because the signs of the two
deviations was opposite and the magnitudes did not follow naive
scaling, hence sfermion mass parameters would have to be chosen in a
very non-universal way. Similarly, it was investigated whether
sizeable contributions $\amu$ and positive shifts of the $W$-boson
mass $M_W$ are compatible. Under broad conditions, the MSSM was found
to be unable \cite{Yang:2022gvz,Bagnaschi:2022qhb,Athron:2022isz}
to accommodate the CDF measurement of the $W$-boson mass
with the very high value $M_W^{\text{CDF}} = 80.4335(94)\text{ GeV}$
\cite{shortauthordoi:10.1126/science.abk1781}, which is however
in tension with other measurements and excluded from the global average  in
Ref.~\cite{ParticleDataGroup:2024cfk}, see also the discussion in Sec.~\ref{sec:EWPO}.

\subsubsection{Baseline SUSY scenarios}
\label{sec:BaselineSUSY}

If the MSSM is realised in nature, it is expected that the soft SUSY
breaking parameters are effectively generated from some underlying
dynamics involving spontaneous SUSY breaking. In this case the soft
breaking parameters become correlated and functions of more
fundamental parameters. Here we collect a few of the most discussed
and simplest such scenarios. They are interesting in their own right
and provide a baseline for phenomenological discussions and for
understanding the literature. It is,
however, important to note that each of them could be modified and
generalised, and that SUSY could also be realised in quite different
ways. Our discussion follows the review Ref.~\cite{Martin:1997ns}, where
further details and original references can be found.

One well known scenario is called the Constrained MSSM (CMSSM) or
minimal supergravity (mSUGRA), see
e.g.~Ref.~\cite{Nilles:1983ge}. Here the fundamental theory is assumed 
to be supergravity, where local SUSY is spontaneously broken in a
hidden sector and where gravitational interactions mediate the SUSY
breaking into the MSSM sector, leading to effective soft SUSY breaking
terms. In the minimal setup it is assumed that the running soft
parameters become universal at a high scale around the GUT or Planck
scale. The CMSSM thus has the free SUSY parameters
\begin{align}
  m_0, m_{1/2}, A_0, \tan\beta,
\end{align}
where $m_{0,1/2}$ correspond to the unified scalar and gaugino masses
at the high scale, and $A_0$ to the unified $A$-parameters. Further
parameters such as the Higgsino mass $\mu$ and the Higgs mass spectrum
are then predicted. As an example, the running soft parameters at the
TeV-scale can then be approximated as \cite{Drees:1995hj}
\begin{align}
  m_{\tilde{l}}^2 &\approx m_0^2 + 0.5 m_{1/2}^2,\\
  m_{\tilde{e}}^2 &\approx m_0^2 + 0.15 m_{1/2}^2,\\
  m_{\tilde{q}}^2 &\approx m_0^2 + 6.5m_{1/2}^2,\\
  m_{\tilde{u}}^2 &\approx m_0^2 + 6m_{1/2}^2,\\
  m_{\tilde{d}}^2 &\approx m_0^2 + 6m_{1/2}^2,\\
  M_{1,2,3}  &\approx \frac{\alpha_{1,2,3}}{\alpha_{\text{GUT}}} m_{1/2},  
\end{align}
where the $\alpha_i$ in the last equation are GUT-normalised
finestructure constants, which result in the gaugino mass pattern
\begin{align}
  M_1:M_2:M_3 &\approx (1:2:6)\,0.4\, m_{1/2}.
\end{align}
This means
that a pattern emerges between the different gaugino and scalar
masses, where the coloured gluinos and squarks are the heaviest sparticles, while the sleptons, Bino and Wino are lighter by fixed
factors.

Phenomenologically, the CMSSM is constrained by the very
strong LHC mass limits on gluinos and squarks, where particularly the
gluino mass limits imply robust lower limits on the Bino and Wino
masses of the order $M_2\approx2M_1\gtrsim 1$ TeV. Further, the
Higgs-boson mass $M_h\approx125 $ GeV requires heavy stops or large
mixing, and the $\mu$-parameter is a dependent quantity. All of these
constraints imply rather large masses and thus constraints on the
contributions to $\amu$, see Fig.~\ref{fig:briefsurveyplot} as well
as on the possibilities to accommodate the dark matter relic density
(DMRD).
The remaining CMSSM parameter space corresponds to scenarios 4, 5 or 6 in
Tab.~\ref{tab:MSSMscenarios}. As a result, the CMSSM is incompatible with very large contributions
to $\amu$, as already found in global fits after the first LHC runs
\cite{Buchmueller:2013rsa,Bechtle:2015nua,Han:2016gvr,GAMBIT:2017snp}. The
more recent update \cite{Chakraborti:2021bmv} 
finds the maximum values
\begin{align}\label{amuCMSSM}
  \Damu^{\text{CMSSM}}\left\{\begin{array}{ll}
 < 2\times10^{-10} & \text{(DMRD fully explained),}\\
 < 4\times10^{-10} & \text{(DMRD not overabundant).}
  \end{array}\right.
\end{align}
In view of the progress on $\amu$ leading to the result $
\DamuFinal$ in Eq.~(\ref{eq:DamuFinal}), the CMSSM prediction
is actually well in agreement with observations and thus constitutes a
viable scenario for SUSY.

In the literature, many extensions of the
CMSSM were proposed and investigated. Some of them were partially motivated by
the previous large deviation $\DamuOld$. For instance,
a simple generalisation is to relax only the unification between the
three gaugino masses, such that $M_{1,2,3}$ become a independent input
parameters while other GUT
constraints still apply. This  is
sufficient to allow large values of $\Damu^{\text{SUSY}}\sim20\times10^{-10}$ together with explaining
the dark matter via Bino-like LSP with slepton coannihilation in the
few-100 GeV mass region
\cite{Akula:2013ioa,Gogoladze:2014cha,Chakrabortty:2015ika,Wang:2018vrr,Aboubrahim:2021xfi,Chakraborti:2021bmv,Li:2021pnt,Ahmed:2021htr,Wang:2021bcx,Wang:2022rfd,Ellis:2024ijt}, even
in the ``no-scale'' special case where $m_0$ is negligible
\cite{Forster:2021vyz}. The LHC and DMDD constraints can also be
fulfilled
\cite{Chakraborti:2021bmv,Li:2021pnt,Ahmed:2021htr,Wang:2021bcx,Forster:2021vyz}, and some scenarios can also
be augmented to SO(10) GUTs where even 3rd generation Yukawa couplings
unify \cite{Aboubrahim:2021phn}.
Similarly,
relaxing the unification between $M_1$ and $M_2$ also allows the
cases of Higgsino-like or Wino-like LSP
\cite{Iwamoto:2021aaf}.
Interestingly, relaxing only the unification between the scalar masses
is not always sufficient to open the parameter space and allow
sizeable $\Damu^{\text{SUSY}}$. E.g.\ the SU(5) GUT motivated unification into four
different scalar masses $m_{10}$, $m_5$, $m_{H_u}$, $m_{H_d}$ for each
kind of GUT multiplet is not
sufficient \cite{Bagnaschi:2016afc}. In contrast, 
again sizeable contributions to $\amu$ are possible provided at least
the 3rd-generation sfermions are disunified from the other sfermions
\cite{Ibe:2013oha,Okada:2016wlm,Hussain:2017fbp,Tran:2018kxv,Ibe:2019jbx,Ellis:2024ijt}, though
Ref.~\cite{Ibe:2019jbx} finds strong constraints from dark matter
unless also gaugino masses are disunified. Similarly, the gaugino and scalar mass patterns
corresponding to so-called flipped SU(5) allow sizeable $\Damu^{\text{SUSY}}$
\cite{Ellis:2021zmg,Lamborn:2021snt,Ellis:2021vpp}.
Further GUT-inspired extensions of the CMSSM with sizeable $\Damu$ are based on the  left-right
symmetric Pati-Salam
gauge group \cite{Belyaev:2016oxy} (here, also flavour symmetry is
considered and light Winos
are preferred), or Pati-Salam symmetry with specific Higgs sector
\cite{Gomez:2022qrb,Gomez:2024dts} (which is compatible
with Bino-like LSP explaining the dark matter relic density).

A second class of basic scenarios is gauge-mediated SUSY breaking
\cite{Giudice:1998bp}.
Though
this also assumes an underlying supergravity theory and a hidden
sector, the interactions producing the soft SUSY breaking terms are
the known and renormalizable $\GSM$ gauge
interactions.  In the minimal setup (mGMSB), a set of $N_5$ messenger fields
which form complete SU(5) GUT multiplets and thus couple to all
quark, lepton and gauge fields is assumed. The basic parameters are
then
\begin{align}
 N_5,  M_{\text{mess}}, F_Z/A_Z, \tan\beta,
\end{align}
where $  M_{\text{mess}}$ is the messenger mass, $F_Z,A_Z$ are \vev{}s
relevant for SUSY breaking in the hidden sector. Computing the
effective MSSM soft SUSY breaking parameters in mGMSB leads to
patterns of scalar masses which differ from the CMSSM case,
while the gaugino mass pattern turns out to be the same,
\begin{align}
  m^2_{\tilde{F}}&=2\left(\frac{F_Z}{A_Z}\right)^2N_5\sum_{i=1,2,3}C_i(F)\left(\frac{\alpha_i}{4\pi}\right)^2
  \qquad(\tilde{F}\in\{\tilde{q},\tilde{u},\tilde{d},\tilde{l},\tilde{e}\}),
  \\
  M_{1,2,3}&=\frac{\alpha_{1,2,3}}{4\pi}\frac{F_Z}{A_Z}N_5.
\end{align}
The values of the Casimir invariants $C_i(F)$ can be  found e.g.~in
Ref.~\cite{Martin:1997ns}. The above equation is imposed for the running
parameters at the scale $M_{\text{mess}}$, and running according to
MSSM $\beta$ functions is used to compute the final values of the soft
breaking parameters at the TeV scale.

A big advantage of mGMSB is that the flavour universality of the gauge
interactions automatically implies that the scalar masses are flavour
universal, thus avoiding large unobserved flavour violating
effects, see Sec.~\ref{sec:SUSYconstraints}. This is in contrast to
the CMSSM scenario where the 
assumption of flavour universality, leading to the parameter $m_0$, is
ad hoc. At the same time, the $A$-parameters turn out to vanish in the
mGMSB scenario. Because of this, however, the mixing left-right mixing in the
stop sector is very small, and the Higgs-boson mass $M_h\approx125 $
GeV can only be accommodated for stop masses of at least around 10
TeV. As a result the slepton masses are very large and contributions
to $\amu$ are tiny in mGMSB. A broad parameter scan in
Ref.~\cite{Ibe:2021cvf} shows that
\begin{align}\label{amumGMSB}
  \Damu^{\text{mGMSB}}<0.5\times10^{-10}
\end{align}
if the Higgs-boson mass is
correctly accommodated in the mGMSB scenario.
This upper limit is even smaller than the one in the CMSSM in
Eq.~(\ref{amuCMSSM}), but it is also in agreement with the current
status given in Eq.~(\ref{eq:DamuFinal}). 
Hence the mGMSB scenario is
viable in view of $\amu$, and we refer to the literature for further
properties of the scenario with respect to dark matter and LHC data.

Given the very rigid correlations in mGMSB and the tiny prediction
(\ref{amumGMSB}) it might seem less promising to obtain larger values in
generalisations of the mGMSB setup. Still,
Refs.~\cite{Bhattacharyya:2018inr,Ibe:2021cvf,Evans:2022oho} have considered generalisations 
of mGMSB where the messenger sector is non-minimal, leading to an
increased freedom in the relationships between squark, slepton and
gaugino masses. The resulting SUSY breaking mechanisms are viable,
can explain the
absence of observed CP-violating or CLFV effects, can lead to
rather light Wino and Higgsino. The scenarios can also lead to
significant $\Damu^{\text{SUSY}}$ and are thus constrained by the small current value
$\DamuFinal$. 
Also, Ref.~\cite{Kim:2024bub} uses standard SU(5) GUT unification
conditions in conjunction with gauge-mediated SUSY breaking and finds viable
parameter space and also potentially sizeable $\Damu^{\text{SUSY}}$. However, it
also quantifies a tension between 
proton decay constraints and $\amu$ where smaller values of $\Damu$ as
preferred now widen
the allowed parameter space for the fundamental SUSY breaking mechanism.

A third basic scenario of SUSY breaking is called anomaly-mediated
SUSY breaking (AMSB) \cite{Randall:1998uk,Giudice:1998xp}. Here
the mediation mechanisms present in the CMSSM and mGMSB are assumed to
be absent, and what remains are universally present contributions
governed by the superconformal anomaly, i.e.~by the existence of
non-vanishing $\beta$ functions. As a result, the gaugino masses
satisfy the pattern
\begin{align}\label{AMSBgauginomasses}
  M_1:M_2:M_3 &= \beta_1:\beta_2:\beta_3\approx
  3.3:1:9.
\end{align}
The importance of the modified gaugino mass patterns has also been stressed
in Ref.~\cite{Choi:2007ka}. In the very minimal AMSB setup,
the slepton masses turn out to be tachyonic, which rules out the setup
\cite{Martin:1997ns}. Extended models have been constructed, however, and
these can lead to interesting spectra where the phenomenology differs
substantially from the cases of the CMSSM or mGMSB.  For instance,
in Ref.~\cite{Yin:2021mls}, AMSB with the gaugino relation
Eq.~(\ref{AMSBgauginomasses}) and resulting Wino-like LSP is
analysed. Changing the minimal setup for the scalars generates a
viable scenario, where the Wino-LSP and the  sleptons can have masses
as low as about 700 GeV and $\Damu^{\text{SUSY}}$ can reach more than
$20\times10^{-10}$. A further generalisation allowing modified gaugino
mass patterns is considered in Ref.~\cite{Jeong:2021qey}, such that
also a Bino-like LSP and very large $\mu$ becomes possible, again
leading to sizeable $\Damu^{\text{SUSY}}$.

In what follows, we will comment on the phenomenology in scenarios
motivated by these and further model-building efforts, but also on the
general MSSM phenomenology.

\subsubsection{Phenomenological results for heavy sparticles in the MSSM}
\label{sec:HeavySUSY}

Here and in the following subsubsections 
we discuss the SUSY $\amu$ phenomenology in more detail. We will
consider models with high-scale constraints as discussed in
Sec.~\ref{sec:BaselineSUSY} as well as the general MSSM, where we only
impose assumptions on parameters mentioned in
Sec.~\ref{sec:MSSMDefinition}.
Here we will begin with heavy SUSY, essentially corresponding to the 
upper right quadrant of the plot in  Fig.~\ref{fig:briefsurveyplot},
delineated by the black lines. 
For smuon masses above 700 GeV and chargino masses above 1100 GeV
 the LHC limits are
trivially fulfilled for arbitrary mass splittings, see
Sec.~\ref{sec:SUSYconstraints}. As a simple guideline,
Fig.~\ref{fig:briefsurveyplot} shows that
\begin{align}\label{DamuSUSYmax1TeV}
  \Damu^{\text{SUSY,max}}(\tan\beta=40,m_{\text{LSP}}\ge1\text{ TeV}) <10\times10^{-10}.
\end{align}

As mentioned in Sec.~\ref{sec:BaselineSUSY}, several of the well-known
SUSY scenarios such as the CMSSM or mGMSB can accommodate the current
LHC and $M_h$ constraints only for rather heavy SUSY particles, and
they thus predict small $\Damu^{\text{SUSY}}$; concrete values are
given in Eqs.~(\ref{amuCMSSM},\ref{amumGMSB}).  These small values are now compatible with the
small current $\DamuFinal$. The observed dark matter can also
be accommodated in such scenarios, such as the CMSSM with Bino-like
LSP with in a so-called funnel region where the Higgs mass
$M_A\approx2m_{\text{LSP}}$ or with Higgsino-like LSP with
$m_{\text{LSP}}\approx$ 1 TeV
\cite{Kowalska:2015kaa,Roszkowski:2017nbc,Kowalska:2018toh}.
In this way, such simple
scenarios re-emerge as well motivated potential realisations of SUSY,
although the recent DMDD bounds from the LZ experiment exclude
significant portions of the parameter space
\cite{LZ:2024zvo}.\footnote{%
Comparing the CMSSM predictions in Ref.~\cite{Kowalska:2018toh} with
the LZ limit appears to essentially exclude the entire Higgsino-like
dark matter region of the CMSSM, although a detailed analysis is
lacking. }

In general, even independently of the CMSSM or similar setups,
the scenarios with Higgsino-like LSP with mass around 1
TeV or Wino-like LSP with mass around 2--3 TeV are very attractive:
These scenarios lead to explanations of the observed dark matter
relic density without further tuning of masses or mass splittings and
can be realised in the constrained MSSM or in more general variants of the MSSM (see
e.g.\ Ref.\ \cite{Roszkowski:2017nbc}).
They are listed as scenarios 5, 7 in Tab.~\ref{tab:MSSMscenarios}.
However, all such scenarios
restrict $\Damu^{\text{SUSY}}$ to very small values (actually much
smaller than visible in Eq.~(\ref{DamuSUSYmax1TeV}) because of dark
matter constraints) and are thus
compatible with observation but not constrained by
$\DamuFinal$.

It is interesting to note that there exist heavy SUSY scenarios with
rather large contributions to $\amu$. They are not visible in the plot
of Fig.~\ref{fig:briefsurveyplot}, but they involve ultra-high values
of $\tan\beta$  or of  $\mu$.
In both cases
the linearity in
  $\tan\beta$ and $\mu$ visible in Eqs.\ (\ref{eq:SUSYMIapprox}) is replaced by a
  saturation resulting from resummed higher-order effects as
  illustrated in Eq.~(\ref{muonmassMSSM1L}). And the vacuum stability
  constraint related to the off-diagonal slepton mass matrix elements and
  illustrated in Eq.~(\ref{VacStabEndo}) implies upper 
  limits on the combination $\mu\tan\beta$. Such
  extreme parameter regions have been studied in several directions.

First, by increasing primarily the Higgsino mass parameter $\mu$, the BLR
contribution to $\amu$ becomes dominant since it is the only one
without Higgsino propagators. It is then possible to analyse the
overall maximum possible value of the BLR contribution by maximising
$\mu$ for any given value of the smuon and Bino masses \cite{Endo:2013lva,Chigusa:2022xpq,Chigusa:2023mqy}. It
turns out that $\Damu^{\text{SUSY}}$
as large as $25\times10^{-10}$ is possible even if the LSP mass is
around 1 TeV. More generally, Ref.~\cite{Chigusa:2023mqy} finds e.g.
\begin{align}\label{SUSYBLRmax}
  \Damu^{\text{SUSY,BLR,max}}&=\left\{
  \begin{array}{ll}
    25\times10^{-10} & \text{ for }m_{L,R}=M_1=1\text{ TeV}\\
    25\times10^{-10} & \text{ for }m_{L,R}=1.2\text{
      TeV, } M_1=600\text{ GeV}\\
    25\times10^{-10} & \text{ for }m_{L,R}=m_{\tilde{\tau}}=M_1=250\text{ GeV}
  \end{array}
  \right.
\end{align}
where in the first two lines the stau masses are assumed to be large
and irrelevant for vacuum stability and the last line is included as a
contrast in case of degenerate stau/slepton masses. It has also been found that a ratio $m_L=m_R$ is
optimal for large contributions; similar limits also exist in the NMSSM \cite{Wang:2021lwi}.
Such a scenario with ultra-large
$\mu$ and TeV-scale Bino has also been investigated in
Ref.~\cite{Gu:2021mjd}, which showed that dark matter can be
accommodated if--- as motivated by gauge-mediated SUSY breaking ---
the MSSM neutralinos decay into a lighter gravitino which constitutes
dark matter.

Second, by increasing $\tan\beta$ to ultra-large values a similar
behaviour can be obtained, provided $\tan\beta$ becomes large enough to compensate a loop
factor such that the muon mass correction $\Delta_\mu$ becomes of
${\cal O}(1)$, see Eqs.~(\ref{eq:sigmamuparts})
\cite{Dobrescu:2010mk,Altmannshofer:2010zt,Bach:2015doa}. The limit
$\tan\beta\to \infty$ exists if the resummation
Eq.~(\ref{muonmassMSSM1L}) is taken into account, but it leads to a
larger variety of contributions than the large-$\mu$ limit.

The actual
$\tan\beta\to\infty$ limit is appealing since it corresponds to
$v_d=0$ and vanishing tree-level muon mass, and thus to one way to
realise radiative muon mass generation in the MSSM
\cite{Bach:2015doa}. The mass is generated via the one-loop coupling of the muon to
the ``wrong'' Higgs \vev $v_u$, described by the quantity
$\Delta_\mu$. Thanks to the resummation in
Eqs.~(\ref{muonmassMSSM1L},\ref{amuSUSYdecomposition}), the final result for $\Damu^{\text{SUSY}}$ in this
scenario becomes a ratio of the form $\text{1-loop}/\text{1-loop}$,
like the generic examples of Sec.~\ref{sec:MuonMass}. Many factors and
sign dependences
cancel between numerator and denominator, and the final result mainly
depends on which of the contributions dominate. 
Ref.~\cite{Bach:2015doa} finds e.g.\ the approximations
\begin{align}\label{amuTBinfinity}
  \Damu^{\text{SUSY},\tan\beta\to\infty}\approx
  \left\{\begin{array}{ll}
-72 \times 10^{-10} \,
\left(\frac{\text{1 TeV}}{M_{\text{SUSY}}}\right)^2
&
\text{ for }|\mu|= |M_2|= |M_1|=m_L=m_R\equiv M_{\text{SUSY}},
\\ 
+37\times10^{-10} \,
\left(\frac{\text{1 TeV}}{M_{\text{SUSY}}}\right)^2
& \text{ for }m_L\gg |\mu|=|M_1|=m_R\equiv M_{\text{SUSY}}.
  \end{array}
\right.
\end{align}
Here the first result corresponds to universally equal SUSY masses and
is negative definite,
the second case is positive definite and corresponds to heavy $m_L$, where the BHR contribution to $\amu$
dominates. For heavy $\mu$ and $\tan\beta\to\infty$, the result
of Eq.~(\ref{SUSYBLRmax}) is recovered. 

The common feature of both the large-$\mu$ and $\tan\beta\to\infty$
limits is the strong modification of the muon mass generation
mechanism. Two further scenarios with different modifications which
also allow large $\Damu^{\text{SUSY}}$ for heavy SUSY masses are
proposed in
Refs.~\cite{Borzumati:1999sp,Crivellin:2010ty,Thalapillil:2014kya,Altmannshofer:2021hfu,Ke:2021kgy}.
The first of these ideas is to realise radiative
muon mass generation by setting the Yukawa coupling $y_\mu=0$ while
keeping $v_d\ne0$
\cite{Borzumati:1999sp,Crivellin:2010ty,Thalapillil:2014kya} and generating $m_\mu$ via loops involving
non-standard soft SUSY breaking parameters.
The second idea is to introduce additional Higgs doublets into the
theory \cite{Altmannshofer:2021hfu}. In this way the $\tau$-lepton
mass is obtained from a different \vev than the muon mass, hence
effectively different $\tan\beta$'s
apply in the two sectors, which allows  having larger effective
$\tan\beta^{\text{eff}}_{\mu}$ in the muon sector than in the tau sector,
alleviating e.g.~vacuum stability constraints while keeping agreement
with the observed Higgs-boson mass \cite{Ke:2021kgy}. All of these
proposals can lead to contributions to $\amu$ of similar magnitude as
Eqs.~(\ref{SUSYBLRmax},\ref{amuTBinfinity}).

\subsubsection{Phenomenological results with light Bino-like LSP in the MSSM}

\begin{figure}[t]
  \begin{subfigure}[t]{0.405\textwidth}
    \centering \includegraphics[scale=.7]{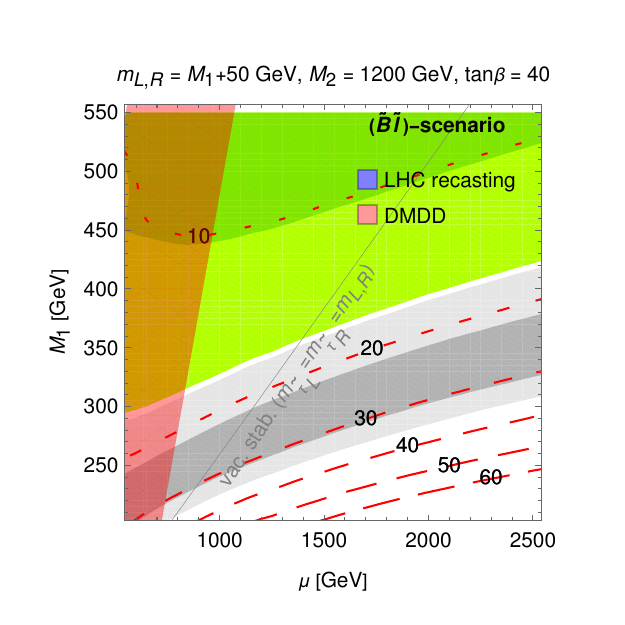}
    \caption{}
    \label{fig:Binosleptonscompresseda}
  \end{subfigure}
  \begin{subfigure}[t]{0.405\textwidth}
    \centering\includegraphics[scale=.7]{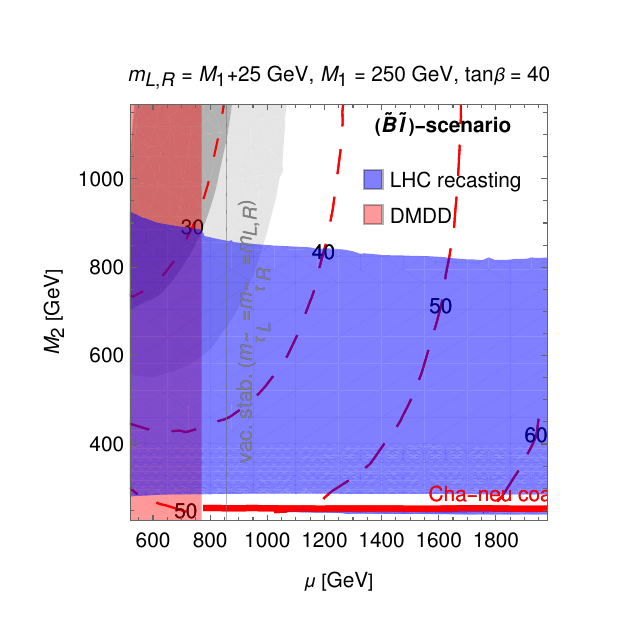}
    \caption{}
    \label{fig:Binosleptonscompressedb}
  \end{subfigure}
  \begin{subfigure}[t]{0.18\textwidth}
    \centering\includegraphics[scale=.7]{Plots/PlotAmuLegend.pdf}
  \end{subfigure}
  \caption{\label{fig:Binosleptonscompressed} 
    MSSM in the $(\tilde{B}\tilde{l})$-scenario with Bino-like LSP and light
    sleptons. In the plots either $M_2=1200$ GeV is fixed
    or $M_1=250$ GeV is fixed, and the remaining chargino and slepton
    masses are varied. For the exact parameter values see the
    plots. The meaning of the red dashed lines, the green/grey coloured
    $\Damu^{\text{SUSY}}$ regions, and the reinterpretation in the
    legend plot are as in Fig.\ \ref{fig:briefsurveyplot}.
    The red shaded regions are excluded by dark matter direct detection
    if the LSP is assumed stable; the blue shaded region corresponds
    to LHC limits, see text for details.
    The red thick
    solid line in the right plot corresponds to the parameter strip
    where chargino--neutralino coannihilation is possible; directly
    below this strip a tiny region is excluded by the LHC-constraint
    from compressed masses, Ref.\ \cite{CMS:2018kag}, but we verified that this
    does not exclude the chargino--neutralino coannihilation region.
    The grey thin line corresponds to the vacuum
    stability constraint of Ref.\ \cite{Endo:2013lva}, see Eq.~(\ref{VacStabEndo}); it applies in case the
    left- and right-handed stau-masses are set equal to the
    smuon/selectron masses and excludes the points to the right,
    i.e.\ with larger $\mu$. 
}
\end{figure}

Next we consider the $\amu$ phenomenology of lighter SUSY particles in
connection with LHC and dark matter data, here focusing on the case of
a Bino-like neutralino LSP. Mass spectra with Bino-like LSP are frequently
studied and well motivated by the gaugino mass constraint $M_1\approx
M_2/2$ of the CMSSM, mGMSB or many of their extensions, see
Sec.~\ref{sec:BaselineSUSY}. In particular, the case of Bino-like LSP
has been analysed in view of LHC run-II data and the significant $\amu$ deviation already
before the Fermilab $g-2$ experiment in
Refs.~\cite{Hagiwara:2017lse,Cox:2018qyi,Abdughani:2019wai,Endo:2020mqz,Chakraborti:2020vjp,Chakraborti:2021kkr}, 
in
Refs.\ \cite{Cox:2018qyi,Abdughani:2019wai,Chakraborti:2020vjp,Chakraborti:2021kkr}
also including dark matter;
after the Fermilab $g-2$ Run-1 result,
Refs.~\cite{Cox:2021nbo,Athron:2021iuf,Chakraborti:2021dli,Chakraborti:2021mbr,Chakraborti:2022vds,Chakraborti:2023pis,Endo:2021zal,VanBeekveld:2021tgn,He:2023lgi}
presented further very detailed general analyses, partially also
including other scenarios.\footnote{%
We refer to Ref.~\cite{Athron:2021iuf} for more detailed comments on
comparisons between these studies.}
Here we summarise the status.

The mass range of interest is $M_1$
in the few-100 GeV region, while some slepton  and/or chargino masses are
below the 700 and 1100 GeV limits used in Sec.~\ref{sec:HeavySUSY} to
characterise heavy SUSY. The Bino-like LSP
is neutral and stable, hence a dark matter candidate. In general,
the considered mass region leads to small
annihilation cross sections and thus to a relic density which is too 
large, unless one of the following coannihilation mechanisms is
possible \cite{Roszkowski:2017nbc}, see also scenarios 1, 2, 3 in
Tab.~\ref{tab:MSSMscenarios}:
\begin{itemize}
\item
  Slepton coannihilation, where one or several sleptons (possibly also
  staus) are very close in mass to the LSP;
\item
  Stau coannihilation, where one or two staus are very close in mass
  (but the sleptons significantly heavier)  to the LSP;
\item
  Wino coannihilation, where the Wino-like neutralino and chargino are
  very close in mass  to the LSP.
\end{itemize}
Intriguingly, in each case the coannihilation partner is hidden from
LHC because of the small mass splitting (exceptions may arise in the
very-low mass region below around 200 GeV where dedicated
compressed-mass searches are relevant). Hence the coannihilation
mechanisms are essentially unconstrained by LHC. The masses of further
SUSY particles, however, are important for $\amu$ and constrained by
LHC, dark matter and other considerations.

The interplay of
constraints is illustrated in Fig~\ref{fig:Binosleptonscompressed}, which 
updates results of Ref.~\cite{Athron:2021iuf} for the Bino-LSP
scenario with additionally light sleptons evading LHC limits. 
The left plot
Fig.~\ref{fig:Binosleptonscompresseda} shows results in the
$\mu$--$M_1$-plane. The Wino mass is fixed to the
rather high value $M_2=1200$ GeV, safely but not too far above the
maximum LHC chargino mass limit (\ref{maximumChamasslimit}); the
mass splitting is fixed to the reference value $m_{\txl, \txr}-M_1=50$
GeV. The right plot 
Fig.\ \ref{fig:Binosleptonscompressedb} shows results in the
$\mu$--$M_2$-plane, while the Bino and slepton masses are fixed to the
rather light values $M_1=250$ GeV and
$m_{\txl, \txr}=275$ GeV. The plots are evaluated for $\tan\beta=40$,
but most quantities except $\amu$ do not strongly depend on
$\tan\beta$.

Along the thick red strip in the right plot, Wino-coannihilation takes
place and generates the correct DMRD. Outside of this strip,  both
plots are not very
sensitive to the precise choice of the slepton  and stau masses, so the plots are
representative for a wider range of values for $m_{\txl, \txr}-M_1$;
in particular it would be possible to finetune the values of the
slepton and stau masses to fit the correct DMRD everywhere 
(but the finetuning  is not  unique and hence not done  \cite{Athron:2021iuf}).

Assuming now that the relic density is correctly explained,
the constraints from direct detection experiments are shown as the
(light) red shaded bands; they exclude a large portion of the parameter space with
small $\mu$, corresponding to the bound given in
Eq.~(\ref{DMDDmulimit}) in Sec.\ \ref{sec:SUSYconstraints}. Hence
the Higgsino mass must be rather large. There is, however, also a potential
upper limit on the Higgsino mass,  shown as the thin solid grey line
in the plots. It corresponds to the vacuum
stability constraint of Ref.\ \cite{Endo:2013lva} on stau-mixing already explained
around Eq.\ (\ref{VacStabEndo}). It excludes the large-$\mu$ region to its right
under the condition that both left- and
right-handed stau masses are as light as the
smuon/selectron masses. This upper limit on $\mu$ thus applies in particular if
stau-coannihilation and $m_{\tilde{\tau}_\txl}\approx m_{\tilde{\tau}_\txr}$
is assumed. Conversely, the region to the right of the grey line is
allowed provided at least one stau is
significantly heavier. Further, in interpreting the
interplay between the DMDD limit and the vacuum stability limit on
$\mu$ one should note that the vacuum stability limit essentially
constrains $\mu\tan\beta$, while the DMDD limit has a mild sensitivity
on $\tan\beta$. Hence for smaller $\tan\beta$, the allowed region for
$\mu$ widens considerably even in case of light staus.

This
discussion also changes in case of negative sign of $M_1$, where there are
cancellations in the DMDD cross section, also allowing smaller $\mu$
\cite{Huang:2014xua,Baum:2021qzx,Baum:2023inl}. The parameter space
considered in these references corresponds to the Wino-coannihilation
region of Fig.\ \ref{fig:Binosleptonscompressedb}, but with smaller
allowed $\mu$, which can alleviate  electroweak finetuning.
We also refer to Ref.~\cite{Acuna:2021rbg} for a study of the
Bino-slepton scenario in a simplified setup without the additional
MSSM fields.

LHC-constraints obtained as described in Ref.~\cite{Athron:2021iuf}
are
displayed by the blue shaded region in the plots. 
The parameter space of the left plot
Fig.\ \ref{fig:Binosleptonscompresseda} is entirely allowed. The right plot
Fig.\ \ref{fig:Binosleptonscompressedb} shows a large excluded
region approximately for $300\text{ GeV}<M_2<900\text{ GeV}$. Here the
channel of Wino-like chargino/neutralino production with decay into
sleptons is relevant. An additional strip of parameter
space at around $M_2\approx M_1-5$ GeV (in which case the mass
eigenvalues satisfy $m_{\chi^\pm_1}-m_{\chi^0_1}\approx15$ GeV) is
excluded by dedicated compressed-mass searches.

Before discussing $\amu$ constraints, we highlight that in the plots  we can identify
the three generally viable  parameter regions mentioned above, corresponding to the three ways to
accommodate dark matter (scenarios 1, 2, 3 in
Tab.~\ref{tab:MSSMscenarios}): Wino coannihilation is possible along the
thick red strip; stau/slepton coannihilation with equal stau and
slepton masses is possible to the left of the grey vacuum stability
constraint, with either very light or very heavy Wino, and slepton
coannihilation with heavy staus is the only option 
to the right of the vacuum stability constraint at very large $\mu$.

On this parameter space, the red dashed lines and the green coloured
regions show the contours of $\amu$ for $\tan\beta=40$ and the allowed region
corresponding to the new result 
$\DamuFinal$. To guide the eye, the $1/2\sigma$ regions corresponding to
the old result $\DamuOld$ are shown in grey. 
The behaviour of $\amu$ in this Bino-LSP scenario with light sleptons is dominated by the WHL and BLR
contributions of
Eqs.\ (\ref{eq:SUSYMIapprox}) and can be well
understood via these approximations. The
WHL contributions dominate in the left plot at large $M_1$ and
very small $\mu$ and in the right plot at $\mu\lesssim1 $ TeV; in these
regions $\Damu^{\text{SUSY}}$ decreases with increasing $\mu$. The BLR contributions
are linearly enhanced by $\mu$ and dominate at large $\mu$ in both
plots. As the plots show very large $\Damu^{\text{SUSY}}$ can be obtained both for large $\mu$,
where the BLR-contribution dominates, and for small $\mu$ with 
WHL-dominance.

The results for $\amu$ illustrate why this parameter
region was considered a promising explanation of the old deviation
(\ref{eq:DamuOld}). Now, the small value of $\DamuFinal$ places tight
bounds on the parameter space. For fixed $\tan\beta=40$, the masses
must be rather large; smaller masses are allowed for smaller values of
$\tan\beta$.
 For instance, the
Wino-coannihilation region in the right plot is compatible with
$\DamuFinal$ only for small $\tan\beta\lesssim15$ at the $2\sigma$
level, as can be read off 
with the help of the legend plot.

As mentioned above, the same parameter region is investigated in
a number of further references, with similar
results. From large scans of the parameter space,
Refs.~\cite{Chakraborti:2020vjp,Chakraborti:2021dli,Chakraborti:2021mbr} find that slepton
as well as Wino coannihilation is possible for LSP masses in the wider range
around
\begin{align}
  m_{\chi^0_1}\approx 130\ldots500\text{ GeV},
\end{align}
and Ref.~\cite{VanBeekveld:2021tgn} argues that the scenario leads to
moderate electroweak finetuning.

If sleptons are heavier than
considered in Fig.~\ref{fig:Binosleptonscompressed}, there are several
relevant considerations. First, the LHC constraints become stronger
and exclude sleptons up to about 600 GeV for realistic scenarios \cite{Endo:2021zal},
slightly less than the limit obtained by LHC for simplified
models and quoted in Sec.~\ref{sec:SUSYconstraints}. Second, there are
significant dark
matter constraints. A viable way to explain dark matter in the general
parameter region with Bino-like
LSP and heavier sleptons is Wino 
coannihilation
\cite{Athron:2021iuf,Chakraborti:2021dli,Chakraborti:2021mbr}. An
alternative way can be
stau coannihilation in case the stau masses are not unified with the
slepton masses. However, there are significant LHC constraints on
chargino masses \cite{Endo:2021zal}, which in case of stau
coannihilation and light stau  require at least one
rather heavy chargino \cite{Athron:2021iuf,Chakraborti:2023pis}. The 
case where the left- and right-handed 
slepton masses are different has been considered in several
references, and particularly
Refs.~\cite{Endo:2021zal,Chakraborti:2022vds} have considered the
extreme case where one of the two slepton chiralities decouples,
$m_\txl\to\infty$ or $m_\txr\to\infty$, such that only the BHL or BHR
contributions to $\amu$ survive. They find that dark matter data
already excludes such scenarios with Bino-like LSP and significant
$\amu$.

The recent model-building literature has put forward a variety of
constructions leading to potentially large $\amu$ with mass spectra of
the kind of Fig.~\ref{fig:Binosleptonscompressed}. First, simple
extensions of baseline scenarios such as the CMSSM
\cite{Akula:2013ioa,Ibe:2013oha,Gogoladze:2014cha,Chakrabortty:2015ika,Okada:2016wlm,Hussain:2017fbp,Tran:2018kxv,Wang:2018vrr,Aboubrahim:2021xfi,Chakraborti:2021bmv,Ellis:2021zmg,Lamborn:2021snt,Ellis:2021vpp,Li:2021pnt,Ahmed:2021htr,Wang:2021bcx,Forster:2021vyz,Gomez:2022qrb,Wang:2022rfd,Ellis:2024ijt}
 or AMSB
\cite{Jeong:2021qey}, mentioned in
Sec.~\ref{sec:BaselineSUSY}, can often lead to Bino-like LSP in
combination  with light 
sleptons and very heavy $\mu$, thus allowing large contributions to
$\amu$ via
BLR- and WHL-contributions. Also, going beyond the baseline scenarios,
dedicated constructions of SUSY breaking and mediation mechanisms were
proposed in
Refs.~\cite{Zhu:2016ncq,Ibe:2019jbx,Yanagida:2017dao,Yanagida:2020jzy,Agashe:2022uih}. 
The common feature of these constructions is the motivation to
naturally explain heavy coloured SUSY particles and the heavy Higgs
mass $M_h$, while allowing light sleptons and light Bino. A by-product
of these constructions tends to be  very large $\mu$ in the  multi-TeV
region, which can lead to large BLR contributions to $\amu$. 

The scenario with large $\mu$ in the multi-TeV region also appears in the context of various
specific model constructions, such as models based on Pati-Salam
symmetry \cite{Belyaev:2016oxy,Gomez:2022qrb,Gomez:2024dts} or on
SO(10) \cite{Aboubrahim:2021phn}, or models with usual GUT constraints but extra vector-like
matter fields \cite{Choudhury:2017acn,Choudhury:2017fuu}.
All mentioned
scenarios are intrinsically motivated; the current constraint
$\DamuFinal$ then imposes stringent lower mass limits, or equivalently
stringent upper $\tan\beta$ limits of the kind visible in
Fig.~\ref{fig:Binosleptonscompressed}.

Quite different scenarios with Bino-like LSP but small $\mu$, small
$M_2$ and heavier sleptons have been considered
e.g.~in
Refs.\ \cite{Hagiwara:2017lse,Altin:2017sxx,Pozzo:2018anw}. Given the
dark matter constraints, these scenarios are now strongly constrained.

\subsubsection{Phenomenological results with light Higgsino- or Wino-like LSP in the MSSM}
\begin{figure}[t]
  \begin{subfigure}[t]{0.405\textwidth}
    \centering\includegraphics[scale=.7]{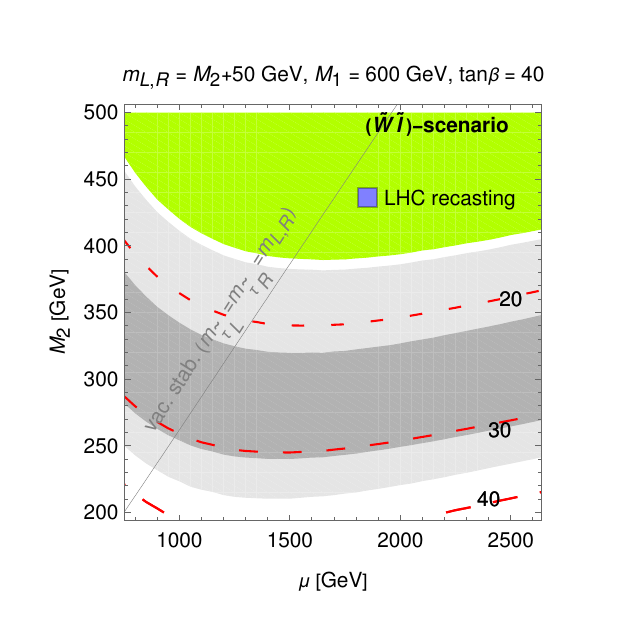}
    \caption{}
    \label{fig:WHsleptonscompresseda}
  \end{subfigure}
  \begin{subfigure}[t]{0.405\textwidth}
    \centering\includegraphics[scale=.7]{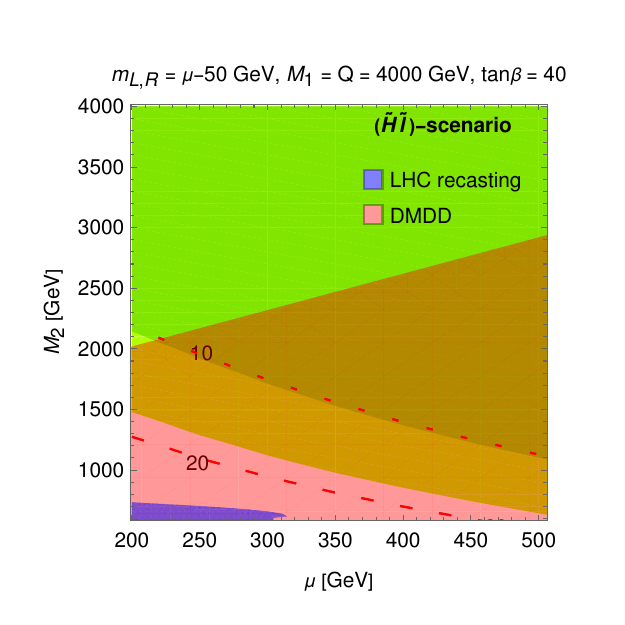}
    \caption{}
    \label{fig:WHsleptonscompressedb}
  \end{subfigure}
  \begin{subfigure}[t]{0.18\textwidth}
    \centering\includegraphics[scale=.7]{Plots/PlotAmuLegend.pdf}
  \end{subfigure}
  \caption{\label{fig:WHsleptonscompressed} MSSM in the (a)
    $(\tilde{W}\tilde{l})$- and (b)
    $(\tilde{H}\tilde{l})$-scenarios with either Wino- or
    Higgsino-like LSP and light sleptons. For parameter values see the
    plots and the text.
The meaning of the red dashed lines, the green/grey coloured
    $\Damu^{\text{SUSY}}$ regions, and the reinterpretation in the
    legend plot are as in Fig.\ \ref{fig:briefsurveyplot}.
    The  red shaded region in the $(\tilde{H}\tilde{l})$-scenario is excluded by dark matter direct detection
    if the LSP is assumed stable; the blue shaded region corresponds
    to LHC limits. The thin solid grey line corresponds to the vacuum
    stability constraint of Ref.\ \cite{Endo:2013lva}; it applies in case the
    left- and right-handed stau-masses are set equal to the
    smuon/selectron masses and excludes the points to its right,
    i.e.\ with larger $\mu$. Different from the other plots, the plot
    (b) has been updated from Ref.~\cite{Athron:2021iuf} with modified
    input parameters to include higher gaugino masses: the running
    parameters at $Q=4$ TeV are $M_1=4 $ TeV, all squark mass
    parameters are set to 4 TeV, and the slepton mass parameter is set
    50 GeV lower than the Higgsino mass. The loop-corrected pole
    masses then correspond to Higgsino-like LSP with slightly heavier sleptons.
}
\end{figure}

Now we focus on the $\amu$ phenomenology in the MSSM with Wino- or
Higgsino-like LSP. For masses above 2 TeV or 1 TeV, respectively, the
full DMRD can be explained, see
Sec.~\ref{sec:SUSYconstraints}. Here we consider lighter masses
considered here, such that the DMRD is below the observed one, see
scenarios 6, 8 in Tab.~\ref{tab:MSSMscenarios}. The underabundant DMRD
is compatible with observations, although additional dark matter
candidates from beyond the MSSM are required. 

We
illustrate the scenarios in Fig.~\ref{fig:WHsleptonscompressed}. To discuss
the parameter choices and results we first note that Wino- or
Higgsino-like LSP are neutralinos which are accompanied by at least
one chargino close in mass. Hence the LSP mass is directly constrained
by dedicated LHC searches for such compressed neutralino--chargino
pairs; above around 200 GeV such constraints become ineffective.

For the sleptons, the LHC constraints exclude a triangular
$m_{\tilde{l}}$--$m_{\chi^0_1}$ region; for LSP masses below around
200 GeV,  either slepton
masses above around 400--500 GeV or close to the LSP are allowed, and
for LSP masses above 200 GeV, the slepton mass limits become weak
\cite{Chakraborti:2021kkr}. Hence the plots in
Fig.~\ref{fig:WHsleptonscompressed} show LSP mass parameters $M_2$, $\mu$ above 200 GeV and choose the
slepton masses $m_{\txl, \txr}$ very close, just 50 GeV above $M_2$ or
$\mu$, respectively, in order to maximise the contributions to
$\amu$. Among all mass hierarchies between the Wino, the Higgsino and
the sleptons, the hierarchy ``Wino$>$sleptons$>$Higgsino'' leads
to the strongest additional LHC limits on the Wino-like heavier chargino mass, which
extends up to around 700 GeV in the plot \cite{Athron:2021iuf}; otherwise the
chargino masses in these scenarios are not strongly constrained by
LHC \cite{GAMBIT:2018gjo}.\footnote{%
The most recent LHC analyses from after the appearance of Refs.~\cite{GAMBIT:2018gjo,Athron:2021iuf} have
not been used for this statement, and their evaluation is beyond the
scope of the present review. However, their impact is not expected to
change the viable parameter space considered here. 
}

However, the LHC limits generally have far less impact on these scenarios than
the DMDD limits from the most recent LZ data, see
Eq.~(\ref{DMDDMlimit}). For the Higgsino-like LSP case, these limits allow
only values of $M_2>2$ TeV in the plot; in the Wino-like LSP case, the
limits are not evaluated but by comparison to previous calculations
\cite{Chakraborti:2021kkr,Athron:2021iuf} are expected not to affect
the displayed parameter region.

In the allowed parameter regions, the SUSY contributions to $\amu$ are
dominated by the WHL contributions, and for very large $\mu$ in case
of Wino-like LSP also the BLR contributions are relevant. In the
Wino-like LSP
case the general behaviour is similar to the Bino-like LSP case shown in
Fig.~\ref{fig:Binosleptonscompressed}, and the displayed parameter region focuses on
the turnover point where the BLR contributions start becoming
important. As is visible, the Wino-like LSP parameter
space is now strongly restricted by the new small result
$\DamuFinal$. For $\tan\beta=40$, agreement at the $2\sigma$-level can be obtained for $M_2>400 \text{
    GeV}$.
For smaller $\mu$ and also for larger $\mu$ than shown in the plot,
$\Damu^{\text{SUSY}}$ increases further;\footnote{%
The DMDD constraints will exclude too small $\mu$, similarly to the
case of the Bino-like LSP shown in
Fig.~\ref{fig:Binosleptonscompressed}. }
hence heavier Wino masses (or
smaller $\tan\beta$) are required by $\DamuFinal$. Another parameter
region which is consistent with the new $\Damu$ constraint  is light
Wino-like LSP in conjunction with significantly heavier sleptons.

In case of Higgsino-like LSP,  only rather small contributions to $\amu$ remain
possible for $\tan\beta=40$. This used to be different but is now
caused by the LZ DMDD constraint which forces $M_2$ to
be huge. This strong constraint is in line with expectations in
Refs.~\cite{Chakraborti:2021mbr,Zhao:2022pnv}. Hence the entire
Higgsino-like LSP parameter region allowed by DMDD constraints is
automatically in agreement with the $\amu$ constraints (as long as
$\tan\beta\lesssim40$, see Sec.~\ref{sec:SUSYamu} for the limit
$\tan\beta\to\infty$). Interestingly, in this parameter scenario it is also
possible  to
explain a slight excess seen at the LHC in $\chi^0_2\chi^\pm_1$
channels \cite{Chakraborti:2024pdn}.

In the model-building literature, scenarios with Wino- or
Higgsino-like LSP have been widely considered.
Specifically light Higgsinos can be generally motivated by electroweak
finetuning \cite{Li:2016ucz,Padley:2015uma,Baer:2018rhs} and according models have been
constructed as simple CMSSM extensions \cite{Iwamoto:2021aaf} or
based on gauge-mediated \cite{Bhattacharyya:2018inr,Ibe:2021cvf,Evans:2022oho},
anomaly-mediated \cite{Chowdhury:2015rja},
gaugino- and/or Higgs-mediated \cite{Harigaya:2015kfa,Cox:2018vsv,Endo:2019bcj}, or based on gravity-mediated  SUSY breaking
and 
focus-point behaviour  \cite{Harigaya:2015jba}.
The models of these references tend to also produce small Wino masses
and allow either Higgsino-like or Wino-like LSP. Now the Higgsino-like
LSP case with small Wino mass is excluded by the LZ DMDD data (unless one assumes the LSP to be
unstable and not to contribute to dark matter which can be motivated
in gauge-mediated SUSY breaking with light gravitino). The Wino-like LSP
regions of those models can remain viable but are now constrained by
DMDD and by $\amu$.

In contrast, Ref.\ \cite{Han:2020exx} has considered a Higgsino-like
LSP scenario with significant mass gap between
the Higgsino-LSP and the two gauginos, motivated within the context of a
model with seesaw mechanism and  SO(10) GUT constraints on the gaugino masses but
non-universal scalar masses. This reference finds only small viable contributions
to $\amu$, which now turns out to be in agreement with observations.

The scenario with Wino-like LSP but very large $\mu$ has been
constructed in Ref.~\cite{Yanagida:2020jzy,Yin:2021mls}  based on 
Higgs-anomaly mediated SUSY-breaking
\cite{Yanagida:2020jzy,Yin:2016shg,Evans:2013uza}; such scenarios
are now strongly constrained by $\amu$, restricting
essentially the
maximum product of $\mu\tan\beta$ as a function of the LSP and slepton
masses.

\subsubsection{SUSY beyond the R-parity conserving MSSM}
\label{sec:BMSSM}
Now we focus on the phenomenology of SUSY models different from the
R-parity conserving MSSM. We will discuss the MRSSM, a distinct
minimal model without $\tan\beta$ enhancement, the NMSSM and further
non-minimal models which can change the LHC and dark matter
phenomenology, and return to the MSSM but without assuming a stable
neutralino-LSP.

The Minimal R-symmetric Supersymmetric Standard Model (MRSSM) is
defined as the minimal model with continuous unbroken U(1)$_R$
symmetry \cite{Kribs:2007ac}. This symmetry does not commute with SUSY and it
assigns R-charges 0 to SM-like fields and R-charges $\pm1$, $\pm2$ to
the additional fields. R-symmetry implies R-parity conservation but
goes beyond it.
Importantly, the different squarks $\tilde{q}_{L,R}$ have charges $\pm1$ such that the LHC
production of e.g.~$\tilde{q}_L\tilde{q}_L^\dagger$ and
$\tilde{q}_L\tilde{q}_R$ is allowed while $\tilde{q}_L\tilde{q}_L$ and
$\tilde{q}_L\tilde{q}_R^\dagger$ are forbidden. This reduces the LHC
squark production cross section and alleviates LHC limits
\cite{Heikinheimo:2011fk,Kribs:2013oda,Diessner:2017ske,Diessner:2019bwv}, and the MRSSM also has additional virtues with
respect to dark matter and Higgs phenomenology \cite{Bertuzzo:2014bwa,Diessner:2014ksa,Diessner:2015iln,Kalinowski:2024uxe}. The MRSSM also dramatically
modifies the $\amu$ and the related CLFV
phenomenology \cite{Kribs:2007ac,Kotlarski:2019muo}. The R-symmetry
forbids the MSSM-like Higgsino and gaugino mass parameters $\mu$,
$M_1$ and $M_2$ but introduces Dirac mass terms involving additional
degrees of freedom. As a result, all $\tan\beta$-enhanced MSSM
contributions illustrated in Eqs.~(\ref{eq:SUSYMIapprox}) vanish.\footnote{This is
different in models which Dirac gauginos without the restrictions of
unbroken R-symmetry \cite{Li:2017fbg}.}

Overall, there is
no $\tan\beta$ enhancement in the MRSSM contributions to leptonic
dipole operators relevant for $\amu$ or $\mu\to e\gamma$ and similar
processes \cite{Kotlarski:2019muo}. There is an enhancement by MRSSM-specific
Yukawa-like couplings, which is however much weaker. As a result, the
MRSSM contributions to $\amu$ are generally small, and in a wide
parameter scan, Ref.~\cite{Kotlarski:2019muo} found at most 
\begin{align}
  \Damu^{\text{MRSSM,max}} &<
  10\times10^{-10} &&\text{if }M_{\text{LOSP}}>300\text{ GeV},
\end{align}
where $M_{\text{LOSP}}$ is the mass of the lightest electrically
charged SUSY particle. Hence the MRSSM was no promising explanation of
the previous deviation $\DamuOld$, but it now emerges as a model which
is generically in full agreement with the new result
$\DamuFinal$. Since the dipole operator has at most a moderate
enhancement, the corresponding correlations of CLFV observables
$\mu\to e\gamma$, $\mu\to3e$, $\mu\to e$ conversion mentioned in
Eqs.~(\ref{dipoledominancemu3e},\ref{dipoledominancemue}) are not
valid. This means that the MRSSM might lead to 
visible signals in the upcoming $\mu\to e$ experiments.

The Next-to-Minimal Supersymmetric Standard Model (NMSSM) extends the
MSSM by a Higgs singlet and its superpartner, a singlino. Although the
singlino can mix with the Bino, Wino and Higgsinos to form neutralino
mass eigenstates, the
additional fields do not modify the contributions to $\amu$
significantly,
though non-negligible modifications can occur if also right-handed
neutrino and sneutrino fields are added \cite{Dao:2022rui}. Hence, the
NMSSM can generally also accommodate sizeable contributions to
$\amu$
of similar behaviour as in the MSSM 
\cite{Tang:2022pxh,Huang:2023zvs,Wang:2021lwi,Abdughani:2021pdc,Cao:2019evo,Cao:2021tuh,Cao:2022ovk,Cao:2022chy,Cao:2024axg,Domingo:2022pde,Kim:2021suj}.
Interestingly, the NMSSM
contains several intriguing possibilities to explain dark matter which
are distinct from the MSSM and which lead to different SUSY mass
patterns. These can evade LHC limits and lead to sizeable
contributions to $\amu$.
First, a singlino-like LSP can accommodate the DMRD for rather small masses and
allow e.g.\ Wino- or Higgsino-like neutralinos or even sneutrinos as
the lightest MSSM-like sparticles. In  these cases, LHC searches for electroweak sparticles
become less effective and many sparticles can be light, leading to
large contributions to $\amu$ via the WHL
diagram \cite{Cao:2021tuh,Cao:2022ovk,Cao:2022chy,Cao:2024axg,Domingo:2022pde,Zhou:2025xol}. Second,
if also right-handed neutrinos are introduced, the sneutrino can be
the LSP and constitute dark matter, and again small masses and large
WHL contributions to $\amu$ are possible \cite{Kim:2021suj}. These
scenarios are attractive from the dark matter and LHC point of view,
but the new result $\DamuFinal$ imposes constraints such as lower mass
limits and/or upper $\tan\beta$ limits. The MSSM plots in
Figs.~\ref{fig:Binosleptonscompressed}, \ref{fig:WHsleptonscompressed} can be
used as an illustration: compared to these plots, in the mentioned
NMSSM scenarios
$\amu$ would behave similarly while LHC and DMDD constraints would be
weakened.

Going beyond even the NMSSM, it is often motivated to consider SUSY
models with extended gauge groups, which lead to $Z'$ gauge bosons and
additional neutralino states, which may have stronger interactions
than the NMSSM singlino. In a series of works
\cite{Frank:2017ohg,Yang:2018guw,Dong:2019iaf,Yang:2020bmh,Zhao:2020trt,Su:2020lrv,Zhang:2021gun,Yang:2021duj,Zhao:2021eaa,
	Wang:2022iaf,Wang:2022wdy,Yang:2023krd}
models of this kind with an extra U(1) gauge group were considered, and
additional possibilities for dark matter were found, as well as
additional two-loop Barr-Zee contributions to $\amu$, which are now
constrained.
Similarly, an 
SO(10)-inspired left-right
symmetric model with extra $U(1)_R \times
U(1)_{B-L}$ gauge group there are
non-MSSM-like dark matter candidates such as a Bino corresponding to
one of the extra U(1) gauge groups \cite{Dehghani:2023lde}. Again this
changes the correlation between LHC, dark matter and $\amu$
phenomenology, and interestingly, the model of
Ref.~\cite{Dehghani:2023lde} tends to prefer small contributions to $\amu$.

Within the MSSM and its extensions, we have so far mostly assumed the
LSP to be neutral and stable and to contribute to dark matter. This
does not need to be true, e.g.~under two motivated conditions,
reviewed e.g.~in Ref.~\cite{Barbier:2004ez}: if
R-parity is not conserved, or if the gravitino, the spin-3/2
superpartner of the graviton in supergravity, is lighter than all MSSM
sparticles and therefore forms a dark matter candidate. 

In case of  R-parity  violation (RPV), there can be vertices such as
muon--electron--sneutrino, or muon--bottom-quark--stop, and
accordingly there can be additional one-loop contributions to
$\amu$. These, however, must be 
small given existing constraints on such lepton-number violating
vertices \cite{Choudhury:2023lbp,Choudhury:2024ggy}. The major impact
on phenomenology derives from modified correlations with other sectors.
For instance, Refs.~\cite{Zheng:2021wnu,Zheng:2022ssr,BhupalDev:2021ipu,Afik:2022vpm} considered models where the RPV couplings
explained deviations observed in B-physics, which 
however have been excluded in the meantime. The studies in
Refs.~\cite{Kpatcha:2019pve,Heinemeyer:2021opc} describe how RPV
can loosen the LHC limits on electroweak sparticles, and the so-called
$\mu\nu$SSM can explain all neutrino masses, can explain dark matter via
gravitinos or axinos, and can dynamically generate the MSSM
$\mu$-parameter. However, interestingly in all these mentioned
references, the motivation for RPV leads to specific viable  mass
ranges; these tend to prefer small $\amu$, hence in particular the
$\mu\nu$SSM remains a viable model in view of $\DamuFinal$.

In a more general context, Ref.~\cite{Chakraborti:2022vds} has
investigated the impact of RPV or light gravitino on the MSSM parameter
space of interest for $\amu$. Neither RPV nor a gravitino LSP changes
$\amu$ as such, but complementary constraints from dark matter and LHC
can be significantly modified. In case of RPV, the LSP is unstable;
hence in the simplest case it does not contribute to dark matter and
all the strong constraints from DMDD as well as requirements for
coannihilation and particular mass splittings are
absent. Similarly, in the worst case the neutralino LSP can decay into
three quarks, forming three jets which are hard to identify at the
LHC, and LHC SUSY search limits are significantly weakened. Hence RPV
can significantly open up the parameter space of the kind 
visible e.g.\ in Figs.~\ref{fig:Binosleptonscompressed},
\ref{fig:WHsleptonscompressed}.  In general, therefore, RPV SUSY could 
be a way to accommodate very large $\Damu$, however, the small current
value $\DamuFinal$ does not provide a strong motivation to consider
such scenarios.

In the MSSM with additional light gravitino, the MSSM-like LSP (which
is then the NLSP) is
unstable since it can decay into the gravitino, which is then the true
LSP. In contrast to the RPV case, the decay into gravitinos is 
seriously constrained by dedicated LHC searches, hence the parameter
space of interest for $\amu$ remains strongly constrained
\cite{Chakraborti:2022vds}.  For instance, gravitino LSP with 
Higgsino- or Wino-like NLSP requires very heavy Higgsino and  Wino
masses above 650--750 GeV. Importantly, a light gravitino LSP also
allows the case where charged
particles such as sleptons or staus are the NLSP, i.e.~the lightest
MSSM-like sparticles. These cases are distinct from what is possible
in the ordinary MSSM, and they allow lighter slepton and Higgsino 
masses \cite{Chakraborti:2022vds}. An example embedding of
such a scenario into a full SU(5) SUSY GUT with gauge-mediated SUSY
breaking is presented in Ref.~\cite{Kim:2024bub}. This reference
finds viable parameter space and also potentially sizeable
$\Damu^{\text{SUSY}}$. However it also quantifies a tension between 
proton decay constraints and $\amu$; smaller values of $\amu$ as
preferred by the new $\DamuFinal$ widen
the allowed parameter space for the fundamental SUSY breaking
mechanism.

\subsubsection{Summary}

In summary, the phenomenology of $\amu$ in SUSY models is very rich
and has led to a vast body of literature. SUSY extensions of the SM
are well motivated in a number of ways, and for many years they have
been considered as some of the most plausible explanations of the
large deviation $\DamuOld$. The new $\amu$ result, together with
recent improvements in dark matter and LHC searches, now sheds a different
light on possible realisations of SUSY.

Heavy SUSY, where all sparticle masses are $\gtrsim1$ TeV, almost
inevitably leads to small $\Damu$ (exceptions such as radiative muon
mass generation are discussed above). This was not preferred by the
previous result but is now fully compatible with $\DamuFinal$. Many
scenarios such as the CMSSM, mGMSB and many of their generalisations
or the generic scenarios with Higgsino- or Wino-like dark matter
require heavy sparticles, either in view of LHC, $M_h$, or dark matter
data. These motivated, and often rather simple scenarios thus
re-emerge as attractive viable SUSY models (to the extent allowed by
other constraints). In the future,
they could only be challenged if an improved SM prediction of $\amu$
firmly deviates from the experimental value.

The MSSM with lighter masses is constrained by LHC and dark matter
data. Interestingly, the three scenarios to explain the dark matter
relic density with Bino-like LSP and coannihilation with sleptons,
staus, or Winos, all evade LHC limits rather generically in the mass
range 200\ldots500 GeV, and even lower LSP masses are possible. For
this case, the
LZ dark matter experiment dramatically constrains
the parameter space, preferring very large values of the Higgsino mass
$\mu$. In this way, the BLR contributions to $\amu$ actually increase,
leading to stringent constraints on the parameter space from
$\DamuFinal$. As discussed and illustrated in the plots, the current $\amu$
constraints translate into lower mass limits and/or upper $\tan\beta$
limits on the scenarios with Bino-like LSP.

Scenarios with Higgsino- or Wino-like LSP with mass in the few-100 GeV
range are also viable, though strongly constrained. The Higgsino-like
LSP scenario, though the dark matter relic density is underabundant,
is very strongly constrained by the LZ experiment. As a result,
Higgsino-like LSP with $\tan\beta\lesssim40$ now automatically
predicts small $\Damu$ in agreement with $\DamuFinal$. In contrast,
the Wino-like LSP scenario could lead to very large $\Damu$; it is
therefore very strongly constrained by $\DamuFinal$.

SUSY beyond the R-parity conserving MSSM can be well motivated and can
change the $\amu$ phenomenology. For instance, the MRSSM has an
additional symmetry, fewer free parameters and automatically predicts
small $\Damu$. The MRSSM, but even more so the NMSSM, also enable additional
options to accommodate dark matter via singlinos; in this way the
strict MSSM interplay between $\amu$ and dark matter is
loosened. Similarly, if
the LSP in the MSSM is unstable because of R-parity violation, the
parameter space with small SUSY masses is opened up, while the case
of LSP-decay into light gravitinos is strongly constrained and prefers
small $\Damu$ in line with the new result $\DamuFinal$.

\subsection{Two-Higgs-doublet models}\label{sec:2HDM}

The 2HDM is among the  most studied and best motivated extensions of the
SM. It replaces the single SM Higgs doublet $\Phi$ by two Higgs
doublets $\Phi_1$, $\Phi_2$ with $Y=\frac{1}{2}$. It
represents one of the simplest ways to extend the scalar field content
and
allows to study
Higgs boson properties, the
nature of electroweak symmetry breaking and the
origin of mass. In fact, the 2HDM changes
two SM sectors in important ways. First, it modifies the
Higgs sector and the
Higgs potential,
leading to
a more complex vacuum structure,
additional physical Higgs bosons,  and
modified couplings of the  Higgs bosons to gauge bosons and
Higgs self couplings.
Second, the 2HDM also
strongly alters the
Yukawa sector, leading to a modified
fermion mass generation mechanism, and to
potentially enhanced couplings of fermions to Higgs bosons and resulting
large effects in flavour physics. For reviews of the theory of the
2HDM we refer to Refs.~\cite{Gunion:1989we,Branco:2011iw}.

As analysed in Secs.~\ref{sec:ChiralityFlips} and \ref{sec:MuonMass}, a
modified muon mass generation mechanism 
is typically accompanied by contributions to muon $g-2$. Indeed it
has been known for a long time
\cite{Chang:2000ii,Cheung:2001hz,Wu:2001vq,Krawczyk:2001pe,Krawczyk:2002df,Larios:2001ma}
that the 2HDM could
provide an explanation of  deviations as large as $\DamuOld$ in
Eq.~\eqref{eq:DamuOld} reported at
the time of the FNAL Run-1 or the earlier BNL measurement.
Since then the 2HDM parameter space has been scrutinised by 
measurements at the LHC, by $\amu$ and by a multitude of complementary
constraints, with important recent new developments.
Especially the complementary constraints have changed the  2HDM $\amu$
phenomenology  
significantly  in recent years, even before the new result
$\DamuFinal$ appeared. 
Here we review
the current status, focusing mainly on the scenarios within the 2HDM
with the potential for sizeable $\Damu$. These were motivated and
considered in view of the previous result $\DamuOld$ and are now
constrained by $\DamuFinal$. For recent in-depth analyses and fits of the
parameter 2HDM space in scenarios with large or small $\Damu$ we
refer to
Refs.~\cite{Eberhardt:2020dat,Atkinson:2021eox,Atkinson:2022pcn,Iguro:2023tbk,Karan:2023kyj,Coutinho:2024zyp}. \newline

The Higgs potential of the 2HDM in general contains three quadratic and seven quartic terms in the Higgs doublets $\Phi_{1,2}$ 
and can be written as \cite{Gunion:2002zf,Branco:2011iw}
\begin{align}\label{eq:2HDM-potential}
	\begin{split}
		V(\Phi_1,\Phi_2) &= m_{11}^2 \Phi_1^\dagger \Phi_1 + m_{22} \Phi_2^\dagger\Phi_2 - \Big( m_{12}^2 \Phi_1^\dagger\Phi_2 +  h.c.\Big) \\
		& + \frac{\lambda_1}{2} (\Phi_1^\dagger\Phi_1)^2 + \frac{\lambda_2}{2}(\Phi_2^\dagger\Phi_2)^2 + 
		\lambda_3 (\Phi_1^\dagger\Phi_1)(\Phi_2^\dagger\Phi_2) + \lambda_4 (\Phi_1^\dagger\Phi_2)(\Phi_2^\dagger\Phi_1)\\
		& + \Big(\frac{\lambda_5}{2} (\Phi_1^\dagger\Phi_2)^2 + \lambda_6 (\Phi_1^\dagger\Phi_1)(\Phi_1^\dagger\Phi_2)
		  + \lambda_7 (\Phi_1^\dagger\Phi_2)(\Phi_2^\dagger\Phi_2) + h.c. \Big).
	\end{split}
\end{align}
If the potential contains a global minimum, both scalar doublets will acquire a vacuum expectation value (\vev) $\ev{\Phi_i}=\frac{1}{\sqrt{2}}(0,v_i)$. 
Non-zero complex  phases of the parameters $m_{12}^2,\lambda_{5,6,7}$
and $v_{1,2}$ are in general possible and correspond to CP violation
in the scalar sector \cite{Lee:1973iz}.
However, the $\amu$ phenomenology is not strongly sensitive to this possibility and we therefore assume real parameters and real \vev{s} $v_{1,2}$
in the following.

It is convenient to introduce the \emph{Higgs basis} by rotating the
scalar fields with $\tan\beta=v_2/v_1$ into two doublets $\Phi_v$, $\Phi_\perp$,
such that the \vev and the unphysical Goldstone bosons 
are contained entirely in the doublet $\Phi_v$. The rotation and the
field components read
\begin{align}
\label{2HDMHiggsbasis}  \begin{pmatrix}	\Phi_v \\ \Phi_\perp\end{pmatrix} = 
	\begin{pmatrix} \cos\beta & \sin\beta \\ -\sin\beta & \cos\beta \end{pmatrix}
	\begin{pmatrix}	\Phi_1 \\ \Phi_2 \end{pmatrix},
	\qquad
	\Phi_v = \begin{pmatrix} G^+ \\ \frac{1}{\sqrt{2}}(v + S_1 + i G^0)	\end{pmatrix}, \qquad 
	\Phi_\perp = \begin{pmatrix} H^+ \\ \frac{1}{\sqrt{2}} (S_2 + i S_3)\end{pmatrix},
\end{align}
where $v\approx 246$ GeV is related to the \vev{s} by $v_1^2 + v_2^2 = v^2$.
The physical Higgs spectrum of the 2HDM  consists
of a charged scalar $H^\pm$ as well as three neutral scalars $h,H$ and
$A$ related to $S_i$ by an orthogonal transformation.   
In the CP conserving case assumed here, the physical CP-odd Higgs $A=S_3$ and
the two physical CP-even Higgs bosons are
\begin{align}
	\begin{pmatrix}	H \\ h 	\end{pmatrix} = \begin{pmatrix} \cos(\alpha-\beta) & \sin(\alpha-\beta) \\ - \sin(\alpha-\beta) & \cos(\alpha-\beta)\end{pmatrix} \begin{pmatrix} S_1 \\ S_2 \end{pmatrix}.
\end{align}
After introducing these mass-eigenstates, the Higgs potential can equivalently be parametrised in terms of the 4 scalar masses 
$M_{h,H,A,H^\pm}$, the mixing angles $\tan\beta$ and $\cos(\alpha-\beta)$ and three of the quartic Higgs couplings
$\lambda_1, \lambda_6$ and $\lambda_7$.
In the so-called alignment limit \cite{Bernon:2015qea} where either $\sin(\alpha-\beta)=0$ or $ \cos(\alpha-\beta)=0$, the
doublet $\Phi_v$ becomes purely SM-like and the field $S_1$ becomes
equal to either $H$ or $h$, which then exactly correspond to the SM Higgs
boson. In the following we will assume that $h$ is approximately or
exactly SM-like, such that $\cos(\alpha-\beta)$ is small or
zero.

If no further symmetries are imposed, all SM fermions are allowed to couple to both of the Higgs doublets, 
resulting in the most general possible 2HDM Yukawa Lagrangian
\begin{align}\label{eq:2HDM-Yukawa-Lag}
	\La \supset - \sum_{a=v,\perp} \Big[ \overline{l_L} Y^l_a e_R
          \Phi_a + \overline{q_L} Y^d_a d_R \Phi_a + \overline{q_L}Y^u_a u_R \tilde\Phi_a \Big] + h.c.
\end{align}
The Yukawa Lagrangian can be equivalently written in terms of the
general basis or the Higgs basis; the corresponding Yukawa matrices
$Y^f_{v,\perp}$ and $Y^f_{1,2}$ are related similarly to the fields as
in Eq.~(\ref{2HDMHiggsbasis}). 
The Yukawa couplings to $\Phi_v$ in the Higgs basis are related to the
tree-level fermion mass matrices as $Y^f_v = \sqrt{2} M_f/v$. The (approximately) SM-like Higgs boson
$h$ couples to fermions via $Y^f_v$, up to corrections of order $\cos(\alpha-\beta)$.
In contrast, the physical Higgs bosons $H$, $A$ and $H^\pm$ couple via
the additional Yukawa matrices $Y^f_\perp$ (in case of $H$ up to
$\cos(\alpha-\beta)$ corrections). These $Y^f_\perp$ are in general
unrelated to the fermion masses and can therefore result in 
large effects in flavour physics. This includes potentially significant
contributions to $\amu$, but also flavour-changing neutral currents
(FCNC) that are strongly constrained by experiment. We refer to the
review \cite{Branco:2011iw} for a survey of ways to avoid too large
FCNC contributions. Broadly speaking, one can impose exact or softly
broken symmetries which force  $Y^f_\perp$  
to be diagonal in the fermion mass basis (such as the Type I, II, X, Y
models or the muon-specific model) or one can postulate more
general patterns of flavour textures of
$Y^f_2$ where FCNC are either completely absent (such as the flavour-aligned 2HDM) or
strongly suppressed (such as various forms of the so-called Type-III
model as discussed in Ref.~\cite{Branco:2011iw}). In the following we
focus on scenarios without FCNC.

\subsubsection{Type I, II, X, Y and Flavour-aligned 2HDM}\label{subsec:FA2HMD}
\begin{table}
	\centering
	\setlength{\arrayrulewidth}{0.3mm}
	\begin{tabular}{|c||c|c|c|c|c|c|}
		\hline
		Model & type I & type II & type X & type Y & inert & FA\\ \hline\hline
		$\zeta_l$ & $\cot\beta$ & -$\tan\beta$ & -$\tan\beta$ &  $\cot\beta$ & 0 & $\lesssim 300$  \\ \hline
		$\zeta_d$ & $\cot\beta$ & -$\tan\beta$ &  $\cot\beta$ & -$\tan\beta$ & 0 & $\lesssim 1$ \\ \hline
		$\zeta_u$ & $\cot\beta$ &  $\cot\beta$ &  $\cot\beta$ &  $\cot\beta$ & 0 & $\lesssim 1$ \\ \hline
	\end{tabular}
	\caption{Values of the Yukawa alignment parameters $\zeta_f$ in the Type-I,II,X,Y and inert 2HDM.}
	\label{tab:2HDM-zeta_f}
\end{table}
The most widely discussed way to avoid FCNC is to introduce a discrete
$\mathds{Z}_2$ symmetry under which the general-basis fields $\Phi_1$ and $\Phi_2$ transform
differently.\footnote{This also forces $m_{12}^2$ and $\lambda_{6,7}$ to vanish. However, $m_{12}^2\neq 0$
breaks the discrete symmetry only softly (leading to finite FCNC at one-loop order) and is therefore often included \cite{Gunion:1989we}.}
This forces each of the fermion fields to couple to only one 
of these scalar doublets, depending on the choice of charge assignment.
Without loss of generality one can assume that the up-type quarks
always couple to $\Phi_2$.
There are then four distinct possibilities classified as type I (all fermions couple
to $\Phi_2$), type II ($d$ and $l$ couple to $\Phi_1$, $u$ to $\Phi_2$), type X 
(or lepton-specific: $l$ couples to $\Phi_1$, $u$ and $d$ to $\Phi_2$) and type Y 
($d$ couples to $\Phi_1$, $u$ and $l$ to $\Phi_2$).
In all these cases, the Yukawa matrices corresponding to the
Higgs basis $Y^f_v$ and $Y^f_\perp$ are then proportional to each
other for each $f=l,d,u$, i.e.\ one can write
\begin{align}\label{eq:FA2HDMrelation}
	Y^l_\perp = \zeta_l Y^l_v, \qquad Y^d_\perp = \zeta_d Y^d_v, \qquad Y^u_\perp = \zeta_u Y^u_v,
\end{align}
where the values of the proportionality factors $\zeta_{l,d,u}$ for the different types depend on $\tan\beta$ and $\cot\beta$ and are listed in Tab.~\ref{tab:2HDM-zeta_f}. 

A generalisation of the type I, II, X, Y models is given by the
flavour-aligned 2HDM (FA2HDM), proposed and
explored in Refs.~\cite{Pich:2009sp,Jung:2010ik}. In the FA2HDM one
assumes the validity of Eq.~\eqref{eq:FA2HDMrelation} with arbitrary, in
general complex, values of the coefficients $\zeta_{l,d,u}$. Such
general values are not necessarily governed by symmetries, and the
general FA2HDM Yukawa structure is not guaranteed to be stable under
renormalisation. But it provides a very useful
unified framework which contains all types with discrete
symmetries as special cases.

We mention in passing that there is a special case of the scenarios
with discrete symmetries where one of the two Higgs doublets receives
no \vev, i.e.\ where $\cot\beta=0$. This is only compatible with
the type I Yukawa couplings in order for all fermions to receive mass.
The resulting model is called the Inert Doublet model. The lightest
BSM scalar is guaranteed to be stable thanks to the unbroken
$\mathds{Z}_2$ symmetry. The Inert Doublet model is therefore of high
interest in view of dark matter. However, since the BSM scalars do not
couple to fermions (unless in generalisations \cite{Wang:2021fkn,Han:2021gfu,Han:2022juu}) it provides only tiny contributions to $\amu$, and
$\amu$ does not constrain the Inert Doublet model parameter
space. Hence we will not discuss this case further.

For reference we also provide the Yukawa Lagrangian written in terms
of mass-eigenstate fields. It follows from 
Eqs.~\eqref{eq:2HDM-Yukawa-Lag}, \eqref{eq:FA2HDMrelation} and
(\ref{2HDMHiggsbasis})
and reads
\begin{align}
	\begin{split}
		\La &\supset - \sum_{f,\mathcal{S}} \Big[ Y_f^\mathcal{S} \mathcal{S} \bar{f} \tfrac{M_f}{v}\PR f \Big ] 
		+ \sqrt{2}H^+ Y_l^{H^\pm} \bar{\nu} \tfrac{M_e}{v} \PR e  \\
		&\hspace{.45cm}+ \sqrt{2} H^+ \bar u \Big[Y^{H^\pm}_d V_\text{CKM}\tfrac{M_d}{v} \PR + 
		Y_u^{H^\pm} \tfrac{M_u}{v}V_\text{CKM}\PL\Big] d + h.c.,
	\end{split}
\end{align}
where $\mathcal{S}\in\{h,iA,H\}$.
The sum over fermion fields runs over the mass eigenstates $f=e, u$ and $d$,
where $M_{e,u,d}$ denote the diagonal fermion mass matrices and generation
indices are suppressed. The Yukawa modifier parameters appearing in
the Lagrangian are given by 
\begin{subequations}\label{eq:2HDM-Yukawa-parameters}
	\begin{gather}
		Y_f^h = \sin(\beta-\alpha) + \cos(\beta-\alpha) \zeta_f \\
		Y_f^H = \cos(\beta-\alpha) - \sin(\beta-\alpha) \zeta_f \\
		Y_{d,l}^A = Y_{d,l}^{H^\pm} = - \zeta_{d,l}, \qquad Y_u^A = Y_u^{H^\pm} = \zeta_u,
	\end{gather}
\end{subequations}
such that the couplings to the BSM Higgs states $H,A,H^\pm$ are
potentially enhanced by the $\zeta_f$, where important values
are collected in Tab.~\ref{tab:2HDM-zeta_f} and the Higgs alignment
limit is given by $\cos(\beta-\alpha)\to0$.

\subsubsection{2HDM contributions to $\amu$}

\begin{figure}
	\centering
	\begin{subfigure}{.59\textwidth}
		\centering
		\includegraphics[width=.8\textwidth]{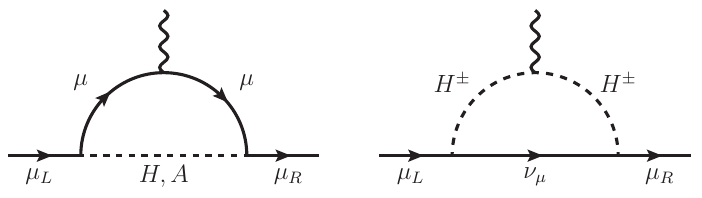}
		\caption{}
		\label{fig:FA2DM-amu-1L}
	\end{subfigure}
	\begin{subfigure}{.4\textwidth}
		\centering
		\includegraphics[width=.7\textwidth]{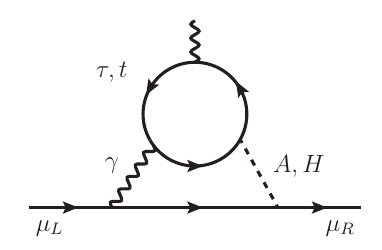}
		\caption{}
		\label{fig:FA2DM-amu-2L}
	\end{subfigure}
	\caption{Example of one- and two-loop contributions of the heavy scalars $H,A$ and $H^\pm$ to $a_\mu$ in the FA2HDM.}
	\label{fig:FA2HDM-amu}
\end{figure}
In the 2HDM significant corrections to the muon magnetic moment can arise at both the one-loop and two-loop order. 
The one-loop diagrams with the exchange of the new Higgs states $H,A,H^\pm$ 
shown in Fig.~\ref{fig:FA2DM-amu-1L} are of the generic form given in Eq.~\eqref{eq:Damu-FS}
\begin{subequations}
	\begin{align}
		\Damu^{1l} &\simeq \frac{1}{8\pi^2} \frac{m_\mu^2}{v^2} \bigg\{ 
		 (Y_\mu^A)^2\frac{m_\mu^2}{M_A^2} \Big[\ln(\tfrac{m_\mu^2}{M_A^2}) + \tfrac{11}{6}\Big]
		-(Y_\mu^H)^2\frac{m_\mu^2 }{M_H^2} \Big[\ln(\tfrac{m_\mu^2}{M_H^2}) + \tfrac{7}{6}\Big]
		-(Y_\mu^{H^\pm})^2\frac{m_\mu^2}{6M_{H^\pm}^2}    \bigg\}.
	\end{align}
\end{subequations}
In the absence of lepton-flavour violating Higgs couplings, these one-loop diagrams are
suppressed by two powers of the muon mass and two additional powers of
the muon Yukawa couplings. Therefore, even if the BSM muon Yukawa coupling is 
enhanced by a large value of $\zeta_l$, the one-loop diagrams are typically subdominant. 

References~\cite{Chang:2000ii,Cheung:2001hz,Wu:2001vq,Krawczyk:2001pe,Krawczyk:2002df,Larios:2001ma}
have stressed that the two-loop Barr-Zee diagrams discussed in
Sec.~\ref{sec:Barr-Zee} are chirally enhanced relative to the one-loop
contributions and can dominate in large parts of the 2HDM parameter
space. The simplest such diagram shown in Fig.~\ref{fig:FA2DM-amu-2L} involves 
 a one-loop
$\gamma\gamma$--Higgs subdiagram with a heavy fermion loop, where one photon and the neutral Higgs boson connects to the muon line; it
results in the contribution (cf. sec.~\ref{sec:Barr-Zee})
\begin{align}
	\Damu^{2l,f} \simeq \frac{\alpha}{8\pi^3} \frac{m_\mu^2}{v^2} N_c^f Q_f^2 \bigg\{ 
	Y_\mu^H Y_f^H \frac{m_f^2}{M_H^2} \F_1\Big(\tfrac{m_f^2}{M_H^2}\Big) + 
	Y_\mu^A Y_f^A\frac{m_f^2}{M_A^2} \F_2\Big(\tfrac{m_f^2}{M_A^2}\Big) \bigg\},
\end{align}
where the loop-functions are given in Eq.~\eqref{eq:BZ-loop-functions}.

Though the above formulas could be generalised, we continue to focus
on the framework of the FA2HDM, which contains the
well-known type I, II, X, Y models as special cases. In this general
setup the qualitative behaviour of the most important contributions
can be well understood in terms of the scalings
\begin{align}\label{eq:FA2HDM-amu-estimate}
  \Damu^{1l}\sim \mp\frac{\zeta^2_l}{8\pi^2} \frac{m_\mu^4}{v^2 M_S^2} , \qquad
  \Damu^{2l,\tau} \sim \pm \zeta_l^2 \frac{ m_\tau^2 }{v^2}\times  \frac{\alpha}{8\pi^3} \frac{m_\mu^2}{M^2_S}, \qquad
  \Damu^{2l,t} \sim -\zeta_l \zeta_u \frac{ m_t^2 }{v^2} \times
  \frac{N_cQ_t^2\alpha}{8\pi^3} \frac{m_\mu^2}{M_S^2}.
\end{align}
Here $\Damu^{1l}$ is the one-loop contribution from one of the
neutral scalars $A,H$, and $\Damu^{2l,\tau}$ and $\Damu^{2l,t}$ correspond to the Barr-Zee diagrams with a $\tau$-loop and top-loop, and we have explicitly inserted the expressions Eq.~\eqref{eq:2HDM-Yukawa-parameters} for the Yukawa
couplings in the flavour-alignment limit.
Furthermore, the behaviour of the loop functions has been approximated by $1/M_S^2$, ignoring logarithmic enhancements. The signs correspond to $S=A$ or $H$, respectively.

These scalings can also be understood in terms of the discussion of
chirality flips presented in sec.~\ref{sec:ChiralityFlips}. In particular,
the factors before $\times$ correspond to the chiral enhancement factor $R_\chi$
of Eq.~\eqref{amuchiralenhancement} (see also Tab.~\ref{tab:estimates}).
For the one-loop contribution, regardless of whether the chirality is flipped
at the external muon or the internal muon, there always appears an
explicit factor of $m_\mu^2$ and two additional powers of BSM muon
Yukawa couplings and consequently $\Damu^{1l} \propto m_\mu^4$. For the
Barr-Zee diagrams, the muon chirality is flipped via the single
coupling of the muon to the BSM Higgs, contributing one power of
$\zeta_l m_\mu/v$. The BSM Higgs also couples to the inner
fermion loop, yielding a factor of the BSM Yukawa coupling of the
inner fermion. As explained in Sec.~\ref{sec:ChiralityFlips}, any
contribution to $\amu$ must be proportional to the
EWSB \vev. In the Barr-Zee diagrams, the \vev enters via the fermion
mass of the inner fermion loop, $m_\tau$ or $m_t$. Taken together,
these considerations fully explain the factors appearing in $\Damu^{2l,\tau}$
and $\Damu^{2l,t}$. 

The estimate in Eq.~\eqref{eq:FA2HDM-amu-estimate} demonstrates how the
Barr-Zee diagrams can dominate over the one-loop diagrams. The
Barr-Zee $\tau$-loop diagram depends on the same Yukawa parameters, is
suppressed by an additional loop factor but enhanced by
$m_\tau^2/m_\mu^2$. The detailed calculations show that it dominates
over the one-loop contribution provided that the Higgs masses
$M_{A,H}$ are larger than around $M_A\gtrsim3$ GeV and $M_H\gtrsim5$
GeV
\cite{Chang:2000ii,Cheung:2001hz,Wu:2001vq,Krawczyk:2001pe,Krawczyk:2002df,Larios:2001ma}. For
lower masses, the one-loop diagram is relatively enhanced by the
logarithm $\ln(M_S/m_\mu)$ and dominates, see also the limit of the
loop function in Eq.~(\ref{LoopfunctionFSlimits}). 
For Higgs masses above these values, the total sign of the one-loop
plus the Barr-Zee $\tau$-loop contribution is positive for
$A$-exchange and negative for $H$-exchange. The Barr-Zee top-loop
contribution involves the even larger enhancement factor
$m_t^2/m_\mu^2$ compared to the one-loop contribution. But it depends on the Yukawa
parameter $\zeta_u$, which is severely constrained as discussed below.

For the following phenomenological discussion in the present section we only
consider BSM Higgs boson masses above around 5 GeV. For contributions
of lighter particles to $\amu$ we refer to the discussion in
Sec.~\ref{sec:LightDarkSector}. Specifically for the 2HDM, the few-GeV mass range
of scalars with significant Yukawa couplings is strongly constrained
by $B$ physics, see e.g.\ \cite{Dedes:2001nx,Dedes:2001hh,Schmidt-Hoberg:2013hba}.
For large Higgs masses $M_A\approx M_H\approx M_{H^\pm}\gtrsim500$ GeV the following simple seminumerical formulas hold
\begin{subequations}\label{eq:2HDM-amu-numerical}
	\begin{align}
		\Damu^{2l,\tau} &\simeq + \bigg(\frac{\zeta_l}{100}\bigg)^2  \bigg\{\frac{5.3 + \ln(x_s)}{x_s^2}\bigg\} \cdot 7.4 \times 10^{-13} \\
		\Damu^{2l,t} &\simeq - \bigg(\frac{\zeta_u\zeta_l}{100}\bigg)\hspace{.085cm} \bigg\{\frac{3.5 + 3\ln(x_s) + \ln(x_s)^2}{x_s^2}\bigg\} \cdot 1.9  \times 10^{-10},
	\end{align}
\end{subequations}
with $x_s=M_S/1000$ GeV. In contrast to the similar seminumerical results for small $M_A\ll M_{H,H^\pm}$ obtained in Ref.~\cite{Cherchiglia:2017uwv},
here the equal scalar mass limit implies a cancellation of the $\ln^2(x_s)$ enhanced terms in the $\tau$-loop contribution, which is consequently
even more suppressed compared to the top-loop correction $\Damu^{2l,t}$.

Finally, besides the dominant contributions from the fermionic Barr-Zee diagrams there are
also bosonic contributions to $\Damu$ at two-loop order. These do not
give rise to additional chiral enhancements and are therefore
generally small.
The  general class of all Barr-Zee type diagrams has been evaluated in
Ref.~\cite{Ilisie:2015tra}. A complete two-loop calculation of $\amu$
in the 2HDM has been obtained and discussed in
Refs.~\cite{Cherchiglia:2016eui,Cherchiglia:2017uwv}, and a corresponding
numerical computer code has been published in Ref.~\cite{Athron:2021evk}.
For a detailed discussion of the behaviour of the full two-loop result
as a function of all parameters in the FA2HDM see
Refs.~\cite{Cherchiglia:2016eui,Cherchiglia:2017uwv}. In general, the additional bosonic corrections can be noticeable
if $\cos(\beta-\alpha)$ differs from the SM-limit and $\zeta_l$ and $M_{H^\pm}-M_H$ become large; but given existing constraints the bosonic corrections are typically
$\lesssim\mathcal{O}(10^{-10})$.

\subsubsection{Constraints from $\amu$ and other observables on the Type I, II, X, Y and Flavour-aligned 2HDM}
\label{sec:FA2HDMpheno}
\begin{figure}
	\centering
	\includegraphics[width=.25\textwidth]{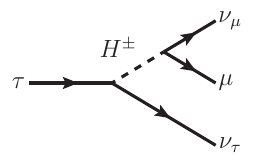}\hspace{.5cm}
	\includegraphics[width=.25\textwidth]{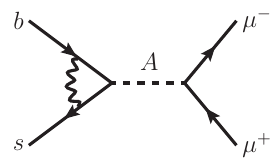}\hspace{.7cm}
	\includegraphics[width=.25\textwidth]{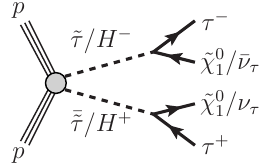}
	\caption{ Illustration of diagrams contributing to low-energy flavour observables and Higgs production at the LHC in the 2HDM.
		\emph{left}: tree-level correction of $H^\pm$ to the $\tau$ decay.
		\emph{middle}: contribution of $A$ to $B_s^0\to \mu^+\mu^-$ with virtual $W$-boson and top-quark in the loop.
		\emph{right}: stau ($H^\pm$) pair production and subsequent decay searched for at the LHC. }
	\label{fig:FA2HDM-constraint-diags}
\end{figure}

\begin{figure}
	\centering
	\includegraphics[height=.3\textwidth]{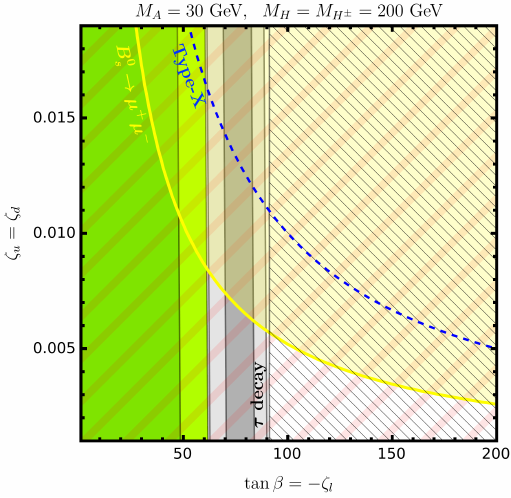}\hfill
	\includegraphics[height=.3\textwidth]{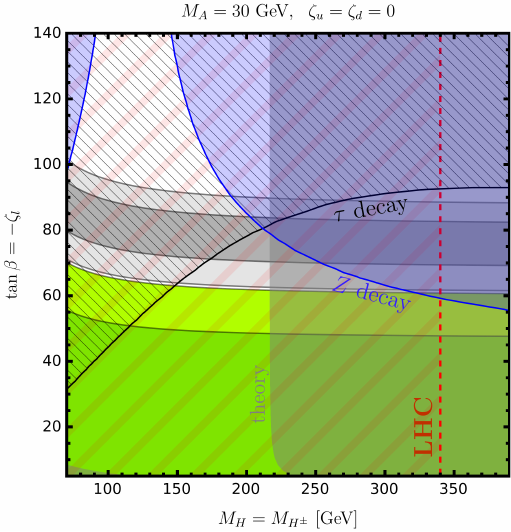}\hfill
	\includegraphics[width=.3\textwidth]{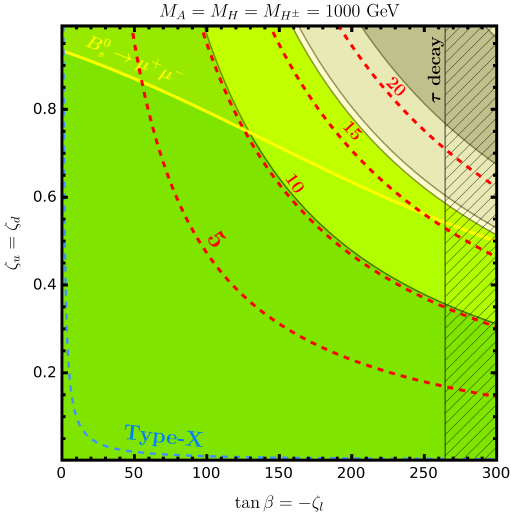}
	\caption{$\Damu$ contour plots for the FA2HDM in 
		the $\tan\beta-\zeta_u$ plane with light $M_A$ \emph{(left)},		
		the $M_{H^\pm}-\tan\beta$ plane with light $M_A$ \emph{(middle)} and 
		the $\tan\beta-\zeta_u$ plane with large scalar masses \emph{(right)}.
		The (light) green region corresponds to the 1$\sigma$ (2$\sigma$) range of $\DamuFinal$,
		and the (light) grey region to the 1$\sigma$ (2$\sigma$) range around $\DamuOld$. In the right plot
		we have included (red, dashed) contour lines for $\Damu\times 10^{10}$.		
		The excluded regions are indicated by their corresponding constraints and were obtained
		using the formulas given in Refs.~\cite{Chun:2016hzs,Iguro:2023tbk}. In particular, the red hatched regions
		are excluded by the $\tilde\tau$ search recasting of Ref.~\cite{Iguro:2023tbk}.}
	\label{fig:2HDM-plots}
\end{figure}

The estimate Eq.~\eqref{eq:FA2HDM-amu-estimate} illustrates in which parameter regions of 
the 2HDM sizeable contributions to $\amu$ are generated. In particular, a large value of $|\zeta_l| \gg 1$
(equivalently large $\tan\beta$ in type II or type X) is required, and
small Higgs masses as well as large $|\zeta_u|$ are additionally preferred.
In the past decade, after the Higgs discovery at the LHC and in
anticipation of the Fermilab $g-2$ measurement, this parameter
space has been extensively scrutinised
\cite{
  Broggio:2014mna,Wang:2014sda,
  Han:2015yys,Abe:2015oca,Chun:2015hsa,Chun:2016hzs,
  Cherchiglia:2016eui,Cherchiglia:2017uwv,Wang:2018hnw,
    Iguro:2019sly,
    Chun:2019oix,Sabatta:2019nfg,
    DelleRose:2020oaa,
    Ghosh:2021jeg,
    Athron:2021iuf,
    Jueid:2021avn,Dey:2021pyn,Atkinson:2022qnl,Kim:2022xuo,Kim:2022hvh}.
Here we discuss the most relevant constraints  identified in the
literature, see Fig.~\ref{fig:FA2HDM-constraint-diags} for sample
diagrams, then we discuss the phenomenological results, see
Fig.~\ref{fig:2HDM-plots}.

First, the possible spectrum of Higgs masses compatible with $m_h\approx125$
GeV is restricted by theoretical constraints on the Higgs potential Eq.~\eqref{eq:2HDM-potential}.
The existence of a global minimum requires \cite{Barroso:2013awa}
\begin{align}
	m_{12}^2\big(m_{11}^2 - m_{22}^2 \sqrt{\lambda_1/\lambda_2}\big)\big(\tan\beta - \sqrt[4]{\lambda_1/\lambda_2}\big) > 0,
\end{align}
and the stability of the vacuum (i.e. $V$ should be bounded from below) implies \cite{Gunion:2002zf}
\begin{align}
	\lambda_{1,2} > 0, \qquad \lambda_3 > - \sqrt{\lambda_1\lambda_2}, \quad \text{and} \quad \lambda_3+\lambda_4 - |\lambda_5| > -\sqrt{\lambda_1\lambda_2}. 
\end{align}
The quartic couplings must also be (approximately) bounded by $|\lambda_i| \lesssim 4\pi$ in order to fulfil the tree-level 
unitarity constraints \cite{Kanemura:1993hm,Ginzburg:2005dt}.
In addition, an important experimental constraint on the 2HDM spectrum
arises from requiring consistency with the $\rho$-parameter and
electroweak precision observables. Ref.~\cite{Broggio:2014mna} showed
that these basic constraints restrict the mass splittings between the
three BSM Higgs states $A$, $H$, $H^\pm$. At least two out of the
three masses must be similar, and the difference $|M_A-M_{H^\pm}|$ is
generally restricted. Two allowed mass patterns therefore are:
\begin{itemize}
	\item  A light $M_A\lesssim100$ GeV requires not too heavy $M_H\approx
		M_{H^\pm}\lesssim300$ GeV. This pattern is exemplified in our
		Fig.~\ref{fig:2HDM-plots} \emph{(left},\emph{middle)}. Here  the
		Barr-Zee diagrams with $\tau$-loop can dominate and potentially lead to
		sizeable, positive $\Damu$.
	\item For $M_H\approx M_{H^\pm}\gtrsim 300$ GeV, $M_A$ must be of the same order as
		$M_{H^\pm}$. This mass pattern is exemplified in our
		Fig.~\ref{fig:2HDM-plots} \emph{(right)}, where we set
		$M_A=M_H=M_{H^\pm}=1000$ GeV. Here all 2HDM contributions to $\amu$
		are suppressed, though $\Damu^{2l,t}$ can still reach values up to $\O(10^{-9})$.
\end{itemize}
While Ref.~\cite{Broggio:2014mna} assumed the quartic potential terms to be $\mathds{Z}_2$ symmetric,
a similar conclusion with slightly larger mass windows was also obtained for the general Higgs potential \cite{Han:2015yys}.\newline

For each pattern of Higgs masses, the values of Yukawa couplings are constrained by flavour and collider physics. 
In the FA2HDM, the important coupling parameter $\zeta_l$ is mainly restricted particularly by
$\tau$-physics, in particular by lepton flavour universality measurements in $\tau$-decays \cite{HFLAV:2022esi},
studied in Refs.~\cite{Krawczyk:2004na,Wang:2014sda,Abe:2015oca,Chun:2016hzs,Cherchiglia:2017uwv,Wang:2018hnw,Iguro:2023tbk}.
The main 2HDM contribution stems from tree-level exchange of the charged Higgs $H^\pm$ shown in Fig.~\ref{fig:FA2HDM-constraint-diags} \emph{(left)},
as well as one-loop diagrams where all new scalars contribute, yielding a correction
\begin{align}
	\Gamma_{\tau\to l \nu\nu} \simeq \Gamma_{\tau\to l \nu\nu}^\text{SM} \big(1 + 2\delta_\text{tree} + 2 \delta_\text{loop}\big),
\end{align}
where the explicit expressions for $\delta_\text{tree,loop}$ together with the current experimental bounds are summarised 
in Ref.~\cite{Iguro:2023tbk}. Complementary constraints arise also from $Z\to\tau\tau$ decays \cite{Denner:1991ie,Abe:2015oca,Chun:2016hzs,Cherchiglia:2017uwv,Wang:2018hnw,Iguro:2023tbk}
and, for $M_A\lesssim20$ GeV, $\tau\tau$-production \cite{Cherchiglia:2017uwv} both of which were measured at LEP \cite{DELPHI:2004bco,ALEPH:2005ab}. 
The resulting upper limits on $|\zeta_l|$ range from around 100 (for
$M_A=20\ldots100$) to around 250 (for $M_A=1000$ GeV). The excluded
regions are shown in Fig.~\ref{fig:2HDM-plots} as the hatched and the
blue regions, respectively, where Fig.~\ref{fig:2HDM-plots}
\emph{(left,right)} only show the stronger of the two limits, while
Fig.~\ref{fig:2HDM-plots} \emph{(middle)} shows both to illustrate
their complementarity.

The quark Yukawa couplings are severely restricted by
$B$-physics, see  Fig.~\ref{fig:FA2HDM-constraint-diags}
\emph{(middle)}, and by LHC Higgs physics. For the models with
discrete symmetry listed in Tab.~\ref{tab:2HDM-zeta_f} this has immediate
conclusions \cite{Broggio:2014mna}. Obviously, in  the type I and Y
models, $\zeta_l$ and $\zeta_u$ are equal and  constraints  on the top
Yukawa coupling translate into a strong suppression of $\zeta_l$ and
hence of $\Delta\amu$. 
In the type II model, a strong constraint on the charged Higgs mass $M_{H^\pm} \gtrsim 580$ GeV arises from 
$\text{BR}(b\to s\gamma)$ \cite{Misiak:2017bgg}.

As a result, Ref.~\cite{Broggio:2014mna} concluded that the type X model is the
only of the $\mathds{Z}_2$-symmetric models which has any potential of
significant contributions to $\amu$. The type X model was then
investigated in more detail also in view of CP violation and the
electron electric dipole moment \cite{Chun:2019oix}, and highly
sensitive modest-energy lepton collider tests via the Yukawa process $e^+e^-\to\gamma^*/Z^*\to\tau\tau A$
were proposed   \cite{Chun:2019sjo}. Within the type X model, also the
``wrong-sign'' Yukawa coupling is of interest, where the SM-like Higgs
coupling modifier $Y_e^h\approx-1$ in
Eq.~(\ref{eq:2HDM-Yukawa-parameters}), and the impact of this
difference has been investigated in
this context in Refs.~\cite{Chun:2015hsa,Chun:2016hzs,Iguro:2019sly,Ghosh:2020tfq,Ghosh:2021jeg}.

In the more general FA2HDM scenario, constraints on the quark Yukawa parameters
$\zeta_{d,u}$ can be quantified individually. This was done in 
Refs.~\cite{Han:2015yys,Cherchiglia:2016eui,Cherchiglia:2017uwv,Athron:2021iuf} and
more recently, based on important experimental updates, in Ref.~\cite{Iguro:2023tbk}. 
The recent measurements of $B^0_s\to\mu^+\mu^-$ by LHCb \cite{LHCb:2021vsc} and CMS \cite{CMS:2022mgd} have shifted the world average
from $\text{BR}(B^0_s\to\mu^+ \mu^-)^\text{2014} = (2.8 \pm 0.7) \times 10^{-9}$ \cite{CMS:2014xfa} 
to \cite{ParticleDataGroup:2024cfk}
\begin{align}
	\text{BR}(B^0_s\to\mu^+ \mu^-)^\text{2023} = (3.34 \pm 0.27) \times 10^{-9},
\end{align}
in significantly better agreement with the SM prediction \cite{Bobeth:2013uxa,Buras:2022wpw}.
As shown in Ref.~\cite{Iguro:2023tbk}, this measurement by far supersedes all previously considered
constraints on the up-type Yukawa couplings. 
In Fig.~\ref{fig:2HDM-plots} \emph{(left)} the corresponding
constraint is shown as the yellow region.
For the Higgs masses chosen in this plot, $B^0_s\to\mu^+ \mu^-$ limits
$\zeta_u$ to around 0.01 and even excludes the type 
X model where $\zeta_u=-1/\zeta_l$.

There also exist a number of direct searches at the LHC that dramatically restrict the possible 
range of BSM Higgs masses. As pointed out in Refs.~\cite{Chun:2015hsa,Iguro:2019sly},
in case of large $\zeta_l$ the heavy Higgs bosons decay dominantly into $\tau$ leptons,
resulting in a significant excess in $\tau$-rich signatures.
Resulting signatures closely resemble those of stau $\tilde\tau$ production in the context of supersymmetric models, 
which have been studied extensively. The similarity is illustrated in
Fig.~\ref{fig:FA2HDM-constraint-diags} \emph{(right)}.
In fact, the CMS search for $\tilde\tau_L$ pair production
\cite{CMS:2022syk} can be directly adapted to the FA2HDM and excludes charged 
scalar masses in the range of $115\leq M_{H^\pm} \leq 340$ GeV, assuming BR$(H^\pm\to\tau^\pm\bar\nu)\simeq 1$~\cite{Iguro:2023tbk}.
In Fig.~\ref{fig:2HDM-plots} \emph{(left,middle)} this constraint is
shown as the red hatched region. In the \emph{(middle)} plot, it is overlaid on top of the parameter space for $M_A=30$ GeV allowed by the $B$-physics, $Z$ and $\tau$ constraints, 
covering precisely the gap where  large $\Delta\amu$ was still
possible. In the \emph{(left)} plot, it excludes the entire shown region.
A proper recasting of the updated stau analyses performed by ATLAS and CMS
\cite{ATLAS:2017qwn,ATLAS:2019gti,ATLAS:2022nrb,CMS:2022syk} using the full Run-2 data set 
was also presented in Ref.~\cite{Iguro:2023tbk} and excludes charged
scalar masses up to more than 500 GeV and also excludes the region
with small scalar masses that is still allowed in Fig.~\ref{fig:2HDM-plots} \emph{(middle)}.

These recent additional $B$-physics and LHC search constraints
change the viable 2HDM parameter space strongly. To illustrate the
difference we note that prior to these constraints,
Refs.~\cite{Cherchiglia:2017uwv,Athron:2021iuf} evaluated the overall 
maximum possible contribution $\Damu$ in the type X and FA2HDM and
found values around
$(30\ldots50)\times10^{-10}$, respectively. In order to achieve such
values, $M_A$ around 30 GeV and maximised $\zeta_{l,u}$ were
needed. Now, with the recent additional constraints this entire region
with light pseudo-scalar, large $|\zeta_l|$  and $\Damu\sim\O(10^{-9})$  considered
in
Refs.~\cite{Broggio:2014mna,Wang:2014sda,Han:2015yys,Abe:2015oca,Chun:2015hsa,Chun:2016hzs,Cherchiglia:2016eui,Cherchiglia:2017uwv,Wang:2018hnw,Iguro:2019sly,Chun:2019oix,Sabatta:2019nfg,DelleRose:2020oaa,Ghosh:2021jeg,Jueid:2021avn,Dey:2021pyn,Atkinson:2022qnl,
    Kim:2022xuo,Kim:2022hvh}
is excluded \cite{Iguro:2023tbk}.

In a recent global fit of the FA2HDM
focusing on light new Higgs bosons, the model parameter space was extensively
investigated  \cite{Coutinho:2024zyp}.
This reference finds allowed parameter regions where light
$M_A$ is viable, and it even finds larger allowed
$\zeta_{l,u}$ than in Ref.~\cite{Iguro:2023tbk} and visible in the
plots of Fig.~\ref{fig:2HDM-plots}  \emph{(left, middle)}. However, it also
confirms that there is no viable possibility for sizeable 
contributions to $\amu$ within all its investigated FA2HDM parameter
space where at least one of the BSM Higgs masses is below
$M_h\approx125$  GeV.

This negative  result is further complemented by a recent dedicated search for the light pseudo-scalar $A$ with masses 
in the range of $M_A=20...90$ GeV performed by ATLAS \cite{ATLAS:2024rzd}, where no significant excess above the SM background was found.
This search considered the process $gg\to A\to\tau\tau$ with single production of $A$ via gluon fusion and
subsequent decay into a $\tau$ pair and provides both a
model-independent upper limit on the production cross-section as well
as a FA2HDM-specific bound 
on $|\zeta_u|\lesssim 0.2$ in the relevant parameter region.\newline

We are thus led to focus on the remaining parameter space for the 2HDM
in case of large lepton Yukawa couplings $\zeta_l$, which
is characterised by large Higgs masses  where the theoretical constraints enforce smaller (relative) mass splittings such that we consider $M_A\approx
M_H\approx M_{H^\pm}\gtrsim500$ GeV.
Since the FA2HDM contributions to the $Z$ decay vanish for
degenerate masses and the corrections to the $\tau$ decay scale as $\zeta_l^2 / M_S^2$,
larger values of $\zeta_l\sim 250$ are allowed in this region. 
At the same time thanks to the heavy masses, $B$-physics 
constraints are much weaker and allow 
up to $\zeta_u\lesssim 0.8$.
In Fig.~\ref{fig:2HDM-plots} \emph{(right)} we show contours for
$\Damu$ in the $\zeta_l$--$\zeta_u$ plane in this heavy mass region,
fixing $M_A=M_H=M_{H^\pm}=1000$ GeV. 
Following Eq.~\eqref{eq:2HDM-amu-numerical} the dominant corrections to $\Damu$ stem from the top loop,
while the $\tau$ contributions are strongly suppressed.
Particularly, if both couplings $\zeta_{l,u}$ are maximised, $\Damu$ can
still reach values up to $10^{-9}$ in the FA2HDM even for Higgs masses
as heavy as 1 TeV.
A similar result has been obtained in a recent global fit
\cite{Karan:2023kyj} with focus on the
heavy-mass region of the FA2HDM. Although $\amu$ was not included as a
constraint in the fit, significant contributions $\Damu$ were found to
be possible.
Conversely, this means that the current result $\DamuFinal$ constrains the FA2HDM parameter
space in this region of very large $\zeta_l\zeta_u$ in a non-trivial way. 
In contrast, in the type X scenario the small value of
$\zeta_u$ prevents significant contributions to $\amu$, such that the
small value $\DamuFinal$ 
does not provide additional constraints on the type X model any more.

\subsubsection{General  2HDM and summary}

Although the FA2HDM is already a very general framework, it is
of interest to study the range of possibilities of more general Yukawa
structures in the 2HDM, for reviews focusing on $\amu$ in this context
see Refs.~\cite{Wang:2022yhm,Iguro:2023tbk}. The major limiting feature of the FA2HDM is
the equality between the muon and $\tau$-lepton Yukawa enhancements
via the common factor $\zeta_l$. Removing this equality immediately
relaxes the constraints related to $\tau$-physics and LHC searches
involving $\tau$ final states.

Refs.~\cite{Botella:2020xzf,Botella:2022rte}  consider the obvious
generalisation where effectively, the single quantity $\zeta_l$ in the
lepton sector is generalised to three parameters
$\zeta_{e,\mu,\tau}$ (equivalent to $n_{e,\mu,\tau}$ in
Refs.~\cite{Botella:2020xzf,Botella:2022rte} and $\kappa_{e,\mu,\tau}$
in Ref.~\cite{Han:2018znu}; the so-called power-alignment of
Ref.~\cite{Li:2020dbg} can be considered as a special case of this
idea).
This does not introduce
FCNC but it allows for 
greater freedom in the Yukawa sector.
In this way, $\tau$-constraints can be relaxed by smaller
$\zeta_\tau$, which in turn allows significantly larger
$\zeta_\mu$. Translating to our conventions, Ref.~\cite{Botella:2022rte} finds 2HDM scenarios with
$\Damu\sim10^{-9}$ of several kinds:
\begin{itemize}
\item
  All $A$, $H$, $H^\pm$ Higgs masses  of order 1 TeV and similar, with
  very large $\zeta_\mu\sim1000$ and significant $\zeta_u$: this region is
  similar to the FA2HDM region shown in Fig.~\ref{fig:2HDM-plots}
  \emph{(right)} and allows sizeable $\Damu$ via top-loop Barr-Zee
  diagrams.
\item
  All $A$, $H$, $H^\pm$ Higgs masses of order 1 TeV,  $M_H<M_A$, with
  very large $\zeta_\mu$: if the mass gap between $M_H$ and $M_A$ is
  sufficiently large, the one-loop diagram with $H$-exchange can
  provide a significant, positive contribution to $\amu$.
\item
  All $A$, $H$, $H^\pm$ Higgs masses between 500\ldots1000 GeV, with
  $M_H<M_A$ and  opposite signs for $\zeta_\mu$ and $\zeta_\tau$: in
  this case the $\tau$-loop Barr-Zee diagram with $H$-exchange is
  positive and can provide a relevant contribution to $\amu$.
\end{itemize}
In addition, Ref.~\cite{Jana:2020pxx} finds viable parameter regions
for additional 2HDM Higgs bosons with very light masses $M_H\lesssim1$
GeV if only Yukawa couplings to electron and muon are sizeable.

The so-called $\mu$-specific 2HDM \cite{Abe:2017jqo} can be regarded as a special case of
the scenario of Refs.~\cite{Botella:2020xzf,Botella:2022rte}, where
the Yukawa pattern follows a discrete symmetry singling out the
muon. In contrast to the usual type I, II, X, Y models, the
$\mu$-specific model features a $\mathds{Z}_4$ symmetry, which
effectively implies type I Yukawa couplings with the exception that
the muon couples to $\Phi_1$ instead of $\Phi_2$, such that
$\zeta_\mu=-\tan\beta$. For $\tan\beta$ of order 1000, and a
sufficient mass splitting $M_H< M_A$, the one-loop diagrams with
$H$-exchange can provide a positive and sizeable contribution to $\amu$
\cite{Abe:2017jqo}. The recent update in Ref.~\cite{Iguro:2023tbk}
however finds that the required Higgs mass splitting is strongly
constrained by perturbativity of the Higgs potential, in particular if
perturbativity up to higher energy scales is required.

A different variant of the 2HDM is given by the $\mu$--$\tau$-specific
model, where the BSM Higgs--lepton Yukawa coupling mediates a
$\mu$--$\tau$ transition. It is possible to control charged lepton
flavour violation while obtaining sizeable contributions to $\amu$. We
refer to
Refs.~\cite{Iltan:2001nk,Wang:2016rvz,Wang:2019ngf,Hou:2021sfl,Hou:2021qmf,Asai:2022uix,Blanke:2022kpi,Benbrik:2022dja,Babu:2022pdn,Iguro:2023tbk}
for discussions of this case and its generalisations.

In summary, the 2HDM $\amu$ phenomenology has changed a significantly
in recent years. The parameter region with low pseudoscalar mass $M_A$
in the type-X or more general FA2HDM with large $\tan\beta$/large
$\zeta_l$  has been scrutinised and is
now excluded as a viable region with large $\Damu$.
Generally now, most viable 2HDM scenarios predict small contributions to $\amu$
within the $1\sigma$ band around the new value $\DamuFinal$.

Still, in exceptional parameter regions the 2HDM can accommodate
sizeable contributions to $\amu$; 
e.g.\ the FA2HDM can reach $\Damu\sim10^{-9}$
in the heavy-mass region with maximised $\zeta_l\zeta_u$ as illustrated
in Fig.~\ref{fig:2HDM-plots} \emph{(right)}, or  the models with
generalised Yukawa patterns mentioned in the previous paragraphs can
reach even larger values.
We stress again that here we  have mainly focused on those 2HDM
  scenarios which have the potential of sizeable contributions to
  $\amu$. These are now constrained in non-trivial ways by the new
  result $\DamuFinal$, but could also have the potential of accommodating
  discrepancies, should such discrepancies emerge in the
  future.

  Independently of $\amu$, there are motivated and simple variants such
  as the 2HDM type I or type II which are very well studied, and we refer
  to Refs.~\cite{Eberhardt:2020dat,Athron:2024rir,
    Atkinson:2021eox,
    Athron:2021auq,
    Atkinson:2022pcn,
    Karan:2023kyj,                 
    Coutinho:2024zyp}
  for recent in-depth 2HDM analyses in
  the light of current experimental data.
  These references carry out systematic and general global fits of
  various types of 2HDM parameter spaces without including $\amu$ as a
  constraint in
  the fits. They find viable regions in the FA2HDM with light new Higgs
  states \cite{Coutinho:2024zyp} or heavy new Higgs states
  \cite{Eberhardt:2020dat,Karan:2023kyj}, viable parameter regions in the general 2HDM
  \cite{Athron:2021auq,Athron:2024rir} and best-fit regions with Higgs masses around
  $1.7\ldots2$ TeV in the type I \cite{Atkinson:2022pcn} and type II \cite{Atkinson:2021eox}
  models. Except for the FA2HDM region mentioned earlier, all viable regions
  found in these references predict
  negligible contributions to $\amu$, which is now in agreement with
  the new result $\DamuFinal$.

Finally, we remark that the 2HDM is frequently considered not in
isolation but as part of a larger framework, e.g.~together with
 additional very light new physics
\cite{Arcadi:2021zdk,Arcadi:2021yyr,Arcadi:2022dmt,Arcadi:2022lpp,Arcadi:2023rbv} or
further scalars
\cite{Chun:2021rtk,Bharadwaj:2024gfo,Chakrabarty:2021ztf,Chakrabarty:2022dai,Chen:2021jok,Herms:2022nhd},
where also dark matter may be explained, together with
vector-like fermions
\cite{Frank:2020smf,Chun:2020uzw,Dermisek:2023tgq,Ferreira:2021gke,Bharadwaj:2021tgp,Dermisek:2020cod,Dermisek:2021ajd,Raju:2022zlv,Raju:2024ram}
(see also Sec.~\ref{sec:VLF+scalar}) or neutrinos \cite{Raju:2024ram},
extra U(1) gauge bosons \cite{Lee:2022sic,Mondal:2021vou} or more
complicated extensions
\cite{Ghosh:2020fdc,Nomura:2020kcw,Nomura:2021adf,Hue:2021xzl,Hernandez:2021xet,Hernandez:2021tii,Hernandez:2021kju,Lee:2021gnw,Liu:2021kug,Tran:2022cwh,Abdallah:2022shy,Duy:2024lbd}.
The 2HDM can also be extended to the 3HDM which offers a variety of potential mechanisms, including
radiative muon mass generation, such that $\Damu$ behaves as discussed
generically in Sec.~\ref{sec:MuonMass} \cite{Hernandez:2021iss}.
All mentioned references focus on $\amu$ in
connection with complementary constraints, and we refer to the
literature for  details of these investigations.

\subsection{Leptoquark models}
\label{sec:LQ}

Leptoquarks by definition are particles with direct couplings to
quarks and leptons that can be assigned baryon and lepton
number. Leptoquarks naturally arise both in the Higgs and the gauge
sectors of unified theories where quarks
and leptons are combined into larger gauge multiplets. More generally
leptoquark models can be viewed as simple extensions of the SM by one
or few new fields which are motivated by the structure of the quark
and lepton sectors and their relationships. Ref.~\cite{Buchmuller:1986zs} has identified 10 possible types of leptoquarks with
renormalizable and gauge invariant couplings.\footnote{An additional
two leptoquarks states were included in the review
\cite{Dorsner:2016wpm}, however these can only couple to right handed
neutrinos that are not present in the standard model.     }  The 10 types have either
spin 0 or spin 1, and they differ by their electroweak gauge quantum
numbers and the resulting possible couplings to up- or down-type left-
or right-handed quark and lepton fields.

Adding leptoquarks to the SM dramatically modifies the flavour
structure and introduces a large number of free parameters governing
the interactions with quarks and leptons of all generations. Depending
on these couplings and on the leptoquark mass, many flavour
observables can be affected, including $\amu$ but also CLFV and quark
flavour observables. In addition, all leptoquarks are strongly
interacting and hence can be produced at the LHC, provided their
masses are accessible. The LHC experiments therefore obtained lower
limits on viable leptoquark masses with values in the range
1~TeV\ldots5~TeV, see the discussion of Sec.~\ref{sec:Collider} and
the summary Sec.~94 of Ref.~\cite{ParticleDataGroup:2024cfk}. The values of the limits depend
on the assumed flavour patterns of couplings and resulting decay
modes. For the scenarios relevant for $\amu$, leptoquark couplings to
muons and heavy quarks are needed, and the appropriate limits are
around 2 TeV. Compared to other scenarios of interest for $\amu$, these are
very high mass limits, but the $\amu$ phenomenology is not very
sensitive to the precise values; hence in the following we focus on
leptoquark masses above  2~TeV.

For an extensive review covering collider and quark and
lepton flavour physics of leptoquarks we refer to
Ref.~\cite{Dorsner:2016wpm}.  Since the appearance of that review,
leptoquark models have been intensely studied in view of experimental
progress on $\amu$ and also on $B$-physics where several results
hinted at a violation of lepton flavour universality \cite{Crivellin:2021sff}.
It turned out that leptoquark models were particularly promising and
intensely studied for
combined explanations of deviations from the SM in $\amu$ and
$B$-physics
\cite{Bauer:2015knc,Das:2016vkr,Popov:2016fzr,Cai:2017wry,Buttazzo:2017ixm,Crivellin:2017zlb,Angelescu:2018tyl,Dorsner:2019itg,DEramo:2020sqv,Babu:2020hun,Crivellin:2020tsz,Gherardi:2020qhc,Lee:2021jdr,Greljo:2021npi,Nomura:2021oeu,Marzocca:2021azj,Zhang:2021dgl,
  Wang:2021uqz,FileviezPerez:2021lkq,Greljo:2021xmg,Chen:2022hle,Crivellin:2022mff,Freitas:2022gqs,Julio:2022ton,Julio:2022bue,Cheung:2022zsb,Dev:2024tto}.\footnote{%
Similarly, leptoquarks were considered as combined explanations of the
deviations from the SM in $\amu$ and the 2022 CDF measurement of $M_W$
\cite{Athron:2022qpo,Bhaskar:2022vgk,Cheung:2022zsb,Chowdhury:2022dps,He:2022zjz}. The
CDF  measurement and more recent measurements of $M_W$ were briefly
discussed in
Sec.~\ref{sec:EWPO}.

}
However,
more recent experimental results have not confirmed those hints for
lepton flavour universality violation in $B$-physics
\cite{LHCb:2022qnv,Capdevila:2023yhq}. 
In the
following we explain the possible leptoquark contributions to $\amu$
and how leptoquark models are constrained by $\amu$ and the most
directly connected observables. At the end of the subsection we also
survey important recent developments related to $B$-physics.

\subsubsection{Leptoquark types with and without chiral enhancements}

The phenomenology of leptoquarks strongly depends on the type
\cite{Buchmuller:1986zs} and the assumed flavour structure of 
couplings. For $\amu$ it is particularly important whether couplings
to left-handed and right-handed leptons are simultaneously
possible. It has been known for a long time
\cite{Chakraverty:2001yg,Mahanta:2001yc,Cheung:2001ip} that in this case 
very large, chirally enhanced contributions to $\amu$ exist. Three out of the 10 types of leptoquarks allow for such
chiral enhancements.\footnote{%
A fourth one, the spin-1 leptoquark $V_1$ is constrained strongly when
it is incorporated into an SU(5) ultraviolet completion and hence
effectively cannot contribute to $\amu$ with significant enhancement
\cite{Biggio:2016wyy}; for phenomenological discussions of $\amu$ in this model see
Ref.~\cite{Bhaskar:2024swq}.  
}
As an illustration we
focus on one of them, denoted as $S_1$ model in
Ref.~\cite{Buchmuller:1986zs}. The Lagrangian terms for the $S_1$
couplings to quarks and leptons read
\begin{align}
\La_{S_1} &= 
-
{\lambda}^{ij}_L~\overline{{q}^{C}_{Li}} i\sigma_2 l_{Lj}\,S_1 
- {\lambda}^{ij}_{R}\overline{{u}_{Ri}^{C}}{e}_{Rj}\,S_1  +h.c.
.
\end{align}
Here the leptoquark $S_1$ has spin 0 and the $\GSM$  gauge quantum numbers $(\overline{\pmb{3}},\pmb{1},1/3)$. 
It is thus a scalar $\SUL$ singlet with electric charge  $Q_{S_1} =
1/3$. These quantum numbers allow gauge invariant couplings  both to
left-handed leptons (and left-handed quarks) and to right-handed
leptons (and right-handed quarks). The couplings are described by two
$3\times3$ matrices in flavour space $\lambda_{L,R}^{ij}$.
We omit here other gauge invariant operators such as diquark
operators, which could mediate proton decay
\cite{Arnold:2013cva,Queiroz:2014pra}.

In writing the Lagrangian above we have assumed that the lepton and
up-type quark fields are already rotated to their mass basis, while
the down-type quark fields inside the doublets need to be rotated to
the mass basis using the CKM matrix
 $V_{\mathrm{CKM}}$, such that the interactions with mass eigenstate
fields become
\begin{align}
  \label{LQLagrangianflavour}
\La_{S_1} \ni 
-\overline{u_i^C} 
\big( \lambda_L^{ij} P_L + \lambda_R^{ij} P_R \big) e_{j} S_1 
+ 
\overline{d_i^C} \big( \lambda_L^{kj} V^{ki}_{\mathrm{CKM}} P_L \big) \nu_{j} S_1  + h.c.
\end{align}
In particular, the left-handed (right-handed) muon couples to the
left-handed (right-handed) up-type quarks. Similarly, the so-called
$R_2$ is   a scalar $\SUL$ doublet
leptoquark which allows couplings of the left-handed (right-handed) muon to
the right-handed (left-handed) up-type quarks; and the $U_1$
leptoquark is  a
spin-1 $\SUL$ singlet which allows couplings of the left-handed (right-handed) muon to
the left-handed (right-handed) down-type quarks. For all other seven
types of leptoquarks, only 
couplings to either left- or right-handed leptons are allowed, but not
both.

The $S_1$ leptoquark model, and similarly $R_2$ and $U_1$, can be
related to our previous discussion of 3-field models in
Sec.~\ref{sec:genericthreefield} and chirality flips in
Sec.~\ref{sec:ChiralityFlips}. Together with the up-type quarks and
their Yukawa couplings to the Higgs, the $S_1$ model precisely
represents a 3-field model of Class I. The left- and
right-handed up-type quarks
correspond to the fermions $\psi$, $\chi$, and the up-type Yukawa
couplings correspond to the Higgs coupling $\lambda_H$ in the
Lagrangian of Eq.~\eqref{eq:LaClassI}. In such models, the chiral
symmetry related to the muon mass, see Eq.~(\ref{muchiralsym}), is not
only broken by the muon Yukawa coupling but also by the product of the
additional couplings as shown in
Eq.~(\ref{3fieldchiralenhancements}). Specifically for the $S_1$ model, the
relevant 
new sources of chirality flips are the combinations
\begin{align}
  \lambda_L^{32}\lambda_R^{32} (y_u)_{33},&&
  \lambda_L^{22}\lambda_R^{22} (y_u)_{22},
\label{LQchiralitybreaking}
\end{align}
which involve the top-quark and the charm-quark Yukawa coupling,
respectively. They can be significantly larger than the muon Yukawa
coupling even if the $\lambda_{L,R}$-couplings are perturbatively
small.

\subsubsection{Leptoquark contributions to $\amu$}

\begin{figure}
	\centering
	\begin{subfigure}{.6\textwidth}
		\centering
		\includegraphics[width=.9\textwidth]{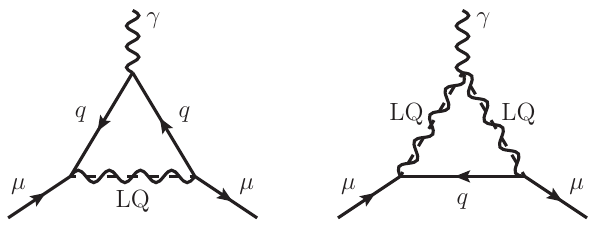}
		\caption{LQ correction to $\Damu$}
		\label{fig:LQ-amu}
	\end{subfigure}\hspace{1cm}
	\begin{subfigure}{.26\textwidth}
		\centering
		\includegraphics[height=.8\textwidth]{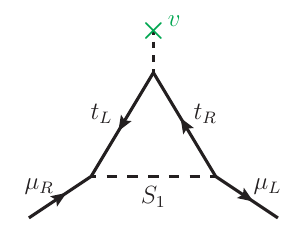}
		\caption{LQ correction to $m_\mu$}
		\label{fig:LQ-mmu}
	\end{subfigure}
  	\caption{One-loop Feynman diagrams with leptoquarks
            (LQ).  \textbf{(a)}: Generic spin-0 or spin-1 LQ one-loop correction to $\Damu$, \textbf{(b)}: $S_1$
  		one-loop contribution to $m_\mu$ with top-quark mass insertion.  }
  	\label{fig:LQdiagrams}
\end{figure}

Leptoquarks can contribute to $\amu$ via 1-loop Feynman diagrams as
shown in Fig.~\ref{fig:LQ-amu}, with a leptoquark and a SM quark in
the loop. For scalar leptoquarks, the
diagrams correspond to the generic diagrams \ref{fig:amu-FV}, for
vector leptoquarks to \ref{fig:amu-FV}. Specifically for the $S_1$
leptoquark, the loop contains up-type quarks like the top quark, and
the couplings allow a chiral enhancement as illustrated in
Fig.~\ref{fig:LQ-mmu}, where the chirality is flipped via
the top-quark mass instead of the muon mass.

The resulting 1-loop contribution to $\amu$ in the $S_1$ model from a
loop with the up-type quark $u_i$ reads
\begin{align}\label{DamuLQ}
\Damu^{S_1} &=
		\frac{N_C}{16\pi^2} \frac{m_\mu^2}{M_{S_1}^2} \bigg(
\frac{m_{u_i}}{m_{\mu}} \Re{\lambda_L^{i2}{}^*\lambda_R^{i2}}
\F^\text{FS}(0, x;-1/3)
+ \big\{|\lambda_L^{i2}|^2+|\lambda_R^{i2}|^2 \big\} \G^\text{FS}(0,x;-1/3)
	\bigg),
\end{align}
where $N_C=3$ is the number of colours and $M_{S_1}$ is the mass of
the leptoquark. In the loop functions, the mass ratio $x = 
m_{u_i}^2/M_{S_1}^2$ has been used, the charge $-Q_{S_1}=-1/3$ needs to
be inserted, while the muon mass has been neglected.
The first term in Eq.~(\ref{DamuLQ}) contains the chiral enhancement factor
ratio $m_{u_i}/m_{\mu}$, resulting from the new sources of chiral
symmetry breaking identified in
Eq.~(\ref{LQchiralitybreaking}). It is illuminating to compare this to Eqs.\
(\ref{eqn:GeneralGM2Contribution}) and (\ref{oneloopexamples}) and the discussion of
  the correlation to the muon mass. We can write the chirally enhanced contributions to
  $m_\mu$ and to $\amu$ in a uniform way as
  \begin{align}
      \frac{\Delta m_\mu^{S_1}}{m_\mu} &=
- C_{\text{BSM}} \times B_0(0,m_{u_i},M_{S_1}),
  &
  \Damu^{S_1} &=
C_{\text{BSM}} \times \F^\text{FS}(0, x;-1/3)\frac{m_\mu^2}{M_{S_1}^2} ,
\end{align}
where the
coefficient governing the magnitude of the contributions  can be written as
\begin{align}
  \label{CBSMLQ}
  C_{\text{BSM}} = \frac{N_C\big[\Re{\lambda_L^{i2}{}^*\lambda_R^{i2}}}{16
    \pi^2}\frac{m_{u_i}\big]}{ m_\mu}\,.
\end{align}
The important factors related to chirality flips and electroweak
symmetry breaking have been written between square brackets, as in Eqs.~(\ref{mmugeneric},\ref{amugeneric}).
Especially in case of the top-quark the ratio $m_t / m_\mu \approx
1600$ can provide a huge chiral  enhancement, which can lead to huge
contributions to $\amu$, but also $\O(1)$ corrections to $m_\mu$. Even in case
of the charm-quark the chiral enhancement can be very large, if the
LQ couplings are large.

The result in Eq.~(\ref{DamuLQ}) can also be used to illustrate the
behaviour of other types of leptoquarks different from $S_1$. For all
three types with chiral enhancement, $S_1$, $R_2$, $U_1$, a term of
the form $m_{u_i,d_i}\lambda_L^{i2*}\lambda_R^{i2}$ exists which has a
strong or very strong chiral enhancement in case the leptoquark
couples muons to charm or top quarks (in case of $S_1,R_2$) or bottom
quarks (in case of $U_1$). In case these chirally enhanced terms are
present, they strongly dominate the result. In case they are zero, the
 $|\lambda_L|^2$- or $|\lambda_R|^2$-terms remain. These have no
chiral enhancement, but at least one term  of this
structure exists for all types of leptoquarks.

The strongest chiral enhancement from the top quark is present if both coupling factors
$\lambda_L^{32}$ and $\lambda_R^{32}$ connecting the muon to the top
quark are non-vanishing. Similarly, the weaker chiral enhancement via the
charm quark is invoked if both coupling factors $\lambda_L^{22}$ and
$\lambda_R^{22}$ relating only 2nd-generation fields are
non-vanishing. If either $\lambda_L$ or $\lambda_R$ vanishes, no
chiral enhancement is possible.

Instructive seminumerical approximations for the
most interesting contributions read
\begin{align}
\text{\emph{top-loop:\quad}}  \Damu^{S_1}
&
\approx
33 \times 10^{-10}
~
\frac{1+0.64 \ln(M_{S_1}{}/2\,{\text{\,TeV}})}{(M_{S_1}{}/2\,{\text{\,TeV}})^2}
~
\frac{ \Re{\lambda_L^{32}{}^*\lambda_R^{32}} }{0.01}
\,,
\label{LQtopnumerical}
\\
\text{\emph{charm-loop:\quad}}\Damu^{S_1}
&
\approx
54 \times 10^{-10}
~
\frac{1+0.14 \ln(M_{S_1}{}/2\,{\text{\,TeV}})}{(M_{S_1}{}/2\,{\text{\,TeV}})^2}
~
\Re{\lambda_L^{22}{}^*\lambda_R^{22}}
\,,
\label{LQcharmnumerical}
\\
\text{\emph{no chiral enhancement:\quad}}
\Damu^{S_1}
&
\approx
8\times10^{-12}
~
\frac{\big|\lambda_{L,R}^{i2} \big|^2\,
}{(M_{S_1}{}/2\,{\text{\,TeV}})^2}.
\label{eq:LQnochiralityflipnumerical}
\end{align}
Here the logarithmic terms arise from the limit of the loop function
(\ref{LoopfunctionFSlimits}) for small fermion and large scalar mass;
the numerical factors are from
Ref.~\cite{Khasianevich:2023duu}, where
the pure one-loop result (\ref{DamuLQ}) has been evaluated using running
quark masses at the TeV-scale and improved by including the
QED-logarithms as discussed in Sec.~\ref{sec:photonic} and as
implemented in {\tt FlexibleSUSY}
\cite{Athron:2014yba,Athron:2017fvs}.

These numbers show that large contributions of
$\Damu\sim20\times10^{-10}$ are easy to obtain in this model, for
perturbative values
of the couplings. Hence leptoquark models with such chiral
enhancements were discussed as  promising explanations of
$\DamuOld$ in Eq.~(\ref{eq:DamuOld})  in the literature
in the Brookhaven era
\cite{Chakraverty:2001yg,Mahanta:2001yc,Cheung:2001ip} and later, partially
motivated by deviations from the SM in $B$-physics,
\cite{Bauer:2015knc,Das:2016vkr,Popov:2016fzr,Cai:2017wry,Buttazzo:2017ixm,Crivellin:2017zlb,Angelescu:2018tyl,Dorsner:2019itg,DEramo:2020sqv,Babu:2020hun,Crivellin:2020tsz,Gherardi:2020qhc,Lee:2021jdr,Nomura:2021oeu,Marzocca:2021azj,Zhang:2021dgl,
	Chen:2022hle,Crivellin:2022mff,Freitas:2022gqs,Julio:2022ton,Julio:2022bue,Wang:2021uqz,FileviezPerez:2021lkq,Dev:2024tto,Greljo:2021xmg,%
  Biggio:2014ela,
  Bauer:2015knc,
    Popov:2016fzr,
  Das:2016vkr,ColuccioLeskow:2016dox,Kowalska:2018ulj,
  Dorsner:2019itg,Crivellin:2020tsz,Gherardi:2020qhc,Babu:2020hun,
  Crivellin:2020mjs,Athron:2021iuf,Fajfer:2021cxa,Khasianevich:2023duu}. More
generally, the $S_1$, $R_2$ and $U_1$ models are three of very few
viable single-field explanations of $\DamuOld$
\cite{Queiroz:2014zfa,Chiu:2014oma,Biggio:2014ela,Biggio:2016wyy,Athron:2021iuf},
where the spin-1 leptoquarks have  been particularly considered in
Refs.\ \cite{Queiroz:2014zfa, Biggio:2016wyy}.

In contrast, leptoquark models without chiral enhancements behave as
Eq.~(\ref{eq:LQnochiralityflipnumerical}), which fits to  the
generic result (\ref{nonenhancedoneloop}). Since the leptoquark masses
are strongly constrained from below, the non-enhanced contributions to $\amu$ are
tiny, even if the couplings $\lambda_{L,R}$ are as large as,
e.g.\ $\sqrt{4\pi}$. Such leptoquarks could not explain any
significant deviation $\Damu$. Conversely, since the new result
$\DamuFinal$ does not show a significant deviation, such leptoquark
models are fully in agreement with $\amu$ data but cannot be
constrained by $\amu$ in a non-trivial way.

\subsubsection{Constraints on leptoquarks from $\amu$ and related
  observables}

\begin{figure}[ht!]
	\centering
	\includegraphics[width=0.32\textwidth]{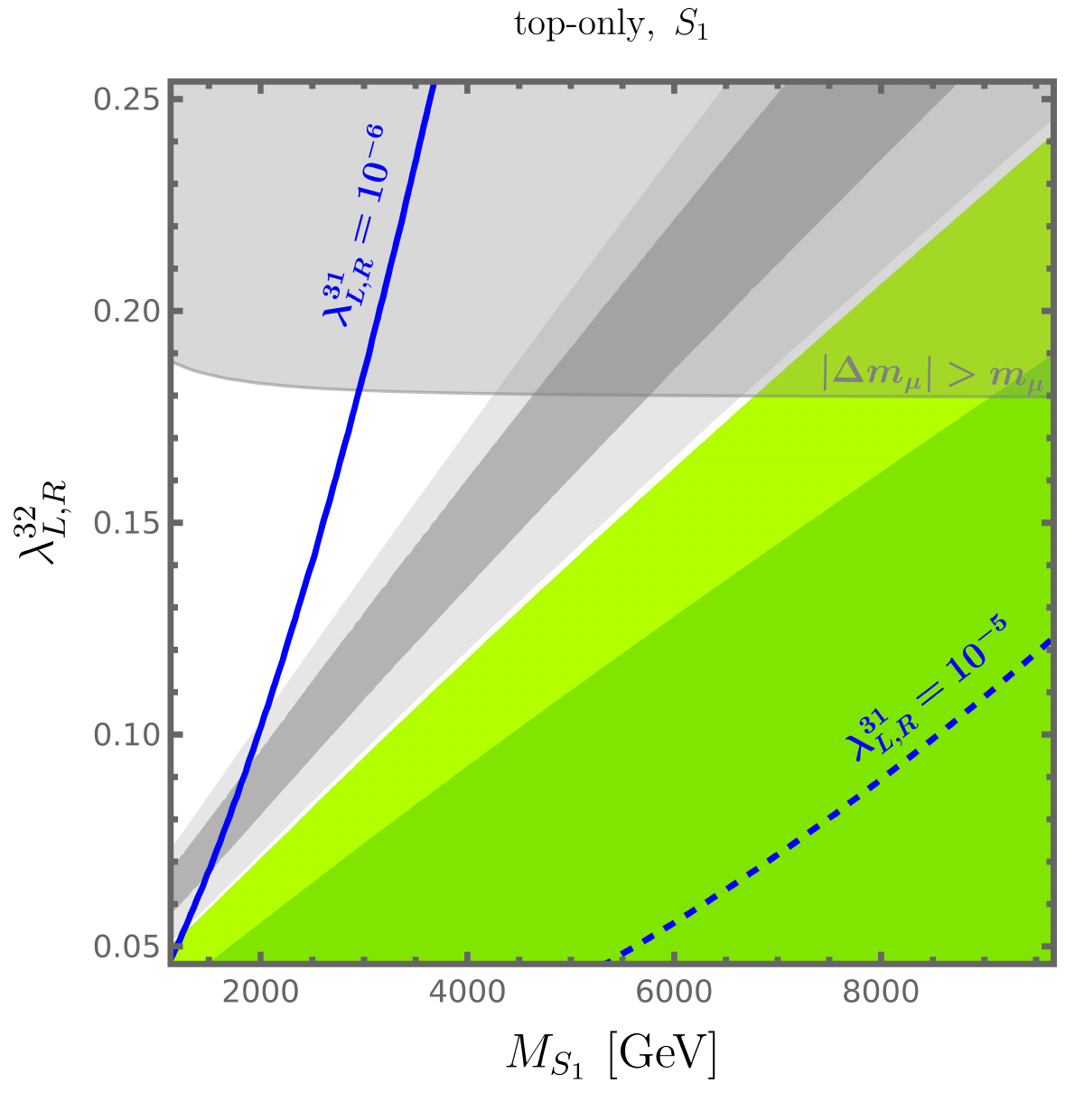}
	\hspace{1.5cm}
	\includegraphics[width=0.32\textwidth]{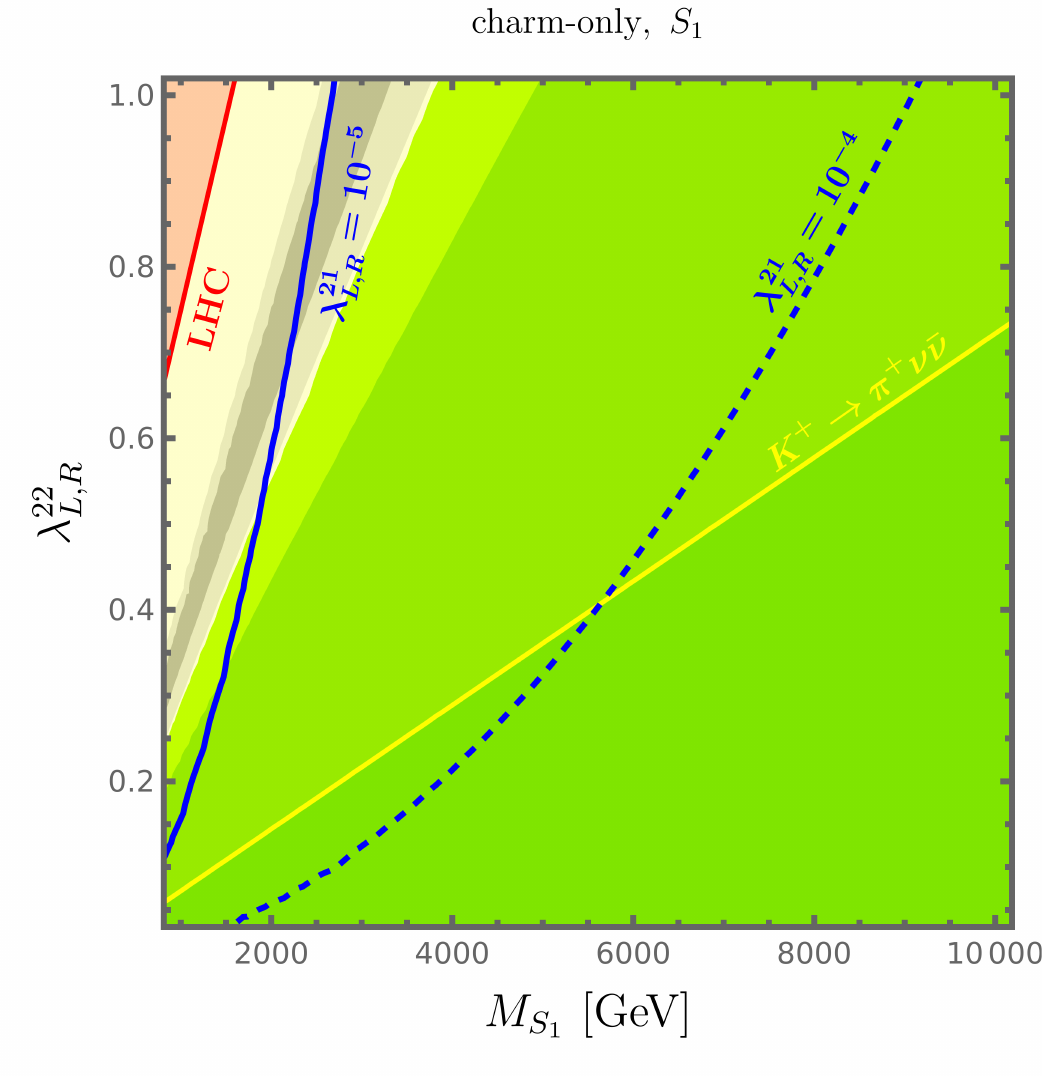}
	\caption{The leptoquark $S_1$ parameter space in 
          mass--coupling planes, in the case of couplings only to the
          top-quark (\emph{left}) or only to the charm-quark
          (\emph{right}). In both cases the left- and right-handed
          couplings are set equal,
          $\lambda_R^{i2}=\lambda_R^{i2}$. The (light) green and grey
          contours correspond to the 1$\sigma$ (2$\sigma$) ranges for
          $\DamuFinal$ and $\DamuOld$, respectively. In the top-only
          scenario, the transparent grey region shows the region
          where the LQ one-loop correction to the muon mass
          contributes more than 100\%.
          In the charm-only scenario, the displayed exclusion regions
          correspond to  BR$(K^+\to\pi^+ \nu\bar\nu)$ (yellow) and LHC
          (red).
          In both cases, blue contour lines indicate the maximum
          values the corresponding electron-couplings
          $\lambda^{i1}_{L,R}$ can take given the bounds from BR$(\mu\to e\gamma)$.
}
	\label{fig:leptoquarkresults}
\end{figure}

As the above approximations show, leptoquark models without chiral enhancements
cannot significantly modify $\amu$; these models are in agreement with but
not constrained by the current  $\amu$
results. In contrast, the  $S_1,R_2,U_1$  leptoquark models with
chiral enhancements  potentially lead to large contributions to $\amu$. In the
following we discuss the   relevant constraints derived from $\amu$
and the most directly related  complementary observables. 

The three cases of leptoquark models with chiral enhancements behave
quite similarly to each other
\cite{Queiroz:2014zfa,Biggio:2016wyy,Kowalska:2018ulj,Dorsner:2019itg,Crivellin:2020tsz,Bigaran:2020jil,Dorsner:2020aaz,Athron:2021iuf}, where the
$S_1$ model is compatible with the   largest 
contributions to $\amu$. Hence we provide explicit results   here
specifically for the $S_1$ case. Representative results are   shown
in the plots  in Fig.~\ref{fig:leptoquarkresults}. The left plot shows
the pure
top-loop case (only $\lambda_{L,R}^{32}$ are non-zero), the right plot
shows  the pure charm-loop case (only $\lambda_{L,R}^{22}$ are
non-zero).

The relevant parameters for the $\Damu^{S_1}$ contributions in
Eqs.~(\ref{LQtopnumerical},\ref{LQcharmnumerical})  are the leptoquark
mass $M_{S_1}$ and the 
leptoquark couplings between the muon and the top- or charm-quark,
$\lambda_{L,R}^{32}$ or $\lambda_{L,R}^{22}$. As mentioned in the
beginning, direct LHC searches for leptoquarks are evaded for
$M_{S_1}\gtrsim2$ TeV, which is what we focus on.
In both plots
the coupling values have been chosen such that $\Damu^{S_1}$
ranges from 0 to the ballpark of
$\DamuOld$, while the LQ masses are well above the basic LHC limit.

  Two of the directly connected observables are the muon mass
  and the muon--Higgs coupling, as discussed in general in
  Secs.~\ref{sec:MuonMass}, \ref{sec:muon-Higgs}. The muon mass is
  particularly interesting here because of the  potentially very large
  chirality flipping parameter in Eq.~(\ref{LQchiralitybreaking}). In
  terms of the
  coefficient $C_{BSM}$ in Eq.~(\ref{CBSMLQ}) its one-loop correction
  in the $\overline{\text{MS}}$ scheme at the scale $\mu=M_{S_1}$ is
  \begin{align}
	\left|\frac{    \Delta m_\mu^{\text{fin}} }{m_\mu}\right| &=
	    C_{BSM}\times|B_0^{\text{fin}}(0;m_{u_i}^2,M_{S_1}^2)|
	    \sim  C_{BSM}.
  \end{align}
  
In Fig.~\ref{fig:leptoquarkresults} the parameter region with
$|\Delta m_\mu|>m_\mu$ is indicated as the transparent grey
region above around $\lambda^{32}_{L,R}\gtrsim0.18$ in the 
top-loop case. This region might be
  considered unplausible in view of finetuning in the muon mass,
  though this is not a strict bound
  \cite{Athron:2021iuf,Stockinger:2022ata}, see also Sec.~\ref{sec:MuonMass}. 
In line with the general relation (\ref{eqn:GeneralGM2Contribution})
and in view of the lower LHC leptoquark mass   limit, this
consideration  cuts off leptoquark contributions to $\Damu$ larger
than
$\DamuOld$ for $S_1$ masses above 5 TeV.\footnote{%
The bound is only visible in the top-only case shown in the left panel
of Fig.~\ref{fig:leptoquarkresults};  essentially the same bound
also exists in the charm-only case (right panel) but is located
outside the region shown in the plot.
}

  The modification of the muon--Higgs coupling is generally correlated with
  contributions to $\amu$ as explained in
  Sec.~\ref{sec:muon-Higgs}. Refs.~\cite{ColuccioLeskow:2016dox,Crivellin:2020tsz,Fajfer:2021cxa}
  and~\cite{Crivellin:2021rbq}  specifically  studied the correlation
  in case of leptoquark $S_1$ and $R_2$ models, and generalised 3-field models as
  defined in Sec.~\ref{sec:genericthreefield}. In all these models, the
  muon--Higgs coupling is modified only at the loop level, and the
  resulting effects are below experimental sensitivities which can be
  reached at the LHC. Hence we do not show constraints from the
  muon--Higgs coupling in Fig.~\ref{fig:leptoquarkresults}.
Similarly, Z-boson couplings to leptons
\cite{Arnan:2019olv,Crivellin:2020mjs} and $h\to\gamma\gamma$ decays \cite{Crivellin:2020ukd} can be modified by
  leptoquark loop contributions,
  but the constraints do not interfere significantly with the
  potential of large leptoquark contributions to $\amu$.

  Very important constraints on the relevant parameters result from
  quark flavour-violating processes. Quark flavour violation is unavoidable because the coupling matrices
  acting in the up-/lepton sector and in the down-/neutrino sector,
  see Eq.~(\ref{LQLagrangianflavour}),
  cannot be simultaneously diagonal. For general  flavour constraints on
  leptoquarks we refer to 
  Refs.~\cite{deBoer:2015boa,Mandal:2019gff}. Here we follow
  Ref.~\cite{Kowalska:2018ulj} and highlight several instructive
  minimal cases related to $\amu$: 
  \begin{itemize}
    \item Charm-philic up-origin case: the  muon--charm couplings
      $\lambda_{L,R}^{22}$ are the only non-vanishing matrix elements
      of $\lambda_{L,R}$: in this case the
      muon-neutrino will inevitably couple to quarks of all
      generations, and the processes $K^+\to\pi^+\nu\bar{\nu}$
      \cite{Kowalska:2018ulj} and
      $\bar{B}\to\bar{K}\nu\bar{\nu}$ \cite{Bauer:2015knc,ColuccioLeskow:2016dox}
      are affected.
    \item
      Top-philic up-origin case: similarly for the  muon--top 
      couplings $\lambda_{L,R}^{32}$: the same processes are affected.
    \item
      Charm-philic down-origin case: the muon-neutrino--strange coupling
      $(V_{\text{CKM}}^T\lambda_L)^{22}$ is assumed to be the only
      non-vanishing matrix element of $(V_{\text{CKM}}^T\lambda_L)$,
      and similarly for $\lambda_R^{22}$: in this case the coupling
      $\lambda_L^{12}$ to the up-quark is generated and the
      $D^0\to\mu\mu$  \cite{Bauer:2015knc,Kowalska:2018ulj} and
      further charm-quark decays such as $D^+\to\pi^+\mu\mu$ \cite{deBoer:2015boa}
      are affected.
  \end{itemize}
One may also study an analogous top-philic down-origin case
\cite{Kowalska:2018ulj} or a case where the leptoquark couples equally
strongly to all lepton generations \cite{Khasianevich:2023duu}, with
analogous conclusions.

  Quantitatively, the $\bar{B}\to\bar{K}\nu\bar{\nu}$ process does not
  constrain the couplings relevant for
  $\Damu^{\text{LQ}}\lesssim20\times10^{-10}$ at all
  \cite{Bauer:2015knc,ColuccioLeskow:2016dox}.
  In contrast, the $D^0\to\mu\mu$
  decay very significantly restrict the allowed ranges  of the
  couplings to the charm- and up-quarks
  \cite{Kowalska:2018ulj,Bauer:2015knc}. For instance, in the charm-philic down-origin
  case,
  $\Damu^{\text{LQ}}$ is bound 
  to be smaller than around $4\times10^{-10}$ \cite{Kowalska:2018ulj} in the
  $S_1$ model, and similarly in the $R_2$ model.

  In the charm-philic up-origin case \cite{Kowalska:2018ulj}, where
  $\lambda_{L,R}^{22}$ are the only non-vanishing matrix elements,
  larger $\Damu^{\text{LQ}}$ are possible. Here, the
  $K^+\to\pi^+\nu\bar{\nu}$ decay directly restricts the 
  $\lambda_L^{22}$ coupling.  In an updated analysis,
  Refs.~\cite{Kowalska:2018ulj,Khasianevich:2023duu} find
  \begin{align}
    \label{LQKaonDecayLimit}
    |\lambda_L^{22}|<0.14\left(\frac{M_{S_1}}{2\,\text{TeV}}\right).
  \end{align}
  This bound is shown as the yellow area in
  Fig.~\ref{fig:leptoquarkresults}, but it is important to note that
  the bound in the plot relies on the equality
  $\lambda_L^{22}=\lambda^{22}_R$ --- larger $\Damu$ would be
  compatible with the $K^+$ decay if   $\lambda_L^{22}<\lambda^{22}_R$.
  
The muon--charm couplings $\lambda_{L,R}^{22}$ are also constrained by
LHC searches in a way that is relevant for $\amu$
\cite{Kowalska:2018ulj,Angelescu:2021lln}. Di-lepton or single-lepton production LHC processes at high
$p_T$ can be affected by leptoquark $t$-channel exchange even for very
heavy leptoquark masses above the direct leptoquark search
limit. Recasting LHC results for such processes results in
constraints which essentially depend on the ratios $\lambda_{L,R}^{22}/M_{S_1}$ to
a good approximation and which are approximately equal for
$\lambda_L^{22}$ and $\lambda_R^{22}$
\cite{Kowalska:2018ulj,Angelescu:2021lln}. E.g.\ for $M_{S_1}=2$ TeV, these
references find upper limits on the couplings between 1.2 and 1.8,
depending on the employed LHC processes. As an approximation to the
limits found in these references we take
\begin{align}
  \label{LQLHClimits}
  |\lambda^{22}_{L,R}| < 0.3 + 0.9\frac{ M_{S_1}}{2\text{ TeV}}.
\end{align}
  In the plot this bound is weaker than the bound
(\ref{LQKaonDecayLimit}) from Kaon decay, but the LHC limits are still important
since they independently apply onto the right-handed coupling
$\lambda_R^{22}$. In 
  Fig.~\ref{fig:leptoquarkresults} we show the LHC limit of
  Ref.~\cite{Kowalska:2018ulj} as the red region.

Turning from quark to lepton flavour violation,
Sec.~\ref{sec:LeptonDipole} has explained the generic correlation
between BSM contributions to  $\amu$ and  to CLFV processes such as
$\mu\to e\gamma$. This correlation is particularly acute in the case
of leptoquarks, which can couple to all quark and all lepton
generations with in principle arbitrary coupling
matrices.\footnote{Similarly, if complex couplings are allowed, there
are strong correlations to electric dipole moments. As discussed in
Sec.~\ref{sec:LeptonDipole}, this correlation is particularly
constraining in case of naive scaling, which is not necessarily the
case for leptoquark models due to the flavour structure of the
couplings $\lambda^{ij}_{L,R}$, which is in general unknown. Here we
assume real couplings and refer to 
the literature for EDM studies in the context of leptoquarks
\cite{Dekens:2018bci}.}

While quark flavour violation is an unavoidable consequence of the
leptoquark interactions, lepton flavour violation (in the limit of
vanishing neutrino masses) can be avoided by specific choices of the
interaction matrices. However, once all coupling matrix elements are
allowed to be non-zero, the constraints are extremely strong.

The most important and most directly related CLFV process is
the $\mu\to e\gamma$ decay. Leptoquark contributions to $\mu\to
e\gamma$ have been computed and discussed in detail
in Refs.~\cite{Benbrik:2008si,Benbrik:2010cf,Arnold:2013cva,Dorsner:2016wpm,Cai:2017wry,Khasianevich:2023duu}.
In line with the generic estimate (\ref{eq:mueg-semi-num}) we can
relate the leptoquark contributions to $\mu\to e\gamma$ and $\amu$ as
\begin{subequations}\label{LQMuegformulas}
  \begin{align}
\text{top-loop:}&&  \text{BR}(\mu\to e\gamma)^{S_1} &=
2\times10^{-4}  \left(\frac{\Damu^{S_1}}{10^{-9}}\right)^2
  \left(
\left| \frac{ \lambda_L^{31}}{\lambda_L^{32}} \right|^2 +
\left| \frac{ \lambda_R^{31}}{\lambda_R^{32}} \right|^2
\right),
\\
\text{charm-loop:}&&  \text{BR}(\mu\to e\gamma)^{S_1} &=
2\times10^{-4}  \left(\frac{\Damu^{S_1}}{10^{-9}}\right)^2
  \left(
  \left| \frac{ \lambda_L^{21}}{\lambda_L^{22}} \right|^2 +
\left| \frac{ \lambda_R^{21}}{\lambda_R^{22}} \right|^2
  \right),
  \end{align}
\end{subequations}
where in each line we only allow either only couplings to the top- or
to the charm-quark, respectively, like in Eq.~\eqref{eq:LQnochiralityflipnumerical}.
Here, in addition to the parameters contributing to $\amu$, also
couplings to the electron, i.e.\ $\lambda_{L,R}^{i1}$, appear. The
appearing ratios of couplings to electrons and muons are very strictly
constrained because of the MEG-II limit on $\text{BR}(\mu\to
e\gamma)$. In particular if the contributions to $\amu$ are
significant and of the order $10^{-9}$, the coupling ratios must be
smaller than around $10^{-5}$.

This implies that significant leptoquark contributions to $\amu$ are
only tenable if the leptoquark couplings are highly non-universal ---
couplings to the muon must be many orders of magnitude larger than
couplings to electrons. This result is illustrated in   Fig.~\ref{fig:leptoquarkresults},
where we show blue contours indicating where the maximum allowed values of
leptoquark couplings to electrons are $10^{-4,-5,-6}$.
Again, this is in line with the generic
conclusions of Sec.~\ref{sec:LeptonDipole}.\footnote{With respect to
the correlation with other lepton dipole observables, we note that leptoquarks were also considered as a potential
explanation of the deviation  of the electron magnetic moment
$a_e$ resulting from the SM value in Eq.~(\ref{aeSM2018}) at the time
  \cite{Crivellin:2018qmi,Bigaran:2020jil,Dorsner:2020aaz,Bigaran:2021kmn,Keung:2021rps,Parashar:2022wrd}\cite{Gherardi:2020qhc}, and that leptoquarks provide a slight variation of
the ``single particle contribution'' explained in Sec.~\ref{sec:LeptonDipole}
with less constraints from $\mu\to e\gamma$ because of the quark
flavour structure allowed in the couplings \cite{Dorsner:2020aaz}.}

Next to  $\mu\to e\gamma$, further CLFV processes such as
$\tau\to\mu\gamma$ and $\mu\to e$ conversion provide important
additional constraints on leptoquark couplings. $\tau\to\mu\gamma$
constrains the couplings to the $\tau$ lepton $\lambda_{L,R}^{i3}$, and
$\mu\to e$ conversion especially constrains the couplings to
1st-generation quarks, $\lambda_{L,R}^{1i}$. In all these cases,
similar statements can be made as for $\mu\to e\gamma$: if
leptoquarks lead to $\Damu\sim10^{-9}$, the 
leptoquark couplings mediating CLFV processes must be several orders
of magnitude smaller than the ones to muons
\cite{Cai:2017wry,Freitas:2022gqs,Khasianevich:2023duu}.

It is instructive to summarise the main constraints on leptoquark
couplings for the specific case of the $S_1$ leptoquark with mass $M_{S_1}\gtrsim2$ TeV,
as follows:
\begin{itemize}
\item
  $\Damu^{S_1}\sim10^{-9}$ is possible via muon--top couplings of
  order $0.1$; the left- and right-handed couplings
  $\lambda_{L,R}^{32}$ can be similar. CLFV constraints require the
  electron--top and $\tau$-lepton--top couplings to be several orders
  of magnitude smaller. This scenario predicts small effects e.g.\ in
  $\bar{B}\to\bar{K}\nu\bar{\nu}$ and  $K^+\to\pi^+\nu\bar{\nu}$ below
  the current sensitivity.
\item
  Contributions of $\Damu^{S_1}\sim10^{-9}$ are even possible for
  leptoquark masses in the multi-TeV region, however these are then
  accompanied by radiative corrections to the muon mass of order
  100\%, which may be considered unnatural.
\item
  $\Damu^{S_1}\sim10^{-9}$ is also possible via muon--charm
  couplings. 
  In this case,   $K^+\to\pi^+\nu\bar{\nu}$ requires a
  strong hierarchy $\lambda^{22}_L\ll\lambda^{22}_R$, and again CLFV
  constraints require the couplings of electron and $\tau$-lepton to
  be much smaller.
\item
  If only muon--charm couplings are allowed and equality
  $\lambda^{22}_L\approx\lambda^{22}_R$ is assumed, the contributions
  to $\amu$ are bound to be below around $10^{-10}$ and thus
  negligible.
\item
  If a generic flavour pattern is assumed where all leptoquark
  couplings $\lambda_{L,R}^{ij}$ are of the same order of magnitude,
  the CLFV constraints from e.g.\ Eq.~(\ref{LQMuegformulas}) limit contributions to
  $\amu$ even stronger to below $10^{-14}$.
\end{itemize}

These statements illustrate why leptoquarks were considered so
frequently in the literature as simple explanations of the old
deviation $\DamuOld$. Still, the parameter regions which produce
sizeable contributions to $\amu$ are rather narrow, requiring specific
types of leptoquark couplings and flavour patterns. The current result
$\DamuFinal$ is compatible with smaller leptoquark contributions to
$\amu$, which could allow e.g.~anarchic leptoquark flavour patterns or
small contributions to the muon mass.
\subsubsection{Brief survey of leptoquark scenarios with specific
  flavour structures}
\label{sec:LQflavour}
The summary above highlights that only leptoquark models with very
non-generic flavour structure have the potential for significant
contributions to $\amu$.
In the past years, leptoquarks were very frequently considered also in
view of several reported $B$-physics ``anomalies'', i.e.\ deviations
between SM predictions and measurements. The most prominent such
anomalies were reported for the ratios of observables $R({K^{(*)}})$
and $R({D^{(*)}})$ related to $b\to s$ and $b\to c$ transitions, respectively.
In the meantime, improved measurements show a reduced tension. Here we
nevertheless provide a brief survey of the literature to illustrate
the range of possible effects that can be caused by leptoquarks and
how observables from other sectors can be related to $\amu$.

The ratios $R({K^{(*)}})$ are
defined  as $R({K^{(*)}})=\Gamma(\bar{B}\to
\bar{K}^{(*)}\mu\mu)/\Gamma(\bar{B}\to \bar{K}^{(*)}ee)$ and
test lepton flavour universality (LFU) in  $b\to s$ transitions.  As reviewed
in 2021 e.g.~in Ref.~\cite{Crivellin:2021sff}, the SM predicts these
ratios to be very close to unity, because gauge couplings to leptons
are universal for all
lepton generations, whereas experiments showed a $>3\sigma$ discrepancy
from the SM prediction.  The  LHCb measurements published in 2022
\cite{LHCb:2022qnv} supersede the previous ones and are in full
agreement with the SM prediction, eliminating the hints for LFU violation
\cite{Fedele:2023rxb,Capdevila:2023yhq}. The ratios $R({D^{(*)}})$ are defined via $b\to c$ decays 
$R({D^{(*)})}=\Gamma(\bar{B}\to
\bar{D}^{(*)}\tau\bar\nu)/\Gamma(\bar{B}\to \bar{D}^{(*)}\ell\bar\nu)$
where $\ell=e,\mu$. For these processes, there still is a certain
disagreement between experiment and SM prediction
\cite{HeavyFlavorAveragingGroupHFLAV:2024ctg},
  \begin{align}
    R(D)^{\text{SM}} &=0.296 \pm 0.004 &
    R(D^*)^{\text{SM}} &=0.254 \pm 0.005,\\
    R(D)^{\text{Exp}} &=0.342 \pm 0.026&
    R(D^*)^{\text{Exp}} &=0.286 \pm 0.012. 
  \end{align}

When the results for $R(K^{(*)})$ and $R(D^{(*)})$ were first reported
to be anomalous, i.e.\ in tension between SM and experiment,
Ref.~\cite{Bauer:2015knc} showed that 
all these anomalies, and a significant contribution to $\amu$ were
simultaneously possible in the $S_1$ leptoquark model, and it was also
shown that this scenario could be embedded into a left-right symmetric
gauge theory \cite{Das:2016vkr}  or into a model of radiative neutrino
mass generation \cite{Popov:2016fzr,Cai:2017wry}. Improved
measurements and constraints from the $B\to K\nu\bar\nu$ however ruled
out simple explanations of the reported anomalies in
$R(K^{(*)})$, $R(D^{(*)})$ and $\amu$ at the time
by just a single leptoquark
\cite{Buttazzo:2017ixm,Crivellin:2017zlb,Angelescu:2018tyl,Dorsner:2019itg}. 

Simultaneous explanations remained possible in more complicated models
such as two leptoquarks $S_1$ and $S_3$ 
\cite{Crivellin:2017zlb,Gherardi:2020qhc,Lee:2021jdr},
three different leptoquarks \cite{Nomura:2021oeu}, the $S_1$
leptoquark plus a charged scalar singlet \cite{Marzocca:2021azj}.
Also simultaneous explanations of the anomalies at the time together
with neutrino masses were considered with   $S_1$ and $\tilde{R}_2$ 
\cite{Zhang:2021dgl,Freitas:2022gqs} or with
$S_1$ $\tilde{R}_2$ and $S_3$ \cite{Chen:2022hle} or with $R_2$ and
$S_3$ leptoquarks \cite{Babu:2020hun}, or with leptoquarks plus other
BSM states \cite{Julio:2022bue,Julio:2022ton,Wang:2021uqz,FileviezPerez:2021lkq,Dev:2024tto}. Reference
\cite{Greljo:2021xmg} also constructs elaborate
models involving
gauge symmetries related
to flavour and include leptoquarks charged under that gauge symmetry
to explain various observations, and Ref.~\cite{Heeck:2022znj} uses such scenarios to
even explain dark matter and neutrino masses at the same time. This
exemplifies that one can build models with deep construction
principles which can lead to large effects in all flavour observables
--- but it remains true that even within such symmetry-based models,
large $\Damu$ and $B$-physics effects arise independently and are not
connected.
The need for such complicated constructions has been generalised into
a no-go theorem \cite{Arcadi:2021cwg} which states that at least four
BSM fields are required to accommodate large $\Damu$, the $B$-physics
anomalies circa 2021, and dark matter.

The discussion shows that if leptoquarks exist in the few-TeV mass
region, their flavour structure is likely highly non-generic. This
point is also visualised graphically with best-fit points in
Ref.~\cite{Freitas:2022gqs}. It motivates studies how non-generic
flavour patterns could emerge from underlying mechanisms.

  Basic ways to generate flavour patterns include the Froggatt-Nielsen
  mechanism \cite{Froggatt:1978nt}, which can be applied to single leptoquark models
  \cite{Hiller:2016kry} or models with $S_1$ and $S_3$ leptoquarks
  \cite{Bordone:2020lnb}, or generalisations of the Minimal Flavour
  Violation (MFV) paradigm \cite{Davidson:2010uu}.
In all these cases the leptoquark couplings to muons, electrons and
$\tau$-leptons  depend on underlying parameters and are thus
correlated. Huge coupling ratios such as the ones needed in the plots
of Fig.~\ref{fig:leptoquarkresults} cannot occur. As a result, the
bounds on CLFV processes imply that only negligibly small
contributions to $\amu$ are possible.

There exist, however, also leptoquark flavour constructions which
allow significant contributions to $\amu$. An appealing example is
provided by the combination of leptoquarks with a flavour-dependent
gauge group such as U(1)$_{L_\mu-L_\tau}$ discussed already in
Sec.~\ref{sec:LmuMinusLtau}. Leptoquarks can be charged under this
gauge group, thus allowing e.g.~only couplings to muons but not to
other leptons. Such leptoquarks, called muoquarks in
Ref.~\cite{Greljo:2021xmg}, cannot mediate CLFV 
processes and hence can have significant couplings to muons and lead
to $\Damu$ contributions of the order $10^{-9}$.

In total, the current status of $\amu$ as well as the flavour
anomalies mentioned here do not show evidence for new physics. While
the fundamental motivation for the existence of leptoquarks related
e.g.~to grand unification remains, current phenomenological
constraints do not prefer particular parameter regions and do not
require leptoquarks at the TeV-scale.

\subsection{Vector-like fermion models}\label{sec:VLF}

Over the past decade many properties of the Higgs boson have been measured with impressive precision and, consequently, 
	reinforced our picture of the SM with exactly three
        generations of chiral fermions. The existence of a fourth,
        sequential chiral generation is conclusively ruled out \cite{Eberhardt:2012gv,Djouadi:2012ae}.
	This statement, based mainly on measurements of $h\to \gamma\gamma$, however, does
	not rule out the existence of  new fermions beyond the SM in
        general. Clearly, introducing new fermions is one of the
        obvious ways to extend the SM. For example, additional
        gauge singlet fermions (``right-handed neutrinos'') with large Majorana masses not only
	avoid the bounds from the Higgs signal strength but could also explain the observed tiny neutrino masses (cf. Sec.~\ref{sec:neutrino_masses}).

	Another possibility, discussed in this section, is the
        existence of vector-like, or Dirac fermions. Unlike the chiral fermions in the SM, both 
	the left- and right-handed chiralities of such fermions by
        definition transform in the same way under the SM gauge group, leading to
	vector-like interactions with the SM gauge bosons. Because of
        this, these \emph{vector-like fermions} (VLF) can also possess
        a gauge invariant Dirac mass
	term before EWSB,
	\begin{align}
	  \La \supset \overline{F_L} M_F F_R + \overline{F_R} M_F^* F_L .
          \label{Eq:VLL_Ldirac}
	\end{align}
	Consequently, the VLF can be arbitrarily heavy and again easily evade the Higgs signal strength bounds without violating perturbativity.
	Historically, VLF have gained  attention in the context of Kaluza-Klein excitations \cite{Kaluza:1921tu,Klein:1926tv} in models with extra dimensions,
	mainly introduced in an attempt to address the hierarchy
        problem \cite{Randall:1999ee,Arkani-Hamed:1998jmv}. Also
        anthropic arguments have been put forward to motivate that VLF
        should be the lightest BSM states, while new scalars should be
        much heavier \cite{Kannike:2011ng,Arkani-Hamed:2021xlp}.
	At the same time, VLF have also been explored phenomenologically in the context of dark matter \cite{Dimopoulos:1990kc,Cheng:2002ej}, collider experiments \cite{Thomas:1998wy,Sher:1995tc,Djouadi:1993pe,Fujikawa:1994we}
	and, after the BNL measurement, also increasingly received
        interest in the context of $\Damu$
        \cite{Kannike:2011ng,Dermisek:2013gta,Falkowski:2013jya,Freitas:2014pua,Queiroz:2014zfa}.

	In general, corrections from VLF first enter $\Damu$ at the two-loop level e.g.\ through Barr-Zee diagrams discussed in Sec.~\ref{sec:Barr-Zee}. Therefore,
	unless additional scalars are present that lead to further enhancements (see Sec.~\ref{sec:VLF+scalar}), their contributions are typically rather moderate.
	However, there is also a special class of VLF, the \emph{vector-like leptons} (VLL), which can mix with the SM leptons at tree-level
	and can lead to new sources of chirality flips.

        The phenomenology of such VLL is rich and distinctive, and the
        new tree-level chirality flips modify
        $\amu$, $m_\mu$, Higgs decays and more generally the lepton
        mass-generation mechanism  in unique ways which differ from other
        scenarios and can lead to large effects. Here we focus mainly on this case, first in the
        pure context of the SM extended by fermions, later by also
        allowing additional scalar fields in Sec.~\ref{sec:VLF+scalar}.

	\subsubsection{Vector-like lepton models and chiral enhancement}\label{subsec:VLL}
	\begin{table}[t]
		\centering
		\begin{tabular}{ r|c|c|c|c|c|c|}
			\cline{2-7}
			& $N$ & $E$ & $L$ & $L_{\frac{3}{2}}$ & $N^a$ & $E^a$\\ \hhline{~======}
			\textbf{Repr.} & $(\bm{1},0)$ & $(\bm{1},-1)$ & $(\bm{2},-\frac{1}{2})$ 
			& $(\bm{2},-\frac{3}{2})$ & $(\bm{3},0)$ & $(\bm{3},-1)$ \\ 
			\textbf{Multiplet} & $N$ & $E$ & \raisebox{.5\totalheight}{\vphantom{\Big(}}$\renewcommand*{\arraystretch}{.8}\mqty(L^0 \\ L^-)$\raisebox{-0.5\totalheight}{\vphantom{\Big(}} &
			$\renewcommand*{\arraystretch}{.8}\mqty(L^- \\ L^{--})$ & $\renewcommand*{\arraystretch}{.8}\mqty(N^0 &\!\!\! \sqrt{2}N^+ \\ \sqrt{2}N^- &\!\!\! -N^0)$ & 
			$\renewcommand*{\arraystretch}{.8}\mqty(E^- &\!\!\! \sqrt{2}E^0 \\ \sqrt{2}E^{--} &\!\! -E^-)$ \\ 
			\rule{0pt}{2.5ex}$-\La_\text{int}$ & $\lambda_L \bar l_L N \tilde\Phi$ & $\lambda_L \bar l_L E \Phi$ & $\lambda_R \bar{L} e_R \Phi$ &
			$\lambda_R \bar{L}_{\frac{3}{2}}e_R\tilde\Phi$ & $\lambda_L\bar{l}_L \sigma^a N^a \tilde\Phi$ & $\lambda_L\bar{l}_L \sigma^a E^a\Phi$ \\ \cline{2-7}
		\end{tabular}\hspace{.5cm}
		\caption{$\GEW$ representation, multiplet structure ($\sigma^a F^a$ for the triplets) 
			and interaction terms for all possible vector-like fermions coupling to SM leptons and the Higgs doublet. 
			The new Yukawa couplings follow the conventions from Sec.~\ref{sec:genericthreefield}.}
		\label{tab:VLL-quantum-numbers}
	\end{table}
        We define VLL as the subclass of VLF which can mix with the SM
        leptons via gauge invariant and renormalisable operators.
	Without the introduction of new scalars, this requirement of non-trivial mixing puts strong restrictions on their possible quantum numbers.
	The only allowed renormalisable kinds of mixing terms are
        either dimension-3 gauge invariant fermion mass terms, in which case the VLL is required to have the same quantum numbers as the SM counterpart,
	or dimension-4 Yukawa couplings of a VLL to a SM lepton and
        the Higgs doublet. Since the mass terms can be diagonalised
        (without breaking  gauge invariance), the relevant mixing stems from the latter. 
	Including the charge-conjugate Higgs field $\tilde\Phi = i\sigma^2 \Phi^*$, there are exactly 6 different possible such Yukawa terms
	(corresponding to model 10--15 in Tab.~\ref{tab:onefieldmodels}),
	\begin{align}
	  \bar{l}_L N \tilde \Phi, \qquad \bar{l}_L E \Phi, \qquad \bar{L} e_R \Phi, \qquad \bar{L}_{\frac{3}{2}} e_R\tilde\Phi, \qquad \bar{l}_L \sigma^a N^a \tilde\Phi, \quad \text{and} \quad \bar{l}_L \sigma^a E^a\Phi,
          \label{Eq:2fieldVLLModels}
	\end{align}
	where the representations and hyper-charges of the Dirac
        fermion fields $N$, $E$, $L$, $L_{\frac{3}{2}}$, $N^a$, $E^a$ are listed in Tab.~\ref{tab:VLL-quantum-numbers} and we have suppressed the SM family indices.
	In the following we denote the couplings of VLL to the
        left-handed SM doublet by $\lambda_L$ and to the right-handed
        singlet by $\lambda_R$.

        In isolation, these new interactions can lead to Feynman
        diagrams as illustrated in Fig.~\ref{fig:1VLL} for the example
        of $L$. The left of the diagrams illustrates one-loop corrections to the
        muon--Higgs interaction, the muon mass and $\amu$.
        For instance, the correction to the muon mass and its
        correlation to the fundamental Yukawa coupling is of the form
	\begin{align}\label{VLL-nonenhanced}
		m_\mu = \frac{y_\mu v}{\sqrt{2}} \bigg[1 + \O\Big(\tfrac{\lambda^2_{L/R} v^2}{M^2}\Big)\bigg].
	\end{align}
        This effect is obviously amplified if several VLL are
        present.
	However, there is no chiral enhancement since each of the
        terms in Eq.~(\ref{Eq:2fieldVLLModels}) on its own does not
        lead to an additional breaking of the
        SM lepton chiral symmetries.

        In the case the SM is extended by several VLL at the same
        time, there is the possibility of a much stronger  enhancement due to
	new sources of chiral symmetry breaking for the SM leptons. Concretely, this chiral enhancement emerges if Yukawa couplings between the VLL are allowed.
	In the minimal case of two VLL this is again allowed for
        precisely 6 different combinations, given by
        \cite{Kannike:2011ng}
	\begin{align}\label{chargedseesawpossibilities}
		L\oplus E, \qquad L\oplus N, \qquad L_{\frac{3}{2}}\oplus E, \qquad L\oplus E^a, \qquad L\oplus N^a, \quad \text{and} \quad L_{\frac{3}{2}}\oplus E^a.
	\end{align}
	The corresponding Yukawa terms are given by\footnote{\label{footnoteVLL}
		Though very similar, the VLL models do not all map directly onto the three-field models discussed in Sec.~\ref{sec:genericthreefield}.
			Instead e.g.~the $L\oplus E$ model is more
                        closely related to Model 6 discussed in Ref.~\cite{Kowalska:2017iqv} and mentioned in
			footnote \ref{note:3field}, however, with the crucial difference that $\phi\equiv\Phi$.
			As a consequence the muon chiral symmetry
                        cannot be extended to the scalar doublet and
                        is indeed broken in the VLL models. For this
                        reason, chiral enhancements are possible and
                        the phenomenological behaviour is similar to
                        the one of the three-field model of Class
                        II. Similar comparisons can be made for all
                        six cases listed in Eq.~\eqref{chargedseesawpossibilities}. }
	\begin{subequations}
		\begin{alignat}{3}
			L\oplus E:& \qquad \La \supset - \bar{L} &&\Big( \lambda_\Phi \PR + \bar{\lambda}_\Phi \PL \Big) E \Phi &&+ h.c. \\
			L\oplus N:& \qquad \La \supset - \bar{L} &&\Big( \lambda_\Phi \PR + \bar{\lambda}_\Phi \PL \Big) N \tilde\Phi &&+ h.c. \\
			L_{\frac{3}{2}}\oplus E:& \qquad \La \supset - \bar{L}_{\frac{3}{2}} &&\Big( \lambda_\Phi \PR + \bar{\lambda}_\Phi \PL \Big) E \tilde\Phi &&+ h.c. \\
			L\oplus E^a:& \qquad \La \supset - \bar{L} &&\Big( \lambda_\Phi \PR + \bar{\lambda}_\Phi \PL \Big) \sigma^a E^a \Phi &&+ h.c. \\
			L\oplus N^a:& \qquad \La \supset - \bar{L} &&\Big( \lambda_\Phi \PR + \bar{\lambda}_\Phi \PL \Big) \sigma^a N^a \tilde\Phi &&+ h.c. \\
			L_{\frac{3}{2}}\oplus E^a:& \qquad \La \supset - \bar{L}_{\frac{3}{2}} &&\Big(
			        \lambda_\Phi \PR + \bar{\lambda}_\Phi \PL \Big) \sigma^a E^a
			        \tilde\Phi \quad &&+ h.c. .
		\end{alignat}
	\end{subequations}
        Fig.~\ref{fig:2VLL} shows sample diagrams  involving these
        terms; the last of these contributes to the muon mass
        similarly to the seesaw mechanism for neutrino masses, hence
        the name charged-seesaw  models
        \cite{Kannike:2011ng}. 
	Notably, the appearing coupling parameters $\lambda_\Phi$ and $\bar\lambda_\Phi$ for the two different chirality combinations are, in general, independent and 
	in combination with $\lambda_L$ and $\lambda_R$ both lead to a
        new, separate breaking of the muon chiral symmetry of the kind
        discussed in Sec.~\ref{sec:ChiralityFlips}. The two
        corresponding order parameters can be written as
	\begin{align}\label{eq:VLL-chiral-enhancement}
		\lambda_R \lambda^*_\Phi \lambda_L, \qquad \text{and} \qquad \lambda_R M^*_L \bar\lambda^*_\Phi M^*_{E/N} \lambda_L .
	\end{align}
	The appearance of the Dirac masses in the second product is
    noteworthy and in particular implies that the $\bar\lambda_\Phi$-term
    dominates the contributions to $\amu$ in the limit of large VLL masses as discussed below.

        The diagrams in Fig.~\ref{fig:2VLL} illustrate the correlated
        contributions of these VLL to the muon mass, the muon--Higgs
        coupling, and to $\amu$. Their analysis makes explicit the
        general arguments presented in Secs.~\ref{sec:ChiralityFlips},
        \ref{sec:MuonMass} and \ref{sec:muon-Higgs}.
        Historically, the analysis of the
        correlations in VLL models of
        Refs.~\cite{Kannike:2011ng,Dermisek:2013gta} preceded the 
        generic analysis presented already in
        Sec.~\ref{sec:muon-Higgs}. Here we first sketch the key
        observations, before providing detailed calculations and results.

        Because of the new source of tree-level chirality flips, the
        muon mass receives two tree-level contributions
        \begin{align}
		  m_\mu^H &= \frac{y_\mu v}{\sqrt2},&
		  m_\mu^{HHH} &\sim  - \frac{\lambda_{Li} \bar\lambda^*_\Phi \lambda_{Rj}
		   v^3}{2\sqrt{2} M_E M_L} ,
		  \label{mHHH}
        \end{align}
        where the first one results from the fundamental Yukawa
        interaction and the second one from the diagram with three
        external Higgs fields in Fig.~\ref{fig:2VLL}.  The exact
        form of its
        explicit result depends on the choice of the VLL model but is
        always of the generic form indicated in Eq.~(\ref{mHHH}),
        involving the new chiral symmetry breaking parameter of
        Eq.~(\ref{eq:VLL-chiral-enhancement}) governed by
        $\bar{\lambda}_\Phi$.  This second 
        contribution corresponds to the Wilson coefficient for the
        dimension-6 operator introduced in Sec.~\ref{sec:muon-Higgs} as
        $m_\mu^{HHH}=-C_{\underset{22}{e\Phi}}v^3/2\sqrt2$. With
        these two tree-level quantities representing the two origins
        of chirality flips, the physical results for the muon mass,
        the muon--Higgs coupling and for $\Damu$ can be written as
        \begin{subequations}\label{previewVLLrelations}
          \begin{align}
  m_\mu &= m_\mu^{H} + m_\mu^{HHH},\\
  \frac{\lambda_{\mu\mu}}{\lambda_{\mu\mu}^{\text{SM}}} &=
  \frac{m_\mu^{H} + 3m_\mu^{HHH}}{m_\mu}=1+\frac{2m_\mu^{HHH}}{m_\mu},
  \label{VLLlambdaSimple}
  \\
  \Damu &= -\mathcal{Q}\frac{m_\mu m_\mu^{HHH}}{8\pi^2 v^2},\label{VLLDamuSimple}
        \end{align}
        \end{subequations}
        in full agreement with the discussions of
        Sec.~\ref{sec:muon-Higgs} and Eq.~(\ref{eq:SMEFT-muon-Higgs}). As announced there, the VLL models
        represent the important class of models, where there is a
        relative loop factor between the new chirality flips and the
        contributions to $\amu$, i.e.~the prefactor $	\kappa =
        \frac{64\pi^2}{\mathcal{Q}}$ and $\mathcal{Q}$ is of the order
        one. In the following we discuss the detailed computations of
        these and further related quantities, provide the explicit
        values of $\mathcal{Q}$ and discuss the phenomenological
        behaviour. In that discussion, non-chirally enhanced 
        corrections such as Eq.~(\ref{VLL-nonenhanced}) will be
        neglected unless explicitly mentioned.

        In order to determine the corrections more precisely one can either follow the EFT discussion presented in Sec.~\ref{sec:muon-Higgs}
	or compute the exact results in the full theory, which we outline here. To start, performing the calculation in the full theory requires the introduction
	of unitary matrices $U_L$ and $U_R$ in order to diagonalise the lepton mass matrix. This matrix results on the one hand from the above Yukawa terms after EWSB, but
	also contains contributions from the Dirac mass terms of the VLL. For example, in the $L\oplus E$ model it is given by
	(the analogous matrices in the other models are listed in Ref.~\cite{Kannike:2011ng})
	\begin{align}\label{eq:VLL-mass-matrix}
		\mathscr{M}^- = \bordermatrix{ & e_{Rj} & L^-_R & E_R \cr	
			\overline{e_{L}}_{i} & y_{eij} \tfrac{v}{\sqrt{2}} & 0 & \lambda_{Li} \tfrac{v}{\sqrt{2}} \cr 
			\overline{L^-_L} & \lambda_{Rj} \tfrac{v}{\sqrt{2}} & M_L & \lambda_\Phi \tfrac{v}{\sqrt{2}} \cr 
			\overline{E_L} & 0 & \bar\lambda_\Phi \tfrac{v}{\sqrt{2}} & M_E }.
	\end{align}
	The unitary matrices are chosen such that
	$U_L^\dagger \mathscr{M}^- U_R =
        \text{diag}(m_e,m_\mu,m_\tau,m_4,m_5) \equiv m_l $ and the
        corresponding mass-eigenstate lepton fields\footnote{%
        Because of the more complicated mixing structure than in the
        SM, we use modified conventions here compared to
        \ref{sec:Conventions}: the fields $\hat{e}^a$ $a=1\ldots5$ correspond
        to mass eigenstate fields, which are linear combinations of
        the SM lepton fields $e_i$, $i=1\ldots3$ and the new fields
        $L^-$ and $E$. A similar notation can be used in the neutrino
        sector and in the other VLL models.}
	are given by $\hat e_{L/R} = U_{L/R}^\dagger (e_i, L^-,
        E)_{L/R}$.	Analytic expressions for $U_{L/R}$ can be
          found e.g. in Ref.~\cite{Dermisek:2013gta} up to order
          $v^2$, and in the other models similar diagonalisation matrices also need to be introduced for the neutral leptons $\hat\nu_{L/R}$ 
	and doubly charged fields $\rho_{L/R}$.
	As a result of going to the mass basis, the couplings to both the
        gauge bosons and the Higgs boson are changed by the unitary transformation. 
	In particular, due to the presence of the Dirac masses, the lepton mass- and Yukawa matrices are no longer simultaneously diagonalised. 
	In the above example we obtain for the  coupling of lepton
        mass eigenstates to the physical Higgs
	\begin{align}
		-\La \supset \bar{\hat{e}}_L U_L^\dagger \begin{pmatrix} y_e & 0 & \lambda_L \\ \lambda_R & 0 & \lambda_\Phi \\ 0 & \bar\lambda_\Phi & 0	\end{pmatrix} U_R \hat e_R \frac{h}{\sqrt{2}}
		= \bar{\hat{e}}_L \frac{1}{v}\Big[m_l - U_L^\dagger\text{diag}(0,M_L,M_E) U_R \Big]\hat e_R h
		\equiv \bar{\hat{e}}_L^a \lambda_{ab} \hat{e}_R^b h.
	\end{align}
	This yields an enhanced tree-level correction to the effective (SM) lepton--Higgs couplings
	\begin{align}\label{eq:VLL-Higgs-coupling}
		\lambda_{ij} \approx \frac{m_i}{v} \delta_{ij} - \frac{\lambda_{Li} \bar\lambda^*_\Phi \lambda_{Rj} v^2}{\sqrt{2} M_E M_L} + \O(v^4),
	\end{align}
	in agreement with Eq.~\eqref{eq:SMEFT-muon-Higgs} and the
        announcement in Eq.~\eqref{VLLlambdaSimple}. As a side note, if one generation of
        VLL couples to multiple generations of SM leptons, LFV Higgs couplings
	are induced at tree-level which are already strongly
        constrained e.g. by LFV Higgs decays like $h\to \tau\mu$ and
        low-energy LFV observables like $\mu\to e$. This behaviour is
        similar, though slightly more complex, to the single-field case discussed in
        Sec.~\ref{sec:LeptonDipole} and Eq.~(\ref{LFVsingleparticleresult}).

	Similarly, because of the mixing between leptons of different $\SUL$ representations, the couplings to the $Z$- and $W$-boson are also no longer diagonal,
	\begin{align}
		\La \supset -\frac{g_2}{2c_W} \bar{\hat e} \slashed{Z} \Big(g^Z_L \PL + g^Z_R \PR \Big) \hat{e} - \frac{g}{\sqrt{2}} \bar{\hat \nu} \slashed{W}^+ \Big(g^W_L \PL + g^W_R\PR\Big) \hat e + h.c.
	\end{align} 
	where $(g^W_{L/R})_{\nu_a b} = (U_{L/R})_{ab}$ and the $Z$ coupling matrices are given by
	\footnote{Note that in the $L\oplus E$ model diagonalisation matrices for the neutrino mass eigenstates are not required, such that $\hat\nu_L=(\nu_{L,i},L^0_L,0)$ and $\hat\nu_R=(0,L^0_R,0)$. 
		For the other combinations this is usually not the case, such that e.g. the $W$-boson couplings also receive a contribution from the neutrino diagonalisation.}
	\begin{subequations}
		\begin{align}
			(g^Z_L)_{ab} &= (s_W^2-\tfrac{1}{2}) \delta_{ab} + \tfrac{1}{2} (U_L^\dagger)_{a5} (U_L)_{5b}, \\
			(g^Z_R)_{ab} &= \qquad~~ s_W^2 \delta_{ab} - \tfrac{1}{2} (U_R^\dagger)_{a4} (U_R)_{4b}.
		\end{align}
	\end{subequations}
	Again the deviation from the SM coupling, in general, introduces LFV couplings at tree-level.
	However, unlike in the case of the chirality-flipping observables (in particular Eq.~\eqref{eq:VLL-Higgs-coupling}) here the corrections are not proportional to either of the combinations
	in Eq.~\eqref{eq:VLL-chiral-enhancement} but are instead given by
	\begin{subequations}\label{eq:VLL-dgZ}
		\begin{align}
			(\delta g^Z_L)_{ij} &=  +\tfrac{1}{2} (U_L^\dagger)_{i5} (U_L)_{5j} \approx +\frac{\lambda^*_{Li}\lambda_{Lj} v^2}{4M_E^2} + \O(v^4) \\
			(\delta g^Z_R)_{ij} &=  -\tfrac{1}{2} (U_R^\dagger)_{i4} (U_R)_{4j} \approx -\frac{\lambda^*_{Ri}\lambda_{Rj} v^2}{4M_L^2} + \O(v^4)
		\end{align}
	\end{subequations}
	This different structure compared to the corrections in the lepton--Higgs coupling is crucial to satisfy both the $Z$-pole precision observables
	and LHC constraints (discussed in more detail below) with non-zero Yukawa couplings to the VLL.

\subsubsection{Vector-like lepton contributions to $a_\mu$}
	\begin{figure}[t]
		\centering
		\begin{subfigure}{.49\textwidth}
			\centering
			\includegraphics[width=.4\textwidth]{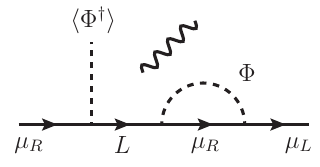}
			\includegraphics[width=.4\textwidth]{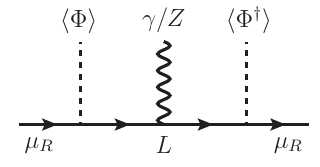}
			\caption{Single VLL without chiral enhancement}
			\label{fig:1VLL}
		\end{subfigure}
		\begin{subfigure}{.49\textwidth}
			\centering
			\includegraphics[width=.4\textwidth]{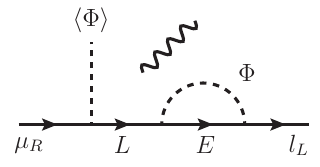}
			\includegraphics[width=.4\textwidth]{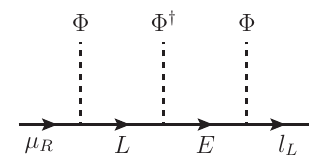}
			\caption{Two VLL combined as in Eq.~(\ref{chargedseesawpossibilities})}
			\label{fig:2VLL}
		\end{subfigure}
		\caption{Sample VLL diagrams illustrating contributions to the muon
                  mass, $\amu$, the muon--Higgs coupling and the
                  dimension-6 operators defined in
                  Sec.~\ref{sec:muon-Higgs}.  The Higgs fields can
                  either be interpreted as \vev{}s or as physical
                  Higgs bosons, and the gauge boson line can
                  optionally be connected to the
                  diagrams. \textbf{(a)}:  Single VLL  contributes to
                  e.g.~$\Damu$ and the effective 
                  $Z\to\mu\mu$ coupling with the same non-enhanced chiral structure.
			\textbf{(b)}: Two  VLL $L\oplus E$
                        provide one-loop contributions to e.g.~$\Damu$ and a tree-level correction to $h\to\mu\mu$ with the same chirally enhanced  structure.  }
		\label{fig:VLL-amu-diagrams}
	\end{figure}	

	Thanks to the mixing, couplings between the muon and the VLL
        arise at tree level, and contributions to $\amu$ arise at the
        one-loop level. They can be obtained immediately using the 
	generic formulas given in Sec.~\ref{sec:genericoneloop} and
        the couplings computed above. The dominant contributions are
        generated by the diagrams with internal VLL and Higgs- or $Z/W$-boson,
	where the chirality flip appears on the VLL line. For example, in the $L\oplus E$ model these are given by\footnote{In the models with $L_{3/2}$ or $E^a$ there will also be contributions from the doubly-charged VLL and $W^+$.}
\begin{subequations}
		\begin{align}
			\Damu^h &= \sum_{a=4,5} \frac{m_\mu m_a}{16\pi^2 M_h^2} \Re{\lambda_{\mu a}\lambda_{a\mu}} \F^\text{FS}\big(0,\tfrac{m_a^2}{M_h^2};0\big) \\
			\Damu^Z &= \sum_{a=4,5} \frac{m_\mu m_a}{16\pi^2 M_Z^2} \Re{(g^Z_L)^*_{a\mu} (g^Z_R)_{a\mu}} \F^\text{FV}\big(0,\tfrac{m_a^2}{M_Z^2};0\big) \\
			\Damu^W &= \frac{m_\mu M_L}{16\pi^2 M_W^2} \Re{(g^W_L)^*_{4\mu} (g^W_R)_{4\mu}}
                        \F^\text{FV}\big(0,\tfrac{M_L^2}{M_W^2};-1\big).
		\end{align}
	\end{subequations}
	In addition, diagrams involving only SM fields are also different from the pure SM since, as discussed above, the mixing also affects the couplings between those fields.
	These effects are however strongly constrained and the
        resulting corrections to $\Damu$ therefore
        suppressed.

	To gain qualitative insight into the above formulas for the case of $M\gg v$ it is useful to consider the mass-insertion diagrams shown in Fig.~\ref{fig:VLL-amu-diagrams}.
	Fig.~\ref{fig:1VLL} illustrates the contribution of a single
        VLL to $\Damu$ (left) as well as a tree-level contribution to
        the muon--$Z$ coupling with the same chiral structure
        (right). For $\Damu$, the required chirality flip and \vev is
        provided by the SM muon Yukawa coupling and $v$, and the
        relevant mass scale in the loop is ${\cal O}(v)$.
	As a consequence, diagrams of this form do not result in chirally enhanced contributions to $\Damu$, but instead give
	\begin{align}\label{eq:amu-VLL-non-enhanced}
		\Damu^\text{VLL,non-enhanced} \sim \frac{m_\mu
                  y_\mu v}{16\pi^2v^2} (\delta g^Z_{L/R})_{\mu\mu},
	\end{align}
	Since the constraints on deviations in the $Z$--$\mu$ coupling by electroweak precision data \cite{Cynolter:2008ea,delAguila:2008pw} and $Z$-pole observables \cite{Freitas:2014pua,Lourenco:2023sde} 
	are already rather stringent, such contributions can result at most in $\Damu^\text{VLL,non-enhanced} \lesssim \O(10^{-12})$. 

	On the other hand, in Fig.~\ref{fig:2VLL} (left) we illustrate the contribution of two virtual VLL. 
	Importantly, this diagram has the same chiral structure as the tree-level diagram Fig.~\ref{fig:2VLL} (right) that generates the $\bar{l}_L \mu_R \Phi (\Phi^\dagger\Phi)$ 
	operator discussed in Sec.~\ref{sec:muon-Higgs}	and which results in the chirally enhanced contributions to $\lambda_{\mu\mu}$ in Eq.~\eqref{eq:VLL-Higgs-coupling}.
	This implies that such diagrams also generate the same chiral
        enhancement in $\Damu$, and the full contributions can be
        estimated by multiplying with the required factors as
        discussed before and in Sec.~\ref{sec:ChiralityFlips},
	\begin{align}\label{eq:VLL-Damu-enhanced}
		\Damu^\text{VLL, enhanced} &\sim \frac{m_\mu}{16\pi^2}
                 \frac{[\lambda_{L2} \bar\lambda^*_\Phi \lambda_{R2} v]}{\sqrt{2} M_E M_L} ,
        \intertext{where the terms collected in the square brackets have the same
        meaning as in Eqs.~(\ref{amugeneric}), or equivalently}
		\Damu^\text{VLL, enhanced} &\sim \frac{m_\mu}{16\pi^2 v} \delta \lambda_{\mu\mu},
	\end{align}
        where $\delta\lambda_{\mu\mu}$ is the correction term on the
        r.h.s.~of Eq.~(\ref{eq:VLL-Higgs-coupling}). This equation
        reproduces the announced result (\ref{VLLDamuSimple}).
	Compared to Eq.~\eqref{eq:amu-VLL-non-enhanced} this
        contribution is larger roughly by a chiral enhancement factor of $\bar\lambda_\Phi / y_\mu \sim \O(1000)$ and can therefore
	comfortably reach large values such as  $\DamuOld$.

        Yet another useful way to understand the chirally enhanced VLL
        contributions to $\amu$ is via the three-field models
        discussed in Sec.~\ref{sec:genericthreefield}. As mentioned in
        footnote \ref{footnoteVLL}, the enhanced
        contribution in Eq.~(\ref{eq:VLL-Damu-enhanced}) behaves
        similarly to the one of the Class II three-field model, see also
        Eq.~(\ref{eq:MIAcontribution}). A similar VLL contribution
        proportional to $\lambda_\Phi$ exists as well and is similar 
        to the Class I model. In the relevant limit where the VLL
        masses $M_F$ are much larger than the SM Higgs-boson mass, the
        limiting behaviour of the three-field model analysed in
        Eq.~(\ref{eq:3-field-limits}) shows that the
        $\lambda_\Phi$-contribution has an additional
        $v^2/M_F^2$-suppression compared to the
        $\bar{\lambda}_\Phi$-term.

        Now we turn to the correlation with the muon--Higgs coupling,
        which plays a unique role in VLL models as previewed in
        Eqs.~(\ref{previewVLLrelations}). Indeed, combining the
        formula for $\Damu$ with the one for the muon--Higgs coupling
        in
        Eq.~\eqref{eq:VLL-Higgs-coupling}  results in a correlation like Eq.~\eqref{eq:Rmumu-ellipse}
	(in the absence of CP violation) \cite{Kannike:2011ng,Dermisek:2013gta,Endo:2020tkb,Dermisek:2022aec,Dermisek:2023nhe},
	\begin{align}\label{eq:Rmumu-VLL}
		R_{\mu\mu}\equiv\frac{\Gamma(h\to\mu\mu)}{\Gamma(h\to\mu\mu)_\text{SM}} \approx \bigg|\frac{\lambda_{\mu\mu}}{\lambda_{\mu\mu}^\text{SM}}\bigg|^2 = \bigg|1 - 0.86 \bigg(\frac{\Damu}{\mathcal{Q} \times 10^{-9}}\bigg)\bigg|^2,
	\end{align}
	where the constants $\mathcal{Q}$ for the different models can
        be obtained from the full computation of all quantities and are
        listed in Tab.~\ref{tab:VLL-coefficients}. As discussed in Sec.~\ref{sec:muon-Higgs} a correlation of this form
	appears in any model where chirally enhanced contributions to $\Damu$ are present. However, in generic theories the corrections to $\lambda_{\mu\mu}$
	appear first at one-loop order \cite{Crivellin:2021rbq}
        resulting in suppressions equivalent to $\mathcal{Q}\sim \O(16\pi^2)$.

	As mentioned in Sec.~\ref{sec:muon-Higgs}, the VLL models are unique in that they predict much smaller values of $\mathcal{Q}\sim \O(1)$ as a consequence of the fact
	that the chirally enhanced corrections to $\lambda_{\mu\mu}$
        arise already at tree level. Therefore, these models exhibit a strong correlation between $R_{\mu\mu}$ and $\Damu$ 
	which aids in discriminating between the different scenarios.
	The only  exception is the $L\oplus N$ model. In this case the operator $(\bar{l}\tilde\Phi e) (\Phi^\dagger\tilde\Phi)$ induced by the chiral structure in Fig.~\ref{fig:2VLL} 
	vanishes due to the anti-symmetric $SU(2)$ index contraction. Similarly, the leading-order corrections to $\Damu$ also vanish \cite{Arkani-Hamed:2021xlp}, such that the model 
	effectively behaves like one without any chiral enhancement. Hence, in this model the correlation is also accidentally weak. 
	We note that allowing for complex couplings results in an additional contribution from $d_\mu$ to the above correlation (cf. Sec.~\ref{sec:muon-Higgs}) 
	and thus also a constraint on the possible CP violation \cite{Cherchiglia:2021syq,Hamaguchi:2022byw,Dermisek:2023nhe}.
		
\begin{figure}
		\centering
		\includegraphics[width=.45\textwidth]{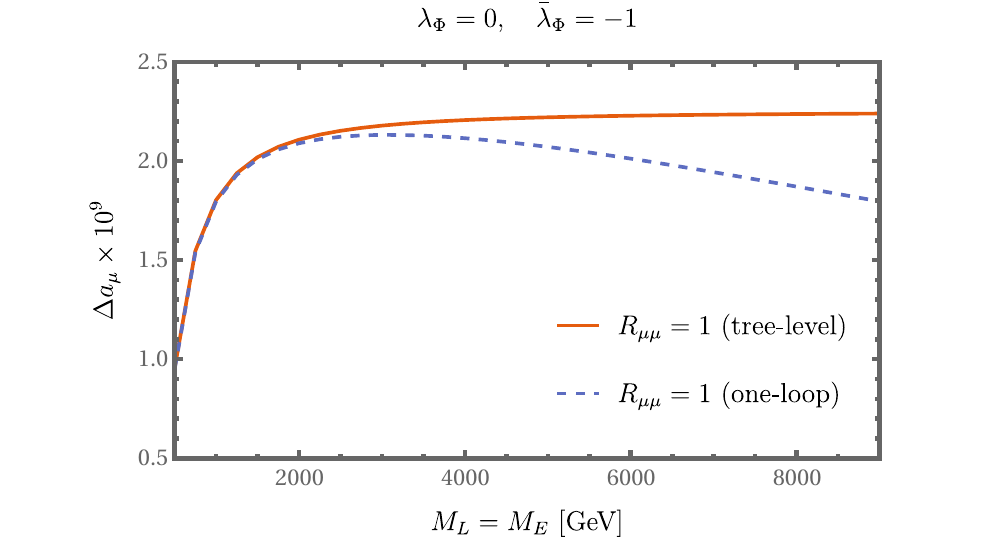}
		\includegraphics[width=.45\textwidth]{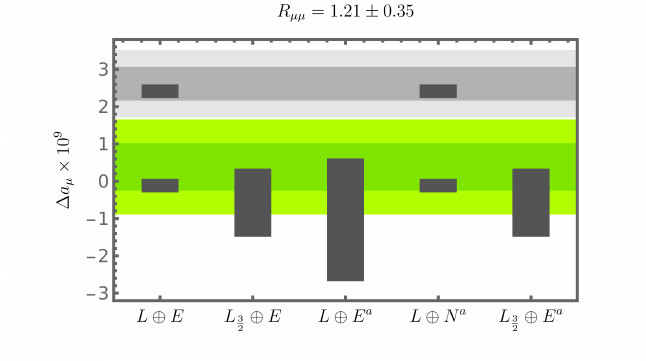}
		\caption{Correlation between $\Damu$ and the
                  muon--Higgs coupling and its impact in the VLL model $L\oplus E$. \emph{left}: Comparison between the value of $\Damu$ required to obtain $R_{\mu\mu}=1$ at tree-level (red) and one-loop (blue). \emph{right}: Ranges of $\Damu$
		allowed by Eq.~\ref{eq:Rmumu-VLL} and the current experimental bounds. The (light) grey band shows the $1\sigma$ (2$\sigma$) range of $\DamuOld$ while 
		the (light) green band shows the 1$\sigma$ (2$\sigma$) range for $\DamuFinal$. }
		\label{fig:VLL-Rmumu-correlation}
	\end{figure}

	The above correlation has an immediate, strong
        phenomenological consequence: In order to achieve
        $R_{\mu\mu}\approx1$ in agreement with LHC measurements,
        $\Damu$ is restricted, and numerically the restriction affects
        exactly the range of interest. One solution is sufficiently
        small $\Damu$, a second solution is where the
        term involving $\Damu$ in Eq.~(\ref{eq:Rmumu-VLL}) becomes
        $\approx-2$. This second, \emph{flipped-sign} solution requires one
        specific non-vanishing and positive value of $\Damu$ in the
        ballpark of ``a few''$\times10^{-9}$. Here we point
        out that higher-order corrections to the correlation Eq.~\eqref{eq:Rmumu-VLL} 
	can have a potentially significant effect on the predicted asymptotic value of $\Damu$ \cite{Mohling:2024qvk}. Fig.~\ref{fig:VLL-Rmumu-correlation} (left) shows a comparison between 
	$\Damu$ (in the $L\oplus E$ model) obtained for $R_{\mu\mu}=1$
        at tree-level (red) and one-loop order\footnote{These results
        are based on consistent one-loop calculations for both
        observables, but including two-loop (i.e.\ next-to-leading
        order) corrections to $\Damu$
	could lead to further shifts of the correlation.} (blue). As predicted, for large masses
        the leading-order result converges to the fixed value
        $\Damu\approx22\times10^{-10}$ dictated by
        Eq.~\eqref{eq:Rmumu-VLL}, which is intriguingly close 
        to the previous deviation $\DamuOld$. The one-loop corrected
        correlation predicts a significantly smaller value and  no longer converges to a fixed value 
	for increasing VLL masses due to the logarithmic enhancement present in the one-loop corrections. This effect is even 
	stronger when non-zero values of $\lambda_\Phi$ are allowed
	since at one-loop both combinations of chiral enhancement from
        Eq.~\eqref{eq:VLL-chiral-enhancement} contribute in a similar
        way, as discussed in Ref.~\cite{Mohling:2024qvk}.
	
	\begin{table}[t]
		\centering
		\begin{tabular}{|c||c|c|c|c|c|c|}
			\hline
			Model & $L\oplus E$ & $L\oplus N$ & $L_{\frac{3}{2}}\oplus E$ & $L\oplus E^a$ & $L\oplus N^a$ & $L_{\frac{3}{2}}\oplus E^a$  \\ \hline
			$\mathcal{Q}$ & $1$ & -- & $5$ & $9$ & $1$ & $5$  \\ \hline
		\end{tabular}
		\caption{Correlation factors for the VLL models.}
		\label{tab:VLL-coefficients}
	\end{table}
	
\subsubsection{Collider searches and precision constraints}
	\begin{figure}[t]
		\centering
		\includegraphics{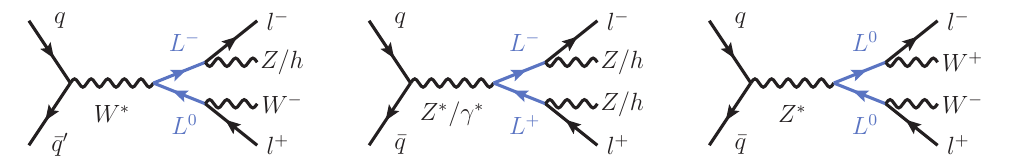}
		\caption{Illustration of pair-production and subsequent decay of a doublet VLL at the LHC.}
		\label{fig:VLL-search}
	\end{figure}
	Since the VLL have only a negligible impact on the Higgs production cross section, the experimental bounds on $R_{\mu\mu}$
	can immediately be obtained from the current measurements of the di-muon Higgs decay at the LHC resulting in \cite{ATLAS:2020fzp,CMS:2022dwd,ParticleDataGroup:2024cfk}
	\begin{align}
		R_{\mu\mu} = 1.21 \pm 0.35.
	\end{align}
	The resulting ranges of $\Damu$ allowed by Eq.~\eqref{eq:Rmumu-VLL} are shown for the different models in Fig.~\ref{fig:VLL-Rmumu-correlation} (right).
	The grey band shows the (1 and 2$\sigma$) region $\Damu = \DamuOld$
          and the green band the new result $\DamuFinal$, which is compatible with zero.
	As discussed, the correlation (\ref{eq:Rmumu-VLL}) for real
        parameters leads to
        two possible solutions. Agreement with the LHC data is
        achieved for either very small values of $\Damu$ or 
        sufficiently large values such that the muon--Higgs coupling
        has the same magnitude but opposite sign (\emph{flipped-sign} case) to the SM. 
	Coincidentally, two of the models ($L\oplus E$ and $L\oplus
        N^a$) predict a large deviation very close to $\DamuOld$ in
        this \emph{flipped-sign} scenario, as already shown in
        Fig.~\ref{fig:VLL-Rmumu-correlation} (left) in more detail. 
	In contrast, the $L_{\frac{3}{2}}\oplus E, ~L\oplus E^a$ and
        $L_{\frac{3}{2}}\oplus E^a$ models with $\mathcal{Q}=5$ or $9$ would require a deviation much larger than $\DamuOld$ to be compatible with $R_{\mu\mu}$.
	Consequently, for these last three models, the \emph{flipped-sign}
        case had already been firmly excluded at the time of the
        previous result $\DamuOld$.

        For the two $L\oplus E$ and $L\oplus N^a$ models, the
        previous result $\DamuOld$ was very well compatible with the
        \emph{flipped-sign} case; now, the new result $\DamuFinal$
        disfavours this option. Even in the case where a future
        non-vanishing deviation is established, however of a smaller
        magnitude of order $\lesssim10^{-9}$, the \emph{flipped-sign} case
        would remain disfavoured or excluded for all these VLL models.

        For all the five models shown in
        Fig.~\ref{fig:VLL-Rmumu-correlation} (right), on the other
        hand,  the normal-sign case predicts small $\Damu$ and is thus
        compatible with $\DamuFinal$. In this case the difference
        between the  five models is the possible range of $\Damu$
        values. In the two models with $\mathcal{Q}=1$, the
        $R_{\mu\mu}$ constraint is  only compatible with very small
        $\Damu$, while the models with $\mathcal{Q}=5$ or $9$ still
        comfortably allow  $|\Damu| \gtrsim \O(5\times 10^{10})$ even
        in the normal-sign case. The 
        current measurement of $\DamuFinal$ thus constrains the
        parameter space of the  $\mathcal{Q}=5,9$ models in  a
        non-trivial way, and a future even more precise determination
        of $\Damu$ could be important to discriminate between the VLL models.
 We finally note that all models predict $R_{\mu\mu} < 1$ for a positive deviation $\Damu$, 
	while the current measurement favours values larger than one. This leads to the asymmetric distribution around $\Damu=0$ seen in Fig.~\ref{fig:VLL-Rmumu-correlation}.

	Besides the indirect tests the LHC also provides very stringent exclusion limits on the VLL masses, significantly improving on the bound of $M \gtrsim 100$ GeV set by LEP \cite{L3:2001xsz}.
	Several searches for direct production of VLL at the LHC were performed by ATLAS \cite{ATLAS:2015qoy,ATLAS:2023sbu,ATLAS:2024mrr}, CMS \cite{CMS:2019hsm,CMS:2023dhv,CMS:2024bni}
	and Refs.~\cite{Kearney:2012zi,Dermisek:2014qca,Freitas:2014pua,Kumar:2015tna,Guedes:2021oqx,Kawamura:2023zuo,Dehghani:2024lgz} focused in particular on scenarios with the SM-like doublet and singlet VLL $L$ and $E$.
	At the LHC these would mainly be pair-produced via exchange of a virtual gauge-boson, i.e. $pp\to \gamma^*/Z^*\to L^+ L^-$, $pp\to Z^* \to L^0 \bar{L}^0$ or $pp\to W^* \to L^- \bar{L}^0$
	in case of the doublet or $pp\to \gamma^*/Z^* \to E^+ E^-$ in case of the singlet. After production these VLL primarily decay into a SM lepton and gauge- or Higgs boson with branching ratios given by
	(in the limit of $M\gg v$) $\text{BR}(L^\pm \to l^\pm + Z/h)\simeq 50\%$, 
	$\text{BR}(L^0 \to l^\pm + W^\mp)\simeq 100\%$ for the doublet as well as $\text{BR}(E^\pm\to l^\pm Z/h)\simeq 25\%$ and $\text{BR}(E^\pm\to \nu W^\pm)\simeq 50\%$ in case of the VLL singlet.
	The resulting processes relevant for the LHC analyses are shown in Fig.~\ref{fig:VLL-search}.
	Recently, searches for VLL in the full run-2 data set have been performed by ATLAS \cite{ATLAS:2024mrr}, CMS \cite{CMS:2024bni} and \cite{Kawamura:2023zuo},
	yielding the most stringent exclusion limits to date. For VLL coupling to exclusively to muons (electrons) [tau] \cite{ATLAS:2024mrr,CMS:2024bni}
	these are given by\footnote{It has been argued \cite{Chala:2020odv} that searches for single production of VLL might be necessary to conclusively rule out VLL below the TeV scale.}
	\begin{align}
		\text{doublet}&: \qquad M_L > 1270~(1220)~[1040]~\text{GeV} \\
		\text{singlet}&: \qquad M_E > 400~~\,(320)~~\,[170]~~\,\text{GeV}.
	\end{align}
	The significantly weaker bounds on VLL singlets result from the smaller pair production cross section compared to the doublet scenario.
	So far, no dedicated searches for $L_{3/2}$ or the triplet VLL have been performed, however, the similar type III see-saw triplet has already been excluded for masses below 1 TeV \cite{ATLAS:2022yhd}.
	Given that the pair production cross section is comparable to or even larger than in the doublet case \cite{Kannike:2011ng}, a recasting could further strengthen this bound.\newline

	Another important constraint on the VLL models stems from EWPO like $Z$-pole measurements \cite{Kannike:2011ng,Dermisek:2013gta}. These result in bounds on the mixing between
	VLL and SM leptons, i.e. in particular the combination Eq.~\eqref{eq:VLL-dgZ}.
	For example, in case of the $L\oplus E$ model the updated
        bounds on the $\mu$--$Z$ coupling from
        Ref.~\cite{Lourenco:2023sde} translate (at tree-level) into \cite{Mohling:2024qvk}
	\begin{align}\label{eq:VLL-Z-limits}
		\bigg|\frac{\lambda_R^\mu v}{M_L}\bigg| \lesssim 0.03, \qquad \bigg(\frac{\lambda_L^\mu v}{M_E}\bigg)^2 + 1.16 \bigg(\frac{\lambda_R^\mu v}{M_L}\bigg)^2 \lesssim 0.001.
	\end{align}
	Similar constraints also hold for the other models.
	Notably, in combination with Eq.~\eqref{eq:VLL-chiral-enhancement} and Eq.~\eqref{eq:Rmumu-VLL} these constraints require either $\bar\lambda_\Phi\approx 0$ (with $\Damu\approx 0$)
	or $|\bar\lambda_\Phi| \gtrsim 1.5$ (with $\Damu\approx \DamuOld$). While the latter case is consistent with perturbativity and other experimental bounds, 
	it is potentially in conflict with the constraints from vacuum
        stability
        \cite{Gopalakrishna:2018uxn,Adhikary:2024esf,Cingiloglu:2024vdh},
        which strengthens the conclusion that the \emph{flipped-sign} case
        is now disfavoured.
	The technical reason is that VLF generically contribute in the same way as the top-quark to the RGE of the quartic Higgs coupling $\lambda$
	\begin{align}
		\beta(\lambda) \sim \frac{1}{4\pi^2} \big(2 \bar\lambda_\Phi^2 \lambda - \bar\lambda_\Phi^4\big)
	\end{align}
	and thus further drive $\lambda(\mu)$ to negative values as $\mu$ increases. Requiring that the quartic Higgs coupling stays positive up to some scale $\Lambda$ therefore
	translates into upper bounds on the VLL Yukawa couplings, which turn out to be rather stringent \cite{Adhikary:2024esf}. In fact, already for moderate choices like
	$\Lambda=100$ TeV, values of $\bar\lambda_\Phi\gtrsim 1$ are essentially excluded.
	This issue is potentially remedied by the introduction of additional scalars \cite{Xiao:2014kba} which will be discussed in the following subsection.\newline

	\begin{figure}
		\centering
		\begin{subfigure}{.24\textwidth}
			\centering
			\includegraphics[width=.8\textwidth]{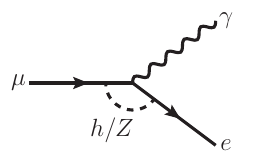}
			\caption{$\mu\to e\gamma$ (chirality flipping)}
			\label{fig:VLL-mu2egamma}
		\end{subfigure}
		\begin{subfigure}{.24\textwidth}
			\centering
			\includegraphics[width=.8\textwidth]{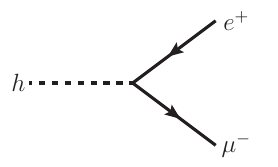}
			\caption{$h\to\mu e$ (chirality flipping)}
			\label{fig:VLL-h2emu}
		\end{subfigure}
		\begin{subfigure}{.24\textwidth}
			\centering
			\includegraphics[width=.8\textwidth]{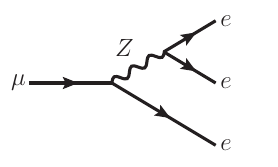}
			\caption{$\mu\to 3e$ (chirality preserving)}
			\label{fig:VLL-mu3e}
		\end{subfigure}
		\begin{subfigure}{.24\textwidth}
			\includegraphics[width=.8\textwidth]{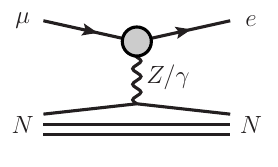}
			\caption{$\mu\to e$ conversion (both)}
		\end{subfigure}
		\caption{Illustration of the leading-order contributions to LFV observables 
			in the VLL models. }
		\label{fig:VLL-mu2e}
	\end{figure}
	
	Lastly, if the VLL couple to multiple SM leptons simultaneously, constraints from LFV observables need to be taken into account.
	The impact of of these constraints on the VLL models has already been studied in several different scenarios.
	Refs.~\cite{Ishiwata:2013gma,Poh:2017tfo} considered in particular constraints from $\mu\to e\gamma$ and $\mu\to 3e$
	on the $L\oplus E$ model, Ref.~\cite{Raby:2017igl} also included a 4th generation of vector-like quarks and
	Refs.~\cite{Falkowski:2013jya, Crivellin:2018qmi} considered extensions to 3 generations of VLL and with constraints from 
	$\mu\to e$ conversions and LFV processes involving also the 3rd generation. VLL besides the SM-like doublet and singlet
	have also been studied in Ref.~\cite{Crivellin:2020ebi} with focus on models without chiral enhancement.
	
	As discussed in Sec.~\ref{sec:LeptonDipole}, if $a_\mu$ receives large contributions, the strongest LFV constraints typically 
	stem from the dipole operators generated at one-loop through diagrams like Fig.~\ref{fig:VLL-mu2egamma}.
	In the VLL models the flavour mixing angles defined in Sec.~\ref{sec:LeptonDipole} are given by
	\begin{align}
		\theta_{2i} = \frac{\lambda_R^i}{\lambda_R^\mu}, \qquad \text{and} \qquad \theta_{i2} = \frac{\lambda_L^i}{\lambda_L^\mu},
	\end{align}
	corresponding to the scenario of single particle contribution Eq.~\eqref{Crivellinsingleparticle}.
	The current CLFV bounds from BR$(\mu\to e\gamma)$ and BR$(\tau\to\mu \gamma)$ in Eq.~\eqref{eq:ltolgamma-bounds} imply
	\begin{align}
		|\theta_{i\mu}|^2 + |\theta_{\mu i}|^2 &\lesssim \bigg(\frac{10^{-9}}{\Delta a_\mu}\bigg)^2 \begin{cases}
			2.6\times 10^{-9} & i = e \\
			0.4 & i = \tau
		\end{cases}
	\end{align}
	If the muon couplings are near the limits \eqref{eq:VLL-Z-limits} for VLL masses around $\O(1~\text{TeV})$ and sizeable $\Damu$,
	the VLL must couple very weakly to the other generations $|\lambda_{L/R}^e| \lesssim 10^{-5}$ and $|\lambda_{L/R}^\tau| \lesssim 0.1$.
	
	It is also interesting to compare these processes to the Higgs decays BR$(h\to\mu + \tau/ e)$ induced by the tree-level diagram in
	Fig.~\ref{fig:VLL-h2emu}. Similar to Eq.~\eqref{eq:VLL-Damu-enhanced}, the LFV Higgs couplings are related to $\Damu$ by
	\begin{align}
	 	\lambda_{ij} \approx - 0.86 \frac{\theta_{ij} \Delta a_\mu}{\mathcal{Q}\times 10^{-9}} \frac{m_\mu}{v} 
	\end{align}
	The relatively large Higgs decay width $\Gamma_h\approx 4\times 10^{-3}$ GeV combined with the current bounds \cite{ParticleDataGroup:2024cfk}
	\begin{subequations}
		\begin{align}
			\text{BR}(h\to\mu e)< 4.4\times 10^{-5}, \qquad \text{and} \qquad
			\text{BR}(h\to\mu\tau) < 1.5 \times 10^{-3}
		\end{align}
	\end{subequations}
	lead to significantly weaker constraint (particularly in for the models with $\mathcal{Q}=9$ or $5$) 
	on the electron couplings compared to Eq.~\eqref{eq:ltolgamma-bounds}
	\begin{align}
		|\theta_{i\mu}|^2 + |\theta_{\mu i}|^2 &\lesssim \bigg(\frac{10^{-9}}{\Delta a_\mu}\bigg)^2 \begin{cases}
			0.5 \mathcal{Q}^2 & i = e \\
			16 \mathcal{Q}^2 & i = \tau
		\end{cases}
	\end{align}
	
	In scenarios without chiral enhancement strong LFV constraints result from $\mu\to e$ conversion and
	$\mu\to 3e$. Numerically (again up to representation dependent $\O(1)$ factors) we have \cite{Kitano:2002mt,Crivellin:2017rmk}
	\begin{align}
		\begin{split}
			\text{BR}(\mu\text{Au}\to e\text{Au}) &= \frac{m_\mu^5}{8 M^4_Z \Gamma^\text{Au}_\text{capt}} \frac{g_2^2}{c_W^2}
			\Big|(1-4s_W^2)V^{(p)}_\text{Au} - V^{(n)}_\text{Au}\Big|^2 \Big( |g^Z_{L12}|^2 + |g^Z_{R12}|^2\Big) \\
			&\sim \frac{7\times 10^{-13}}{(M/1~\text{TeV})^4} \Big(\Big|\frac{\lambda_L^e\lambda_L^\mu}{10^{-5}}\Big|^2 
			+ \Big|\frac{\lambda_R^e\lambda_R^\mu}{10^{-5}}\Big|^2\Big)
		\end{split}
	\end{align}
	and similarly for $\mu\to 3e$
	\begin{align}
		\text{BR}(\mu\to 3e) = \frac{\alpha m_\mu^5}{128\pi^2 M_Z^4 \Gamma_\mu} \frac{s_W^2}{c_W^2} \Big( |g^Z_{L12}|^2 + |g^Z_{R12}|^2\Big)
		\sim
		\frac{4.2\times 10^{-13}}{(M/1~\text{TeV})^4} \Big(\Big|\frac{\lambda_L^e\lambda_L^\mu}{5\cdot 10^{-5}}\Big|^2 
		+ \Big|\frac{\lambda_R^e\lambda_R^\mu}{5\cdot 10^{-5}}\Big|^2\Big).
	\end{align}
	Again, these require highly non-universal couplings to the SM leptons and,
	as mentioned in Sec.~\ref{sec:LeptonDipole}, the bounds for both $\mu\to 3e$ and $\mu\to e$ conversions are even expected to improve by several order of 
	magnitude in the near future.

\subsubsection{Vector-like fermions with extended scalar sectors}\label{sec:VLF+scalar}
	So far, we have highlighted several motivations for studying
        extended fermion sectors and, e.g.~in Sec.~\ref{sec:2HDM}, also extended scalar sectors.
	Here we want to discuss how a combination of both VLF and new
        scalar sectors can lead to appealing BSM scenarios which can offer
        solutions to open questions and lead to a diverse phenomenology.
	Notably, there are several mutual benefits which motivate both extended scalar sectors from the perspective of VLF and vice versa. 
	On the one hand, additional scalars help to weaken the strong bounds on VLF imposed by vacuum stability \cite{Xiao:2014kba,Arsenault:2022xty,Cingiloglu:2023ylm,Cingiloglu:2024vdh,Adhikary:2024esf}
	which can significantly increase the high energy validity of such models. They also open up further possibilities for including dark matter 
	\cite{ColuccioLeskow:2014epd,Kawamura:2020qxo,Huang:2020ris,Jana:2020joi,Saez:2021qta,Kawamura:2022uft,Ko:2021lpx,Wojcik:2022woa}, 
	neutrino masses (e.g.~in case of triplet scalar) \cite{Wang:2012gm,Bahrami:2014ska,DeJesus:2020yqx,Chakrabarty:2020jro,Jana:2020joi} or even explanations of the entire SM flavour structure 
	\cite{Hernandez:2021xet,Bonilla:2021ize,Hernandez:2021iss,CarcamoHernandez:2023wzf}.
	On the other hand, additional fermions in theories with extended scalar sectors can free up otherwise excluded parameter space \cite{Garg:2013rba,Bahrami:2014ska,Raju:2022zlv}
	and aid in the explanation of tentative anomalies \cite{Arnan:2019uhr,Nagao:2022oin,Brune:2022rlo}. 
	
	Among the examples above, one of the most well studied scenarios is  the addition of new fermion content in the 2HDM.
	Such models not only fit naturally into the context of SUSY, where additional VLF can solve the little hierarchy problem and provide (at least in part) 
	the large required corrections to the Higgs mass \cite{Moroi:1991mg,Martin:2009bg,Graham:2009gy}, but also offer a rich phenomenology, especially in the context of $\Delta a_\mu$.	
	For simplicity we consider only the case of 2HDM+$L\oplus E$ for concrete examples and results, although most of the discussion will also apply for
	other combinations of VLL and more generic VLF. \newline
		
	Similar to the usual Yukawa interaction terms in the 2HDM (cf. Eq.~\eqref{eq:2HDM-Yukawa-Lag}), the VLF generically couple to both Higgs doublets
	\begin{align}
		\La \supset - \sum_{a=1,2} \bigg\{ \bar{l}_L Y^l_a e_R +  \bar{L} \lambda_R^a e_R + \bar{l}_L \lambda^a_L E + 
		\bar{L} \Big(\lambda^a_\Phi \PR + \bar\lambda^a_\Phi \PL \Big) E\bigg\} \Phi_a + h.c.,
	\end{align}
	In order to avoid large FCNC (in the limit $\lambda_{L/R}\to 0$) one can  impose a $\mathds{Z}_2$ symmetry under which $\Phi_1\to -\Phi_1$ and $\Phi_2\to \Phi_2$. Like in the usual 2HDM, this also forces the VLF to couple exclusively to one of Higgs doublets, depending on their charge assignments. Usually, $L$ is chosen to transform like $l_L$ and
	$E$ like $e_R$, such that both the SM leptons and VLL couple only to $\Phi_1$. This scenario was extensively studied in Refs.~\cite{Frank:2020smf,Chun:2020uzw,Dermisek:2020cod,Ferreira:2021gke,Dermisek:2021ajd}
	and corresponds to type-II/X assignments for both the VLL and SM leptons. However, other choices are also possible. For example, taking the opposite charge assignments for
	the VLL implies that the VLL mix with the SM leptons only via
        $\phi_2$, leading to a relative $\cot\beta$ suppression of
        these terms.

        For the usual type-II/X assignments
	the resulting lepton mass matrix is given by
	\begin{align}
		\mathscr{M}^- = \bordermatrix{ & e_{Rj} & L^-_R & E_R \cr	
			\overline{e_{L}}_{i} & (Y^l_1)_{ij} \frac{v_1}{\sqrt{2}} & 0 & \lambda^1_{Li} \frac{v_1}{\sqrt{2}} \cr 
			\overline{L^-_L} & \lambda^1_{Rj} \frac{v_1}{\sqrt{2}} & M_L & \lambda^1_\Phi \frac{v_1}{\sqrt{2}} \cr 
			\overline{E_L} & 0 & \bar\lambda^1_\Phi \frac{v_1}{\sqrt{2}} & M_E }.
	\end{align}
	It is the same matrix as in Eq.~\eqref{eq:VLL-mass-matrix} except for the appearance of $v_1$ instead of $v$. Introducing 
	mass diagonalisation matrices $U_{L/R}$ like in Sec.~\ref{subsec:VLL} 
	therefore yields the same expressions for the gauge-coupling matrices with $v$ replaced by $v_1$.
	The Yukawa Lagrangian in the mass basis is given by
	\begin{align}
		\La \supset - \bar{\hat e}_a \Big[\lambda^h_{ab} h + \lambda^H_{ab} H +
		\lambda^A_{ab} iA\Big] \PR \hat{e}_b 
		- \sqrt{2} H^+ \bar{\hat\nu}_a \lambda^{H^+}_{ab} \PR \hat{e}_b		
		- \sqrt{2} H^- \bar{\hat e}_a \lambda^{H^-}_{ab} \PR \hat{\nu}_b
		+ h.c.
	\end{align}
	where the Yukawa coupling matrices are listed in Ref.~\cite{Dermisek:2021ajd}. In the case where mixing between the VLL and SM leptons is neglected 
	($\lambda^a_{L/R}=0$), the SM couplings reduce to the
        expressions discussed for the pure 2HDM in Sec.~\ref{subsec:FA2HMD}.
	
	Like in case of the usual 2HDM-II/X the couplings $\lambda_{ab}^{H/A/H^\pm}$ to the new scalars 
	are enhanced by	a factor of $\tan\beta$ compared to the coupling to the SM-like Higgs $h$.
	However, because $v_1=v\cos\beta$ the leading-order tree-level corrections to the coupling matrices are 
	suppressed by additional factors of $\cos\beta$ compared to Eq.~\eqref{eq:VLL-Higgs-coupling} or \eqref{eq:VLL-dgZ}.
	In particular, the tree-level correction to the muon--Higgs coupling is
	\begin{align}
		\lambda^h_{\mu\mu} \simeq \frac{m_\mu}{v} - \frac{\lambda_{L\mu}^1\bar\lambda_\Phi^1\lambda_{R\mu}^1 v^2}{\sqrt{2} M_E M_L} \cos^3\beta,
	\end{align}	
	such that the correction term is multiplied by $\cos^3\beta$
        compared to its counterpart with SM Higgs sector in
        Eq.~\eqref{eq:VLL-Higgs-coupling}.
    \newline
    
\begin{figure}[t]
		\centering
		\begin{subfigure}{.44\textwidth}
			\centering
			\includegraphics[width=.44\textwidth]{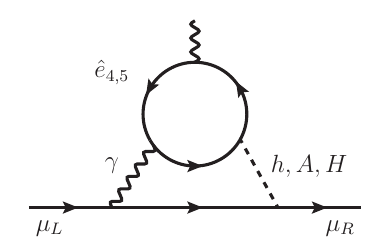}
			\includegraphics[width=.44\textwidth]{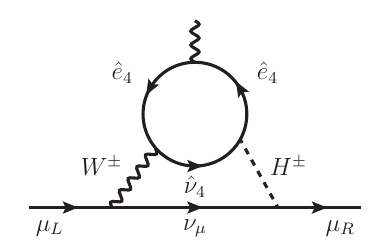}
			\caption{}
			\label{fig:2HDM+VLF-BZ}
		\end{subfigure}
		\begin{subfigure}{.44\textwidth}
			\centering
			\includegraphics[width=.44\textwidth]{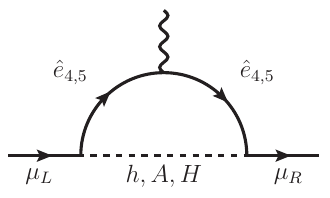}
			\includegraphics[width=.44\textwidth]{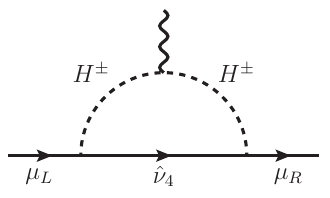}
			\caption{}
			\label{fig:2HDM+VLL-one-loop}
		\end{subfigure}
		\caption{\textbf{(a)}: Two-loop Barr-Zee contributions from generic charged VLF in the 2HDM+VLF. \textbf{(b)}: One-loop contributions to $\Damu$ from VLL and scalars in the 2HDM+VLL. }
		\label{fig:VLL-2HDM-amu-diags}
	\end{figure}

	As discussed in Sec.~\ref{sec:2HDM}, the dominant contributions to $\Damu$ in the 2HDM arise through Barr-Zee diagrams at the two-loop level. Therefore any additional charged
	fermion with significant couplings to the Higgs further increases $\Damu$. This effect was studied e.g. in Refs. \cite{Frank:2020smf, Chun:2020uzw, Ferreira:2021gke} who considered additional VLL while neglecting 
	mixing with the SM leptons. In this case, the VLF will contribute through Barr-Zee diagrams like those shown in 
	Fig.~\ref{fig:2HDM+VLF-BZ} with either exchange of a neutral scalar and a photon discussed in Sec.~\ref{sec:Barr-Zee} 
	or charged scalar and $W$-boson.
	In both cases only the diagonal couplings
        $\lambda^S_{aa}$ in mass-eigenstate basis contribute, such that
	the resulting $\Damu$ behaves similarly to the contribution
        from a heavier $\tau$-lepton with the difference that the VLL
        mass is unrelated to its Yukawa coupling thanks to the Dirac
        mass term. As a result, the $\Damu$ contribution is strongly
        suppressed unless there is significant mixing (i.e.\ large
        Yukawa coupling) between the VLL.
	Specifically, this requires not too heavy VLL with small mass splittings $M_L\sim M_E\equiv M_F$.
	At the same time, the flavour and LHC constraints on the scalar masses discussed in Sec.~\ref{sec:2HDM} 
	still apply and require small mass splittings also between the scalars $M_H\sim M_A\equiv M_S$.
	Therefore, the dominant contribution  from the neutral scalar exchange results in\footnote{The analogous 
		expression for the BZ diagram with $H^\pm$ exchange is slightly more involved and can be found e.g. in Refs.~\cite{Frank:2020smf,Ferreira:2021gke}}
	\begin{align}
		\Damu^\text{VLF,BZ} 
		= \frac{\sin^2\beta}{(M_S/1000~\text{GeV})^2} \bigg\{\big(|\lambda^1_\Phi|^2 + |\bar\lambda^1_\Phi|^2\big) \F_1\Big(\tfrac{M_F^2}{M_S^2}\Big) 
		+ \big(|\lambda^1_\Phi|^2 - |\bar\lambda^1_\Phi|^2\big) \F_2\Big(\tfrac{M_F^2}{M_S^2}\Big)  \bigg\} \cdot 3.3 \times 10^{-13}
	\end{align}
	Compared to the $\tau$ contribution Eq.~\eqref{eq:2HDM-amu-numerical} this is enhanced roughly by a factor
	of $500(\lambda_1^2-6\bar\lambda_1^2)\cos^2\beta$. However, because of the additional factor of 
	$\cos^2\beta$ together with the theoretical constrains on the Yukawa couplings discussed in the previous section,
	the VLF Barr-Zee contribution is still smaller than that from the top-loop. Additionally, if $\bar\lambda^1_\Phi$ is large enough
	the contribution will even be negative.
	
	The situation changes once mixing between the VLL and the muon is permitted. In this case, both the VLF and new scalars will 
	contribute to $\Damu$ already at one-loop order, resulting in a large effect from both the chiral and $\tan\beta$ enhancement
	\cite{Dermisek:2020cod,Dermisek:2021ajd}.
	The resulting correction can be obtained directly from the generic formulas in Sec.~\ref{sec:genericoneloop}. In particular,
	to discuss the qualitative behaviour we can consider the simple case where $M_F\sim M_S$, in this case the contributions to $\Damu$ reduce to
	\cite{Dermisek:2021ajd}
	\begin{align}
		\Damu^{\text{VLF,1L}} \simeq -\frac{m_\mu v}{16\pi^2} \frac{\lambda_{L\mu}^1\bar\lambda^1_\Phi\lambda_{R\mu}^1}{\sqrt2 M_E M_L} \Big(1 + \tan^2\beta\Big)\cos^3\beta.
	\end{align}
	Here the terms enhanced by $\tan^2\beta$ stem from the contributions of the VLF and new scalars shown in Fig.~\ref{fig:2HDM+VLL-one-loop}, 
	while the non-enhanced contributions stem from diagrams involving the VLF and the SM Higgs/gauge-bosons which are analogous to Eq.~\eqref{eq:VLL-Damu-enhanced}.
	Consequently, a similar correlation as in Eq.~\eqref{eq:Rmumu-VLL} arises
	\begin{align}
		R_{\mu\mu} \approx \bigg|1 - \frac{0.87}{1+\tan^2\beta} \bigg(\frac{\Damu}{10^{-9}}\bigg)\bigg|^2.
	\end{align}
	Because of the additional $\tan^2\beta$ suppression, a much wider range of $\Damu$ is consistent with the current measurement on $R_{\mu\mu}$.
	Like Eq.~\eqref{eq:Rmumu-VLL} this correlation also generalises to the other VLL combinations with modified coefficients that
	where computed in Ref.~\cite{Dermisek:2023tgq} .
	In conclusion, the combination of VLL and 2HDM allows larger $\Damu$ than either of the models on their own. Conversely, such models
   	are now severely restricted by the new result $\DamuFinal$.
	
\subsection{Muon $g-2$ and neutrino mass}
\label{sec:neutrino_mass_gm2}

In Sec.~\ref{sec:neutrino_masses} we have already explained that neutrino masses and neutrino oscillations are firmly established. Together with dark matter, they constitute the strongest direct evidence for BSM physics, and the origin of neutrino masses remains one of the most pressing open questions in fundamental physics. 
In that section we also introduced and explained a large number of theoretical proposals to generate non-degenerate neutrino masses and a possible correlation between neutrino masses and $\amu$ in the framework of SMEFT. Generically, there does not have to be a strong correlation between neutrino masses or neutrino mass mechanisms with contributions to $\amu$. 

However, there is a subset of motivated neutrino mass models where correlations with $\amu$ and sizeable corrections $\Damu \sim \mathcal{O}(10 ^{-9})$ can arise. These models were sometimes proposed as potential explanations of the previous deviation $\DamuOld$. Now the new result $\DamuFinal$ constitutes a constraint on the parameter space of such models. In this section we present a survey of a set of neutrino mass models of this kind, where sizeable contributions to $\amu$ are in principle possible.

What makes this section particularly complementary to earlier sections is the diverse nature of the relevant BSM ideas. The models involve new scalars, new fermions, and/or new gauge bosons. Some of them are motivated not only by the quest to understand neutrino masses but also by dark matter or the desire to explain all quark and lepton masses simultaneously with neutrino masses. Many involve heavy BSM states, some very light new particles. For this reason the present section will make use of results and insights of many of the previous sections, but at the same time it will mostly limit the discussion to qualitative statements. 

To set the stage, Sec.~\ref{sec:neutrino_masses} has sketched six different well-known neutrino mass models. In all cases the tiny active neutrino masses are explained by heavy BSM particles that can be integrated out. These heavy particles are either fermions or bosons. 
Three of the models are the seesaw type I, II, III models, which generate the neutrino mass terms at tree level. In the Zee, Babu, and Ma (or scotogenic) models the Majorana mass terms are generated via one- or two-loop Feynman diagrams.
The three seesaw models and the loop-generation models are summarised in Tab.~\ref{table:seesaw} and~\ref{table:ZBM}, respectively.
For general reviews of radiative neutrino mass generation we also refer to Refs.~\cite{Cai:2017jrq,Babu:2019mfe,Adolf:2023wfw}.

\subsubsection{Extensions of the scotogenic model}
\label{sec:extenofscotogenicmodel}

We begin with the scotogenic model (or Ma model)~\cite{Ma:2006km} and extensions thereof and focus on five particular extensions that represent the wide range of possibilities for neutrino masses, particle content and $\Damu$ contributions. Tab.~\ref{table:scotoextension} collects the models and their most important properties. The key feature of these models is the simultaneous explanation of neutrino masses and dark matter via the same new states. All such models contain right-handed neutrino singlets $N_i$, where we suppress the chirality subscript $R$.\footnote{%
Throughout this section, we always use a compact notation and identify $N\equiv N_R$ for right-handed fermion singlets.
}
with gauge invariant Majorana mass terms, a scalar doublet $\eta = (\eta ^+, \eta ^0)^T$ which may be assigned lepton number, and they involve a (fundamental or effective) quartic scalar interaction $\boldsymbol{\lambda}_5(\Phi^\dagger\eta)^2$ which breaks lepton number conservation and governs the magnitude of the loop-generated neutrino masses.\footnote{%
The generalisation for an arbitrary number of sterile (right-handed) neutrinos and inert scalar doublets with zero vacuum expectation value has been discussed in Ref.~\cite{Escribano:2020iqq}.} 
The new fields are odd under a $\mathds{Z}_2$ symmetry such that the lightest new state is a stable WIMP dark matter candidate.

The original scotogenic model has been described in Sec.~\ref{sec:neutrino_masses} and approximations for the neutrino masses and $\Damu$ were given. Two simple numerical estimates are
\begin{align}
  \label{eq:scotonmassestimation}
  M_\nu
  &
  \sim
  0.1\,
  \frac{\boldsymbol{\lambda}_5}{M/\text{100 GeV}}
  \left(\frac{\mathbf{h}}{10^{-5}}\right)^2\text{ eV},
  &
  \Damu &
  \sim  10^{-19} \left(\frac{\text{100 GeV}}{M}\right)^2
  \left(\frac{\mathbf{h}}{10^{-5}}\right)^2,
\end{align}
where $M$ is assumed to be the common mass scale of the new states. Hence the contributions to $\amu$ are extremely tiny if the sub-eV neutrino masses are explained with $\boldsymbol{\lambda}_5 \sim 1$. For very small $\boldsymbol{\lambda}_5$, the Yukawa couplings $\mathbf{h}$ and the resulting $\Damu$ could be larger, but constraints from lepton-flavour violating observables like $\mu \to e \gamma$ rule out sizeable contributions $\Damu\sim10^{-9}$. Models S1-3 contain tree-level quartic coupling terms whereas $\boldsymbol{\lambda}_5$ is induced at the one- or 
two-loop level in S4 and S5 respectively.

\begin{table}
  \centering
  \begin{tabular}{| c | c | c l | c l | c |}
    \hline
    \multirow{2}{*}{\textbf{Model}} & \multirow{2}{*}{\makecell{\textbf{Symmetry}\\\textbf{Groups}}} & \multicolumn{4}{c|}{\textbf{New Fields}} & \multirow{2}{*}{\textbf{Comments}} \\\cline{3-6}
    & & \multicolumn{2}{c|}{Fermions} & \multicolumn{2}{c|}{Scalars} & \\\hline\hline  
    S1 	& $\mathcal{G}_\text{EW}; \mathds{Z}_2$ & $N_{1,2,3}$ & $(\bm1,0;-)$ 
    & \makecell[tc]{$\eta$ \\ $h^+$} & \makecell[tl]{$(\bm2,\frac{1}{2};-)$ \\ $(\bm1,1;-)$} & 
    \makecell[tp{6cm}]{Scotogenic-Zee model \\  chirally enhanced $\Damu$~\cite{Dcruz:2022dao}.}\\\hline
    S2 	& \makecell[tc]{$\mathcal{G}_\text{EW};~\mathds{Z}_2,$ \\ $\text{U}(1)_{L_\mu-L_\tau}$} 
    & \makecell[ct]{$N_{e}$\\$N_{\mu}$\\$N_{\tau}$} & \makecell[lt]{$(\bm1,0;-,0)$\\$(\bm1,0;-,+1)$\\$(\bm1,0;-,-1)$} 
    & \makecell[ct]{$\eta$\\$\Phi_1$\\$\Phi_2$} & \makecell[lt]{$(\bm2,\frac{1}{2};-,0)$\\$(\bm1,0;+,1)$\\$(\bm1,0;+,2)$} &
    \makecell*[tp{6cm}]{Scotogenic  U(1)$_{L_\mu -L_\tau}$
      model,\\ $Z_{\mu\tau}$- and
      $\eta^+$-contributions to $\amu$.\\ $\boldsymbol{\lambda}_5 \ge 10 ^{-8}$ favoured by CLFV~\cite{Borah:2021khc}.} \\\hline
    S3 	& \makecell[tc]{$\mathcal{G}_\text{EW};~\mathds{Z}_2,$ \\ $\text{U}(1)_{L_\mu-L_\tau}$} 
    & \makecell[ct]{$N_{e}$\\$N_{\mu}$\\$N_{\tau}$ \\ $L'$} & \makecell[lt]{$(\bm1,0;-,0)$\\$(\bm1,0;-,+1)$\\$(\bm1,0;-,-1)$ \\ $(\bm2,-\frac{1}{2};-,0)$} 
    & \makecell[ct]{$\eta$\\$\varphi$\\$\chi^-$} & \makecell[lt]{$(\bm2,\frac{1}{2};-,0)$\\$(\bm1,0;+,1)$\\$(\bm1,-1;-,1)$} &
    \makecell[tp{6cm}]{scotogenic  U(1)$_{L_\mu -L_\tau}$ + VLL.\\ $\DamuOld$ obtained for $\mathbf{g}_\mu \sim 1.5$, $\mathbf{h} \sim -1$ and $M_{\chi ^\pm} \sim 160$ GeV~\cite{Kang:2021jmi}.} \\\hline
    S4 	& \makecell[tc]{$\mathcal{G}_\text{EW};~\mathds{Z}_2,$ \\ $\text{U}(1)_{\chi}$}  
    & \makecell[tc]{$N_{1,2,3}$ \\ $N_{0}$ \\ $X$} & \makecell[tl]{$(\bm1,0;-,1)$ \\ $(\bm1,0;+,2)$ \\ $(\bm2,-\frac{1}{2};-,1)$} 
    & $H_I$ & $(\bm2,\frac{1}{2};-,1)$ &  
    \makecell[tp{6cm}]{Two-loop generation of neutrino mass via
      vector-like fermion; chirally enhanced $\Damu$~\cite{Chen:2020ptg}.} \\\hline
    S5 	& \makecell[tc]{$\mathcal{G}_\text{EW};~\mathds{Z}_2,$ \\ $\text{U(1)$^\prime$}$}  
    & $N_{1,2}$ & $(\bm1,0;-,0)$
    & \makecell[tc]{$\eta$ \\ $\varphi$ \\ $\rho$ \\ $\zeta$ \\ $\sigma$} & \makecell[tl]{$(\bm2,\frac{1}{2};-,3)$ \\ $(\bm1,0;-,3)$ \\ $(\bm1,0;-,-1)$ \\ $(\bm1,0;-,0)$ \\ $(\bm1,0;+,\frac{1}{2})$} &  
    \makecell[tp{6cm}]{Three-loop generation of neutrino mass via
      extended Inert Doublet model~\cite{Abada:2022dvm}, non-enhanced
      $\Damu$ like in minimal scotogenic model.} \\\hline
  \end{tabular}
  \caption{List of extended scotogenic models discussed in this
    section, all of which explain neutrino masses and dark matter via
    the same new fields. Again, we identify $N \equiv N_R$ for right-handed
    fermion singlets. $\mathcal{G}_\text{EW}=\GEW$. All fields here
    are singlets under $\SUc$. In S2 and S3, U$(1)_{L_\mu-L_\tau}$ is
    an additional gauge group, while $\mathds{Z}_2$, U$(1)_\chi$ and
    U(1)$^\prime$ are only global symmetries of the respective
    models. The order of representation labels after each field corresponds to the order of the symmetry groups. }
  \label{table:scotoextension}
\end{table}
 
The model S1 in Tab.~\ref{table:scotoextension} is called the scotogenic-Zee model~\cite{Dcruz:2022dao} and is the combination of two one-loop neutrino mass generation models. It combines the scalar doublet $\eta$ and three singlet right-handed neutrinos  $N_{1,2,3}$ of the scotogenic model with one charged  $\mathds{Z}_2$-odd scalar singlet $h^+$ as in the Zee model. 
The neutrino masses are generated as in the scotogenic model by the $\mathds{Z}_2$-odd scalar doublet $\eta$, while $\Damu$ contributions arise from the charged scalar loops involving states that are mixtures of $\eta ^+$ and $h^+$.
In terms of mass-insertion diagrams, the contributions to $M_\nu$ and $\amu$ in this model can be approximated as 
\begin{align}
  \label{eq:gm2scotoZee}
  M_\nu^{\text{S1}} &\sim
  \begin{gathered}
    \includegraphics[scale=.8]{{Diagrams/fig.nmass.Scotogenic.pdf}}
  \end{gathered} \,,
  &
  \Damu^{\text{S1}} & \sim
  \begin{gathered}
    \includegraphics[scale=.8]{{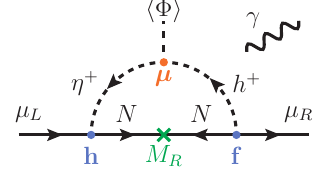}}\,,
  \end{gathered}
\end{align}
where the parameters and the external Higgs \vev insertions are explicitly shown.
The Feynman diagrams in~\eqref{eq:gm2scotoZee} are derived from the triple-scalar interaction term $\boldsymbol{\mu}\tilde{\Phi} ^\dagger \eta h^-$ and the Yukawa interaction terms involving the right-chiral neutrinos like $\mathbf{h}^\dagger_{ij} {\overline{l_{Li}}} \tilde{\eta} N_{j}$ and $\mathbf{f}_{ij} {\overline{e _{Ri}}} h ^- N_{j} ^C$, as well as the typical scotogenic $\boldsymbol{\lambda}_5$ interaction term.

In this model the $\Damu$ contributions are chirally enhanced and resemble the three-field model of Class III in Eq.~\eqref{eq:MIAcontribution}, where the fermion mass insertion $m_\psi$ is the Majorana neutrino mass term $M_R$ and the scalar mixing term $a_\Phi$ corresponds to the trilinear $\eta$--$h^+$--Higgs interaction term. 
In this way the sign of $\Damu$ depends on the relative signs of the model parameters and can be positive or negative, and significant enhancements of $\Damu$ are possible. At the same time the neutrino masses depend on the additional lepton-number violating parameter $\boldsymbol{\lambda}_5$ which can be naturally small.
Ref.~\cite{Dcruz:2022dao} finds indeed that this scenario can describe the neutrino masses and mixing, satisfy bounds from lepton flavour violation, but produce sizeable $\Damu$ as large as $\DamuOld$. Such scenarios are now constrained by the new result $\DamuFinal$ in a non-trivial way.

The two scotogenic extensions S2 and S3 of Tab.~\ref{table:scotoextension} both involve an additional gauge group and associated $Z'$ gauge boson. The chosen gauge group in both cases is U(1)$_{L_\mu - L_\tau}$ where the $Q_{L_\mu -L_\tau}$ charges are $(0,1,-1)$ for both left- and right-handed $(e, \mu, \tau)$ and also for the corresponding neutrinos, including the right-handed neutrinos $N_{e,\mu,\tau}$.
This gauge group has already been discussed in Sec.~\ref{sec:LmuMinusLtau} and it is one of the possible anomaly-free additional U(1) gauge groups and is compatible with a $Z'\equiv Z_{\mu\tau}$ in the MeV--GeV mass region.

For neutrino-mass generation the gauge group is of particular interest since it imposes particular flavour patterns on the Majorana neutrino mass terms, such that $N_eN_e$ and $N_\mu N_\tau$ are allowed while other combinations are forbidden.
Apart from these restrictions, the neutrino masses arise similarly to the original scotogenic model in Eqs.~\eqref{eq:scotonmassestimation} and~\eqref{eq:cdim5ZBscoto}, but the number of particles running inside the loop and their masses vary. The new Yukawa couplings and quartic couplings of the scalar fields play a crucial role.  

Model S2 from Ref.~\cite{Borah:2021khc} contains, apart from the scotogenic field content,  two neutral scalar singlets which have different $U(1)_{L_\mu -L_\tau}$ charges, are even under $\mathds{Z}_2$ and have non-vanishing vacuum expectation values, $\langle \Phi_1 \rangle = v_1$, $\langle \Phi_2 \rangle = v_2$. These break $U(1)_{L_\mu - L_\tau}$ symmetry and the gauge boson gains the mass $M_{Z_{\mu\tau}} = g_{\mu\tau}\sqrt{2(v_1 ^2 + 4 v_2 ^2)}$. The neutrino mass takes the typical scotogenic neutrino mass form given in Eq.~\eqref{eq:cdim5ZBscoto}. In contrast to model S1, in this model S2 the field quantum numbers prevent mass mixing of any of the new fields via couplings to the SM Higgs field. Hence there are no chirally enhanced contributions to $\amu$. The dominant $\Damu$ contributions arise from the Feynman diagram with the $Z_{\mu\tau}$ and/or from the diagram with right-handed neutrinos and charged scalars, schematically
\begin{align}
  \label{eq:gm2ScotoLmutau}
  \Damu^{\text{S2}} &\sim
  \begin{gathered}
    \includegraphics[scale=.8]{{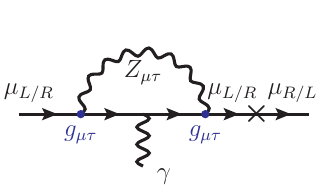}}
  \end{gathered}
  \quad+
    \quad
  \begin{gathered}
    \includegraphics[scale=.8]{{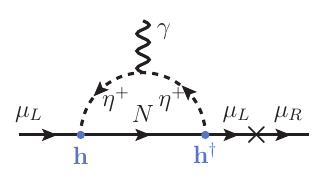}}\,,
  \end{gathered}
\end{align}
where sample assignments of chiralities and the non-enhanced chirality flips at the external lines have also been indicated. The $Z_{\mu\tau}$ contribution is positive and behaves as discussed in Sec.~\ref{sec:LmuMinusLtau}; the FS-type contribution is negative and behaves as the non-enhanced two-field models of Sec.~\ref{sec:MinimalBSM} and Tab.~\ref{tab:twofieldmodels}. 
For this reason the model is automatically in agreement with the new result $\DamuFinal$ except in the small $Z_{\mu\tau}$ parameter region disfavoured in the plot in Fig.~\ref{fig:LmuminusLtau}, see also Eq.~\eqref{DamuLmuLtaumax}.

Model S3~\cite{Kang:2021jmi} realises the U(1)$_{L_\mu-L_\tau}$ symmetry differently. In addition to the scotogenic field content it adds a vector-like fermion $L'$, a charged scalar $\chi ^-$ and one neutral scalar field with nonzero \vev, $\langle\varphi \rangle \equiv \frac{v_\varphi}{\sqrt{2}}$ that gives mass to the $Z_{\mu\tau}$.
All additional fields are odd under $\mathds{Z}_2$ except $\varphi$. The neutrino mass is generated in the same way as in the scotogenic model, though there are now five instead of three  neutral right-handed states. These are three singlet right-handed neutrinos and two new neutral right-handed fields from the vector-like fermions ${L'_L} ^C$ and $L'_R$. The $5\times5$ mixing is constrained by the U(1)$_{L_\mu - L_\tau}$-charges of the fields and results in mass eigenvalues $M_k$.
In this model S3, mass mixing via the SM Higgs \vev\ is possible between specific fermions, e.g.\ between $N_e$ and $L'$, such that contributions to $\amu$ similar to the Class I/II three-field models of Sec.~\ref{sec:genericthreefield} are possible, see also Eq.~\eqref{eq:MIAcontribution}. The corresponding mass-insertion diagram is 
\begin{align}
  \label{eq:gm2ScotoLmutauVLL}
  \Damu^{\text{S3}}
  &\sim
  \begin{gathered}
    \includegraphics[scale=.8]{{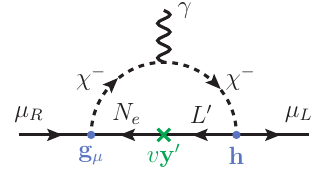}}\,,
  \end{gathered}
\end{align}
where the mass insertion is explicitly indicated.
As discussed in Ref.~\cite{Kang:2021jmi}, these contributions can be sizeable and easily dominate over the non-chirally enhanced contributions from e.g.~the $Z_{\mu\tau}$-loop. This reference identifies parameter regions where dark matter and neutrino masses are explained and bounds from CLFV are avoided. For normal hierarchy of neutrino masses, $\Damu$ could be as large as $\mathcal{O}(10^{-9})$. Hence the parameter regions with normal ordering are now constrained by the new result $\DamuFinal$. For inverted hierarchy of neutrino masses, Ref.~\cite{Kang:2021jmi}  finds that $\Damu$ is at most of order $\mathcal{O}(10^{-12})$. Hence this case is not constrained by $\DamuFinal$ but automatically compatible with it.

The extended scotogenic models S4 and S5 do not involve an extra gauge
group but represent a different idea. Since the neutrino masses are
proportional to the quartic scalar coupling
$\boldsymbol{\lambda}_5(\Phi^\dagger\eta)^2$, their smallness might be
related to the fact that this quartic scalar is forbidden  at tree
level but allowed only via loops.\footnote{%
A different way to generate neutrino masses at the multiloop level is
discussed in Ref.~\cite{Hernandez:2021uxx}, also in connection with $\amu$. In that
reference, the neutrino masses are generated similarly to the
scotogenic mechanism, however the right-handed neutrino Majorana
masses are generated via one-loop diagrams. This leads to the so-called
inverse seesaw mechanism. The model considered in Ref.~\cite{Hernandez:2021uxx} is
based on left-right symmetry and also contains vector-like leptons
that can lead to sizeable corrections to $\amu$.
}
In model S4, the quartic scalar coupling arises at one-loop order, resulting in two-loop neutrino masses; in model S5, the quartic scalar coupling arises at two-loop order and the neutrino masses are therefore three-loop suppressed.
These models are thus well motivated by neutrino mass considerations, but the fields appearing in the loop generating the effective quartic scalar interactions might also enter diagrams that contribute to $\amu$.

Model S4 has been defined in Ref.~\cite{Chen:2020ptg}, and a representative diagram corresponding to the neutrino mass dimension-5 operator is given by
\begin{align}
  \label{eq:twoloopnmass}
  M_\nu^{\text{S4}}
  &\sim
  \begin{gathered}
      \includegraphics[scale=.8]{{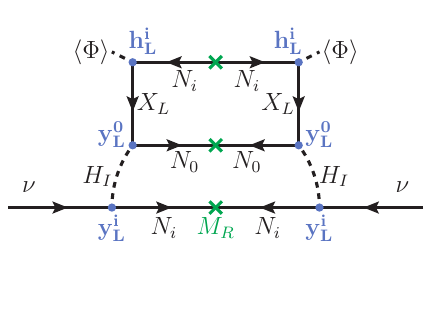}}\,,
  \end{gathered}
\end{align}
where again the Higgs \vev insertions, the field assignments and the
relevant model parameters are explicitly shown and the crosses denote fermion mass term insertions.
In this model the typical scotogenic inert doublet is called $H_I$, and the model contains an additional $\mathds{Z}_2$-even fermion singlet $N_0$ and a $\mathds{Z}_2$-odd vector-like doublet $X$. The couplings are constrained by a global U(1)$_\chi$ symmetry that forbids a fundamental $(\Phi^\dagger H_I)^2$ vertex. However, this symmetry is softly broken such that the effective quartic scalar coupling that is comparable to $\boldsymbol{\lambda}_5$  is generated via a fermion loop composed of $X_L ^0$, $N_i$ and $N_0$ and the effective coupling of $(\Phi^\dagger H_I)^2$ is governed by the product of the four Yukawa-like couplings and the two fermion mass insertions. With this one-loop effective coupling, the two-loop diagram in Eq.~\eqref{eq:twoloopnmass} can generate non-zero neutrino masses.

The fields generating the effective $\boldsymbol{\lambda}_5$ coupling have quantum numbers which simultaneously allow chirally enhanced contributions to $\amu$. Specifically the singlets $N_i$ and the doublet $X$ can mix via the SM Higgs \vev. Thus the $\Damu$ contributions resemble the ones of the Class I/II three-field models discussed in Sec.~\ref{sec:genericthreefield}, like in case of model S3.
A representative mass-insertion diagram for $\Damu$ in this model is
\begin{align}
  \label{eq:gm2twoloopnmass}
  \Damu^{\text{S4}}
  &\sim
  \begin{gathered}
    \includegraphics[scale=.8]{{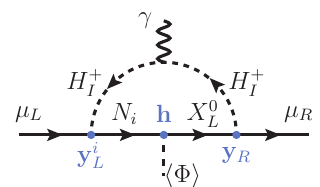}}\,.
  \end{gathered}
\end{align}
The model can accommodate the measured neutrino data and dark matter for masses in the range
100\ldots1000 GeV and couplings of the order $10^{-2\ldots-1}$.
For such parameter choices, Ref.~\cite{Kang:2021jmi} finds values of $\Damu$ between zero and about $30\times10^{-10}$. Hence the model is now significantly constrained by $\DamuFinal$.

Model S5 was investigated in Ref.~\cite{Abada:2022dvm}. The model contains one less right-handed neutrino compared to the original scotogenic model, but it adds four additional scalar singlets. It also requires an additional global U(1)$^\prime$ symmetry under which  the SM quarks and leptons have charges $\frac{1}{3}$ and $-3$, respectively.
The U(1)$^\prime$ charges of the additional fields are specified in Tab.~\ref{table:scotoextension}; the new $\sigma$ scalar acquires a \vev and breaks U(1)$^\prime$ spontaneously. Thanks to this symmetry and the charge assignments, the neutrino masses can only arise at the three-loop level.
An example three-loop mass-insertion diagram generating neutrino masses is 
\begin{align}
  \label{eq:gm2threeloopnmass}
  M_\nu^{\text{S5}}
  &\sim
  \begin{gathered}
    \includegraphics[scale=.8]{{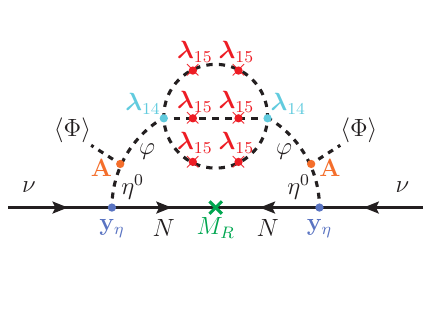}}
  \end{gathered}
\end{align}
where the Higgs \vev{}s  and the parameters are shown but some field
labels are suppressed. For further discussions focusing on the
neutrino phenomenology in models with three-loop neutrino
mass generation we refer to
Refs.~\cite{Krauss:2002px,Aoki:2008av,Kajiyama:2013lja,Ahriche:2014cda,Chen:2014ska,Jin:2014glp,Hernandez:2021iss}.

Important Lagrangian terms and their coupling constants are the triple coupling term $\mathbf{A}\eta ^\dagger \Phi\varphi$ and two quartic coupling terms $\boldsymbol{\lambda}_{14}\varphi \rho ^3$ and $\boldsymbol{\lambda}_{15} \rho\zeta\sigma ^2 $.
The $\boldsymbol{\lambda}_{15}$ coupling leads to $\rho$--$\zeta$ mass
insertions proportional to
$\boldsymbol{\lambda}_{15}\langle\sigma\rangle^2$ shown in the
three-loop diagram as red crosses, and the $\mathbf{A}$ coupling leads to two couplings to the SM Higgs \vev in the diagram and thus to the generation of the dimension-5 neutrino mass operator.

Despite the mass mixing of $\eta$--$\varphi$ via the SM Higgs \vev, there are no chirally enhanced contributions to $\amu$ in this model because the scalar field $\varphi$ cannot couple to the right-handed muon. For this reason, the contribution to $\amu$ is similar to the one in the original scotogenic model, i.e.\ it is negative and has a small magnitude which is essentially automatically in agreement with the current $\DamuFinal$. Ref.~\cite{Abada:2022dvm} studies further extensions of the model by vector-like leptons. With such additional particles, the contributions to $\amu$ could be larger, however the current result  $\DamuFinal$ does not provide a motivation for such extensions.
Rather, the model S5 exemplifies the idea that the neutrino masses could be small because of fundamental symmetries that prevent lower-order diagrams and lead to very strong loop suppressions of neutrino masses. In the model, the mass scale of the relevant particles can be $\O(1\text{ TeV})$ while the relevant couplings are unsuppressed and CLFV constraints are avoided, and dark matter can be explained via the new fermions or new scalars.

\subsubsection{Neutrino mass models based on other mechanisms}

\begin{table}
	\centering
	\begin{tabular}{| c | c | c l | c l | p{7cm} |}
		\hline
		\multirow{2}{*}{\textbf{Model}} & \multirow{2}{*}{\makecell{\textbf{Symmetry}\\\textbf{Groups}}} & \multicolumn{4}{c|}{\textbf{New Fields}} & \multirow{2}{*}{\textbf{Comments}} \\\cline{3-6}
		& & \multicolumn{2}{c|}{Fermions} & \multicolumn{2}{c|}{Scalars} & \\\hline\hline  
		E1 	& $\mathcal{G}_\text{EW}$ &  & 
		& \makecell[tc]{$h^+$ \\ $\Phi_2$} & \makecell[tl]{$(\bm1,1)$ \\ $(\bm2,\frac{1}{2})$} & 
		\makecell[tp{7cm}]{Zee model, see
                  Sec.~\ref{sec:neutrino_masses},
                  Eq.~\eqref{LZee}. $\Damu$ as in non-flavour-aligned 2HDM.} \\\hline
		E2 	& $\mathcal{G}_\text{SM};~\mathds{Z}_2$
		& $N$ & $(\bm1,\bm1,0;-)$
		& \makecell[ct]{$\eta_{1,2}$\\$S$\\$\phi$} & \makecell[lt]{$(\bm1,\bm2,\frac{1}{2};-)$\\$(\bm3,\bm2,\frac{1}{6};-)$\\$(\bm1,\bm1,-1;-)$} &
		\makecell[tp{7cm}]{Similar to scotogenic-Zee model
                  plus leptoquarks (LQ); neutrino mass and dark matter as in scotogenic,
                  $\Damu$ as in scotogenic-Zee, LQ could
                  accommodate $B$-anomalies~\cite{Cepedello:2022xgb}.} \\\hline
		E3 	& $\mathcal{G}_\text{SM}$
		&  & 
		& \makecell[ct]{$S_1$\\$\tilde{R}_2$\\$S_3$} & \makecell[lt]{$(\bar{\bm3},\bm1,\frac{1}{3})$\\$(\bm3,\bm2,\frac{1}{6})$\\$(\bar{\bm3},\bm3,\frac{1}{3})$} &
		\makecell[tp{7cm}]{Extension by three LQs,
                  neutrino mass via LQ loop, chiral
                  enhancement for $\Damu$ via LQ/quark loops~\cite{Chen:2022hle}.} \\\hline
		E4 	&$\mathcal{G}_\text{EW};~\mathds{Z}_2$
		& $F_{1,2,3}$ & $(\bm1,-1;-)$
		& \makecell[tc]{$\phi_1$ \\ $\phi_2$ \\ $\eta$} & \makecell[lt]{$(\bm2,\frac{1}{2};-)$ \\ $(\bm2,\frac{3}{2};-)$ \\ $(\bm1,0;-)$} &  
		\makecell[tp{7cm}]{Neutrino mass via charged
                  vector-like fermions $F_i$, DM candidate and chiral
                  enhancement for $\Damu$ via new scalars~\cite{Jana:2020joi}.}\\\hline
		E5 	& $\mathcal{G}_\text{EW};~\mathds{Z}_2$
		& $N_{1,2}$ & $(\bm1,0;-)$
		& \makecell[ct]{$\varphi^+$ \\ $\Phi_2$} & \makecell[ct]{$(\bm1,1;-)$ \\ $(\bm2,\frac{1}{2};-)$} &  
		\makecell[tp{7cm}]{$\nu$-philic 2HDM (or seesaw type IB)~\cite{Cherchiglia:2023utd},
		one-loop Majorana neutrino mass generation as in
                scotogenic model,
                chiral enhancement for $\amu$ as in E4 or S1.}\\\hline
	\end{tabular}
	\caption{List of non-scotogenic models with radiative neutrino mass generation discussed in the text. We identify $N\equiv N_R$ for right-handed fermion singlets. $\mathcal{G}_\text{EW}$ denotes the EW gauge group as in Tab.~\ref{table:scotoextension} and $\mathcal{G}_\text{SM}$ the full SM gauge group, and $\Phi$ is the SM Higgs doublet. The representation label after each field corresponds to the order of the symmetry groups.}
	\label{table:moreextension}
\end{table}

Here we illustrate the wide range of possible neutrino mass models further. Tab.~\ref{table:moreextension} provides an overview of selected examples, all of which differ in various ways from the examples of Tab.~\ref{table:scotoextension}.

First, model E1 is the original Zee model already discussed in Sec.~\ref{sec:neutrino_masses}. Since this model actually is the 2HDM, further extended by an additional charged scalar singlet, the phenomenology of $\amu$ can be obtained from studying the 2HDM. However, because of the desire to explain neutrino masses, the preferred and typically considered values of the Yukawa matrices by which the quarks and leptons couple to the second Higgs doublet are very different in the Zee model than in the usually considered 2HDM.  Ref.~\cite{Barman:2021xeq} has carried out an extensive study of $\amu$ in the Zee model, taking into account constraints from neutrino masses, lepton flavour violation and from collider searches. The investigated textures of the Yukawa matrices in generation space are very different from the ones of e.g.~the flavour-aligned 2HDM of Sec.~\ref{sec:FA2HDMpheno}. Instead, the authors assume
that the second Higgs doublet only couples to leptons but with different strengths to the electron, muon and tau, and lepton-flavour violating Yukawa couplings are allowed and needed to reproduce the neutrino masses. With such patterns, sizeable $\Damu$ contributions are possible~\cite{Barman:2021xeq}.

Models E2 and E3 in Tab.~\ref{table:moreextension} are models with leptoquarks and thus contain coloured BSM particles. The structure of the two models is quite different, though. Model E2~\cite{Cepedello:2022xgb} may still be regarded as an extended scotogenic model, where the $\mathds{Z}_2$-odd sector is enlarged substantially to a dark sector. The dark sector contains one right-handed neutrino (instead of three) and two inert scalar doublets (instead of
one), and in addition it contains a scalar singlet $\phi$ and a
doublet leptoquark $S$ with hypercharge $\frac{1}{6}$ which is
identical to $\tilde{R}_2$ in model E3. Apart from the leptoquark it
also has similarities to the scotogenic-Zee model S1. In
Ref.~\cite{Cepedello:2022xgb} the model was motivated as a potential
explanation of a set of several hints for BSM physics from $B$-physics
and from $\amu$ that were present around 2020--2022 as already
mentioned in Sec.~\ref{sec:LQflavour}. Though these hints have
substantially reduced in the meantime it may be of interest to discuss
model E2 as a representative example trying to accommodate neutrino
masses, dark matter, and large effects in $\Damu$ and $B$-physics at
the same time.

In model E2, the neutrino masses are explained as in the scotogenic or the scotogenic-Zee models, while $\Damu$ takes the same form as in the scotogenic-Zee model S1. The leptoquark plays a role for $B$-physics observables, and dark matter can be explained by the lightest $\mathds{Z}_2$-odd particle which may be $N$ or, more favourably, one of the neutral $\eta_a$-components.
The fact that the model contains such a large number of new particles is a reflection of the no-go theorem of Ref.~\cite{Arcadi:2021cwg} also mentioned in Sec.~\ref{sec:LQflavour} which states that at least four BSM fields are required to accommodate dark matter, sizeable $\Damu$ and the $B$-physics anomalies of the time.

Model E3~\cite{Chen:2022hle} involves three different leptoquarks, one singlet $S_1$, one doublet $\tilde{R}_2$ and one triplet $S_3$. Its key feature is that it does not introduce any other, non-leptoquark particles. Hence the neutrino masses are explained in a different way, via leptoquark--quark loops.  A sample mass-insertion diagram is
\begin{align}
  \label{eq:nmassLQ}
  M_\nu^{\text{E3}}
  &\sim
  \begin{gathered}
    \includegraphics[scale=.8]{{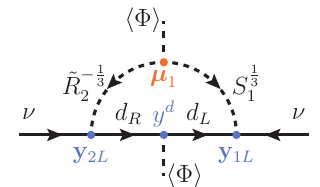}}\,,
  \end{gathered}
\end{align}
which contains leptoquarks $S_1^{\frac13}$ and $\tilde{R}_2^{-\frac13}$ with electric charge $\pm\frac13$.
The loop diagrams generating $M_\nu$ must involve lepton-number violating mass mixing of the different leptoquarks types  induced by couplings to the SM Higgs \vev, either $S_1 - {\tilde{R}}_2$ as shown in Eq.~\eqref{eq:nmassLQ} and/or $S_3 -{\tilde{R}}_2$. A second Higgs insertion happens at the down-quark line, such that  Eq.~\eqref{eq:nmassLQ} corresponds to neutrino mass generation via the dimension-5 operator in Eq.~\eqref{eq:nmassDim5}, like in the other neutrino mass models.
This model E3 has already been discussed in Sec.~\ref{sec:LQflavour}, together with a variety of further models with similar goals where leptoquarks were used to explain flavour anomalies. Similarly to many other such models, model E3 can contribute to $\amu$ particularly via the leptoquark $S_1$, which leads to chirally enhanced contributions. These contributions are now constrained by $\DamuFinal$.

Model E4 is a model without leptoquarks and without right-handed neutrinos. It instead extends the fermion sector by three charged vector-like leptons $F_i$ and the scalar sector by two scalar doublets $\phi_{1,2}$ with different hypercharges and one neutral scalar singlet $\eta_3$. The model is one of the examples mentioned in Sec.~\ref{sec:VLF+scalar} where vector-like fermions with extended scalar sectors can lead to appealing phenomenology including explanations of neutrino masses and dark matter. 

In model E4 specifically, the dark matter candidate is stable thanks to the unbroken $\mathds{Z}_2$ symmetry, like in the scotogenic model.
The neutrino mass is generated from loop diagrams of the form
\begin{align}
  \label{eq:nmassVLLF}
  M _\nu^{\text{E4}}
  &\sim
  \begin{gathered}
    \includegraphics[scale=.8]{{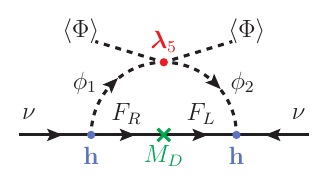}}\,,
  \end{gathered}
\end{align}
which contains the vector-like leptons $F_i$ and the two scalar doublets $\phi_{1,2}$ inside the loop. Here $\phi_1$ couples with the SM lepton doublet $l_L$ and  $F_R$, while $\phi_2$ couples with $l_L$ and $F_L^C$. In the loop generating the neutrino-masses, the $F_i$ undergo a chirality flip via their Dirac mass, and the $\phi_1$--$\phi_2$ transition is governed by a quartic scalar coupling between $\phi_{1,2}$ and the SM Higgs field of the form $\phi_1^\dagger\tilde{\Phi} \Phi^\dagger\phi_2$, which is analogous to the $\boldsymbol{\lambda}_5$ coupling in the scotogenic model and which may be considered lepton-number violating. The entire structure of this loop diagram is thus similar to the one in the scotogenic model shown in Fig.~\ref{fig:nmassScoto}, except for the charges of the involved fields.

The singlet scalar $\eta_3$ in the model can couple with the SM lepton singlet $e_R$ and the left-chiral vector-like singlet fermion $F_L$, and  a mass mixing of $\eta_3$--$\phi_1$ via a coupling to the SM Higgs \vev is allowed by the quantum numbers. In this way, the contributions to $\amu$ are dominated by the diagram
\begin{align}
  \label{eq:gm2nmassVLLF}
  \Damu^{\text{E4}}
  &\sim
  \begin{gathered}
    \includegraphics[scale=.8]{{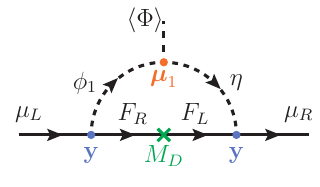}}
  \end{gathered}
\end{align}
and are thus chirally enhanced. The structure of this diagram and the enhancement is analogous to the one in the Class III three-field model, see Eq.~\eqref{eq:MIAcontribution}. 
Ref.~\cite{Jana:2020joi} shows that the model is representative of a wider class of models explaining the neutrino masses and dark matter. In each such model, there is a $\mathds{Z}_2$-odd dark  sector comprised  of charged vector-like fermions and at least three different scalars. Two of the scalars appear in the neutrino-mass loop, and another set of two scalars appears in the $\amu$-loop. It is possible to accommodate neutrino masses and dark matter for BSM masses in the 100\ldots1000 GeV region. Now, the new result $\DamuFinal$ restricts the values of the chirality-flipping parameters.

With model E5, we return to the simpler case of tree-level generation of neutrino masses. This model is an example of several variations of the original seesaw type I mechanism studied in Ref.~\cite{Cherchiglia:2023utd}, which differ in the additional field content and the associated symmetries. The basic idea is that two Higgs doublets  $\Phi_{1,2}$  with non-vanishing \vev{}s exist, where $\Phi_2$ is SM-like but the neutrino masses can only arise due to coupling to $\Phi_1$, thanks to certain discrete symmetries. Our choice E5 corresponds to the neutrinophilic 2HDM considered in the reference, where the neutrino masses are  given by
\begin{align}
  \label{eq:nmassnu2HDM}
  M _\nu^{\text{E5}}
  &\sim
  \begin{gathered}
    \includegraphics[scale=.8]{{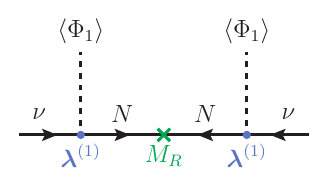}}
  \end{gathered}
\end{align}
and are therefore proportional to $\langle\Phi_1\rangle^2$. In contrast, all other SM fermion masses arise from couplings to the SM-like $\langle\Phi_2\rangle$. In another variant called type IB the neutrino masses are proportional to $\langle\Phi_1\rangle\langle\Phi_2\rangle$. The smallness of neutrino masses is then related to the smallness of the \vev $\langle\Phi_1\rangle$. 
Similarly to the scotogenic-Zee model S1, a charged singlet scalar $\varphi^+$ is introduced which participates in a triple scalar coupling with the both Higgs doublets. Hence contributions to $\amu$ arise via
\begin{align}
  \label{eq:gm2nmassnu2HDM}
  \Damu^{\text{E5}}
  &\sim
  \begin{gathered}
    \includegraphics[scale=.8]{{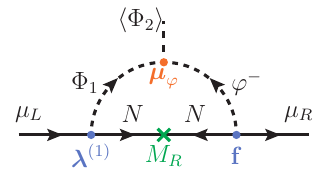}}\,,
  \end{gathered}
\end{align}
which is essentially the same as in models S1 and E4 or in the Class III three-field models discussed before. The chiral enhancement is proportional to the Majorana neutrino mass term and the triple scalar coupling. This model can accommodate neutrino masses with TeV-scale $M_R$ and can at the same time lead to sizeable $\Damu$ via the chiral enhancement. With the current $\DamuFinal$, there is an upper limit on the chiral enhancement.
CLFV constraints can also be important here. As is typical for neutrino mass models, contributions to $\amu$ are inevitably linked to contributions to processes such as $\mu\to e\gamma$, see Sec.~\ref{sec:LeptonDipole} and Eq.~\eqref{eq:mueg-semi-num}. For model E5, the CLFV constraints are strong if $\Damu\sim10^{-9}$. The constraints can be avoided by special regions in parameter space where e.g.~orthogonality conditions between the Yukawa couplings such as $(\mathbf{f}_{1\mu},\mathbf{f}_{2\mu})\propto(\boldsymbol{\lambda}_{2e},-\boldsymbol{\lambda}_{1e})$ hold. If, on the other hand, $\Damu$ is much smaller, as allowed by the current $\DamuFinal$, the CLFV relations between the Yukawa couplings are relaxed. In the type Ib model also investigated in Ref.~\cite{Cherchiglia:2023utd}, it turns out that CLFV constraints entirely exclude sizeable contributions to $\amu$ of the order $10^{-9}$. 

In summary, the previous examples illustrate the great diversity of neutrino mass
models and the richness of potential extensions of the SM,
with new gauge bosons, new vector-like or chiral fermions, and/or new
scalars, including leptoquarks. The models
S1--S5, and E2, E4 can explain dark matter via some of the new
fermions or bosons. The majority of the discussed models lead to
chirally enhanced contributions to 
$\amu$ often of the types discussed in previous sections; model S2 can
contribute to $\amu$ via a light $Z_{\mu\tau}$ 
gauge boson. All of the discussed models explain neutrino masses via 
loops, some at the multi-loop level; they thus naturally allow smaller
mass scales than tree-level neutrino mass models.

All neutrino mass
models predict CLFV effects, although the correlations with $\amu$,
expressed e.g.~via Eq.~\eqref{eq:mueg-semi-num}, are highly
model-specific. In several models, TeV-scale new particles and
sizeable contributions to $\amu$
are compatible with CLFV constraints. Such models are now
non-trivially constrained by the
current $\DamuFinal$. Conversely, the current $\DamuFinal$ is
generally compatible with neutrino mass models that can only produce
small corrections  to
$\amu$, including the more straightforward models discussed in Sec.~\ref{sec:neutrino_masses}.

\section{Summary}

Recently, the Fermilab muon $g-2$ experiment has released its final result that surpasses its design precision and is in full agreement   with earlier measurements, but with a significantly reduced uncertainty. Similarly, the muon $g-2$ Theory Initiative has published an updated SM prediction, however with a shifted value that dramatically transforms the implications of this measurement. After more than 20 years with a tantalising and persisting discrepancy, there is now instead full agreement between SM and experiment, leading to upper limits on possible BSM contributions. Nevertheless, BSM physics must exist, e.g.~in view of dark matter, neutrino masses or other fundamental open questions.

The anomalous magnetic moment of the muon $\amu$ is a key observable for studying BSM physics, with connections to many other observables and relevant effects in a large variety of scenarios. Through significant work in the past years, it is now very well understood how different BSM scenarios can contribute to $\amu$ and how these contributions are related to other, complementary observables. 

The present review aims to provide a comprehensive summary of the impact of $\amu$ for BSM physics and a guide to the literature on the topic. It emphasises model-independent, qualitative and generic results and interrelations to other observables such as dark matter, neutrino masses and collider constraints but also more closely connected observables such as the muon mass, the muon--Higgs coupling and other dipole observables. It also presents quantitative evaluations in specific models, taking into account many constraints. With its brief but self-contained introductions to complementary observables and a range of BSM scenarios, the review may also be useful for researchers interested in BSM phenomenology independently of $\amu$.

Here we summarise a few examples that illustrate the range of  possible results and interconnections.
\begin{itemize}
\item In many models with a light dark sector, there is a
  complementarity between constraints from $\amu$, direct searches and
  other low-energy constraints or even cosmology, see e.g.~Fig.~\ref{fig:ALPsresult}, where $\amu$, $a_e$ and BESIII constraints on the ALPs parameter space are shown. 
\item In SUSY models, there is a strong interplay between $\amu$, LHC
  and dark matter, see e.g.~Fig.~\ref{fig:Binosleptonscompressed}. LHC
  experiments allow   the Bino--slepton and Bino--wino coannihilation
  regions, but dark matter constraints require large Higgsino mass, which enhances $\amu$; the current $\amu$ constraints then limit $\tan\beta$ to be rather small.
\item In leptoquark models with chiral enhancement, $\Damu$ is correlated with contributions to the muon mass and to CLFV processes, see e.g.~Fig.~\ref{fig:leptoquarkresults}. Large contributions to $\amu$ require $\O(1)$ corrections to the muon mass and are only viable in case of highly non-universal flavour patterns of leptoquark couplings. The current small $\DamuFinal$ allows leptoquark parameter regions with less extreme properties.
\item In models with vector-like leptons and seesaw-like tree-level contributions to the muon mass, $\Damu$ is correlated with the muon--Higgs coupling. The combination of current constraints now conclusively rule out several such VLL scenarios, while for other scenarios the parameter space is strongly constrained, see e.g.~Fig.~\ref{fig:VLL-Rmumu-correlation}~\emph{(right)}.
\item In models with radiative muon mass generation, the correlation between $\Delta m_\mu$ and $\Damu$ leads to a fixed contribution to $\amu$, see e.g.~Eq.~(\ref{amuradmmu}), which currently implies a lower mass limit of around 3 TeV on such models.
\end{itemize}
In addition to these examples we also refer to our summaries of many models in Table~\ref{tab:estimates}, the plot in Fig.~\ref{Fig:mass_reach_gm2VScoll}, and the tables in Sec.~\ref{sec:MinimalBSM}. Furthermore, the sections on dark matter and neutrino masses, Secs.~\ref{sec:neutrino_masses}, \ref{sec:DarkMatter} and \ref{sec:neutrino_mass_gm2} discuss many minimal as well as highly elaborate scenarios that show a great diversity in their structure and in their contributions to $\amu$.

Generally, there are examples of BSM scenarios which inevitably predict large $\Damu$ and which are therefore now excluded by $\DamuFinal$. There also exist many BSM models which can accommodate large or small $\Damu$. Especially models with chiral enhancements and/or dark sector particles are often of this kind. The parameter spaces of such models are significantly constrained by $\amu$.

Finally, it is worth noting that several well-known BSM
scenarios are not able to explain deviations as large as $\DamuOld$ in view of constraints related to dark matter,
direct searches, $B$-physics or other observables --- many such
scenarios now re-emerge as viable. Examples  
include minimal SUSY models such as the CMSSM or Higgsino-like LSP dark
matter, the 2HDM of type I or II, 
the minimal dark photon, or many straightforward neutrino mass models.
  More generally, the connection to
CLFV implies that models with generic BSM flavour patterns are only
compatible with $\Damu\lesssim10^{-14}$, which is now viable.

The current agreement between the SM prediction and the final Fermilab measurement may mark a new era for $\amu$ and BSM physics. However, the experimental   precision of $\amu$ is now about four times better than the current SM precision, which hopefully can be further improved. With such an improvement, the constraints on BSM physics might sharpen, or a discrepancy and signal for non-vanishing BSM contributions might
  re-emerge. At the same time, progress on complementary observables
  such as dark matter or CLFV searches and Higgs-coupling measurements
  is expected. The review was organised with these prospects in mind and
  should be a useful resource 
  accompanying future developments.

\section*{Acknowledgments}

We thank our many collaborators on topics related to the muon magnetic
moment and BSM physics, as well as many experimental and theory
colleagues from the Fermilab muon $g-2$ collaboration and the $g-2$
Theory Initiative, for fruitful collaborations, stimulating
discussions and insightful  comments. 
We are particularly grateful to
Rodolfo Capdevilla,
Uwe Hernandez Acosta,
Martin Hoferichter,
Wojciech Kotlarski, 
Gordon Krnjaic, 
Alberto Lusiani,
Stefan M\"uller, 
Ren\'e Reimann, 
Enrico Sessolo and
Adrian Signer
for discussions on specific parts of the review.
KM, DS and HSK acknowledge funding by
DFG grants STO 876/7-2, 10-1,
and by the
EU Horizon 2020 Marie Sk\l{}odowska-Curie RISE Grant Agreement
No.\ 101006726, and thank Fermilab, where
part of this work was carried out, for the hospitality.
The work of PA is supported by NNSFC Key Projects grant No.\ 12335005 and the supporting fund for foreign experts grant wgxz2022021. The Feynman diagrams were created
using Axodraw \cite{Collins:2016aya}.

\appendix

\section{Conventions}\label{sec:Conventions}
Here we collect the conventions used throughout the review.

We use the Minkowski metric tensor with mostly-minus metric $(g^{\mu\nu}) = (g_{\mu\nu}) = \text{diag}(1,-1,-1,-1)$.
We denote 3-vectors by bold symbols, such that 4-vectors are combined as $x^\mu = (x^0, \bm{x})$ and $x_\mu=g_{\mu\nu}x^\nu=(x_0,-\bm{x})$. 
The 4-gradient is defined as $\partial_\mu=(\partial_0,\bm{\nabla})$, such that $\partial_\mu x_\nu = g_{\mu\nu}$. 
The Minkowski-space Fourier transform is given by
\begin{align}
	A(x) = \int \frac{d^4q}{(2\pi)^4} \tilde A(q) e^{-iq\cdot x}.
\end{align}
For convenience we list the Pauli matrices
\begin{align}
	\renewcommand{\arraystretch}{.7}
	\sigma^1 = \begin{pmatrix} 0 & 1 \\ 1 & 0 \end{pmatrix}, \qquad
	\sigma^2 = \begin{pmatrix} 0 & -i \\ i & 0 \end{pmatrix}, \qquad
	\sigma^3 = \begin{pmatrix} 1 & 0 \\ 0 & -1 \end{pmatrix}, \qquad
\end{align}
which fulfil the (anti-)commutator relations $[\sigma^i,\sigma^j] = 2i\epsilon_{ijk}\sigma^k$ and $\{\sigma^i,\sigma^j\}=2\delta^{ij}$.
The $\gamma$-matrices satisfy 
$\{\gamma^\mu,\gamma^\nu\}=2g^{\mu\nu}$, and we use the usual
notation $\slashed{p}\equiv p_\mu \gamma^\mu$. 
For 4-spinors
$\Psi$ we generally use the decomposition $ \Psi=\Psi_L+\Psi_R$ where
$  \Psi_{L,R}  = P_{L,R}\Psi$ with chiral projectors
$  P_{L,R} =\frac12(1\mp\gamma_5)$,  and we denote charge conjugated
spinors as 
$  \Psi^C  = C\bar\Psi^T $ with $  C = i\gamma^0\gamma^2$.
In cases where explicit
representations are needed we use the Weyl (chiral) representation. It 
can be written as
\begin{align}
	\renewcommand{\arraystretch}{.7}
	\gamma^\mu = \begin{pmatrix} 0 & \sigma^\mu \\ \bar\sigma^\mu & 0 \end{pmatrix}, \qquad
	\gamma^5 = i\gamma^0\gamma^1\gamma^2\gamma^3 = \begin{pmatrix} -\mathds{1} & 0 \\ 0 & \mathds{1} \end{pmatrix}
\end{align}
in terms of $\sigma^\mu=(\mathds{1},\bm\sigma)$ and
$\bar\sigma^\mu=(\mathds{1},-\bm\sigma)$.
The 4-spinors $\Psi$ in this representation could be written in terms of
a left-handed Weyl 2-spinor $\psi_{L,\alpha}$  and a right-handed Weyl
2-spinor $\bar\psi_{R}^{\dot\beta}$ such that
$\Psi={\psi_{L,\alpha}\choose\bar\psi_{R}^{\dot\beta}}$, where
the corresponding  charge-conjugated Dirac spinor becomes
$\Psi^C={\psi_{R,\alpha}\choose\bar\psi_{L}^{\dot\beta}}$ and we
follow the notation of e.g.~Ref.~\cite{Belusca-Maito:2023wah} and
refer to Ref.~\cite{Dreiner:2008tw} for a general review of
2-component spinors.
We will however mostly use representation-independent 4-spinor notation.

\begin{table}
	\centering
	\setlength{\Cwidth}{2cm}
	\begin{tabular}{|c||C|C|C|C|C|C|}
		\hline
		\textbf{Standard Model} &\multicolumn{2}{c|}{Leptons} & \multicolumn{3}{c|}{Quarks} & Higgs \\ \hline\hline
		\rule{0pt}{2.5ex} Field name & ${l_{Li}}$ & ${e_{Ri}}$ & ${q_{Li}}$ & ${u_{Ri}}$  & ${d_{Ri}}$& $\Phi$ \\ \hline
		\rule{0pt}{2.5ex} Representation & $(\bm{1},\bm{2},-\frac{1}{2})$ & $(\bm{1},\bm{1},-1)$ & 
		$(\bm{3},\bm{2},\frac{1}{6})$ & $(\bm{3},\bm{1},\frac{2}{3})$ & $(\bm{3},\bm{1},-\frac{1}{3})$ & $(\bm{1},\bm{2},\frac{1}{2})$ \\\hline\hline
		Gauge Fields & \multicolumn{6}{c|}{\begin{tabular*}{12cm}{@{\extracolsep{\fill}}cc@{\extracolsep{\fill}}c@{\extracolsep{\fill}}c@{\extracolsep{\fill}}c@{\extracolsep{\fill}}}
				&$\SUc: G_\mu^a$ & $\SUL: W_\mu^b$ & $\UY: B_\mu$ &
		\end{tabular*}}\\\hline
	\end{tabular}
	\caption{\label{table:SMparticles}The definition of the Standard
		Model gauge multiplets. $l_{Li}$ denote the left-handed
		lepton doublets and $q_{Li}$ the left-handed quark
		doublets. $u_{Ri}$ and $d_{Ri}$ denote the right-handed up- and
		down-type quark singlets respectively, and $e_{Ri}$ denote
		the right-handed charged lepton singlets. $\Phi$ is the
		Higgs doublet. The index $L/R$ denotes the chirality of the fermion
		field, the generation indices are denoted by
		$i=1,2,3$. Colour and $\SUL$ indices are suppressed
		here and whenever possible without risk of confusion. }
\end{table}

For the Standard Model, our notation of the fundamental fields
together with their quantum numbers is given in
Tab.~\ref{table:SMparticles}. 
We refer to the review Ref.~\cite{Denner:2019vbn} for
details of the SM Lagrangian including its quantisation,
renormalisation and applications, and we refer to Ref.~\cite{Romao:2012pq} for a useful
comparison of different conventions.
In our conventions, the $\GSM$ covariant derivative is given by
\begin{align}\label{eq:SM-cov-derivative-sym}
	D_\mu = \partial_\mu + i \big(g_1 Y B_\mu + g_2 T^i W^i_\mu + g_3 t^a G^a_\mu\big)
\end{align}
with the gauge couplings $g_{1,2,3}$, where we identify $g_s\equiv g_3$.
In the fundamental representation, the $\SUL$ generators are given by $T^i=\sigma^i/2$ and the $\SUc$ generators by $t^a = \lambda^a/2$ where $\lambda^a$ denotes the Gell-Mann matrices
and fulfil the commutation relation $[\lambda^a,\lambda^b]=2if_{abc}\lambda^c$.
The SM field-strength tensors are defined as
\begin{align}\label{eq:SM-field-strengths}
	B_{\mu\nu} = \partial_\mu B_\nu - \partial_\nu B_\mu, \qquad
	W^i_{\mu\nu} = \partial_\mu W^i_\nu - \partial_\nu W^i_\mu - g_2 \epsilon_{ijk} W^j_\mu W^k_\nu, \qquad
	G^a_{\mu\nu}  = \partial_\mu G^a_\nu - \partial_\nu G^a_\mu - g_s f_{abc} G^b_\mu G^c_\nu.
\end{align}
The  Lagrangian of the SM can be decomposed as
\begin{align}
	\La_\text{SM} = \La_\text{gauge} + \La_\text{Fermion} + \La_\text{Higgs} + \La_\text{Yukawa}.
\end{align}
We specify the Higgs and Yukawa part as
\begin{subequations}
	\begin{align}
		\La_{\text{Higgs}}&=
		(D_\mu\Phi)^\dagger (D^\mu\Phi) + \mu^2|\Phi|^2 - \lambda|\Phi|^4,\\
		\La_{\text{Yukawa}}&=
		-(y_e)_{kj}\overline{l_{Lk}}\Phi e_{Rj}
		-(y_d)_{kj}\overline{q_{Lk}}\Phi d_{Rj}
		-(y_u)_{kj}\overline{q_{Lk}}\tilde{\Phi} u_{Rj}+\text{h.c.},
	\end{align}
\end{subequations}
while the remaining terms follow unambiguously from gauge invariance.
Here, the SU(2)$_L$ lepton, quark and Higgs doublets are decomposed as
\begin{align}\label{SU2decomposition}
	\renewcommand{\arraystretch}{.8}
	l_{Li} = \mqty(\nu_{Li}\\ e_{Li}), \qquad
	q_{Li} = \mqty(u_{Li}\\ V_\text{CKM}^{ij}d_{Lj}), \qquad
	\Phi = \mqty(G^+\\ \frac{1}{\sqrt{2}}\big(v + h + iG^0\big)), \qquad
	\tilde{\Phi}=i\sigma^2\Phi^* \equiv \mqty(\frac{1}{\sqrt2}(v+h-iG)\\-G^-)
\end{align}
where $G^-=(G^+)^\dagger$ and we have split off the \vev of the Higgs field with the
normalisation $v\approx 246$ GeV. 
Furthermore, without loss of generality we take the lepton and up-type quark Yukawa matrices $y_e$ and $y_u$ to be diagonal, while the down-type Yukawa matrix $y_d$ is diagonalised
by the CKM matrix $V_\text{CKM}$.

If appropriate, we use explicit names 
\begin{align}
\nu_i&=\{\nu_e,\nu_\mu,\nu_\tau\},
&
e_i&=\{e,\mu,\tau\},
&
u_i&=\{u,c,t\},
&
d_i&=\{d,s,b\}
\end{align}
for the individual generations of fermion mass eigenstates
$f_i=f_{Li}+f_{Ri}$. We also adopt corresponding short-hand notations
with obvious meanings, such as $y_\mu \equiv (y_e)_{22}$ or
$y_t=  (y_u)_{33}$. In certain BSM scenarios, the Yukawa sector is
modified, and in such cases the notation will be adapted as
appropriate.
The gauge boson mass-eigenstate fields are introduced as
\begin{align}
	\renewcommand{\arraystretch}{.8}
	\mqty(A_\mu\\ Z_\mu) = \mqty(c_W&s_W\\-s_W&c_W)	\mqty(B_\mu\\ W^3_\mu), \qquad
	W^\pm_\mu = \frac{1}{\sqrt{2}} \big(W^1_\mu \mp i W^2_\mu\big)
\end{align}
where $c_W \equiv \cos\theta_W$ and $s_W\equiv \sin\theta_W$ denote the cosine and sine of the weak mixing angle. 
For reference, the tree-level masses of the fermions, Higgs and gauge-bosons are given by
\begin{align}
	m_f=\frac{y_f v}{\sqrt{2}}, \qquad M_h = \sqrt{2\lambda}\,v = \sqrt{2}\,\mu, \qquad M_W = c_W M_Z = \frac{g_2 v}{2}.
\end{align}
In terms of the mass-eigenstate fields the covariant derivative Eq.~\eqref{eq:SM-cov-derivative-sym} becomes
\begin{align}
	D_\mu = \partial_\mu + i eQ A_\mu + i \frac{g_2}{c_W} \big(T^3-Q s_W^2\big) Z_\mu + i \frac{g_2}{\sqrt{2}} \big(T^+ W^+_\mu + T^- W^-_\mu\big) + i g_s t^a G^a_\mu,
\end{align}
where $T^\pm=T^1\pm iT^2$, and the electromagnetic gauge coupling and
electric charge operator are given by
\begin{align}
e&=g_1 c_W = g_2 s_W,
&
Q&=T^3 + Y.
\end{align}
The electric charge eigenvalues of $Q$ are $\{0,-1,\frac{2}{3},-\frac{1}{3}\}$ for neutrinos $\nu_i$, charged leptons $e_i$, 
up-type quarks $u_i$ and down-type quarks $d_i$.

The SM covariant derivative contains the QED covariant derivative which, together with the electromagnetic field strength tensor, is given by
\begin{align}\label{eq:QED-covariant-derivative}
	D_\mu^\text{QED} &= \partial_\mu + i eQ A_\mu,
        \\
         F_{\mu\nu} &= \partial_\mu A_\nu - \partial_\nu A_\mu.
\end{align}

\printbibliography

\end{document}